LHAASO

高海拔宇宙线观测站

# The Large High Altitude Air Shower Observatory

Science Book

2021 Edition

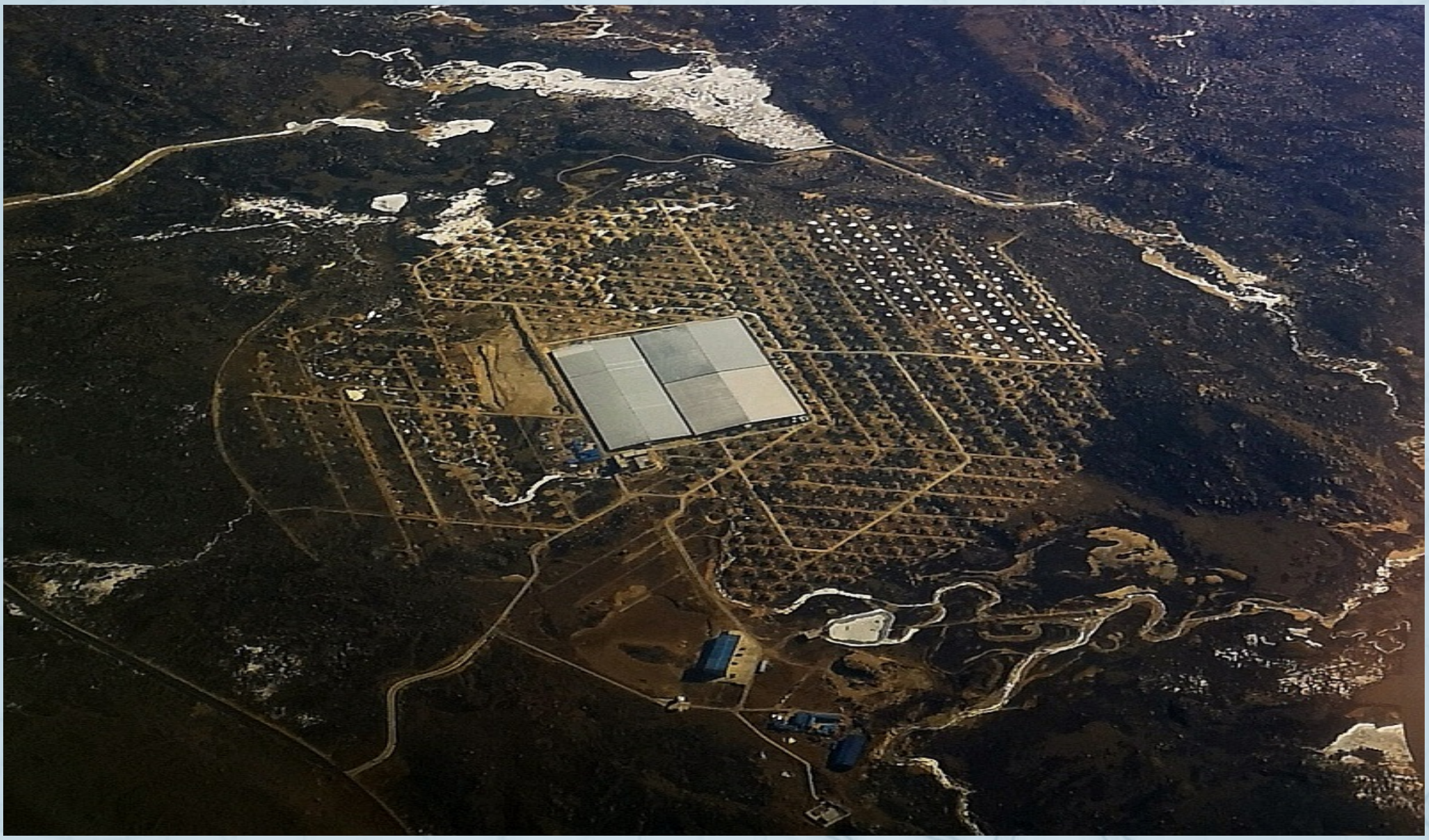

LHAASO

February 16, 2022

Chief Editors: Zhen Cao（曹臻）, Domenico della Volpe,
Siming Liu（刘四明）

Editors: Xiaojun Bi（毕效军）, Yang Chen（陈阳）,
Benedetto D’Ettorre Piazzoli, Li Feng（封莉）,
Huanyu Jia（贾焕玉）, Zhuo Li（黎卓）,
Xinhua Ma（马欣华）, Xiangyu Wang（王祥玉）,
Xiao Zhang（张潇）

External [Internal] Referees: Xiushu Qie（郄秀书）, [Hongbo Hu（胡红波）,
Zhuo Li（黎卓）, Xinhua Ma（马欣华）,
Alejandro Sáiz, Ruizhi Yang（杨睿智）]

## Preface

Since the science white paper of the Large High Altitude Air Shower Observatory (LHAASO) published on arXiv in 2019 [e-Print: 1905.02773 (astro-ph.HE)], LHAASO has completed the transition from a project to an operational gamma-ray astronomical observatory and a fully functioning cosmic ray (CR) air shower detection complex with multiple techniques measuring many shower components. LHAASO is a new generation multi-component facility located in Daocheng, Sichuan province of China, at an altitude of 4410 meters. It aims at measuring with unprecedented sensitivity the spectrum, composition, and anisotropy of cosmic rays in the energy range between $10^{12}$ and $10^{18}$ eV, and acting simultaneously as a wide aperture (one stereoradiant) continuosly-operating gamma-ray telescope in the energy range between $10^{11}$ and $10^{15}$ eV with the designed sensitivity of 1.3% of the Crab Unit (CU) above 100 TeV. During its construction, the partially completed square kilometer array (KM2A) of LHAASO has detected 12 PeVatrons with Ultra High Energy (UHE, $\geq$100 TeV) photon emission, opening the era of UHE gamma-ray astronomy. These Pevatrons are roughly uniformly distributed in the Milky Way implying their ubiquity although only the first group of them having UHE photon fluxes greater than 0.5 CU have been unveiled. LHAASO's capability of measuring simultaneously different shower components (electrons, muons, and Cherenkov/fluorescence ligh), will allow it to investigate the origin, acceleration, and propagation of CR through a measurement of energy spectrum, elemental composition, and anisotropy with unprecedented resolution. The remarkable sensitivity of LHAASO will play a key role in CR physics and gamma-ray astronomy for a general and comprehensive exploration of the high energy universe and will allow important studies of fundamental physics (such as indirect dark matter search, Lorentz invariance violation, quantum gravity) and solar and heliospheric physics. Plenty of science topics will be explored and deeply studied with the LHAASO data in 2020s. LHAASO actually records a few billions of CR showers per day above 100 GeV, allowing us to carry out many researches beyond the UHE gamma-ray astronomy.

To summarize the instrumental features and outline the prospective of scientific researches with the LHAASO experiment, the LHAASO Collaboration organized an editorial working group and finished all editorial work of this science book. Thanks to the editorial team led by Siming Liu and Domenico della Volpe for two years of hard working and the Chinese Physics C (CPC) for the strong technical supports, the book now is finally published in CPC as a special issue in 2022, going on-line immediately. I would like to take this opportunity to thank all LHAASO detector teams for providing the detector descriptions. I also appreciate the reviewing work provided by the LHAASO Collaboration members. Finally, I am grateful to all authors who made this book significantly enhanced in both contents and presentation. Among all editors, I would express my great respects to Benedetto D'Ettorre Piazolli who proof-read almost all chapters and sections.

I wish that this book will bring more scientists into researches with LHAASO and will stimulate more creative studies utilizing the LHAASO experiment.

Zhen Cao
Spokesperson of the LHAASO Collaboration

**Contents**

# Chapter 1 LHAASO Instruments and Detector technology*

Xin-Hua Ma(马欣华)[1,2★¶†] Yu-Jiang Bi(毕玉江)[1,2¶] Zhen Cao(曹臻)[1,2,3¶] Ming-Jun Chen(陈明君)[1,2¶]
Song-Zhan Chen(陈松战)[1,2¶] Yao-Dong Cheng(程耀东)[1,2,3¶] Guang-Hua Gong(龚光华)[4¶]
Min-Hao Gu(顾旻皓)[1¶] Hui-Hai He(何会海)[1,2,3¶] Chao Hou(侯超)[1,2¶] Wen-Hao Huang(黄文昊)[5¶]
Xing-Tao Huang(黄性涛)[5¶] Cheng Liu(刘成)[1,2¶] Oleg Shchegolev[6,7¶] Xiang-Dong Sheng(盛祥东)[1,2¶]
Yuri Stenkin[6,7¶] Chao-Yong Wu(吴超勇)[1,2¶] Han-Rong Wu(吴含荣)[1,2¶] Sha Wu(武莎)[1,2¶] Gang Xiao(肖刚)[1,2¶]
Zhi-Guo Yao(姚志国)[1,2¶] Shou-Shan Zhang(张寿山)[1,2¶] Yi Zhang(张毅)[1,8¶] Xiong Zuo(左雄)[1,2¶]

[1]Key Laboratory of Particle Astrophyics & Experimental Physics Division & Computing Center, Institute of High Energy Physics, Chinese Academy of Sciences, Beijing 100049, China
[2]TIANFU Cosmic Ray Research Center, Chengdu, Sichuan, China
[3]University of Chinese Academy of Sciences, Beijing 100049, China
[4]Department of Engineering Physics, Tsinghua University, Beijing 100084, China
[5]Institute of Frontier and Interdisciplinary Science, Shandong University, Qingdao 266237, Shandong, China
[6]Institute for Nuclear Research of Russian Academy of Sciences, Moscow 117312, Russia
[7]Moscow Institute of Physics and Technology, Moscow 141700, Russia
[8]Key Laboratory of Dark Matter and Space Astronomy, Purple Mountain Observatory, Chinese Academy of Sciences, Nanjing 210023, Jiangsu, China

**Abstract:** The Large High Altitude Air Shower Observatory (LHAASO) (Fig. 1) is located at Mt. Haizi (4410 m a.s.l., 600 g/cm$^2$, 29° 21’ 27.56” N, 100° 08’ 19.66” E) in Daocheng, Sichuan province, P.R. China. LHAASO consists of 1.3 km$^2$ array (KM2A) of electromagnetic particle detectors (ED) and muon detectors (MD), a water Cherenkov detector array (WCDA) with a total active area of 78,000 m$^2$, 18 wide field-of-view air Cherenkov telescopes (WFCTA) and a newly proposed electron-neutron detector array (ENDA) covering 10,000 m$^2$. Each detector is synchronized with all the other through a clock synchronization network based on the White Rabbit protocol. The observatory includes an IT center which comprises the data acquisition system and trigger system, the data analysis facility. In this Chapter, all the above-mentioned components of LHAASO as well as infrastructure are described.

## I. KM2A

The KM2A focuses on discovering galactic cosmic ray sources by searching for galactic gamma ray sources above 30 TeV in the northern sky and measuring primary cosmic rays in the energy range from 10 TeV to 100 PeV. The observation of a gamma-ray source with spectrum extending above 100 TeV represents an unambiguous identification of the ‘PeVatrons’, the sources of cosmic rays for energies up to the knee region of the cosmic ray spectrum.

The Crab Nebula, one of the most luminous sources of very high energy (VHE) gamma rays, is considered as a reference for detector design. The gamma ray spectrum of the Crab Nebula can be expressed as follows:

$$\frac{dN_\gamma}{dE} = 3.0\times10^{-11}E^{-2.59}\text{cm}^{-2}\text{s}^{-1}\text{TeV}^{-1}. \quad (1)$$

Above 100 TeV, the flux of the Crab-like source is 0.45 gamma rays per km$^2$ per hour. Considering the location of LHAASO, the effective observation on the Crab-like source, i.e., the source with same position and same spectral shape with Crab Nebula, is 6.5 hours per day. One square kilometer array will detect about 1000 gamma rays above 100 TeV from the Crab-like source

Received 2 December 2021; Accepted 3 December 2021; Published online 24 January 2022

* Supported by National Natural Science Foundation in China (NSFC) (U2031103, U1831208, 11805209, 11775233), NSFC for Distinguished Young Scholars (12025502), the Science and Technology Department of Sichuan Province (2021YFSY0031) and the International Partnership Program of Chinese Academy of Sciences (113111KYSB20170055)

† E-mail: maxh@ihep.ac.cn, corresponding author

★ Editors ¶ Contributors. All authors contribute equally to the work.

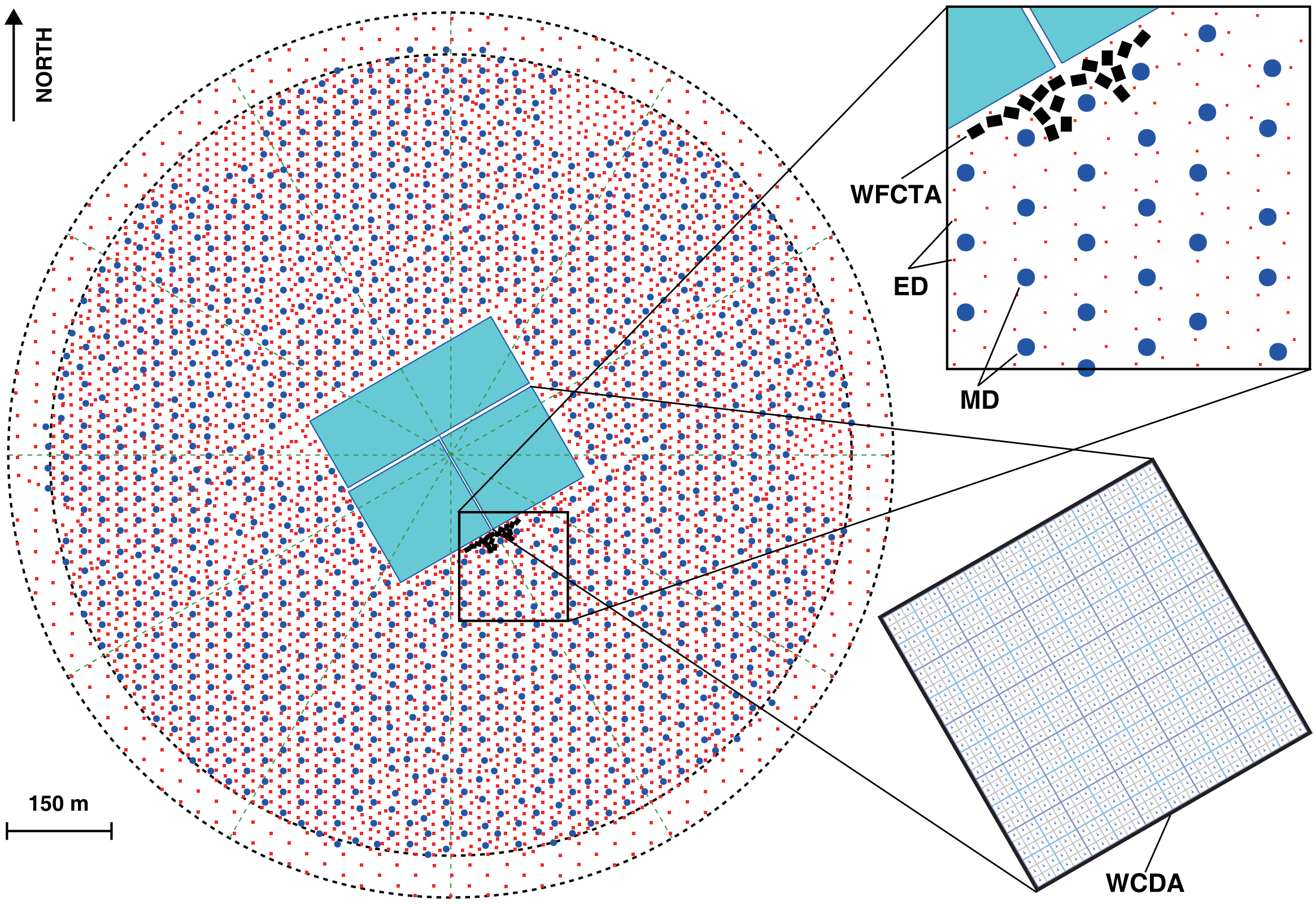

**Fig. 1.** (color online) Layout of LHAASO.

and then 10 gamma rays from a source with a flux of 1% of the Crab-like source in one year.

Therefore to have reasonable effective area for this fluxes, it is necessary to instrument a geometrical area of about 1 km$^2$ to measure extensive air showers (EAS) generated by cosmic rays in atmosphere. In order to achieve threshold energy of 30 TeV and angular resolution of few tenths of degree, the area has to be equipped with electromagnetic detectors of about 1 m$^2$ arranged on a triangle grid with spacing of 15 m. At the same time, above 50 TeV, it is essential to have a background-free sample of gamma rays, i.e. not contaminated by the cosmic nuclei. This will be achieved in LHAASO, by a precise measurement of the muon component of the shower, which will be extremely effective in suppressing the background. An array on muon detectors (MD), each having a surface area of 36 m$^2$, is also deployed on the same areas arranged on a triangle grid of 30 m spacing [1], to form together with the ED array, the KM2A. For gamma ray detection, KM2A has following performance: effective area can reach 0.8 km$^2$ at 30 TeV, angular resolution is 0.4° at 30 TeV and 0.2° at 1 PeV, core position resolution is 7 m at 30 TeV and 2 m at 1 PeV, and energy resolution is 28% at 30 TeV (Fig. 2). KM2A will reach a sensitivity of about 1% of a Crab-like source for energy of 100 TeV in one year of observation [1, 2].

### A. ED

The ED array is uniformly distributed over 1.3 km$^2$. It is divided into two parts: the central array consisting of 4911 EDs with 15 m spacing in a circle area with radius of 575 m, and the guarding ring array consisting of 305 EDs with 30 m spacing surrounding the central one with outer radius of 635 m. The whole array will work for more than 20 years with good stability under severe environmental condition, with large temperature variation (±25°C annually), low atmospheric pressure (0.6 atm) and high humidity [3]. In addition, In order to ensure that the statistical error and dynamic range meet the design requirements, EDs should have a uniformity within 10%. The signal attenuation is required to be less than 20% in 10 years due to detector aging. The main performance of ED are listed in Table 1.

#### 1. *Construction of ED*

A detection sensitive unit of ED (Fig. 3) consists of a plastic scintillator tile (BC408), 12 wavelength shifting (WLS) fibers (BCF92SC, single cladding), a layer of tyvek wrapper (1082D) and black cloth [4]. The tile has high detection efficiency for charged particles, quick response time and good stability. The tile size is 100 cm × 25 cm × 1 cm (length × width × thickness) (Fig. 3).

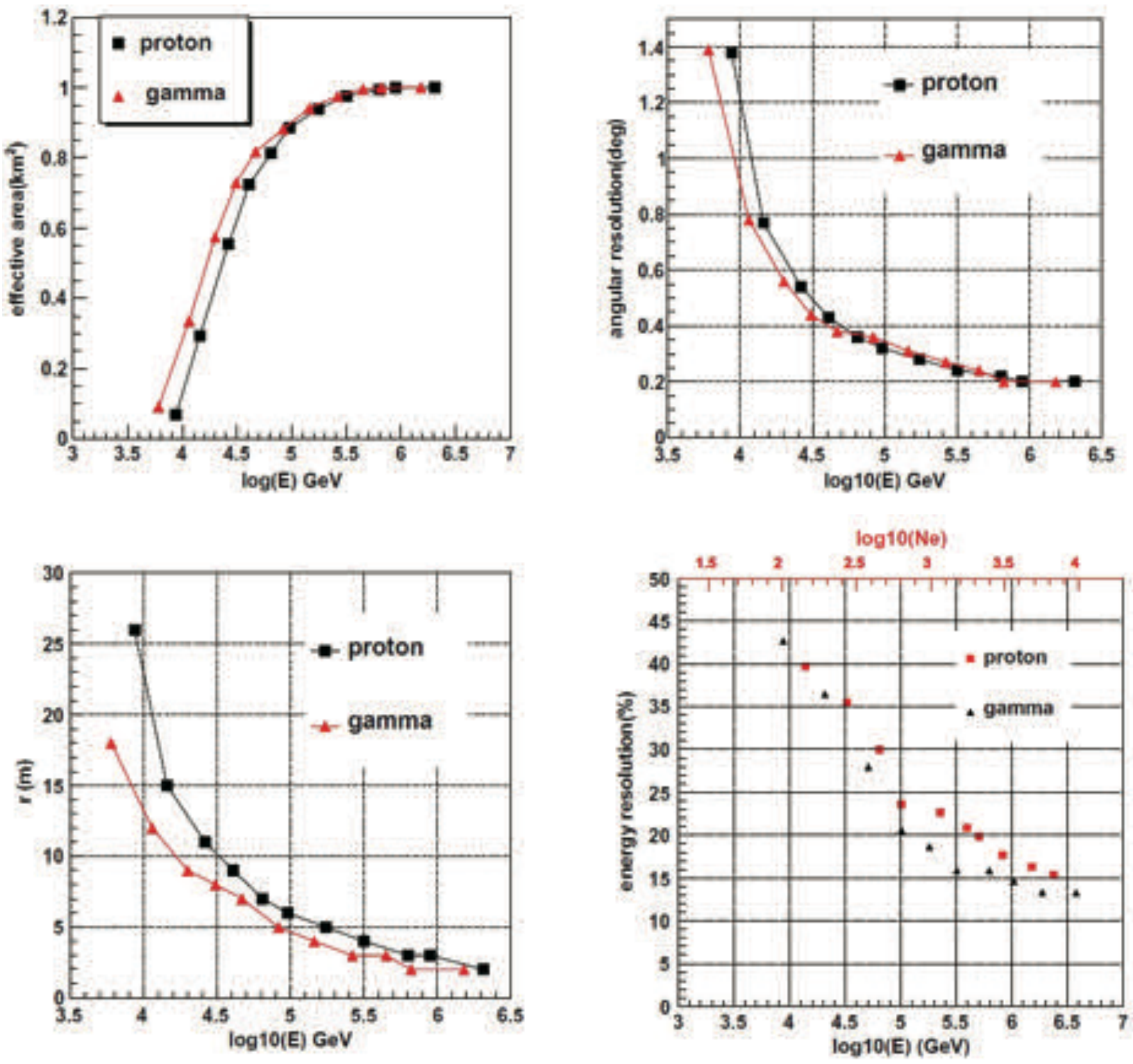

**Fig. 2.** (color online) Effective area (upper left), Angular resolution (upper right), Core position resolution (lower left) and Energy resolution (lower right) of KM2A varies with primary energy [1].

**Table 1.** The main characteristic and performance parameters of ED.

| Item | Value |
|---|---|
| Sensitive area | 1 $m^2$ |
| Secondary electromagnetic particles measured | $\mu^{\pm}, e^{\pm}, \gamma$ |
| Dynamic range | 1-$10^4$ particles/$m^2$ |
| Particles counting resolution | 25% @ single particle, 5% @ $10^4$ particles. |
| Time resolution | <2 ns |
| Total detection efficiency for charged particles | >95% |
| Counting rate at nominal working gain and threshold | <2 kHz |

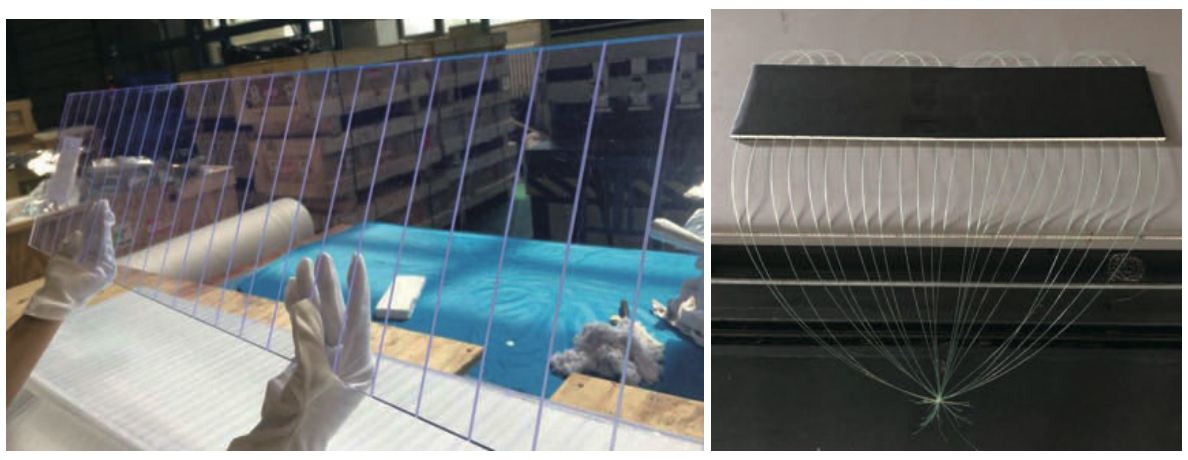

**Fig. 3.** (color online) Photos of ED. Left: One tile of scintillator. Right: A detection sensitive unit.

Twelve fibers with length of 2.7 m and diameter of 1.5 mm are routed in 24 grooves on the tile. The groove section is shaped like a reverse Ω with depth of 1.5 mm and diameter of 1.7 mm. Each fiber passes in the tile twice through two grooves. The separation between fibers is 4.16 cm. Each tile is wrapped with a layer of Tyvek to improve photon collection efficiency. The tile assembly is covered with black cloth to avoid external and inter-tile optical crosstalk.

One ED is composed of four detection sensitive units, a photomultiplier tube (PMT) (XP3960, 9 linear focused dynode stages, 1.5 inch end window), an electronics system, a power supply system and a shell (Fig. 4). Four detection sensitive units are installed into the steel shell symmetrically to cover the 1 $m^2$ detection area [4] (Fig. 5). All ends of 96 fibers are bunched together to be polished and coupled directly to the PMT photocathode. In order to increase detection efficiency for secondary $\gamma$ and absorb charged particles with low energy to improve the angular resolution, a 5 mm thick lead layer is placed on the surface of each ED to convert $\gamma$ into electron-positron pairs.

To build the ED array, more than 5,000 PMTs, 20,000 plastic scintillators and about 720 km WLS fibers are used. To ensure homogeneous response of detectors, for each component a batch test system has been built. Test results of scintillators and fibers show that the light output uniformity between scintillators is below 5% and light attenuation length of fibers is 3 m with 10% differences. The assembled detectors are all tested and qualified before they are sent to the LHAASO site for installation (Fig. 6).

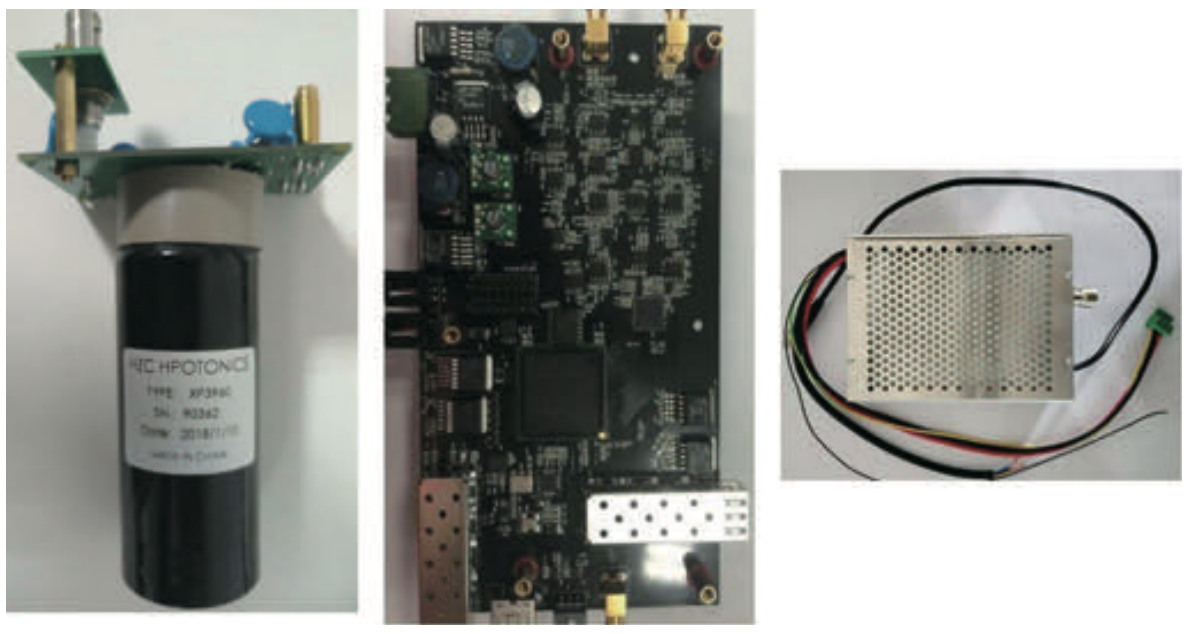

**Fig. 4.** (color online) ED Detector details. The PMT used (left), The readout electronics (middle) and Power supply system (right).

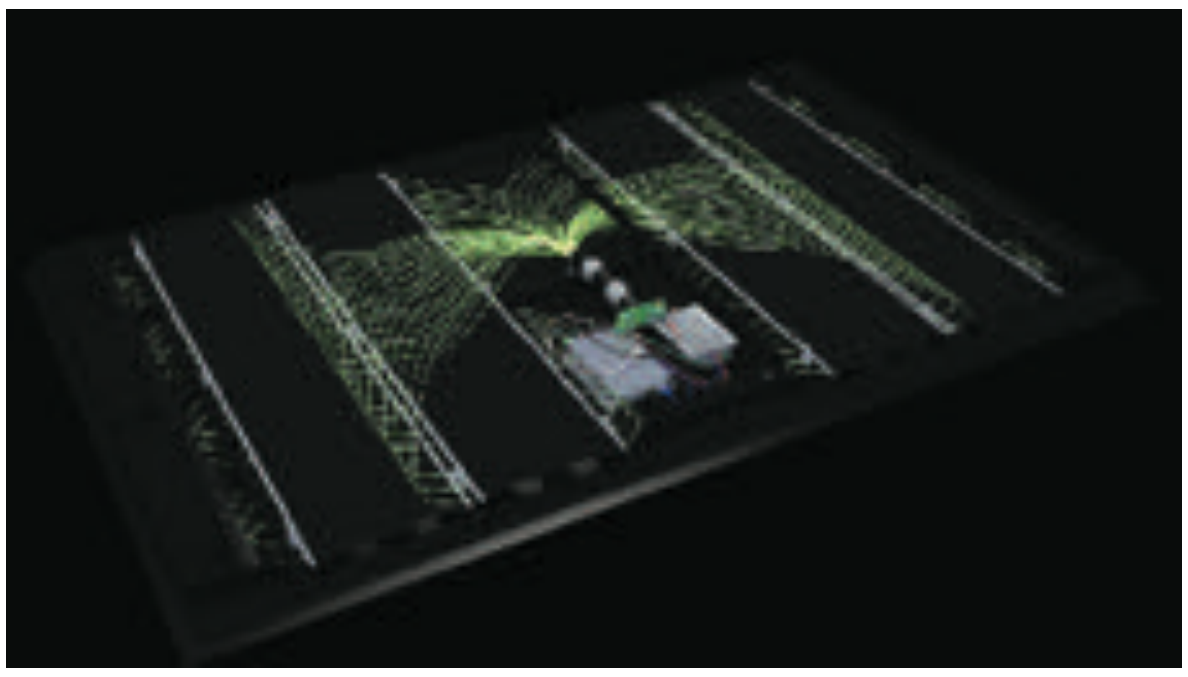

**Fig. 5.** (color online) Photo of an assembled ED sensitive unit.

To achieve the large dynamic range required (4 orders of magnitude), an optimal design of the voltage divider with anode and the 6th dynode readout has been developed [5]. Ratio of anode current to dynode current is around 110. The temperature coefficient is less than 0.2%/°C. PMT is wrapped in a magnetic shielding film of permalloy to reduce the influence of geomagnetic field.

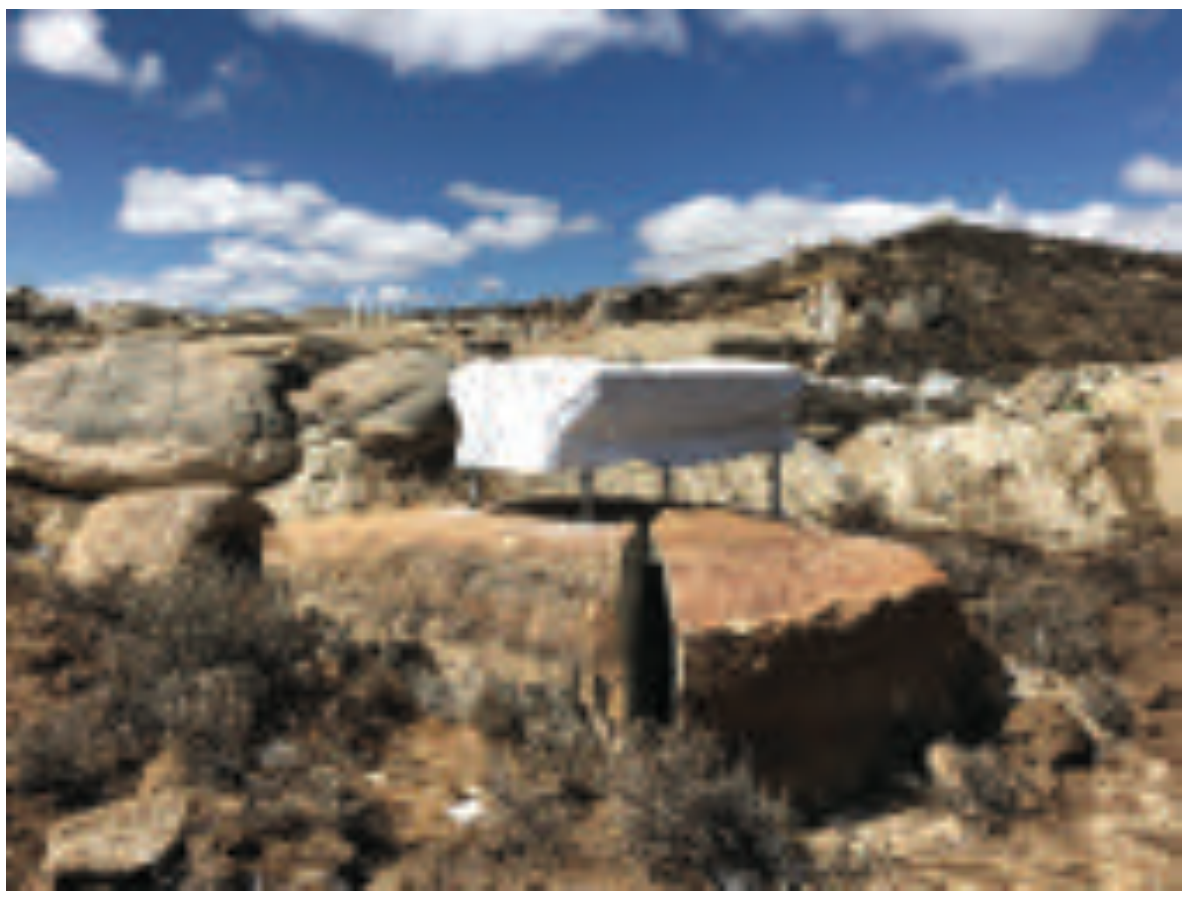

**Fig. 6.** (color online) Photo of an installed ED.

The ED electronics system adopts a trigger-less, independent, front-end digitization scheme (see Fig. 7) where each detector has a local front-end electronics (FEE) for signal digitalization and data transmission to the Data Acquisition (DAQ) system. Data contains arrival time information of hits, accurate charge information, slow control, register feedback, etc. Anode charge measurement ranges from 0.64 pC to 256 pC, and for dynode ranges from 0.32 pC to 128 pC. Charge accuracy is 3% for 1 particle and 1% for $10^4$ particles, and time accuracy is 0.5 ns. Stable operation are guaranteed for temperature ranging from −35°C to 55 °C. With help of the White Rabbit clock system, the FEEs of the different EDs are synchronized among them with a precision of 500 ps. The ED FEEs also allow to measure charge with high precision on the full scale array. The PMT and FEE are powered by the same power supply, but the high voltage of PMT can be adjusted independently. Together with detection units, electronic system and power supply system, a temperature and humidity monitoring system is also installed inside the detector shell [4].

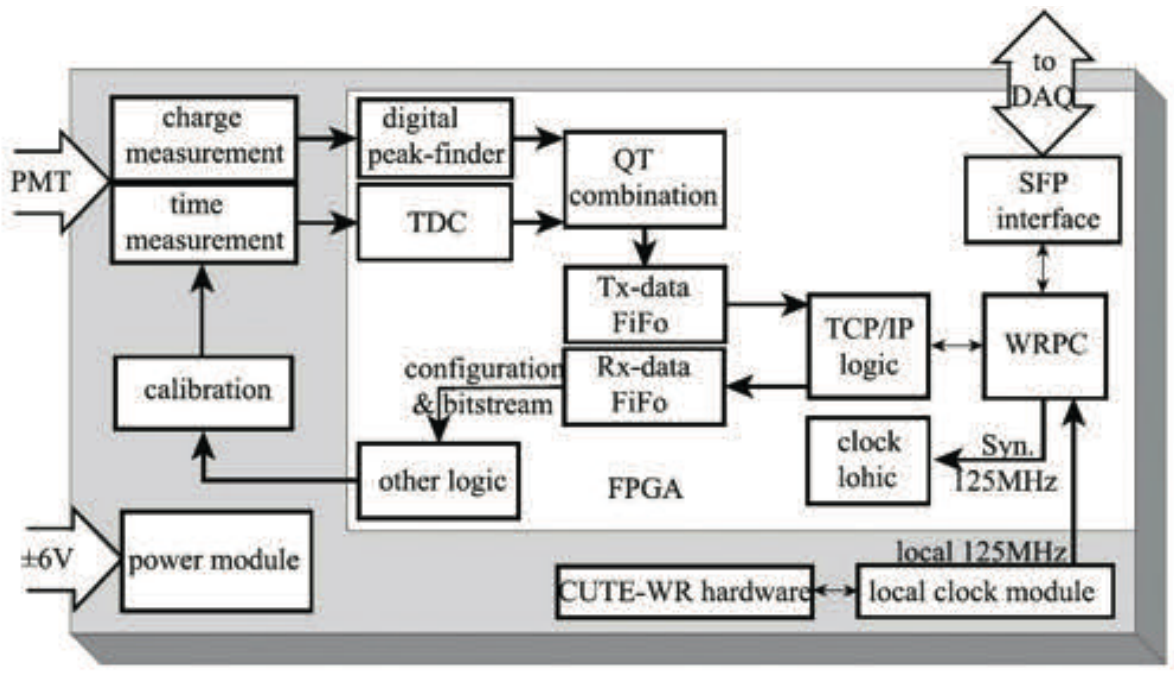

**Fig. 7.** Block diagram of electronics system of ED [6].

## 2. *Performance test of ED*

In order to test the overall performance of ED, Shandong University sets up an ED batch test system (Fig. 8). A telescope system made of scintillator layers is used as a trigger to measure the time of arrival of nearly vertical single cosmic muons. The triggered muon tracks are reconstructed with a gas detector. The reconstructed cosmic-muons allow to scan the ED surface and measure its performance with pixel size of 5 cm × 5 cm [7]. The performance test results of one batch of EDs are listed in Table 2.

Single particle charge spectrum is acquired and fitted with the convolution of Landau with a Gaussian function [4] (Fig. 9). Photo-electrons (p.e.s) are proportional to the charge measured, which is extracted from the peak. Uni-formity of Single particle peak of one ED is shown in Fig. 10. For the time resolution measurement, hit times of ED are compared with the one of the telescope trigger system. The width of distribution of difference between them includes time resolutions of ED and the telescope.

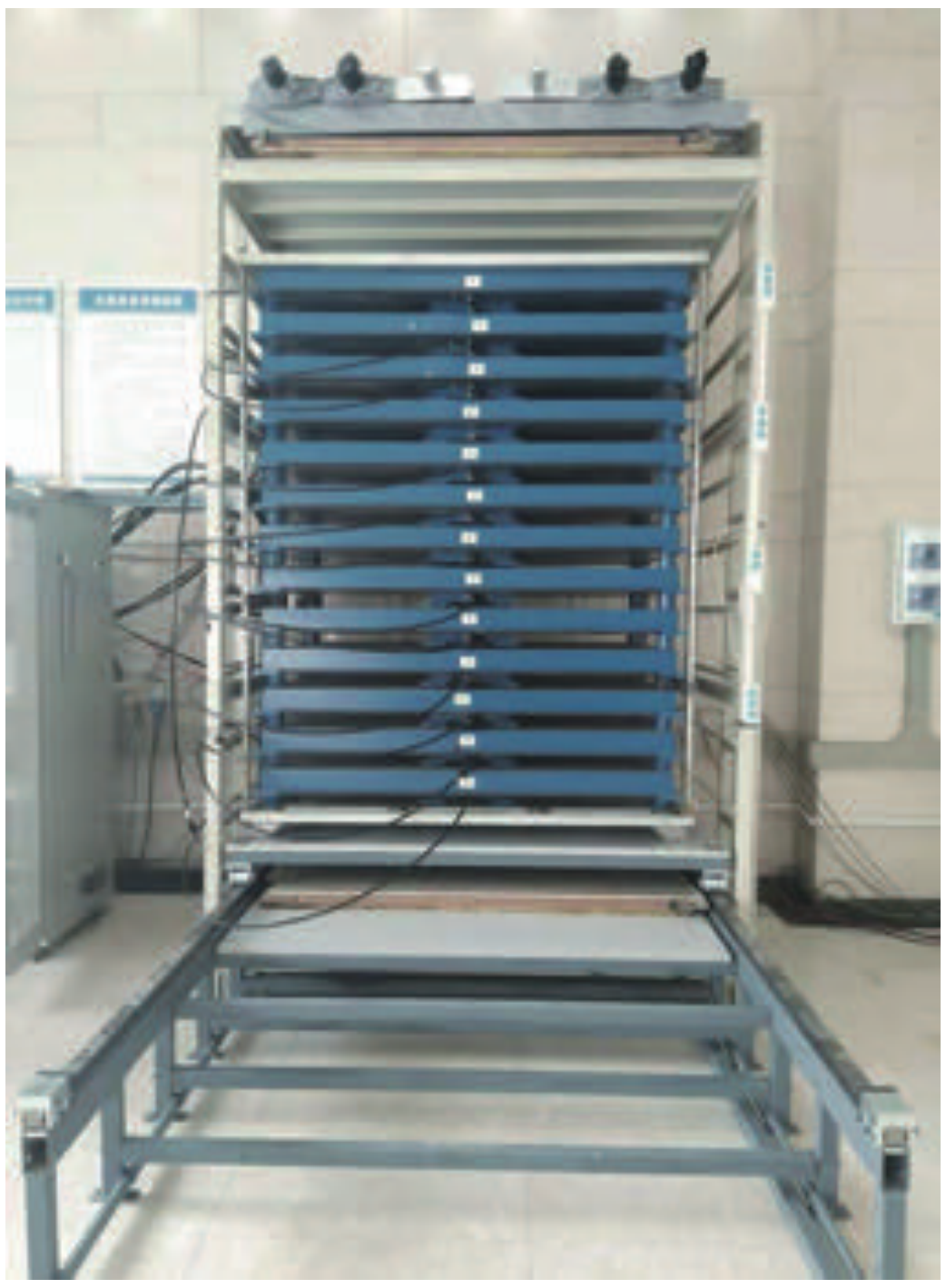

**Fig. 8.** (color online) Photo of the ED batch test system.

**Table 2.** Performance test results of the unit ED.

| Detector performance | Requirement | Batch test results |
|---|---|---|
| p.e.s (no Lead plate) | >20 | 20.6 ± 1.8 |
| Single muon spectral resolution | <25% | (12.7 ± 1.7) % |
| Time resolution | <2 ns | 1.62 ± 0.06 ns |
| Detection efficiency | >95% | (97.8±0.4) |

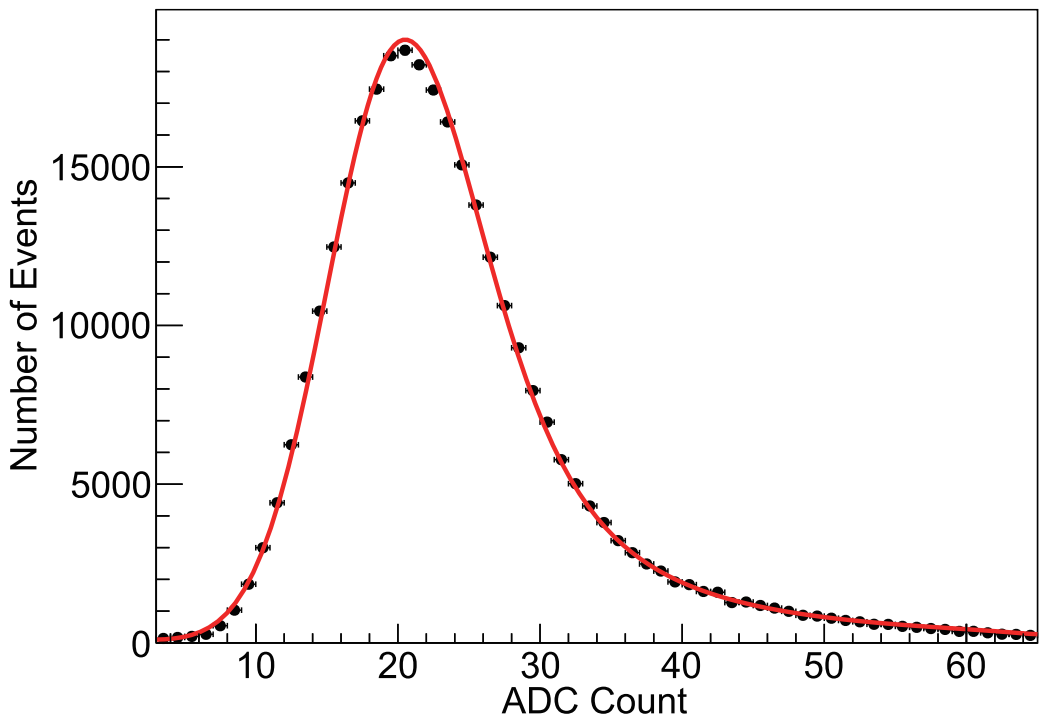

**Fig. 9.** (color online) Single particle charge spectrum (black dots) of one ED fitted with the convolution of Landau with Gaussian function (red line) [8].

Time resolution of the telescope is known and therefore the ED time resolution can be extracted (see Fig. 11). The detection efficiency is measured by the telescope system with muons triggering (Fig. 12). Counting rate refers to the signal rate that exceeds threshold of electronics, which is mainly contributed by the minimum ionization particle (M.I.P.) signal, dark noise of PMT and back-ground radiation. Counting rate is measured below 1 kHz in a laboratory at low altitude. At the LHAASO site, including background and signal, the average counting rate of more than two thousand EDs is (1.7 ± 0.3) kHz.

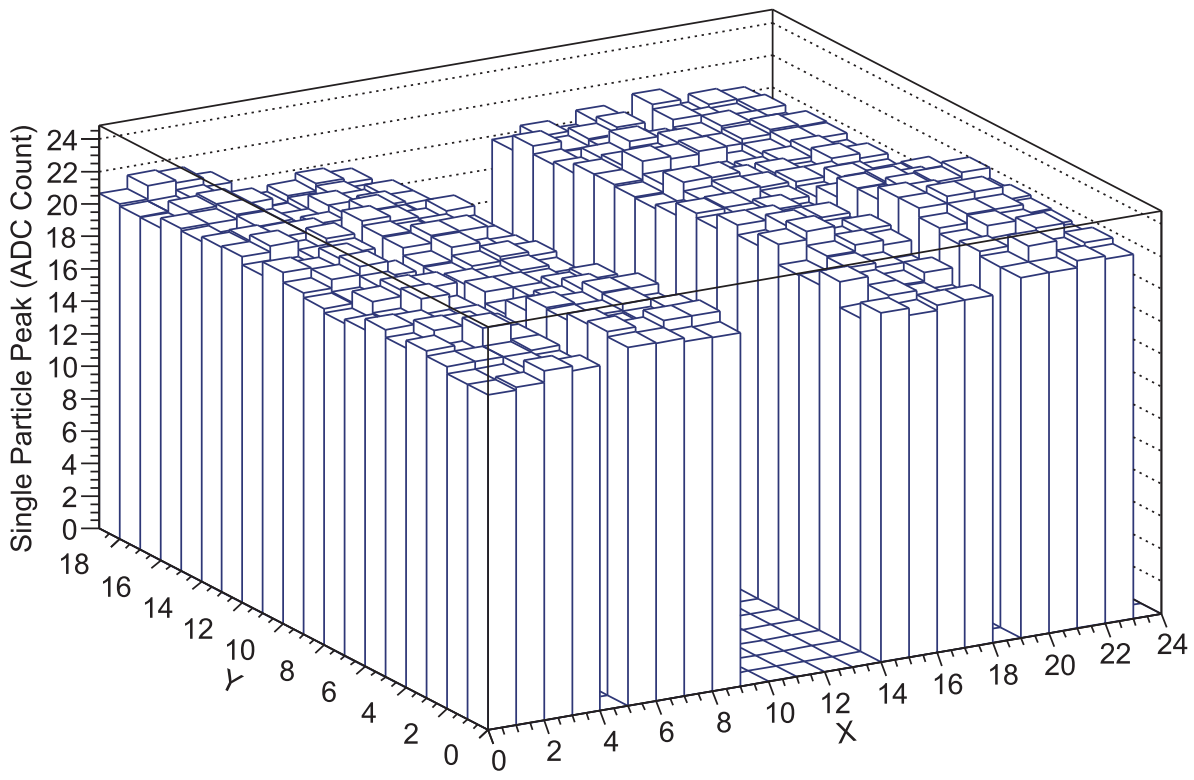

**Fig. 10.** (color online) Scanning result of Single particle peak of one ED [8].

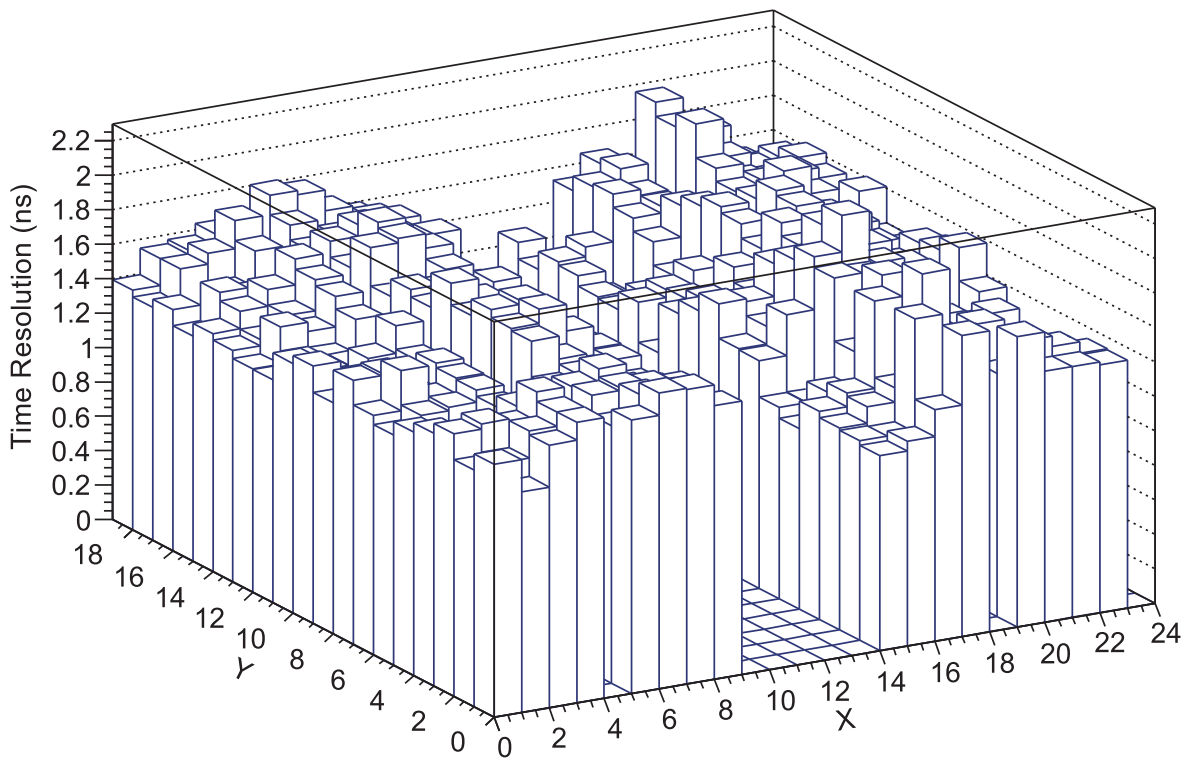

**Fig. 11.** (color online) Scanning result of time resolution of one ED [8].

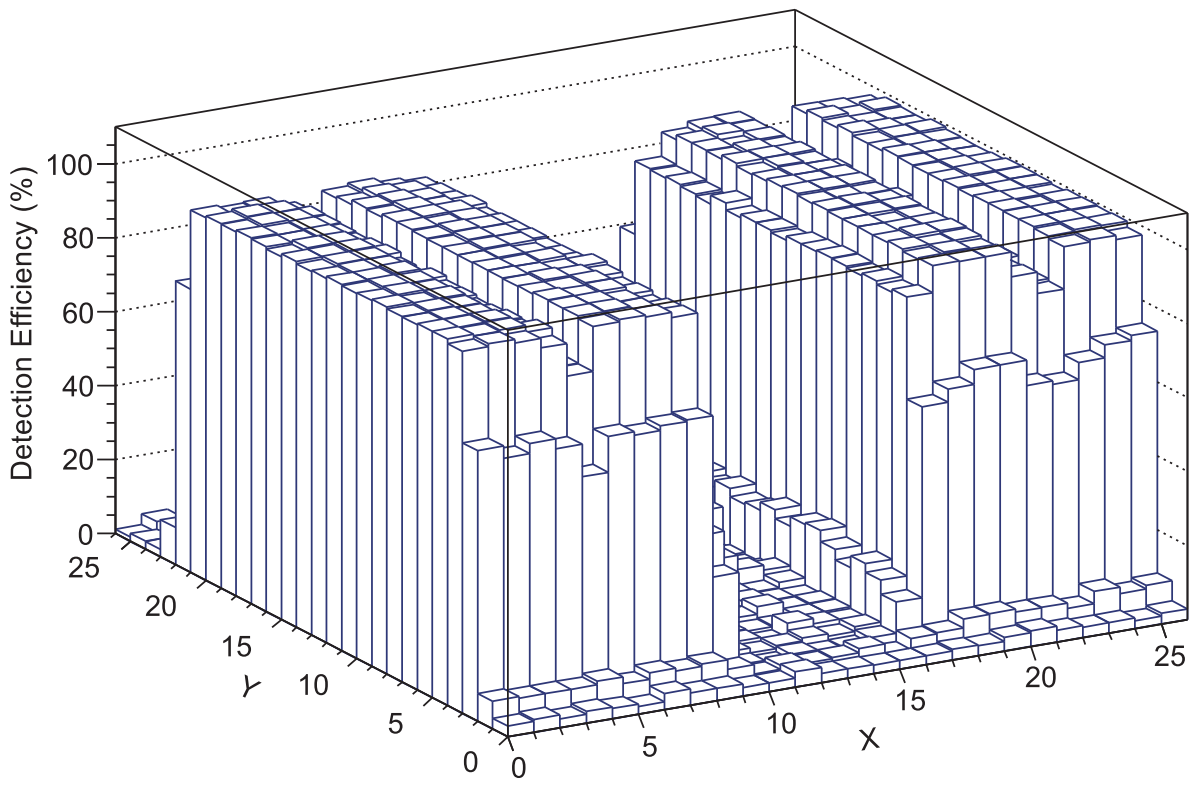

**Fig. 12.** (color online) Scanning result of detection efficiency of one ED [8].

### 3. *On-site performance of the ED array*

In August 2021, all 5216 EDs have been installed and science operation started (Fig. 13). Performance of the full array meets the design requirements. All timestamps are accurately synchronized to within 1 ns [9]. Anode and dynode output charge distributions are consistent with expectation and the overlapping of anode and dynode current reading allow to scan the full dynamic range (Fig. 14). All EDs run stably (Fig. 15) and trigger rate of the array stays constant over time (Fig. 16). Secondary-particle fronts of EAS are reconstructed and detected smoothly (see Fig. 17), and also the high significance of moon shadow is obtained (Fig. 18).

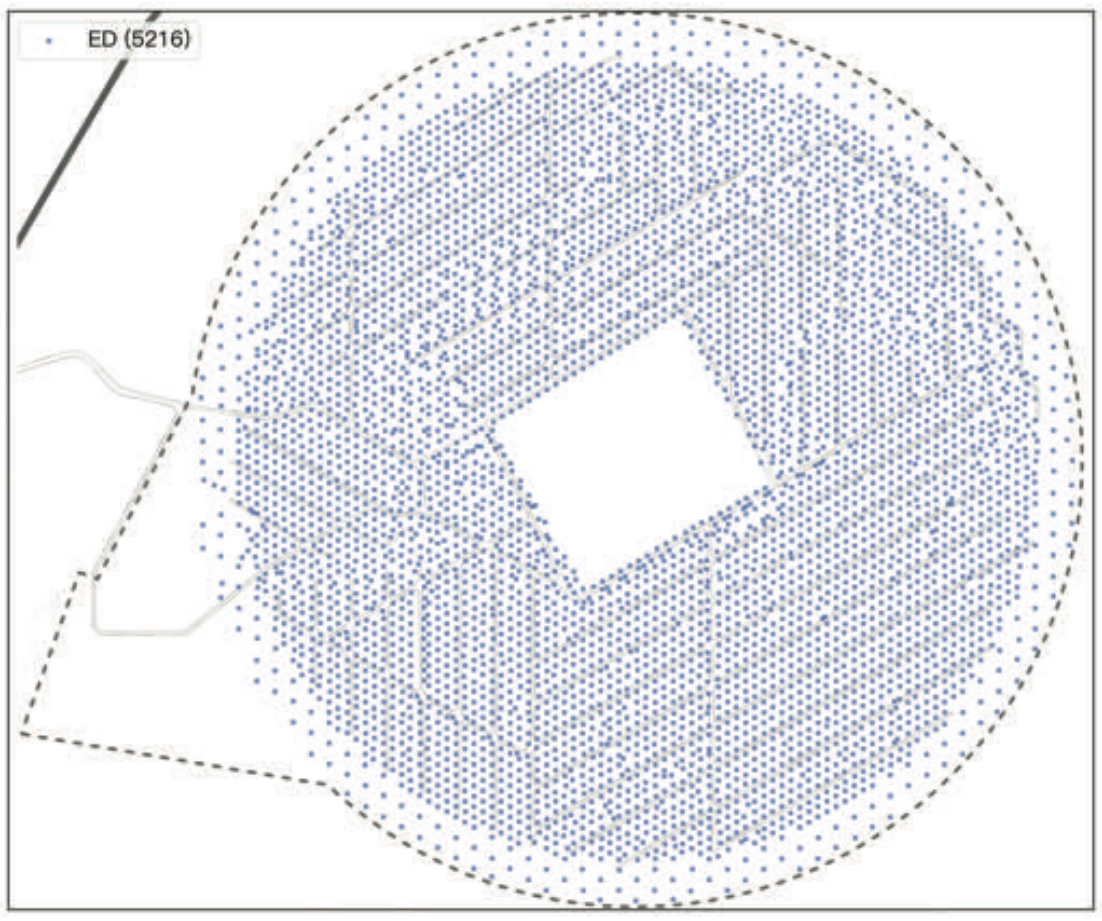

**Fig. 13.** (color online) The layout of the full EDs.

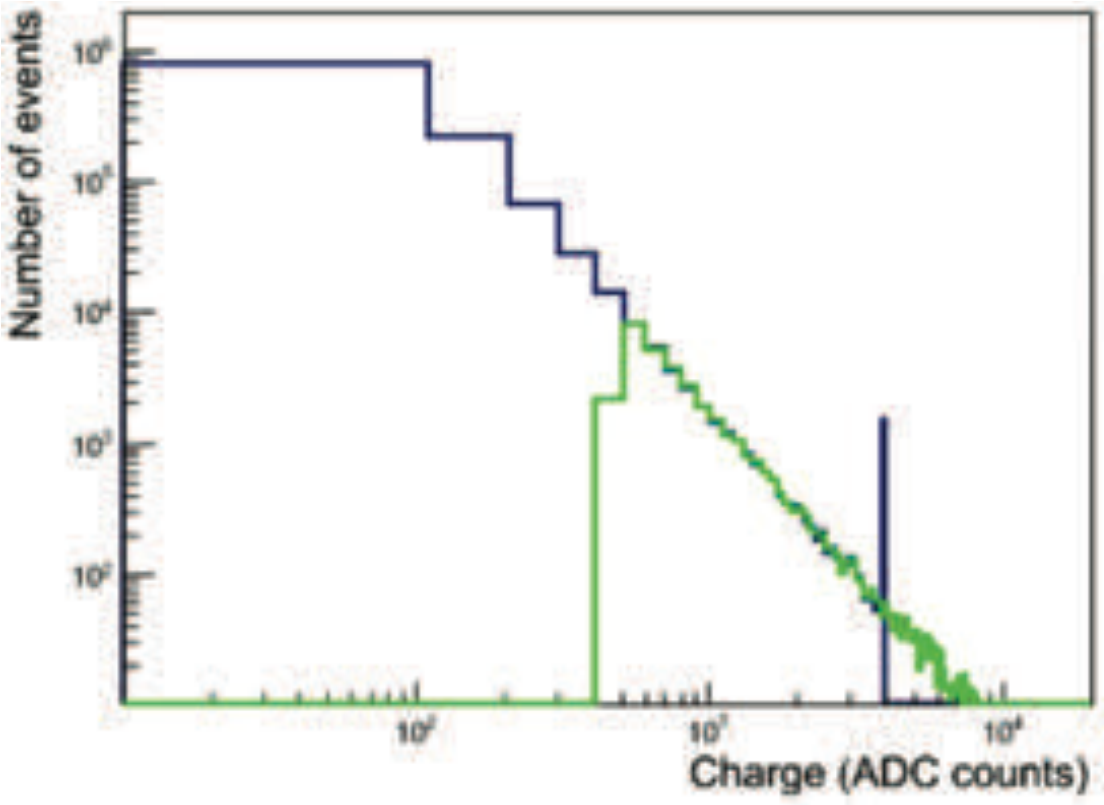

**Fig. 14.** (color online) ED output charge distribution of anode (blue line), which is saturated at 4095, and dynode (green line). (Provided by Hong-Kui Lv.)

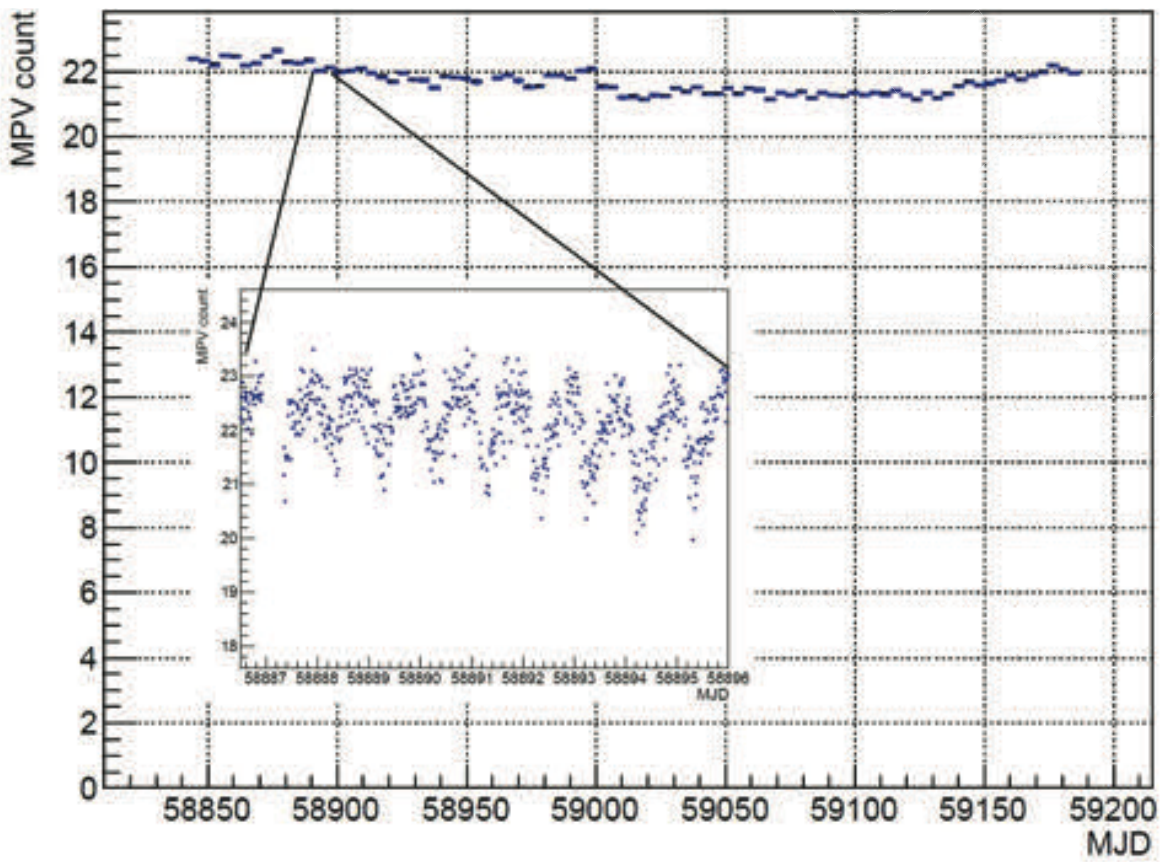

**Fig. 15.** (color online) Long term stability of maximum probability value (MPV) of anode output charge distributions of one ED as examples (after temperature compensation).

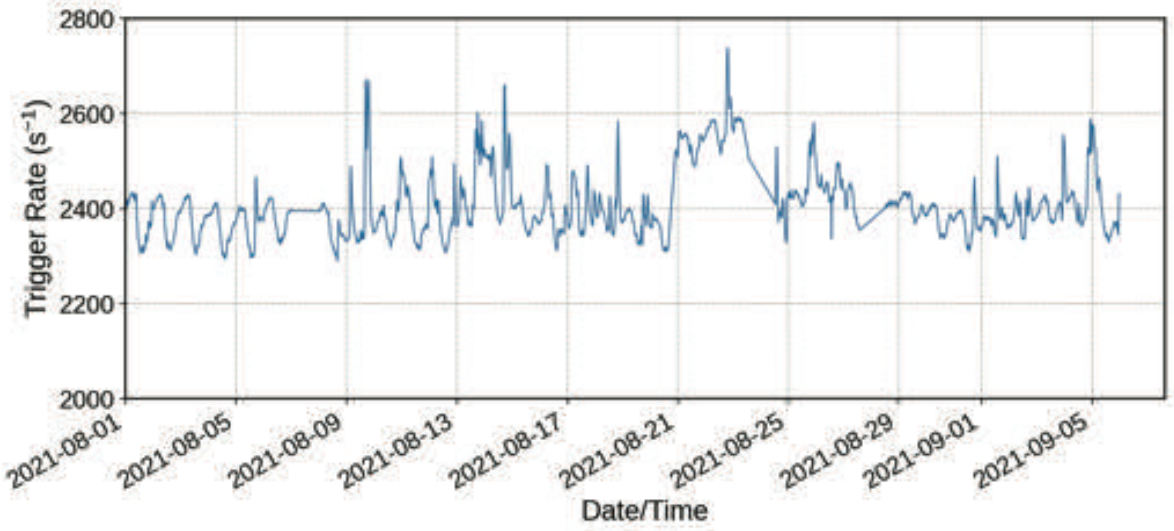

**Fig. 16.** (color online) Trigger rate variation of the ED array. (Provided by Xiao-Peng Zhang.)

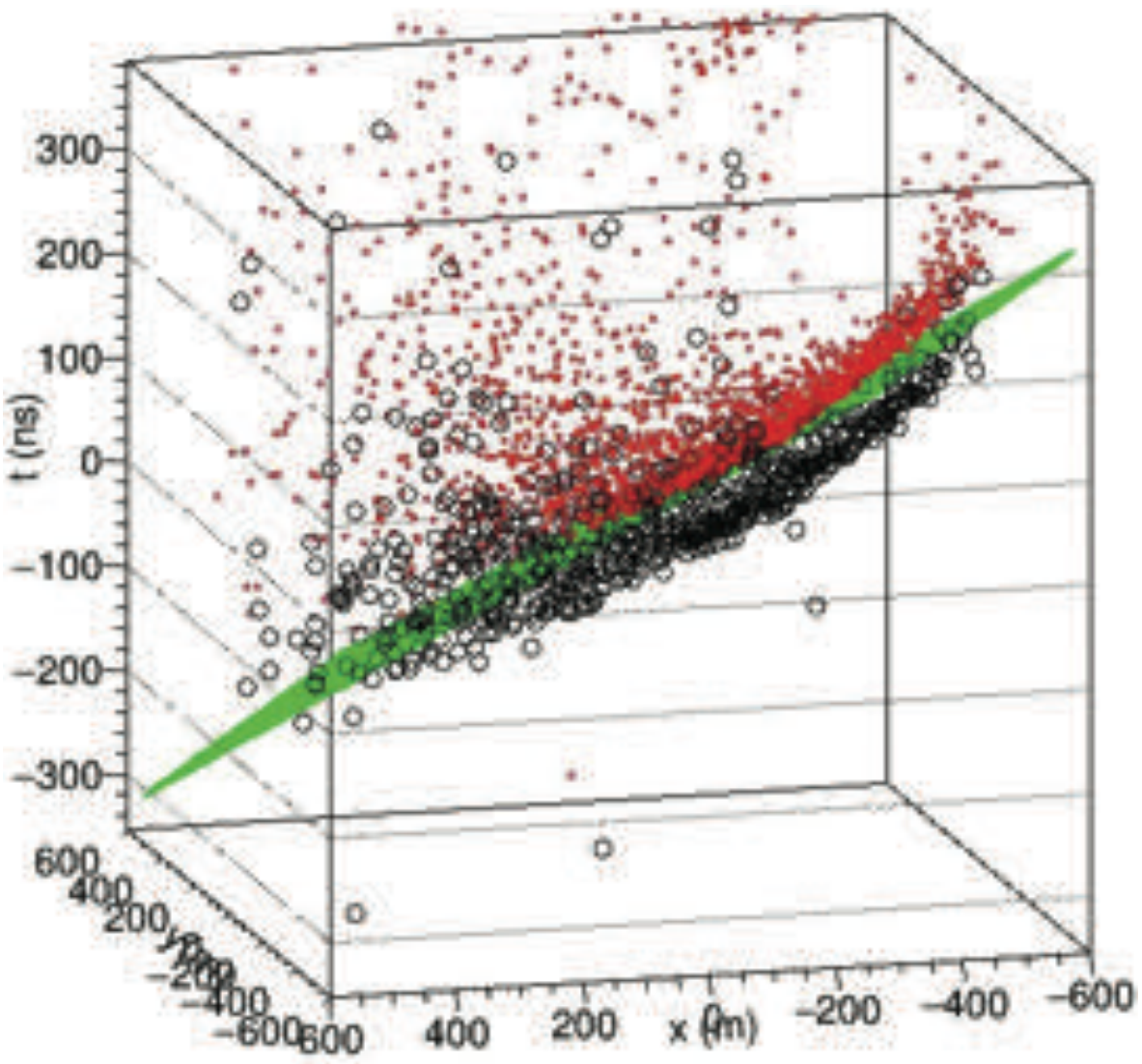

**Fig. 17.** (color online) Scattering plot of Secondary particles of an EAS detected by the ED array. The green plane is EAS front plane. The red squares are the fired ED. The black circles are the fired MD. The second front is caused by MD's dead time. (Provided by Jia Liu.)

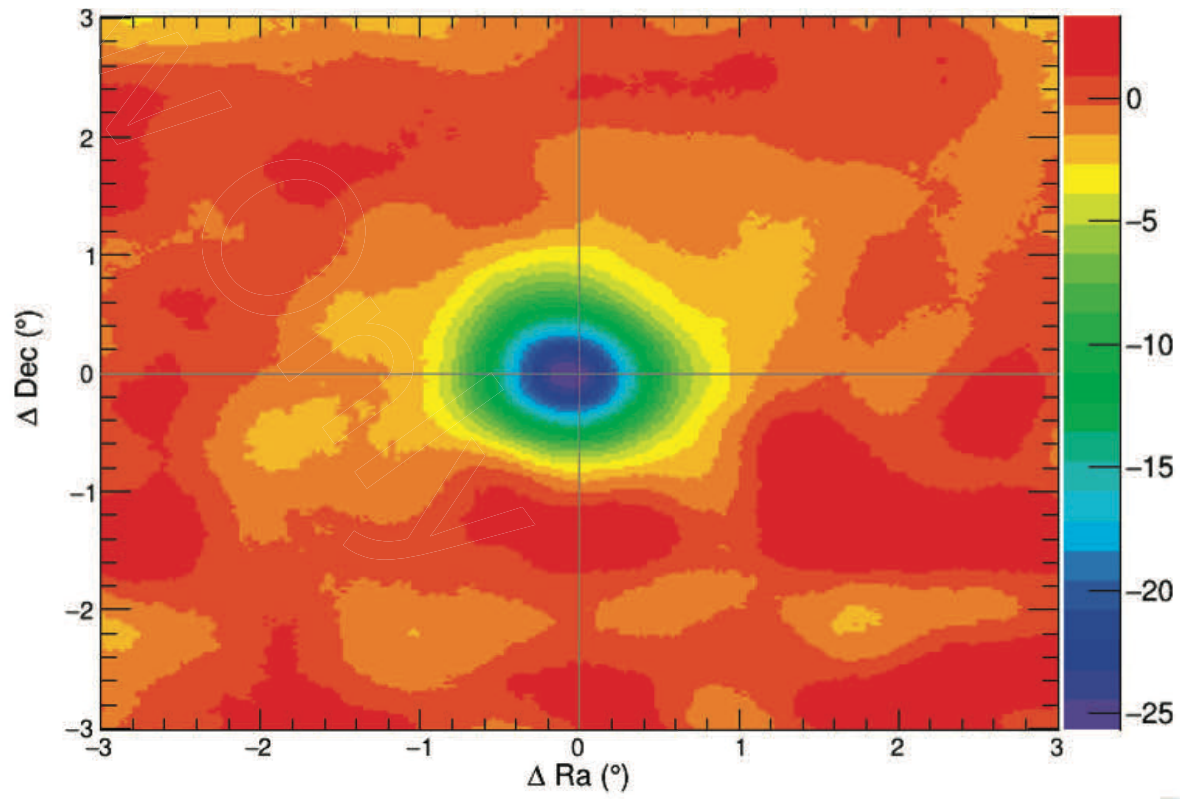

**Fig. 18.** (color online) Moon shadow of 26 days by using the ED array. Δ*Ra* is difference between right ascension of moon shadow and one of the real moon direction, and Δ*Dec* is difference between declination of moon shadow and one of the real moon direction.

## B. MD

The MD array is composed of 1188 water Cherenkov tanks deployed in a grid with a spacing of 30 m (Fig. 19).

### 1. *Construction of MD*

To achieve the main physical purposes, MD should meet the requirements shown in Table 3.

MD area is optimized to 36 $m^2$ by balancing the total area and spacing of MDs [1]. To accurately measure the number of muons detected, the MD should have good detection efficiency (~95%) and signal charge resolution (~25%, defined as the sigma/mean of gaussian fitting of the signal charge distribution) of single muons. Purity of number of muons detected in cosmic ray EAS should be 95%. With total area of > 40000 $m^2$, total background muon rate of MDs could be > $10^7$ Hz, which corresponds to 1 muon in 100 ns. Time resolution is designed as ~ 10 ns because MDs do not participate in triggering and reconstruction of EAS. The lower boundary of MD's dynamic range is set to be single muon signals, while the upper boundary, $10^4$ muons, corresponds to number of muons near EAS cores of the highest energy (100 PeV) cosmic rays to be detected. Finally, long-term stability is also very important since LHAASO will operate for more than 10 years, in which signal attenuation should be less than 20%.

Soil is the cheapest material to shield EAS electromagnetic components and reduce the so-called punch-through effect. EAS electromagnetic components gener-ate sub-showers in the shielding soil, and the products (mainly gamma rays, $e^+$ and $e^-$) have a certain probability to penetrate the soil and to be misidentified as muon. The water Cherenkov detector acts as a calorimeter by measuring the total number of Cherenkov photons yielded by a particle in water, which is approximately proportional to the total deposited energy of particle in water. If the total punch-through energy of EAS electromagnetic components is smaller than the total deposited energy of muons in water, shielding soil can provide further rejection power to EAS electromagnetic components. A thicker the overburden soil results in lower the punch-through effect and higher muon threshold energy. The energy distribution of muons generated by cosmic ray background (Fig. 20) shows that most of muons have energies ~ 1 GeV, thus the overburden soil can be as thick as 2.5 m. On the other hand, if the underneath water Cherenkov detector is deep enough, punch-through electromagnetic components deposit all energy in the water, while energetic muons pass through water with deposited energy proportional to the water depth. The specifications of the design are listed in Table 4.

**Table 3.** Requirements of MD.

| Item | Value |
|---|---|
| Area | 36 $m^2$ |
| Water depth | 1.2 m |
| Overburden | 2.5 m |
| Water transparency (att. len.) | ≥ 30 m (400 nm) |
| Liner reflection coefficient | ≥ 95% |
| Dynamic range | 1-$10^4$ muons |
| Resolution | 25% at 1 muon; < 5% at $10^4$ muons |
| Time resolution | ≤ 10 ns |
| Aging (≤20%) | ≥ 10 year |

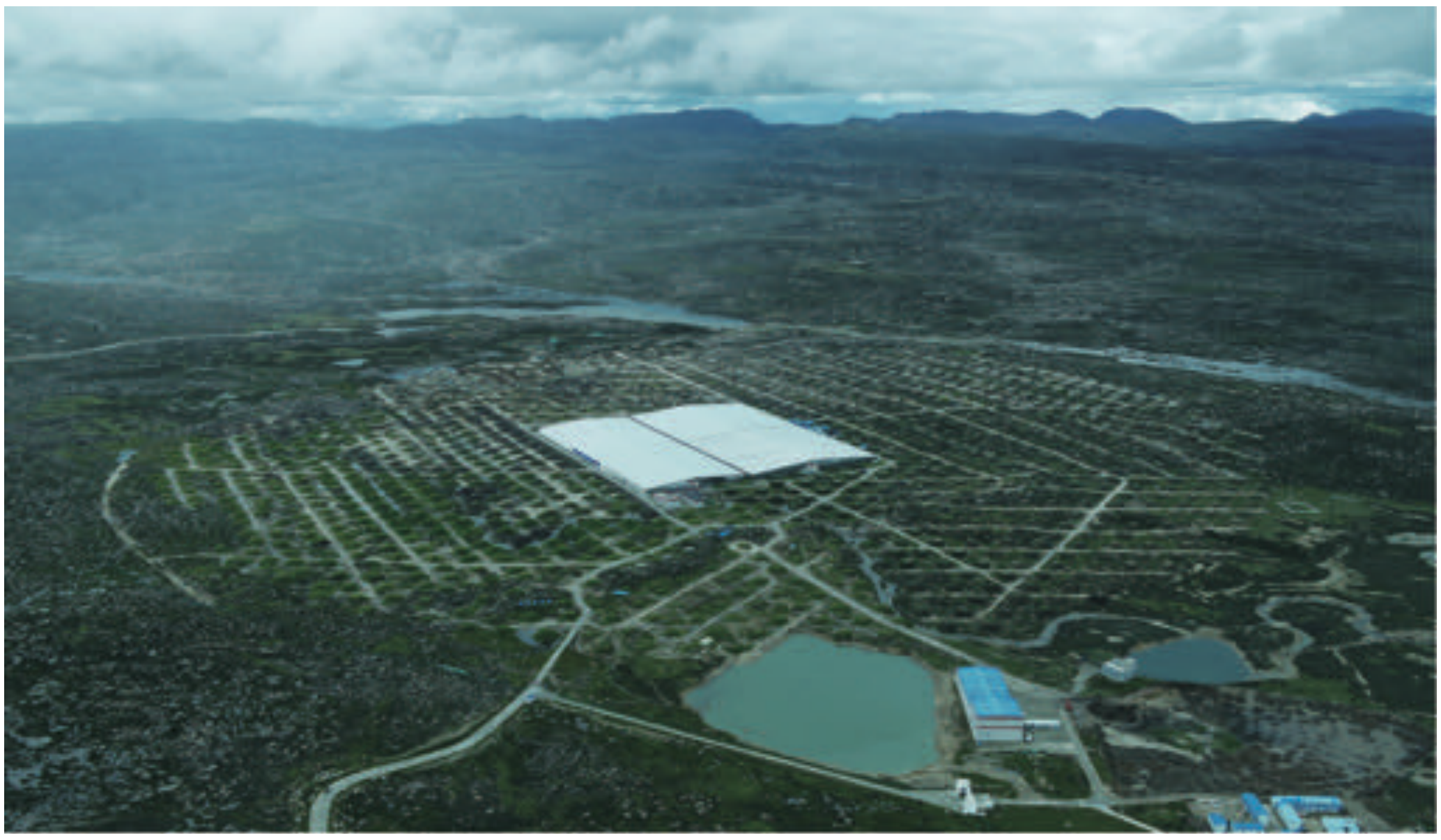

**Fig. 19.** (color online) Photo of LHAASO in August 2021. WCDA is the building at the center. The soil bumps are MDs. The white or green small boxes are EDs. The blue boxes beside WCDA are WFCTs.

A MD consists of concrete tank, thermal layer, water liner, pure water and PMT [10] (Fig. 21). Each cylindrical concrete tank contains a water liner with diameter of 6.8 m and height of 1.2 m to enclose pure water. An eight-inch PMT sits at top center of the liner and looks downwards through a highly transparent window into the water. The liner provides a sealed container for 44 tons of ultra-pure water at least for 10 years, protecting the water from contamination and inhibits bacteriological activities, as well as a high reflectivity inner surface for the Cherenkov photons. An apparatus is developed to measure the attenuation length of water or any other liquid directly [11].

The liner is mechanically supported by the concrete tank and surrounded by the thermal layer. Composition of the liner is (Fig. 22): the inner layer is a complex film of two sheets of DuPont Tyvek 1082D and a sheet of PE in the outer, which has an excellent diffuse reflectivity of Cherenkov light; one middle layer is two independent sheets of LDPE (Low-Density Polyethylene), which seals ultra-pure water; the other middle layer is two independent sheets of EVOH (Ethylene vinyl alcohol copolymer), which is an excellent oxygen barrier; the outer layer is PE woven cloth, which protects the whole inner layers. Liners are manufactured in a food-grade production workshop in order to keep it clean. Each complete liner is tested by high sensitivity SF6 tracer gas leak checking [12].

On the top center of the liner are two valves and PET dome window fitted for the installation of the PMT. The dome is custom-formed to match the PMT front face. The PMT is optically coupled to the dome window using an optical coupling compound (RTV6136A(B)), which eventually ensures a highly transparency for the Cherenkov photons.

The MD electronics board is designed as shown in Fig. 23. Main functions of the electronics system include:

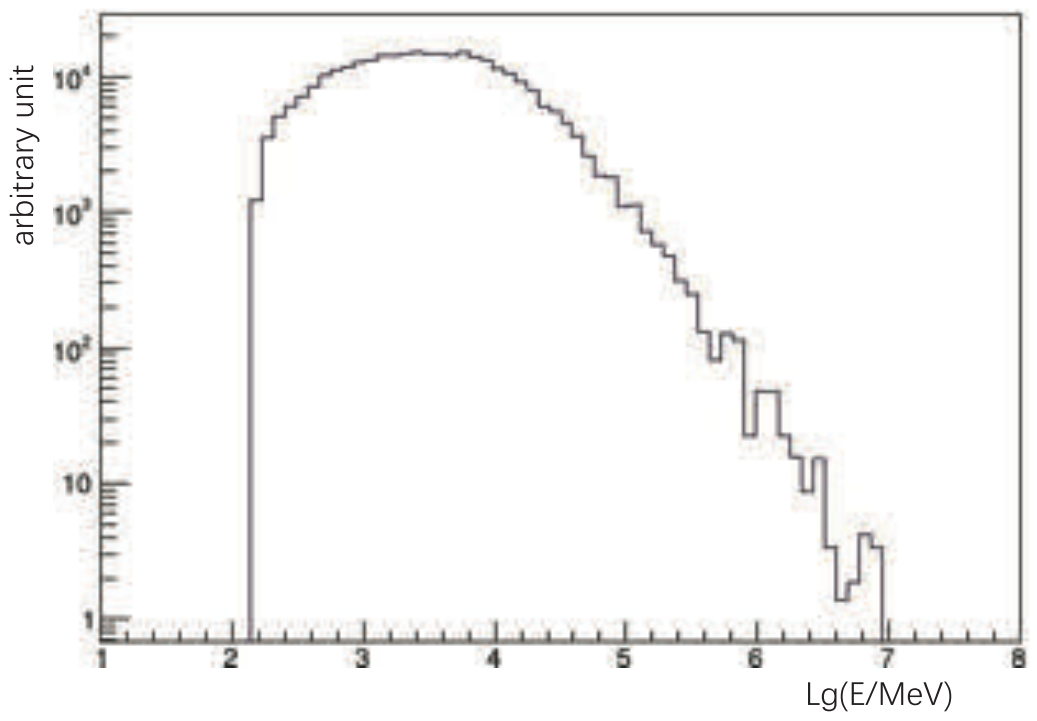

**Fig. 20.** (color online) Energy (E) distribution of muons generated by cosmic ray background along trajectory of the Crab Nebula. (Provided by Ying-Ying Guo.)

**Table 4.** Design specifications of the MD electronic board.

| Item | Value |
|---|---|
| Dynamic range | 1-$10^4$ muons |
| Resolution | 12% at 1 muon; < 5% at $10^4$ muons |
| Time resolution | 4 ns |
| Single Channel hit rate | 10 kHz |
| Threshold | 0.25 muon (Adjustable) |
| Operating temperature | −30° to 40° |

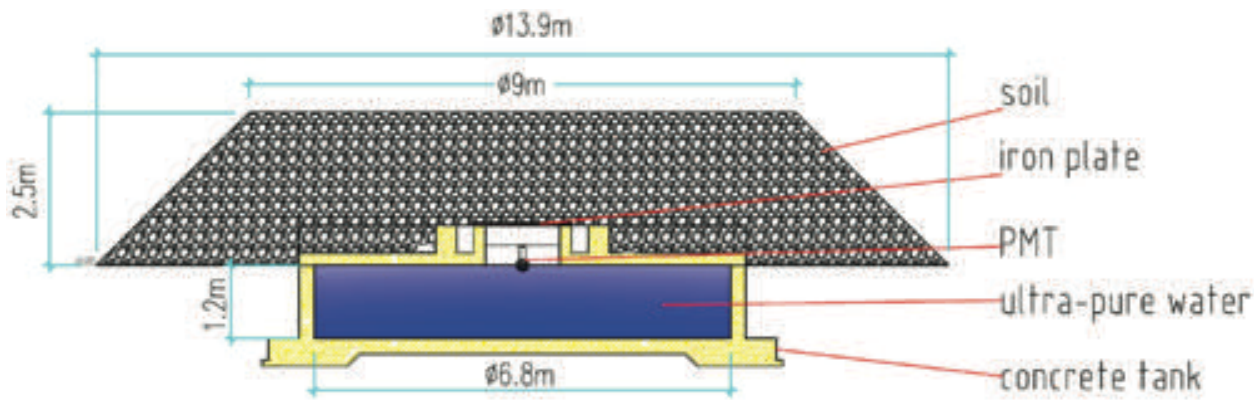

**Fig. 21.** (color online) Schematic of MD [10]. It mainly consist of three parts: probing medium, structural support and shielding layer.

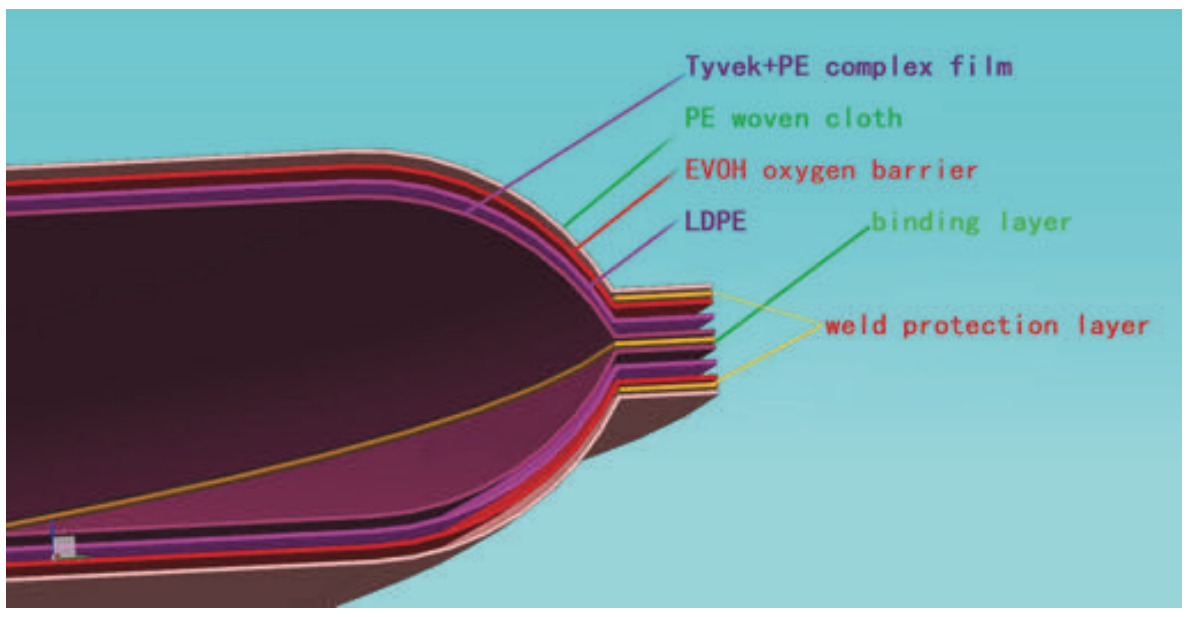

**Fig. 22.** (color online) Composition of the water liner.

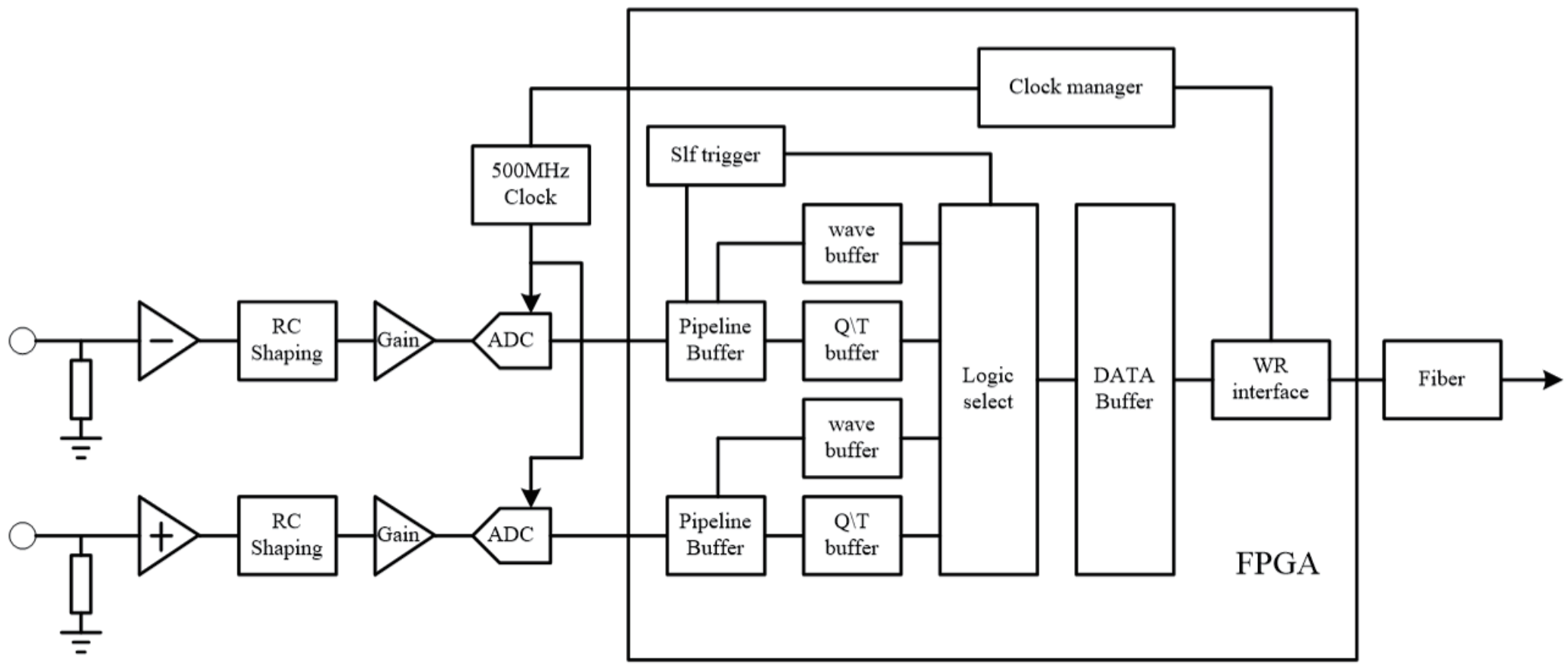

**Fig. 23.** Design block diagram of the MD electronic board. (Provided by Jin-Fan Chang.)

1) To measure the charge from signals from the anode and the dynode of the PMT; 2) To measure the arrival time of the muon signal; 3) To record the waveform of the muon signals periodically.

The signal is triggered when the amplitude of the signal is over the threshold. The time stamp is generated for this triggered event. The signals from the anode and the dynode of PMT are then digitized with two independent flash analog-to-digital-converters (FADCs). By using the recorded waveforms, charges and other parameters are calculated. All data is finally assembled and packaged into a TCP packet and is transmitted to the host computer through the White Rabbit network.

### 2. *Test of MD*

The average single muon waveform of MD is shown in Fig. 24. When a high energy muon passes through the water, it will generate a large number of photons, and these photons will be gradually absorbed by water and reflected by inner layer of the liner, then a part of the photons will reach the PMT by one reflection or several reflections to form a signal. The trailing edge of the wave can be fitted with an exponential function, in which decay time is determined by water absorption length of ultra-pure water and reflectivity of the liner [13]. A new method using muon decay events to calibrate and monitor MDs has been established in an earlier study [14]. Fig. 25 is single muon spectrum of one MD which shows that one a vertical and through muon typically generates about 70 photoelectrons in the detector.

The results of one LHAASO-KM2A prototype array, which is about 1% of the full one, presenting a combined measurement of EAS muons and electromagnetic particles, indicates that the muon content can be studied for EAS with energies from tens of TeV to tens of PeV. The results are compatible with the prediction of Monte Carlo simulations and no obvious excess is observed when taking into account that the mass composition increases above 1 PeV. The results also support a transition from light to heavy elements for cosmic rays in the 'knee' region [15].

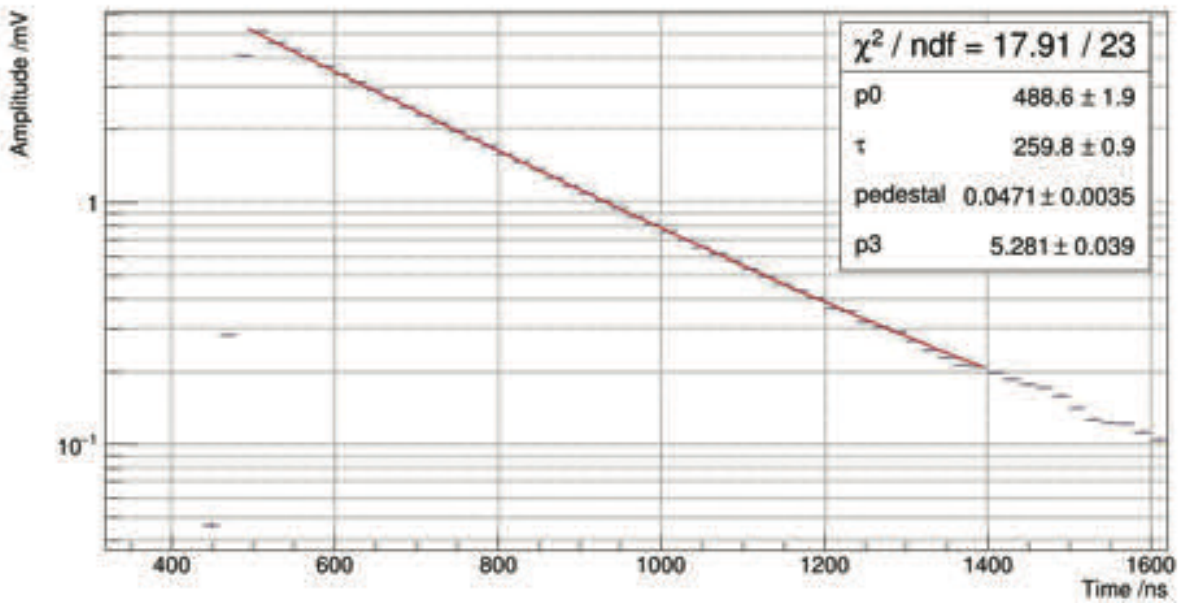

**Fig. 24.** (color online) One average single muon waveform. The trailing edge is fitted with an exponential function.

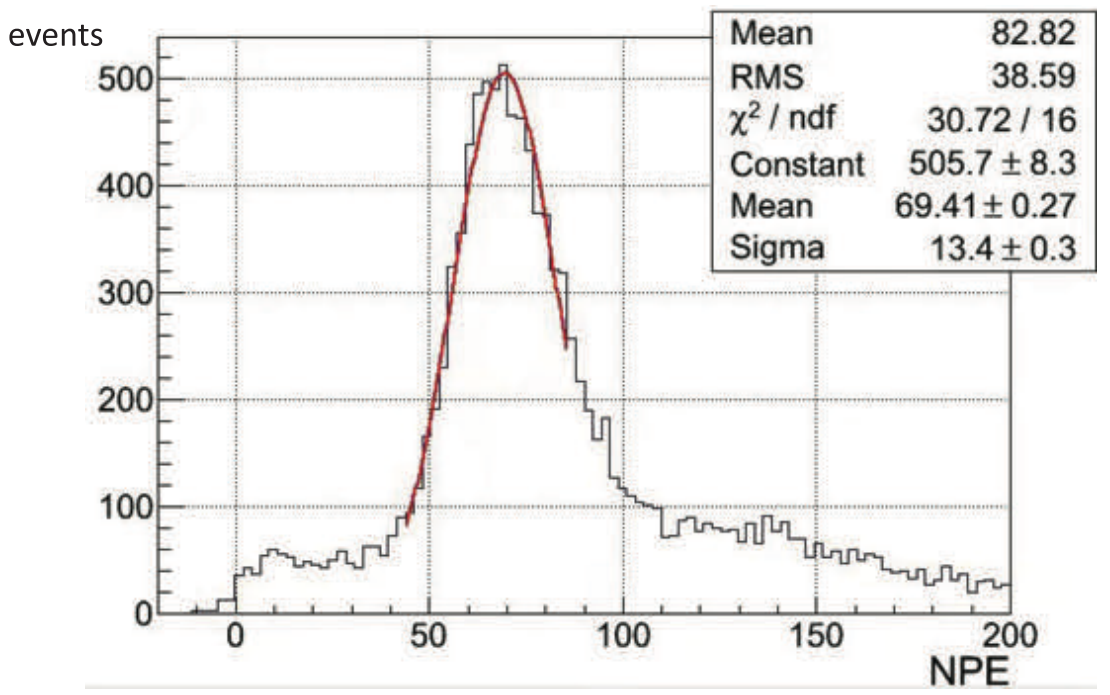

**Fig. 25.** (color online) Single muon spectrum of one detector with gaussian fitting (red line).

## II. WCDA

WCDA is a survey instrument sensitive to gamma rays with energies between 100 GeV - 30 TeV. In the energy region of TeV it can attain the world best survey sensitivity of < 0.1 flux intensity of Crab nebula. Besides observation of Galactic gamma ray sources, WCDA has the discover potential and is sensitive in monitoring extragalactic variant sources, e.g., GRBs, AGNs. WCDA will give an observation with high significance for studying the origin and acceleration of cosmic rays, multi-waveband radiation mechanisms of GRBs, AGNs, EBL, intergalactic magnetic field, galactic evolution, etc. Meanwhile, the WCDA has an advantage for observing extended sources. The important requirements of WCDA including effective area, energy resolution, angular resolution and Q factor varying with energy is shown in Fig. 26.

WCDA covering an area of 78,000 $m^2$, is compose of 3,120 water Cherenkov detectors (WCDs), divided into 3 separate arrays (Fig. 27). Each array is a single water pool with water depth of 4.4 m. Two pools with effective area of 150 m × 150 m contain 900 WCDs individually. The third pool with area of 300 m × 110 m contains 1,320 WCDs.

WCD has a 5 m × 5 m area separated by black plastic curtains with area of 3.3 m × 4.7 m vertically hung in water to absorb scattered light (Fig. 28). Side walls and bottom of the pools are lined with 2 mm thickness HDPE film to keep water from leaking outside. To guarantee an attenuation length of longer than 15 m for the near-ultraviolet light, a water purification and recirculation system is operated. The water recirculation system is composed of two fine filters (5 μm and 1 μm), an ultra-fine filter (0.2 μm), and UV lamps (254 nm and 185 nm) (Fig. 29). It uses a micro-filtration system to remove fine granulation material in water, while using UV185 to degrade Total Organic Carbon (TOC) and using UV254 to kill bacteria in water. Finally, the 0.2 μm filter is used to remove the bacterial carcass, thereby effectively ensuring the cleanness of water. Water purified by this system can reach transparency (attenuation length) above 15 m for light of around 400 nm wavelength. And flow of recirculation system is about one volume/month.

The first pool has 900 WCDs (Fig. 30). Each WCD is equipped by an 8-inch PMT for timing and pulse height in low range and a 1.5-inch PMT for pulse height in high range at the center of WCD and 4 m beneath water surface [16]. The pool measures EAS directions with angular resolution better than 0.2° at 10 TeV and 1.0° at 600 GeV. This is based on time resolution of 0.5 ns of Front-End Electronics (FEE) connected to the large PMTs, watching upwards for direct Cherenkov light only. Dynamic range of WCD is enlarged very much by using the small PMT. This configuration enables measurement of detailed particle density distribution in EAS cores without significant saturation even for energetic EAS up to 10 PeV, and reaches core location resolution better than 3 m over a wide energy range. This is designed for identification of cosmic ray composition and measurement of energy spectrum.

At low energies, EAS is small so that total number of particles may be fully contained in the pools, and the secondary particles generate faint Cherenkov signals in WCDs even near EAS cores. In order to enhance gamma ray detection sensitivity, enlarging the sensitive photocathode of the PMT is considered as one effective way to catch the faint signals. LHAASO's upgrading design is to replace the 8-inch PMTs by 20-inch PMTs in the other two pools with total area of 55,500 $m^2$ (Fig. 30 and Fig. 31). The customized design of the 20-inch PMTs using multi-channel-plate (MCP) instead of the traditional dyn-odes enables good uniformity between PMTs as well as Transit Time Spreads (TTS) less than 7 ns and Cathode Transit Time Distribution (CTTD) less than 2 ns. The photocathode is a factor of 6.25 larger than the 8-inch one so that dynamic range is also shrunk by the same factor. In order to compensate loss of energy dynamic range, a 3-inch PMT is installed beside this large PMT in WCD, read out only pulse height by a simplified version of FEE covering 3 orders of amplitudes in number of photoelectrons. Table 5 shows some basic specifications of WCDA, and Fig. 32 is the general schematic of electronics system.

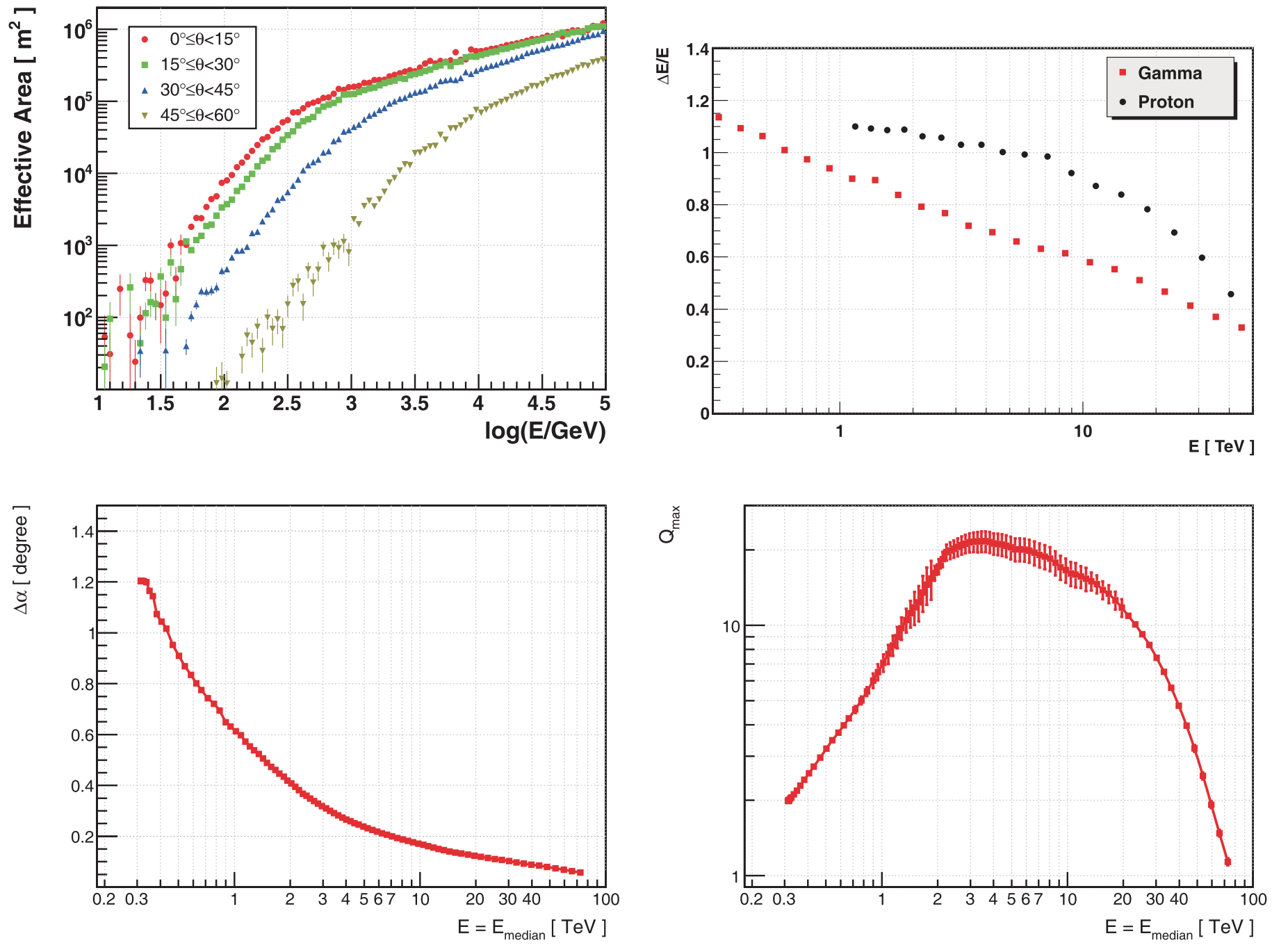

**Fig. 26.** (color online) Effective area (upper left), energy resolution (upper right), angular resolution (lower left) and $Q$ factor (lower right) of WCDA varies with primary energy.

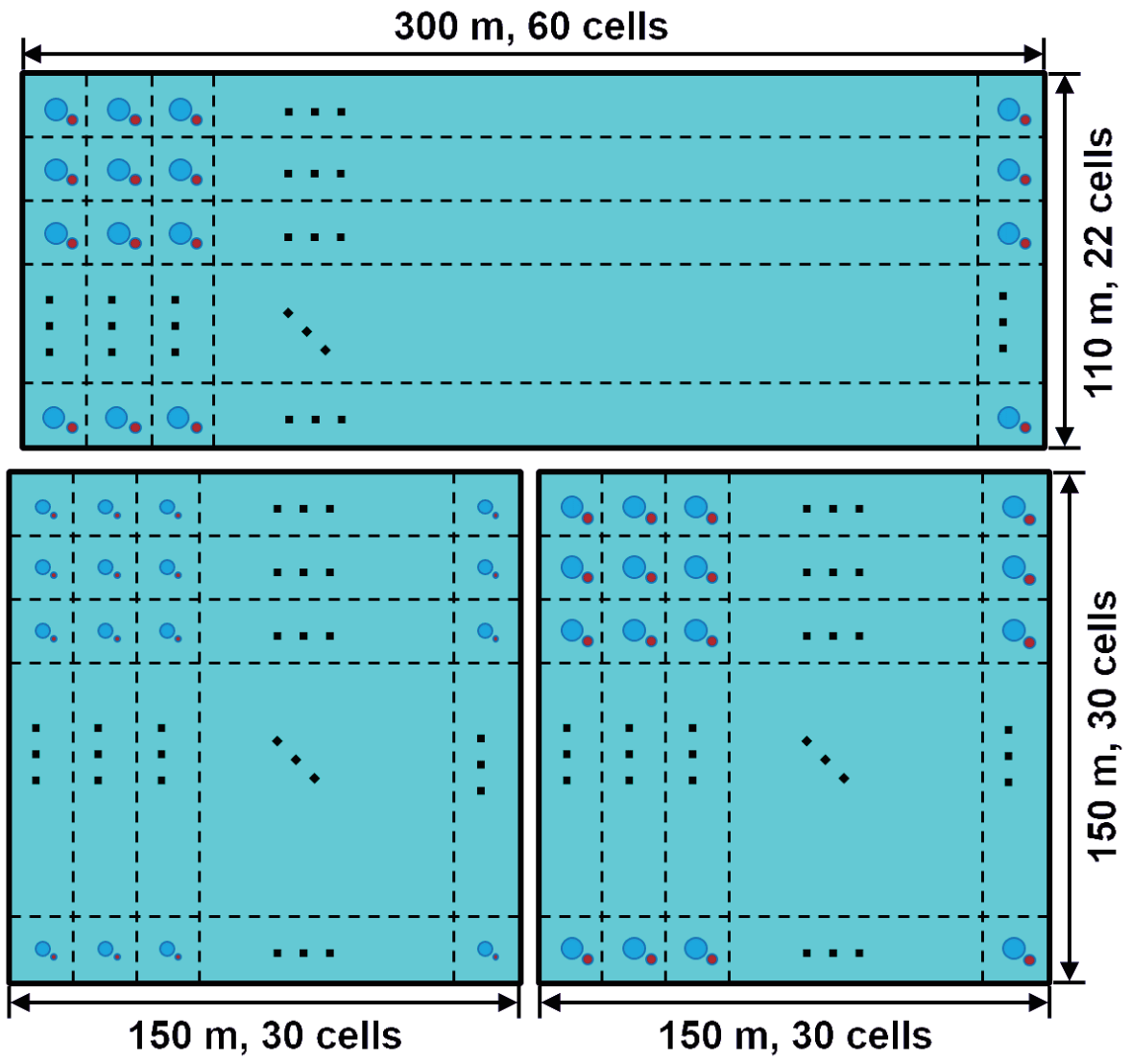

**Fig. 27.** (color online) Layout of WCDA.

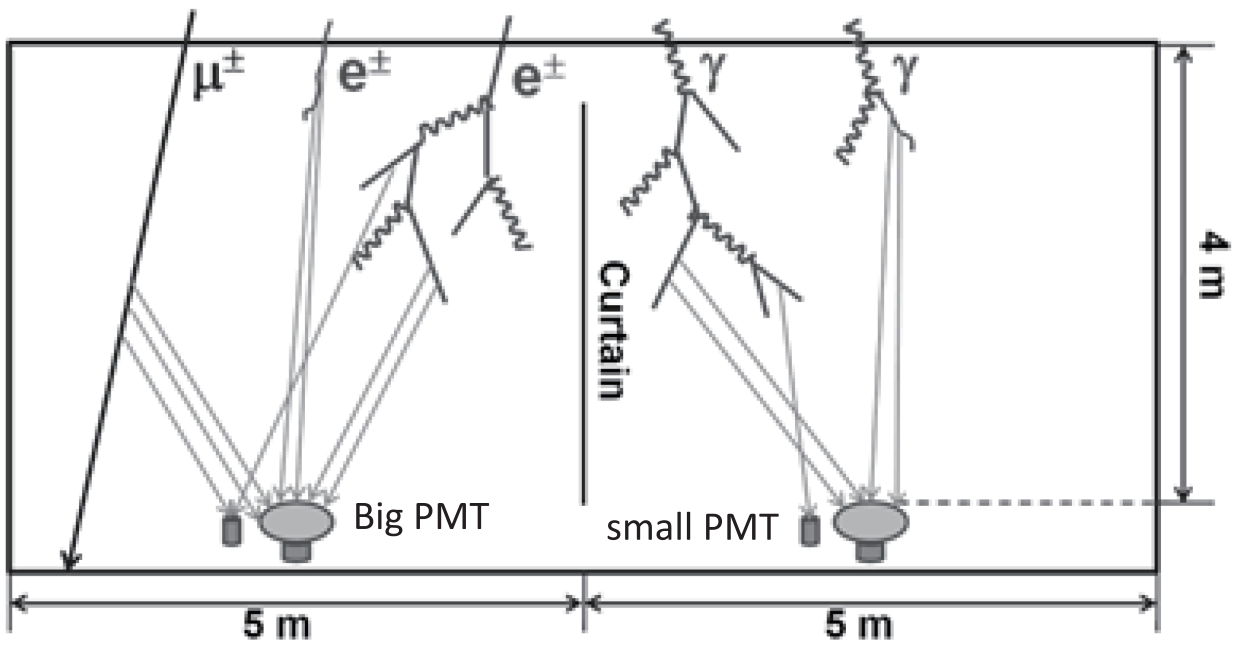

**Fig. 28.** Sketch map of WCD.

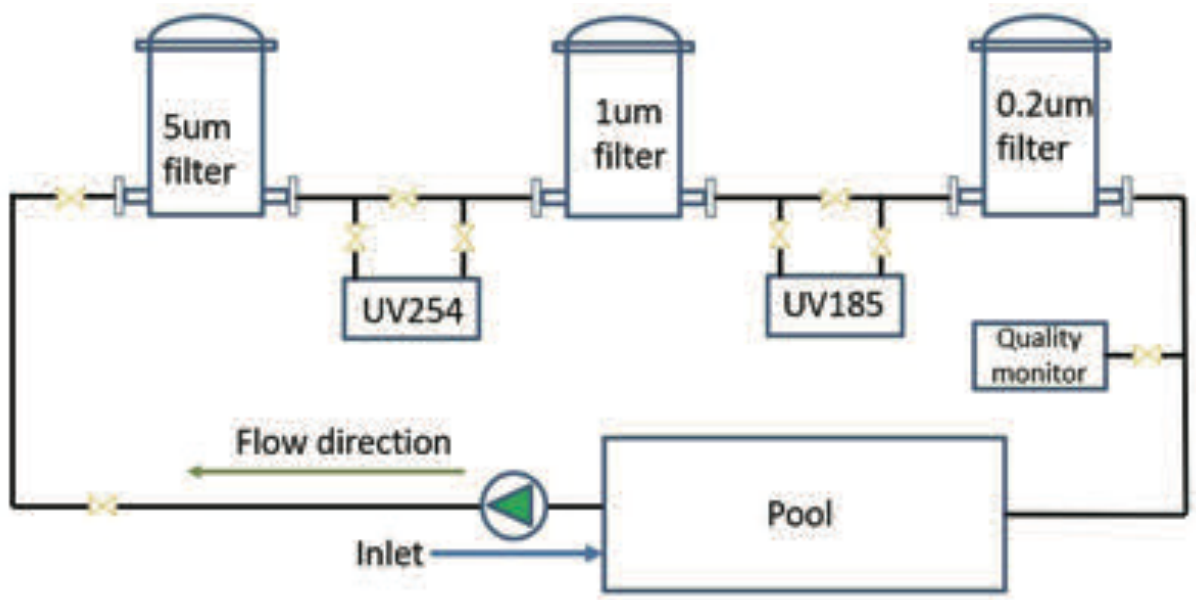

**Fig. 29.** (color online) Schematics of water recirculation system of WCDA.

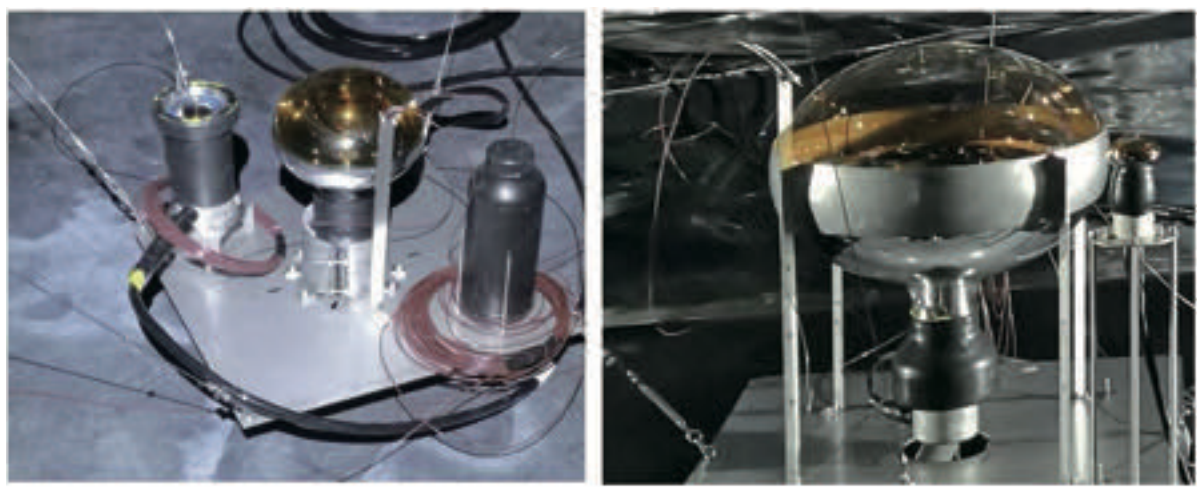

**Fig. 30.** (color online) Photos of WCDs installed in the first pool (left) and the second pool (right).

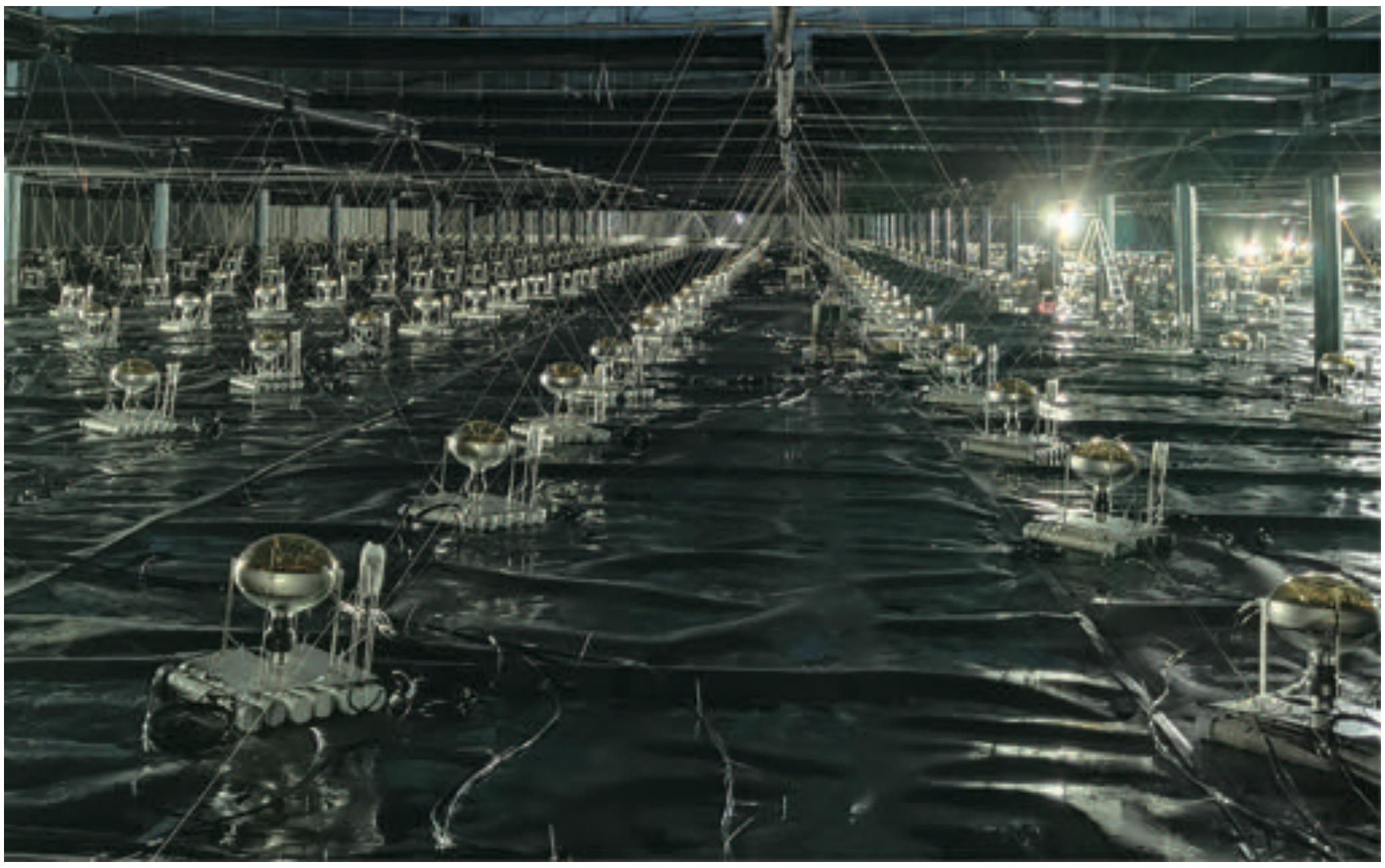

**Fig. 31.** (color online) Photo inside one hall of WCDA without water and curtains.

By using small PMTs, cosmic ray events above 30 TeV are mainly measured, and the event rate is only a few hertz. Therefore, the requirement of time measurement for readout electronics is not high as long as it can meet requirements of cosmic ray composition separation. Moreover, dynamic range meets the requirement of 200,000 (3,000) photoelectrons in the 1st (2nd/3rd) pool, and the target requirements of the readout electronics system are as following Table 6.

**Table 5.** Specifications of WCDA.

| Items | Value |
|---|---|
| Total area | $7.8\times10^4$ m$^2$ |
| Total cells | $3.12\times10^3$ |
| Cell area | 25 m$^2$ |
| Effective water depth | 4 m |
| Water transparency | >15 m (400 nm) |
| Precision of time measurement | 0.5 ns |
| Time resolution | <2 ns |
| Accuracy of charge calibration | <2% |
| Accuracy of time calibration | <0.2 ns |
| Charge resolution(Large PMTs) | 50% at 1 p.e.<br>8-in PMT: 5 % at $4\times10^3$ p.e.s<br>20-in PMT: 3 % at $1.8\times10^3$ p.e.s |
| Dynamic range | 1 - $4\times10^3$ (8"), 20 - $2\times10^4$ (1.5"),<br>1 - $1.8\times10^3$ (20"), 1 - $3\times10^3$ (3") PEs |
| Angular resolution | <$0.4^0$ (at 2 TeV) |

During operation of WCD, time calibration and charge calibration will be proceeded respectively. The time calibration is realized by an optical fiber system [17,18]. On the upper side of every large PMT, two optical fibers with approximately 40 m in length are installed (Fig. 33). All the fibers in a cluster, which contains 6×6 cells, are bundled together, illuminated by a uniform light source consisting of an array of LEDs. These LEDs are powered by the same circuit which can guarantee a fast rise-time (<10 ns) of the LED light pulse output. For every cluster, beside one set of fibers are used for the intra-cluster calibration, another set of fiber are shared by neighboring clusters for cross calibration. With a proper selection and design of the fibers, LEDs, the drivers, and the cross-calibration method, all the PMTs in the pool can achieve a precision < 0.2 ns for the time offset measurement.

Charge of the large PMT signals is calibrated in two ranges: the single photoelectron spectrum of the counting rate and the signal of nearly vertical muons hitting the PMT photocathode [19,20]. For the former, owing to the high single rate of PMT in water, it is possible to fit charge distribution, which is peaked at the single photoelectron position, so that gain of each PMT is obtained.

**Table 6.** Specifications of WCDA small Size PMT Readout.

| Design Requirements | the 1st pool | the 2nd/3rd pool |
|---|---|---|
| Time measurement resolution | <20 ns | <20 ns |
| Dynamic range of time measurement | 2 µs | 2 µs |
| Dynamic range of charge measurement | 20 ~ $2\times10^4$ p.e.s | 1 ~ $3\times10^3$ p.e.s |
| Accuracy of charge measurement | 10 % at 50 p.e.s, <5 % at > 500 p.e.s | 30 % at single p.e., <5 % at > 10 p.e.s |
| Channel Number | 900 | $2.22\times10^3$ |

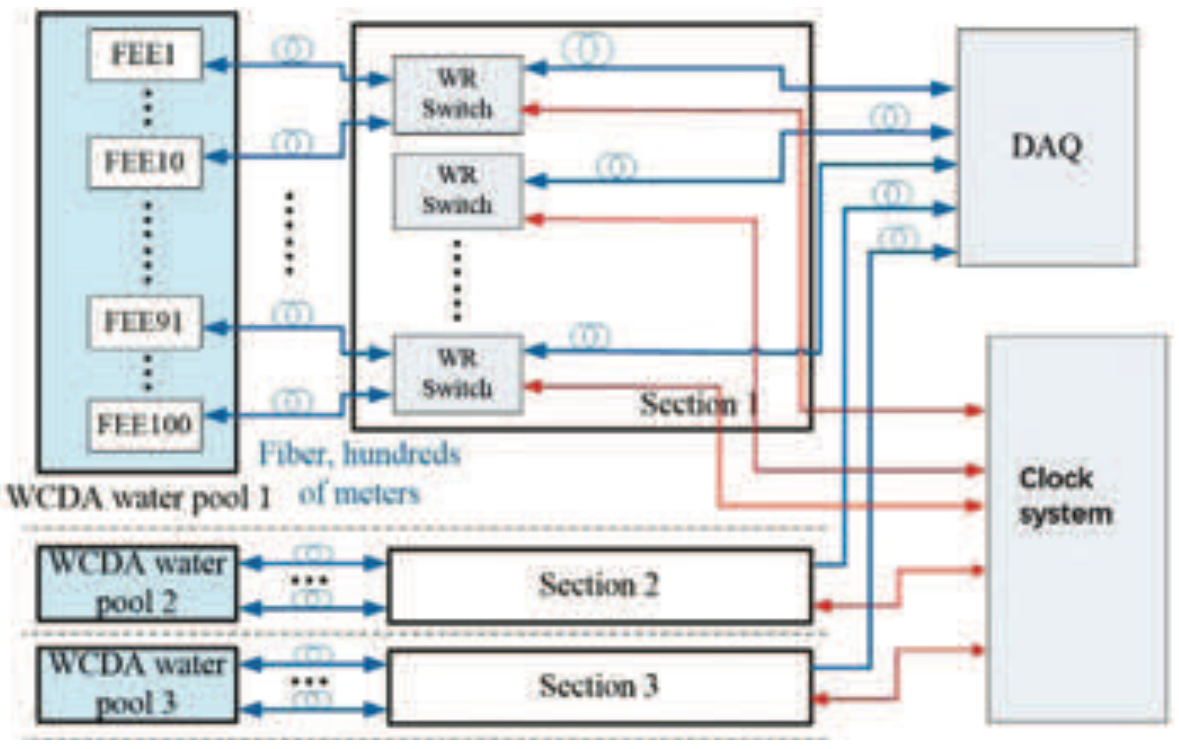

**Fig. 32.** (color online) Schematic of WCDA electronics system. (Provided by Lei Zhao.)

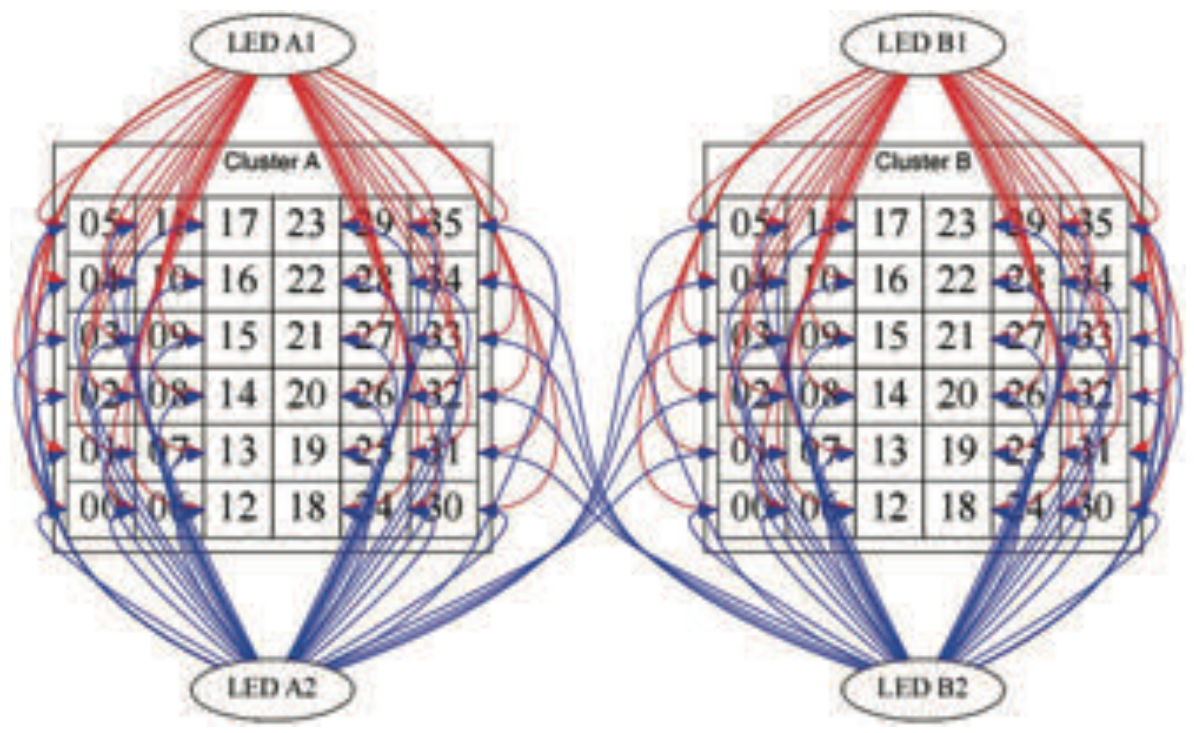

**Fig. 33.** (color online) Schematic of optical fibers layout of time calibration of WCDA.

For the latter, thanks to the enhancement of number of photoelectrons from muons hitting on the photocathode, the peak position of these signals can be distinctly revealed after several hours running. This method can calibrate PMT charge in precision of 2%, involving all the effects, such as quantum efficiency, geometrical effect, collection efficiency, and electronics systm. Besides, the single particle peak in the charge distribution is used to monitor and measure attenuation length of water with a precision of 6% [21].

## III. WFCTA

### A. Configurations of WFCTA

With the advantage of high altitude, LHAASO possesses multiple instruments capable of measuring the complete and 3D observations on the longitudinal and transverse developments of EAS, and enhances greatly accuracy in energy determination and power in identifying the cosmic ray component. The total p.e.s measured by WFCTA can be used to measure the cosmic-ray energy. The shape of the EAS Cherenkov image is observed by WFCTA, including the ratio between the major and minor image axes and the angular distance from the image centroid to the arrival direction of the shower, are closely related to the position when the EAS is developed longitudinally to the maximum. Energy at the core region recorded by WCDA contains the hadron information in the early development of EAS. Muon component recorded by MD reflects the hadron information in EAS. WFCTA is an array of Cherenkov telescopes, a movable design idea is adopted, different array configurations are used for the different observation stages to realize the accurate measurement of component energy spectrum of $10^{13}$ - $10^{18}$ eV, and the energy scale transfer from the direct space measurement to the extremely high energies [22]. In total, three stages are employed by the WFCTA to perform the component and energy spectrum measurement of cosmic rays.

In the first observation stage, WFCTA aims to observe the primary energy range from 10 TeV to 300 TeV, one Cherenkov telescope will point at the zenith angle of 0°.

In the second observation stage, WFCTA aims to observe the primary energy range from 50 TeV to 10 PeV, 6 Cherenkov telescopes will point at the zenith angle of 30° and each telescope points at different azimuth angles, in order to cover the sky with zenith angle from 22° to 38° and azimuth angle from 0° to 156°.

In the third observation stage, WFCTA aims to observe the primary energy range from 1 PeV to 100 PeV. To make sure that the position of shower maximum is far from the observation surface, the main axis of 18 Cherenkov telescopes will point at 45° in zenith angle. Cherenkov telescopes will point at different azimuth angles in order to cover 0° to 360° in azimuth.

### B. Properties of WFCTA

WFCTA is composed of 18 wide field Cherenkov telescopes (WFCTs). Each WFCT (Fig. 34) consists of an array of 32 × 32 Silicon photomultipliers (SiPMs), a 5 $m^2$ spherical aluminized mirror, a portable and adjustable container, power supply system and slow control system. It has a field of view (FOV) of $16° \times 16°$ with a pixel size of approximately $0.5° \times 0.5°$. Mirrors are mounted at the back of the container. The SiPM camera is located at the focal plane which is 2870 mm away from the center of the mirror. Power supply system and slow control system are also installed inside the container with size of 4.2 m × 2.5 m × 2.6 m. The container is mounted on a truck frame with a pitching rotation system that allows the container to be lifted up-and-down manually, from 0° to 90° with a step of 0.1°. The mirror is made of 20 hexagon-shaped segments with side length of 297.1 mm and 5 half hexagon-shaped segments. The radius of each segment curvature is 5800 mm with a tolerance of 10 mm. The reflectivity of the mirror is ≥ 82% for wavelength from 300 nm to 900 nm.

The SiPM camera consists of 64 sub-clusters. Each sub-cluster (Fig. 35) is composed by 4 × 4 SiPMs, each SiPM coupled with a light funnel (Winston cone), a preamplifier board, a voltage and temperature compensation loop board, two analogue boards, a digital board and a power regulator board. The SiPM used is a square Hamamatsu S14466 with photosensitive area of 15 mm × 15 mm. Each SiPM has 360,000 Geiger-Avalanche Photo Diode (APD) cells with size 25 μm [23]. The optical spot size is about 25.8 mm×25.8 mm. A light funnel is used to funnel the photons of the optical spot on the SiPM. Each funnel has an entrance area of 25.8 mm × 25.8 mm (including the funnel walls thickness) and an exit area of 15 mm × 15 mm to match the SiPM area (Fig. 35). SiPM signals are amplified by the pre-amplifier, and then are split into two channels, high gain and low gain, to get a good linearity over a wide dynamic range, to cover the corresponding 3.5 decades in the charge measurement. Finally, the signals are digitized by 50-MHz, 12 bits flash analog-to-digital-converters (FADCs) in the digital boards (DBs) and are collected by FPGAs to do further processing: single channel trigger, event trigger, signals transmission and storage, etc. (Fig. 36). A temperature sensor is mounted on the back of each SiPM chip. The sensor is needed by the voltage and temperature compensation loop used to stabilize the SiPM gain against temperature variations. An optical filter window that allows light between 300 nm and 550 nm passing through with a transmission of 92% is mounted in the front of the SiPM-based camera. The window improves the signal to noise ratio of the camera because the sky background light wavelength is dominated by wavelength ≥ 550 nm where SiPMs are still sensitive enough. The power consumption of each camera is about 720 W. The camera is cooled by an internal self-circulating air-cooling system.

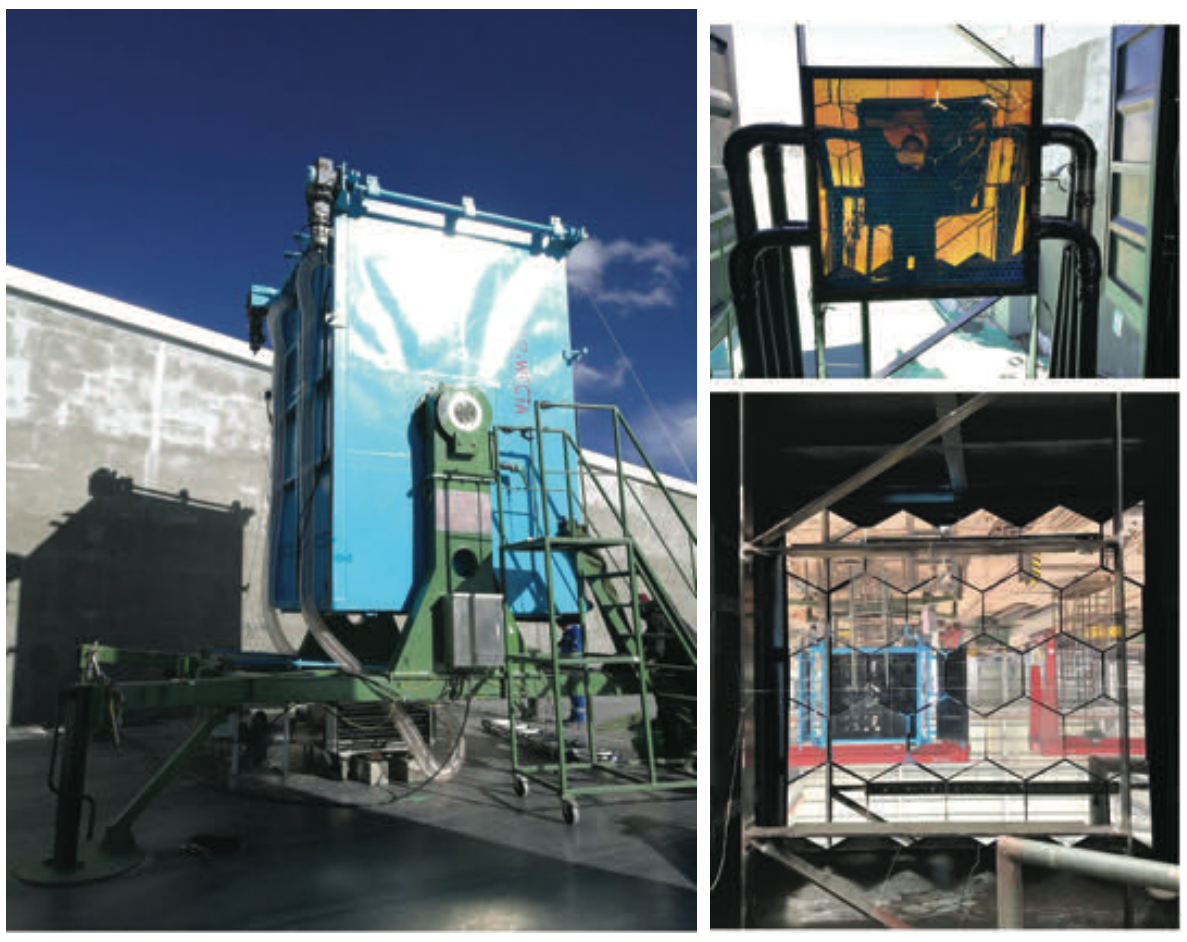

**Fig. 34.** (color online) Left panel: A photo of the telescope. Right top panel: Photos of the SiPM camera. Right bottom panel: mirror.

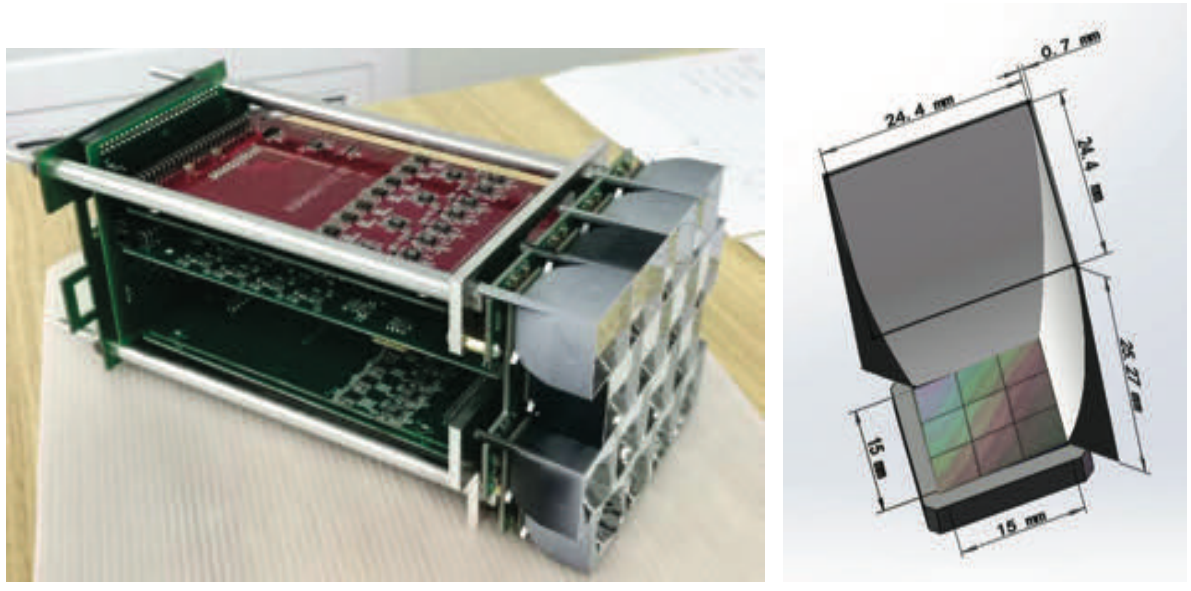

**Fig. 35.** (color online) Left panel: A photo of one sub-cluster. Right Panel: A half light funnel (Winston cone) and an SiPM.

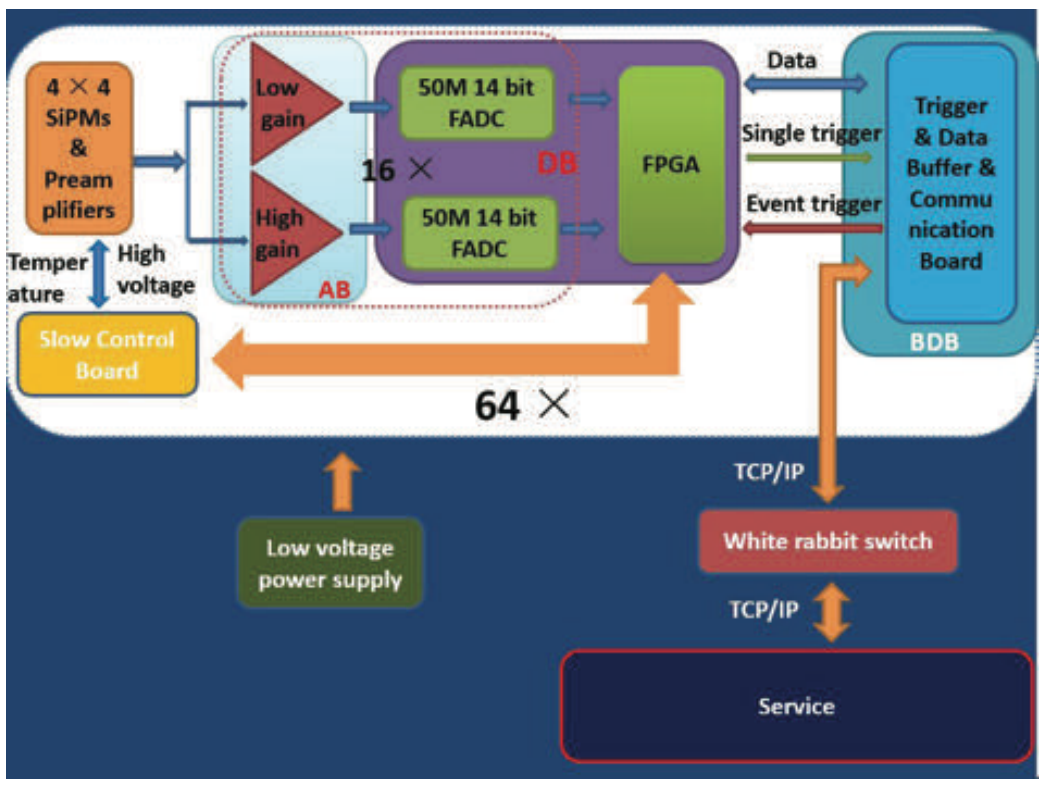

**Fig. 36.** (color online) The architecture of SiPM camera readout electronics.

## C. Performances and results of test run

The telescopes were installed at the southwest corner of the first pool of WCDA and can undergo night observations with and without moonlight, even ran through the entire night with full moon. One coincidence event between the telescope and WCDA is shown in Fig. 37 (Fig. 38).

To monitor the gain of the SiPM camera, six UV-LEDs are mounted at the center of the mirror. The LED driver and six LEDs are installed inside a container that maintains a constant temperature with a self-adjustable heater to be 30.7 °C±0.1°C. The instability of LEDs is less than 0.5% at different temperature [24]. A diffuser is put in front of the LED to make the LED light uniform. One of LEDs is turned on every night. All LEDs are calibrated by the absolute calibration device once a year.

Two 750 W blowers are used for the SiPM camera cooling. The temperature distribution of the SiPM camera is shown in Fig. 39. The difference between the highest and lowest temperatures is about 11 °C±0.2°C. The temperature distribution on the camera is within the adjustable range of bias voltage and temperature compensation loops. The ambient temperature began to drop, and the temperature of SiPM also decreased at night. The ambient temperature rises, and the temperature of SiPM rises as well during daytime. The temperature difference at night is about 10°C. High voltage and temperature compensation loops can keep the SiPM gain stable at different temperature environments.

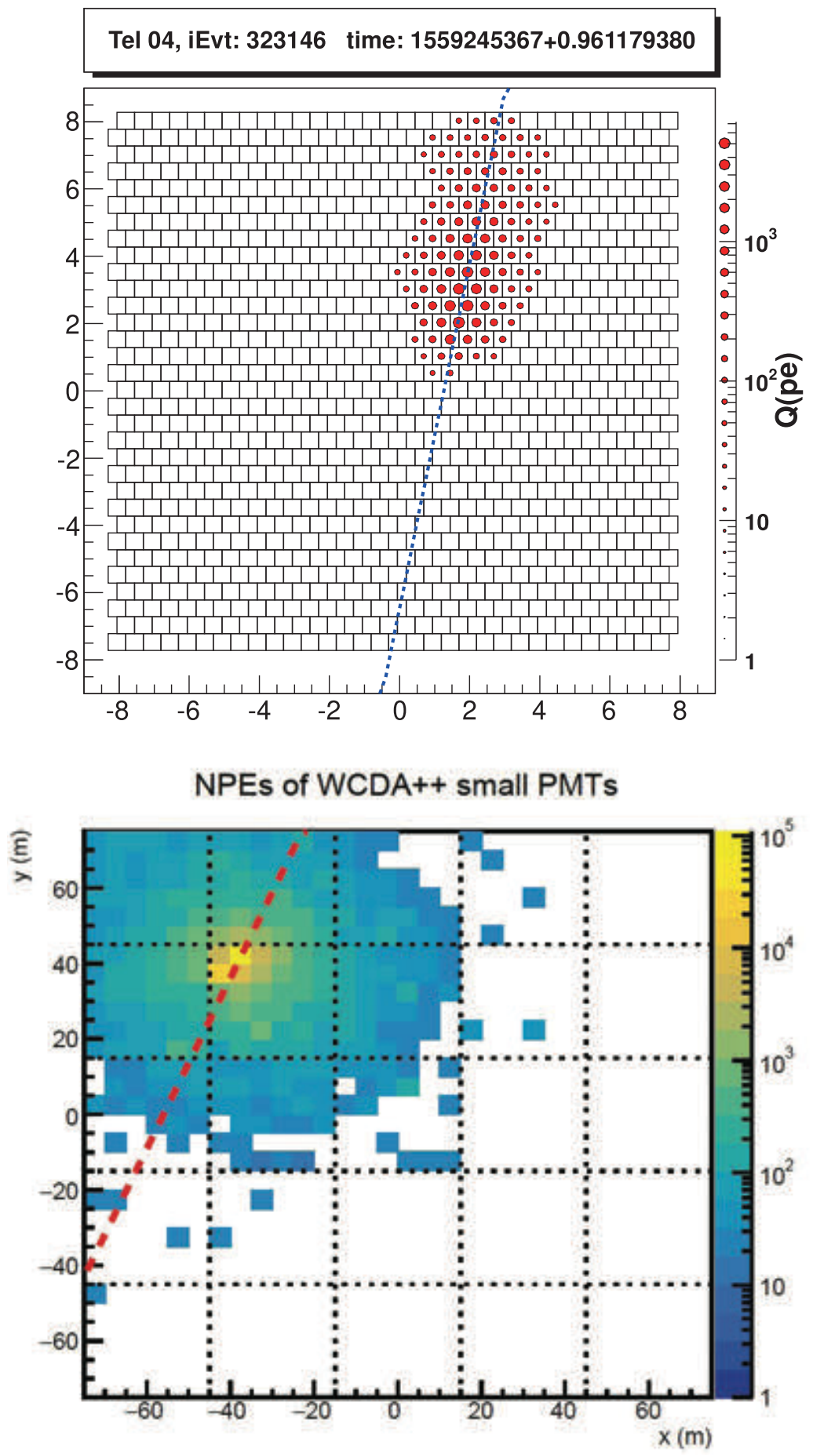

**Fig. 37.** (color online) A coincidence event of two telescoopes and the WCDA first pool. Upper panel: a Cherenkov event of the first Cherenkov telescope. Lower panel: an event of the WCDA first pool using the 1.5-inchs PMT. The dotted lines in (a) and (b) are the shower detector plane reconstructed by the Cherenkov telescope. (Provided by Zhi-Yong You.)

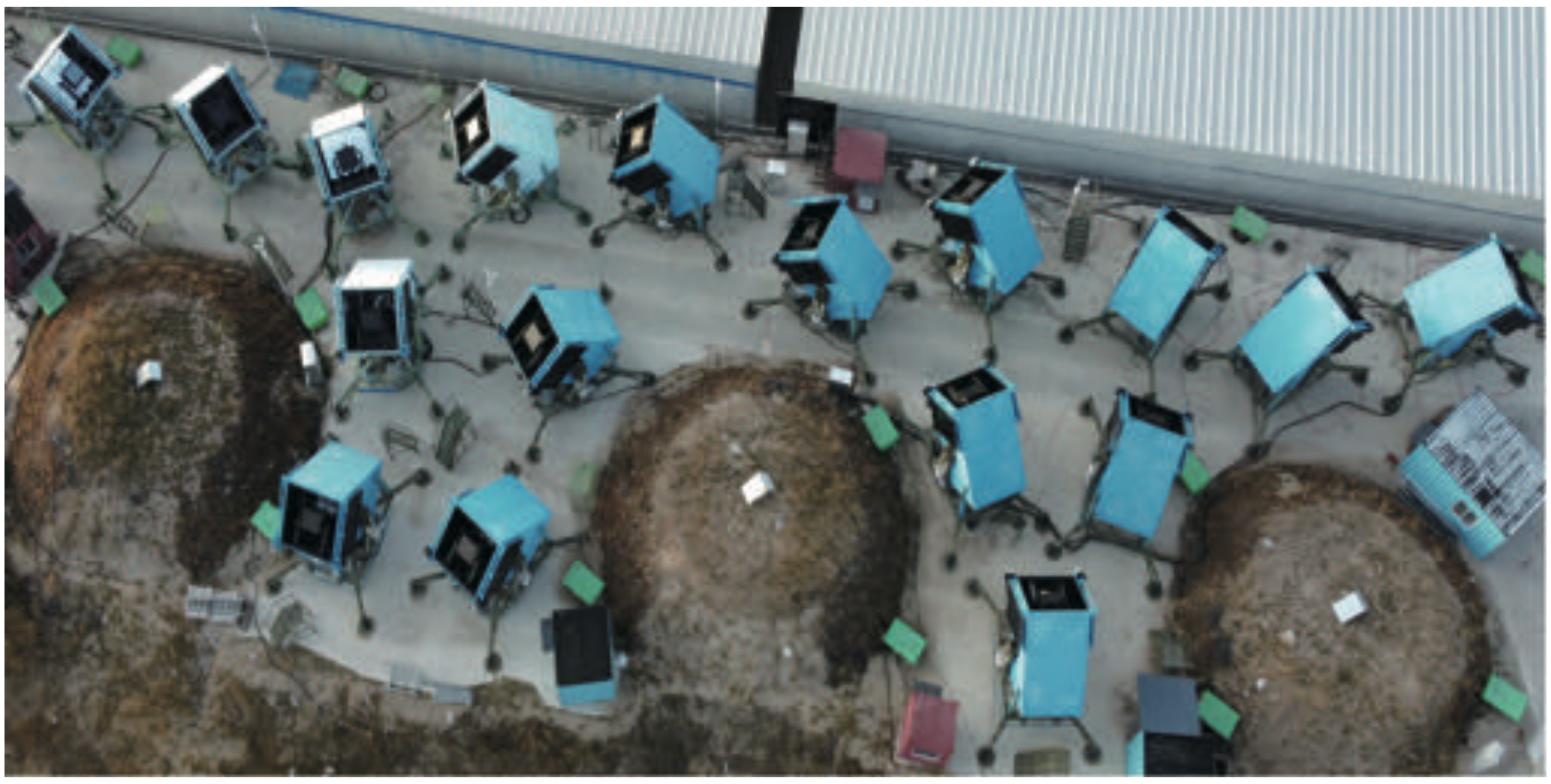

**Fig. 38.** (color online) Bird view of WFCTA.

The gain of SiPM camera is stable in moonless nights. A resistor of 250 Ω is put between the input hight voltage and SiPM cathode to protect the SiPM from damage under the strong light condition. The night sky background light induces a DC current in the SiPM, which induces a voltage drop on the bias resistor. Accordingly, the supply voltage, and then the gain, of the SiPM varies with the intensity of the sky background light. On April 19, 2019, the Cherenkov telescope was operated with the full moon. The space angle between one pixel (pixel 6) of the SiPM camera and the moon as a function of time is shown in Fig. 40. The baseline of SiPM is modulated by moonlight, the smaller the angle between SiPM and the moon, the higher the baseline. The gain of SiPM is linearly related to the intensity of moonlight on SiPM. The variation of SiPM gain is about 4% during the full moon night. Baseline monitoring values can be used to correct the SiPM gain reduction due to moonlight in off-line analysis. After correction, the SiPM gain is constant during the whole moon night (Fig. 41).

Fourteen telescopes are successfully run at LHAASO site. More than 500 million events coincident with the WCDA first pool have been collected. The SiPM-based Cherenkov telescope has been shown to operate steadily on a full-moon night and achieve a longer duty cycle than a PMT-based Cherenkov telescope. More detailed per-formance studies are underway, to validate the telescopes against the design requirements. Eighteen telescopes (full WFCTA array) will be operated at LHAASO site in May 2021.

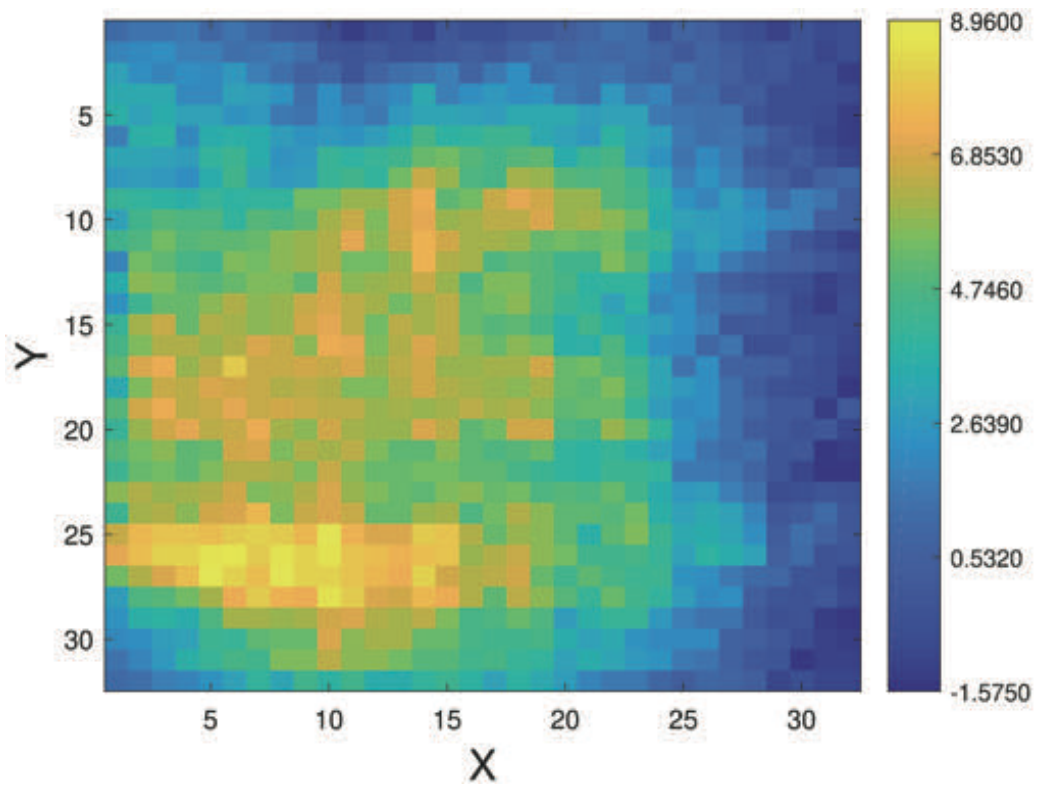

**Fig. 39.** (color online) Two-dimensional temperature distribution of the SiPM camera is shown. The temperature of SiPMs near the air inlet is lower than that of SiPMs near the air outlet. (Provided by Ming-Jie Yang.)

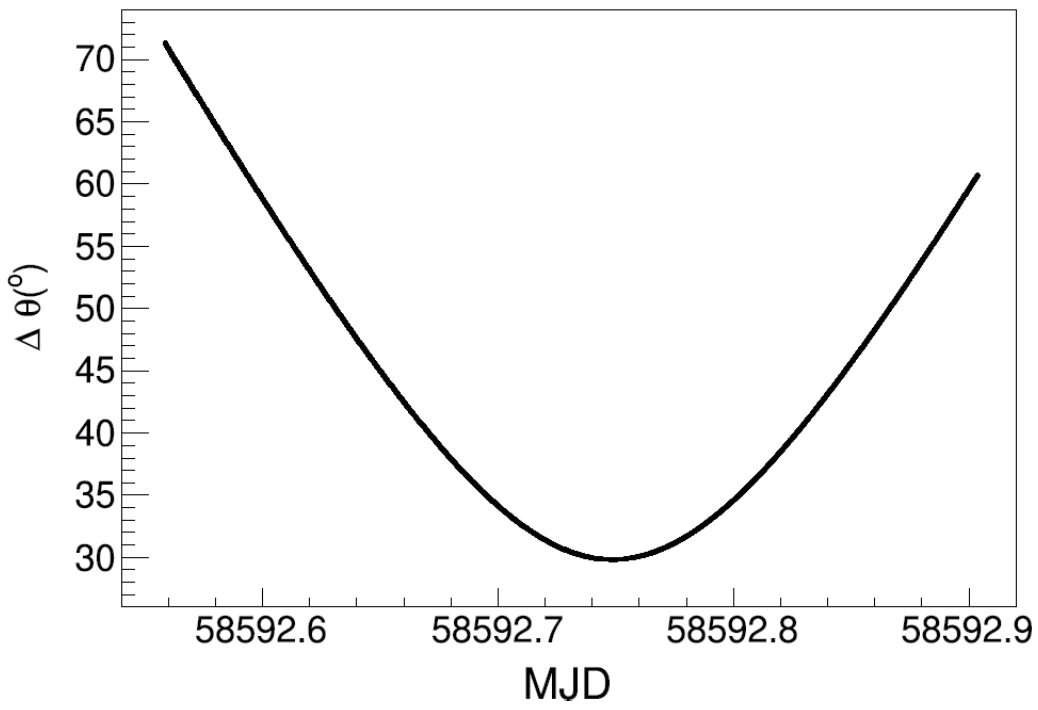

**Fig. 40.** The space angle between one pixel (pixel 6) of the SiPM camera and the moon as a function of time. (Provided by Li-Qiao Yin.)

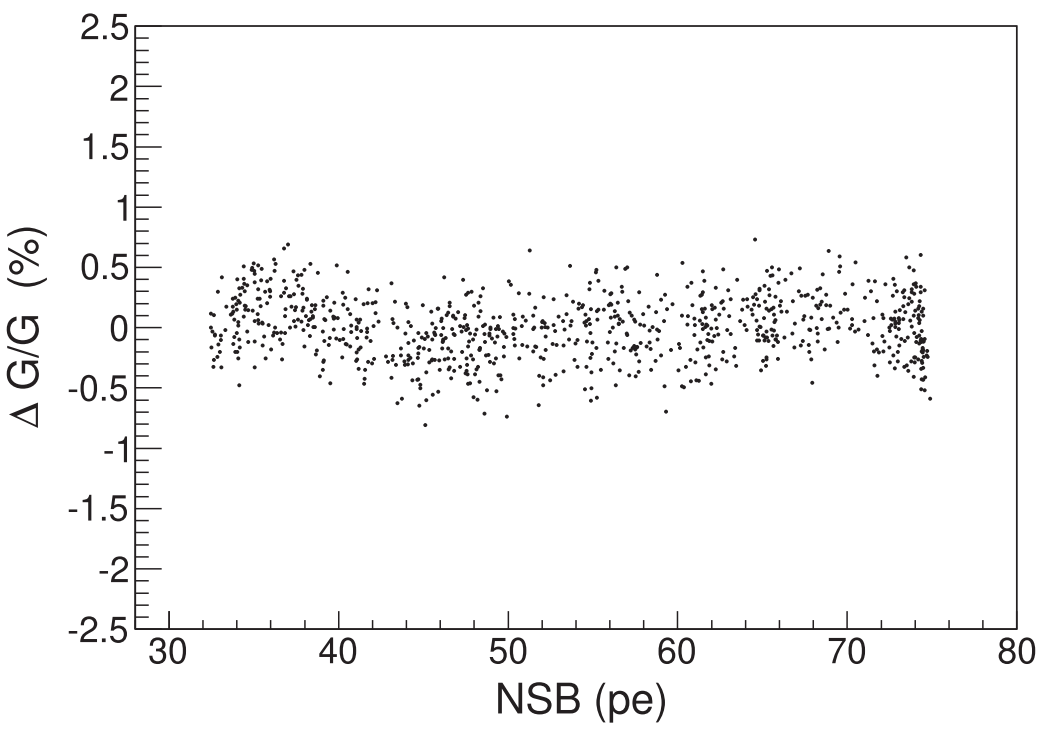

**Fig. 41.** After correction, the gain of the pixel as a function of the night sky background (NSB) in the full moon night. (Provided by Li-Qiao Yin.)

## IV. ENDA

### A. Motivation

When arriving at Earth, high energy cosmic rays interact with the air nuclei thus inducing EAS. They consist of a core of high energy hadrons that continuously feed the electromagnetic part of the shower through fast decays of neutral pions, kaons and etas. Nucleons and other high energy hadrons contribute to the hadronic cascade. High energy hadrons, which constitute the EAS skeleton as G.T. Zatsepin called it, carry important information for multi-parameter correlation studies. For example, hadronic observables, primarily the hadron number/electron number correlation, depend on the nature of the particle inducing the shower [25-27]. Thus, the detection of high energy hadrons, addressed to improve the discrimination power in these analysis, is highly advisable. A way to deal with this problem avoiding the usage of huge and expensive HCALs was brought out in [26]. In these papers the detection of thermal neutrons generated by EAS hadrons is proposed. It is well known that hadrons interacting with ambient matter (air, building, ground, etc.) produce evaporation neutrons due to nuclei disintegration. The neutrons have no charge and lose energy only by scattering. If the medium is a good moderator, i.e., the absorption cross section is much less than the scattering cross section, the neutrons lose energy via scattering down to the thermal ones (moderation process) and then live in the matter until capture. Evaporation neutrons need about 0.5 ms to be thermalized in rock (concrete). Neutrons are generated abundantly, up to 2 orders of magnitude more than parent hadrons [28]. A large fraction of the evaporation neutrons thermalize, so that recording thermal neutrons can be exploited to reconstruct the hadron content in the shower [25]. This approach looks very promising for measurements carried out at high altitude. Indeed, since the hadron content in EAS increases with approaching shower maximum, an abundant production of thermal neutrons can be predicted for experiments at 4 km a.s.l., about a factor 10 higher than that at sea level for showers in the PeV energy region [28]. These considerations suggested the development of a simple and cheap thermal neutron detector, to be deployed over a large area, as 'hadron counter' in EAS experiments at mountain level. This idea led to the development of the EN-detector, based on a mixture of the well-known inorganic scintillator ZnS(Ag) with $^6$LiF, capable of recording both thermal neutrons and relativistic charged particles [29,30]

### B. EN-detector

Of the isotopes used as neutron capture material we used $^6$Li and $^{10}$B. The reactions of neutron capture are:

$$\begin{aligned} & n+{}^6\text{Li}\to{}^3\text{H}+{}^4\text{He}+4.79\ \text{MeV}\ (940\ \text{barns}) \\ & n+{}^{10}\text{B}\to{}^7\text{Li}^*+{}^4\text{He} \\ & \qquad\to{}^7\text{Li}+{}^4\text{He}+2.3\ \text{MeV}\ (93\ \%\ ) \\ & \qquad\to{}^7\text{Li}+{}^4\text{He}+2.8\ \text{MeV}\ (7\%)\ (3980\ \text{barns}). \end{aligned}$$

$^6$Li releases the highest energy during the reaction, but it is a rare material of nuclear fission so that its purchase is strongly limited by government and the price is quite high. Although capturing neutron with lower released energy than $^6$Li, $^{10}$B has larger cross section. Moreover, natural Boron contains 19% of $^{10}$B versus only 7% of $^6$Li in natural Lithium. This and large crossection allowed us to make a natural boron compound comparable in neutron capture efficiency with lithium ones enriched with $^6$Li up to 90%.

A novel type of ZnS(Ag) scintillator alloyed with $B_2O_3$ (65% and 35% respectively) with the $^{10}$B isotope about 19% is developed and produced by Luminofor company (Russia) in collaboration with INR. Powder of ZnS(Ag) and $B_2O_3$ alloy is deposited in optical silicon rubber. The effective thickness of the scintillator layer is 50 mg/cm$^2$.

The design of a typical EN-detector is shown in Fig. 42. The scintillator of 0.35 m$^2$ area is mounted inside a black cylindrical polyethylene (PE) 200-l tank which is used as the detector housing. The scintillator is placed at the tank bottom to a distance of 35 cm from the photomultiplier (PMT) photocathode. A 4"-PMT (Beijing Hamamatsu CR-165) is mounted on the tank lid. A light reflecting cone made of foiled PE foam of 5-mm thickness is used to improve the light collection. As a result, ~ 60 photoelectrons per neutron capture are collected. The efficiency for thermal neutron detection in our scintillator was found experimentally by neutron absorption in the scintillator layer to be about 20%. The peculiar characteristics of the EN-detector output, that are weak and fast signals from charged particles compared to high amplitude, slow and delayed signals from thermal neutron capture, make it well suitable for its use in the framework of EAS experiments. Due to the thin layer of the scintillator, charged particles deposit on average only 60 keV against 2.3 or 2.7 MeV deposited during the neutron capture. A very high $\alpha/e\approx 1$, that is the ratio of the light produced by $\alpha$ particles to the light produced by electrons of the same energy, is the main scintillator feature. This feature allows to collect enough light using only one PMT viewing 0.35 m$^2$ scintillator layer. The charge collection time of a signal due to a neutron capture is several μs, while the characteristic time of the fast emission induced by charged particles is about 40 ns. Note that all signals are integrated with a time of 1 μs and then digitized with a FADC whose resolution is equal to 1 V / 1024 ch = 1 mV/ch. As an example, we show in Fig. 43 the pulses recorded in a high energy EAS event. The first big peak is generated by the large amount of charged particles of the shower front while the smaller delayed signals are generated by thermal neutrons. Thus, the pulse height of the first signal can be used to measure the charged particle density while the delayed signals measured in a time gate of 20 ms give the number of captured thermal neutrons. EN-detectors are also capable to continuously measure natural neutron flux variations connected with weather condition, seasonal effect, solar activity and earthquakes, etc. For this purpose the detector has additional output with 20 μs time of light collection that allows us to more effectively separate neutron pulses.

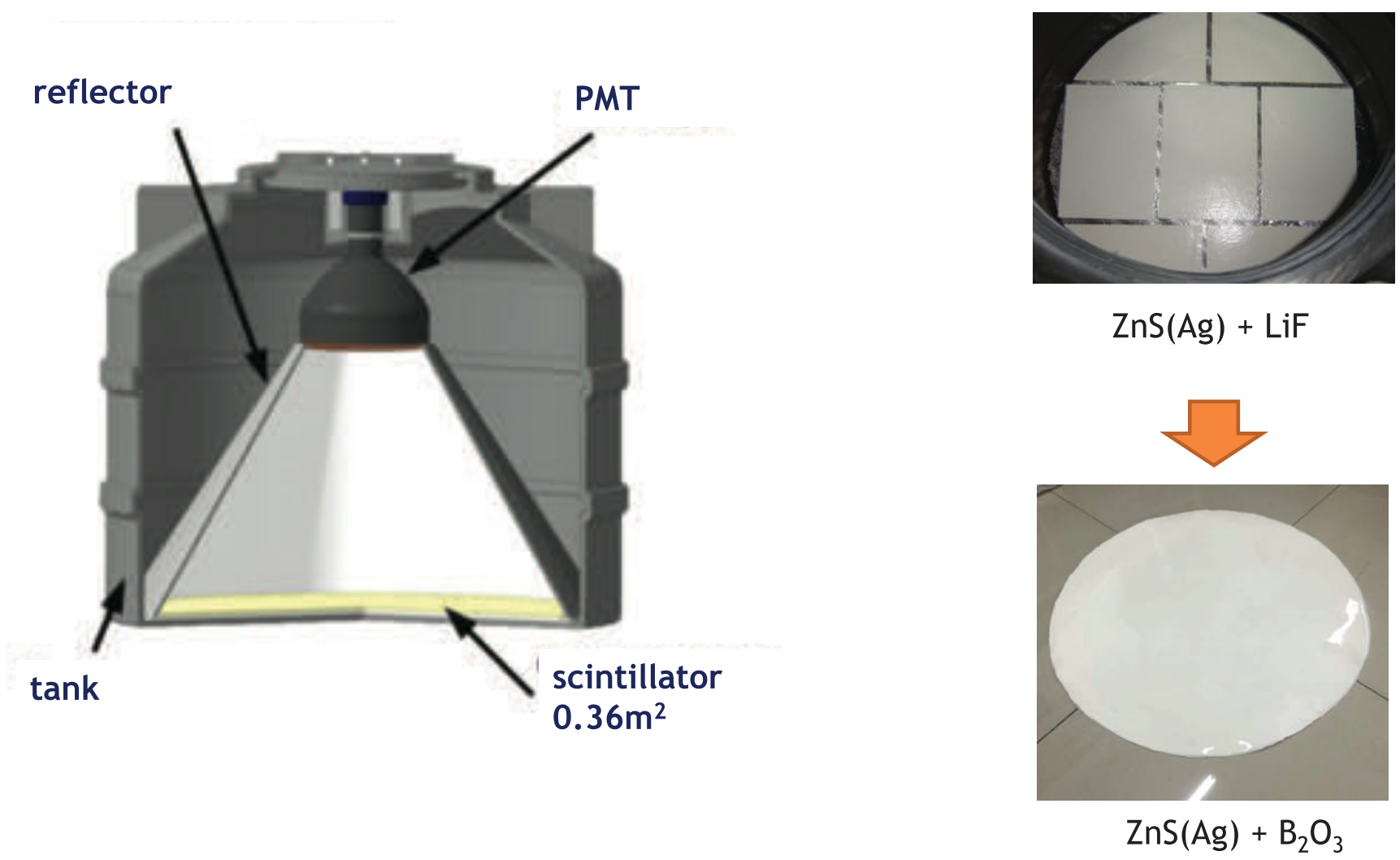

**Fig. 42.** (color online) Left: Scheme of the EN-detector. Right top: Photo of the ZnS(Ag)+LiF scintillator used in PRISMA-YBJ. Right bottom: Photo of the ZnS(Ag)+$B_2O_3$ scintillator used in LHAASO-ENDA.

Front-end electronics (FEE) of each cluster consists of PMT voltage dividers, DIU (Discriminator-Integrator Unit: charge sensitive preamlifier, discriminator and integrator) and IU (Integrator Unit: charge sensitive preamplifier and integrator), 16-channel programmable high voltage (HV) power supply and low voltage (LV) power supply. DIU, IU and HV power supply are shown on Fig. 44.

Voltage dividers have 3 signal readouts from: the 8th and the 5th dynodes and anode. 8th dynode is used for energy deposit measurements, a coincidence selection and counting number of secondary neutrons. the 5th dynode is needed to expand the dynamic range. The 5th dynode pulse will be used in the case of high energy event and big energy deposit in detector when the 8th dynode is saturated and the 5th one is not. The anode's current readout is used only for tests and calibrations. Our working HV range for CR-165 is 700 - 900 V.

DIU and IU are similar to one used in prototypes [31-33]. DIU and IU are supplied by LV -12 and +12 V. DIU has 1 input and 4 outputs. Signal from the 8th dynode comes to the input. Then one readout provides amplified pulse integrated with time of $t$ = 1 μs, the second one gives amplified pulse integrated with time of $t$ = 22 μs. The last two outputs are NIM and TTL pulses in the case of input signal is above the discriminator threshold. 1 $\mu s$ integrated pulse is used for EAS measurements and 22 $\mu s$ integrated pulse is used for monitoring of neutron background variations. NIM and TTL pulses were used in prototypes and aren't used in ENDA. IU has one input for 5th dynode's signal and one output with 1 $\mu s$ time integration. Both devices (DIU and IU) have a linear dynamic range up to 3 V.

HV power supply (produced by Mantigora, Russia) is a stable 16-channel programmable device connected to PC through ethernet. Each channel's range is 400 - 2000

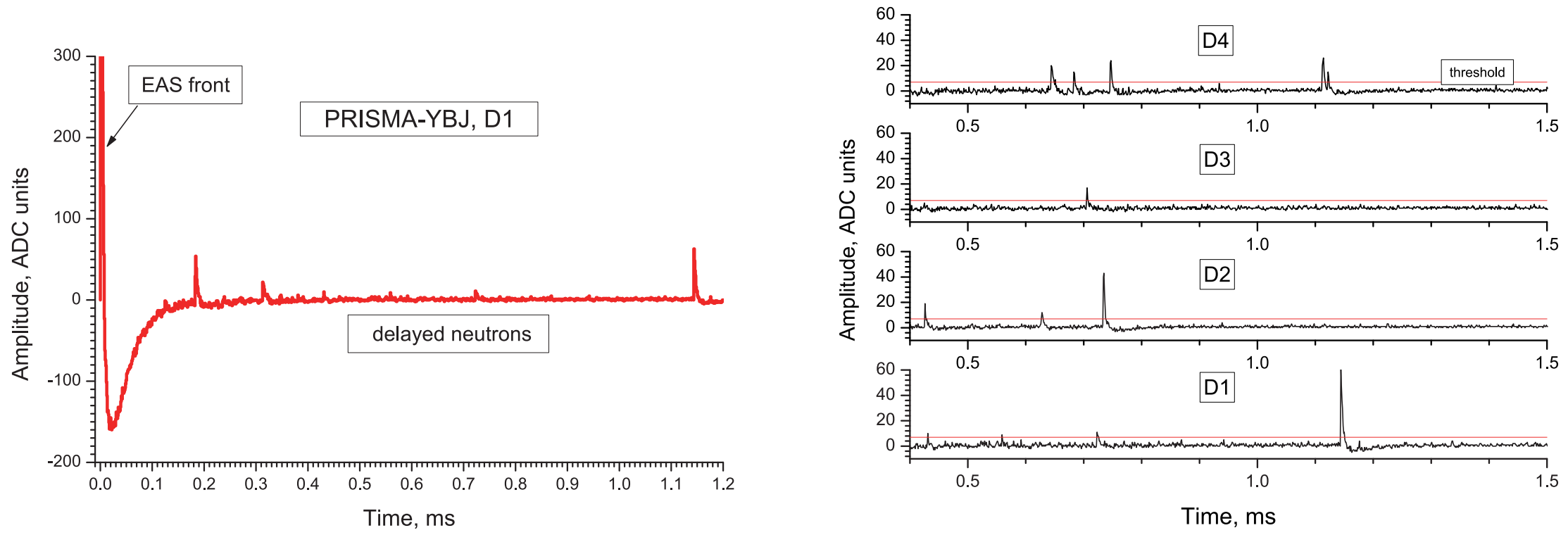

**Fig. 43.** (color online) The shape of the signal from the neutron detector at PRISMA-YBJ.Left: the pulse from 0 to 1.2 ms. The large peak in the first bin is generated by the EAS electrons. Right: the pulses from 0.5 to 1.5 ms. The small delayed peaks following the first peak are generated by thermal neutrons having very wide temporal distribution due to low velocity.

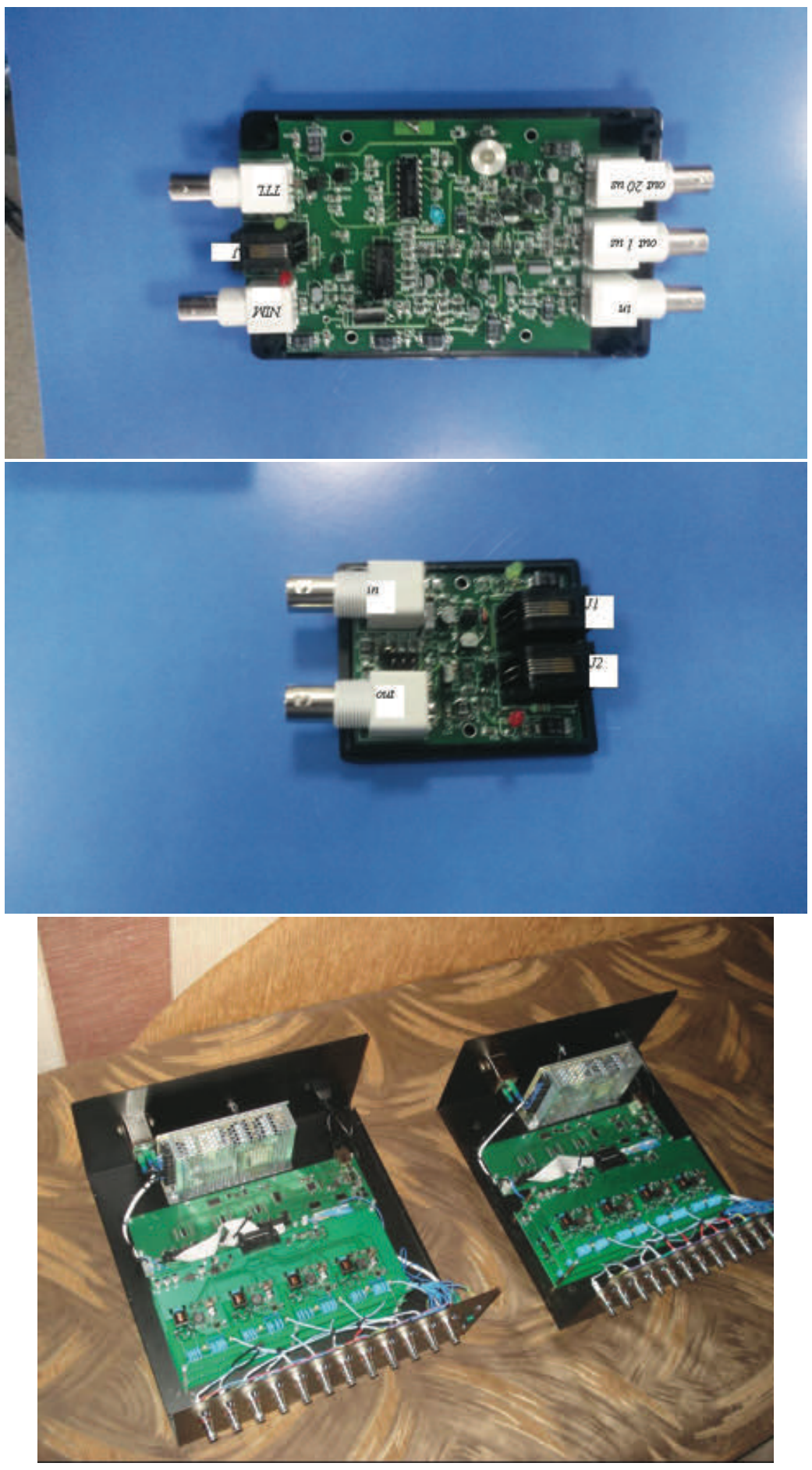

**Fig. 44.** (color online) Front-end electronics. Upper panel: DIU, Middle panel: IU, Lower panel: HV power supply.

V. LV power supply (produced by Dahua, China) has 2 channels providing +12 and -12 V around ground level for DIU and IU.

Data acquisition system (DAQ) of each cluster consists of 32-channel FADC connected to a PC via optical cable. FADC produced by the Sichuan University digitizes each pulse above the threshold. Each digitized pulse has 102 points: 37 points of 20 ns step and 65 points of 1 μs step. Also each pulse has a time tag with 20 ns resolution. The software checks if there is a coincidence of at least 2 detectors in a time gate of 1 μs and if so it counts pulses after that in a time gate of 20 ms analyzing time tags. The first 16 channels of FADC are used for the 8th dynodes and other 16 for the 5th dynodes.

EN-detectors are calibrated with neutron pulse height spectrum. Example of calibration spectra of different detectors with the neutron source Cf-252 are shown on Fig. 45.

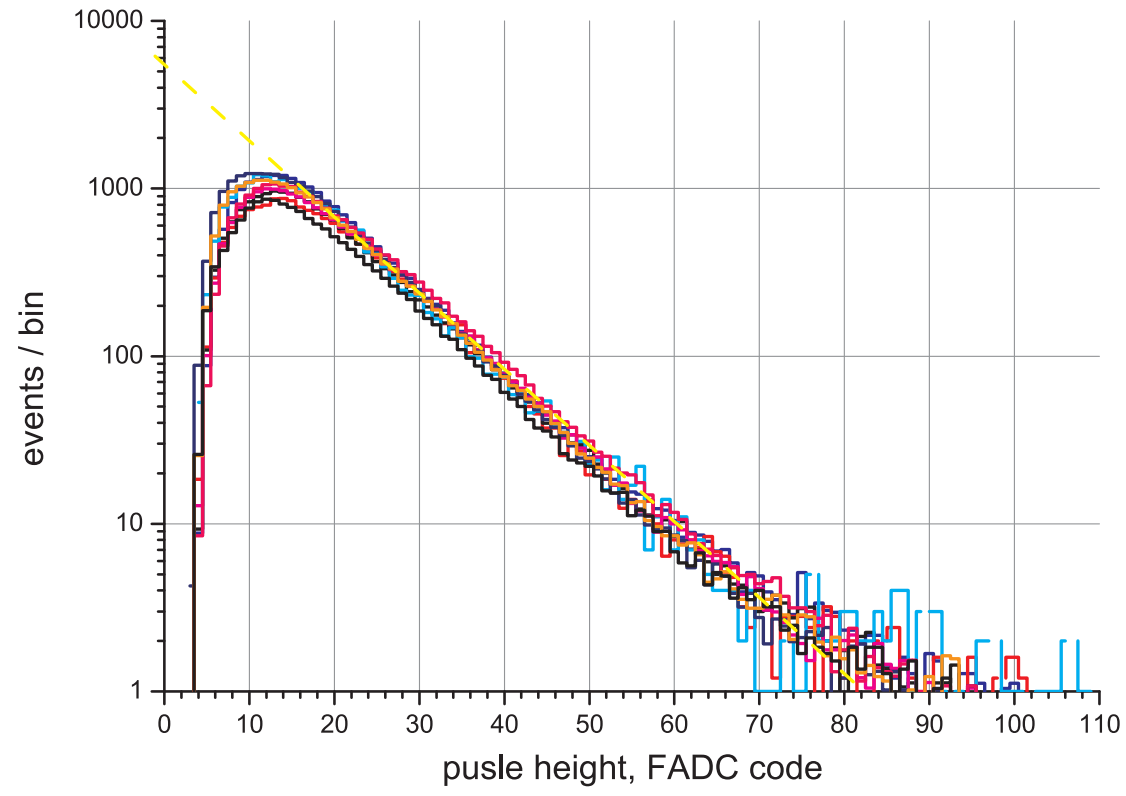

**Fig. 45.** (color online) Neutron pulse height spectra. Different colors are of different detectors.

ZnS(Ag) is not transparent to its own light and because of that neutron pulse height spectrum is exponential:

$$\frac{dN}{dA} = C \cdot \exp\left(-\frac{A}{A_0}\right).$$

In this case the parameter $A_0 = \langle A \rangle$ and is propor-tional to detector gain. EN-detectors are calibrated to set $A_0 \approx 9$ mV. In this instance we have signal from 1 MIP (Minimum Ionizing Particle) equal to ≈ 0.7 mV. Also the number of photoelectrons is ≈ 7 p.e./mV. In addition, the difference between integrals of neutron pulse height spectra (i.e. neutron counting rate) is < 10 %.

### C. Array performance

A prototype array of 32 EN-detectors (PRISMA-32) and P-INR of 16 EN-detectors are now running in Moscow [31,33]. In order to check the performance of this detector at a high altitude site, a small array composed of four EN-detectors (PRISMA-YBJ) has been installed inside the hall hosting the ARGO-YBJ experiment at the Yangbajing Cosmic Ray Observatory (Tibet, China, 4300 m a.s.l., 606 g/cm$^2$). The two arrays (PRISMA-YBJ and ARGO-YBJ) operated together, and coincident events have been analyzed to gather information on the PRISMA-YBJ performance [32]. In order to check the performance of the new type of EN-detectors at a high altitude site, in February 2017 we built an array of 16 EN-detectors based on boron compound scintillator (ENDA-16-YBJ) at Tibet University (TU) in Lhasa, Tibet, China (3700 m a.s.l.), and then moved it to YBJ at the end of 2018 [34, 35]. Up to now, ENDA has totally 66 detectors (ENDA-64 and the other two as backup), ready for deploying inside LHAASO to make a hybrid detection of cosmic ray spectrum above 100 TeV (Fig. 46 left).

ENDA-64 array performance was studied with simulations using GEANT4.10 and CORSIKA7.56 codes. ENDA energy thresholds providing EAS recording efficiency above 90% are ≈ 350 TeV and ≈ 1 PeV for primary protons and iron nuclei respectively. Energy thresholds are shown in Fig. 47. Simulated age parameter distributions for primary protons and iron are presented in Fig. 48.

Primary energy resolution was calculated for protons with fixed energies of $10^{15}$ and $10^{16}$ eV and is ≈ 30% and ≈ 15% respectively and shown in Fig. 49. For energy reconstruction the multi-variable analysis including electron and neutron components was used. Some of these parameters are measured (number of neutrons and electrons), some of them are reconstructed (age parameter, shower size, zenith angle). The analysis technique based on neutron and electron component measurement is described in [36]. Also the distribution of the number of recorded neutrons is presented in Fig. 50.

The core location error is less than 3 m for energies above 1 PeV. The angular resolution of ENDA-64 is about 4 degrees. Also the primary mass reconstruction capabilities of the array were studied by simulations. Using the method described in [35] the primary masses were reconstructed for protons, nitrogen and iron nuclei with energies above 1 PeV and the results are shown in Fig. 51.

After achieving good results, ENDA will be extended up to 400 detectors with array area of 10000 m$^2$ (Fig. 46 right) to study energy spectrum and mass composition at the energy region of $10^{17}$ eV in conjunction with LHAASO.

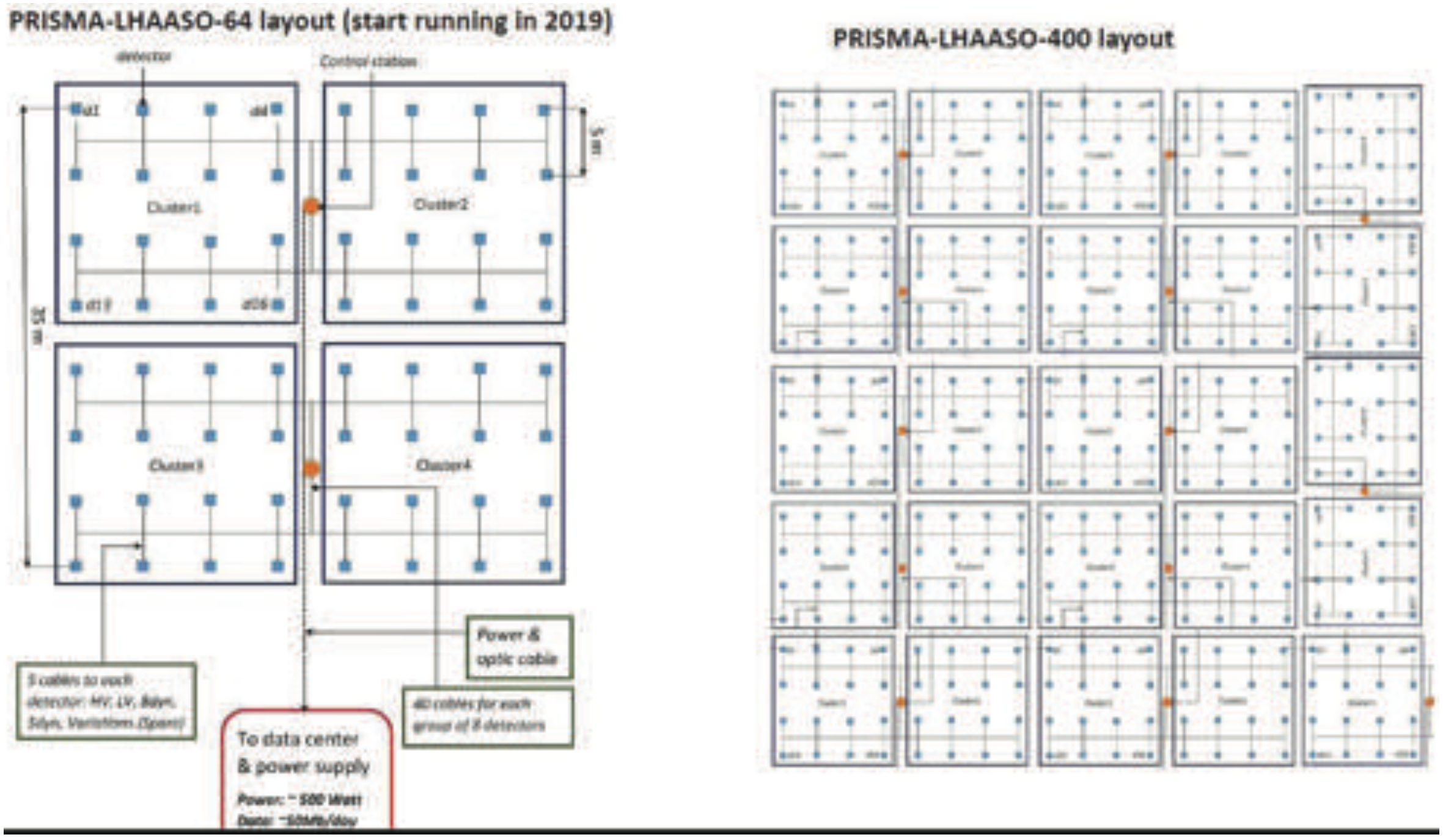

**Fig. 46.** (color online) Configuration of ENDA-64 (left) and ENDA-400 (right).

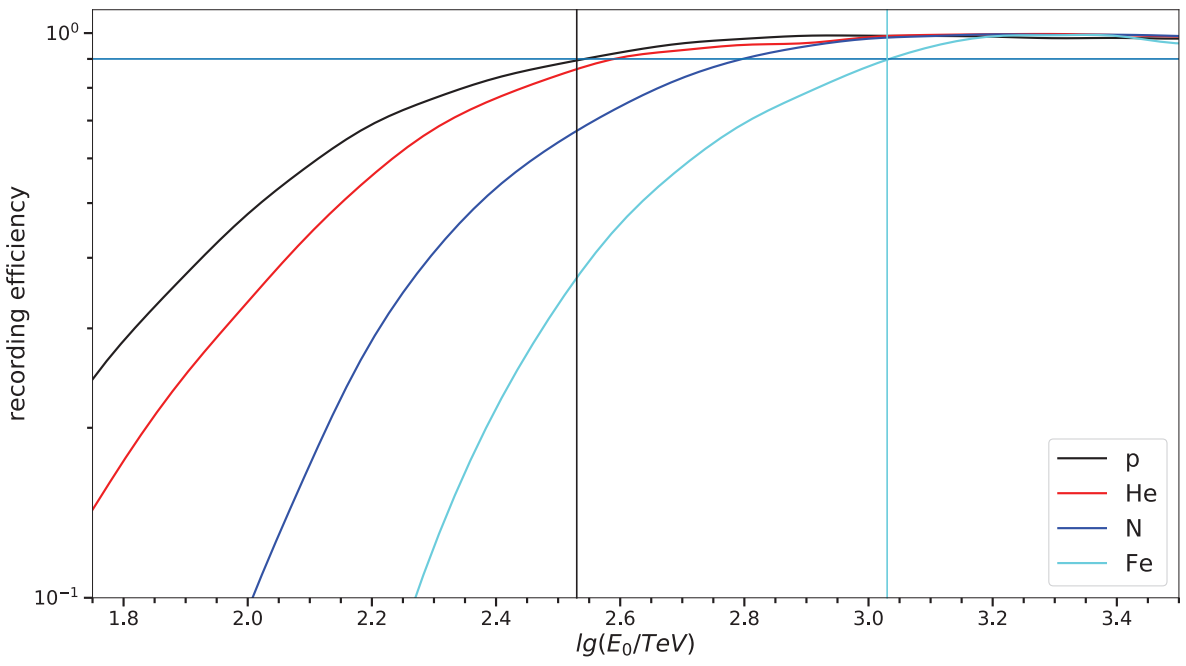

**Fig. 47.** (color online) ENDA EAS recording efficiency for different primaries.

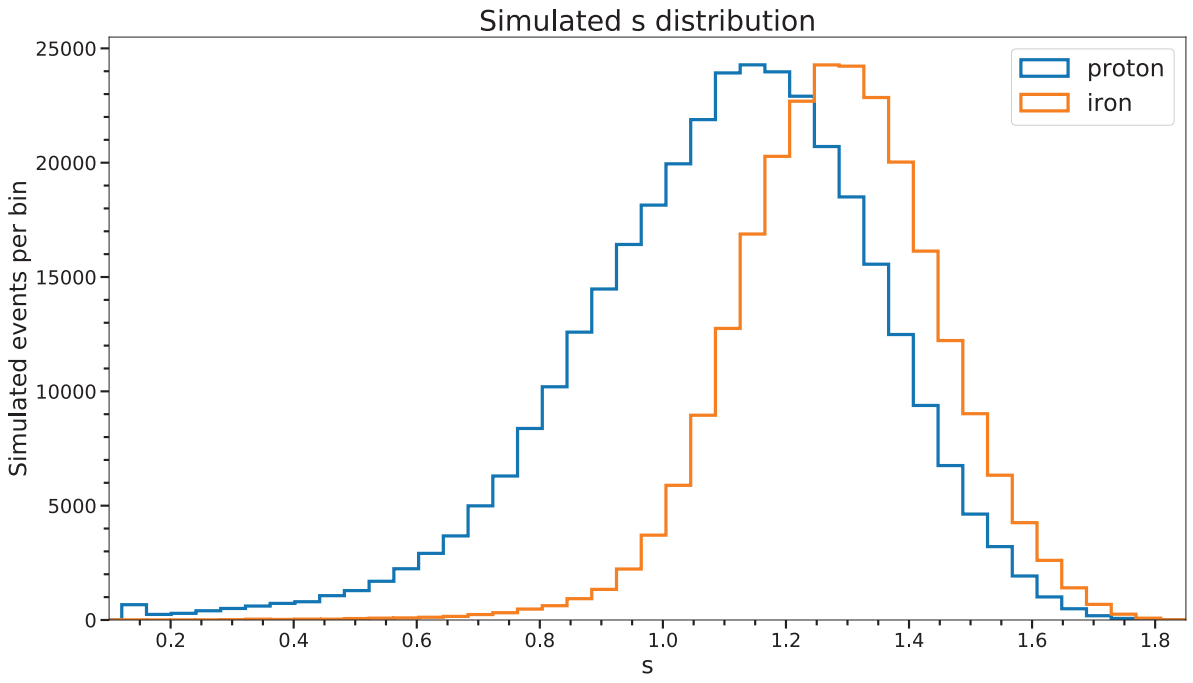

**Fig. 48.** (color online) Simulated age parameter distribution for primary protons and iron nuclei.

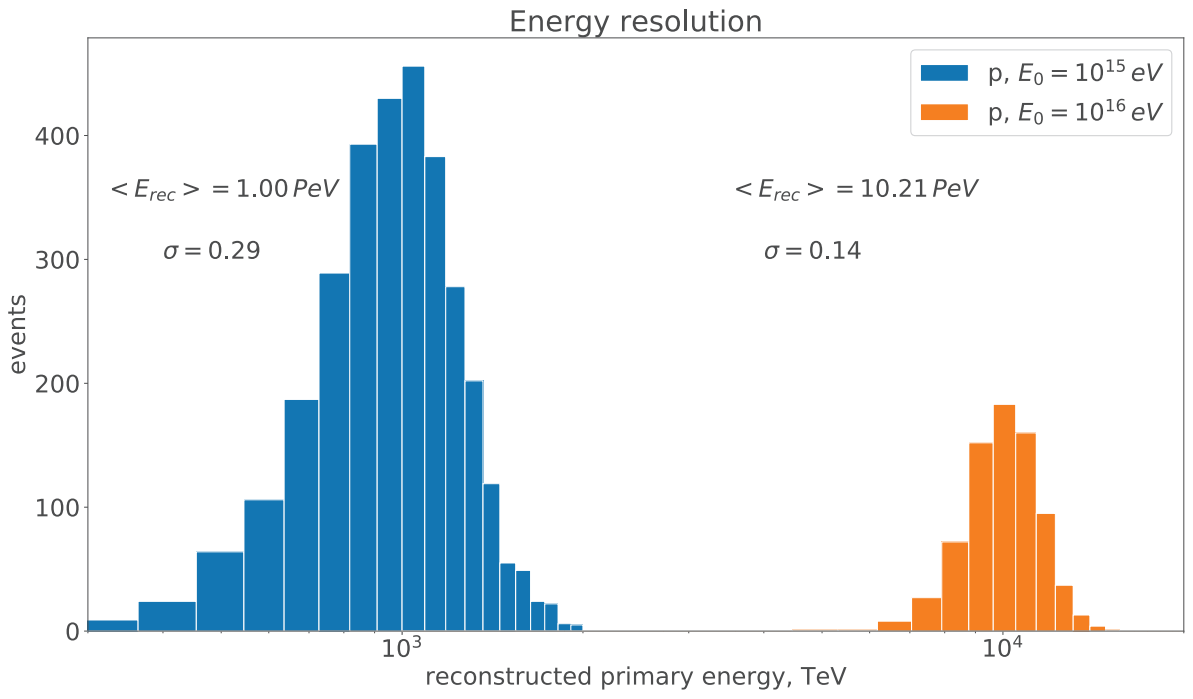

**Fig. 49.** (color online) Simulated primary energy resolution for primary protons with fixed energies of $10^{15}$ and $10^{16}$ eV.

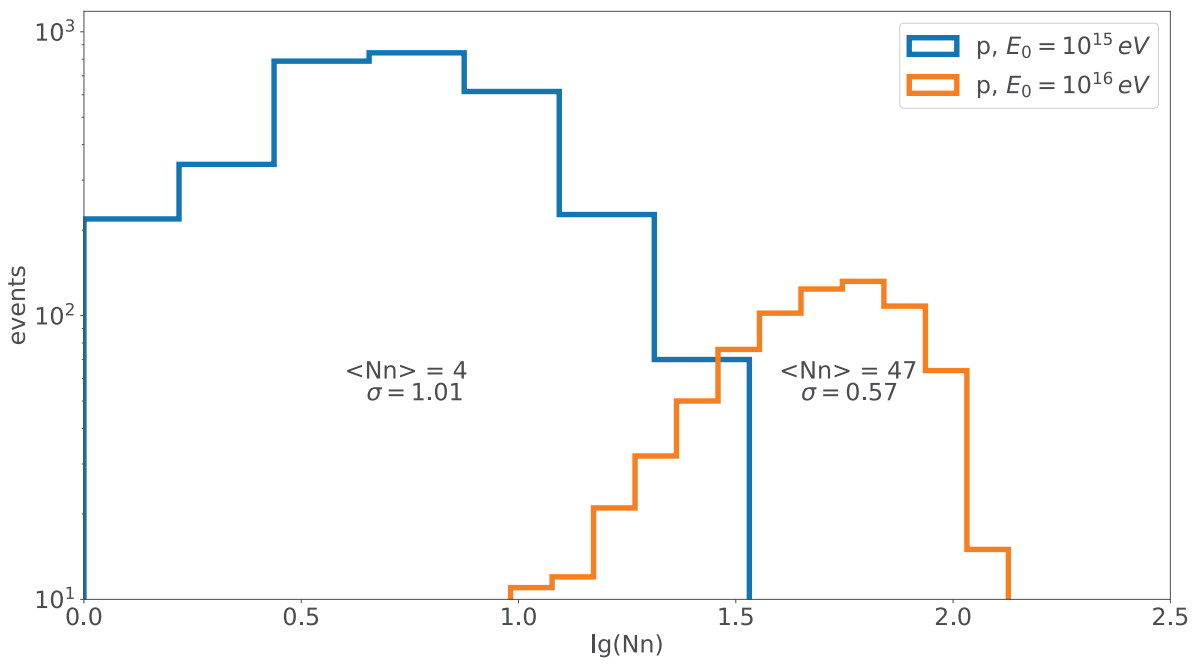

**Fig. 50.** (color online) Simulated neutron number distribution for primary protons with fixed energies of $10^{15}$ and $10^{16}$ eV).

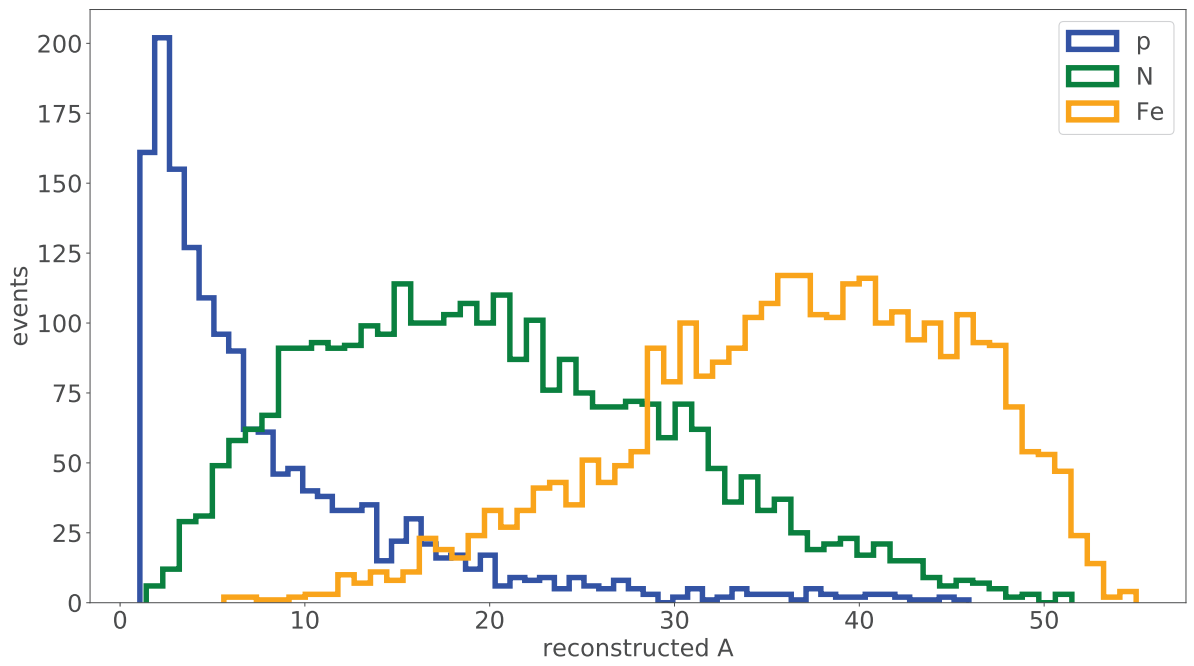

**Fig. 51.** (color online) Distribution of reconstructed masses for primary protons, nitrogen and iron nuclei with energies above 1 PeV.

## V. CLOCK SYNCHRONIZATION NETWORK WITH THE WHITE RABBIT PROTOCOL

### A. Timing system requirement for LHAASO project

In LHAASO, cosmic ray arrival direction is reconstructed using the secondary particles in the air shower front by measuring their arrival time difference among KM2A. To guarantee the angular resolution of reconstruction, a 500 ps RMS overall synchronization precision must be achieved among the thousands of wildly spread detectors.

Due to the large quantity of detector nodes, a combined timing and data link could be more cost effective than the conventional solution of using dedicated clock link and data link, which also greatly simplifies the cable deployments.

Since the most part of the timing network is installed and worked in the unmanned field at very high altitude with simple mechanical housing, strong requirements are also raised for covering wide temperature range, reliability, easy of replacement and prevention from dust.

### B. Introduction to WR technology

WR (White Rabbit) is a technology originally proposed by CERN and GSI and developed by tens of institutes and universities from all over the world. It aims to provide sub nano-second synchronization accuracy and pico-seconds precision over few thousands nodes in the distance of up to 20 km distance. It is compatible with Ethernet standard 1000Base-Lx with additional enhancements which include:

● Sync-E: With the CDR (clock and data recover) circuit that widely used on high-speed serial communication links, the carrier frequency is transferred between the master and slave that a common frequency can be syntonized among the whole network.

● IEEE 1588 (also known as PTPv2) that can syn-chronize the time-counter between the master and slave by exchanging IP packets that contain their local time-counter values. The initial offset and the link delay caused by the physical link can be measured and corrected.

● DDMTD (Digital Dual Mixer Time Difference) that measures the phase difference of two frequencies and then improves the time counter values of the PTP from cycle ticks which is normally 8 nano-seconds to sub pico-second. With the PTP protocol, the time difference from master to slave are measured and calculated to pico-seconds precision, the integer cycle tick are easily corrected by adjusting the slave's local count value and the fractional part are further compensated by tuning the phase of a VCXO (voltage controlled oscillator) output.

To avoid the bi-directional delay asymmetry caused by physical link length difference, WR uses single fiber with wave multiplex for upstream and downstream. For certain types of fiber, the speed difference of two wavelengths can be calibrated.

WR has the advantage of using one single fiber for both Ethernet data transmission up to 1Gbps and sub-ns synchronization, that makes it an ideal solution for LHAASO timing system. The scalability, long distance features are also of importance.

### C. The temperature effect of WR link and correction method

The LHAASO timing system will be running in a very wide temperature range and in particular large day-night temperature variation that could lead to the electronics delay variation and deteriorate the synchronization. The temperature effect is much more obvious and important for cosmic detection experiment such as LHAASO than the originally proposed WR deployment location inside the underground tunnels such as CERN and GSI.

The temperature effect of WR links comes from two parts: the fixed delay from the electronics component and PCB traces on WR nodes and the asymmetric factor defined by the speed difference of the two optical wavelengths. The influence of temperature to WR synchronization are tested with the setup shown below in Fig. 52.

WR nodes and fiber boundle are placed in climate chamber while the synchronization performance are monitored by oscilloscope. Test setups (a), (b) and (c) are designed to measure the temperature effect of fixed delay for WR node, the fiber asymmetric factor and the SFP modules respectively.

The results in Fig. 53. show that

a) the node has an obvious temperature effect which

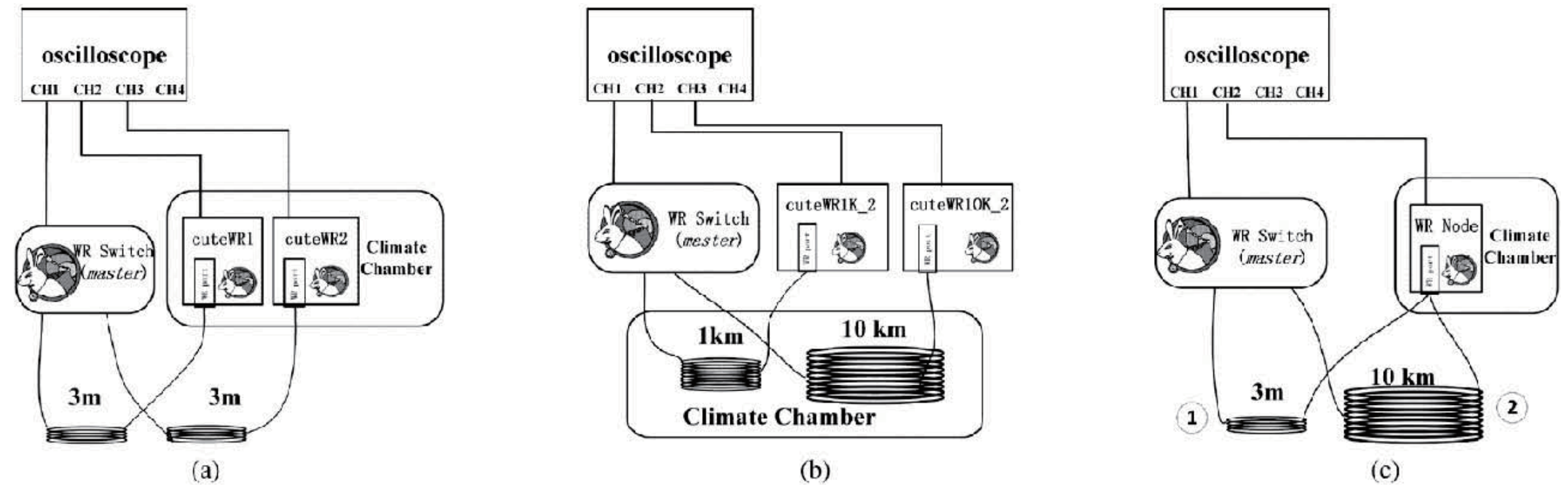

**Fig. 52.** Temperature test of WR devices and links.

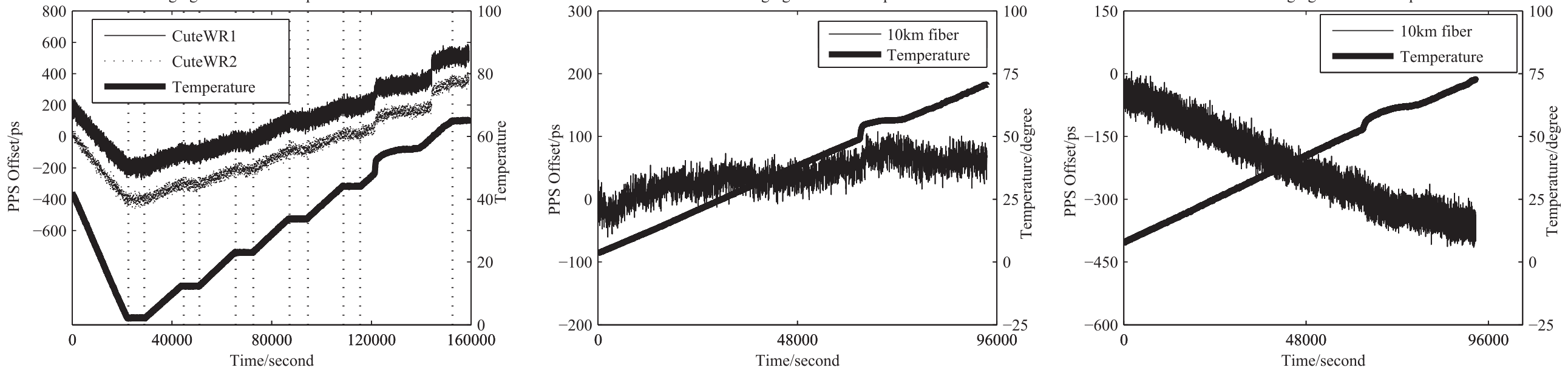

**Fig. 53.** Results of WR Temperature test [37].

is quite linear and can be easily corrected once the temperature is measured;

b) the optical fiber is observed to have a slight dependence of temperature. It is caused by the variation of asymmetric factor that is noticeable only if the fiber is as long as tens of kilometers.

c) the optical wavelength spectrum of the laser used inside the SFP modules shifts slightly with temperature, which is also only noticeable for long fibers.

Since the longest fiber length used in LHAASO is about 500 meters, the contribution from factor b) and c) are negligible. A simple linear temperature correction model is built and applied to the deployed system. The synchronization precision is improved from 700 ps to 120 ps even if the ambient temperature varied over a big range of 65 °C.

### D. The WR components designed for LHAASO

WR network consists of WR switches and WR nodes. In LHAASO experiment the WR node circuit and functionality is integrated inside the detector electronics.

The WR switch (WRS) is the key component of WR network, that has 18 WR ports to support cascade topology. In the cascaded network, thousands of WR devices can be linked. The WR switch is developed by the WR collaboration and produced by commercial companies. LHAASO will need about 550 WRS, most of them will be installed in electronics huts that are distributed over a field of 1.3 $km^2$ without insulation.

The standard WRS type has few flaws that prevent it from direct deployment in LHAASO:

1) the heat dissipation relies heavily on the two fans on the back panel. The dusty environment of LHAASO reduces the lifetime of fans to only a few months. Without fan, the temperature of key components, especially the FPGA kernel, will increase dramatically and may exceed its allowed range and permanent damage may happen.

2) the management port of the switch is RJ45 form that requires a copper Ethernet cable. Since LHAASO needs 550 WRS distributed in a square kilometer range, a dedicated Ethernet cable network with different level of Ethernet repeater/switch to extend the cable connection distance is very expensive and complicated.

3) the panel connectors are not suitable for dusty environment.

A special type of WRS has been developed for LHAASO which overcomes or improves the standard type accordingly, as shown in Fig. 54.

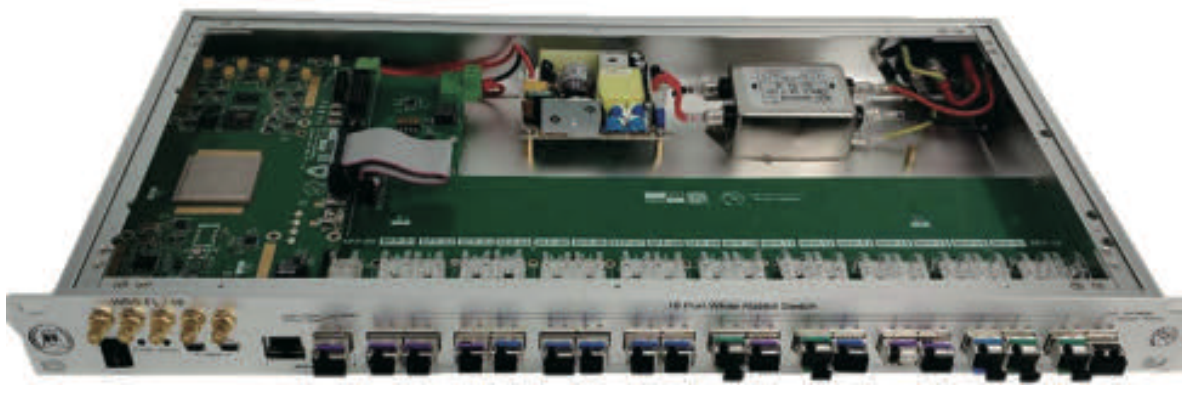

**Fig. 54.** (color online) Fan-less WR switch.

The heat dissipation has been greatly improved: with a CNC customized heat sink that covers the whole PCB attached to the mechanical upper cover via a Thermal conductive rubber gasket, the heat generated from the components are conducted to the heat sink and then to the mechanical box.

A special USB-to-Ethernet converter is used to provide another optical management port besides the existing RJ45 one, thus a single multi-core optical cable can be used for each switch with one dedicated fiber for WR link and one dedicated fiber for WRS management.

Besides, all the electronics components are replaced with industrial grade for wider temp. range, the front panel connectors are replaced with dusty proven types.

The mechanical and installation structure and internal wiring are also optimized, a mass production and QA procedure has been established.

### E. The WR network deployment for LHAASO

About 550 Level-3 PDC (power distribution cabinet) are widely located in the experiment field. Each PDC provides 220V AC power cable and optical fiber to maximum 15 KM2A detector units. One WRS is installed inside each L3 PDC, an optical cable with 4 fibers links the PDC to 1#PDR (power distribution room) where all those 550 WRS are further gathered and managed. To arrange the large amount of ingress optical cables, a dedicated ODF (optical distribution frame) rack provides optical inter-connect patch panels that separate the external optical cable and internal fibers. The devices in the 1#PDR includes 38 L2 and L1 WRS, GPS, Rubidum, WRS management switches, management server and few auxiliary switches, which are organized in 5 racks. The inter-connections among those racks are shown in Fig. 55.

The WRS installed in the L3 PDC and the racks in the 1#PDR are shown respectively in Fig. 56.

## VI. THE DATA ACQUISITION SYSTEM AND TRIGGER SYSTEM

The data acquisition system (DAQ) resides between electronics system and off-line analysis system. It is mainly a software system running on computing nodes, responsible for data collection from electronics, software trigger, and event storage. KM2A and WCDA do not have event trigger at the on-line electronics level, and by Front-End Electronics(FEE) signals are digitalized and packed as "hit packet" sent to DAQ. Each telescope of WFCTA has a hardware trigger built in each camera so that data are packed as event sent to DAQ. All Electronics systems communicate with DAQ through a fiber network. Part of the network also used to synchronize all far end nodes attached to the detectors by running White Rabbit Protocol over the entire network.

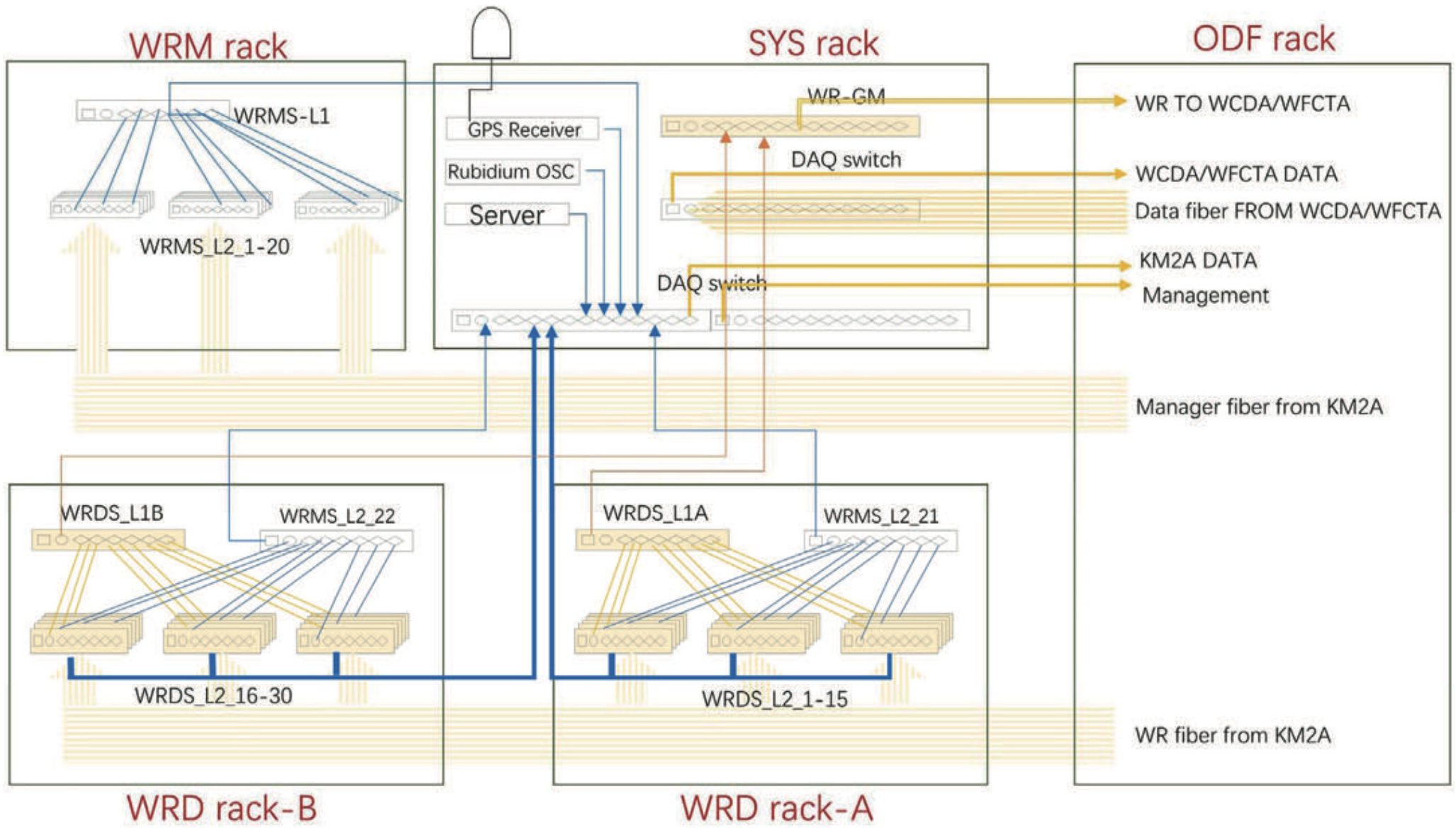

**Fig. 55.** (color online) Connections among the racks in 1#PDR.

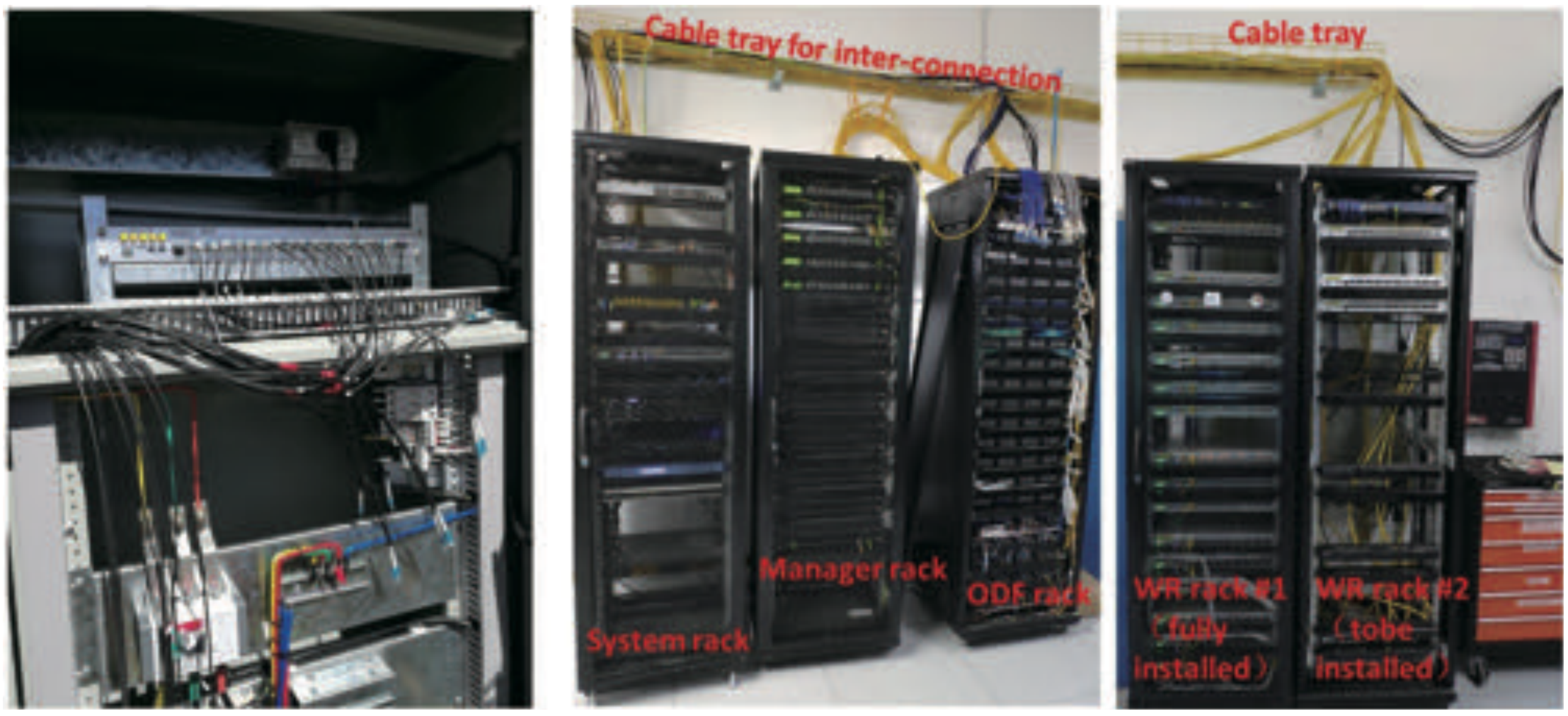

**Fig. 56.** (color online) WRS installation in 1#PDR.

### A. The data acquisition system

In LHAASO the readout electronics and the DAQ system are connected by a standard Gigabit Ethernet system based on White Rabbit switches, communicating with TCP/IP protocol. The Ethernet system can buffer more than 10 seconds of data. The DAQ system is required to collect all these data from readout electronics, perform software trigger and store events to disks. The data readout rate of LHAASO is 5 GB/s, and the data storage rate is 300 MB/s.

The hardware architecture of DAQ system is shown in Fig. 57. DAQ switches are deployed to the power distribution room located at the center of KM2A array and near WCDA. The Ethernet system connects to these DAQ switches through 1 Gbps electrical link. 10 Gbps fibre optical links are deployed between all DAQ switches.

Readout nodes are used to collect and buffer data from electronics. CPUs on both readout and compute nodes can be assigned to perform online processing tasks including software trigger. More than 1760 CPU cores will be assigned for the purpose of online data processing.

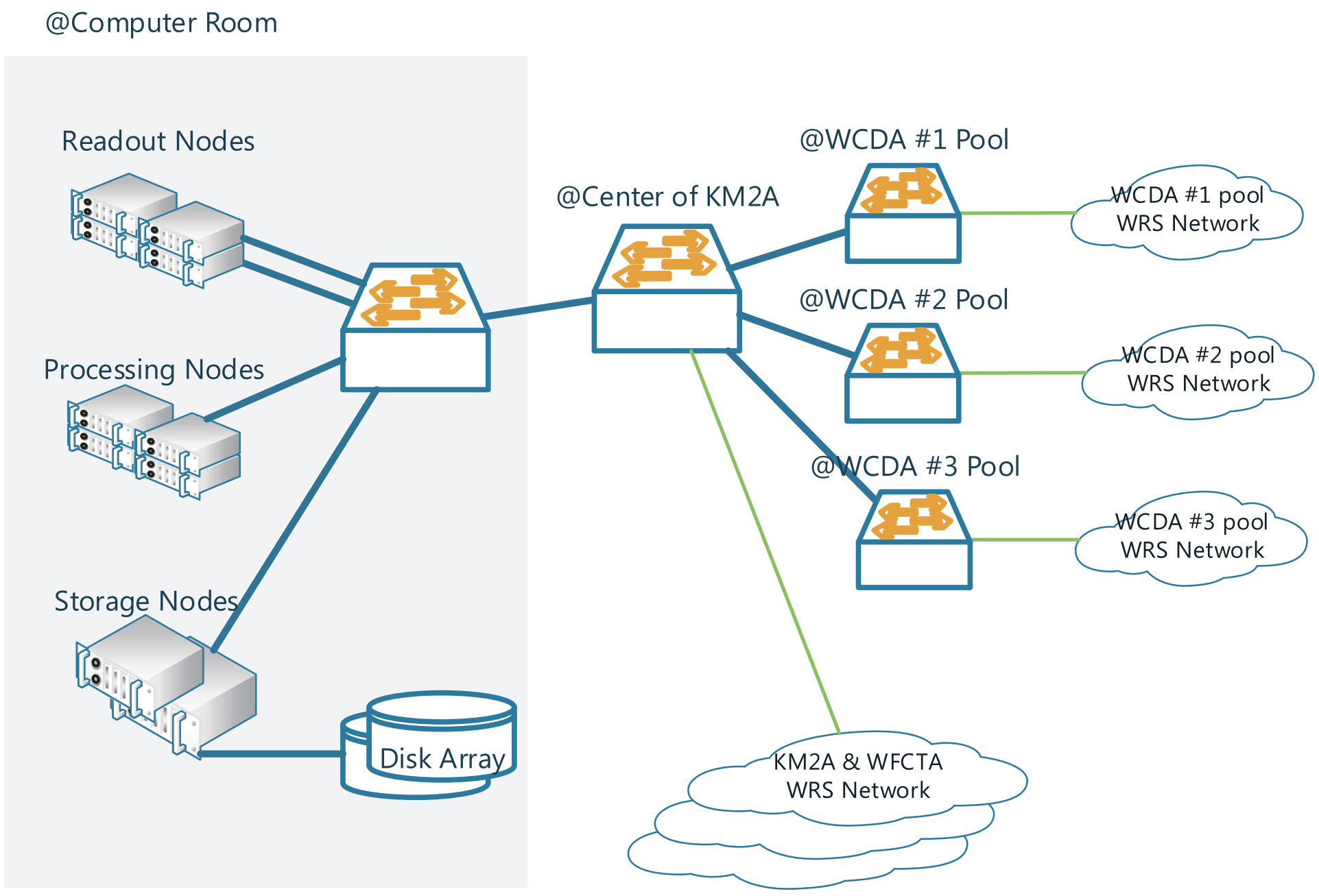

**Fig. 57.** (color online) DAQ system hardware.

Disk arrays are used for online storage. The capacity of online storage system will be 200 TB, that can buffer event data for more than a week before transferring to Beijing.

The DAQ software of LHAASO can be divided into two layers as shown in Fig. 58: online framework and data flow applications. The online framework is concerned with the common functions of a distributed DAQ software such as run control, process management and information sharing. The data flow applications are responsible for all the processing of physics data, receiving and transporting the data to storage.

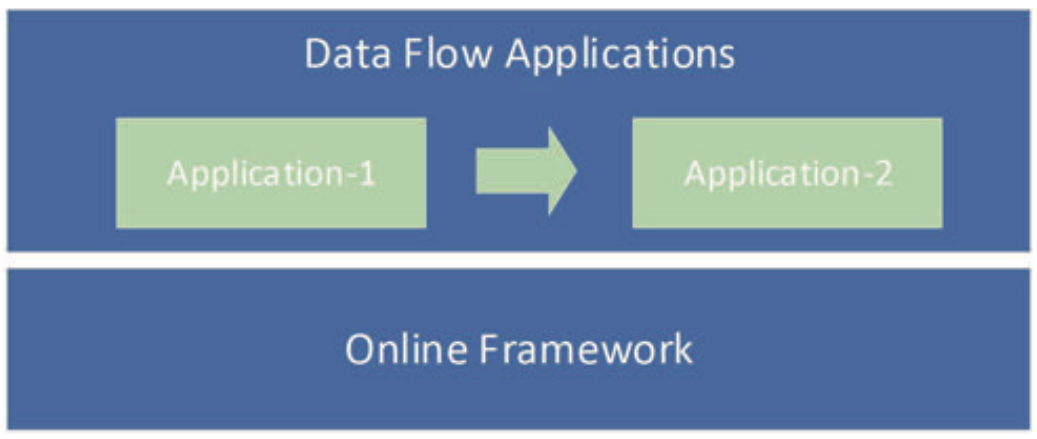

**Fig. 58.** (color online) Architecture of DAQ software.

● Online Framework

The online framework consists of services, interfaces and some scripts. The services (written in Python) running in the background is responsible for starting and stopping the data flow application processes, supervising status of applications and synchronizing some of the run parameters to a persistent database. The interfaces (written in C++) are compiled as libraries providing common functions such as information publishing and subscribing, configuration files accessing and so on.

Design of the online framework is mainly based on Redis, which is an in-memory database and can also be used as a message broker. Current status and parameters of DAQ system are all stored in Redis.

● Data Flow Applications

A message passing interface is implemented based on ZeroMQ library, used for the transmission of both control messages and data fragments between data flow applications. Based on the online framework and message passing interface, mainly five kinds of data flow applications are implemented, shown in Fig. 59.

With the global timing system, every data fragment from electronics is tagged by a time stamp with the precision of less than one nanosecond. To implement software trigger, data from all electronics of some detector array within a specific period of time need to be packed together as a data fragment, which is called a Full Timed Fragment (FTF). The FTF is fed as input for the trigger algorithm.

A readout application receives the stream of data from the electronics and packs them as Subsystem Timed Fragments (STFs). Then, under the coordination of the Data Flow Manager (DFM) application, each Builder can collect all STFs of readout applications and form a FTF. FTFs are transferred to Processors, where software trigger algorithm is performed. The DFM application is responsible for the load balance of Processors. After trigger, events are transferred to Event Store (ES) application for storage.

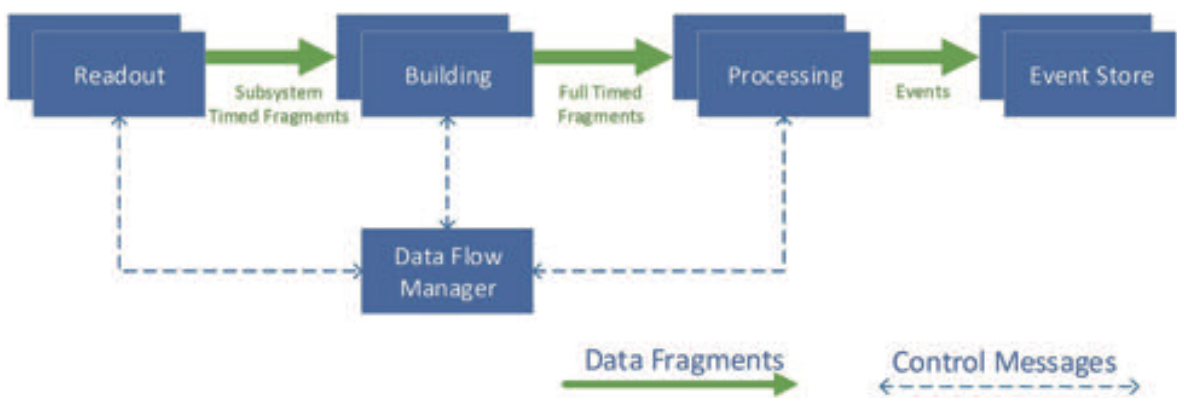

**Fig. 59.** (color online) Data flow applications of DAQ software.

### B. The trigger system

Design of the trigger system aims to remove amount of background and select real events which meet requirements of scientific motivations of LHAASO. All arrays of LHAASO have the following common characteristics:

● One detector transforms light signal into electronic pulse signal via photosensitive device (PMT or SiPM) so that type of signal is unitary and convenient to process.

● Each detector has level 0 (L0) trigger, that is, if the signal of one PMT is greater than a threshold, a hit is generated.

● By use of the clock synchronization network with the White Rabbit protocol, each detector or cluster records arrival time of one hit or one event with hardware trigger level 1 (L1) and transfers data by use of fibers.

The LHAASO trigger framework is presented in Fig. 60. LHAASO events are recorded in three modes: trigger mode (TM), low multiplicity mode (LM) and single particle mode (SM). About TM, KM2A and WCDA take method of "triggerless" front end digitization, that is, the electronics system does not include common hardware trigger system, but each front end electronics (FEE) generates L0 with independent threshold and uses the clock synchronization network to record L0 time. Charge and time information of hits is packed and transferred to the data acquisition system where the higher level triggers (L1 and L2) are performed and finally the selected events are stored into the data files. WFCTA and ENDA are separated into clusters and each cluster has arrival time of L1 recorded by the clock synchronization network. At the data acquisition system, L2 are performed and finally the selected events are stored into the data files. The modes are described in detail as follows:

1. WCDA-TM [38]:

WL0: PMT outputs a steady signal with a normalized amplitude lasting for 250 ns.

tWL1: All the PMT signals in every trigger cluster are counted and summed in the falling edge of a pipeline clock of 25 ns. In any trigger cluster number of fired detectors $\geq 11$ in a time window of 2,000 ns around the trigger time are stored into the event.

tWL2: Pre-reconstruction method with fast searching algorithm are performed to remove noise. Once the trigger formed, all hits of WCDA++ and KM2A in a coincidence time window are collected together with the trigger time and the event so called "tWEvent" is recorded into the data file.

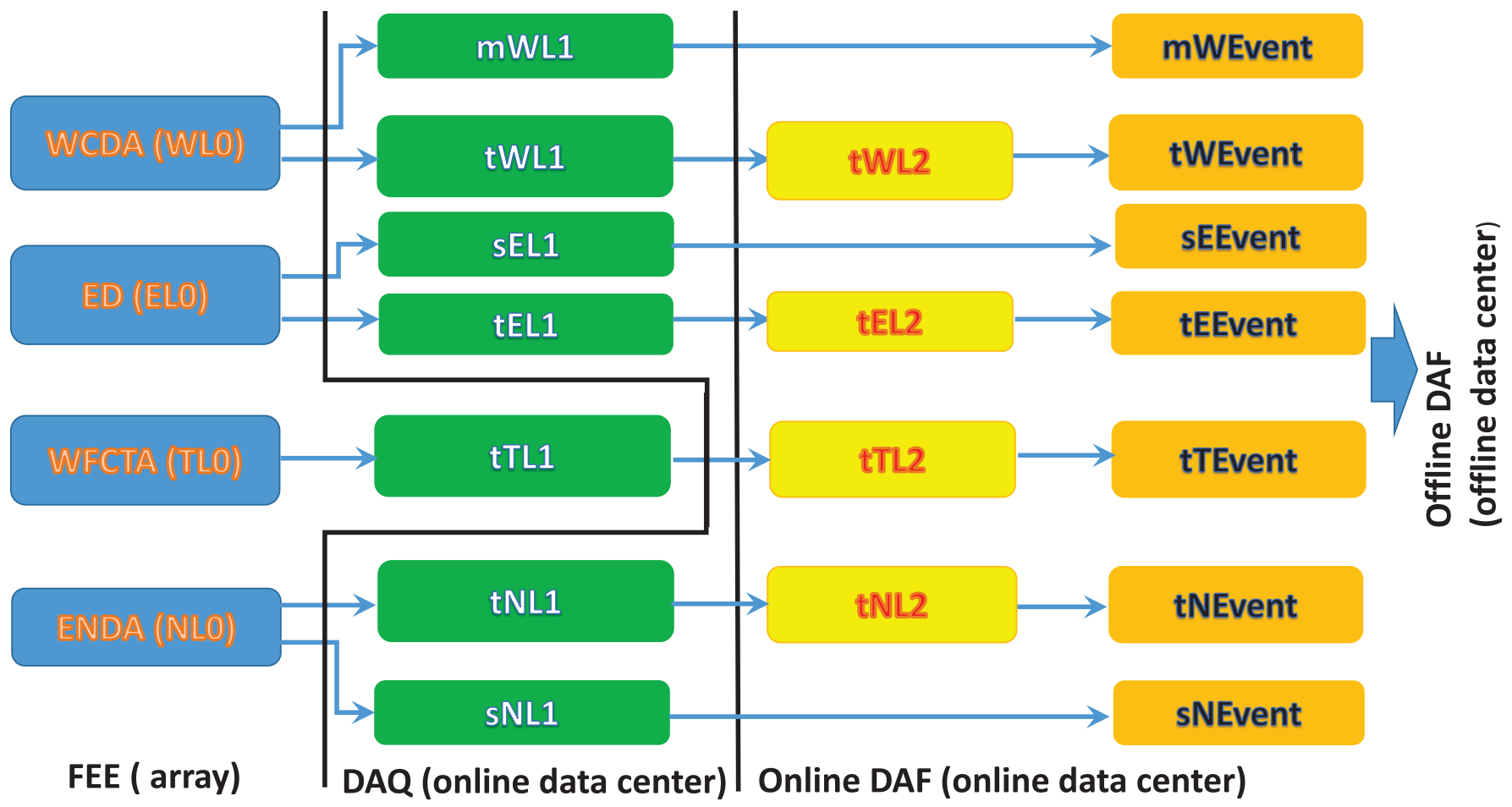

**Fig. 60.** (color online) LHAASO trigger framework.

2. WCDA-LM:

mWL1: Using GRB alert from the Gamma-ray Coordinates Network (GCN) system as a trigger, WCDA makes a follow-up searching for the GRB signals. When a GCN alert arrives, the DAQ system should collect all raw data of WCDA electronics around the alerting time point. The time window of this GRB data is about one to two hours and can be configured. Using the GRB location as their arrival direction, shower events are reconstructed. The method can be used to reduce noise for low multiplicity events to improve significance of the GRB signals. Corresponding event rate as "mWEvent" is recorded into the data file.

3. KM2A-TM [39]:

EL0: Signal of one PMT generates a single channel trigger if it is greater than threshold.

tEL1: In trigger time window 300 ns, any ≥ 6 ED are fired. The hits in sampling time window 10 μs are recorded.

tEL2: A narrowed time window and space window in the vicinity of the shower front are taken into account to reduce background.

Once the trigger formed, all hits of MD, WCDA and WCDA++ are collected and the event so called "tEEvent" is recorded into the data file.

4. KM2A-SM:

sEL1: each ED records count rate for thunderstorm study. Event rate as "sEEvent" is recorded into the data file.

5. WFCTA-TM [40]:

TL0: Signal of one SiPM generates a single channel trigger if it is greater than threshold. It is calculated in FPGA on DBs using signal-to-noise ratio (SNR) algorithm.

tTL1: The telescope trigger in BDB is generated as pattern is recognized. The single channel trigger A pattern recognition technique is used in the second level trigger. Round patterns (e.g., one pixel surrounded by others) are for Cherenkov events and line patterns (e.g., a straight line formed by 4 aligned pixels) are for the fluorescence.

tTL2: The event trigger in the entire array of telescopes are selected. One of the telescopes can trigger the neighbor telescopes. The adjacent telescopes trigger each other through off-line matching which raises the threshold slightly higher.

Once the trigger formed, all hits of WCDA, WCDA++, KM2A and ENDA are selected and the event so called "tTEvent" is recorded into the data file.

6. ENDA-TM [34]:

NL0: Signal of one PMT generates a single channel trigger if it is greater than threshold.

tNL1: Any 1 out of 16 detectors in one cluster starts FADCs.

tNL2: Three flags for event classification:

(1) M1, in a trigger time window of 1 μs, if at least 2 detectors generating the first level trigger;

(2) M2, if the delivered total charge corresponds to more than 250 particles;

(3) M3, if the total number of recorded neutrons is higher than or equal to 3.

During offline data analysis, the coincident events with the other detectors are searched for and the event so called "tNEvent" is recorded into the data file.

7. ENDA-SM:

sNL1: each detector records neutrons count rate and charged particle count rate per 1 minute as "sNEvent".

In brief, the different trigger modes include different contributions from the arrays, and collect different measurement variables from the arrays (Table 7). It leads that the different trigger modes have very different event rates and data rates (Table 8). Consequently, one physical mo-tivation may be correlated with several event types, or one event type may contribute to several physical motivations (Fig. 61). Moreover, the trigger modes on the basis of software control in DAQ are so flexible that, if any new physical motivation is proposed, the present trigger modes can be adjusted conveniently, and even a new trigger mode can be supplemented into the online software.

**Table 7.** LHAASO trigger table. Cross: the detectors take the trigger and the hits of the detectors are read out. Circle: Only the hits of the detectors are read out under the trigger.

| | WC-DA | WCD-A++ | ED | MD | WFC-TA | EN-DA |
|---|---|---|---|---|---|---|
| mWL1 | × | | | | | |
| tWL2 | × | ○ | ○ | ○ | | |
| sEL1 | | | × | | | |
| tEL2 | ○ | ○ | × | ○ | | |
| tTL2 | ○ | ○ | ○ | ○ | × | ○ |
| tNL2 | ○ | ○ | ○ | ○ | ○ | × |
| sNL1 | | | | | | × |

**Table 8.** LHAASO trigger event rate and data rate.

| | count rate of one detector (kHz) | data rate after trigger (MB/s) |
|---|---|---|
| WCDA | 63 | 430 (L1) 50 (L2) |
| KM2A | 1.5 | 10 (L1) |
| WFCTA | <0.2 | 15 (L1) |
| ENDA | $\sim 1 \times 10^{-3}$ | <1(L1) |

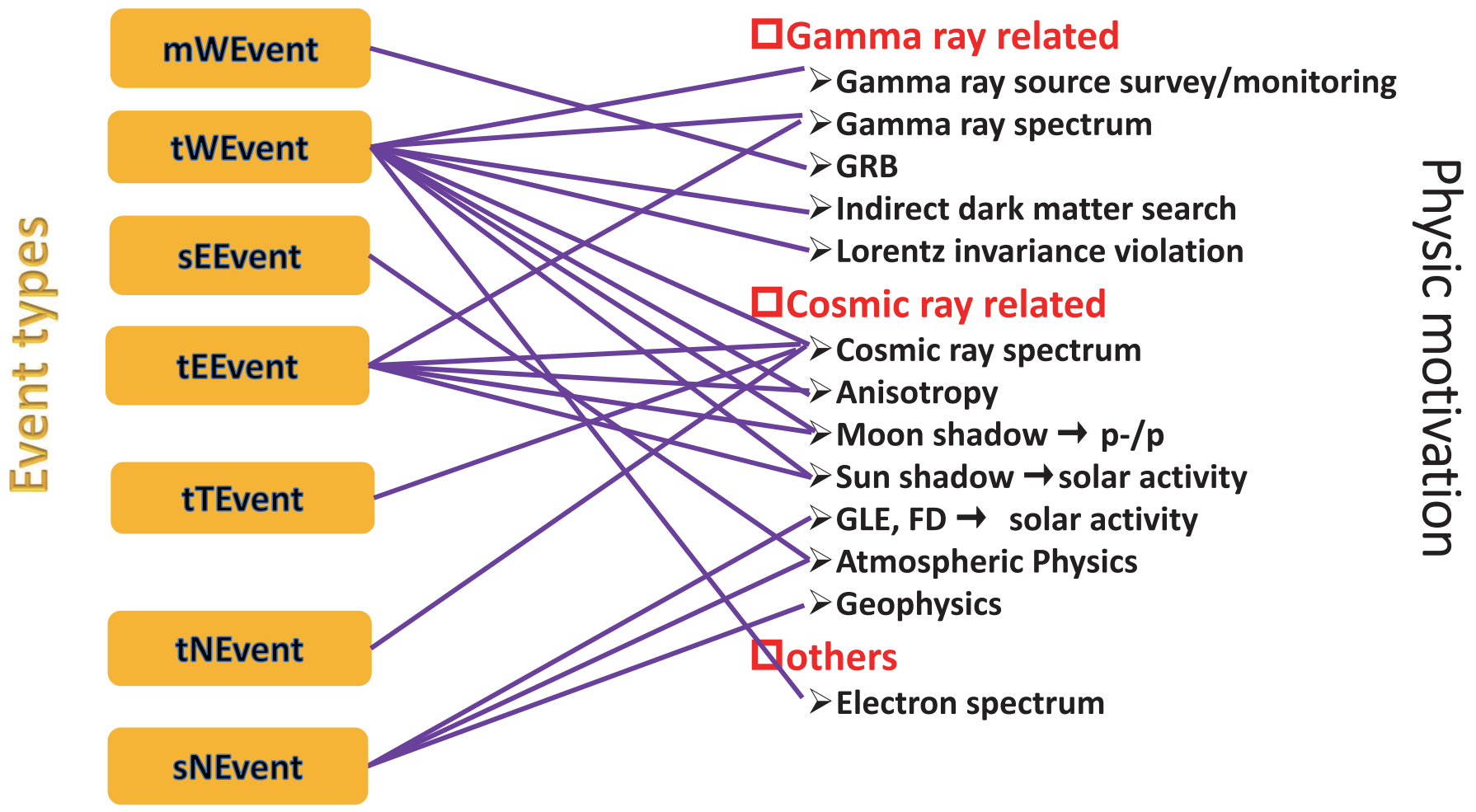

**Fig. 61.** (color online) LHAASO event types and physical motivation.

## VII. THE OFFLINE SOFTWARE FRAMEWORK (LodeStar)

LHAASO Offline Data Processing Software Framework (LodeStar) is developed for the LHAASO offline data processing, including Monte Carlo (MC) data production, experimental data processing as well as physics analysis.The typical data processing is to loop over all events with some calculations applied on the event data. The usual procedure for one event during event looping can be divided into three steps: read in an event, process the event, and write out the results. LodeStar consists of an underlying framework and experiment specific applications such as event generator, detector simulation, electronics simulation, calibration, reconstruction and physics analysis tools.

The framework plays very important roles for the performance and functionalities of LodeStar. It provides most of the common functionalities for the offline data processing, such as data management, data processing controlling, data input/output, common services and tools, as well as friendly user interfaces. The framework also builds the unified software environments and computing platform for the developers, and defines the standard interfaces between different applications. Therefore many developers can synchronously develop their data processing applications. All applications developed with standard interfaces can be smoothly plugged into the framework.

### A. Architecture

The architecture of LodeStar is shown in Fig. 62. LodeStar consists of three projects: External Interface (EI), SNiPER [41] and Offline.

● EI

The EI project is to build the interfaces for external libraries, such as Corsika, Geant4, ROOT, Python, etc. Therefore SNiPER and Offline can easily use external libraries via the interfaces defined in the EI project.

● SNiPER

SNiPER (Software for Non-collider Physics ExpeRiment) is the underlying framework of LodeStar to provide the common functionalities, including data processing controlling, event data management, standard interfaces for different components and applications, and some common services. It is designed for non-collider physics experiments from scratch following the Object-Oriented technology, and developed with bi-language, C++ and Python. So far, It has been successfully used by the JUNO (Jiangmen Undergound Neutrino Observatory) as the foundation to develop its whole offline software system.

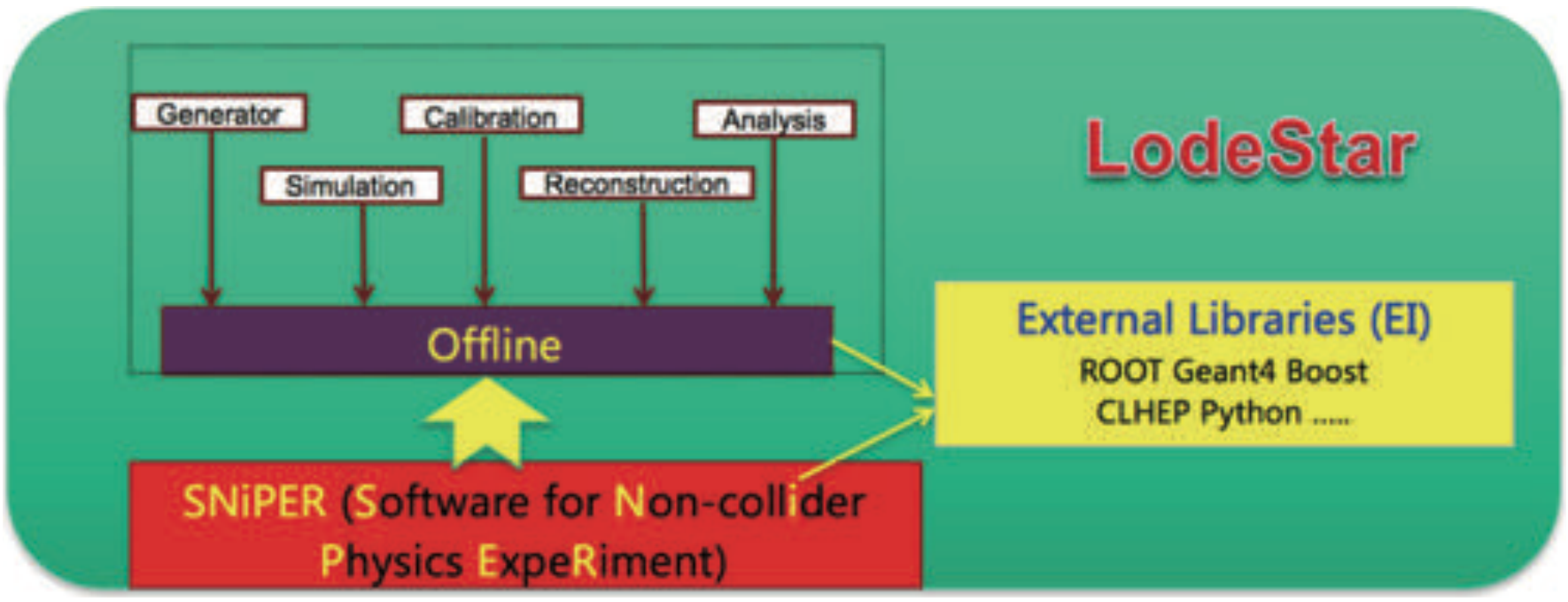

**Fig. 62.** (color online) The architecture of the LHAASO Offline Software System.

● Offline

Based on the interfaces and functionalities provided by SNiPER, the Offline project is developed according to LHAASO specific requirements. It contains two types of packages: one is to define the event data for different data processing stages, the other is for event data processing including data input/output, detector description, detector response simulation, digitization, calibration, reconstruction and physics analysis.

### B. Key components

From the users point of view, the key components of LodeStar are described as below.

● Algorithm

Algorithm is a basic unit for data processing and provides the interfaces for users and developers to develop their new specific algorithms which can be dynamically loaded and executed by LodeStar. Detector simulation, calibration and reconstruction are all implemented as Algorithms.

● Service

Service provides a specific functionality needed during event data processing. For example, users can write data into ROOT files via the RootWriter service, or access the database via the DatabaseSvc, or access to the detector geometry via the GeometrySvc. Services can be invoked by users' Algorithms or the framework.

● Tool

Tool is similar with Service, also a dynamically loadable element, but it belongs to an specific Algorithm and helps the Algorithm to organize the code more flexibly.

● Task

Task is a lightweight application manager, as shown in Fig. 63. It is a controller for the event looping and the entrance of a job. Task manages Algorithms, Services, and sub-tasks and controls the execution of all Algorithms belonging to it according the order of the Algorithms when they are created.

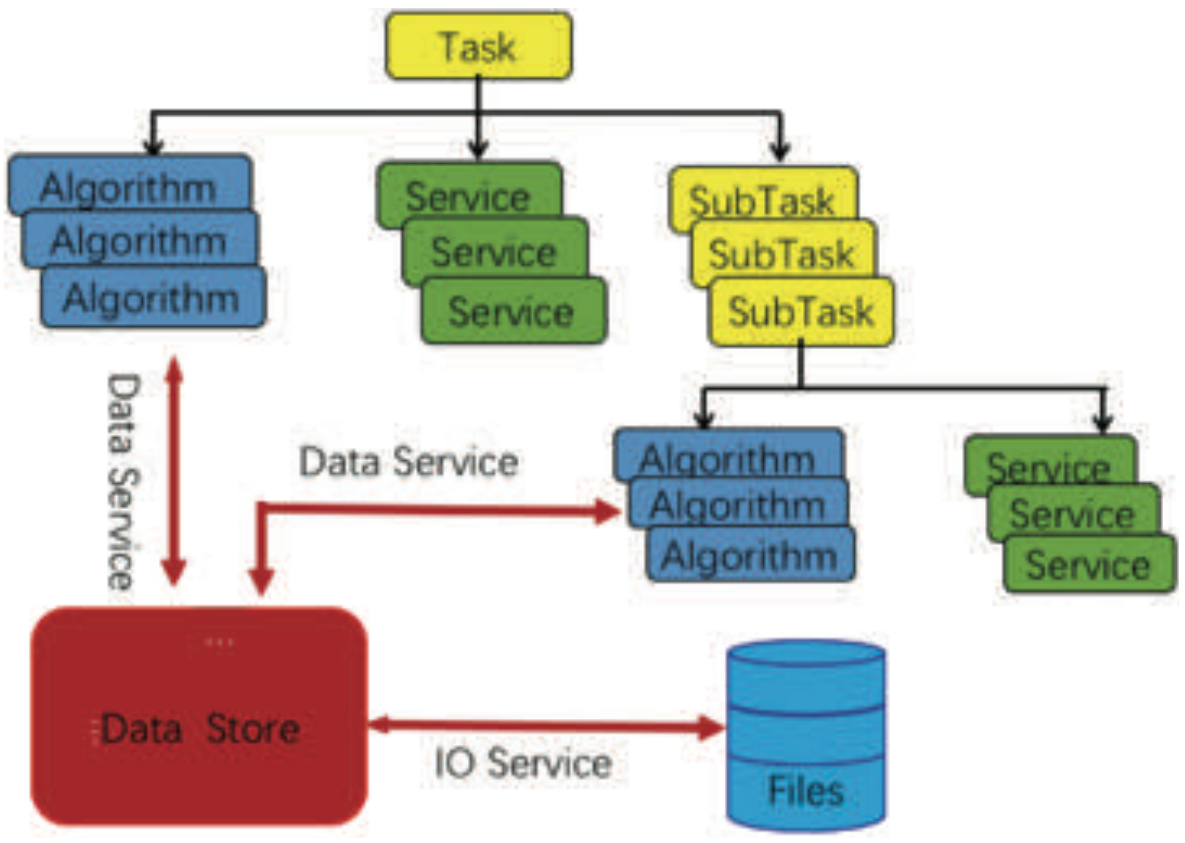

**Fig. 63.** (color online) Main Components of LodeStar.

● Event Data Model

Event data model (EDM) defines the event information at the different data processing stages and build the correlations between them. LHAASO EDM classes are implemented based on the ROOT TObject. Fig. 64 shows the schema of the LHAASO EDM. At each data processing stage, the event data consists of two parts: HeaderObject and EventObject. The HeaderObject only contains some "tag" information in order to speed up event selection, the EventObject holds all information of this event. SmartRef is deployed to build the relationship between HeaderObject and EventObject and provides the data lazy-loading functions.

● Data Store

Data Store is the place in memory holding the current event data in order to be easily accessed during the event processing. The event data in different processing stages for different sub-detectors can exist at the same time in the Data Store, and each type of data has one unique path. Users can use the unique path to retrieve the event in-formation and put the new event data back to the Data Store. All the event data in the Data Store can be read/written from/to data files by the data input/output service.

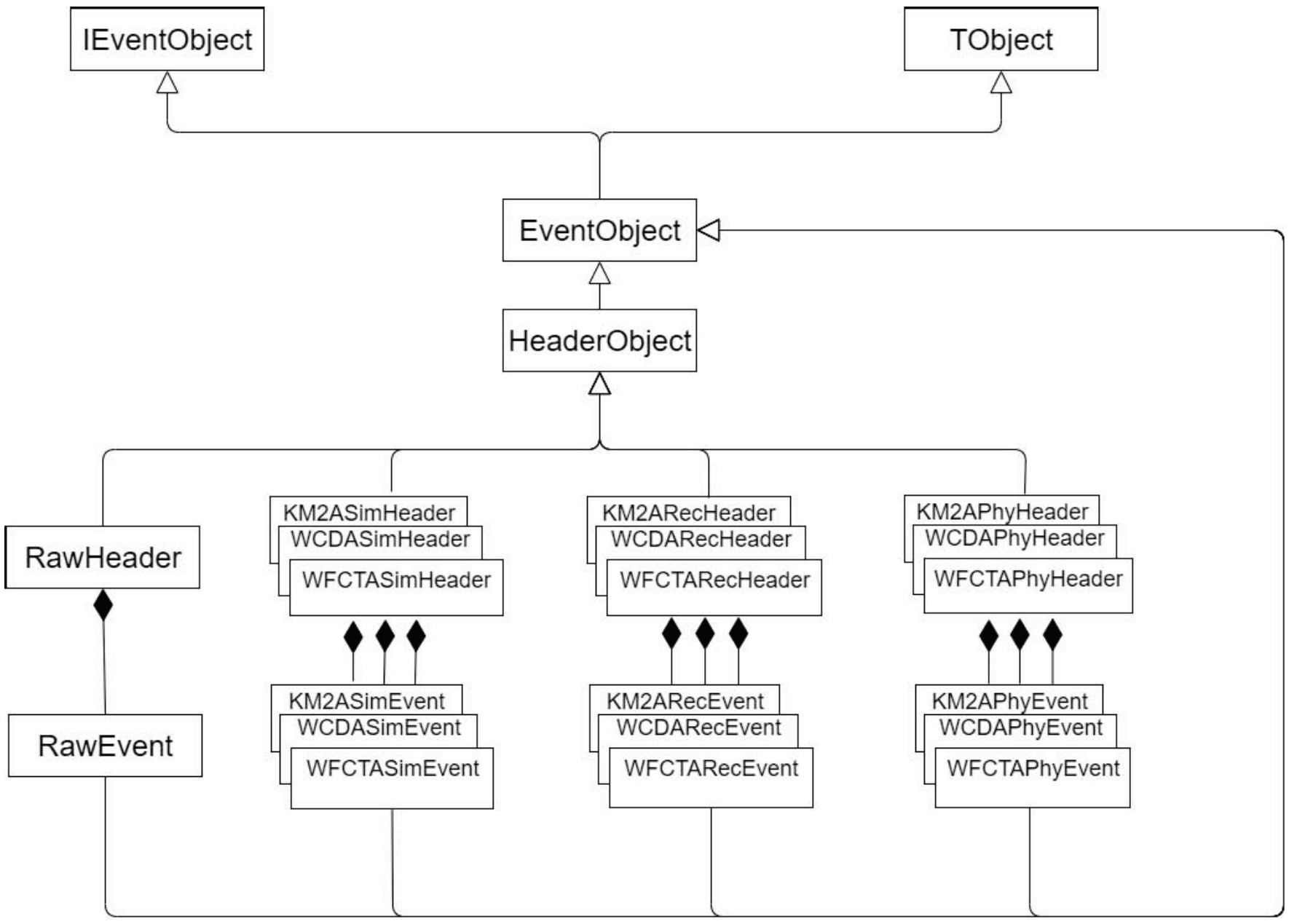

**Fig. 64.** Schema of LHAASO Event Data Model.

● Detector Simulation

LodeStar manages detector simulation with a Task as illustrated in Fig. 65. A dedicated Algorithm, DetSimAlg, is developed for simulation of all sub-detectors (including KM2A and WCDA). A dedicated service, G4Svc, provides the way to launch Geant4 within LodeStar, and a user-end service, DetSimFactory, is developed to set up and organize all the Geant4 related classes. User-end Tools, classified as AnalysisElement are provided for defining the user actions and the physics list for Geant4.

● Property

Property is a configurable variable. Users can customize its value in the job configuration to avoid hardcoding. All Algorithms, Services, Tools and Tasks can have their own Properties by declaring a variable as a Property in their constructors. Then all these Properties can be configured in the job configuration python file.

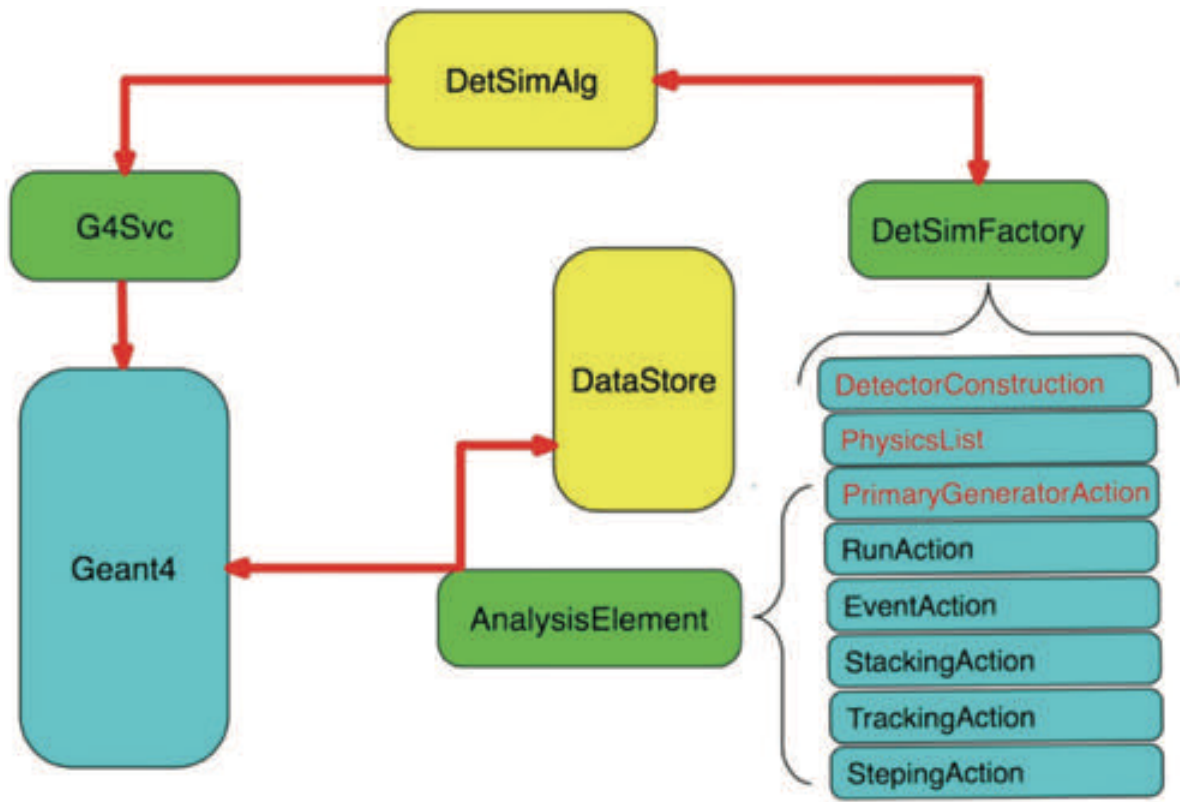

**Fig. 65.** (color online) Detector simulation system.

### C. Status and outlook

Now LodeStar switches from the development to the optimization. LodeStar's most important functionalities have been implemented as described above, and it is being released one or two versions every year. The detector simulations including KM2A, WCDA and WFCTA, have been integrated into LodeStar. The physics analysis framework has been also built up based on the LodeStar in order to speed up the whole LHAASO data processing.

## VIII. THE DATA ANALYSIS FACILITY

### A. Overview

The data analysis facility for LHAASO is designed to build a integrated platform including a large-scale data storage and computing system to support the running of analysis software and finally provides physicists with a uniform experiment data analysis environment. The relation between LHAASO data analysis facility and other subsystems of LHAASO is shown in Fig. 66.

The data analysis facility provides fundamental computing and network services, such as high performance computing (HPC), massive data storage, high-speed transmission network, distributed data sharing and ana-lysis system to support detector monitoring, data transmission, storage, decoding, calibration, reconstruction and analysis. The facility is composed of a small on-site data center (DC) at Daocheng, a large data center at Beijing, high-speed fiber network from Daocheng to Beijing, remote operation center (ROC) and IT Service platform and so forth.

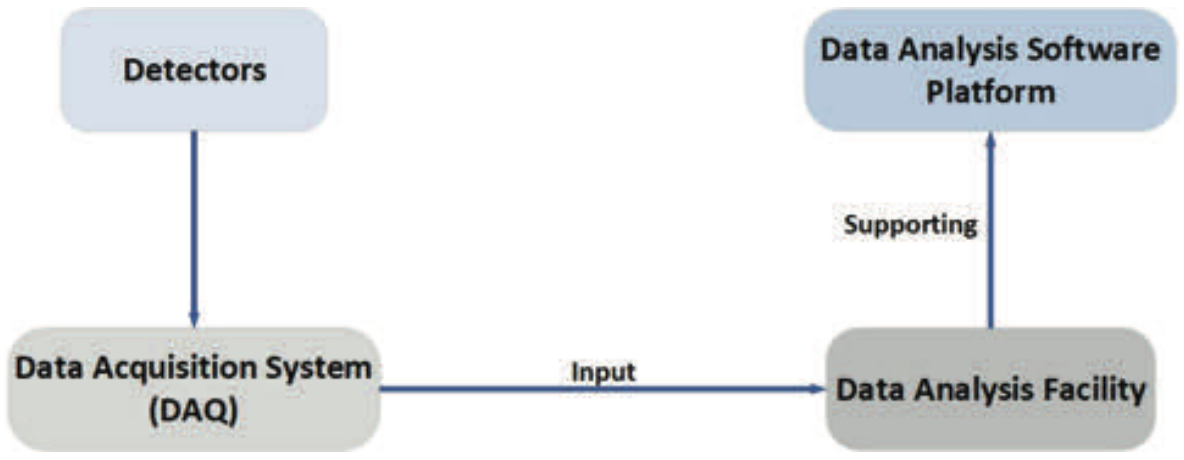

**Fig. 66.** (color online) The architecture of LHAASO data analysis facility.

The small on-site DC is designed to pre-process and compress data of WCDA from DAQ, and keep the processed data about one month, as well as equipped with a HPC cluster, a massive disk storage system (EOS), a job scheduling system (HTCondor), a WAN data transfer system (Spade) and a synthetic monitoring tool. Meanwhile the large DC at Beijing is designed for data storage and analysis, supporting physical simulation, data reconstruction, physics analysis and long-term data preservation, equipped with a large HPC cluster, a disk storage system, a large tape library system, a data backup system and so on.

LHAASO produces about 9 PB raw data annually, which is stored and processed on the data analysis facility. The components and current status of the data analysis facility are as follows:

- One large tape library equipped with 20 LTO7/LTO8 drivers and more than 25 PB tapes.

- One advanced tape storage system with the ability to manage 200PB tape capacity.

- One distributed disk storage system with the capacity of 25 PB and aggregate performance of several tens of GB per second.

- One high throughput computing system with more than fifteen thousand modern CPU cores deployed in Daocheng and Beijing data centers.

- a dedicated network link between LHAASO observatory and Beijing data center with the bandwidth over 2.4 Gps.

The data analysis facility consists of many subsystems, such as the front-end login system, the massive storage system, the tape library and the computing cluster, as shown in Fig. 67. We will discuss some subsystems in the following.

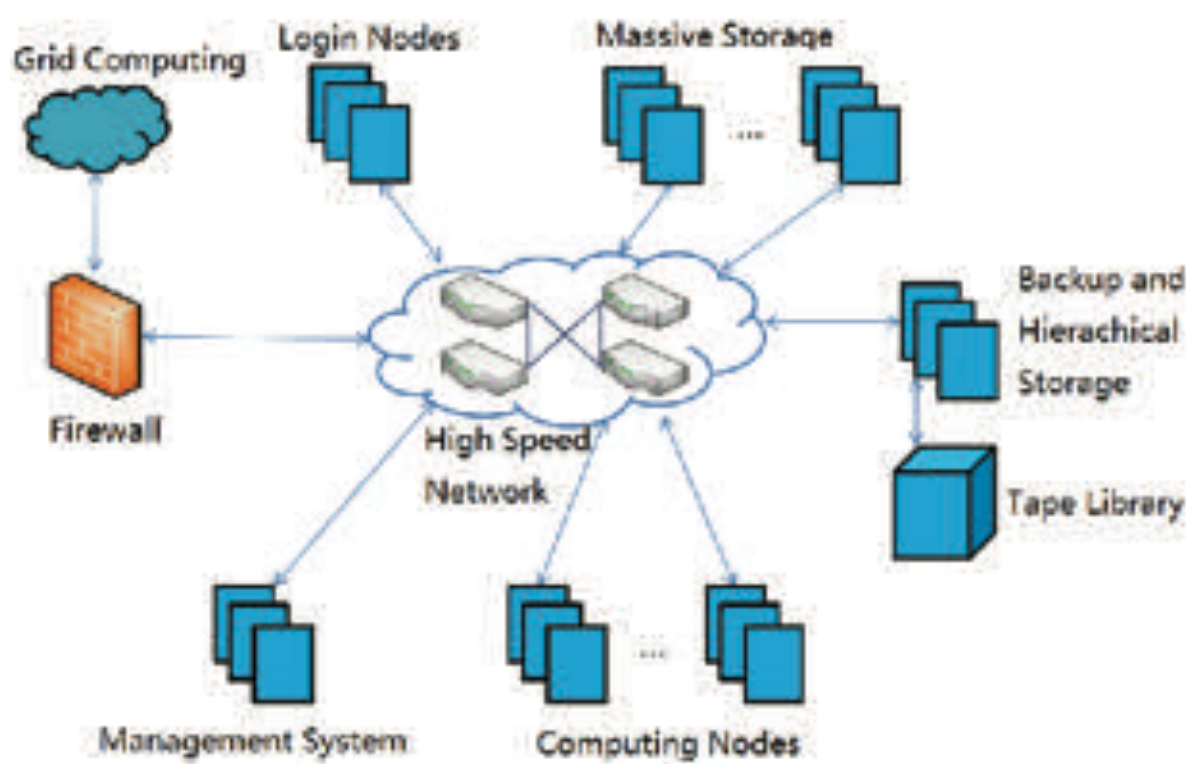

**Fig. 67.** (color online) The LHAASO data analysis facility.

### B. Front-end login system

The front-end login system provides access services to users directly, such as login service, HPC service, web service and massive storage service. It has a single entry point, connecting a dynamical scalable and load-balanced cluster based on DNS load-balancing, which ensure users log into the cluster and do further work from wherever they are.

### C. Account management system

There are a large number of domestic and foreign users in the LHAASO collaboration, and unified management for user information, authentication and management for users' HOME directories are necessary. For LHAASO, AFS filesystem is used for user directory management, and unified authentication combined with Kerberos. AFS is a open-source distributed filesystem, supporting many kinds of OS, and is highly secure via combination of authentication database and ACL.

### D. Computing cluster system

Computing cluster is widely used in high energy physics data analysis facility, and IHEP also adopt the technology. There are two computing clusters for LHAASO, a small one at Daocheng with about 5,000 CPU Cores for preliminary pre- processing and compressing data from DAQ, and a large one with up to 10,000 CPU Cores for offline data analysis for LHAASO, such as data simulation, reconstruction and analysis.

The job sc792 heduling system integrates CPU resources on a large mount of computing nodes, and manage jobs submitted by different users according to some scheduling policies. Common job scheduling system include PBS, HTCondor, LSF, SGE, etc.. HTCondor is an open source scheduling systems, supporting High

Throughput Computing (HTC) especially, with the ability of job controlling and scheduling, resource monitoring and management, and is widely used in high energy physics filed. One home-made job scheduling tool - hepjob is developed based on HTCondor to simplify the procedure of job submission, and has come into service in both computing clusters for LHAASO and other experiments.

### E. Virtual computing platform

In consideration of the difficulties of routine maintenance for the on-site data center at Daocheng Haizi Mount, we adopt the virtualization technology based on Openstack and Kubernetes to build the computing cluster at Daocheng. The virtual computing platform includes the virtual computing cluster, login cluster and management services. All login nodes and management servers are providing services in the format of containers, and the container cluster is managed by Kubernetes, and the system availability is highly increased by taking full advantage of the container's load-balancing and fault-tolerant mechanism.

The virtual computing cluster includes management and scheduling of virtual computing resources, automatic management for virtual machines, and well-defined interfaces for cluster scaling. HTCondor is used as the job scheduling system as well as for the physical cluster.

### F. Disk storage system - EOS

Generally, computing and data storage systems are separated in high energy physics, and the same with LHAASO. There are many kinds of distributed filesystem such as Lustre, EOS which we deploy as the storage filesystem 812 for LHAASO. EOS is a distributed filesystem developed by CERN, which is widely used for many experiments at CERN, such as CMS, ATLAS, ALICE, LHCb, and stores more than 3 billions of files with the capacity larger than 300 PB. Ths EOS system can meet the requirements for data storage and management for LHAASO very well. EOS consists of management service (MGM), message queue (MQ) and file storage service (FST). In general, the EOS is connected to the computing cluster via high-speed fiber network, the computing nodes on which EOS is mounted on via FUSE/FUSEX, and the analysis programs can access files on EOS as if on native filesystem, or just access files via xrootd directly.

Currently two EOS instances are deployed for LHAASO, one in the on-site DC at Daocheng to support data quick pre-processing, compressing and storage, while another one in the large DC at Beijing to provide storage service for data reconstruction, simulation and analysis. The small EOS at Daocheng currently store 57 millions of files with the capacity of 4.5 PB, and the large one at Beijing has a capacity of about 20 PB and stores up to 210 millions of files so far. Both will be expanded continuously in the future.

### G. Tape storage system - castor

LHAASO will produce about 9 PB/year raw data as mentioned above, and massive storage techniques are necessary to store data and ensure data safety. The tertiary storage model - "tape - disk - memory" is a universally adopted storage policy, especially for HEP. Castor is a hierarchical storage management system, developed by CERN, designed for file transmission between disk system and tape library, and has been put into service at IHEP for BESIII, YBJ, and DYB as well as LHAASO experiment.

The Castor system meets the requirements of data backup and retrieval for LHAASO because of the scalable architeture and we have deployed one Castor instance for LHAASO. Currently there are 20 LTO7 tape drives and more than 4200 LTO7 tapes with a capacity up to 25 PB. Data from Daocheng is firstly stored into the large EOS instance at Beijing, and then be migrated to the tape library for data security. User can easily retrieve files from tape library using some commands provided by Castor, while individual file backup to the tape library is restricted. And a new tape library system - CTA developed by CERN will replace Castor in the near future. CTA is modernly designed and cooperates with EOS tightly.

### H. Network system

Experiment data on the on-site DC at Daocheng is needed to be transferred to the large DC at Beijing, requiring high-speed network, stable transmission system and network security and safety. The network between on-site data center and IHEP is rented from telecom carriers with bandwidth up to 2.4 Gbps, quite enough for stable transmission of experiment data from Daocheng to Beijing, remote monitoring, and on-site working, etc. Massive data is transferred from Daocheng to Beijing and stored in EOS system then permanently into tape storage system, which requires a stable and effcient transmission system and the Spade transfer system is adopted for this task. The Spade system is widely used for data transmission, like DYB and JUNO, and has been adopted to transfer experiment data for LHAASO since 2018.

## IX. THE INFRASTRUCTURE OF THE HIGH ALTITUDE LABORATORY

### A. Laboratory sites

LHAASO has two sites in Daocheng, Sichuan province, China (see Fig. 68). One is in Mt. Haizi and the other is in the downtown of Daocheng. The distance between these two sites is about 50 km. The site in Mt.

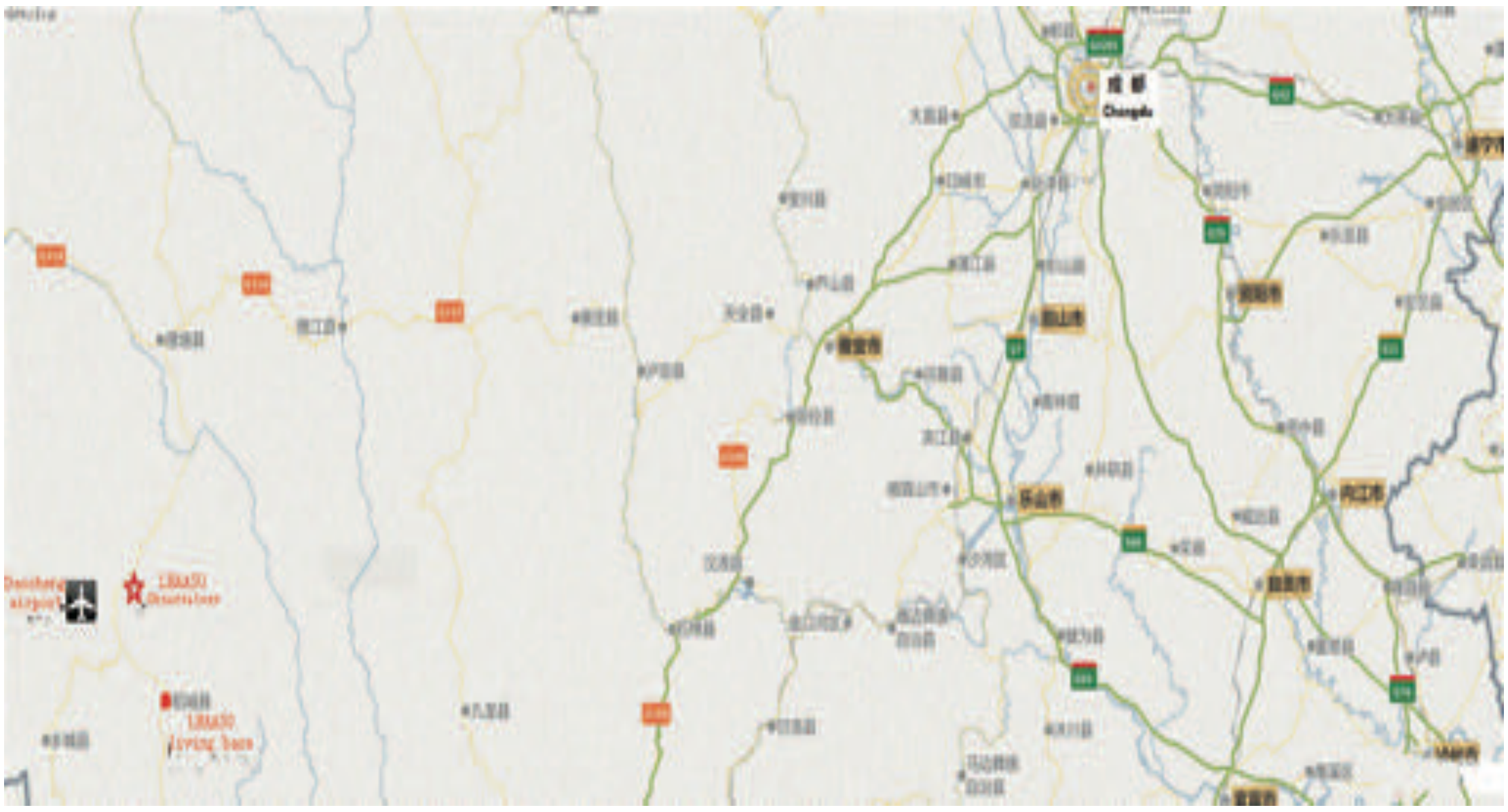

**Fig. 68.** (color online) The two sites of LHAASO (marked on the map from https://map.baidu.com).

Haizi at 4410 m a.s.l. is the observatory where all detectors are installed in arrays. The total area of the observatory is about 1.35 km$^2$. The observatory is about 10 km away from the Daocheng airport. The other site in the Daocheng downtown is the living base at 3750 m a.s.l. The total area of the living base is about 13,000 m$^2$.

### B. Facilities in the site

**1) The living base**

The total area of buildings is about 7000 m$^2$ distributed in 5 buildings in the living base (see Fig. 69). The four-floor building is the dormitory. There are 31 rooms with a capacity of 50 beds in the dormitory. All rooms are equipped with permanent heating system. The heat source is the hot spring water provided by a local company. To allow the guests good sleep in the night at the altitude of 3750 m a.s.l., extra oxygen is supplied in the rooms by a centralized oxygen generator in the living base. The equivalent altitude in the rooms could be reduced down to 2700 m a.s.l. according to the concentration of oxygen in the air. There are fifteen offices, a computer room and one exhibition room on the second floor of two other buildings in the living base. Meeting room, eight labs, kitchen, dining room, gym, utility houses and warehouses are all distributed in the first floor of all buildings.

**2) The observatory**

The bird view of the observatory is in Fig. 70. The area where the observatory is located was wild field before LHAASO started. By connecting to the 35 kV national power grid 30 km away, a dedicated transformer station nearby the observatory has been built. The capacity of the 35 kV transformer is 4000 kVA. Two 10 kV power cables are run from the transformer station to the two power distributing stations where the electric voltage will be further transformed to 380 V and 220 V. The No. 1 power distributing station is located beside the WCDA pools which are located at the central area of the observatory. It mainly powers the WCDA, WFCTA and most ED and MD. The No. 2 power distributing station is located in the assembling hall and provides electric power to the whole assembling hall, including the computer center, the water pumping station, ED and MD detectors nearby.

In the LHAASO project, the WCDA needs about 400 thousand tons of pure water to fill all three pools up and the MD needs about 50 thousand tons of ultrapure water to fill up all the 1171 tanks. The original water is pumped from a small lake which is connected the nature river about 300 m away. The capacity of the lake is about 15 thousand tons of original water. The water will be simply filtered by a filter bag to reduce dust in the pumping station before being pumped to the water purification factories. The water flux of 500 ton/hour, 160 ton/hour and 40 ton/hour can be supplied according to the requirement at different phases. The water pipe between the pumping station and the purification factories is about 1000 m long. The pipe is buried 1.7 m deep under the surface for prevention from freezing in winters. The record frozen depth of the local soil is about 1.7 m.

The total length of roads in the observatory is 21.9 km including 5.9 km paved. The width of all roads is 4 m. The roads look like a web on the ground which are designed to reach every MD detector in the array.

The assembling hall is a one-floor building with the size of 80 m × 24 m. The area reserved for the WFCTA telescope assembling is about 600 m$^2$ with a high roof of 15 m. The roof of the rest part of the building is 8.5 m high. The No.2 power distributing station and the computing center are located in the hall as well with their own roof at 5.5 m.

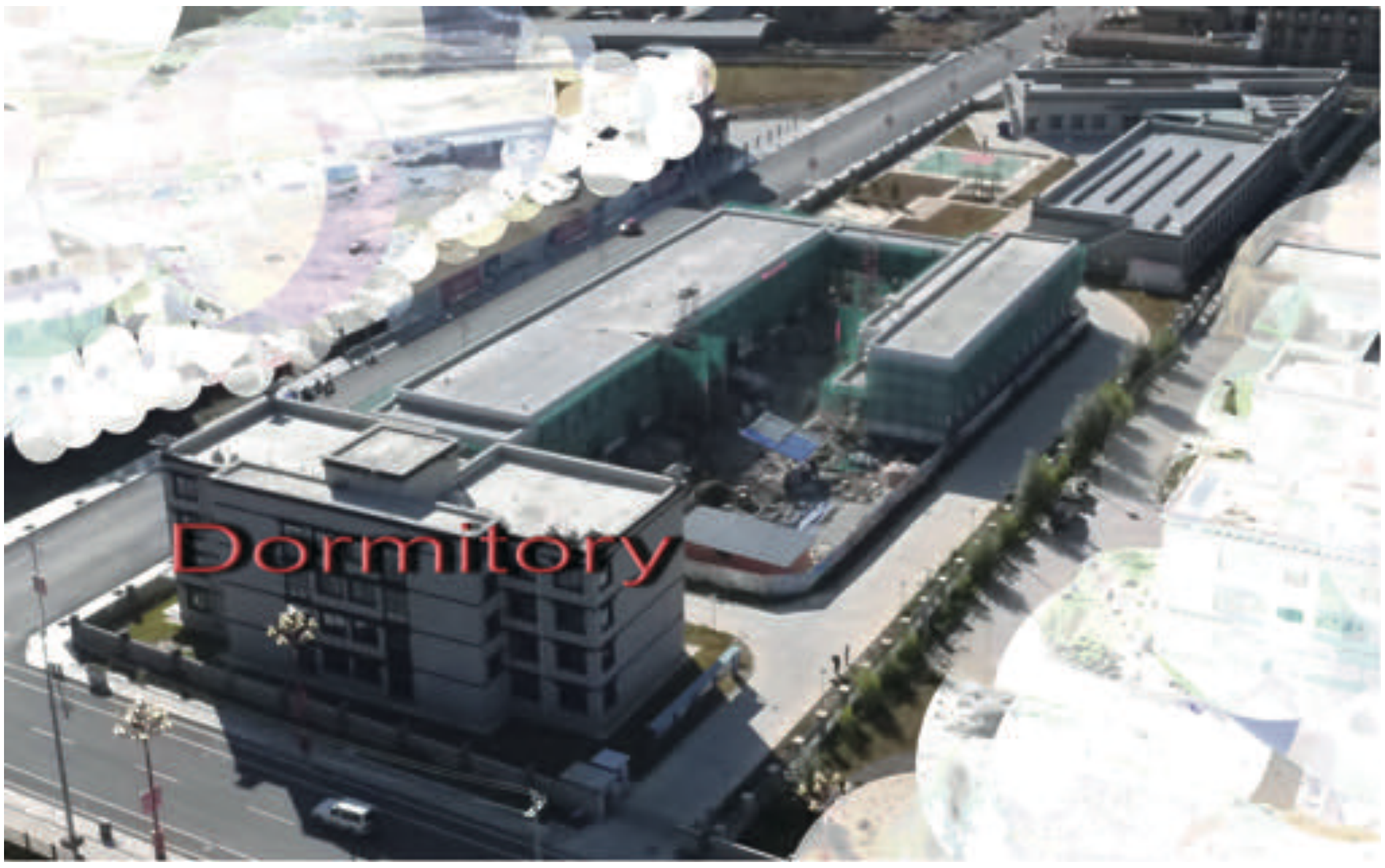

**Fig. 69.** (color online) The living base of LHAASO (2019).

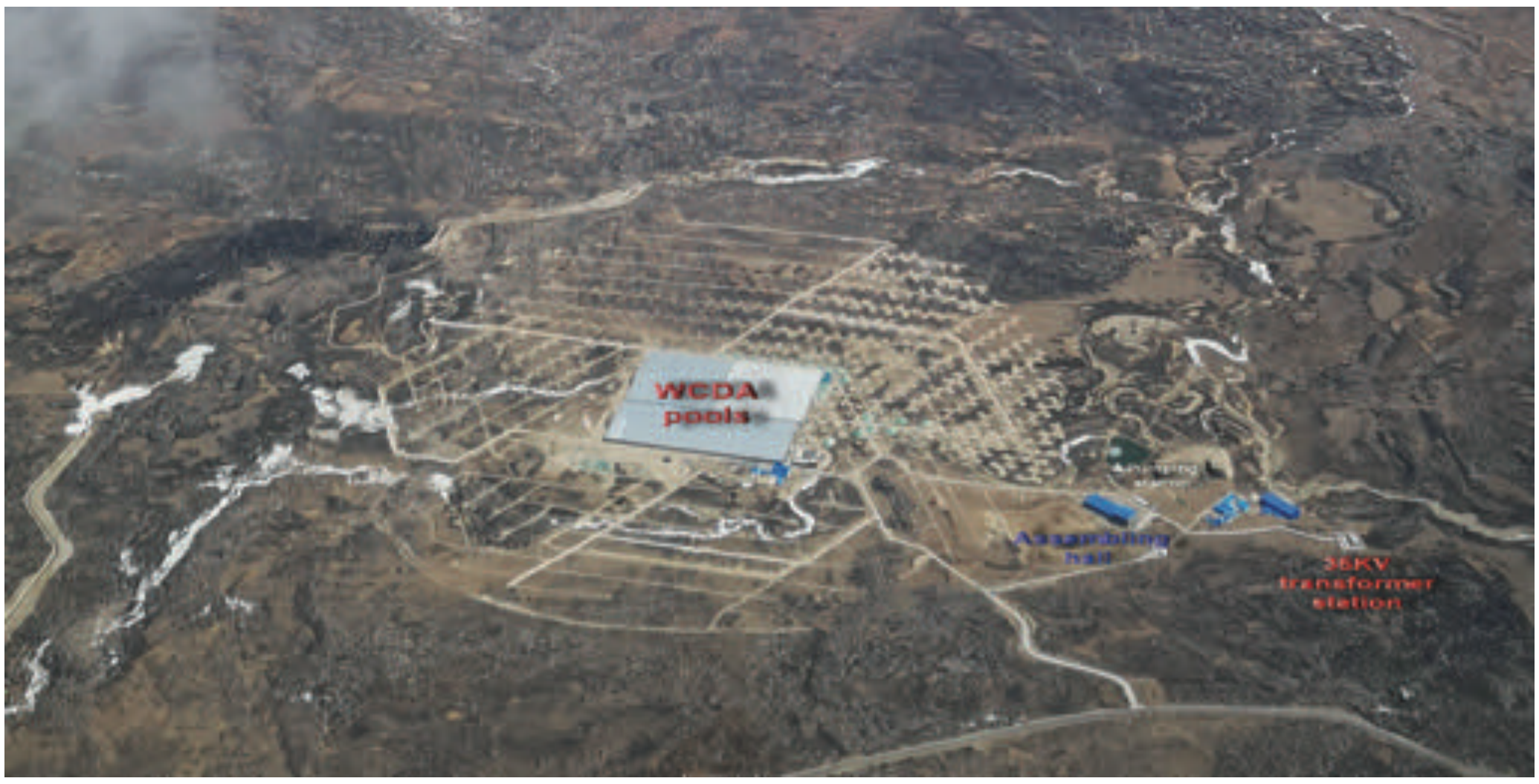

**Fig. 70.** (color online) The bird view of LHAASO observatory in 2019. Assembling hall, pumping station, 35KV transformer station and WCDA pools are marked.

The No.1 power distributing station is a single building of 390 $m^2$ with the UPS power supplier of 600 kVA and a dedicated room reserved for the central timing and White Rabbit switches together.

The water purification system is distributed in two factories for the ultrapure water (428 $m^2$) and pure water (372 $m^2$), respectively (see Fig. 71). They are located near the center of the array around the northwest corner of No.3 WCDA pool. The highest output of the pure water is 180 tons/hour and it can be adjusted to about 40 tons/hour in the running phase if needed. The characters of pure water are TOC (Total Organic Carbon) < 0.2 ppm, turbidity < 0.1 NTU (Nephelometric Turbidity Unit), number of particles (diameter > 1 μm) < 100 /ml. The highest output of the ultrapure water is 20 tons/hour. The characters of ultrapure water are dissolved oxygen < 10 ppb, resistivity ≥ 18 MΩcm (25 °C), the rest characters meet the EW-I standards in " GB/T 11446.1-2013 semiconductor grade water ".

Four sets of water recycling and filtering systems are installed outside the WCDA pools at the same level as the bottom of the WCDA pools. Water naturally flowing out from the WCDA pools are pumped back after filtered and sterilized with UV light through the walls of the pools. Pool 1 and 2 have one for each respectively, while pool 3 has two of them located at both ends of the pool which is 300 m long.

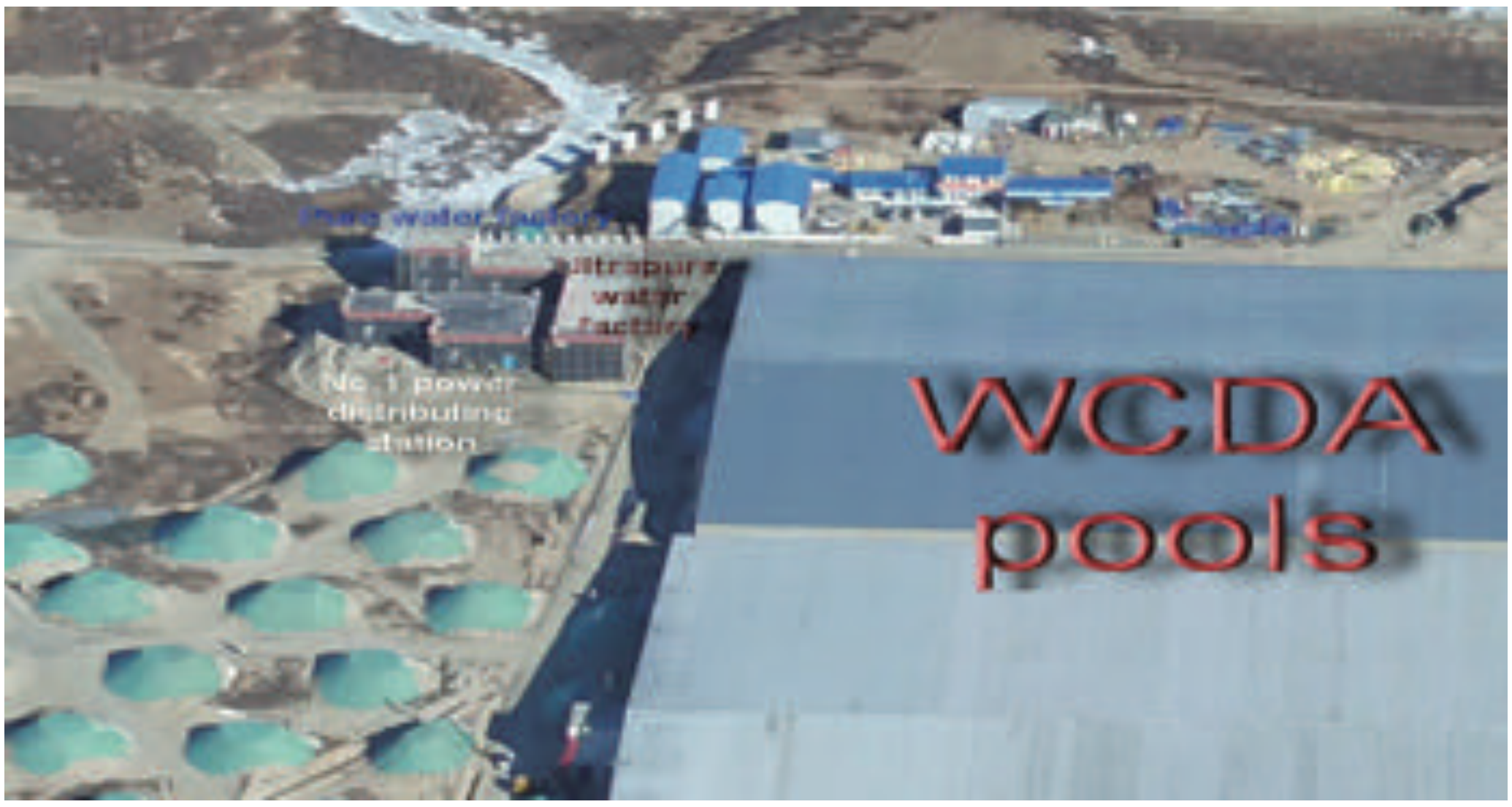

**Fig. 71.** (color online) The water factories and No.1 power distributing station of LHAASO observatory (2019).

Four control rooms are arranged on the second floor of those buildings. Central timing, clock distributing WR switches, data collecting network, high voltage power suppliers and water transparency monitoring system are hosted in the control room. The solo entrance of each pool with a door way equipped with air blowers is also located in the control room.

The WCDA pools are located at the center of the site. There are 3 pools. The pool 1 and 2 are both 150 m × 150 m. Pool 3 is 300 m × 110 m. The two smaller pools are at south-east and the bigger one is at north-west. The gap between the pool 1 and 2 is 2 m and they are 5 m apart from pool 3. Height of concrete walls of the pools is 5 m for holding the water in the pools. They all are covered by double layered roofs of steel sheets. The insulation layer which was made of polystyrene between the steel sheets is 120 mm thick. To prevent from oxidization of the steel roof and thus the pollution of water in the pool, all the inner surfaces of steel roof, structures and walls are completely painted with polyuria. The pillars have HDPE pipes covering their lower parts in water and the top ends of the pipes have been well sealed off. The bottom and the inner walls of the pools are fully covered by HDPE film with the thickness of 2 mm to prevent from water leakage. The HDPE film is conjoined with the pipes by soldering. Two layers of GCL (Geosynthetic Clay Liner) are put underneath the HDPE film at the bottom to protect the film from any possible damage due to the roughness of the ground inside the pools. They have a function of stopping further enlargement of any leakage once it happened after the water is filled up. The water leakage rate is designed to be less than 0.01% of the total capacity of each pool per day. To avoid freezing inside the pools, the walls of pools are insulated with polystyrene foam 110 mm thick. The foam is installed all the way from the top of walls down to the base 2.5 m under the ground. It has been observed during the winter starting from Nov. 2019 that the water temperature inside the pool drops 0.02 degree per day.

# Chapter 2 Galactic Gamma-ray Sources*

Editors: Yang CHEN, Xiao ZHANG

Contributors: Yang CHEN[1,10], Xiao-jun BI[2,10], Kun FANG[2,10], Yi-Qing GUO[2,10], Ye LIU[3,10], P. H. Thomas TAM[4,10], S. VERNETTO[5,10], Zhongxiang WANG[6,10], Rui-zhi YANG[7,8,9,10], and Xiao ZHANG[1,10]

[1] School of Astronomy and Space Science, Nanjing University, Nanjing 210023, China
[2] Key Laboratory of Particle Astrophysics, Institute of High Energy Physics, CAS, P.O. Box 918, 100049 Beijing, China
[3] School of Management Science and Engineering, Hebei University of Economics and Business, Shijiazhuang, Hebei 050061, China
[4] School of Physics and Astronomy, Sun Yat-Sen University, Guangzhou 510275, China
[5] Istituto Nazionale di Astrofisica, OATO, Torino, Italy
[6] Key Laboratory for Research in Galaxies and Cosmology, Shanghai Astronomical Observatory, Chinese Academy of Sciences, 80 Nandan Road, Shanghai 200030, China
[7] Department of Astronomy, School of Physical Sciences, University of Science and Technology of China, Hefei, Anhui 230026, China
[8] CAS Key Laboratory for Research in Galaxies and Cosmology, University of Science and Technology of China, Hefei, Anhui 230026, China
[9] School of Astronomy and Space Science, University of Science and Technology of China, Hefei, Anhui 230026, China
[10] All authors contribute equally to the work.

**Abstract:** In the $\gamma$-ray sky, the highest fluxes come from Galactic sources: supernova remnants (SNRs), pulsars and pulsar wind nebulae, star forming regions, binaries and micro-quasars, giant molecular clouds, Galactic center, and the large extended area around the Galactic plane. The radiation mechanisms of $\gamma$-ray emission and the physics of the emitting particles, such as the origin, acceleration, and propagation, are of very high astrophysical significance. A variety of theoretical models have been suggested for the relevant physics and emission with energies $E \geq 10^{14}$ eV are expected to be crucial in testing them. In particular, this energy band is a direct window to test at which maximum energy a particle can be accelerated in the Galactic sources and whether the most probable source candidates such as Galactic center and SNRs are "PeVatrons".

Designed aiming at the very high energy (VHE, >100 GeV) observation, LHAASO will be a very powerful instrument in these astrophysical studies. Over the past decade, great advances have been made in the VHE $\gamma$-ray astronomy. More than 170 VHE $\gamma$-ray sources have been observed, and among them, 42 Galactic sources fall in the LHAASO field-of-view. With a sensitivity of 10 milli-Crab, LHAASO can not only provide accurate spectrum for the known $\gamma$-ray sources, but also search new TeV $\gamma$-ray sources [1]. In the following sub-sections, the observation of all the Galactic sources with LHAASO will be discussed in details.

## 1 Introduction

During the last twenty years, the achievements in Gamma Ray Astronomy both in the GeV energy range with space borne instruments (like EGRET, AGILE and *Fermi*) and in the TeV region with ground based detectors (like Whipple, HESS, MAGIC, VERITAS, MILAGRO and ARGO-YBJ), produced extraordinary advances in high energy astrophysics, with the detection of a large number of sources (more than 5000 in the Fermi catalogue), about 170 of which emitting up to TeV energies.

† E-mail: xiaozhang@nju.edu.cn, ygchen@nju.edu.cn

* Supported by Key Research and Development Project (Grant Nos. 2018YFA0404204, 2016YFA0400804), the National Natural Science Foundation of China (11905043, 11803011, 11773014, 11633007, U1931204, U1731136), the Original Innovation Program of the Chinese Academy of Sciences (E085021002), Guangdong Major Project of Basic and Applied Basic Research (2019B030302001), and science research grant from the China Manned Space Project (CMS-CSST-2021-B09).

These sources belong to different classes of objects, both Galactic, such as pulsars, pulsar wind nebulae (PWNe), supernova remnants (SNRs), compact binary systems, etc., and extragalactic, such as active galactic nuclei (AGNs) and $\gamma$-ray bursts, in which the emission of high energy photons can be produced by different mechanisms. All these results are deeply modifying our understanding of the "non-thermal Universe". The field is extremely dynamic: the observations continuously provide new results, often unexpected and surprising, while the theory makes efforts to clarify the structure of these sources and the mechanisms operating in the acceleration regions.

In this scenario a strong interest is addressed to the development of new instruments capable of making more precise observations, with a better sensitivity and in a more extended energy range. This interest brought to new ambitious projects, like CTA (Cherenkov Telescope Array [2]), HAWC (High Altitude Water Cherenkov [3]), HiSCORE (Hundred Square-km Cosmic Origin Explorer [4]), and LHAASO.

Most results of TeV Gamma Ray Astronomy have been obtained with Imaging Atmospheric Cherenkov Telescopes (IACT), directional instruments with a field of view limited to a few square degrees, that can make observations only in clear and moonless nights. These are obvious limitations in a field of research aimed to discover unknown sources, and where most of objects have variable emissions, in some case explosive and unpredictable as $\gamma$ bursts, whose emission can last only a few tens of seconds.

Air shower detectors, detecting the secondary particles of $\gamma$-ray induced showers reaching the ground, do not have these limitations, since they can continuously observe the whole overhead sky. Air shower detectors like MILAGRO and ARGO-YBJ, even if with a lower sensitivity with respect to Cherenkov telescopes, have obtained relevant achievements. Starting from their results, new instruments based on the same concepts, but with a much higher sensitivity, like HAWC and LHAASO, have been designed to complement in a natural way the observation of future IACTs.

As pointed out before, the fundamental idea of LHAASO is the development of an instrument capable of measuring the cosmic ray (CR) spectrum, composition and anisotropy in a wide energy range ($\sim$1-$10^5$ TeV) and at the same time to be a sensitive gamma ray telescope at energies from $\sim$300 GeV to 1 PeV. In particular, the LHAASO design makes the detector particularly competitive in the $\gamma$-ray energy range above a few tens of TeVs, an energy region almost completely unexplored. Only a few sources have been observed to emit photons at these energies, and the data are affected by large uncertainties since the sensitivity of current instruments is not enough to determine clearly the spectral shape.

Actually, $\gamma$-ray astronomy at and above 100 TeV is of extreme importance since it is related to one of the most puzzling aspect of high-energy astrophysics: the identification of CR sources. Presently there is a general consensus that CRs with energy up to the so-called "knee" of the spectrum (2-4 PeV), and probably even up to 10-100 PeV, are accelerated inside our Galaxy. SNRs are long since believed the most likely sources, because the shock wave of the expanding shell could provide the conditions suitable to accelerate particles to relativistic energies, and secondly, the frequency of supernova explosions and their energy release could provide the total energy budget of CRs in the Galaxy [5].

This idea, however, is still lacking clear experimental evidence. Actually TeV $\gamma$-rays have been observed from a number of SNRs, demonstrating that in SNRs some kind of acceleration occurs. However the question whether TeV $\gamma$-rays are produced by the decay of $\pi^0$ from protons or nuclei interactions, or by a population of relativistic electrons via Inverse Compton scattering or bremsstrahlung, still needs a conclusive answer.

AGILE and Fermi observed GeV photons from two mid-aged SNRs (W44 and IC443) showing the typical spectrum feature around 1 GeV (the so called "$\pi^0$ bump", due to the decay of $\pi^0$) related to hadronic interactions [6]. This important measurement however does not demonstrate the capability of SNRs to accelerate CRs up to the knee and above. A key observation would be the detection of $\gamma$-rays of energy as high as a factor $10-30$ times less than the maximum energy of Galactic CRs. The observation of a $\gamma$-ray power law spectrum with no break up to 100 TeV would be a sufficient condition to prove the hadronic nature of the interaction, since the Inverse Compton scattering at these energies is strongly suppressed by the Klein-Nishina effect.

Recently IceCube reported a first piece of evidence of neutrinos of astrophysics origin of energy $0.4-1$ PeV [7]. The nature of such a flux has been the subject of intense discussions and different hypotheses have been expressed about the Galactic or extra-galactic origin of the signal. If neutrinos were extra-galactic the $\gamma$-rays generated by the same processes would be absorbed by the CMB through pair production and would not be observable at Earth, but in case of a Galactic origin of a fraction of the measured neutrino flux, it would be important to detect a photon signal of comparable energy.

LHAASO, thanks to the large area of the array KM2A and the high capability of background rejection, can reach sensitivities above 30 TeV about 100 times higher than those of current instruments, offering the possibility to monitor for the first time the $\gamma$-ray sky up to PeV energies.

### 1.1 LHAASO sensitivity to gamma-rays

LHAASO can study $\gamma$-ray sources over almost 4 decades of energy. Figure 1 shows the differential sensitivity in one year measurement, obtained by simulating the response of the detector to $\gamma$-ray flux from a source like the Crab nebula one. In the same figure the Crab nebula spectrum is also shown as a reference flux. The LHAASO sensitivity curve is the combination of two components, the first relative to the water Cherenkov detector (WCDA), operating in the energy range $\sim$0.3 - 10 TeV, the second relative to the KM2A array, sensitive to energies above 10 TeV [8].

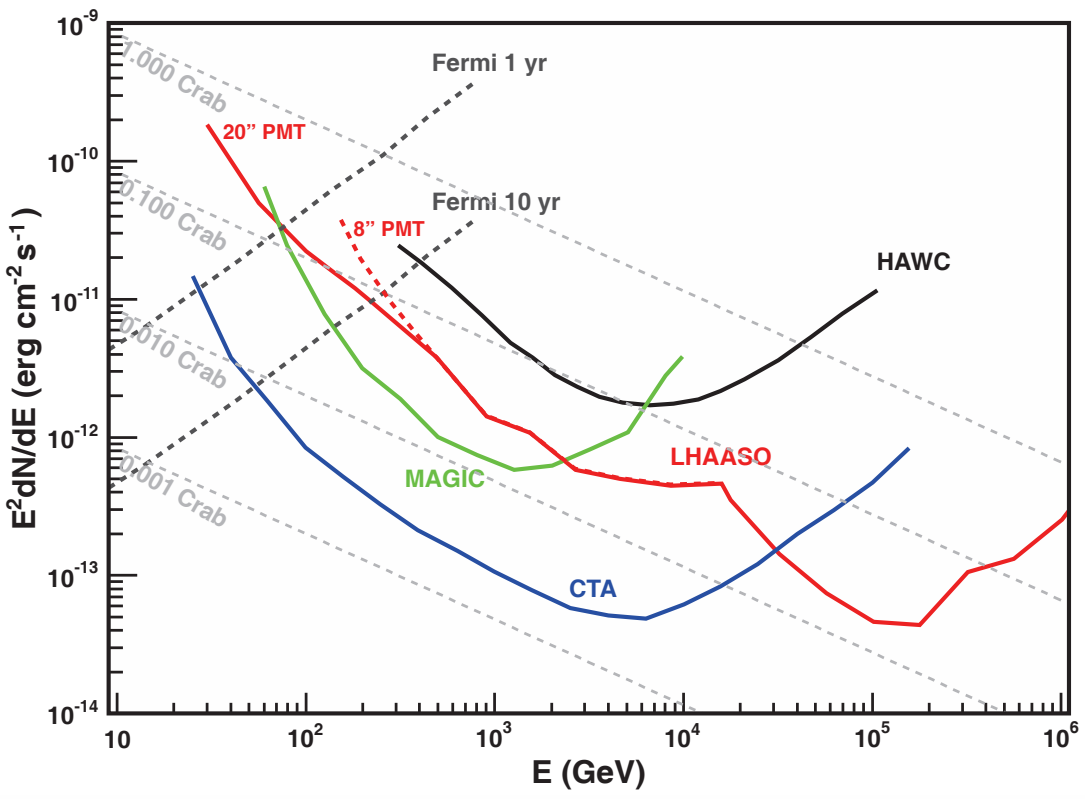

Fig. 1. Differential sensitivity (multiplied by $E^2$) of LHAASO to a Crab-like point $\gamma$-ray sources compared to other experiments. The Crab nebula data obtained by different detectors [9] are taken into account, and the spectral index of -2.6 is extrapolated and extended to 1 PeV.

According to simulations, the minimum $\gamma$-ray flux detectable by LHAASO is less than 3% of the Crab flux in the energy range $\sim$1-5 TeV and about 10% Crab around 100 TeV. In Figure 1 the sensitivity curves of other detectors (some in operation, some in project) are also reported. It has to be noted, however, that for a general convention the sensitivity of air shower detectors is reported for one year of operation, while that of Cherenkov telescopes is relative to 50 hours of "on source" measurement. Note that EAS arrays observe every day all the sources in the field of view for a fixed time interval depending on the source declination, while IACTs observe only one source at the time, and only in the season of the year when the source culminates during night time.

The differences in observation times for which the sensitivity curves are evaluated makes the comparison of different detectors not so straightforward. To evaluate the effective performance of different instruments, one must first determine the type of the observation to be done (sky survey, single source follow-up, observation of a flare/burst, etc.). In the observation of a single source during a flare, for example, lasting a certain number of hours, one must consider the sensitivity curves for *that* observation time. This correction however is not simply obtained by shifting the curves by an amount proportional to the square root of time, because some energy regions can be background free. Due to the different background regime, the sensitivity curves can change shape with different observation time. Decreasing (increasing) the time with respect to the time used in the figure, the background also decreases (increases) and the measurement can be background free at a lower (higher) energy.

Actually, the two techniques - Cherenkov Telescopes and EAS array - are complementary, each of them exploring different aspects of the gamma ray emission. Below 10 TeV, observing a single source, a telescope array as CTA has a higher sensitivity compared to EAS arrays like HAWC and LHAASO. Thanks to the better angular and energy resolution, a Cherenkov telescope can study more in detail the source morphology and spectral features. EAS arrays however have the possibility to monitor a source all days of the year, that in case of AGNs or variable sources in general, it's a clear advantage. Moreover, thanks to the large field of view, they have a much bigger chance to catch unpredictable transient events like flares.

Concerning LHAASO-WCDA and HAWC, their geographical positions (China and Mexico, respectively) allow the observation of the same source at different times during the day, increasing the covering time.

At higher energies LHAASO-KM2A is clearly the most sensitive instrument. According to Fig. 1, at 30 TeV the LHAASO sensitivity is comparable to that of CTA-South and 4 times better than that of CTA-North. Above this energy the sensitivity rapidly increases. The minimum observable flux at 100 TeV is $\sim 3\times10^{-18}$ photons s$^{-1}$ cm$^{-2}$ TeV$^{-1}$, about a factor $\sim$13 (65) lower than that of CTA-South (CTA-North).

At 1 PeV the minimum flux is $\sim 10^{-19}$ photons s$^{-1}$ cm$^{-2}$ TeV$^{-1}$. At the same energy, the combined air shower/neutrino detector Ice-Top/Ice-Cube, located at the South Pole, reports a minimum observable $\gamma$-ray flux ranging from $\sim 10^{-19}$ to $10^{-17}$ photons s$^{-1}$ cm$^{-2}$ TeV$^{-1}$ (depending on the source declination) for sources on the Galactic plane in 5 years of measurements [10]. It has to be noted, however, that at these energies the observations can be seriously hampered by pair production with the CMB photons, that can affect the fluxes of Galactic sources with a distance larger than a few kpc.

Using proton and $\gamma$ fluxes from the Crab direction (with the zenith within $8^\circ \sim 45^\circ$ and assumping a spectrum index $\gamma = 2.7$), the simulated efficiency of $\gamma$s and protons that pass the hadron rejection cuts is shown on

Fig. 2. The simulation includes the ED, MD, and WCDA components of the LHAASO. Above 10 TeV the measurement of the $\mu$ component in the showers allows a very efficient rejection of CR showers. According to simulations the fraction of CRs that survive the discrimination cuts is 0.01 and 0.004 % at 10 and 30 TeV, respectively, while above ∼150 TeV, it is found to be less than 0.0001%. This means that above ∼150 TeV the study of the $\gamma$-ray emission from a point source can be considered as background free, because after applying the rejection procedure the expected background in a cone around a source is less than one event in one year. The flux has been evaluated by dividing each energy decade into 4 bins, and requiring a statistical significance of 5 standard deviations per bin and a minimum number of 10 events. In the background free regime, only the later constraint is adopted. One should note that in the background free regime, the sensitivity increases linearly with the observation time instead of the square root of time, as in presence of background, and the energy threshold for the background free regime changes with the observation time.

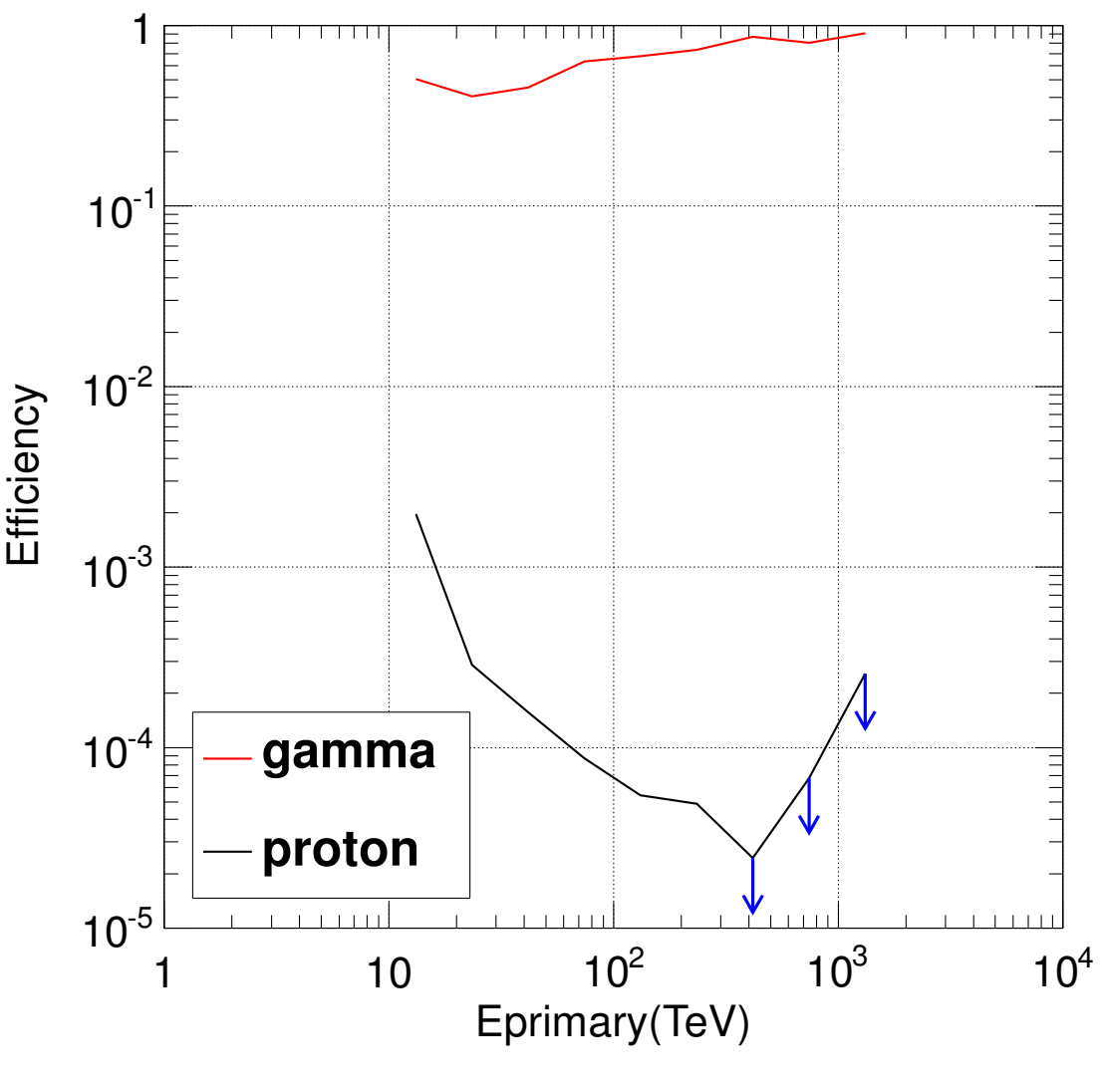

Fig. 2. Simulated LHAASO detection efficiency of $\gamma$s and protons: the number of events normalized to a year of flux from the Crab.

### 1.2 LHAASO and sky survey

One of the most interesting aspect of LHAASO is the large field of view (FOV) and the capability to monitor every day a consistent fraction of the sky. In principle the FOV can include all the sky above the horizon, but the sensitivity decreases at large zenith angles.

Considering only the region of the sky that culminates at zenith angles smaller than 40°, every day LHAASO (located at latitude 29° North) can survey the declination band from -11° to +69° (about 56% of the whole sky) that includes the Galactic plane in the longitude interval from +20° to +225°. Most of this region will be observed by LHAASO with unprecedented sensitivity. For the most energetic events, the extension of the field of view to larger zenith angles will increase the sky coverage allowing observations close to the Galactic center. Fig. 3 shows the observation time per day as a function of the source declination, for different values of the maximum zenith angle.

In the past, the air shower detectors ARGO-YBJ and Milagro have surveyed about the same region of the sky visible by LHAASO, at energies above 0.3-1 TeV and ∼10 TeV respectively, with a sensitivity of about 0.3 Crab units [11, 12]. The new EAS array HAWC, in full operation since 2015, has a sensitivity ∼4 times lower than that expected for LHAASO in the 1-10 TeV region, but more than 100 times lower at 100 TeV. Concerning Cherenkov telescopes, their limited field of view and duty cycle prevent a survey of large regions of the sky. In the past a fraction of the Galactic plane have been surveyed by IACTs with an excellent sensitivity in the TeV energy range. HESS performed a survey of the Galactic plane between longitude -110° and 65° in the latitude band ±3.5° with a sensitivity of ∼0.02 Crab units at energies above 100 GeV [13], that led to the discovery of more than 60 sources, while VERITAS surveyed the Cygnus region between longitude 67° and 82° with a sensitivity of about 0.04 Crab units [14].

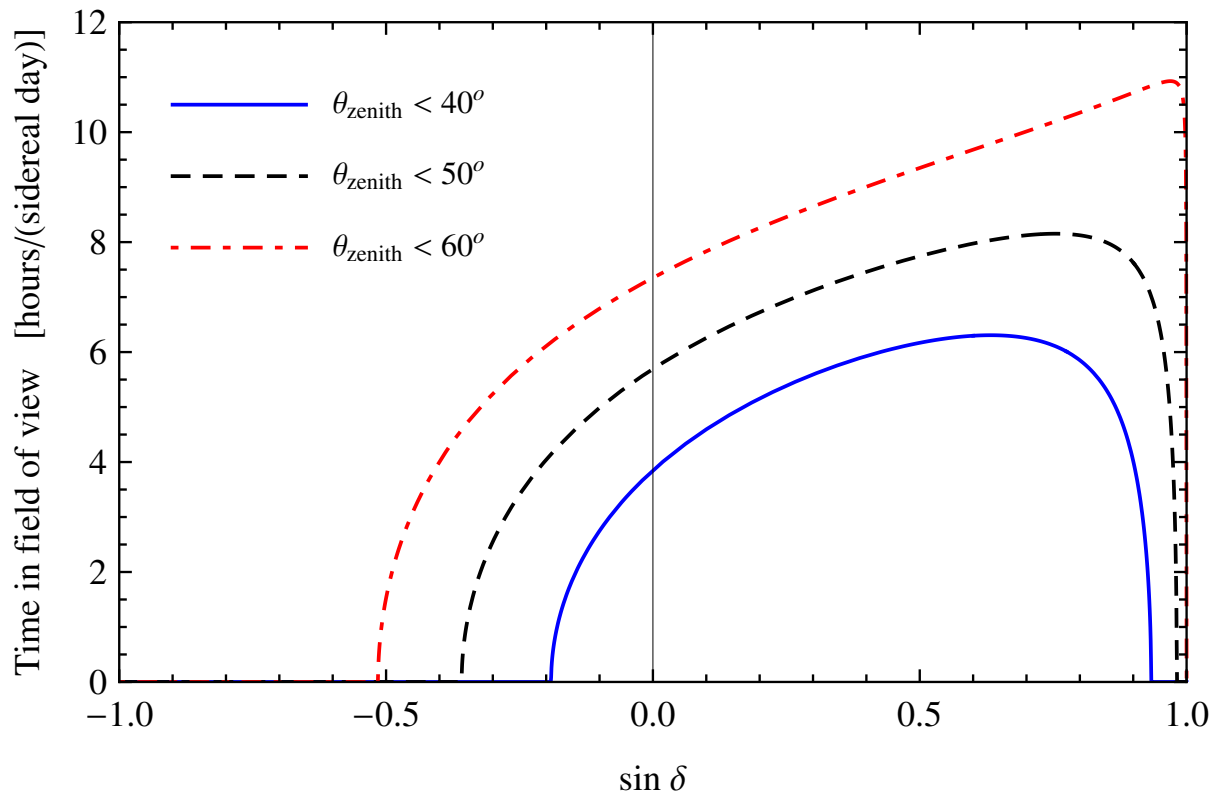

Fig. 3. Observation time (hours) per day as a function of the source declination, for 3 values of the maximum zenith angle. The area under the curves is proportional to the total exposure (observation time × solid angle).

It is interesting to compare the performance in sky survey of LHAASO and the future array CTA. Let's consider a survey of the Galactic plane in a longitude interval of 200° and a latitude band from -4° to +4°. A detector like CTA, with its limited field of view, must scan the whole region with different pointings. The number of pointings determines the maximum observation time that can be dedicated to any location. Assuming a field of view of 5° radius and a decrease of sensitivity of about 50% at a distance of 3° from the center (according to the design of SSTs, the CTA-South small area telescopes planned to work at the highest energies), a reasonable step for pointings could be 4°. With this step, 100 pointings are necessary to cover the entire region. Since in one year a Cherenkov telescope can make observations for a total time of ~1300 hours (assuming the use of the silicon photomultipliers that allow the data taking also in presence of the Moon), every source can be observed for ~13 hours. At 1 TeV, the CTA-South sensitivity in 13 hours is still higher than that of LHAASO in one year. At ~25 TeV LHAASO starts to become more sensitive than CTA. Above 30 TeV, the CTA-South sensitivity is no more limited by the background but by the number of detected events (that must be at least equal to 10), hence it must be rescaled linearly with the time. According to this rough estimation, LHAASO would be ~4 and 50 times more sensitive than CTA-South at 30 and 100 TeV, respectively.

The LHAASO performance is more impressive in case of an *all sky* survey, where assuming a region of 7 steradians to be scanned, the number of CTA pointings would be as large as ~1600 and every location would be observed for less than one hour. In this case the LHAASO sensitivity would be more then ~60 and 800 times higher than that of CTA-South for energies of 30 and 100 TeV, respectively.

Finally, in the comparison with CTA-North (that will be located in the Canary island of La Palma at about the same latitude of LHAASO and will observe about the same sky), LHAASO will gain a further factor 4-5 due to the lower sensitivity of the Northern array with respect to the Southern one. Furthermore, it has to be noted that the CTA-North telescopes will have a field of view with a radius not larger than 4° and consequently the number of pointings necessary to cover the region to be scanned will be larger by at least 40% with respect to the value used above, decreasing correspondingly the observation time and the sensitivity.

### 1.3 Galactic gamma-ray astronomy

According to the online TeV source catalogue TeVCat [15] at the time of writing the number of known sources is 169. Among them, 60% belong to our Galaxy and 40% are extragalactic (mostly AGNs of blazar type). About 1/3 of Galactic sources are still unidentified, 1/3 are PWNe, and the remaining are SNRs, compact binary systems and massive star clusters. Note the the sensitivity of the current instruments allowed the detection of three "Galactic" sources inside an extragalactic object, the Large Magellanic Cloud.

The sky map in Fig. 4 shows the position of all the sources in Galactic coordinates. The sky region that culminates at zenith angle smaller than 40°, delimited by green lines in the figure, includes 84 objects, 23 Galactic, 47 extragalactic, and 14 still unidentified. All the unidentified sources but one, lay on the Galactic plane, being probably Galactic objects that cannot be identified due to the number of possible associations in their positional error box.

The spectrum of the Galactic sources has been generally measured in the energy range from a few hundreds GeV to 10-20 TeV, and for most of them it is consistent with a power-law behavior. The precise measurement at higher energies would be of extreme interest to understand the emission mechanisms of $\gamma$-rays, that for most of the sources is still not understood, and will surely help in the source identification.

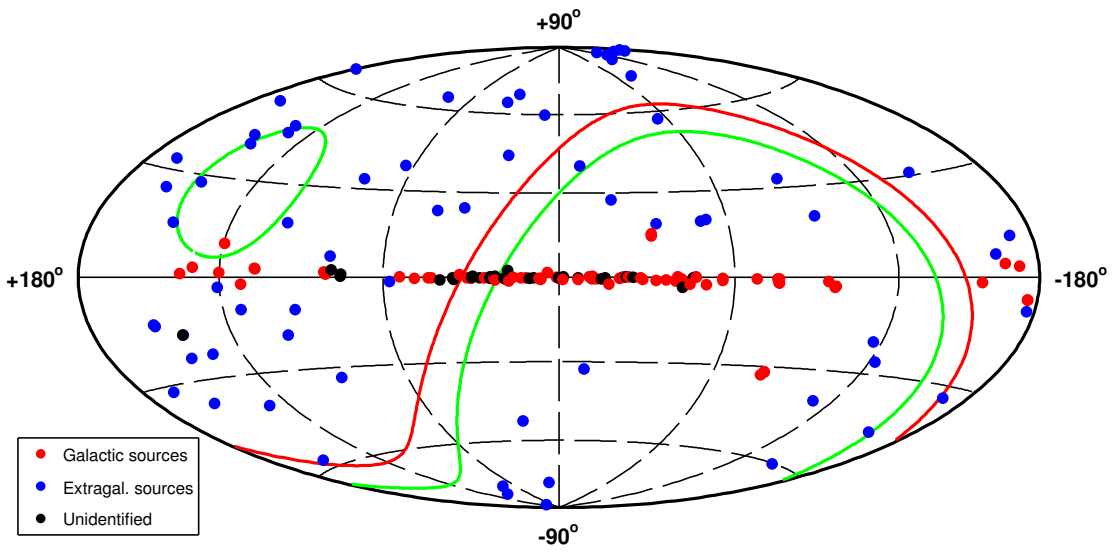

Fig. 4. Sky map in Galactic coordinates, showing the positions of the known TeV sources. The red line represents the celestial equator. The green lines limits the region of the sky that culminates at zenith angles smaller than 40° at the LHAASO site. The sources are indicated by different colors according to their type: Galactic, extragalactic, unidentified (note that the three sources denoted as "Galactic" around the position r.a.= 83° and dec.= -69° are actually in the Large Magellanic Cloud).

So far, only six sources have data above 30 TeV. They are all Galactic and are among the most luminous objects of the TeV sky: the SNR RX J1713.7-3946, the PWNe Crab and Vela-X, and the three MILAGRO extended sources MGRO J2031+41, MGRO J2019+37 (actually resolved in two different sources by VERITAS), and MGRO J1908+06, all them probably PWNe, too. Their spectrum above 30 TeV is however known with large uncertainties.

PWNe are the most common type of Galactic sources. They are believed to be the product of the ultra-relativistic $e^{\pm}$ wind emitted by young pulsars with large spin-down rates, interacting with magnetic and radiation fields around the pulsar. Other leptons can also be accelerated in the shock produced in the collision of the wind with the environment matter. All these relativistic leptons produce synchrotron and Inverse Compton (IC) radiation.

The Crab nebula, the most luminous TeV source and the first to be detected at TeV energies at the beginning of the Cherenkov telescopes era in 1989, is the most famous example of this class of objects. Its spectral energy distribution (SED) shows a double-humped structure. The first one, extending from radio waves to ∼1 GeV, is due to synchrotron emission, the second one, peaking at ∼100 GeV is the product of IC scattering of electrons off low energy photons (synchrotron, thermal and CMB photons). The SED is well defined up to 10-20 TeV. Above this energy is not precisely known. Fig. 5 shows the high energy Crab spectrum measured by different ground based experiments [16–20]. Even considering the large error bars, a disagreement is evident among the higher energy data. The HEGRA spectrum is a power law with a weak steepening above 10 TeV whereas MAGIC and HESS measurements show a more evident spectral curvature in all the energy range considered. The precise measurement of the high energy emission, the "end" of the spectrum, would bring important information on the particle acceleration and the magnetic and radiation fields in the pulsar environment, constraining some parameters that the lower energy spectrum alone cannot determine unambiguously.

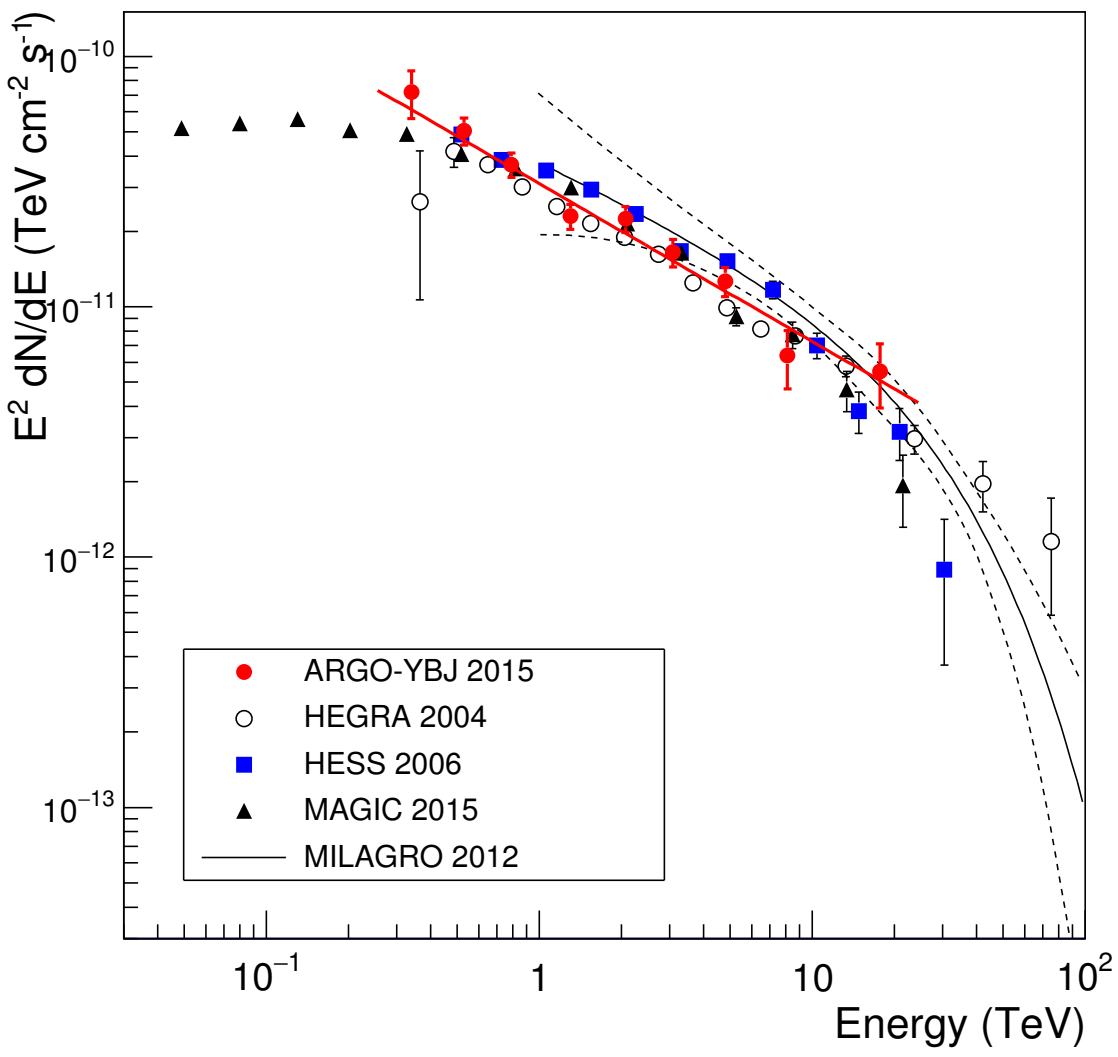

Fig. 5. Energy spectrum of the Crab nebula measured by different experiments.

The high energy measurement would also be of great importance in understanding the intriguing phenomena of the Crab nebula flares. Since long considered as the "standard candle" for $\gamma$-ray astronomy, the Crab nebula has unexpectedly shown a variable behavior in the 100 MeV-1 GeV energy range, with strong flares lasting hours/days [21–23], and rate variations on time scales of months [24], that are still waiting for a shared interpretation. During flares, the SED shows a new hard component above 100 MeV, generally interpreted as synchrotron emission of a new population of electrons accelerated to energy up to $10^{15}$ eV, whose origin is still not understood. A TeV flux enhancement in coincidence with GeV flares have been reported by ARGO-YBJ [25], but with low statistical significance, and has not been confirmed by later measurements by the more sensitive Cherenkov experiments [26, 27]. The question of the possible existence of an Inverse Compton emission associated to the GeV flares remains open, in particular in the energy region around and above 100 TeV, where the IC emission is more likely to occur. LHAASO is the most suitable detector for such a study, due to the high sensitivity at these energies and to the possibility of observing the Crab nebula for 5-6 hours every day of the year.

Besides the Crab nebula, LHAASO can perform accurate spectra measurements above 30 TeV for most of the known TeV Galactic sources visible from its location. To give a quantitative idea of the LHAASO capabilities, it is useful to compare the detector sensitivity with the fluxes of such sources.

Out of 84 sources crossing the detector field of view with a zenith angle less than 40°, 23 are associated with known Galactic objects, while 13, even if not yet associated with certitude with a source, lay on the Galactic plane, and can be reasonably considered Galactic too. For 35 out of these 36 Galactic sources the flux has been measured and reported in [15] and for 24 of them a spectral index is available, ranging from 1.75 to 3.1, with an average value of 2.4.

Fig. 6 shows the spectra of 35 objects extrapolated to 1 PeV (with the same spectral index measured in the TeV region) compared to the LHAASO one-year sensitivity. The spectral index has been set to 2.5 for the sources without an available spectral measurement. It should be specified that for a correct comparison the LHAASO sensitivity should be calculated for each source using its individual spectrum, angular extension and declination, while in the figure the sensitivity refers to a Crab-like source. The spectra extrapolations are clearly unrealistic, since the real spectra likely would show steepening or cutoffs at some energy, but the purpose of the figure is to show that the flux of almost all the considered sources is above the LHAASO sensitivity. LHAASO can study in detail the behavior of the higher energy emission of most of the sources, down to fluxes of $\sim 3\times10^{-18}$ photons $s^{-1}$ $cm^{-2}$ $TeV^{-1}$, at 100 TeV in one year of measurement. These high energy data are likely to play a crucial role for the understanding of the properties of the sources.

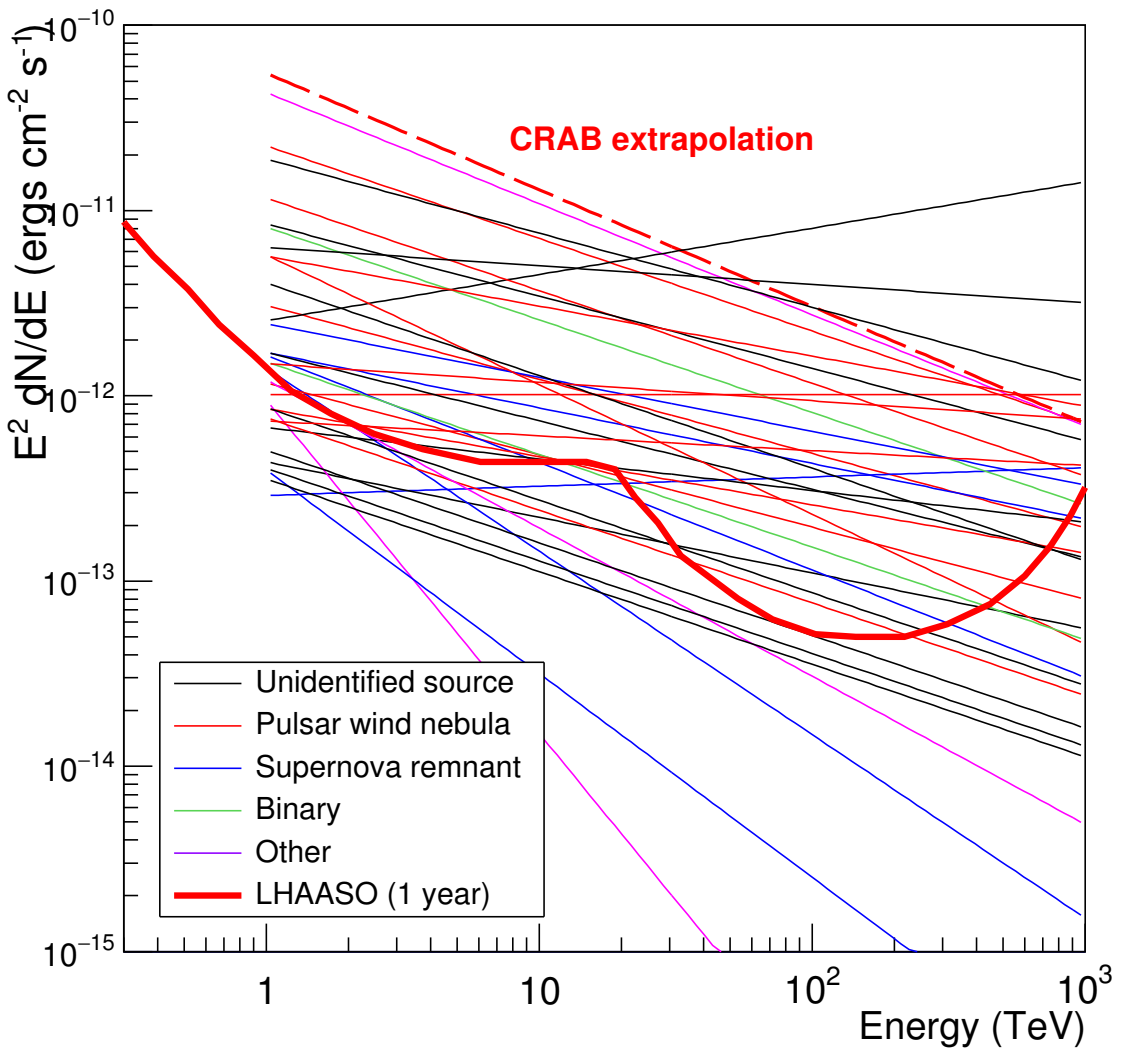

Fig. 6. Differential spectra (multiplied by $E^2$) of the TeV $\gamma$-ray sources visible by LHAASO extrapolated to 1 PeV, compared to the LHAASO sensitivity. The dashed red line represents the Crab nebula flux, as measured by ARGO-YBJ [16] extrapolated to 1 PeV as a power law.

Among Galactic sources, shell SNRs are probably the most interesting to be studied at high energy because the detection of an emission above 100 TeV could be the footprint of hadronic acceleration. In general, from an emission of hadronic origin, one expects a $\gamma$-ray spectrum showing the "$\pi^0$ bump" followed by a power law with a slope consistent with parents spectrum slope up to 50-100 TeV, or even more, depending on the parent nuclei maximum energy. A leptonic emission (Inverse Compton scattering of electrons with a power law spectrum) would produce a flatter power law $\gamma$-ray spectrum, but with a gradual steepening due to the Klein-Nishina effect. The position of the break depends on the energy of the target photons. For example, electrons with a spectral index of -2.2, scattering off CMB photons, would produce a $\gamma$-ray spectrum of index -1.6 in the Thomson regime, that gradually steepens up to -3.2 in the Klein Nishina regime. At 100 TeV the flux is already suppressed by a factor of 3 with respect to the extrapolation of the spectrum before the break.

Actually, in a SNR one could expect a combination of the two emissions, leptonic and hadronic, with different weights depending on many parameters, such as the density of target material for hadronic interaction, the magnetic field strength, the age of the Supernova, etc., which make difficult to identify the emission origin. However, the observation of a spectrum extending above 100 TeV would be a strong indication of a hadronic emission.

So far, only one remnant, SNR RX J1713.7-3946, has been detected above 30 TeV (actually, the spectrum reaches almost 100 TeV [28]). In this case the spectrum steepens above a few TeV and does not show the "$\pi^0$ bump", being more consistent with a leptonic emission [29]. All other TeV SNRs have data up to 15 TeV at maximum, Based on the new data, RX J0852.0-4622 [30], Cas A [31], and RCW 86 [32] have a high-energy cutoff around few TeV.

In the LHAASO field of view there are six shell SNRs (Tycho [33], CAS A [34], W51 [35], IC443 [36], W49B [37] and SNR G106.3+2.7 [38]). The measured spectra show a power law behavior without any cutoff up to the maximum energy reached by the current instruments, that ranges from ~2 to 15 TeV for the sources considered. It should be noted that a recent result of VERITAS [39] shows an updated spectrum of Tycho, steeper than the one reported in the figure, which would make the LHAASO measurement more challenging for this source.

Besides the observation of known sources, given the LHAASO capabilities in sky survey, new Galactic sources will likely be discovered at high energy, since objects with fluxes at 1 TeV below the current instruments sensitivity but with hard spectra (i.e. spectral index <2) would be easily detectable by LHAASO above ~10 TeV.

## 2 Supernova Remnants

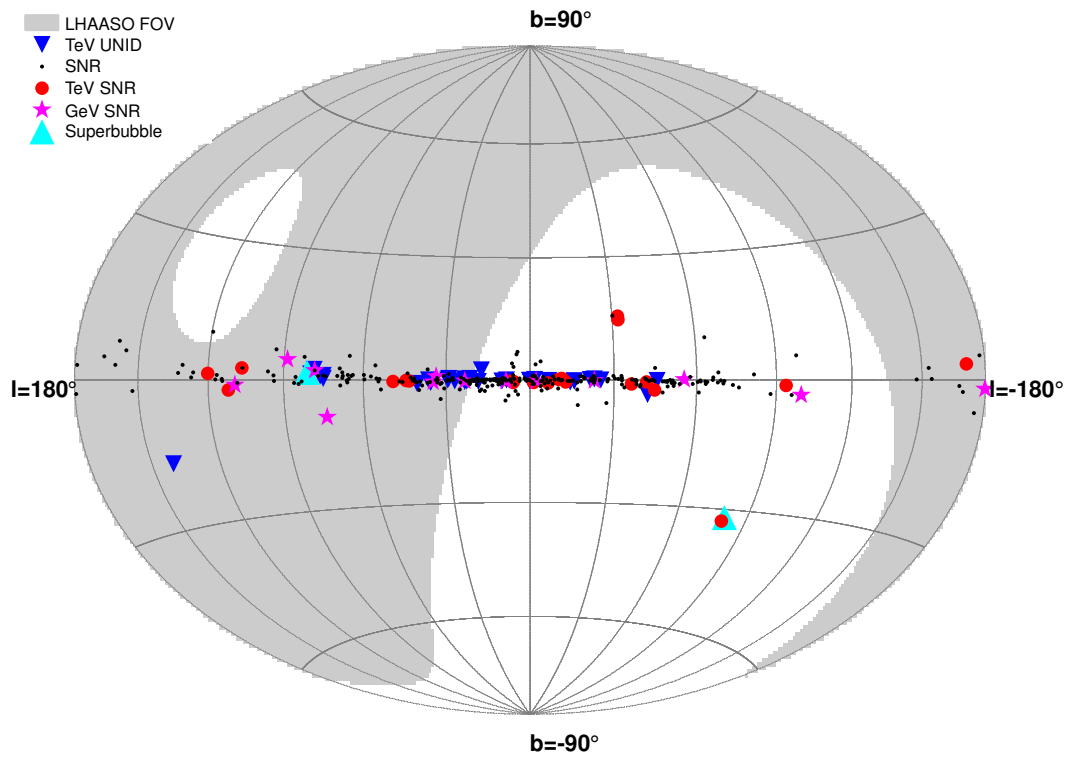

Fig. 7. Locations of SNRs and unidentified TeV $\gamma$-ray sources in Galactic coordinates, compared with the field-of-view of LHAASO (grey region) [1]. Black dots represent SNRs from Green[1], red filled circles and magenta stars show TeV and GeV $\gamma$-ray SNRs[1] [40], blue triangles represent the unidentified TeV $\gamma$-ray sources, and cyan triangles represent two super-bubbles which were detected in TeV $\gamma$-ray bands.

### 2.1 Gamma-ray observation of SNRs

Among Galactic $\gamma$-ray point sources, SNRs are considered to be one of the most plausible candidates for acceleration of cosmic-rays up to PeV energies [41–43].

According to Dave Green's Galactic SNR catalogue , 295 SNRs have been detected up to now. Most of these SNRs are only detected in low energy bands [44]. In the GeV energies range , the Fermi-LAT collaboration reported their first SNR catalog based on three year's survey data, in which 12 firm identifications and 11 possible associations with SNRs were found [40]. In the TeV energies range , there have been at least 23 SNRs or SNR candidates detected up to now, 10 of which are also GeV $\gamma$-ray emitters . Furthermore, there are 34 unidentified TeV $\gamma$-ray sources which do not have clear counterparts in other wavelengths. Unlike the Fermi unidentified sources, which are expected to be dominantly constituted by active galactic nuclei [45], most of the unidentified TeV sources are located in the Galactic plane (see Fig. 1) and could be potential SNRs. Fig.7 illustrates the locations of those sources (symbol) and their visibility by LHAASO (shaded region). In total, 92 out of 295 SNRs in Green Catalog, 6 GeV SNRs or SNR candidates, 2 TeV SNRs and 6 GeV-TeV SNRs are in the field of view of LHAASO. Moreover, 17 TeV unidentified sources locate in the field of view of LHAASO.

Moreover, it has been found that some SNRs could emit TeV $\gamma$-rays while in GeV energy band there was no detection, such as G106.3+2.7 and HESS J1912+101. G106.3+2.7 was first observed by DRAO at radio energy range [46]. In 2000, Pineault & Joncas confirmed the object as an SNR, with an estimated age of 1.3 Myr and distance of 12 kpc [47]. The pulsar PSR J2229+6114 is located at the northern edge of the remnant's head and it is associated with boomerang-shaped radio and X-ray emitting wind nebula. At GeV energy band, the EGRET source 3EG J2227+6122 is compatible with the pulsar position, as well as the main bulk of the radio remnant [48]. At TeV energy band, VERITAS reported the total flux from the SNR G106.3+2.7 above 1 TeV is about $\sim$5% of the Crab Nebula in 2009 [38]. HESS J1912+101 is plausibly associated with the PSR J1913+1011, which is detected by H.E.S.S. experiment. The integral flux between 1-10 TeV is 10% of the Crab Nebula and the measured energy spectrum can be described by a power-law with a photon index $\sim$ 2.7. From the current observations on these two TeV SNRs, we can conclude that LHAASO might detect a number of SNRs compared to conservative predictions based on the current SNR catalogs.

### 2.2 Hadronic or leptonic origin of the gamma-ray emission

Generally, there are two types of scenarios for the production of high-energy $\gamma$-rays from SNRs: the leptonic interaction via IC scattering of background photos by relativistic electrons and hadronic interaction via decay of neutral pions produced by inelastic collisions of relativistic ions with ions in the background plasma [49–52].

Up to now, the evidence for efficient leptonic acceleration in SNRs is clearly established [53, 54]; however, the question of whether SNRs are efficient hadron accelerators is more difficult to answer. The recent observation of $\gamma$-ray spectrum for W44 and IC443 by Fermi shows that accelerated protons and nuclei via hadronic interactions with ambient gas and subsequent $\pi^0$ decays into $\gamma$-rays [6, 55]. But at TeV band, the current observations [36, 56, 57], is not enough to identify the radiation mechanism. With the wide FOV, LHAASO is suitable not only to measure their SEDs but also carry out morphological investigations on those sources at high energies.

Young SNRs, typified by Tycho and Cassiopeia A (Cas A), are believed to be energetic accelerators of relativistic particles. Tycho's SNR, which appeared in 1572 [58], has been observed from radio to TeV $\gamma$-ray band [33, 59–64]. At the GeV range, Fermi-LAT reported a $5\sigma$ detection of GeV $\gamma$-ray emission from Tycho, which can be described by a power-law with a photon index 2.3$\pm$0.2 [65]. At the TeV range, VERITAS found that the total flux of Tycho above 1 TeV is $\sim$ 0.9% of Crab Nebula and the spectral index between 1 TeV and 10 TeV is about 1.95$\pm$0.51 in 2011. But in 2015, the spectral index turns to 2.92$\pm$0.42 [39]. If the spectral index is about 2 up to 10 TeV as VERITAS reported in 2011, it implies that the corresponding spectrum of primary protons extends without a significant steepening or a cutoff to at least several hundred TeV [5, 66]. Due to the large uncertainties of the data sets of Fermi and VERITAS, the energy spectrum from GeV to TeV can be described by a broad range of function, which is not enough to constrain the origin of high energy $\gamma$-ray emission.

Cassionpeia A (Cas A), which might appear in 1680, is the youngest of the historical Galactic SNRs [55, 67]. It is one of the best studied objects with both thermal and non-thermal broad-band emission ranging from radio wavelengths to TeV $\gamma$-rays [34, 67–69]. TeV $\gamma$-ray observations revealed a rather modest $\gamma$-ray flux, compared to the synchrotron radio through X-ray emission, which further strengthens the argument for a rather high magnetic field. In the GeV range, Fermi-LAT observation suggests that leptonic model can not fit the turnover well at low energy because the bremsstrahlung component that is dominant over IC below 1 GeV has a steep spectrum, and hadronic emission describing the $\gamma$-ray spectrum by a broken power-law is preferred. However, because the observed TeV $\gamma$-ray fluxes have large sta-tistical uncertainties, it can not be judged yet whether the TeV $\gamma$-rays are generated by interactions of accelerated protons and nuclei with the ambient gas or by electrons through bremsstrahlung and IC scattering. And the maximum energy of the observed TeV $\gamma$-ray is only several TeV, the question whether Cas A accelerates particles to PeV energy is still open.

‡http://www.mrao.cam.ac.uk/surveys/snrs/

§http://tevcat.uchicago.edu/

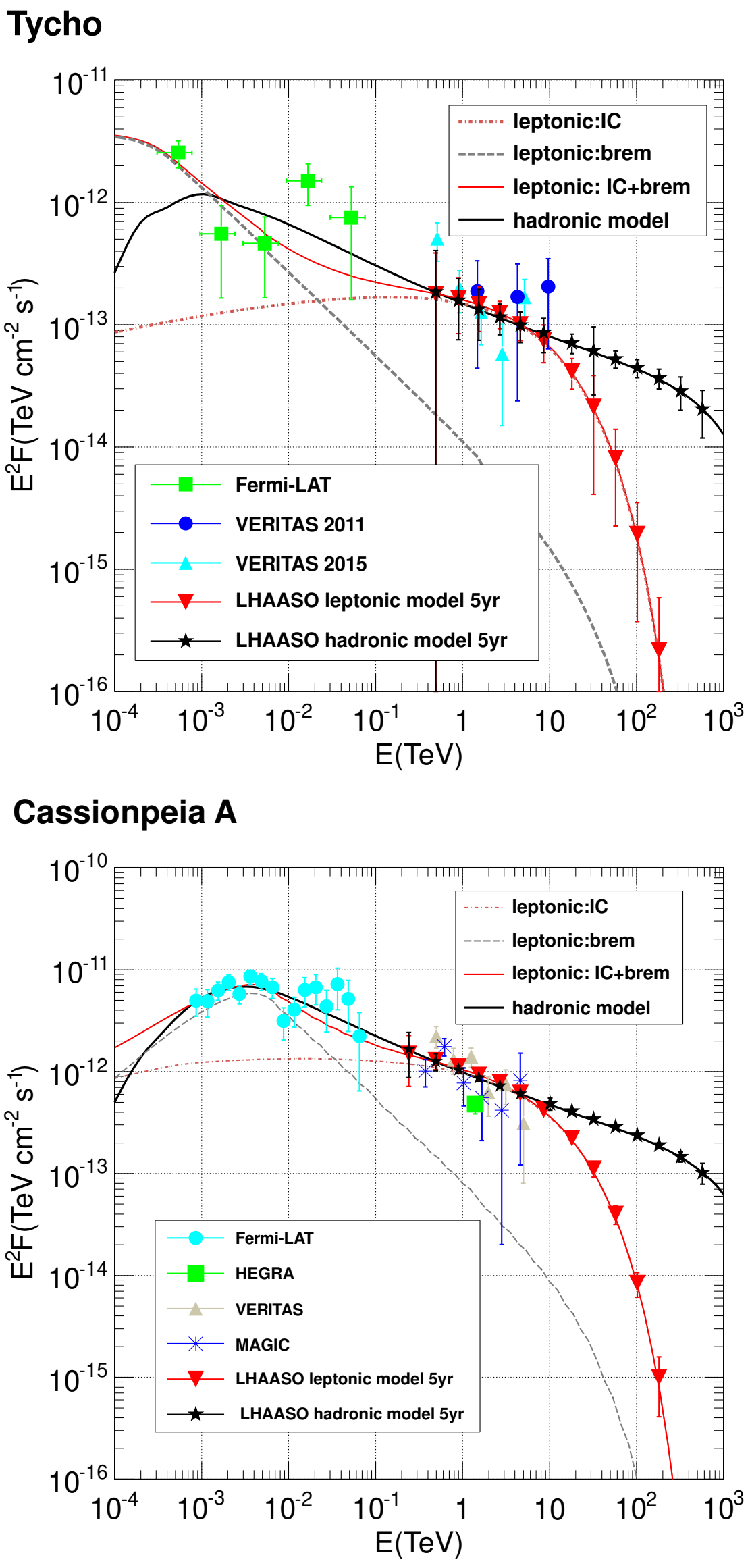

Fig. 8. Expectation of the LHAASO project on the historical SNR spectrum [1].

At the LHAASO site, the effective observation time is 6.2 hours per day for Tycho and 6.8 hours per day for Cas A with zenith angle less than 45°. Tycho culminates with a zenith angle of 34° and Cas A culminates with a zenith angle of 29°. The expected spectrum of Cas A from 0.3 TeV to 1 PeV is shown in Fig.8, we can see that from 300 GeV to 500 TeV, the statistic error of data obtained by LHAASO will be less than 10%. Due to the Klein-Nishina effect, the spectrum dominated by electrons is much softer than the hadronic emission above 10 TeV, and the expected result of LHAASO with a low statistic errors can give a reasonable explanation on the high energy range. These estimations show that the LHAASO observation would be just sufficient not only to give the final judgement for the hadronic/leptonic models but also to confirm whether the historical SNRs are PeVatrons or not.

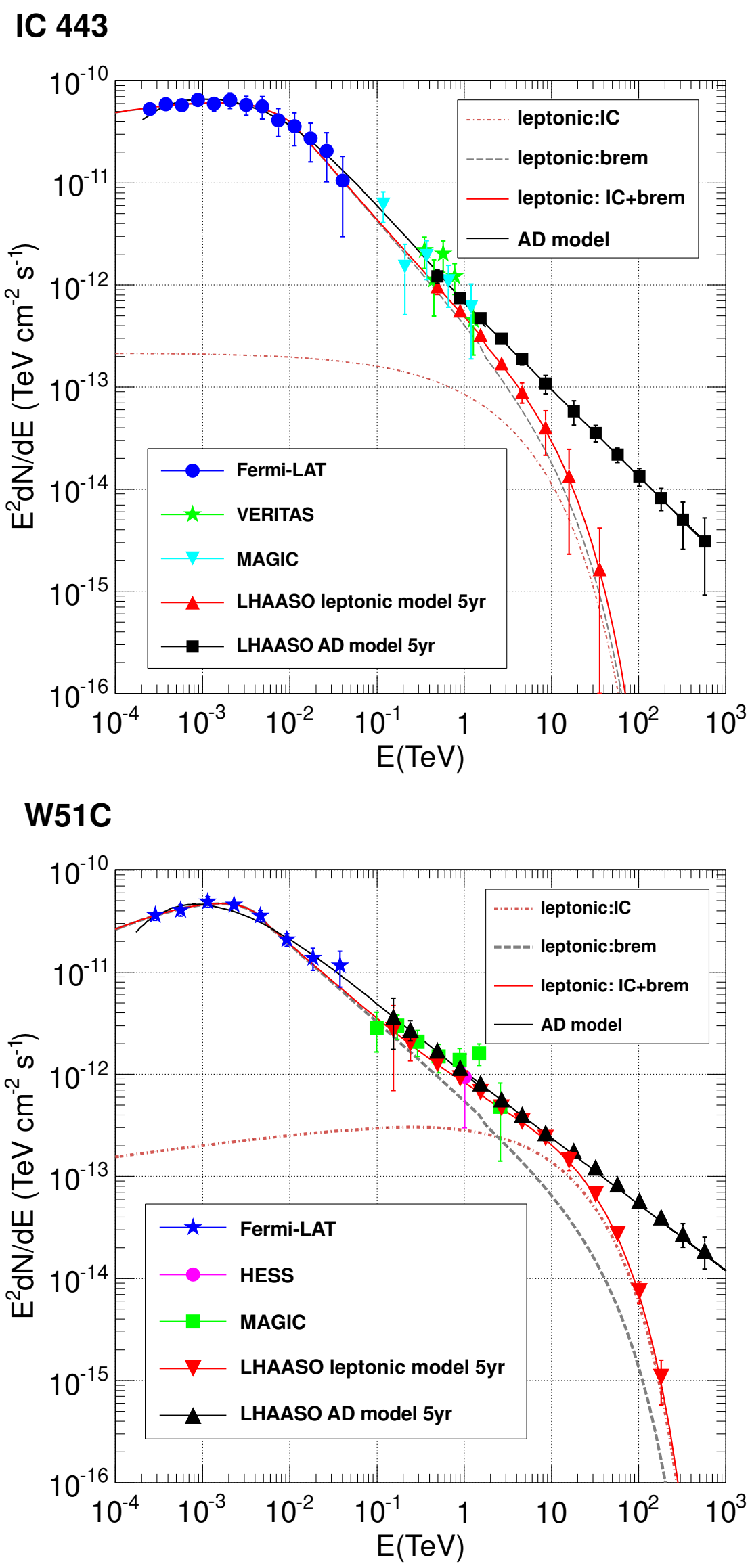

Fig. 9. Expectation of the LHAASO project on SNRs interaction with molecular clouds spectrum [1].

Middle-aged SNRs that are associated with $\gamma$-ray emission are usually in interaction with molecular clouds and feature hadronic emission in $\gamma$-rays. As one of the well studied middle-aged SNRs, IC 443 possesses strong molecular line emission regions that makes it a case for an SNR interacting with molecular clouds. The X-ray emission of IC 443 is primarily thermal and peaked towards the interior of the northeast shell, indicating that IC 443 is a thermal composite or mixed-morphology SNR. Fermi [70] in the GeV band and VERITAS [36], and MAGIC [71] in the TeV band detected the $\gamma$-ray emission from IC 443 and obtained the spectra up to 1 TeV, but there is not yet observation at higher energies, which is very important for determination on $\gamma$-ray emission mechanism.

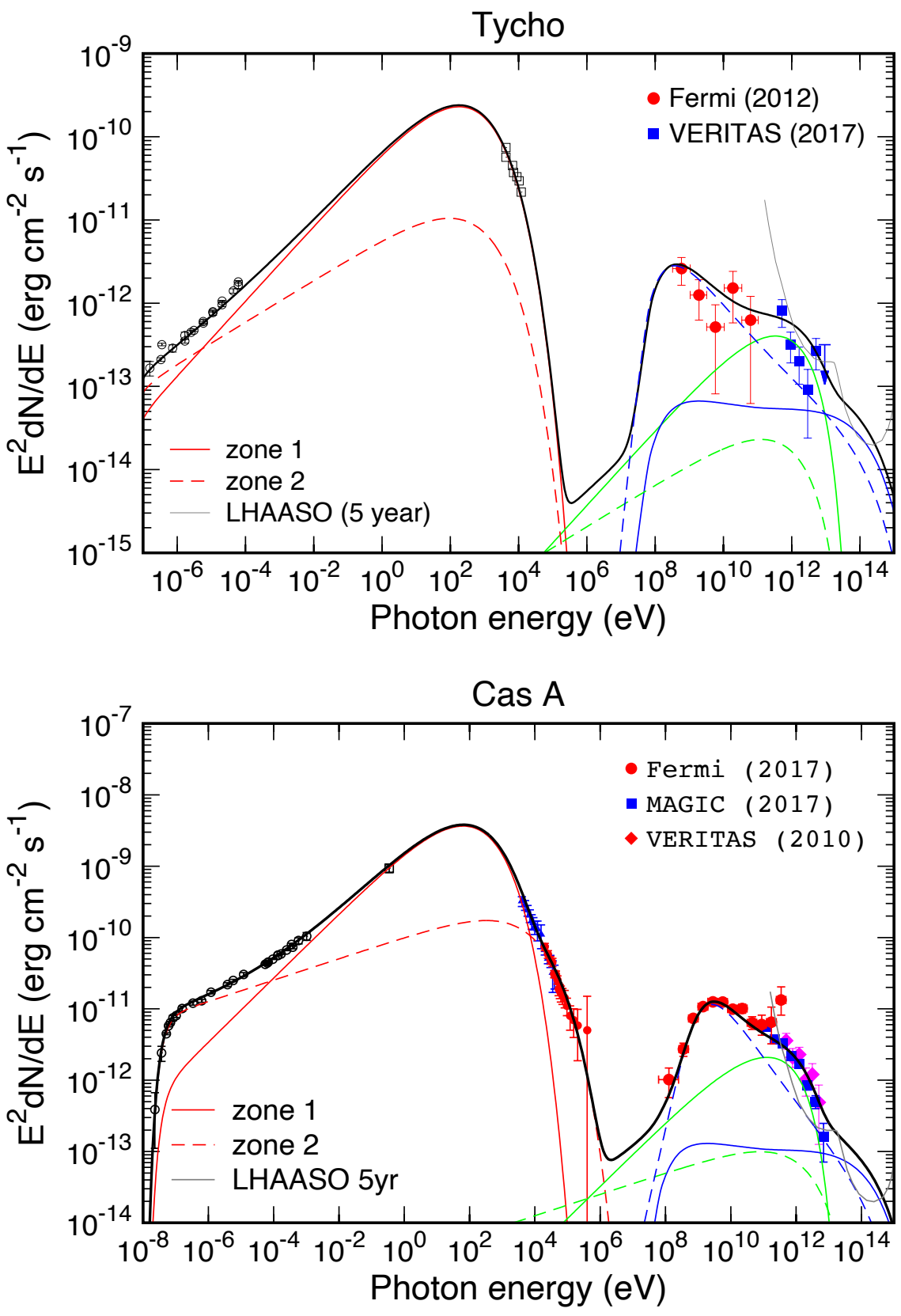

Fig. 10. SED of SNR Cas A (top) and Tycho (bottom). The black solid line represents the total emission from zone1 (solid) and zone 2 (dashed) with components: synchrotron (red), IC (green), and p-p collision (blue).

The middle-aged SNR W51C (G49.2-0.7) also interacts with the molecular clouds. The W51 region was heavily studied as it is known to host several objects. It contains three main components: two star-forming regions W51A and W51B surrounded by very giant molecular cloud, and SNR W51C. W51C is a radio-bright SNR at a distance of 6 kpc from Earth with an estimated age of $\sim 3 \times 10^4$ yr [72]. A smaller distance of 4.3 kpc is also suggested based on the HI absorption results [73], which is supported by the simulation of the radio morphology [74]. W51C is visible in X-rays showing both a shell type and center filled morphology. Shocked atomic and molecular gases have been observed, providing direct evidence on the interaction of W51C shock with a large molecular cloud [35, 75]. The GeV spectral result provided by Fermi indicates that leptonic model is difficult to explain $\gamma$-rays production and the most reasonable explanation is that hadronic interaction taking place at the shocked shell of W51C emits GeV $\gamma$-rays [76]. Moreover, MAGIC and H.E.S.S. also indicates the $\gamma$-ray emission from W51C tends to be dominated by $\pi^0$-decay up to several TeV [35, 75, 77]. But this still has uncertainties for the acceleration mechanism above 10 TeV.

At the LHAASO site, the effective observation time is 6.53 hours per day for IC 443 and 6.0 hours per day for W51C with zenith angle less than 45°. IC 443 culminates with a zenith angle of 8° and W51C culminates with a zenith angle of 16°. The expectation of LHAASO is given in Fig.9, compared with the measurement of Fermi, MAGIC and VERITAS. From 300 GeV to 500 TeV, the statistic error of data obtained by LHAASO will be less than 10%. The discrepancy between the expectations from the two models will reach more than 5 sigma above 20 TeV. It indicates that LHAASO will make a great contribution to the acceleration measurement in the TeV range, providing the final judgement on leptonic or hadronic origin.

### 2.3 Are SNRs PeVatrons?

LHAASO will be powerful in showing whether Galactic SNRs are PeVatrons or not. Whether young SNRs are PeVatrons or nor may have an effect on their $\gamma$-ray spectra. With 158h of high quality data, MAGIC collaboration [31] updated the TeV gamma-ray spectrum of SNR Cas A and revealed a high-energy cutoff of 3.5 TeV with 4.6$\sigma$ significance. This spectral feature seems to rule out Cas A as a PeV particle accelerator if the TeV $\gamma$-ray emission has a hadronic origin. However, the cutoff also can be explained by the leptonic process in a two-zone model [78]. In this model, the electrons accelerated by the forward shock (zone 1) dominantly contribute the TeV $\gamma$-rays via the IC process, while the GeV $\gamma$-rays are mainly produced by the protons accelerated by the inward/reverse shock (zone 2) (see Fig. 10). Thus, the proton spectrum does not need a cutoff, implying that Cas A can still be treated as a PeVatron. Moreover, the hadronic $\gamma$-rays from zone 1 can dominate the hundreds of TeV range if the total energy in the relativistic protons accelerated by the forward shock reaches the order of $10^{48}$ erg, which is also sufficient to supply the high-energy component of CR ions in the frame of SNR origin of Galactic CRs [79]. This two-zone model also could be applied to the Tycho SNR and explain the very soft TeV spectrum observed by VERITAS (see the right panel of Fig. 10). The spectral data obtained with LHAASO can thus be used to determine the maximum energy to that energetic protons can be accelerated by the SNR shock wave.

Using the hadronic interaction model for the diffusive protons [51, 52], Fig. 11 shows the expected hadronic spectra of the middle-aged SNRs W28, W41, W51C, and CTB37A for a proton energy cutoff at 3PeV, which are within the detection ability at the TeV photon energy for 5-yr LHAASO observation.

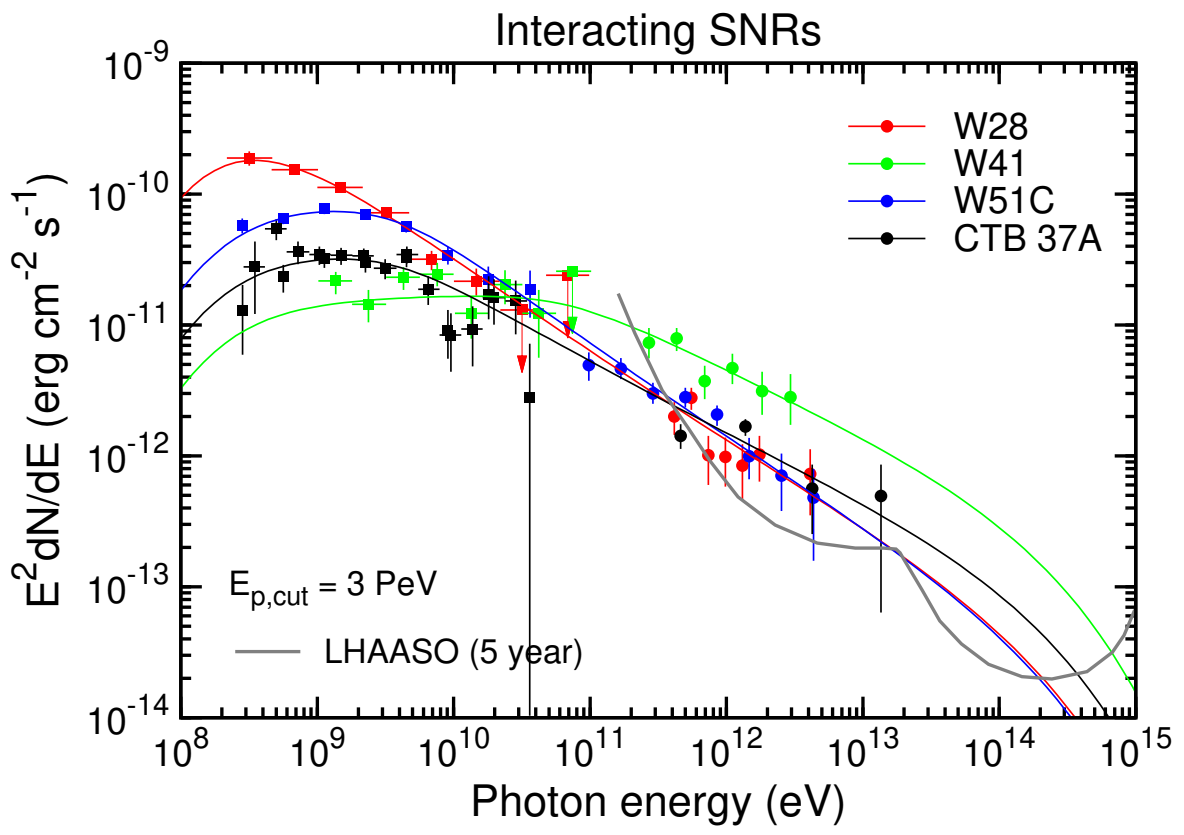

Fig. 11. Hadronic emission spectra expected for four SNRs that interact with molecular clouds using the diffusive proton model [51, 52].

## 3 Star-forming Regions

Star-forming regions are the factories of stars, containing young OB stars and related super-bubbles with strong collective stellar winds. The wind shocks and turbulence created by the collective stellar winds can accelerate particles to the relativistic regime. So they are the potential CR sources. On the one hand, the recent measurements of $^{60}$Fe abundance in CRs [80] indicate that a substantial fraction of CRs could be accelerated in young OB star clusters and related super-bubbles. Furthermore, the measurements of the Galactic diffuse $\gamma$-ray emission show that the CRs have a similar radial distribution as OB stars rather than SNRs [81, 82]. On the other hand, super-bubbles do have sufficient kinetic energy, supplied by supernova explosions therein or collective stellar winds, to provide the flux of the locally measured CRs [83]. Meanwhile, theses objects should be visible in $\gamma$-rays due to the freshly accelerated CRs interacting with ambient gas. In this regard a principal question is whether these objects can operate also as PeVatrons, i.e. whether they can provide the bulk of the locally observed CRs up to the so-called knee around 1 PeV. The most straightforward and unambiguous answer to this question would be the detection of $\gamma$-rays with a hard energy spectrum extending to energies well beyond 10 TeV.

### 3.1 Cygnus region

The Cygnus region of the Galactic plane is the famous region in the northern sky for the complex features observed in radio, infrared, X-rays, and $\gamma$-rays. It contains a high density interstellar medium and is rich in potential CR acceleration sites such as Wolf-Rayet stars, OB associations, and SNRs. This region is home of a number of GeV $\gamma$-ray sources detected by Fermi-LAT [84] and several noteworthy TeV $\gamma$-ray sources detected by Milagro, ARGO-YBJ in the past decade. The Cygnus Cocoon, located in the star-forming region of Cygnus X, is interpreted as a cocoon of freshly accelerated CRs related to the Cygnus super-bubble. The extended TeV $\gamma$-ray source ARGO J2031+4157 (or MGRO J2031+41) is positionally consistent with the Cygnus Cocoon discovered by Fermi-LAT at GeV energies in the Cygnus super-bubble, and another TeV source MGRO J2019+37 is a mysterious source only being detected by MILAGRO [85, 86] above 20 TeV and VERITAS [14] above 1 TeV. The reason for the hard SED from such a spatially extended region is totally unknown. The discovery of this kind of sources and the more detailed multi-wavelength spectroscopic investigations can be an efficient way to explain the radiation mechanism of them.

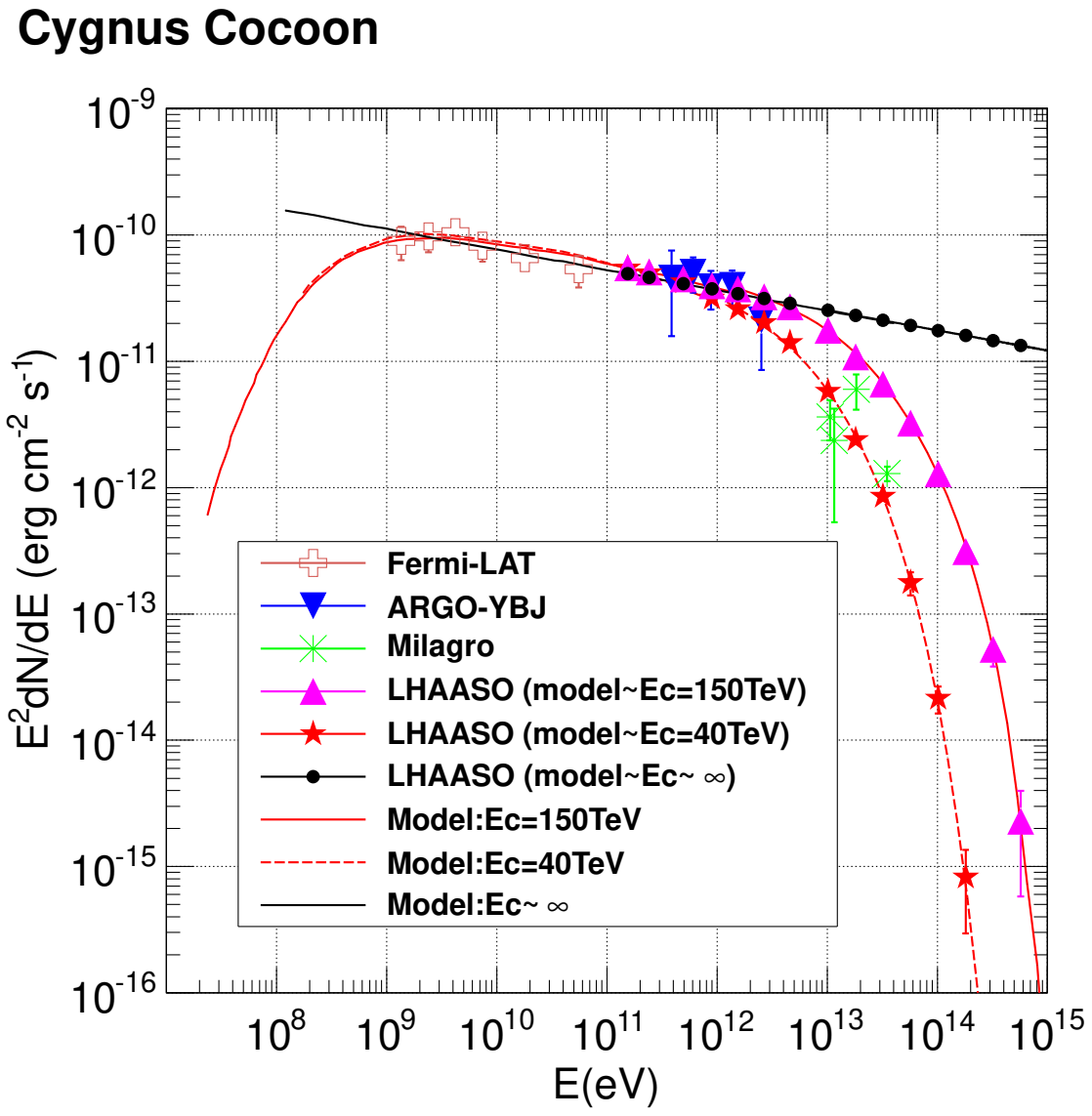

Fig. 12. Expectation of the LHAASO project on Cygnus Cocoon by using one year MC data [1], compared with the measurement of Fermi-LAT [87], ARGO-YBJ [88], Milagro[85, 86].

Figure 12 shows all the spectral measurements by Fermi-LAT [87], ARGO-YBJ [88], Milagro [86], and the expectation results with LHAASO. One year observation of LHAASO will be sufficient to give a judgement on the different energy cutoff models from 300 GeV to several hundred TeV. It will provide important information for investigating the particle acceleration within the superbubble.

### 3.2 W49A: a Galactic mini-starburst

As a part of the W49 complex [89], the powerful thermal radio continuum source W49A is one of the brightest Galactic giant radio H II regions ($\sim 10^7 L_\odot$) and is identified as an active star-forming region. It is located in a giant molecular cloud with a total mass of $\sim 10^6 M_\odot$ [90, 91] and is the best Galactic analog to the starburst phenomenon seen in other galaxies. This region contains $\sim$ 40 ultracompact H II regions, each hosting at least one massive star (earlier than B3) [92], and the brightest water maser cluster in our Galaxy [93]. Based on the proper motion of the strong $H_2O$ masers, the distance is estimated to be 11.4±1.2 kpc [94]. These massive stars can output a copious amount of kinetic energy via stellar winds, which may be sufficient to accelerated CRs. Two expanding shells as well as remnants of two gas ejections were found in W49A [95]. The shells may be driven by the massive stars and have a total kinetic energy of $\sim 10^{49}$ ergs. The gas ejections are likely to have the same origin as the expanding shells and a total energy of $\sim 10^{50}$ ergs. All these observational results make it as a likely potential $\gamma$-ray source. Indeed, the observations of HESS telescopes toward the direction of W49A reveal an excess of TeV $\gamma$-rays with a significance of more than $4.4\sigma$ [37], although the GeV emission has not been reported.

However, another star-forming region NGC 3603 was detected by Fermi-LAT as an extended source with radius of 1.1° at a significance level of more than $\sim 10\sigma$ [96]. Although NGC 3603 is not located in the field of view of LHAASO, its properties in the GeV band may give some clues to explore the TeV $\gamma$-rays for the other star-forming regions. The spectrum of NGC 3603 in energy range from 1 to 250 GeV didn't show any sign of cutoff and can be well fitted with a single power law with a photon index of $\Gamma \approx 2.3$, indicating the existence of the particles with multi-TeV energies at least. In Figure 13, the Fermi-LAT data are modeled via the hadronic scenario with different proton cutoff energy. As can be seen, LHAASO observation toward the other star-forming regions, including W49A, may help us to answer whether energy particles can be accelerated up to PeV in the star-forming regions.

Further consideration of LHAASO targets of candidate PeVatrons harbored in star forming regions will be given in §**7.3**.

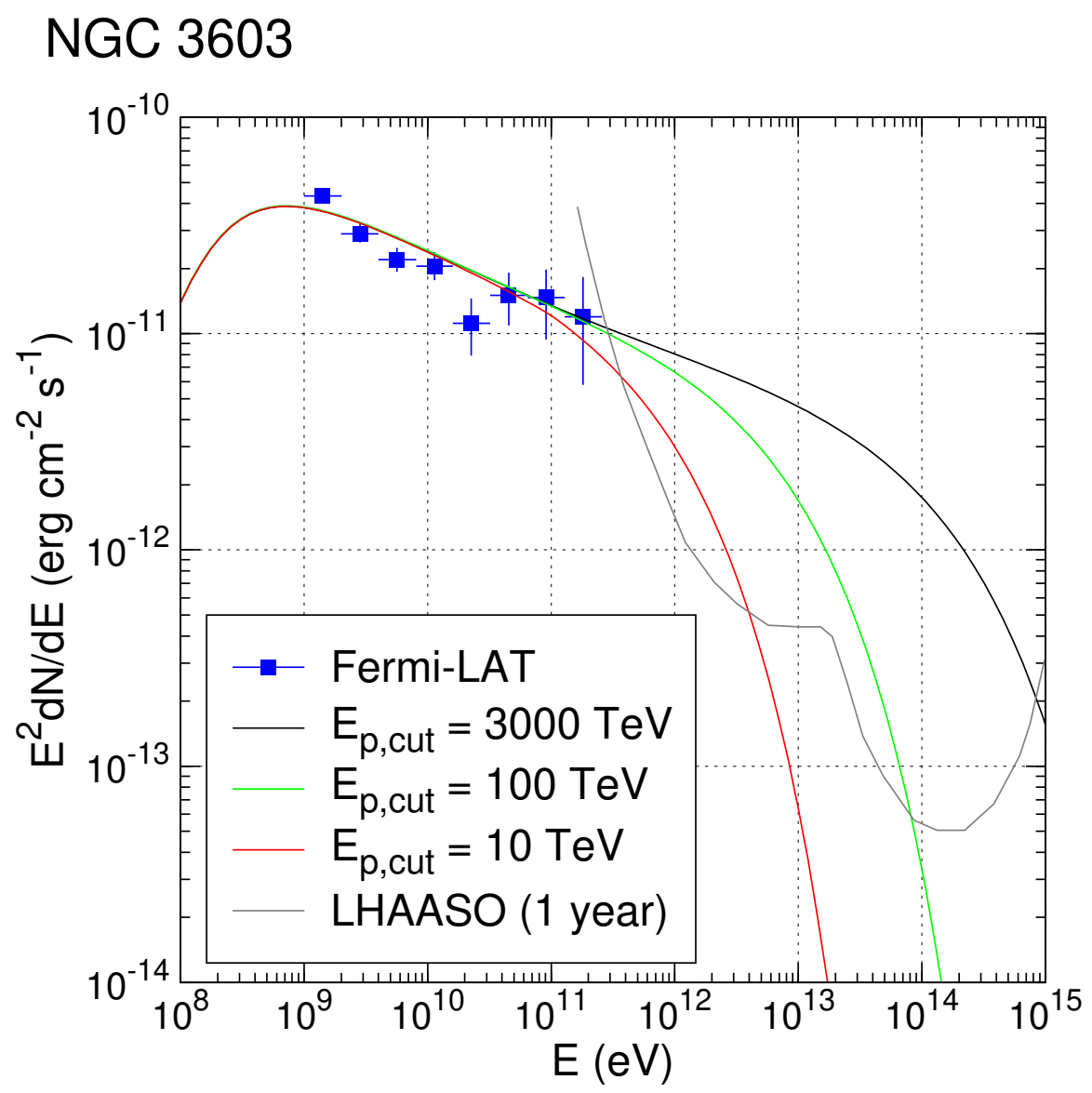

Fig. 13. Modeling the Fermi-LAT data of source NGC 3603 [96] with the different proton cutoff energies: 3000 TeV (black), 100 TeV (green) and 10 TeV (red), compared with the LHAASO's sensitivity curve (gray).

## 4 Pulsars and Pulsar Wind Nebulae

### 4.1 High-energy TeV emission from pulsars

Thanks to *Fermi Gamma-ray Space Telescope (Fermi)*, which was launched in 2008 June, we have learned from its observations that pulsars are the dominant $\gamma$-ray $0.1-100$ GeV sources in our Galaxy [40]. Thus far more than 200 pulsars have been detected by *Fermi*, and from the studies we now know that pulsars generally have $\gamma$-ray emission described by a power law with exponential cutoff at several GeV. Such a spectral shape matches the theoretical expectations, as the emission arises due to curvature radiation from the magnetosphere (near the magnetic poles) of a pulsar (e.g., [97]). It was certainly a surprise when 100 GeV pulsed emission from the Crab pulsar was detected by VERITAS [98], and recently the MAGIC Collaboration has recorded the photons with energies up to 1.5 TeV [99]. In addition, pulsed photons above 50 GeV from the Vela pulsar were also detected [100]. Is such high energy emission only seen from the brightest, young pulsars? Not really! In a recent paper, [101] has reported the detection of up to 200 GeV photons from an old, so-called millisecond pulsar (MSP; they spin rapidly, at periods of several milliseconds).

The detection of photons above 100 GeV challenges the theoretical understanding of the pulsar emission mechanisms, because all the pulsar emission models predict a cutoff in the curvature radiation of pulsars as large as ∼100 GeV. Currently the IC scattering process in the outer magnetosphere or the pulsar wind region is considered to produce the pulsed emission detected in the ≥10 GeV band from the Crab pulsar (see, e.g., [102, 103]). Alternatively, a non-stationary outer-gap scenario has also been proposed recently [104], in which the observed spectrum of a pulsar is the superposition of emission from the variable outer gap structures.

LHAASO will certainly explore the high-energy TeV emission from pulsars, helping by finding a full sample of them and setting constraints for theoretical modeling. We note that high-energy $\gamma$-ray emission is seen from 27 pulsars, as reported in the first *Fermi* catalog of sources above 10 GeV [105]. Among them 20 sources were found to have pulsed $\gamma$-ray emission in the >10 GeV band, including 17 young pulsars and three MSPs. These sources could be good targets for LHAASO.

### 4.2 Pulsar wind nebulae

Pulsars are powered by their fast rotation, and most of the rotational energy of a pulsar is released in a form of the pulsar wind (see, e.g., [106]). The high-energy, relativistic particles in the pulsar wind interact with the ambient medium around a pulsar forming a terminal wind shock. Particles at the shock emanate synchrotron radiation, making the PWN bright from radio to X-ray energies. At GeV and TeV $\gamma$-ray energies, it is believed that the IC scattering process gives rise to emission, with Lorentz factor of $\sim 10^6$ electrons up-scattering background infrared photons to GeV/TeV range. The modeling of a broad-band spectrum of a PWN thus allows us to study its particle population, magnetic field, and dynamical evolution (after the birth of the pulsar; e.g., [107–109]). Especially, when a PWN enters the last evolutionary stage in which the pulsar escaped from its parent SNR or its parent SNR almost disappeared, the original PWN will first dim in the low energy band, leaving an extended TeV emission region called "TeV halo" [110–112] (see §**4.3** for more details). Thus far, more than 30 PWNe or candidates have been detected at TeV energies, and *Fermi* has been able to detect a few of them [113]. Particularly, the first astrophysical source detected above 100 TeV is the prototype PWN Crab, which was achieved by the Tibet AS+MD [114] and HAWC [115] experiments. More recently, HAWC collaboration released a ≥56 TeV catalog [116] containing nine sources which all are likely related to pulsars and may be the so-called TeV halos. With LHAASO's great sensitivity around 100 TeV and large-sky area monitoring capability, it is conceivable that more PWNe or halos will be detected, allowing to obtain a full sample of them in the northern sky.

Apart from SNRs, PWNe are also believed to be a kind of Galactic CR source. According to the Hillas criteria [117], the particles with energy below the knee energy can be effectively trapped by the magnetic fields of PWNe. Thus, PWNe could store a large amount of energy in relativistic protons if pulsars or PWNe could continuously produce energetic protons. Based on the outer magnetospheric gap model, Cheng et al. [118] pointed out that the Crab pulsar can produce relativistic protons if $\vec{\Omega}\cdot\vec{\mu}>0$, where $\vec{\Omega}$ and $\vec{\mu}$ are the angular velocity and magnetic moment of the star, respectively. It is recently suggested that the PWNe inside SNRs can further accelerate the relativistic protons that were accelerated by the SNR shocks up to an energy of order 1 PeV, and hence such PWNe may also be PeVatrons [119]. If a PWN is located in a dense environment and contains relativistic protons, the hadronic emission from the energetic protons may have a significant contribution to the GeV-TeV $\gamma$-rays [120]. Indeed, the lepton-hadronic model has been applied to some PWNe to explain their broadband spectra, e.g. in the cases of Vela X [121], G54.1+0.3 [122], DA 495 [123], and G106.3+2.7 [124, 125]. Of course, it is also pointed out that protons can only take away the spin-down energy of pulsars with a very small fraction [126, 127]. With LHAASO's great capacity of detecting $\gamma$-rays up to energy of ∼100 TeV, it may help us testing the proton acceleration in PWNe.

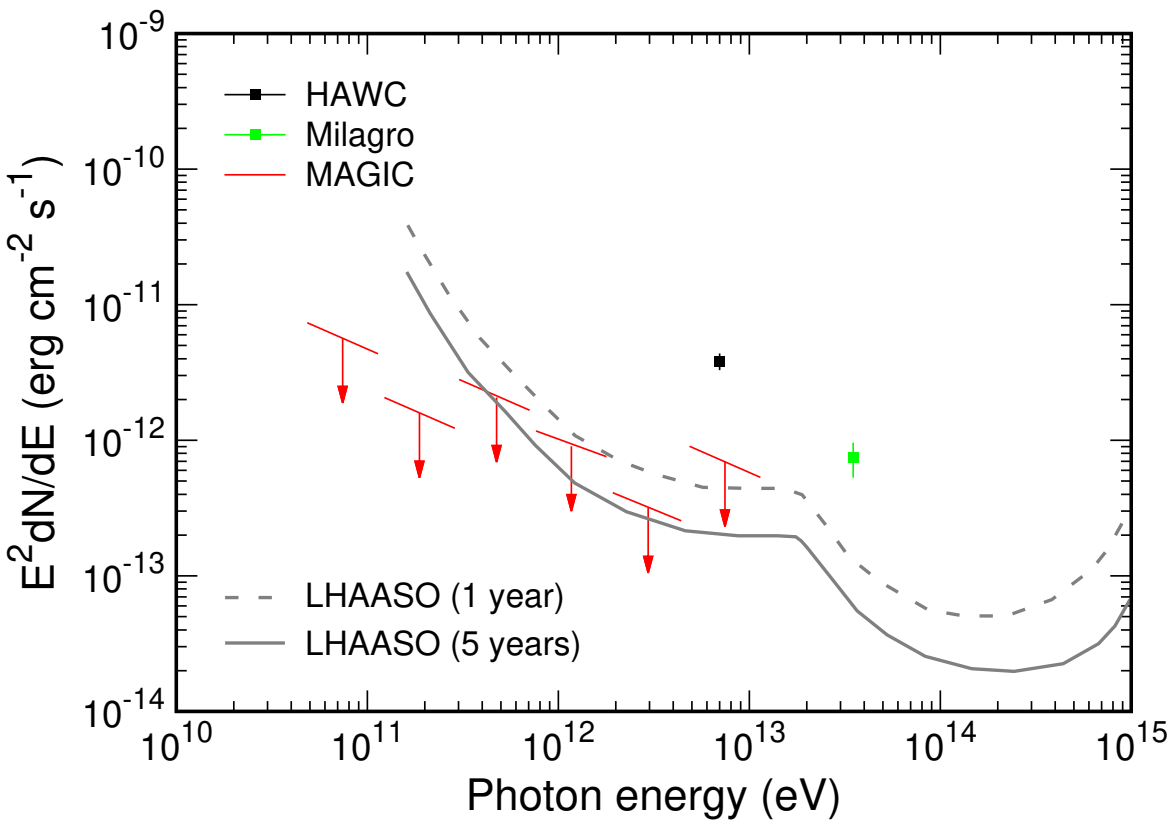

Fig. 14. The spectrum of the nebula around the Geminga pulsar measured by Milagro [128], HAWC [129] and MAGIC [130].

Space experiments (PAMELA [131], Fermi [132], AMS-02 [133]) have revealed an excess of high-energy positrons relative to the standard predictions for secondary production in the interstellar medium (ISM). In order to explain this positron excess, it can be confirmed that significant quantities of TeV positrons should be produced within the local volume (the surrounding ∼ kpc), but the source of positrons is still in debates. Pulsars and/or PWNe are widely suggested to be the domi-nant sources of the local population of TeV electrons and positrons, which can account for the observed positron excess [134–138]. Among the known pulsars, Geminga (PSR J0633+1746) and B0656+14 (PSR J0659+1414) are the potential sources due to their short distance to us. These two pulsars are relatively young (370 and 110 kyrs, respectively) and are located within a few hundred parsecs of the solar system ($250^{+230}_{-80}$ and $280^{+30}_{-30}$ pc, respectively [139]). The electrons and positrons released by the pulsars can diffuse into the surrounding medium and produce $\gamma$-rays. Indeed, the extended TeV $\gamma$-ray emission (2°–3° radius) surrounding the Geminga pulsar, a prototype TeV halo, has been reported by Milagro [128] and HAWC [129], although the observations by the MAGIC telescopes show no significant detection above 50 GeV [130]. The extended TeV $\gamma$-ray emission from B0656+14 also has been detected by HAWC [129]. Based on the HAWC results, Hooper et al. [138] calculate the expected contributions from the two pulsars to the local positron spectrum via fitting the $\gamma$-ray spectrum and conclude that pulsars are likely sources of the local TeV positrons. In Figure 14, the observed results for Geminga and the LHAASO's sensitivity are shown. As can be seen, LHAASO has the ability to accurately measure the $\gamma$-ray spectrum from 200 GeV to 100 TeV, which will give more stronger constraints on the properties of these pulsars and test the pulsar scenario of the positron excess, thus settling the dispute between the MAGIC and HAWC observations.

¶it was called "TeV PWN" in the early literature.

Table 1. The top 30 bright middle-aged pulsars within the field of view of LHAASO. The parameters of the pulsars are given by the ATNF catalog

| NAME | RA (°) | Dec (°) | $l$ (°) | $b$ (°) | $r$ (kpc) | $t$ (100 kyr) | $\dot{E}$ ($10^{34}$erg s$^{-1}$) | $\dot{E}/r^2$ ($10^{34}$ergs$^{-1}$kpc$^{-2}$) | Comments* |
|---|---|---|---|---|---|---|---|---|---|
| J0633+1746 | 98.5 | 17.8 | 195.1 | 4.3 | 0.19 | 3.42 | 3.25 | 90.03 | Geminga, detected by HAWC |
| B0656+14 | 105.0 | 14.2 | 201.1 | 8.3 | 0.29 | 1.11 | 3.80 | 45.18 | detected by HAWC |
| B1951+32 | 298.2 | 32.9 | 68.8 | 2.8 | 3.00 | 1.07 | 374 | 41.56 | with X-ray PWN, missed in TeV |
| J1954+2836 | 298.6 | 28.6 | 65.2 | 0.4 | 1.96 | 0.69 | 105 | 27.33 | detected by Milagro |
| J1740+1000 | 265.1 | 10.0 | 34.0 | 20.3 | 1.23 | 1.14 | 23.2 | 15.33 | with X-ray PWN, missed by HAWC |
| J1913+1011 | 288.3 | 10.2 | 44.5 | -0.2 | 4.61 | 1.69 | 287 | 13.50 | detected by HESS,YBJ,HAWC |
| J1836+5925 | 279.1 | 59.4 | 88.9 | 25.0 | 0.30 | 18.3 | 1.14 | 12.67 | missed in TeV |
| J2032+4127 | 308.1 | 41.5 | 80.2 | 1.0 | 1.33 | 2.01 | 15.2 | 8.59 | detected in X-ray,TeV |
| J1928+1746 | 292.2 | 17.8 | 52.9 | 0.1 | 4.34 | 0.83 | 160 | 8.49 | detected by HAWC? |
| J1831-0952 | 277.9 | -9.9 | 21.9 | -0.1 | 3.68 | 1.28 | 108 | 7.97 | detected by HESS,HAWC |
| B0114+58 | 19.4 | 59.2 | 126.3 | -3.5 | 1.77 | 2.75 | 22.1 | 7.05 | |
| J0633+0632 | 98.4 | 6.5 | 205.1 | -0.9 | 1.35 | 0.59 | 11.9 | 6.53 | detected by HAWC |
| J0248+6021 | 42.1 | 60.4 | 136.9 | 0.7 | 2.00 | 0.62 | 21.3 | 5.33 | |
| B0355+54 | 59.7 | 54.2 | 148.2 | 0.8 | 1.00 | 5.64 | 4.54 | 4.54 | the Mushroom X-ray PWN |
| J1938+2213 | 294.6 | 22.2 | 57.9 | 0.3 | 3.42 | 0.62 | 36.6 | 3.13 | |
| J0538+2817 | 84.6 | 28.3 | 179.7 | -1.7 | 1.30 | 6.18 | 4.94 | 2.92 | X-ray PWN, missed by HAWC? |
| B1830-08 | 278.4 | -8.5 | 23.4 | 0.1 | 4.50 | 1.47 | 58.4 | 2.88 | with X-ray PWN |
| J2043+2740 | 310.9 | 27.7 | 70.6 | -9.2 | 1.48 | 12.0 | 5.64 | 2.57 | |
| J2021+4026 | 305.4 | 40.4 | 78.2 | 2.1 | 2.15 | 0.77 | 11.6 | 2.51 | detected in X-ray,TeV |
| J1857+0143 | 284.4 | 1.7 | 35.2 | -0.6 | 4.57 | 0.71 | 45.1 | 2.16 | detected by HESS,HAWC |
| B0611+22 | 93.6 | 22.5 | 188.8 | 2.4 | 1.74 | 0.89 | 6.24 | 2.06 | |
| J1841-0345 | 280.4 | -3.8 | 28.4 | 0.4 | 3.78 | 0.56 | 26.9 | 1.88 | |
| J1913+0904 | 288.3 | 9.1 | 43.5 | -0.7 | 3.00 | 1.47 | 16.0 | 1.78 | |
| B0540+23 | 85.8 | 23.5 | 184.4 | -3.3 | 1.56 | 2.53 | 4.09 | 1.68 | detected by HAWC |
| J1846+0919 | 281.6 | 9.3 | 40.7 | 5.3 | 1.53 | 3.60 | 3.41 | 1.46 | |
| J0611+1436 | 92.8 | 14.6 | 195.4 | -2.0 | 0.89 | 10.7 | 0.80 | 1.01 | |
| J0357+3205 | 59.5 | 32.1 | 162.8 | -16.0 | 0.83 | 5.40 | 0.59 | 0.85 | missed by ASgamma |
| J1838-0549 | 279.7 | -5.8 | 26.3 | 0.2 | 4.06 | 1.12 | 10.1 | 0.61 | |
| B0919+06 | 140.6 | 6.6 | 225.4 | 36.4 | 1.10 | 4.97 | 0.68 | 0.56 | |
| J1835-0944 | 278.9 | -9.7 | 22.5 | -1.0 | 4.22 | 5.25 | 5.64 | 0.32 | |

### 4.3 TeV gamma-ray halos

In continuation of the above discussion of TeV halos, more specific issues are detailed here.

#### 4.3.1 Very slow-diffusion halos around pulsars

In late 2017, the HAWC collaboration reported the spatially resolved measurement of the $\gamma$-ray halos around the Geminga pulsar and PSR B0656+14 [129]. Unlike $\gamma$-ray PWNe, which are confined structures around young pulsars, these two $\gamma$-ray halos should be generated by free electrons and positrons ($e^{\pm}$s) diffusing out from the corresponding PWNe. The Geminga pulsar and PSR B0656+14 are middle-aged pulsars (0.1–1 Myr). According to the evolution model of PWNe [106], the original PWNe of these pulsars were already broken long time ago, and now these pulsars are traveling in the ISM, driving bow-shock PWNe with scales much smaller than the TeV halos. This is consistent with the X-ray observations [140, 141]. In the ISM, the observed multi-TeV $\gamma$-rays are mainly emitted by $e^{\pm}$s through IC scattering of the homogeneous CMB, so the $\gamma$-ray spatial profile should unambiguously indicate the diffusion speed of the escaping $e^{\pm}$s. This may be so far the most straightforward measurement of CR propagation in local regions of the ISM.

‖It is difficult for imaging atmospheric Cherenkov telescopes (IACTs) to detect the large extended sources.

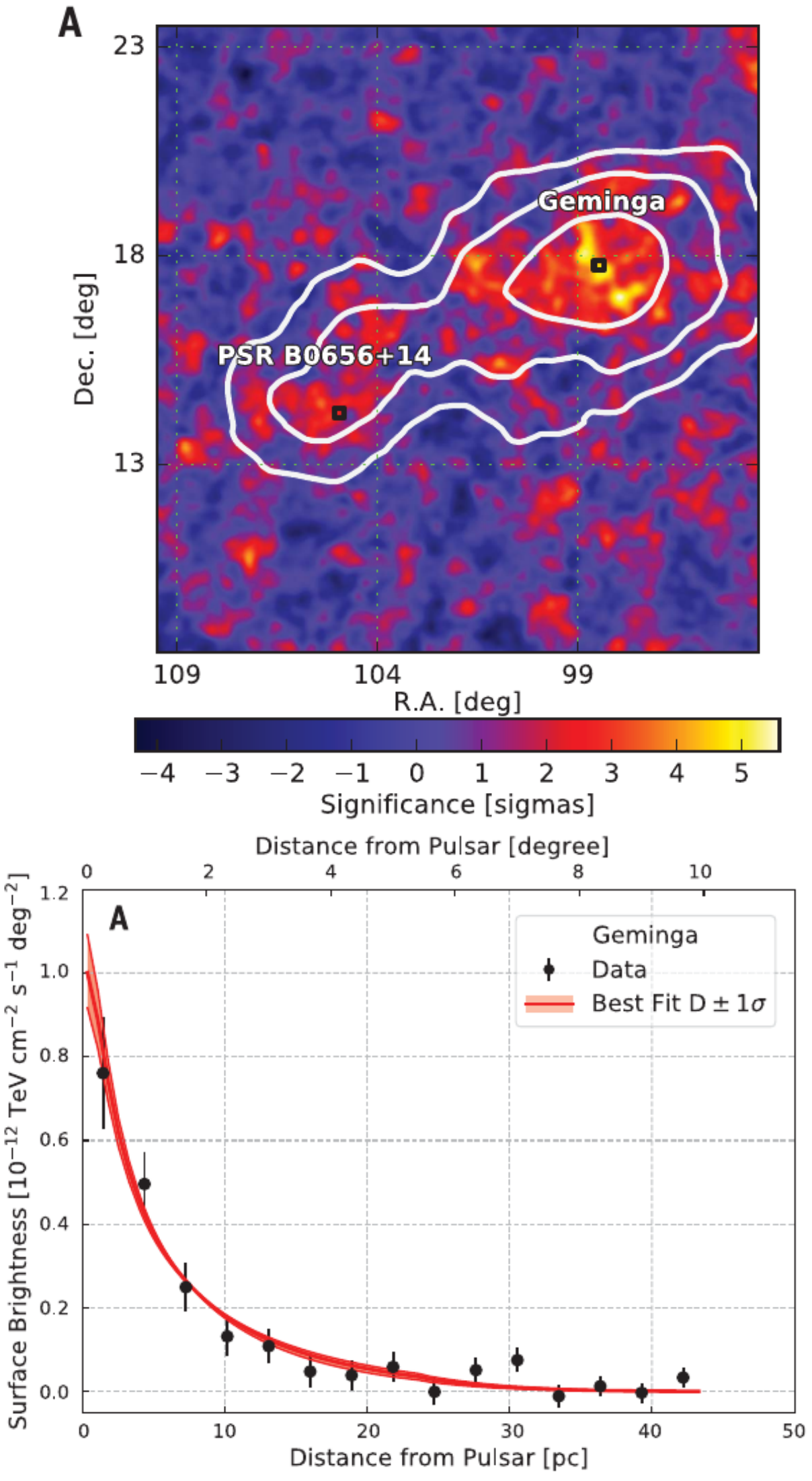

Fig. 15. Top: HAWC significance map for the region around Geminga and PSR B0656+14. Bottom: surface brightness profile around Geminga. Both adopted from [129].

However, the diffusion coefficient derived from these halos is several hundred times smaller than that indicated by the CR boron to carbon ratio (B/C) [142]. The latter is currently the most important indicator of the global CR propagation in the Galaxy. With the diffusion coefficient derived from B/C, the CR antiproton spectrum and the Galatic diffuse $\gamma$-rays can be well reproduced (e.g. [143]). Thus, the slow diffusion around the two pulsars must not be representative in the whole Galaxy. Then what picture would it be? Does the slow diffusion only happen in the nearby region of pulsars? Or is it common in the Galactic disk? The answer of this question could be vital for investigating the origin of CRs, for example, the origin of the positron excess.

Geminga and PSR B0656+14 are generally believed to be the most likely sources of the positron excess [134, 136, 144], due to their close ranges and the capability of accelerating high-energy $e^{\pm}$s. However, if the slow diffusion is everywhere in the local environment of the solar system, positrons released by these pulsars can hardly arrive at Earth [129]. In this case, a very close pulsar is required to explain the positron excess [145–147]. In contrast, if the slow-diffusion regions are only limited in the vicinity of the pulsars, Geminga may still be a probable source of the positron excess [138], and even more optimistic than in the former fast-diffusion case [148].

There are also efforts in studying the possible origin of the slow-diffusion halos. The most straightfoward explanation may be the self-generated case, in which the slow diffusion is induced by the escaping $e^{\pm}$s from the pulsars through streaming instability [149]. While considering the proper motion of Geminga [150], the slow-diffusion halo must be a newly formed structure. The injection power in the recent age of Geminga is too small to significantly suppress the diffusion coefficient [151]. In contrast to the self-generated scenario, the slow diffusion region could be preexisting [151]. If the progenitor of Geminga was in a rarefied environment, the current scale of the associated SNR of Geminga could be large enough to include the Geminga pulsar and the TeV halo inside. Although the SNR is too old to be visible at present, it could be energetic enough to leave a turbulent environment for Geminga, which explains the slow diffusion. Besides, it is also possible that the TeV halos are not interpreted by a turbulent environment, but by the anisotropic diffusion of $e^{\pm}$s along the local regular magnetic field lines which are required to be aligned with the line of sight towards the pulsars [152]. All these interpretations should be tested by a larger sample of TeV halos.

#### 4.3.2 Significance of TeV-halo Observation

Currently, new TeV halos around pulsars (such as PSR J0633+0632 and PSR B0540+23) are detected by HAWC with high significance [153]. TeV halos are thus very likely to be common around middle-aged pulsars. Observation of this type of sources has significant mean-

ings:

- As mentioned in §4.3.1, TeV halos around middle-aged pulsars are ideal tools for studying the CR propagation. Firstly, the background level in the energy range of multi-TeV is much lower than that in low-energy $\gamma$-ray observations. This helps us to obtain clear $\gamma$-ray profiles around sources, which carry the information of particle propagation. Secondly, the observed TeV $\gamma$-rays are generated by very young $e^{\pm}$s. For example, 20 TeV $\gamma$-rays are emitted by $\sim 100$ TeV $e^{\pm}$s, the lifetime of which is only $\sim 7$ kyr. Thus, the $\gamma$-ray spatial distributions unambiguously indicate the *present* MHD status of the ISM around the pulsars. This means an evolution-independent model is enough to explain the data, in which we do not need to consider some complicated factors like the proper motion of the pulsars and the evolution of the ISM. Thirdly, compared with young SNRs, middle-aged pulsars should be better targets for the study of CR propagation. For the latter, the high-energy $e^{\pm}$ are accelerated by pulsars or PWNe, the sizes of which are significantly smaller than those of the TeV halos. So the propagation zone will not be mixed with the acceleration zone, and a clear pattern of CR propagation can be obtained.

- TeV halos could be indicators for invisible pulsars [110]. Pulsars are lighthouse-like emitters. We cannot detect the pulsed signal from pulsars if their electromagnetic emission is not beamed towards Earth. In this case, extended TeV halos could be implications for those misaligned pulsars. This is especially meaningful for relatively old pulsars whose host SNRs already faded. Follow-up observations in multiwavelengths would then provide further evidence about the origin of the halos.

- We can derive the $e^{\pm}$ injection spectra from the observation of TeV halos. The injection spectra include the information of particle acceleration and escaping in PWNe. Current observations show that even relatively old sources like Geminga can still accelerate particles to at least $\sim$100 TeV. Energy spectra of more halos will be measured in the future, which may reveal the relationship between the acceleration limit of PWNe and other parameters, such as the age of the pulsars.

#### 4.3.3 Candidate Objects for LHAASO

With its high sensitivity, wide energy range, and large field of view, LHAASO is very competent to perform deep investigation of TeV halos. In Table 1, we list the top 30 bright middle-aged pulsars in the ATNF catalog [154], which could be the targets of the future observation. Of course, we only choose those within the field of view of LHAASO ($-10^{\circ} < \mathrm{Dec} < 70^{\circ}$). The pulsars are ordered by $\dot{E}/r^2$, where $\dot{E}$ is the current spin-down luminosity of the pulsar and $r$ is the distance to the pulsar. Obviously, the two halos reported by HAWC [129] correspond to the two brightest pulsars. The pulsar age $t$ in the table is limited between 50 kyr and 10 Myr. As we mentioned in the beginning, what we focus on here are the halos generated by the free $e^{\pm}$s diffusing in the ISM, rather than the confined PWNe. For pulsars younger than $\sim 50$ kyr, the original PWNe may not be totally broken, which might be confused with the free diffusion halos. There is also discussion about the criterion of TeV halos [112].

It can be found from the "comment column" of Table 1 that current TeV experiments have already measured the top 10 bright sources. However, $\gamma$-ray halos are not significantly detected for three of them—PSR B1951+32, PSR J1740+1000, and PSR J1836+5925. We note that all of these three pulsars are relatively far from the Galactic plane (146 pc for B1951+32, 420 pc for J1740+1000, and 126 pc for J1836+5925). In fact, all the pulsars with height larger than 100 pc (from Galactic plane) in Table 1 are not detected with TeV halos so far. Does it imply that the slow-diffusion region is a very thin disk in the Galaxy which is possibly left by myriad old SNRs? Or is the non-detection due to other reasons? The future measurement of LHAASO may give an explanation to it.

## 5 Gamma-ray binaries

A new class of high-mass X-ray binaries (HMXBs) have been discovered as strong $\gamma$-ray emitters: PSR B1259−63, LS 5039, LS I +61° 303, HESS J0632+057, and 1FGL J1018.6−5856 (see [155] for a review). Other recent candidates such as PSR J2032+4127 have also been reported [156]. These $\gamma$-ray binaries contain a compact object orbiting an OB companion star, emitting non-thermal emission from radio to TeV $\gamma$-rays that are modulated on the orbital period. Studying the emissions from $\gamma$-ray binaries can probe the surroundings of compact objects at AU scale, which is a largely unexplored distance scale. The complexity of the immediate environment of $\gamma$-ray binaries also shed light on physical processes that are poorly understood.

The detection of very high energy (VHE) $\gamma$-rays (above 100 GeV) by the current imaging atmospheric Cherenkov Telescopes (IACTs) from all known $\gamma$-ray binaries, gives hint for very efficient particle acceleration in these systems. Indeed, there is no lack of particle acceleration sites for $\gamma$-ray binaries: the interaction of the pulsar wind (for those the compact object is a pulsar) with the strong wind of a massive star, accretion onto a compact object and/or jet activities (similar to micro-quasars), and a relativistic outflow interacting with the ISM at a larger scale. Micro-quasars or interacting stellar binaries are also observed to emit $\gamma$-rays above 60 MeV, e.g., in Cygnus X-1 [157], Cygnus X-3 [158], V 404 Cyg [159], and Eta Carinae [160].

$\gamma$-ray binaries, such as LS 5039, have a very high efficiency of particle acceleration. The very good sensitivity of LHAASO in the energy band of 10–100 TeV and above will allow us to probe the acceleration mechanism, the magnetic field strength, stellar wind densities, and short-term variability of the acceleration and/or radiation regions. This is because the opacity and orbital dependence of $\gamma$-$\gamma$ absorption, and the angular dependence of the IC emission, or other sources of variability, are less important in this energy range than in sub-TeV energy band. In addition, the spectrum of the emission also depends on whether the accelerated particles are leptons or hadrons.

LHAASO, being an excellent all-sky detector at the TeV to multi-TeV energies, are a good monitor of the TeV transient sky, including transient phenomena related to $\gamma$-ray binaries.

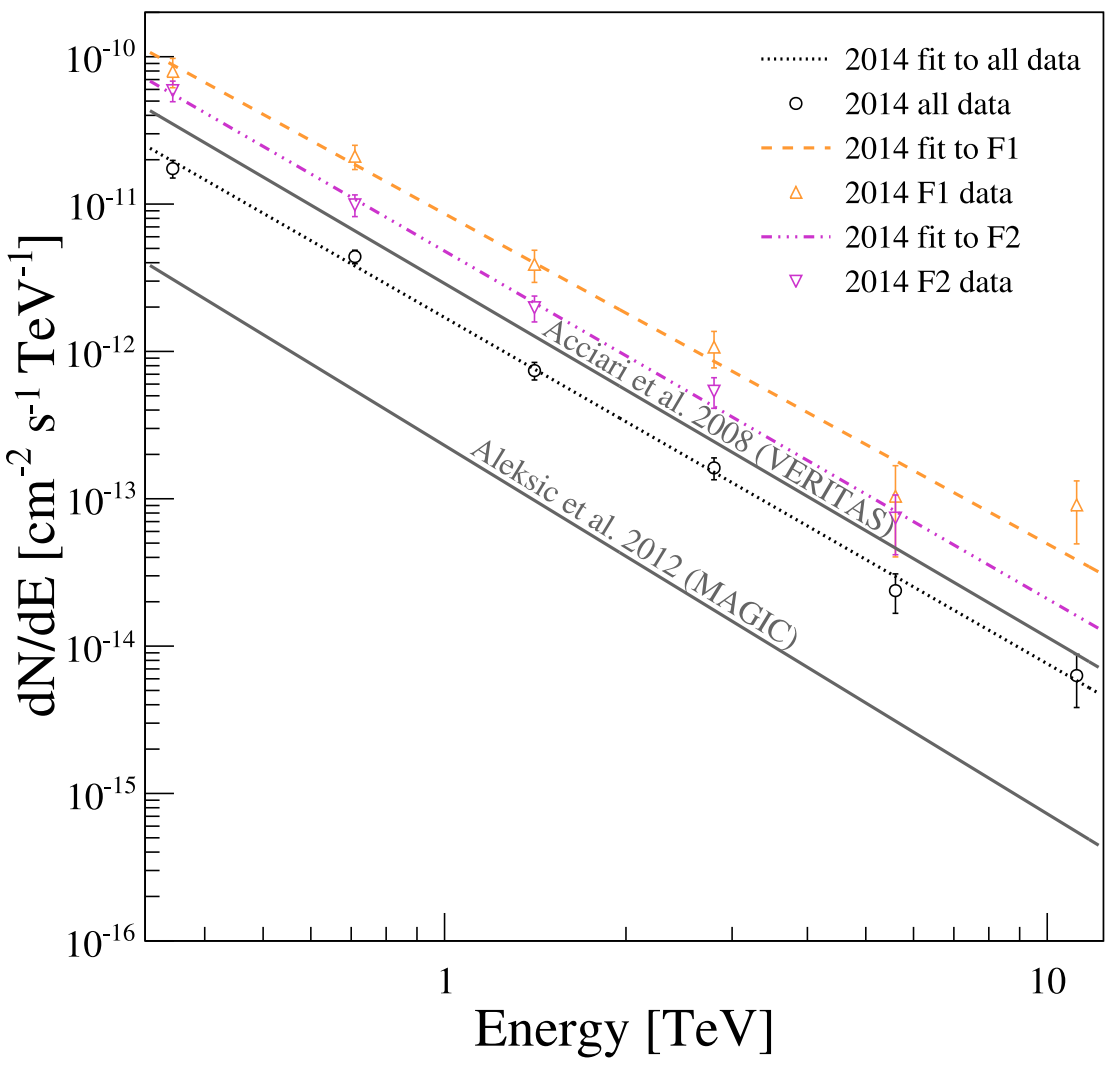

Fig. 16. Differential spectra of LS I +61° 303 during a flaring period from the VERITAS observations in 2014, together with those average spectra in previous publications (from [161]).

For $\gamma$-ray binaries, the most surprising transient behavior came from the GeV observations of PSR B1259-63. During late 2010 to early 2011, the Fermi-LAT observed PSR B1259-63 through a periastron passage, for the first time since its launch in 2008. Before and during the passage, the LAT detected a weak emission above 100 MeV. Unexpectedly, a GeV flare occurred 30 days after the passage, with a flux about an order of magnitude higher than the pre-periastron value. The flare continued about three months after the periastron passage [162, 163]. It turned out that the GeV flare was seen again in 2014 periastron at a similar orbital phase as in 2011. The major obstacle to understand the GeV flare is that it occurred at an orbital phase well after the second/post-periastron disk crossing, and did not correspond to any activities in other wavelengths as of 2011. Although PSR B1259-63, visible only from southern hemisphere, is not visible to LHAASO, this highlights the possibility that any VHE emission from $\gamma$-ray binaries can be unpredictable and transient, which is best probed by an all-sky detector like LHAASO.

In fact, unexpected 'flares' of VHE emission was already seen before. LS I +61° 303 is one of the most studied $\gamma$-ray binary but the nature of its compact object is still under debate because of the poorly constrained mass of the compact object and the inclination angle of the system. Radio to $\gamma$-ray emission are all modulated at the orbital period (26.5 days) and even at the super-orbital period of 1667±8 days. VERITAS observations of LS I +61° 303 clearly observed VHE flares in two consecutive orbits in similar orbital phase (October and November 2014; [161]). Figure 16 shows that the 0.3–20 TeV flux of the VHE flare is about a factor of 2–5 above that of the average flux measured previously, and the flare spectrum does not show any cut-off up to 20 TeV. With the planned sensitivity of LHAASO, it is possible that the VHE emission can be seen by LHAASO, if such elevated TeV level remains for months.

Although leptonic scenario prevails to explain the multi-wavelength emissions from $\gamma$-ray binaries, if hadrons are also accelerated in the complicated binary environment, they might also contribute to >10 TeV emission. An observational 'evidence' for hadronic emission is a low-significance neutrino signal (pre-trial $p$-value is 0.087) from HESS J0632+057 reported by the IceCube collaboration [164]. Although this signal is fully compatible with the background fluctuation after taking the trial factor into account, if similar events are detected in the future, it could increase the likelihood of a >10 TeV emission from accelerated hadrons.

Chances are that there are more $\gamma$-ray binaries to be discovered, based on the fact that known $\gamma$-ray binaries tend to be nearby. Paredes et al. [165] estimate that the total number of $\gamma$-ray binaries in our Galaxy is about 50, but this number can depend on the duty cycle of $\gamma$-ray emission: VHE emission in HESS J0632+057, LS I +61° 303, and PSR B1259−63 is strongly dependent on orbital phase and in some sources the orbital periods can be (very) long, e.g., the 30–50-year orbital period binary pulsar PSR J2032+4127 has only been recently discovered by long-term monitoring (i.e., years) by the Fermi-LAT. With its very large field of view at all times, LHAASO will be the best instrument to observe known and yet-to-discover $\gamma$-ray binaries at energies above 100 GeV.

# 6 The Galactic Center

## 6.1 Galactic center as a high energy emission source

It is well known that the Galactic Center (GC), with a supermassive black hole ($\sim 4{\times}10^6\,M_\odot$), is a good laboratory for the study of high energy astrophysical phenomena. Currently, the overall behavior of the GC is quite silent except some continuous weak activities. Transient X-ray events with a 2–10 keV energy output up to $10^{35}\,\mathrm{ergs}^{-1}$ are observed from the GC on a regular basis, as well as transient events at MeV/GeV energies. Flares from the X-ray binaries located in the GC region can reach luminosities up to $10^{37}\,\mathrm{ergs}^{-1}$. However, there are sufficient evidences to prove that the GC had violent activities in the past, such as X-ray outbursts [166] and the Fermi-Bubbles [167]. During the violent activities, the accretion of stars and gas by the supermassive black hole could be effective to accelerate particles. The maximum energy that protons can achieve by diffusive shock acceleration is [168]

$$E_{\max}\sim eBR\approx 10^{14}\left(\frac{B}{\mathrm{G}}\right)\left(\frac{M}{4\times10^6 M_\odot}\right)\left(\frac{R}{10R_g}\right)\ \mathrm{eV} \quad (1)$$

where $B$ is the magnetic field and $R$ is the size of the acceleration region. As in [168], we assume the acceleration takes place within 10 Schwarzschild radii ($R_g\sim10^{12}$ cm) of the black hole. To accelerate protons to above ~PeV requires magnetic field strength of tens of G in the acceleration region [169, 170]. Such a condition could be reached in the very central region of the GC [168, 171]. On the other hand, if the acceleration takes place in larger regions, the required magnetic field could be smaller. When the accelerated CRs diffuse out of the GC, hadronic interaction with the ISM will happen and produce similar amount of $\gamma$-rays and neutrinos. The observations of high energy $\gamma$-ray emissions can shed new light on the acceleration mechanism at the GC. In fact, with the state of art technologies, current $\gamma$-ray observations have provided unprecedented sensitivity in studying the acceleration activities in the GC.

## 6.2 Gamma-ray emission of the GC

The very high energy $\gamma$-rays from hundreds of GeV to tens of TeV in the direction of the GC have been observed by several atmospheric Cherenkov telescopes such as CANGAROO [172], VERITAS [173, 174], HESS [175–178], and MAGIC [179]. The diffusive $\gamma$-ray emission is also observed at Galactic Center Disk (GCD) range by HESS experiment [176]. More interesting thing is that the map of the central molecular zone as seen in $\gamma$-rays demonstrates a strong correlation between the brightness distribution of very-high-energy $\gamma$-rays and the locations of massive gas-rich complexes [180]. This points towards a hadronic origin of the diffuse emission, where the $\gamma$-rays result from the interactions of relativistic protons with the ambient gas.

Figure 17 shows the spectra of very-high-energy $\gamma$-rays for the diffuse emission of GC. The best-fit to the data found that the spectrum with power law index ~2.3 can extend the energies up to tens of TeV, without any indication of a cutoff or a break. It is suggested that such a $\gamma$-ray spectrum, arising from hadronic interactions, is detected in general for the first time. Since these $\gamma$-rays result from the decay of neutral pions produced by p-p interactions, the derivation of such hard power-law spectrum implies that the spectrum of the parent protons should extend to energies close to 1 PeV. Simultaneously, the spectral index at TeV energy range for the GC point source is the same as that of the diffusive one, which may possibly share the same origin: GC supermassive black hole. The result supports that the $\gamma$-ray emissions come from ~PeV energy protons and the most plausible accelerator is the GC [180].

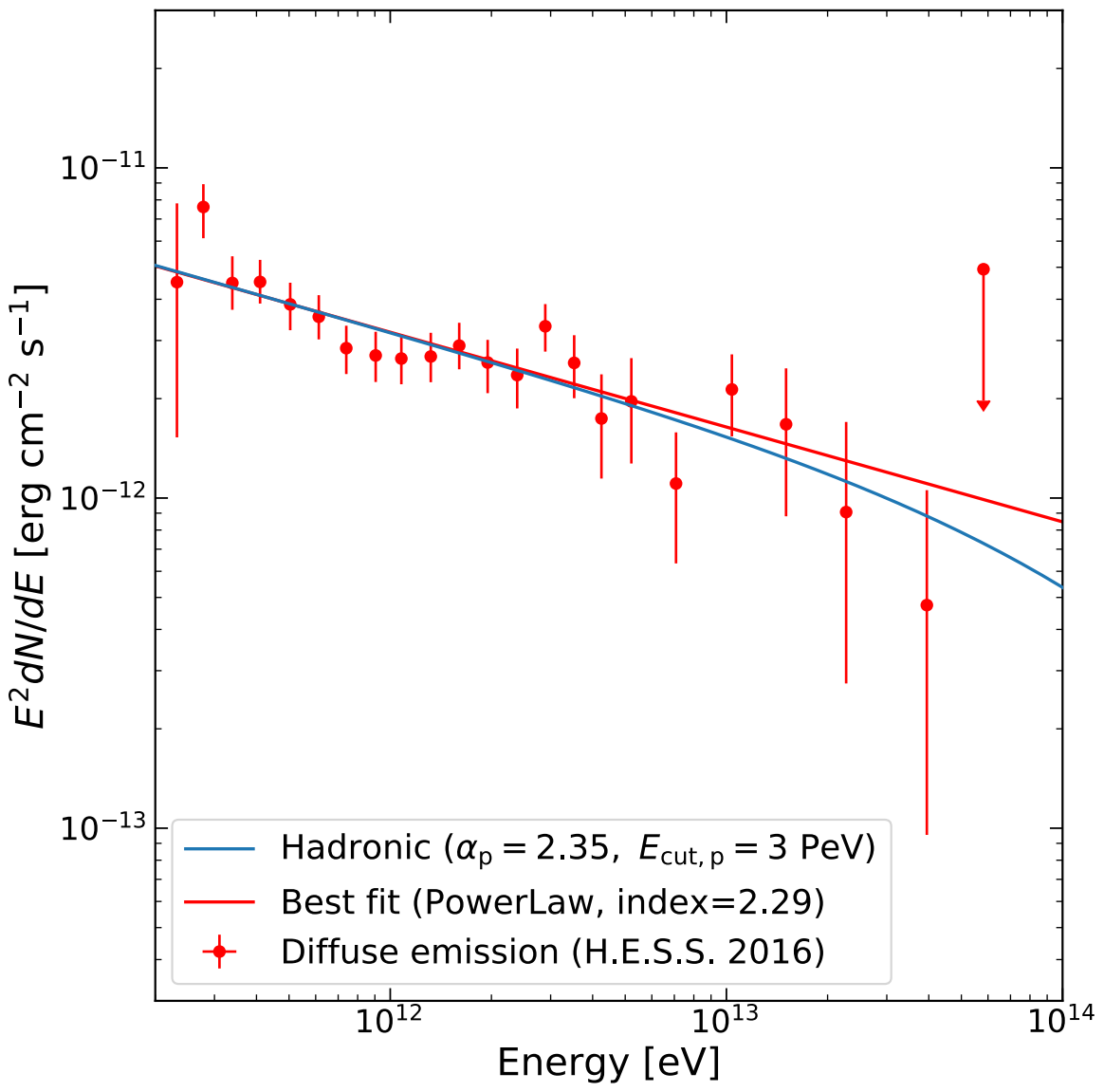

Fig. 17. The very-high-energy $\gamma$-ray spectra for the diffuse emission of GC (Data are adopted from [180]).

However, the $\gamma$-ray emission from the point source in GC has a broken power law spectrum at tens of TeV. The best fit of the cut-off can be described by exponential function in high energy [181]. While adopting the traditional model of ISRF, the absorption effect is too small to explain the observed cut-off spectrum of HESS J1745-290 [181]. The alternative solution attributes it to the intrinsic cut-off, which characterizes the acceleration limit of the flaring event with the critical energy $E_c \sim 200$ TeV for protons. Let's look into the diffuse $\gamma$-ray emission at GC region. The uncertainty at tens of TeV in the $\gamma$-ray spectrum leads to the poor ability to discriminate the different energy cutoff of protons. It is to say that the observation of $\gamma$-ray emission at ∼100 TeV energy will play a very important role to determine the acceleration ability of GC in the future.

### 6.3 The LHAASO sensitivity at 100 TeV energy range

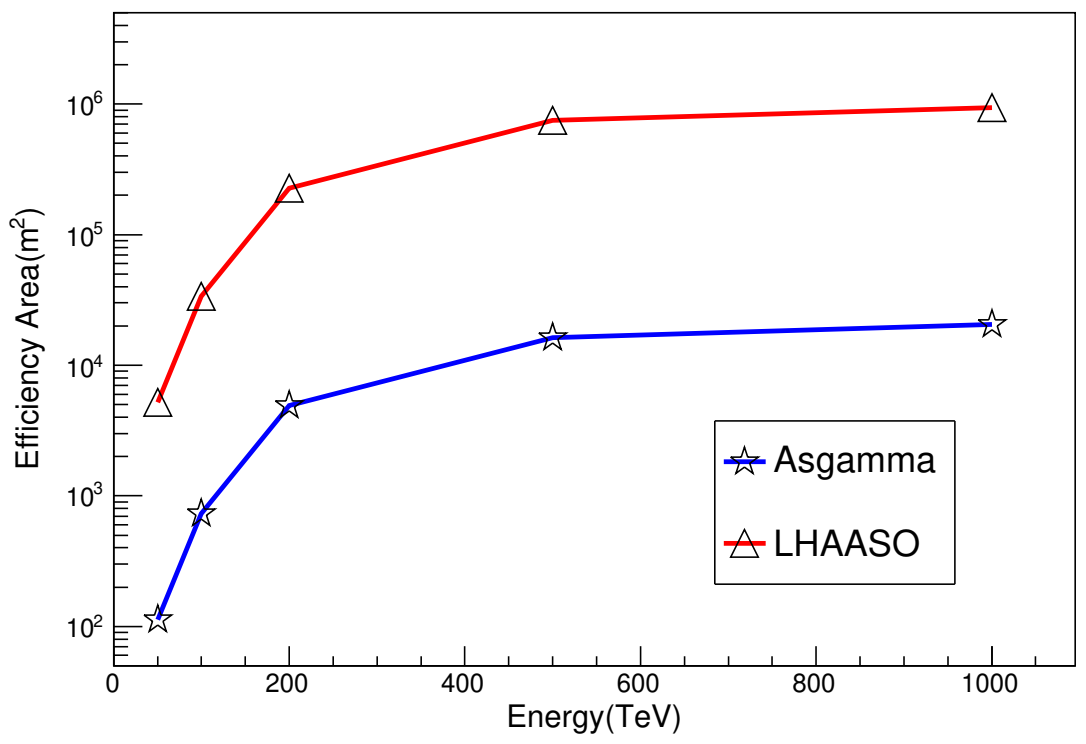

Fig. 18. The effective area of LHAASO for $\gamma$-rays from GC direction.

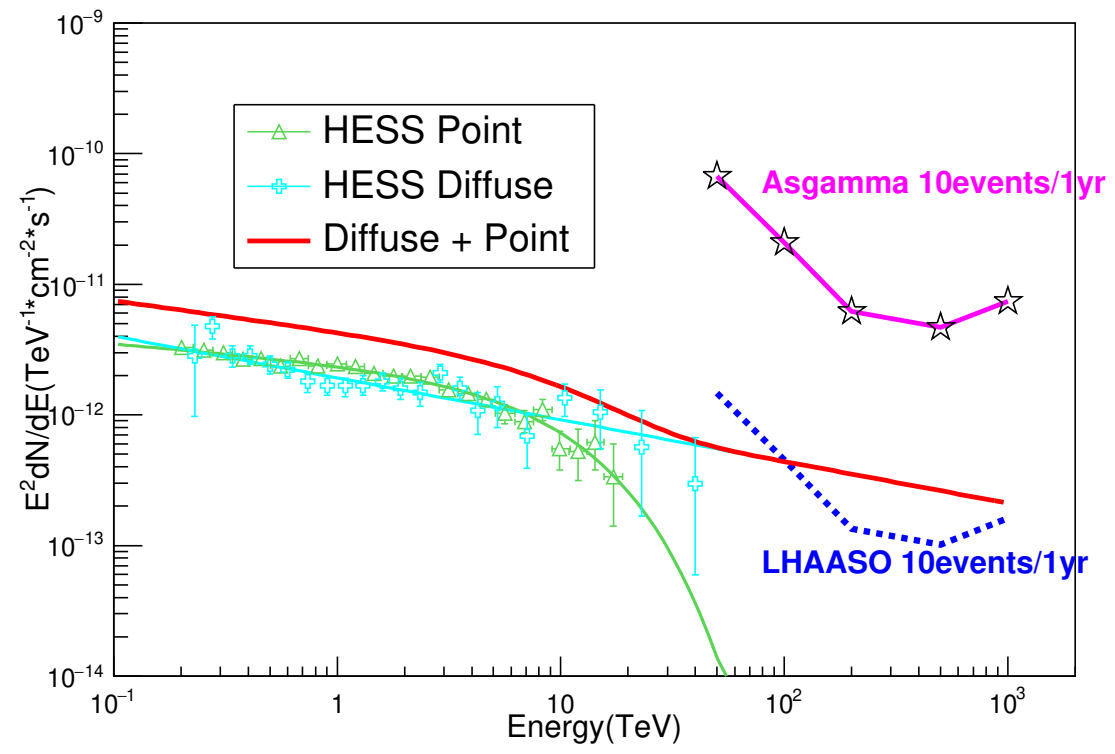

Fig. 19. The sensitivity of LHAASO for $\gamma$-rays from GC direction.

The problem is that the GC in LHAASO field of view is with the zenith angle of ∼65°, which will seriously reduce the sensitivity of LHAASO. So the special analysis technology for wide field of view should develop to study the $\gamma$-ray emission from the GC region based on the simulation. The air shower development in the atmosphere has been generated with the CORSIKA v7.405 code [8]. The electromagnetic interactions are described by the EGS4 package while the hadronic interactions are reproduced by the QGSJET model. The low-energy hadronic interactions are described by the FLUKA package. CR spectra have been simulated in the energy range from 10 TeV to 10 PeV. About 8-yrs showers have been sampled in the zenith angle interval from 55° to 70°. For $\gamma$-rays, we produce 20000 events at every energy point including: 50, 100, 200, 500, 1000 TeV. The experimental conditions (trigger logic, time resolution, electronic noises, etc.) have been taken into account via a GEANT4-based fast simulation code and analyzed with the same reconstruction code.

The event selection is performed for the reconstructed simulation data. With the specified cut the backgrounds of CRs can be almost totally rejected. Fig. 18 is the effective area of KM2A array. It can reach $\sim 5\times10^3$ m$^2$ at 50 TeV, $\sim 3\times10^4$ m$^2$ at 100 TeV and larger than $\sim 2\times10^5$ m$^2$ above 200 TeV. Owing to the zero background, 10 $\gamma$-ray events detected can be defined 5 $\sigma$ level, Fig. 19 shows the sensitivity of LHAASO with one year observation. It is obvious that the LHAASO have enough sensitivity to observe this source at above 100 TeV. However, if the protons can not be accelerated to ∼PeV, LHAASO can not have enough sensitivity.

### 6.4 Short summary

Galactic CRs can reach energies of ∼ PeV. The first PeV accelerator, GC, has been evidenced by HESS experiment based on the observation of $\gamma$-ray emission at tens of TeV. However, the uncertainty at tens of TeV for the spectrum of $\gamma$-rays leads to the poor ability to discriminate the different energy cutoff of protons. We employ the MC simulation to examine the LHAASO sensitivity to Galactic center at 100 TeV energy range and see that the LHAASO has enough sensitivity with one year observation to detect this source at above 100 TeV if the proton can be accelerated to PeV energy. On the contrary, if the maximum energy is ∼200 TeV, the LHAASO can not have enough sensitivity to detect it.

## 7 Giant Molecular clouds

A giant molecular cloud (GMC) has a typical mass of $10^5$ solar mass and a density of more than 100 cm$^{-3}$. The molecular gas in GMCs can be observed and measured via molecular lines, such as the rotational transition lines of CO. Furthermore the infrared emission from the dust inside GMCs provides an alternative way to study the gas contents. GMCs are the birth place of young stars and thus also harbor HII regions and bubble-like structure. GMCs are also regarded as $\gamma$-ray emitters. The main $\gamma$-ray production mechanisms inside GMCs are the decay of neutral pions produced in the collision of CR nuclei with the ambient gas, IC of relativistic electrons on background radiation fields, and bremsstrahlung of relativistic electrons. Due to the high gas density, pion-decay dominates the other mechanisms above about 100 MeV[182]. In the energy range of LHAASO, the IC and bremsstrahlung are further suppressed due to the high energy cutoff at several TeV observed in the CR electron spectrum [183]. The dominance of pion-decay mechanism in $\gamma$-ray production makes it an ideal place to measure CR density beyond the solar system. Several famous GMCs locate inside the field of view (FOV) of LHAASO. Their positions, mass, and distances are listed in Table 2. The predicted $\gamma$-ray flux from GMCs are proportional to the value $M/d^2$, which are also listed in Table 2.

Table 2. Properties of the GMCs in the FOV of LHAASO. The estimated distance and position are obtained from Dame et al, 1987. The mass values listed in the second column are calculated from the CfA maps (see [184] for detail).

| Region | $M$ | $D$ | $l$ | $b$ | $M/d^2$ | size |
|---|---|---|---|---|---|---|
| | [$10^5$ $M_\odot$] | [pc] | [°] | [°] | [($10^5$ $M_\odot$/kpc$^2$)] | [arcdeg$^2$] |
| $\rho$ Oph | 0.12 | 165 | 356 | +18 | 8.4 | 68 |
| Orion B | 0.78 | 500 | 205 | −14 | 3.9 | 22 |
| Orion A | 1.2 | 500 | 213 | −18 | 5.2 | 28 |
| Mon R2 | 1.1 | 830 | 214 | −12 | 1.7 | 19 |
| Taurus | 0.30 | 140 | 170 | −16 | 15.0 | 101 |
| Polaris flare | 0.055 | 230 | 130 | +26 | 0.96 | 40 |

### 7.1 GMCs as CR calorimeter

The current paradigm of CRs postulates that, because of the effective mixture of CRs during their propagation in the interstellar magnetic fields, the CR density locally measured in the neighborhood of Earth should correctly describe the average density of CRs throughout the Galactic disk [185]. However, small variations of CRs on large (kpc) scales do not exclude significant fluctuations on smaller scales, particularly in the proximity of young CR accelerators. Therefore, it is not obvious that the locally measured component of CRs can be taken as an undisputed representative of the whole Galactic population of relativistic particles. In particular, it is possible that the flux of local CRs might be dominated by the contribution of a few nearby sources. However, the density of CRs in different parts of the Galaxy can be probed uniquely through observations of $\gamma$-rays from GMCs [186–188]. On GeV band the investigations in this regard have already been done on the nearby GMCs in Gould belt [184, 189–191] as well as on Sgr B complex in Galactic center [192]. But on TeV band the GMCs are still left undetected. One reason for the non-detection of GMCs is the extended size of these objects and the limited FOV of Imaging Atmospheric Cherenkov Telescope (IACT). In contrast, the high sensitivity and large FOV of LHAASO will provide a unique opportunity to detect such objects and measure the CR density in TeV-PeV band in different positions of the Galaxy. To show the detection prospect we plot the predicted $\gamma$-ray flux as well as the LHAASO sensitivities for a typical GMCs with a $M/d^2$ value of $10^6$ ($M_\odot$/kpc$^2$) in Figure 20. It should be mentioned that the sensitivities for extended sources are estimated as $F_{\rm ext}=F_{\rm PS}(\Omega_{\rm ext}/\Omega_{\rm PSF})^{1/2}$, where $F_{\rm ext}$ and $F_{\rm PS}$ are sensitivities for the extended source and point source, respectively, and $\Omega_{\rm ext}$ and $\Omega_{\rm PSF}$ are the angular size of extended source and point spread function, respectively. Thus the detection capacity of GMCs depends on their angular size. Indeed, the GMCs show filamentary morphology and the $\gamma$-ray emission region is much smaller than that listed in Table 2. Thus the estimation of LHAASO sensitivities in Figure 20 is very conservative and should be regarded as an upper limit.

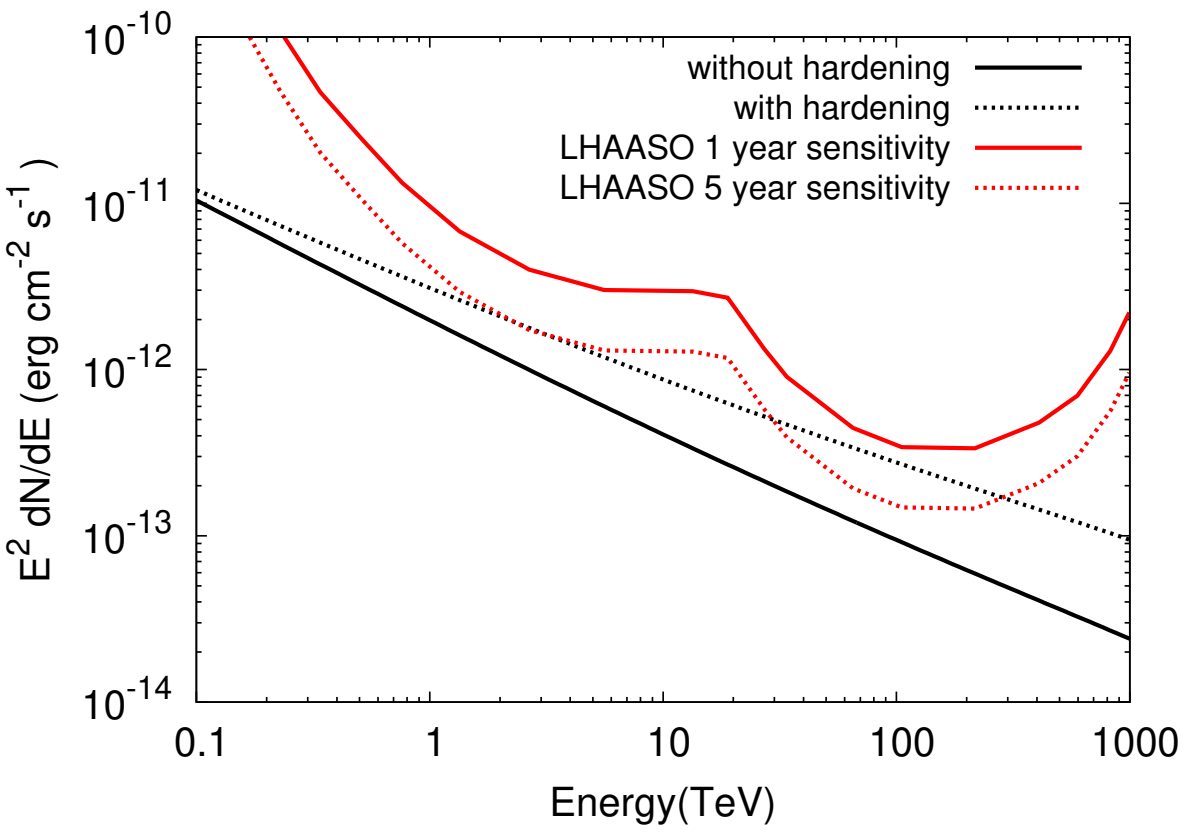

Fig. 20. The $\gamma$-rays flux produced in a molecular clouds with a M/d$^2$ value of $10^6$ (M$_\odot$/kpc$^2$), the angular size is 20 arcdeg$^2$. The CR spectrum measured by AMS-02 extrapolated to 10 PeV with and without a hardening are used in deriving the $\gamma$-ray flux. The LHAASO sensitivity are estimated by considering the source extension.

In addition to the absolute CR fluxes at different positions of the Galaxy, it would also possible to measure the spectral property of CRs using the $\gamma$-ray observations on GMCs. Recently a hardening in CR spectrum above 200 GeV was reported by several observations [193–195]. This effect can be observed in the $\gamma$-ray flux in the nearby GMCs given the hardening extends to more than 100 TeV. To illustrate the effect we plot in Figure 20 the predicted $\gamma$-ray flux in GMCs with and without such a hardening. Furthermore, the $\gamma$-ray above 100 TeV are already produced by CRs with the energy close to the *knee*. Thus, LHAASO observation of $\gamma$-rays from GMCs in this energy range will provide an alternative method in measuring the CR spectral property near the *knee*.

### 7.2 Young stellar associations inside GMCs

Young star associations and corresponding super bubbles are considered to be the origin of a substantial fraction of Galactic CRs [196, 197]. Fermi LAT has detected a cocoon like structure near the young star association of Cygnus OB2 with a hard spectrum and argue that this is produced by fresh accelerated CRs [87]. The GMCs harbored various young star associations and young HII regions. For example, the Orion Nebula Cluster (ONC) in the Orion A molecular cloud and NGC 2024 in the Orion B molecular cloud are the two largest clusters in the youngest subgroup of Orion OB1, with ages less than 2 Myr [198]. These young star clusters are also potential accelerating sites of the CRs. Although these young accelerators are not observed in GeV band, one can not exclude the possibility that they would dominate in multi-TeV ranges, due to their hard spectra. In this case the CR density inside GMCs are contaminated by the embedded acceleration and GMCs can no longer be regarded as CR calorimeters. Furthermore, if the hard spectra in these young structures are detected in multi-TeV energy range, this would be a strong evidence for the existence of *PeVatron*, which will be discussed in detail in an independent section (§**7.3**).

### 7.3 PeVatrons

The hard spectrum in multi-TeV range without cutoff is considered as the sign of hadronic origin of the emissions. This is because the Klein-Nishina (KN) effects will introduce a break in the spectrum of IC scattering off CMB photons at this energy range, even if there is no cutoff in electron spectrum. Thus such a hard spectrum can only be produced by CRs protons with energy up to PeV. This argument has been adopted for the PeVatrons in Galactic center observed by H.E.S.S. [199] (also see §6).

As a result, all the hard TeV sources without detected high energy cutoff can be regarded as candidates for PeVatrons. Several famous young SNRs, such as Cassiopeia A and Tycho, are detected by Air Cherenkov telescopes without cutoff up to several TeV. As already discussed in the section "SNRs" (see §**2**, §**2.3**) such objects should be regarded as PeVatron candidates. Along with young SNRs, the unidentified TeV sources without cutoff should also be examined. One recent example is the H.E.S.S detection of hard spectra up to more than 20 TeV without cutoff in the source HESS J1641-463 [200]. However, the limited statistics cannot rule out a cutoff at higher energy caused by KN effects. By comparison, the much higher sensitivity of LHAASO at the energy range of 10–100 TeV provides an ideal window to study the spectral property of the PeVatron candidate. Although HESS J1641-463 is located beyond the LHAASO FOV, there are still a few unidentified Galactic source in the northern sky with hard spectra.

One remarkable example is TeV J2032+4130 in Cygnus region (§**3.1**), which is also related with the Fermi Cygnus cocoon [87]. The hard spectra (index of $-2$) and non-detection of cutoff at TeV range has been reported by VERITAS [14]. Furthermore, the study on Fermi Cygnus cocoon reveals that the Cygnus region indeed harbors CR acceleration site and fresh CRs. The Cygnus region, as well as other star-forming regions (see §**3**), is a very promising target to hunt for PeVatrons.

Another interesting source is HESS J1848-018. H.E.S.S measurement has revealed a spectral index of $-2.8$ [201], which makes it unlike a PeVatron. However, the recent HAWC observations [202] reveal a much higher flux at high energy and thus a harder spectra. The difference may comes from the diffusive nature of this source. The source is spatially correlated with the star forming region W43, which has a similar environment as that of the Cygnus cocoon (§**3.1**). We note that, at GeV range, the Cygnus cocoon also has a spatial extension of more than 3 degrees. Indeed, if the CRs are accelerated in the super-bubbles surrounding the young star clusters, the $\gamma$-ray emission should be diffuse due to the low ambient density in the cavities. Such diffuse structure can hardly be detected by IACT due to the very limited FOV. LHAASO, however, with much larger FOV and continuous exposure, has the capability to detect such structures.

In conclusion, in addition to the strong indication of the Galactic center (§6), the hard unassociated TeV sources noted here, SNRs (§**2**), PWNe (§**4.2**), and star-forming regions (§**3**) considered in the previous sections can be Galactic PeVatron candidates. Whether high energy cutoff is present at dozens of TeV is crucial to identify the PeVatron nature of these sources. The energy range of LHAASO is perfectly suitable to study their spectral features. On the other hand, the PeVatrons can also be diffusive rather than compact, and such kind of sources can hardly be detected by the former IACT but would be very promising to be detected by LHAASO.

## 8 Diffuse Galactic Gamma-Rays

It is recognized that the $\gamma$-rays above 100 MeV chiefly spring from the diffuse emission. Three major mechanisms are thought to be responsible for the creation of $\gamma$-rays, and they are respectively[203]: the decay of neutral pions which are generated through the inelastic collisions between CRs (mostly protons and heliums) and ISM, the IC scattering of high energy electrons off interstellar radiation field, as well as the bremsstrahlung of CR electrons with interstellar gas. Each process is dominant in different parts of the $\gamma$-ray spectrum.

Observation of these diffuse emission is beneficial in acquiring the following knowledge: 1.) spatial distributions of hadronic and leptonic components of CRs, 2.) origin and propagation of cosmic CRs in the Galaxy, 3.) composition and allocation of interstellar medium, and 4.) large-scale distribution of Galactic magnetic field and turbulence. Moreover as the Galactic diffuse emission often represents the natural background to many different signals, a thorough understanding of diffuse Galactic $\gamma$-ray emission (DGE) is also essential for deducing the spectra of other components of the diffuse emission, unveiling the undiscovered $\gamma$-ray sources, enhancing the measurement accuracy of the position and spectral energy distribution (SED) of Galactic or extra-galactic point/extended sources and even searching for the sign of dark matter annihilation or decay.

### 8.1 Progresses on the observations of Galactic diffuse gamma-rays

The observation of diffuse $\gamma$-rays started with the OSO-III satellite in 1968[204]. The measurements have been dramatically ameliorated during the surveys of SAS-2[205], COS-B[206], COMPTEL[207, 208], HEAO 1[209] and EGRET[210]. With the launch of a new generation telescope, Fermi Large Area Telescope (LAT), it maps the $\gamma$-ray sky up to a few hundreds of GeV with unprecedented accuracy[211, 212], which deepens our understanding of the generation and propagation of Galactic CRs. In lower energies, the SPI instrument on INTEGRAL observatory has extended the observations of CR-induced diffuse emissions into the hard X-ray range[213, 214]. As for the higher energies, subject to very low flux and limited area of space-based detector, the observations above TeV have been carried out principally on ground-based instruments, such as Whipple[215], HEGRA[216], Milagro[85], HESS[217], ARGO-YBJ[218] and so on.

Higher-quality data enable us to model the DGE based on CR transport and interactions in magnetic halo[211, 212, 219–221]. In the GeV energy range, the EGRET data show a significant excess in all directions (called "GeV excess") with respect to the predictions supposing the same CR spectrum in the Galaxy as that at Earth. But this excess has not been confirmed by the following observation of Fermi-LAT at intermediate Galactic latitudes[223]. Up to now, the DGE model generated by the numerical package GALPROP well conforms the observations at both high and intermediate latitudes published by the 21-month Fermi-LAT survey[211]. But in the Galactic plane the models all underestimate the data above a few GeV, especially toward the inner Galaxy. This has been reconfirmed in the renewed measurements by Fermi-LAT[212]. Possible explanations include the contribution from the unresolved point source population such as pulsars, SNRs, PWNe, spectral variations of CRs or even dark matter annihilation/decay[224, 225]. Recently Guo et al.[226] suggest that a hard CR component within the Galactic plane can self-consistently explain the excess of diffuse $\gamma$-rays at the inner Galaxy, the observed B/C and $\bar{p}/p$ ratio. For the diffuse TeV $\gamma$-rays, the Milagro telescope made the first observation towards the Galactic disk and corroborated the existence of diffuse TeV $\gamma$-ray emission[227, 228]. In the Galactic plane, the Cygnus region inhabits abundant CR sources and large column density of matter, and is recognized as the brightest $\gamma$-ray region in the entire northern sky[228]. Milagro telescope performed the observations of Cygnus region and found the diffuse TeV $\gamma$-ray emission[85]. Subsequently ARGO-YBJ experiment carried out similar observation as well[218], whose data agree well with the measurements of Fermi-LAT at lower energies. Meanwhile, HESS telescope array also performed surveys at both Galactic plane[217] and center[176, 180].

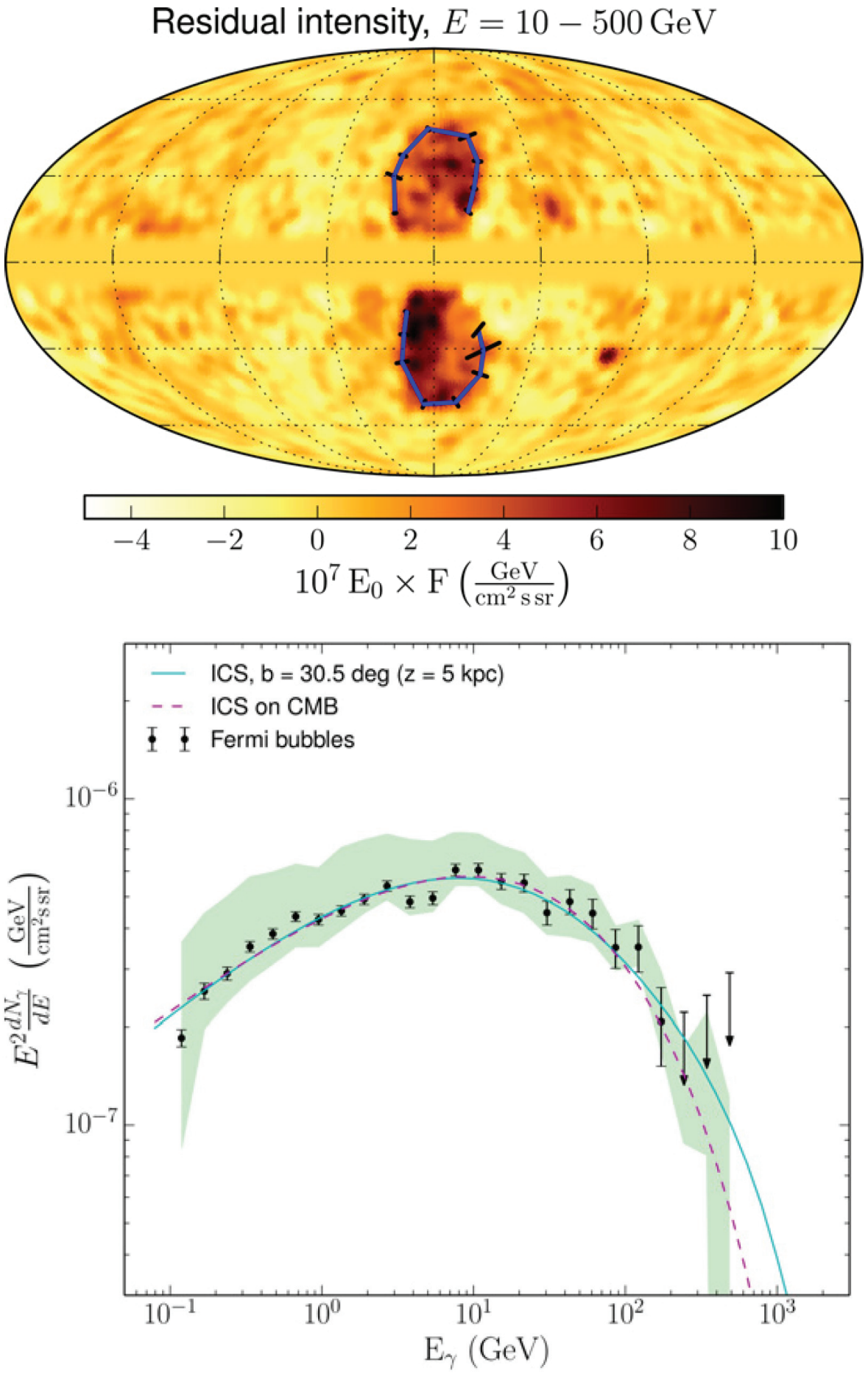

Fig. 21. Image (top) and Spectral energy distribution (bottom) of Fermi bubbles [222].

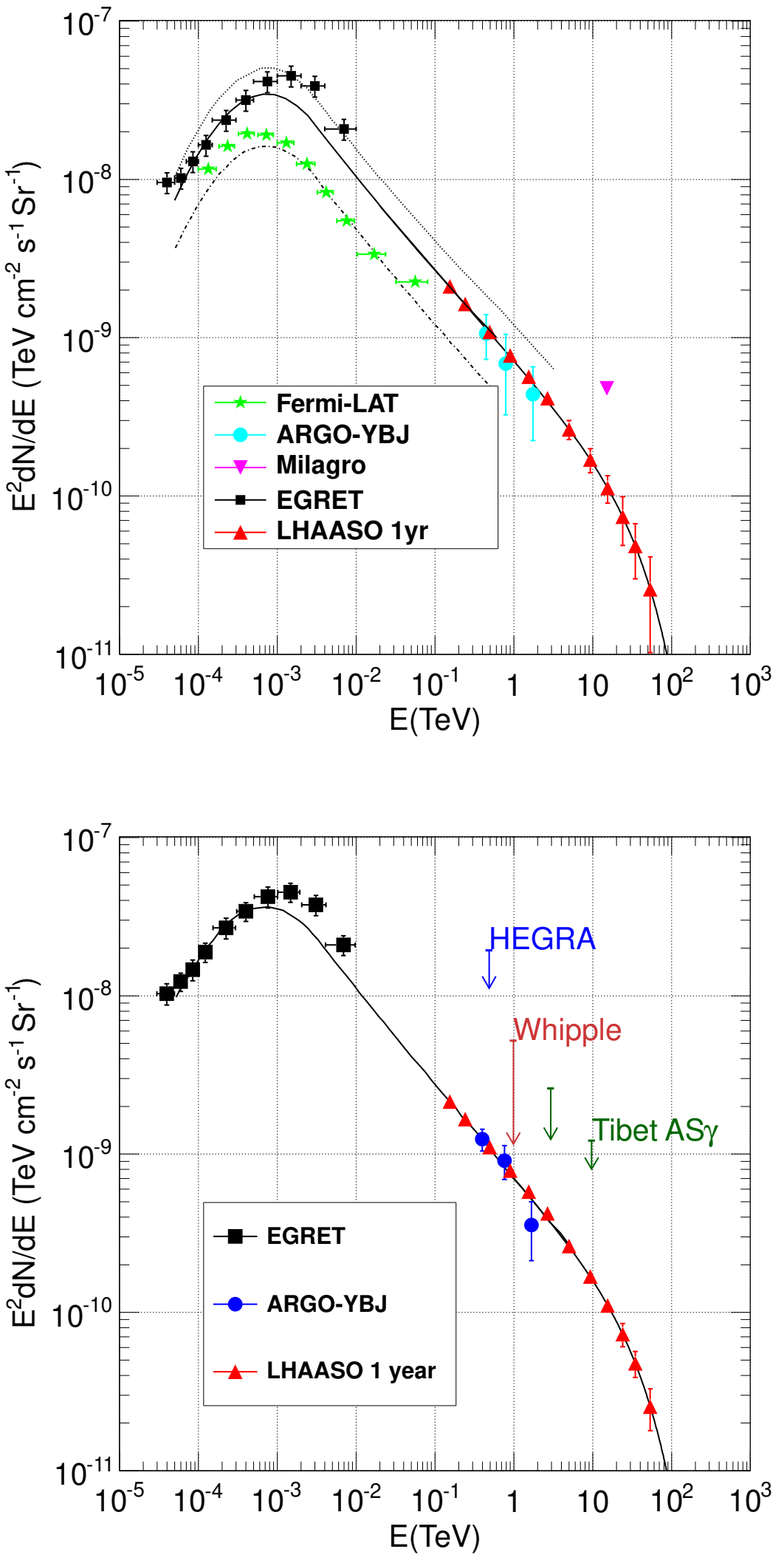

Fig. 22. The predicted DGE flux respectively at Cygnus region (top) and $25° < l < 100°$, $|b| < 5°$ (bottom) observed by LHAASO experiment.

Probably, the most spectacular discovery about the extended emission in recent years is the so-called Fermi Bubbles[167, 222]. The Fermi bubbles are two giant lobes, roughly symmetrically distributed at two sides of the Galactic center. Each bubble owns an oval emission region with sharp edge, which extends over several kiloparsecs beyond the Galactic plane. Compared with the diffuse $\gamma$-rays, the Fermi bubbles have a visibly harder $\gamma$-ray spectrum with index $\sim -2$, see Figure 21. So far the origin of Fermi bubbles is still under debate. Many theories have been proposed including jet radiation of massive black hole at the Galactic center, shock wave from accretion events of the central black hole, shock wave from supernova explosions near the Galactic center and so forth (see [229] and references therein).

### 8.2 The outlook of LHAASO project on Galactic diffuse gamma-rays above 30 TeV

Nevertheless the above intriguing findings only reach to energies around tens of TeV at most, while for the higher energy, i.e. 100 TeV $\gamma$-rays, the observations so far are still poor. One main part of the LHAASO project, KM2A (one KM$^2$ Array), is designed to observe the $\gamma$-rays above 30 TeV. The large detection area and high capability of background rejection enable the sensitivities of LHAASO experiment to reach $\sim$100 times higher than that of current instruments above 30 TeV[8, 230]. It will also be the first time to monitor $\gamma$-ray sky at PeV energies.

**1. The observations of diffuse $\gamma$-rays above tens of TeV** LHAASO project plans to map the DGE above a few hundreds GeV throughout the Galaxy with high sensitivity. It is going to perform an unbiased sky survey of the northern sky with a detection threshold of $\sim$0.03 Crab unit at TeV energies and $\sim$0.1 Crab around 100 TeV by one year operation, respectively, which is capable of continuously surveying the $\gamma$-ray sky from 100 GeV to 1 PeV. For the LHAASO sensitivity to the DGE flux, it can be evaluated roughly according to the point source sensitivity multiplied by a correction factor $f = (\Omega_{\rm PSF}\Omega_{\rm GP})^{-1/2}$, in which $\Omega_{\rm PSF}$ is the observation angular window, related to the detector point spread function (PSF), and $\Omega_{\rm GP}$ is the solid angle of a certain region in the Galactic plane. Therefore according to the above rough evaluation, after one year observation towards the longitude interval 25°–100°, the 5-$\sigma$ minimum flux detectable by LHAASO can reach as low as $F_{\rm min} \sim 7\times10^{-16}$ photons cm$^{-2}$ s$^{-1}$ eV$^{-1}$ sr$^{-1}$ at 100 TeV, which is about 6 times lower than the extrapolation of the Fermi-DGE model at the same energy[8, 230]. Figure 22 shows the predicted DGE flux observed by one quarter LHAASO project after one year run, at Cygnus region (top) and $25° < l < 100°$, $|b| < 5°$ (bottom), respectively. It can be seen that LHAASO is very sensitive to 100 TeV $\gamma$-ray photons. Hence we could expect a better measurements for the DGE at this energy range, especially the presence of exponential cutoff at TeV energies.

**2. Diffuse $\gamma$-ray constraints to the origin of the Fermi bubbles**

Part of Fermi bubbles are in LHAASO's wide field of view. If the $\gamma$-rays stem from the interaction between the CRs and gas within the bubbles, the spectral index is anticipated to be harder, thus the spectra of $\gamma$-rays could extend to 100 TeV. According to the sensitivity of LHAASO, it can make precise measurement between 10–100 TeV and thus offer the support to the acceleration mechanism of CRs within Fermi bubbles and hadronic origin of $\gamma$-rays. Figure 23 shows the energy spectrum of the Fermi bubbles extrapolated according to the hadronic model of [231] (black solid line) and the integral sensitivity of one quarter LHAASO project (red solid line).

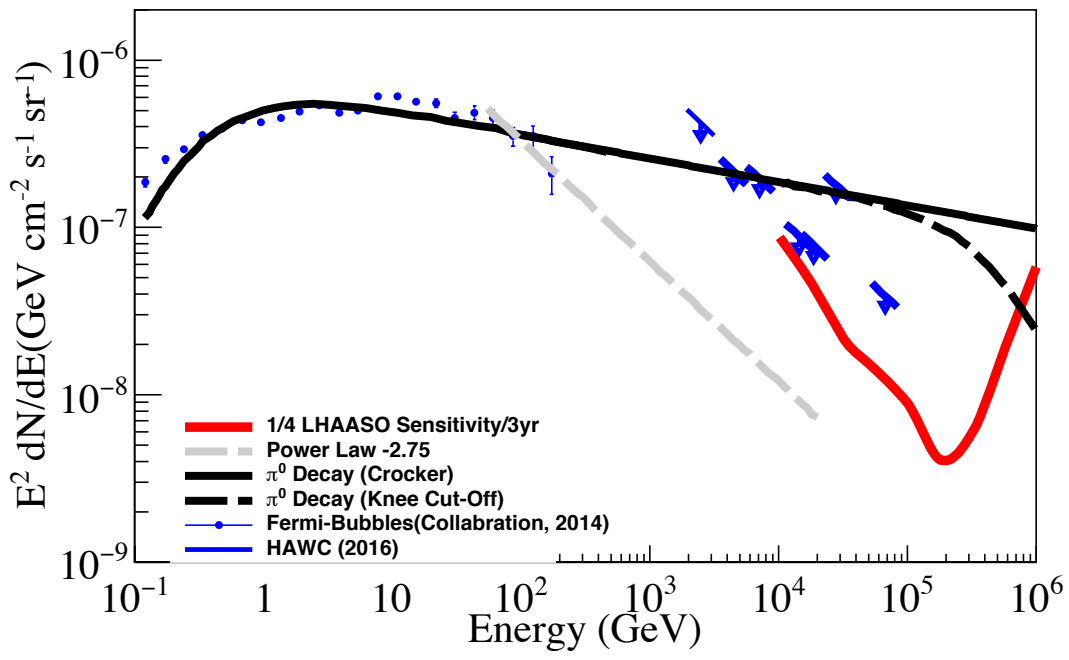

Fig. 23. The extrapolated energy spectrum of the Fermi bubbles in hadronic model (black solid line) according to [231] and the integral sensitivity of one quarter LHAASO project (red solid line).

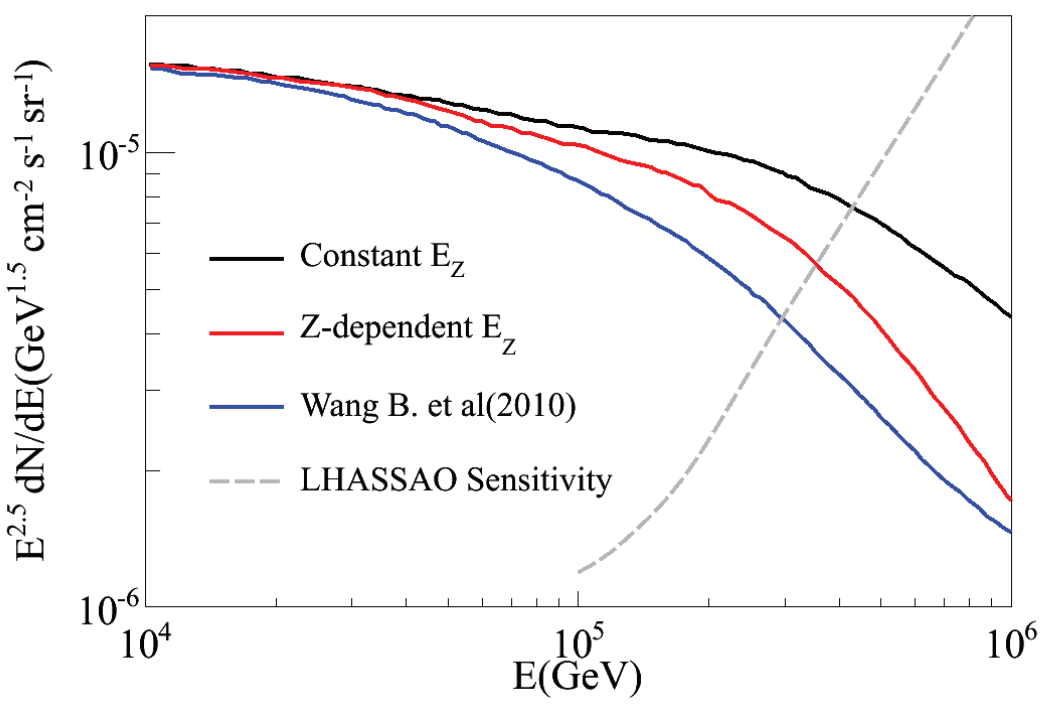

Fig. 24. The $\gamma$-ray spectra predicted by different knee models[232] and the LHAASO sensitivity (grey dash line).

**3. Diffuse $\gamma$-ray constraints to the CR knee region** The origin of the CR knee has been a mystery since its discovery. So far there are various of models proposed to explain the break of all particle spectrum at the knee region. However, due to the large uncertainties of the measurement of individual composition, it is hard to further testify these hypotheses. In [232], Guo et al. argue that the different models about the knee region could generate distinct DGE spectra, in which a knee-like structure also appears at about hundreds of TeV due to the different CR compositions around PeV energies. Thus the measurement of the DGE at hundreds of TeV could be used to distinguish the models of the knee region. Figure 24 shows the $\gamma$-ray spectra predicted by different knee models, and the grey dashed line is the expected LHAASO sensitivity.

### 8.3 Diffuse gamma-ray constraints to Galactic neutrino flux

The IceCube collaboration has recently reported the discovery of high-energy extraterrestrial neutrino flux[233]. After two years' operation, the IceCube experiment has observed 28 neutrino events between 30 TeV and 1.2 PeV, which is far above the 10.6 events evaluated from conventional atmospheric background. It declares that we have entered into a new era of neutrino astronomy. The interactions between CRs and interstellar medium could also generate neutrinos as well as $\gamma$-rays. Thus the neutrinos detected by the IceCube may partly originate from the Galaxy. The DGE could effectively impose restrictions on the origin of the Galactic neutrinos and the contribution of the Galactic neutrino flux[232, 234–237]. The measurement of the DGE above hundreds of TeV by the coming LHAASO can provide more stringent constraints.

### 8.4 Short summary

$\gamma$-ray astrophysics has made a remarkable progress. Especially recent observations of the DGE obtained by both space- and ground-based instruments have significantly changed and deepened our understanding of the origin and transport of the CRs in the Galaxy. While already being investigated at GeV energies over several decades, assessments of the DGE at TeV energies remain sparse and lots of terra incognita is going to be uncovered in the future. LHAASO ground array is promising to provide more detailed observations of VHE DGE above tens of TeV and open a new window at PeV energies.

# Chapter 3 Extra-galactic Gamma-Ray Sources *

Editor: Xiang-Yu Wang[1]

Contributors: Zhen Cao[2], Xiao-Jun Bi[2], Piero Vallania[3], Xiang-Yu Wang[1], Hanrong Wu[2], Dahai Yan[4], Qiang Yuan[5]

[1] School of Astronomy and Space Sciences, Nanjing University, Nanjing 210093, China

[2] Key Laboratory of Particle Astrophysics, Institute of High Energy Physics, CAS, P.O. Box 918, 100049 Beijing, China

[3] INAF - Osservatorio Astrofisico di Torino, Strada Osservatorio 20, 10025 Pino Torinese (TO), Italy and INFN Sezione di Torino, Via P. Giuria 1, 10125 Torino, Italy

[4]Key Laboratory for the Structure and Evolution of Celestial Objects, Yunnan Observatories, Chinese Academy of Sciences, Kunming 650011, People's Republic of China

[5]Key Laboratory of Dark Matter and Space Astronomy, Purple Mountain Observatory, Chinese Academy of Sciences, Nanjing 210023, China

**Abstract:** Extra-galactic gamma-ray sources, such as gamma-ray bursts, active galactic nuclei, starburst galaxies, are interesting and important targets for LHAASO observations. In this chapter, the prospects of detecting these sources with LHAASO and their physical implications are studied. The upgrade plan for the Water Cherenkov Detector Array (WCDA), which aims to enhance the detectability of relatively lower energy photons, is also presented. In addition, a study on constraining the extragalactic background light with LHAASO observation of blazars is presented.

## 1 Prospects for Gamma Ray Bursts detection with LHAASO

Executive summary The LHAASO (Large High Altitude Air Shower Observatory) experiment, currently under design, is planned to be installed in the Sichuan Province (China) at 4410 m a.s.l. with the aim of studying the highest energy gamma-ray sources and cosmic rays in the wide energy range from hundreds of GeV to hundreds of TeV. Among its different components, optimized to study different energy regions, the WCDA (Water Cherenkov Detector Array) will be one of the most important detectors. Three ponds, for a total surface of 78,000 $m^2$, will be equipped with 3120 PMTs to detect the Cherenkov light produced by ultra-relativistic particles. Each PMT will monitor a volume cell of 5x5x4.5 $m^3$. Data (signal amplitude, with a threshold set at 1 pe level, and arrival time) from each PMT are collected and sent to a DAQ system able to build and record events with all multiplicities starting from a single PMT fired. For small numbers, the primary energy for gammas corresponds to a few GeV, overlapping with the actual satellite detectors energy range. In this paper, the scheme to calculate the expected rate and typology of GRBs detectable in follow-up mode with LHAASO is described and discussed.

### 1.1 Introduction

Gamma Ray Bursts are among the most powerful sources in the sky, with an energy spectrum extending from radio to gamma rays of tens of GeV (for a review see e.g. [1, 2]). They occur with a frequency of a few per day, and originate from the entire universe.

GRBs are divided into two classes depending on their duration. The short ones last up to 2 seconds and show a harder spectrum with a mean peak energy of 490 keV. It is believed that their origin is due to the merger of two compact objects such as neutron stars or blacks holes. Long GRBs have durations greater than 2 seconds with a softer spectrum and a peak at about 160 keV. In this case it is thought that the origin is due to the collapse of the nucleus of a type Ib/c Supernova, and in fact, the coincidence between these two phenomena has already been observed in many cases. The shape of the spectrum is well described in most cases by the Band function, characterized by two power laws smoothly connected. Although the majority of the ejection is concentrated in the energy region between keV and MeV, some photons have been observed up to tens of GeV using detectors in space on board the CGRO satellite and more recently Fermi and AGILE.

Until now all the experimental data in the MeV - GeV energy range were obtained from detectors onboard satellites, but due to their small size and the rapid fall

1) E-mail: xywang@nju.edu.cn

* Supported by the National Key R D program of China under the grant 2018YFA0404203, NSFC grants 11625312 and China Manned Spaced Project (CMS-CSST-2021-B11)

of GRB energy spectra they hardly cover the energy region above 1 GeV. The detection on ground can be done by two kinds of detectors able to provide a much larger effective area: telescopes for the atmospheric Cherenkov light and EAS arrays.

With their enormous size, the Cherenkov telescope recently installed at HESS and those planned for the CTA observatory allow the detection of gamma rays with an energy threshold as low as 20 GeV. However, the necessity of working during nights with clear skies and no or few moon light limits the efficiency to 10-15 %. Furthermore, apart very seldom serendipitous observations or specific pointing strategies to cover a wider sky region, the field of view of less than about 5° prevents the study of short GRBs and of the prompt phase of long GRBs, since the repointing requires a minimum time of about 100 seconds. So far, all major Cherenkov detectors (MAGIC, HESS, VERITAS) have tried to point the GRBs following the afterglow but without success. In the case of CTA it is expected a coincident detection of 0.5 - 2 GRBs per year [3] depending on the assumed high energy spectral features, satellite alert rate and array performance.

The EAS arrays have on the contrary a large field of view (nominally 2 sr, limited only by the atmospheric absorption) and high efficiency (up to 100%), but the need to reveal enough secondary particles to reconstruct the arrival direction and energy of the primary increases the threshold to around 100 GeV.

An alternative mode consists in the measurement of the counting rates of the detectors at time intervals of the order of a fraction of second ("single particle technique") [4–6], and then in the search for an excess in coincidence with a GRB detected by a satellite. With this technique it is not possible to measure the arrival direction of the excess, but the threshold energy can be lowered to about 1 GeV. Both techniques have been used by various extensive air shower arrays, such as EASTOP, Chacaltaya, Milagro and more recently ARGO-YBJ which has studied the richest sample of GRBs ever analyzed by a detector on ground (over 200 events) [7]. Even in this case there has been no clear detection.

The HAWC experiment, an extensive air shower array with an area of 22,000 $m^2$ fully operating in Mexico since spring 2015 at an altitude of 4100 m a. s. l., has made a detailed study on the possibility of detecting GRBs with both techniques, estimating an overall detection rate of 1.55/y for short GRBs and 0.25/y for long ones [8, 9]. The shower technique was found to be preferred with the idea of lowering the threshold energy down to 50 GeV. In this paper a method similar to that used for CTA and HAWC to calculate the expected rate and features of detectable GRBs has been applied to LHAASO.

### 1.2 The LHAASO experiment

The LHAASO experiment, planned to be installed in the Sichuan Province (P.R. of China) at 4410 m a.s.l., is currently under design to study cosmic rays and photons in the energy range 0.1 - 1000 TeV. This very wide interval is obtained combining different air shower detection techniques covering different energy windows. At the lower end, from 100 GeV to 30 TeV, the Water Cherenkov Detector Array (WCDA) is one of the major components of LHAASO. It is made of three ponds, covering a total surface of 78,000 $m^2$. Each pond is divided into 900 cells ($5 \times 5$ $m^2$ each, with a depth of 4.5 m) seen by one PMT located at the cell floor centre and looking up to detect the direct Cherenkov light produced by the relativistic particles of the showers. In order to maximize the detector performance a large simulation campaign has been carried out to optimize both the cell dimensions and depth and the number of PMTs for each cell. The results show that a higher PMT density, obtained with both smaller cells and higher number of PMTs per cell gives a better performance in terms of angular resolution and sensitivity, but weighting these results with a cost estimate the $5 \times 5$ $m^2$ cells seen by a unique PMT offer the most effective layout [10]. Besides simulations, a prototype water Cherenkov detector has been built and operated in Beijing and an engineering array corresponding to 1% of one pool ($3 \times 3$ cells equipped with one 8" Hamamatsu R5912 PMT each) has been implemented at the ARGO-YBJ site (Yangbajing Cosmic Ray Laboratory, 4300 m a.s.l., P.R. of China). The measured counting rate was about 35 kHz for each cell, with an expected minimum of 12.5 kHz given by cosmic rays. Since LHAASO will be located at a similar height, we foresee a counting rate close to this value. This very high single counting rate does not allow a simple majority but requires a topological one, with different trigger levels. The basic element is given by a $3 \times 3$ cells matrix, whose signal is collected by a custom FEE and sent to a station where a suitable trigger is generated and the corresponding data are recorded. This quite new approach is called "trigger-less" and allows the maximum DAQ flexibility. For example, overlapping the clusters (corresponding to $12 \times 12$ cells) by shifting them of 30 m and requiring a coincidence of at least 12 PMTs within 250 ns in any cluster, a trigger rate of 70 KHz is expected. In the search for GRBs, this approach is particularly effective. For very low multiplicities, starting from 3, the number of random coincidences does not allow the reconstruction of the arrival direction, and moreover the huge amount of events prevents the storage of data. However, if an on-line alert is provided by satellites, as for the case of Cherenkov telescopes, the DAQ can switch to this very low multiplicity mode for a limited amount of time ($\sim$ minutes) and knowing the arrival direction the random coincidences can be strongly suppressed. Even if for these very low multiplicity events (~ 3-10 hits) the angular resolution is very poor (~ 10-15°) and the primary energy is very badly reconstructed, the background is highly reduced with respect to the single particle technique, where the contribution comes from the whole sky. Providing a buffer to store continuously some hundreds of seconds of low multiplicity data, the GRBs can be followed since the beginning covering the delay of the alert transmission. To sum up, the GRB search will be done by LHAASO using the WCDA data in three different ways, depending on multiplicity:

- for n=1,2 or slightly higher number of particles, the DAQ will simply count the number of events in a fixed time window with the corresponding multiplicity and the search will be done in "scaler mode", looking for a statistical excess in the counting rate of all the PMTs in the detector;
- for n$\gtrsim$10 the events are reconstructed one by one and an excess is looked for in the GRB direction. Since all these data are recorded, this search can be done off-line with unlimited GRB duration;
- for the intermediate multiplicities, data are recorded for a fixed time window before and after the real-time alert given by satellites. If successful, this technique will cover for the first time for EAS arrays the energy region between a few GeV and 100 GeV with some directional information.

In order to evaluate the rate and typology of GRBs detectable by LHAASO, several ingredients must be laid together and precisely a GRB model, a parametrization as a function of energy of the detector performance (effective area and angular resolution) and some hypotheses on the expected external trigger rate. All these items will be presented and discussed in the following sections.

### 1.3 The GRB model

To compare our results with CTA and mainly HAWC, we decided to use the same approach found in [3] and [11]. In these papers, a set of pseudo-GRBs has been simulated sampling their features from the experimental ones measured by Fermi (GBM and LAT) and Swift (BAT) satellites. At first, we assume that the external trigger will be given by Fermi-GBM. For each parameter, a different distribution has been considered for short ($T_{90} \leq 2$ s) and long ($T_{90} > 2$ s) GRBs, and no correlation among them has been considered.

### 1.4 The high energy spectrum

We suppose a high energy emission in the 1-1000 GeV energy range as a simple power law with fixed spectral index and no intrinsic cutoff, that will be given exclusively by Extragalactic Background Light (EBL) absorption. To quantify this high energy contribution, we used the correlation between the fluencies measured by GBM and LAT, respectively in the low and high energy bands. Figure 1 shows this correlation, updated with all the 21 GRBs with fluence calculated in the same time window. The points are very scattered, and for short GRBs (red squares) only 3 events were measured by both detectors. Nevertheless we followed the assumptions made in [12] that the LAT fluence in the 100 MeV-10 GeV energy range is about 10% of the GBM fluence in the 10 keV-1 MeV energy range for long GRBs, while for short ones the amount is 100%.

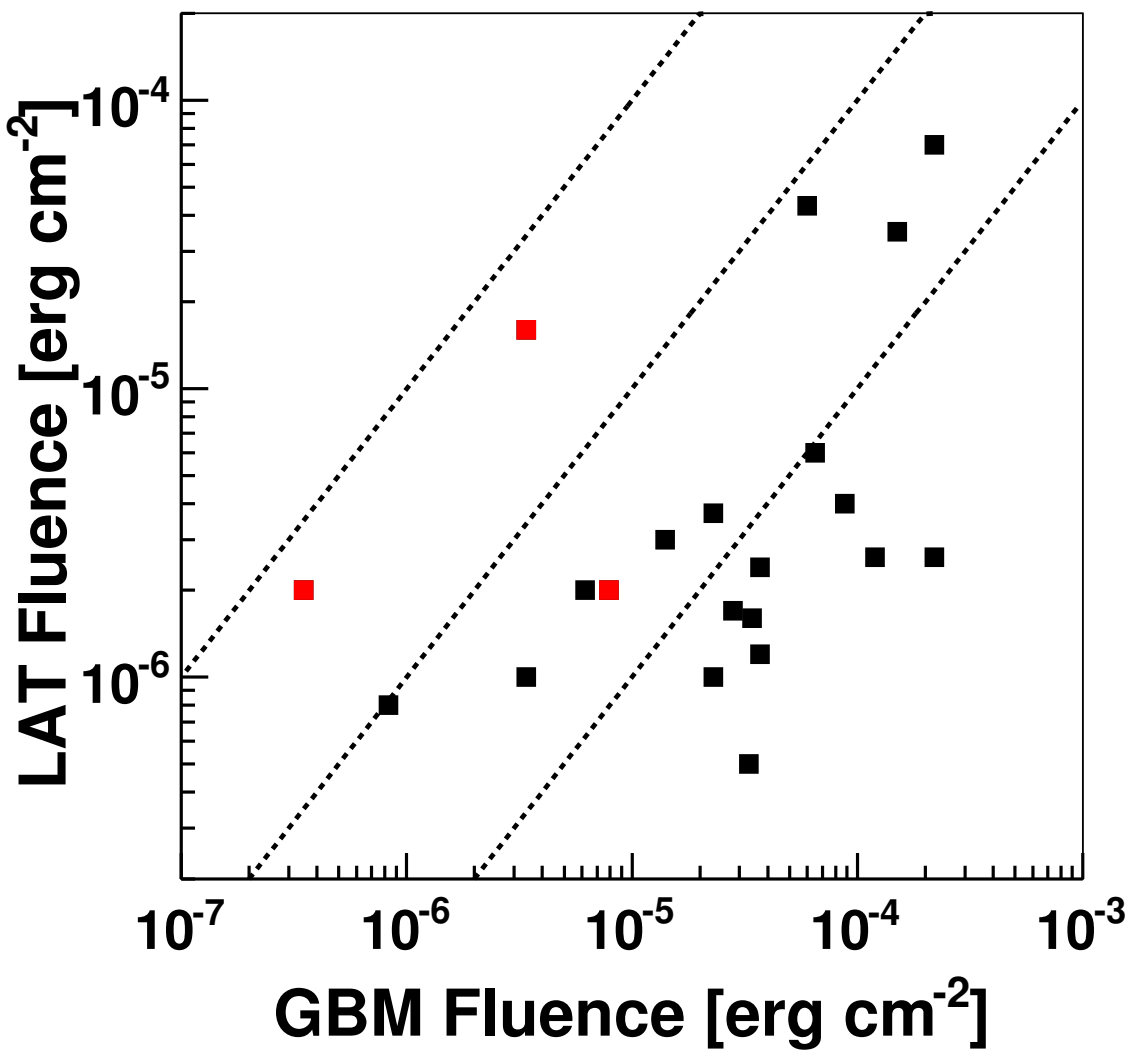

Fig. 1. A comparison of the LAT and GBM fluencies in the [0.1-10] GeV and [10-1000] keV range respectively. Black (red) squares are for long (short) GRBs; dashed lines indicate LAT-GBM fluence ratio of 0.1, 1.0, 10.0 (bottom to top).

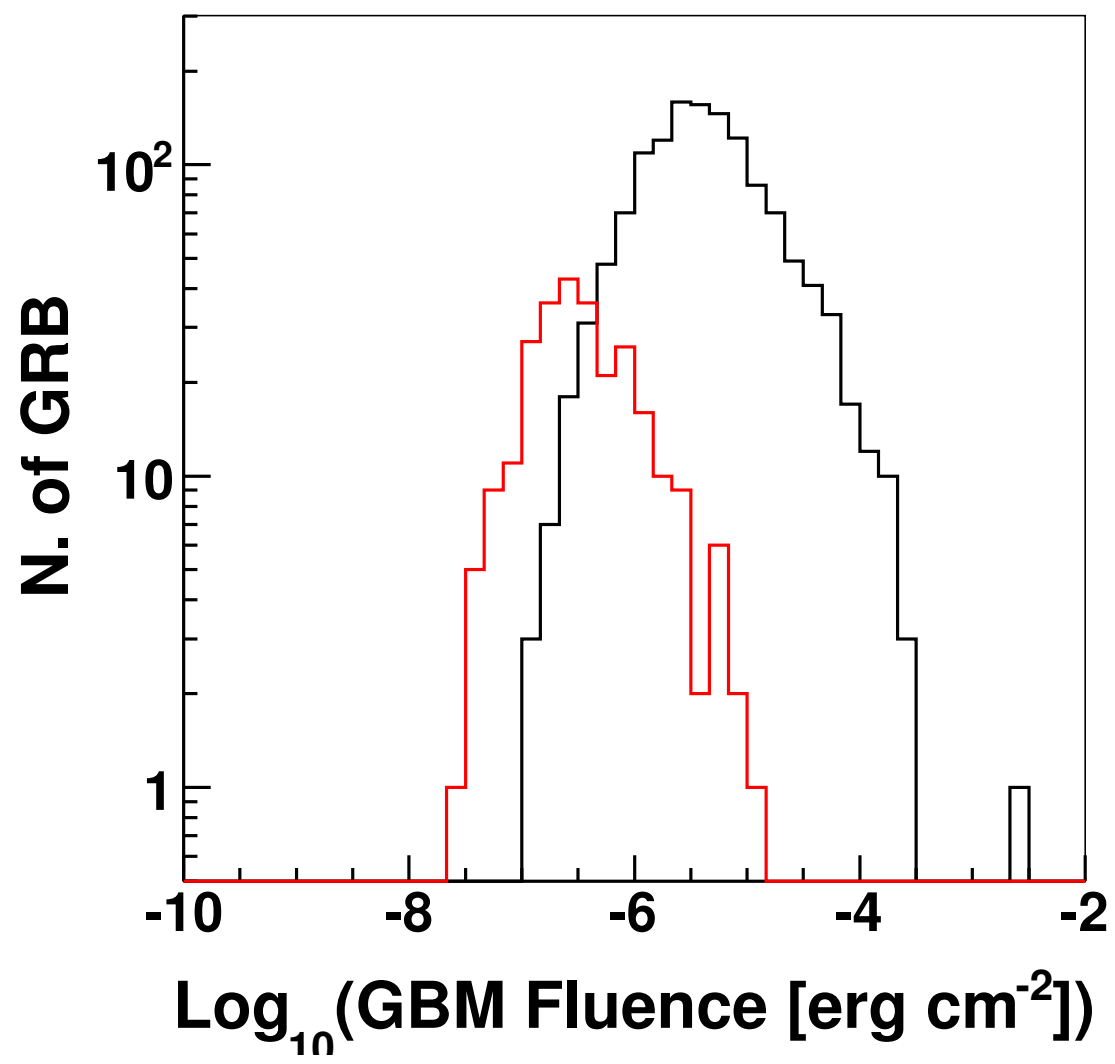

Fig. 2. Distribution of measured GBM fluencies for Long (black) and Short (red) GRBs in the [10-1000] MeV range.

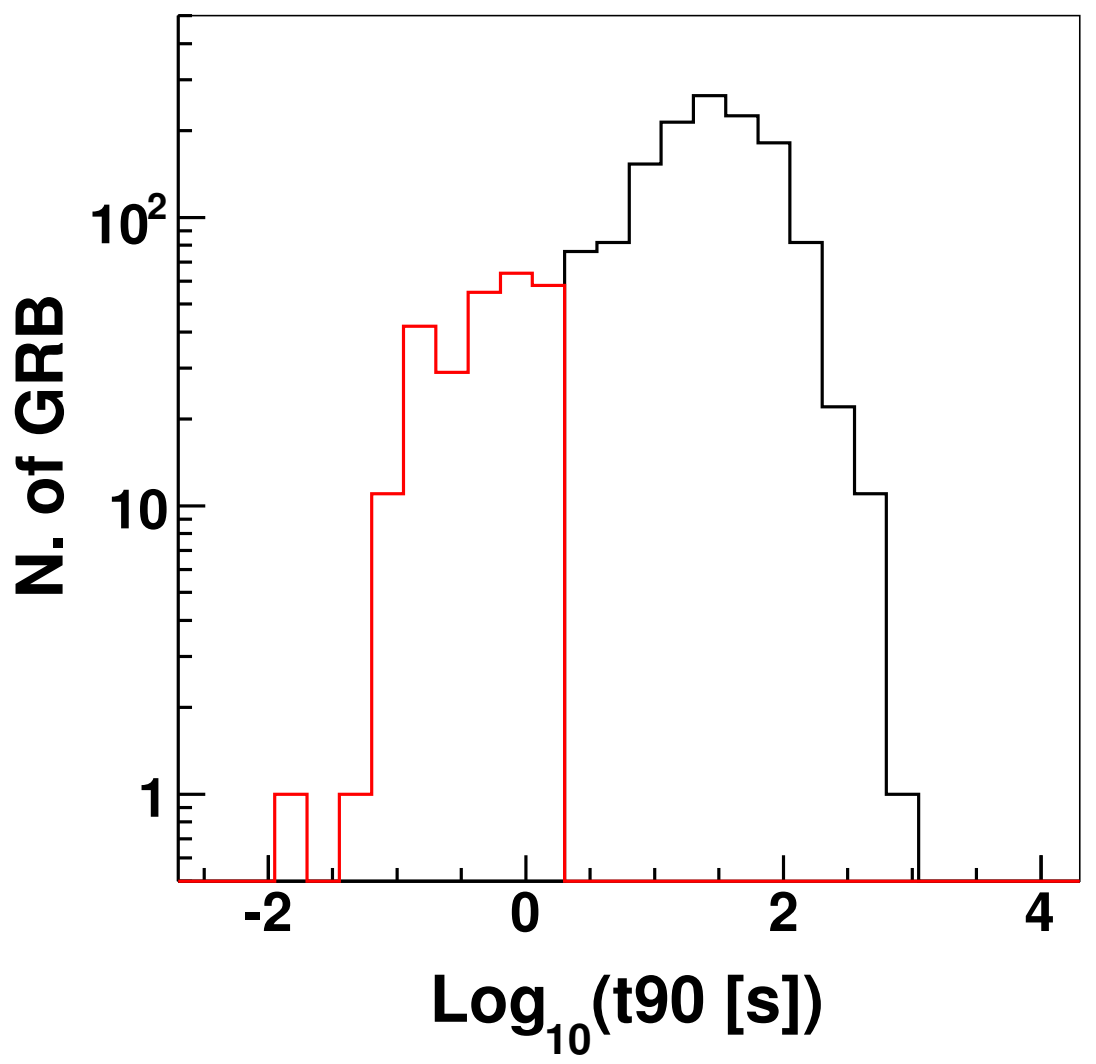

Fig. 3. Distribution of $T_{90}$ durations for Long (black) and Short (red) GRBs detected by Fermi-GBM.

Figure 2 shows the low energy fluence measured by GBM (red: short GRBs; black: long GRBs). For the high energy emission of our pseudo-GRBs we sampled from these distributions a fluence that is reported to the 100 MeV-10 GeV energy region using the quoted percentages. Since the fluence distribution for long GRBs is shifted towards larger values by about a factor of 10 with respect to the short ones, the high energy scaling produces a close distribution for short and long GRBs. For the high energy spectrum we used a spectral index $\alpha$ = -2, since for long GRBs with an additional power law this is the mean value measured by LAT. For short GRBs we used $\alpha$ = -1.6, the same value used in [11]. The assumption that all short and long GRBs have an additional high energy power law with spectral index -1.6 and -2, respectively, and without any cutoff in all the considered energy range is quite raw and optimistic, but in any case it allows us to compare our results with the expected sensitivity of HAWC.

### 1.5 The light curve

As pointed out by Ghisellini et al. in [13], the GRB light curve can be modeled as a constant flux during the $T_{90}$ measured by GBM, followed by a power law fall-off with index $\gamma$=1.5. Due to its expected higher sensitivity and to the fact that it will lose the prompt phase of most GRBs, CTA made this assumption in [3] to estimate the rate of detectable GRBs. We decided instead to follow the approach used for HAWC, i.e., we sampled the $T_{90}$ distribution showed in Figure 3 obtained by Fermi-GBM for long GRBs (black line), while we used a fixed GRB duration of 2 s for short GRBs.

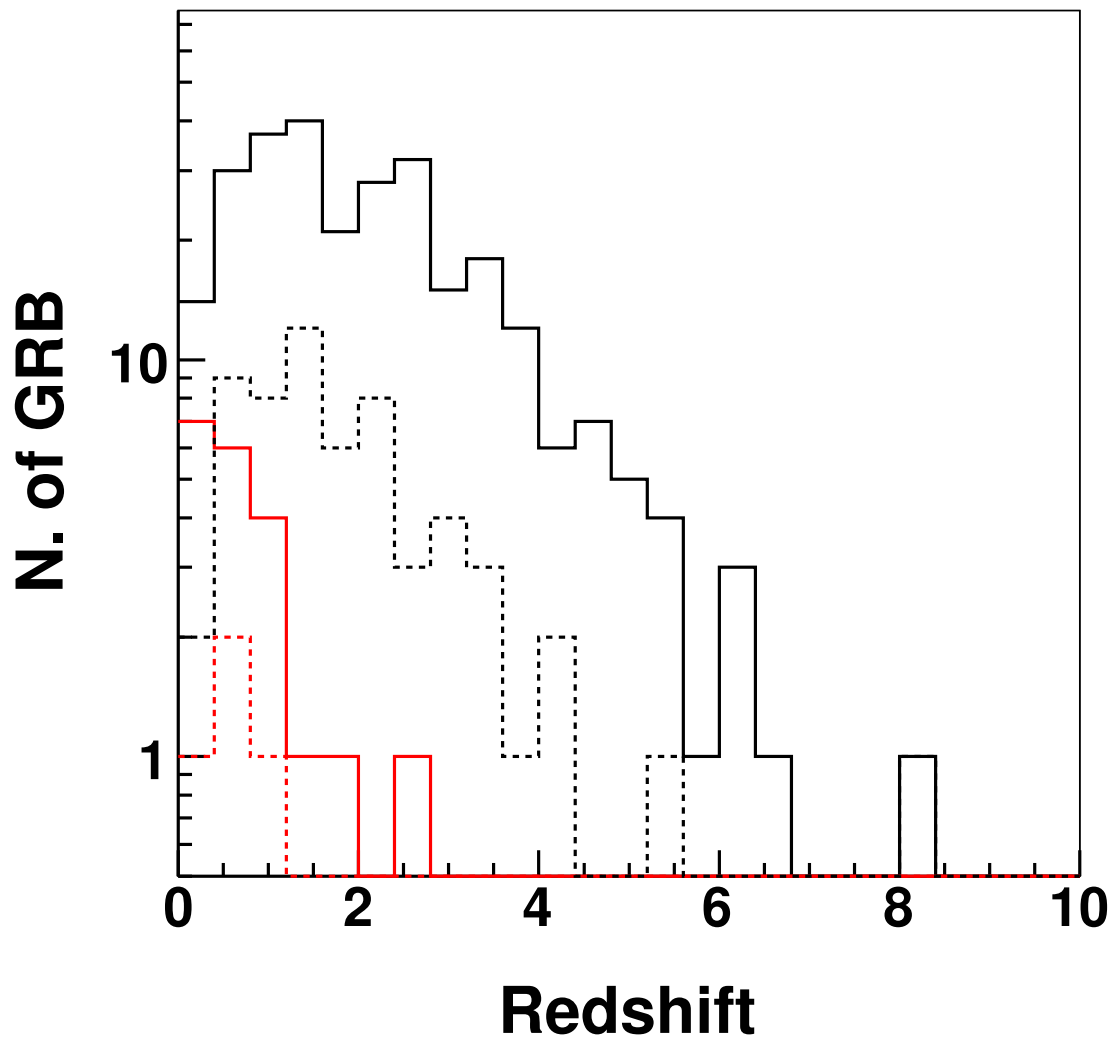

Fig. 4. Distribution of redshift for Long (black) and Short (red) GRBs detected by Fermi-GBM. Dashed lines show the subsamples of Long and Short GRBs observed simultaneously by Fermi-GBM and Swift-BAT detectors.

### 1.6 The EBL absorption

The interaction of very high energy photons with the EBL produces e-pairs and thus a quite sharp spectral cutoff. This absorption depends on the redshift and GRBs are cosmological objects, with a mean value $z\simeq2$. Many models of EBL attenuation have been published in the last decades, with a general trend towards an increase of transparency due to the observation of very high energy photons at larger redshifts [14]. In this work we used the model by Kneiske et al. [15]. Since the energy resolution of the Fermi-GBM instrument does not allow the detection of clear spectral lines, the GRB distance is obtained sampling the redshift distributions measured by Swift-BAT. Figure 4 shows these distributions for long (black line) and short (red line) GRBs. We apply the EBL cutoff starting from z=0.1 and up to 1 TeV, the maximum energy after which the source spectrum is totally absorbed. This choice is due to the fact that z=0.1 corresponds roughly to a cutoff energy of 1 TeV in our model. An important point to be checked is that the higher sensitivity of Swift-BAT with respect to Fermi-GBM could distort the redshift distribution, so we selected the subsample of GRBs detected by both. The corresponding distribution (dashed black and red lines for long and short GRBs respectively) is also shown in Figure 4 and does not show significant deviations from the whole data set.

### 1.7 The detector performance for the different configurations

The sensitivity of an EAS array to any gamma-ray source and in particular to GRBs is given by the angular resolution and the effective area for primary photons. Both of these depend on the primary energy and on the zenith angle (however, the dependency of the angular resolution on the zenith angle is small). The expected performance of the detector is evaluated by means of a detailed Monte Carlo simulation that reproduces the development of gamma-ray showers in the atmosphere and the interaction of the secondary EAS particles with the detector. For each pseudo-GRB the expected signal $S$ is calculated integrating from 1 GeV to 1 TeV:

$$S=\int_{1\ GeV}^{1\ TeV} S_\gamma(E)\times EBL(E,z)\times A^\gamma_{eff}(E,\theta)\times T_{90}\ dE \quad (1)$$

where $S_\gamma(E)$ is the sampled GRB spectrum, $EBL(E,z)$ the EBL absorption, $A^\gamma_{eff}(E,\theta)$ the photons effective area and $T_{90}$ the burst duration. The peak energy $E_{peak}$ is defined as the energy corresponding to the maximum of the signal function before integration. For GRBs, $E_{peak}$ is typically less than 100 GeV.

The zenith angle of the pseudo-GRB is randomly chosen in the range from 0 to 50 degrees, with a uniform distribution in the corresponding solid angle. To calculate the expected background $B$, the same Monte Carlo simulation is run for primary protons, obtaining an effective area $A^p_{eff}(E,\theta)$ as a function of energy and zenith angle. The expected background $B$ is calculated integrating in the same energy range 1GeV-1 TeV:

$$B=\int_{1\ GeV}^{1\ TeV} S_p(E)\times A^p_{eff}(E,\theta)\times T_{90}\times\Omega(E_{peak})\ dE \quad (2)$$

where $S_p(E)$ is the cosmic ray spectrum and $\Omega(E_{peak})$ the solid angle corresponding to the angular resolution for $E=E_{peak}$. As angular resolution, we use the $\Psi_{70}$ aperture that maximizes the signal to noise (i.e. S/$\sqrt{(B)}$) ratio keeping 71.5% of the signal with an aperture of 1.58 $\sigma$.

For the cosmic ray spectrum, all the primary nuclei from p to Fe have been simulated and then grouped into five mass sets (p, He, CNO, NeMgSi, Fe). As a first step, the effective area has been obtained considering a cosmic ray flux made by only protons, normalized to obtain a counting rate equivalent to that produced by all the five mass groups using the Hörandel primary composition [16]. For each primary particle (in our case photons and protons) this simulation procedure, that is very CPU-consuming requiring the generation of a huge amount of events, must be repeated for each trigger condition and several zenith angles. Figure 5 shows the angular resolution for internal events ($r <$ 160 m) and trigger multiplicity $N_{fit} \geq 20$, where $N_{fit}$ corresponds to the total number of hits for each event after an iterative cleaning procedure to discard the random hits on the basis of the temporal features of the shower front. Figure 6 shows the effective areas for gammas (left) and protons (right) in steps of 15° for zenith angle from 0° to 60° and trigger multiplicity $N_{fit}\geq$ 10. These results have been obtained using CORSIKA [17] for the development of EASs from gamma rays and protons, and a custom software derived by the Milagro one for the detector response. The complete set of simulations for the different trigger conditions is currently undergoing.

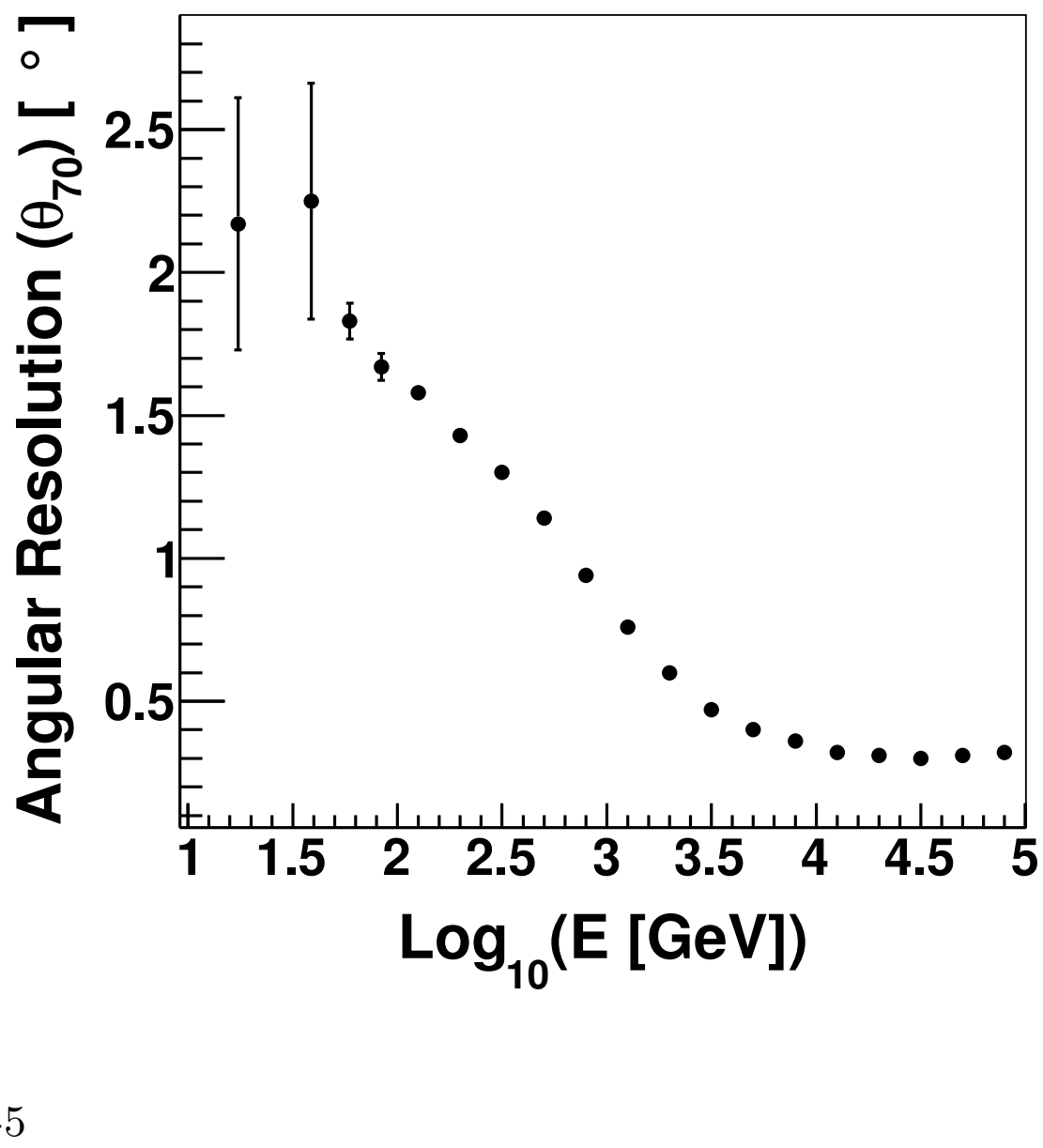

Fig. 5. Expected angular resolution of the WCDA for internal events and $N_{fit} \geq 20$ (see text).

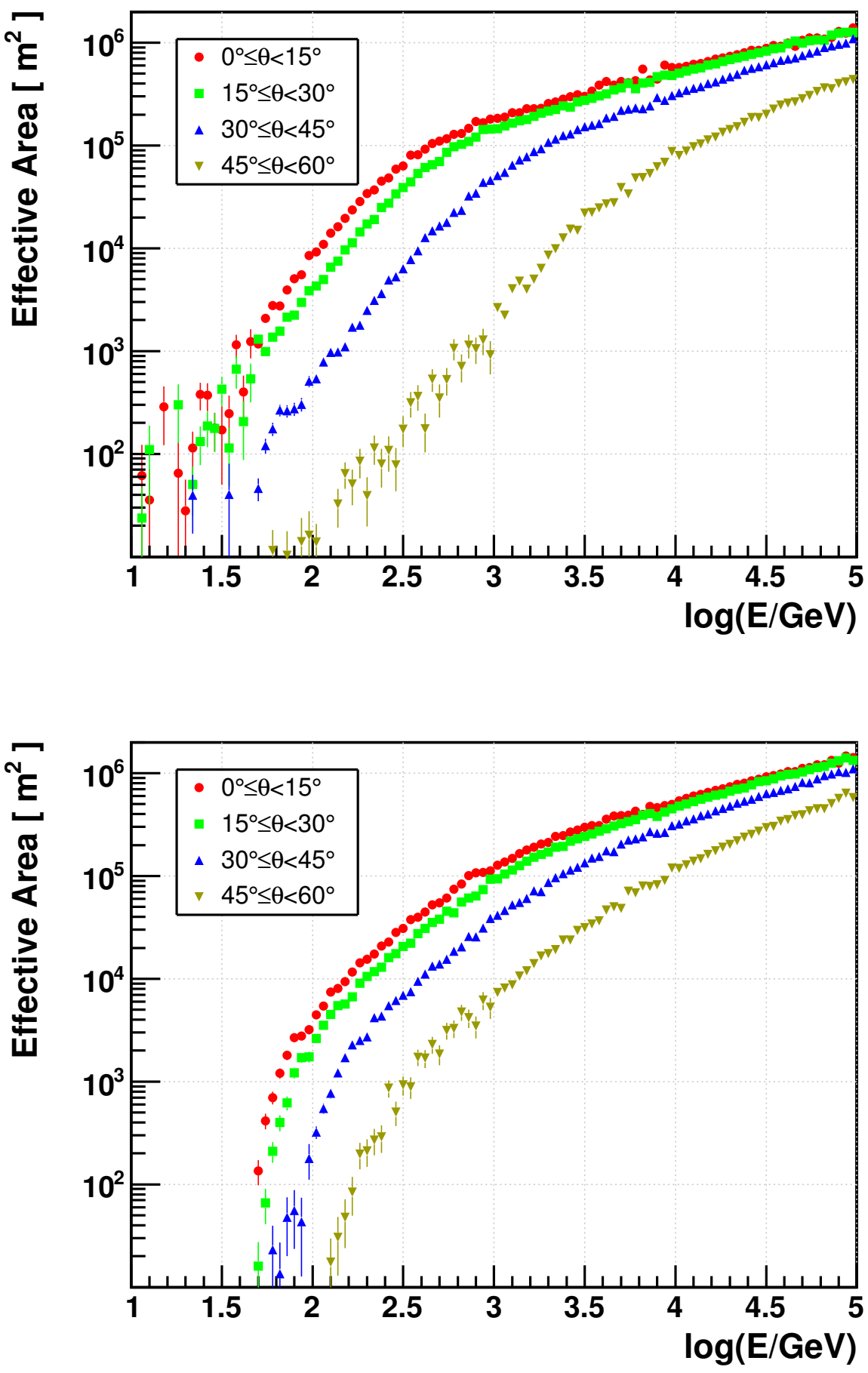

Fig. 6. Effective areas of the WCDA for gamma rays (top) and protons (bottom) for four zenith windows and $N_{fit} \geq 10$ (see text)).

### 1.8 The detector threshold and external trigger rate

The confidence level of a detection is obtained requiring a signal greater than a given number of background fluctuations. A value of 5 s.d. has been set as the detector threshold, and to properly calculate the signal significance, equation (17) of [18] was used. Once the fraction of detectable GRBs has been derived, an external trigger rate must be provided. According to [19], Fermi/GBM detected GRBs with a mean detection rate of $\sim$ 250 yr$^{-1}$ in its f.o.v. of 8.74 sr. This corresponds to more than 700 GRBs yr$^{-1}$ from the whole sky taking into account that GBM has a duty cycle of about 50%. The LHAASO angular acceptance up to 50° in zenith angle is 2.24 sr, with a full sky coverage of 17.9%. The expected trigger rate in GBM follow-up observations is thus 45 yr$^{-1}$, while in independent mode we foresee $\sim$ 130 "GBM-like" GRBs per year. These values will be used to normalize in time our pseudo-GRBs data set.

### 1.9 Discussion and conclusions

CGRO/EGRET in the past and recently Fermi/LAT have clearly demonstrated the emission of GeV photons from GRBs. Nevertheless, this VHE emission is quite unusual and the presence of a hard power-law contribution to the spectrum has not been confirmed for all the GRBs seen by LAT. Moreover, the extrapolation to the GeV region is made over several orders of magnitude, with a fixed ratio between low and high energy fluencies that roughly fits reality. The expected fraction of detectable GRBs is largely dependent on the adopted GRB model, and for this reason we decided to use as much as possible the same assumptions made by CTA and mainly by HAWC to make the results comparable. Presently, our GRB model is defined together with all the calculation details. The determination of the effective area and angular resolution for gamma rays and protons and for the different trigger conditions is currently under way and the very first results on the GRBs detectability and typology for some trigger conditions by LHAASO-WCDA are under check.

## 2 Upgraded WCDA: A Tool for Transient Phenomena and Multi-messenger Astronomy

Executive summary: Large High Altitude Air Shower Observatory (LHAASO) has been building the 2nd and 3rd water Cherenkov detectors, denoted as WCDA-2 and WCDA-3, using Multi Channel Plate (MCP) staffed 20" Photo Multiplier Tube (PMT) instead of the 8" traditional PMTs with dynodes as originally proposed. This will enhance its sensitivity at energies around 100 GeV. The effective area for gamma ray detection will reach to 1800 m$^2$ and differential sensitivity to 0.2 Crab Unit (CU, the gamma ray flux from the Crab Nebula at corresponding energy) around 50 GeV. It will be the very useful survey detection for transient phenomena around 50 GeV in the northern sky. LHAASO is expected to play an important role in the multi-messenger astronomy after the WCDA-2 is put in operation by the end of 2019. WCDA-3 is expected to be finished by the end of 2020.

### 2.1 Motivation for the Upgrade

The first multi-messenger observation of Gravitational Wave (GW) event GW170817 by LIGO and VIRGO GW observatories together with many other Electro-Magnetic (EM) wave observations, such as FERMI[20] and the multi-wavelength campaign on the possible EM partner of the very high energy neutrino IC-170922A detected by the IceCube experiment[21] are very significant progresses in astroparticle physics in past years. They opened new windows for exploring the high energy phenomena in the universe. This, however, becomes a challenge to LHAASO with its original proposal which is designed to target the high energy (above several hundred GeV) gamma ray sources and charged particles at even higher energies up to a few EeV. At energies lower than 300 GeV, the gamma ray detection sensitivity is not sufficient to detect those sources which is typically faint. In order to enhance the sensitivity below 100 GeV, we proposed to enlarge the sensitive area of the photo cathode of the water Cherenkov detector in the LHAASO array. In this section, we are going to briefly describe the experiments with original WCDA as well as the upgrading plan in §**2.2**, and the performance of WCDA in gamma ray astronomy with the the enhancement at low energy region in §**2.3**. The whole upgrading plan is briefly summarized in §**2.4**.

### 2.2 LHAASO-WCDA Experiment and the Upgrading Plan

LHAASO is a multipurpose complex of Extensive Air Shower (EAS) detection observatory consisting of four major detector arrays. In the center of the observatory, the WCDA is divided into 3 independent water pools, namely two smaller pools with the area of 150 $m$×150 $m$ each, denoted as WCDA-1 and WCDA-2, and the larger one with the area of 300 $m$×110 $m$, denoted as WCDA-3. The layout and orientation of WCDA is shown in Figure 7. WCDA-1, in south-west, is already put in scientific operation since April 2019.

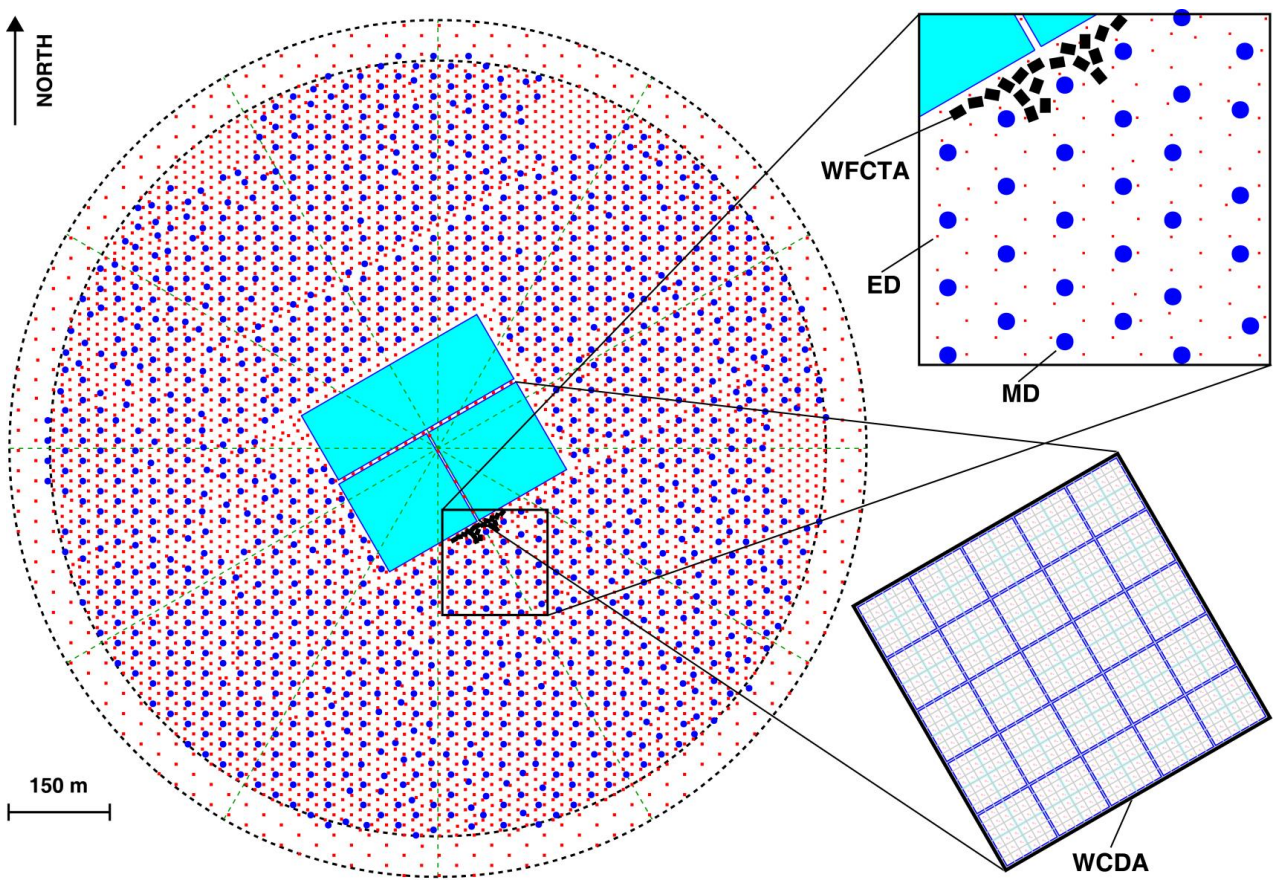

Fig. 7. The layout and orientation of the water Cherenkov detector array (WCDA, rectangle in the center) in LHAASO. In south-west, WCDA-1 of 150m×150m is put in operation. In south-east, WCDA-2 with the same size as WCDA-1 is illustrated with the pup-up square on the right-hand-side. It is divided into 900 detector units. 36 units form a cluster with a junction box with all Front-End Electronics (FEE), digitization, power suppliers and calibration LED-drivers in. The bigger one in north, WCDA-3, has 1320 units covering an area of 300m×110m.

WCDA-1 has 900 units, equipped by a large (8") PMT for timing and a small (1.5") PMT for pulse size in each unit. WCDA-1 measures shower directions with a resolution better than 0.2° above 10 TeV and 1.0° above 600 GeV. The small PMTs enlarge the dynamic range of the detector very much. This enables the measurement of the detailed particle density distribution in the shower cores without saturation even for energetic showers up to 10 PeV and achievement of the core location resolution better than 3 $m$. This is designed for the identification of the primary particle species in the cosmic ray composition and spectrum measurements.

Low energy showers are small in terms of total number of particles that reach to the pools, therefore the total Cherenkov signal generated by those secondary particles in every detector unit is faint, even for units being near the cores of the showers. In order to enhance the gamma ray detecting sensitivity at low energies, enlarging the sensitive photo-cathode area of the PMT in the same size unit could be one effective way to catch the faint signals. LHAASO has made a upgrading plan along with this approach, namely to replace the 8" PMTs by 20" PMTs in the rest WCDA-2 and WCDA-3 with the active area of 55,500 $m^2$ in total. To gain the enlarging factor of 6.25 in terms of the photo cathode area versus the 8" tube, the dynamic range is also shrunk by the same factor. This is the price to pay. In order to compensate the loss, a 3" PMT is installed beside the large PMT in each unit, read out only for the pulse size by a simplified version of the Front-End Electronics (FEE) covering 4 orders of magnitudes of number of photo-electrons.

### 2.3 Performances and Prospects for Gamma Ray Astronomy

Gamma ray induced showers are different from showers induced by charged Cosmic Ray (CR) nuclei in terms of the hits distribution in the pool. In general, the later is more spread out than the former ones as shown in Figure 8, where two Monte Carlo (MC) simulated events due to 1 TeV gamma ray and 2 TeV proton are compared with each other. Even more significantly, the CR events have many "hot spots", the populated hits, outside the core region, while the gamma ray events are much cleaner beyond some distance from the core. This enables us to identify the gamma events out of the CR background, nevertheless they are much more ($10^{4\sim5}$) than gamma ray signals even if within a very small angular region defined by the point-spread-function (PSF) near the sources. By eliminating the events which have the most populated hit in the outer region, 45 $m$ from the core, greater than certain number of photoelectrons, $N_{th}$, the CR background will be suppressed to very low in the nearby region of sources. Making balancing between the elimination of the background CR events and the loss the $\gamma$-like signal events to maximize the sensitivity. It is found that $N_{th}$ increases with the number of hits that are involved in the shower front fit, denoted as $N_{fit}$ which measures the shower energy, i.e. proportional to $N_{fit}/logN_{fit}$. Shown in the Figure 8, one $\gamma$-like event from the Crab Nebula direction measured by WCDA-1 is selected according to the algorithm, together with a typical event that has similar number of hits to the $\gamma$-like event as a comparison. Other parameters characterize the distribution of hits in the whole pool area, such as hit density in the outer region, are used in the identification of gamma ray events. This results an effective area for gamma ray detection of about 230 $m^2$ at 50 GeV and 30,000 $m^2$ at 1 TeV, respectively, if all 3 pools were equipped by 8" PMTs. The corresponding sensitivity of the gamma ray point-like source detection is plotted in Figure 9 as a function of gamma ray energy.

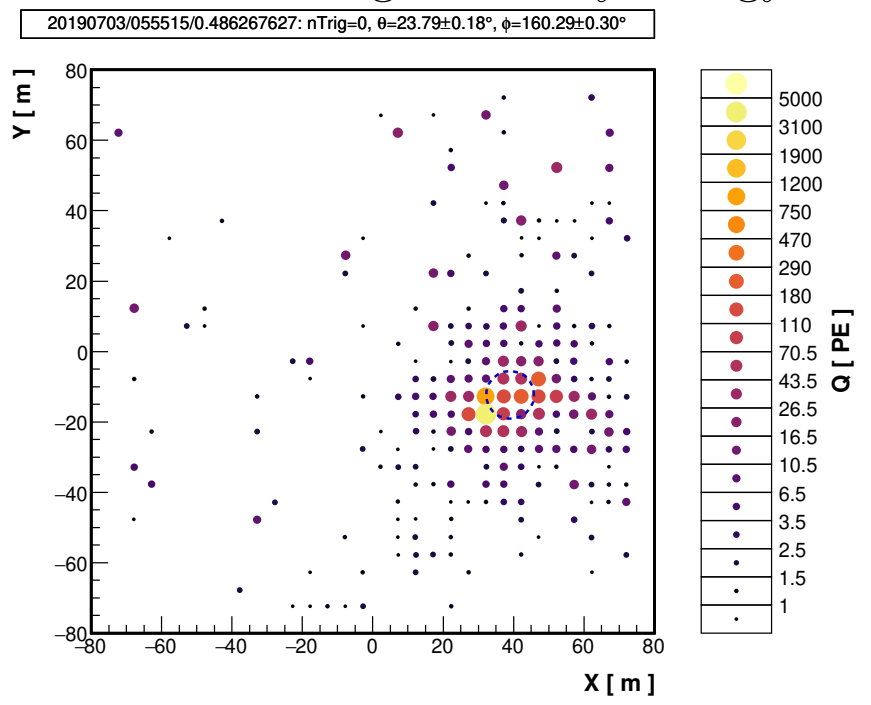

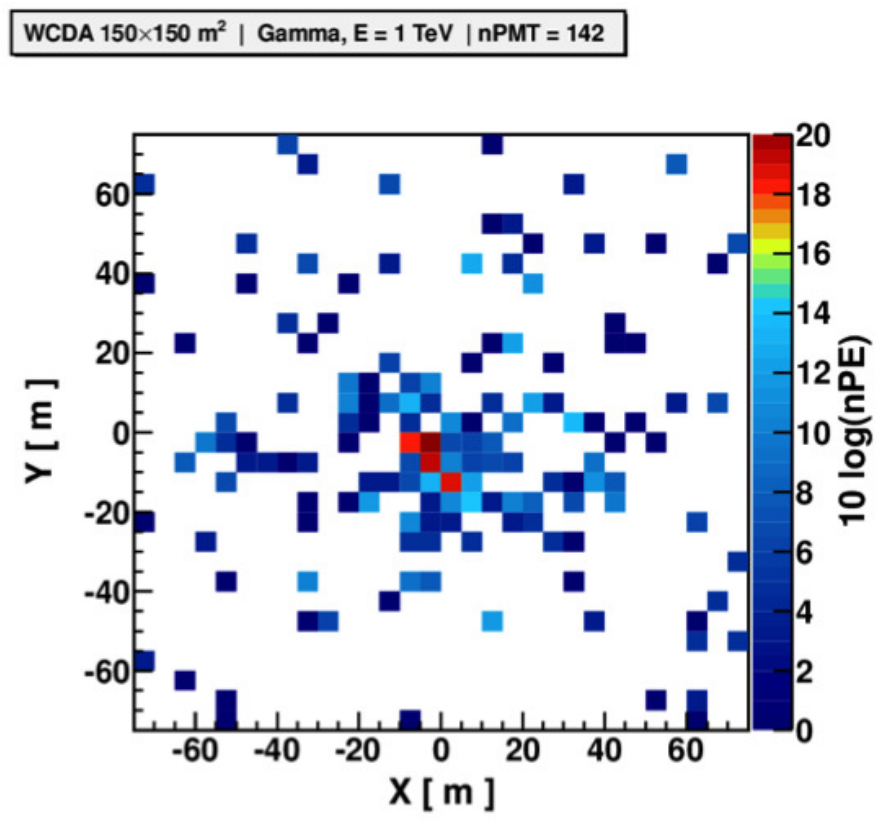

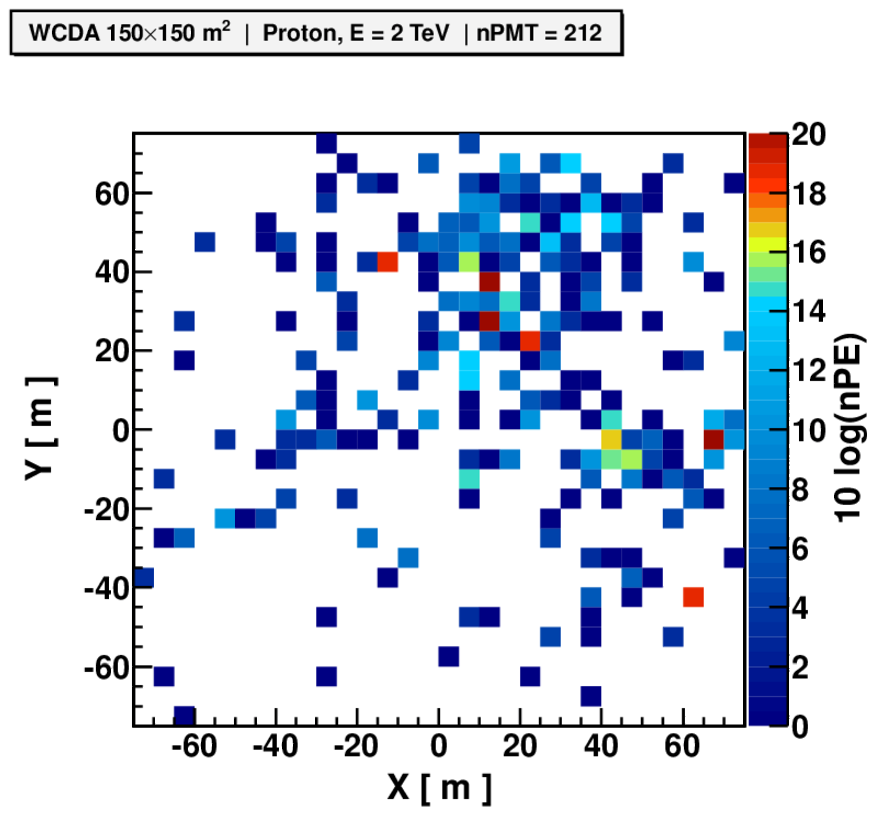

Fig. 8. The top panel shows an event recorded by WCDA-1 representing $\gamma$-like from Crab Nebula direction (left) according to the algorithm of $\gamma/p$ discrimination. This algorithm was tested and optimized using simulated events. The middle panel shows a simulated event induced by 1 TeV gamma ray. As a comparison, a 2 TeV proton event is also showed in the bottom panel which looks much more spread out than the events induced by gamma rays.

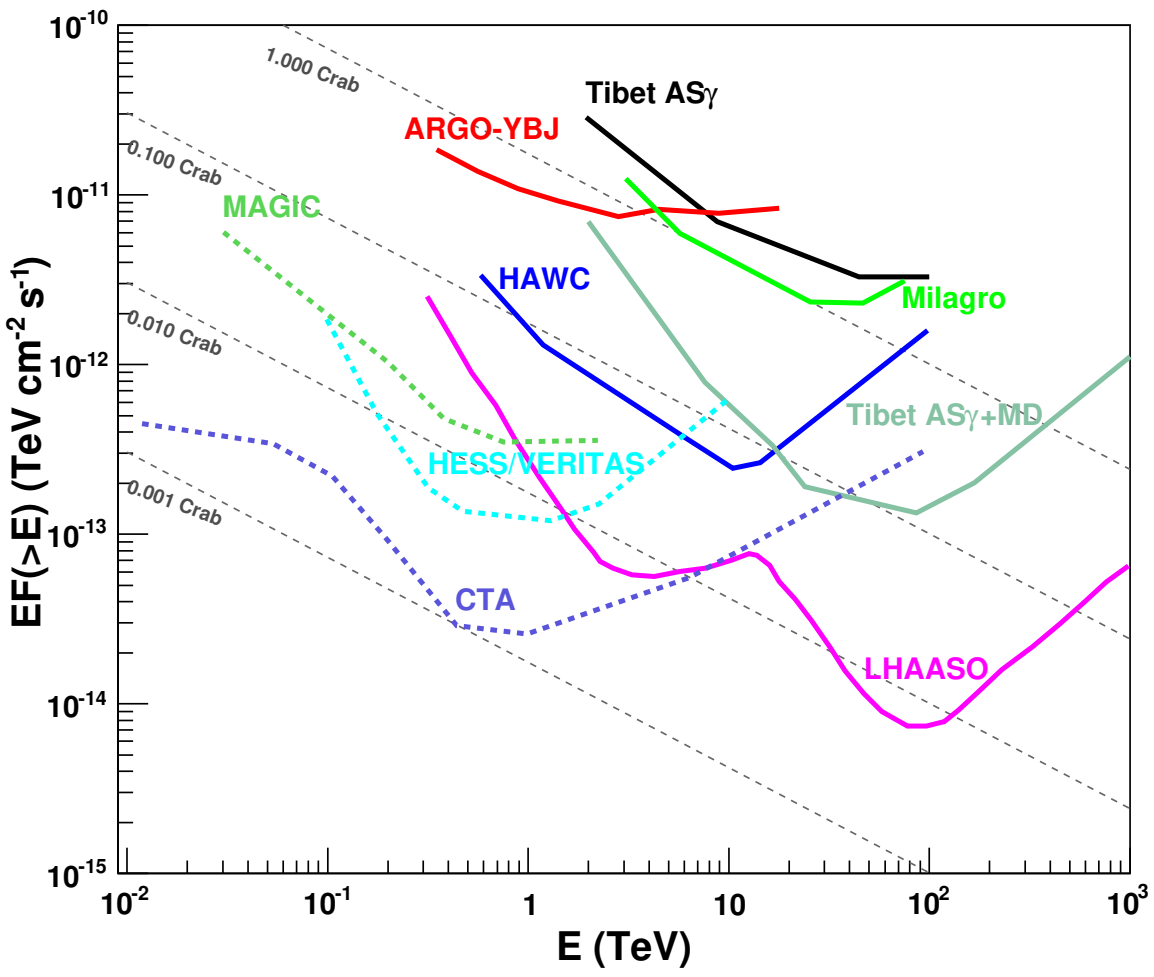

Fig. 9. The integrated sensitivity of LHAASO (in pink) comparing with other experiments. The part of the curve below 10 TeV is the contribution of WCDA optimized with the cuts described in text.

With this configuration, LHAASO has a survey power for discovering all sources that are brighter than 7 mini crab unit above 1 TeV. Taking into account the wide field of view of ∼1/7 of the entire sky and the constant exposure time of 24 hours, LHAASO is very significant in finding new sources. It is actually estimated that about 40 new Active Galactic Nuclei (AGN)[21] could be discovered within one year after it is fully operated. It is also expected that LHAASO will make a deeper survey inside our galaxy comparing with what has been done by the HAWC experiment[22]. LHAASO will play as a wide FoV monitor for any kind of transient phenomena, such as AGN flares and Gamma Ray Bursts (GRB), as well.

**1. Enhancement at Low Energy for follow-up observation of GRBs and monitoring of AGN flares** With the upgraded configuration using 20” PMTs in WCDA-2 and WCDA-3, the effective area at energies below 300 GeV is significantly enlarged, i.e. reaches to 1,800 $m^2$ at 50 GeV and 44,000 $m^2$ at 1 TeV. The corresponding differential sensitivity around 50 GeV is expected to be 0.2 CU per a quart decade of energy which is compatible with the space borne detector FERMI-LAT, as shown in Figure 10. The difference, however, is that the effective area is a factor of 1,800 larger than the later. This means that more than 1000 photons are expected to be recorded if the gamma ray burst event GRB090510 happened again in the field of view of LHAASO. In event GRB130427A, FERMI-LAT recorded a single gamma photon at 95 GeV[23]. If such GRBs with high energy photons happened again, they should be detected by LHAASO given the LHAASO sensitivity. This opens a window for the multi-wavelength campaign in a much convenient way because of the clock-round operation of LHAASO. With any global alarm for transient phenomena, such as GRB, it is easy for LHAASO to recall the data in the window in which the alarm was ringing. Not only the status of the source at $T_0$ can be observed, but also it is in principle possible to find any pre-burst emission of gamma rays if there were.

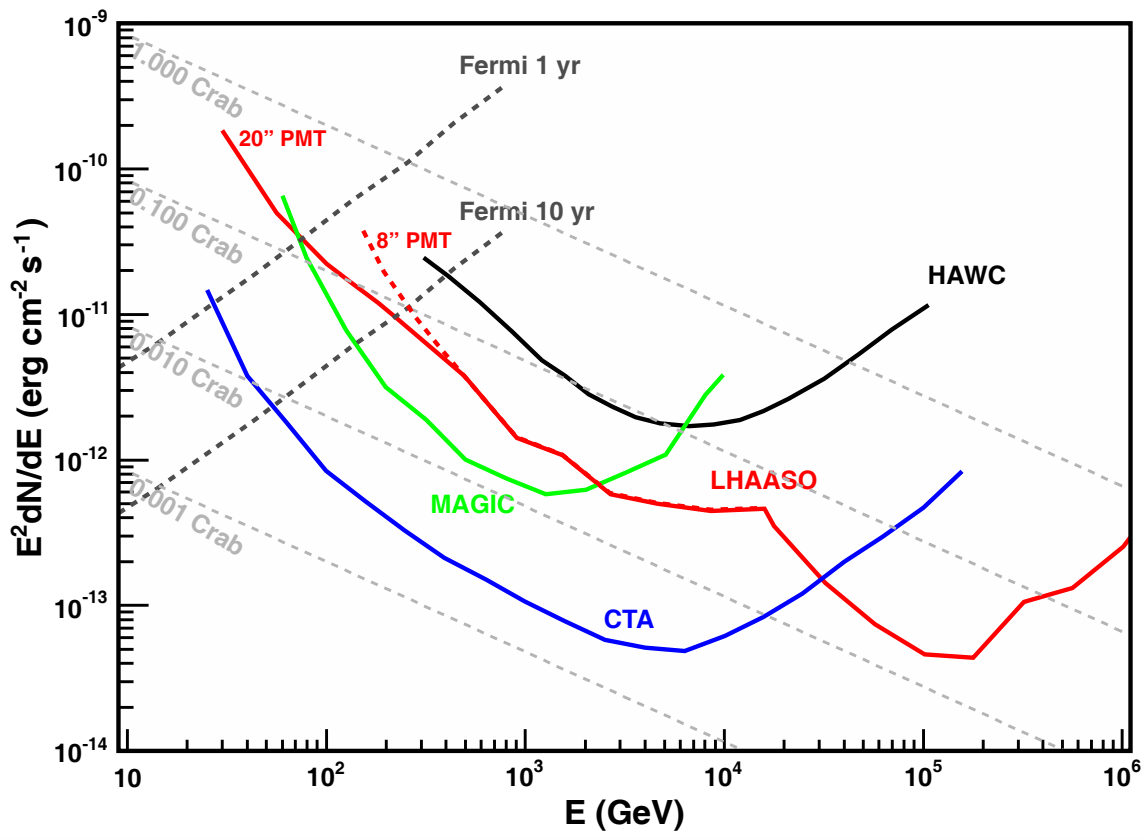

Fig. 10. The differential sensitivity of LHAASO (in red) comparing with other experiments. Below 300 GeV, both estimates with 8” PMT configuration (dashed line) and 20” PMT configuration (solid line) are plotted. It is noticed that WCDA with 20” PMTs is almost as same sensitive as FERMI/LAT at 70 GeV.

With such a sensitivity, LHAASO will be a transient phenomenon finder as well. An alarm trigger algorithm is going to be operated to constantly watch all interested AGNs in LHAASO’s Field of View (FoV) for any excess in various time windows, e.g. from few seconds to hours. It is useful for monitoring any AGN flare, e.g. if its emission level rises to be greater than 1 CU within a hour, an alarm will be broadcast to the whole community.

**2. Prospects for Multi-messenger Exploring** Investigating sources with multi-messengers is very powerful in viewing of inside mechanism of high energy phenomena in the universe, particularly for possible common origins of the messengers, such us neutrinos, gamma rays, charged particles and gravitational waves. However, to identify the sources and verify the association, all corresponding detectors are required for sufficient sensitivities. As an example, we investigated the possible association between the ultra high energy muon neutrino event IC-170922A detected by the IceCube experiment[21] and the blazer TXS 0506+056 which had a faint flare within 20 days after the neutrino event in multiple wavelength bands, including X-ray (SWIFT), gamma ray (FERMI-LAT) and very high energy gamma ray (MAGIC). The Spectral Energy Distribution (SED) of the blazer during the flare is reported in Ref. [21] and is quoted here in Figure 11 over a very wide energy range. Also shown in the figure, sensitivity curves of several experiments, including HAWC, HESS, VERITAS and upgraded LHAASO. According to this, LHAASO will play a significant role in such multi-messenger observation by covering an important energy range starting from as low as 30 GeV to few hundred TeV.

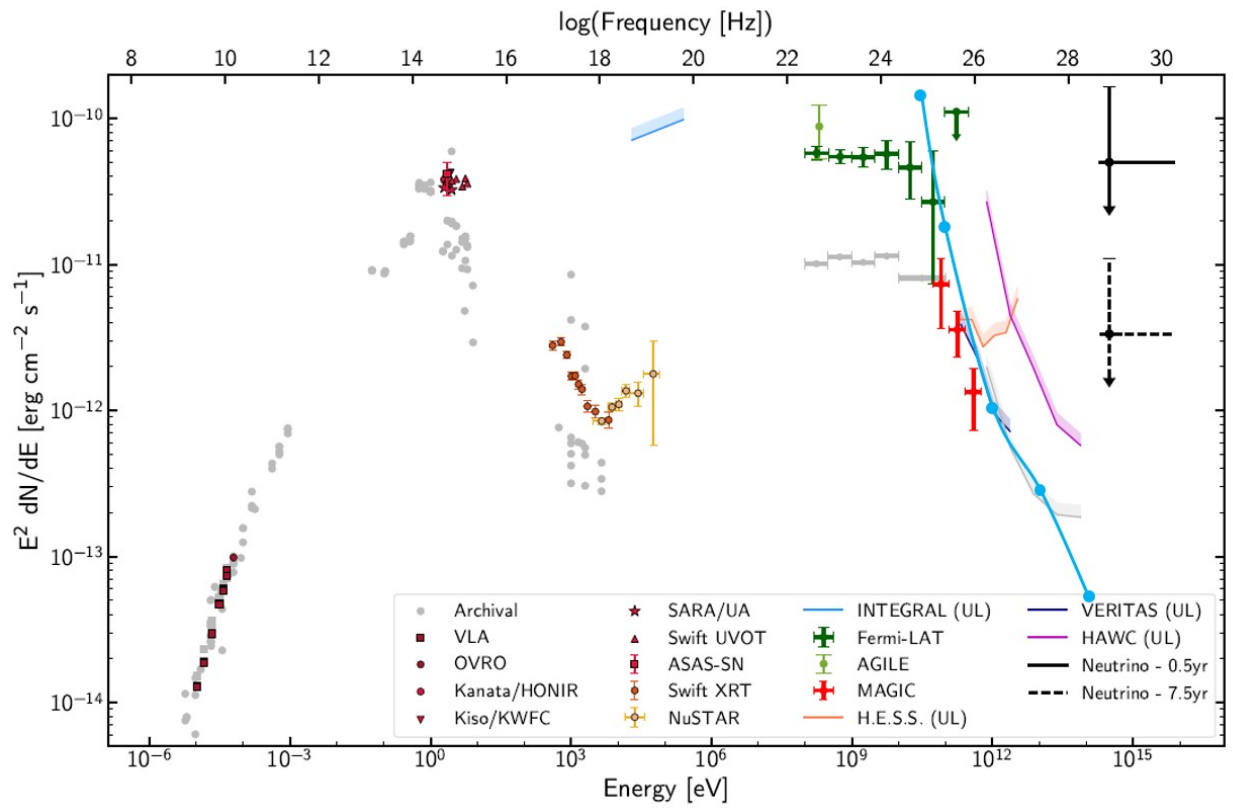

Fig. 11. The SED of TXS 0506+056 during the flare within 20 days after the neutrino event IC-170922A. It is reported in Ref. [21] including sensitivity curves of HAWC, HESS and VERITAS experiments. The upgraded LHAASO sensitivity is also plotted in the same figure (light blue) from 30 GeV to 100 TeV.

For other blazars including BL Lac objects and flat-spectrum radio quasars, the most extreme subclass of AGNs, LHAASO has a full duty cycle and a large FoV to monitor flares of them continuously. In Figure 12, the flare on Mrk501 for 35 days is measured by Swift, FERMI-LAT and ARGO-YBJ (See Ref.[24]). It clearly differs from the steady emission which fits well with the Synchrotron Self-Compton (SSC) model. Assuming the similar flare occurred again, the prediction for LHAASO's observation is plotted in Figure 12 and LHAASO will give an accurate spectrum at TeV region which can be the key to explain the radiation mechanism of flare. LHAASO not only serves as a global alarm system for the high energy flares, but also opens a great opportunity to identify the emitting mechanism during the flares.

Beyond the traditional studies of the extra-galactic gamma ray emissions, the potential of LHAASO in exploring new physics such as intergalactic magnetic field detection and Lorentz invariance tests, has been discussed in depth elsewhere (See Ref.[25]).

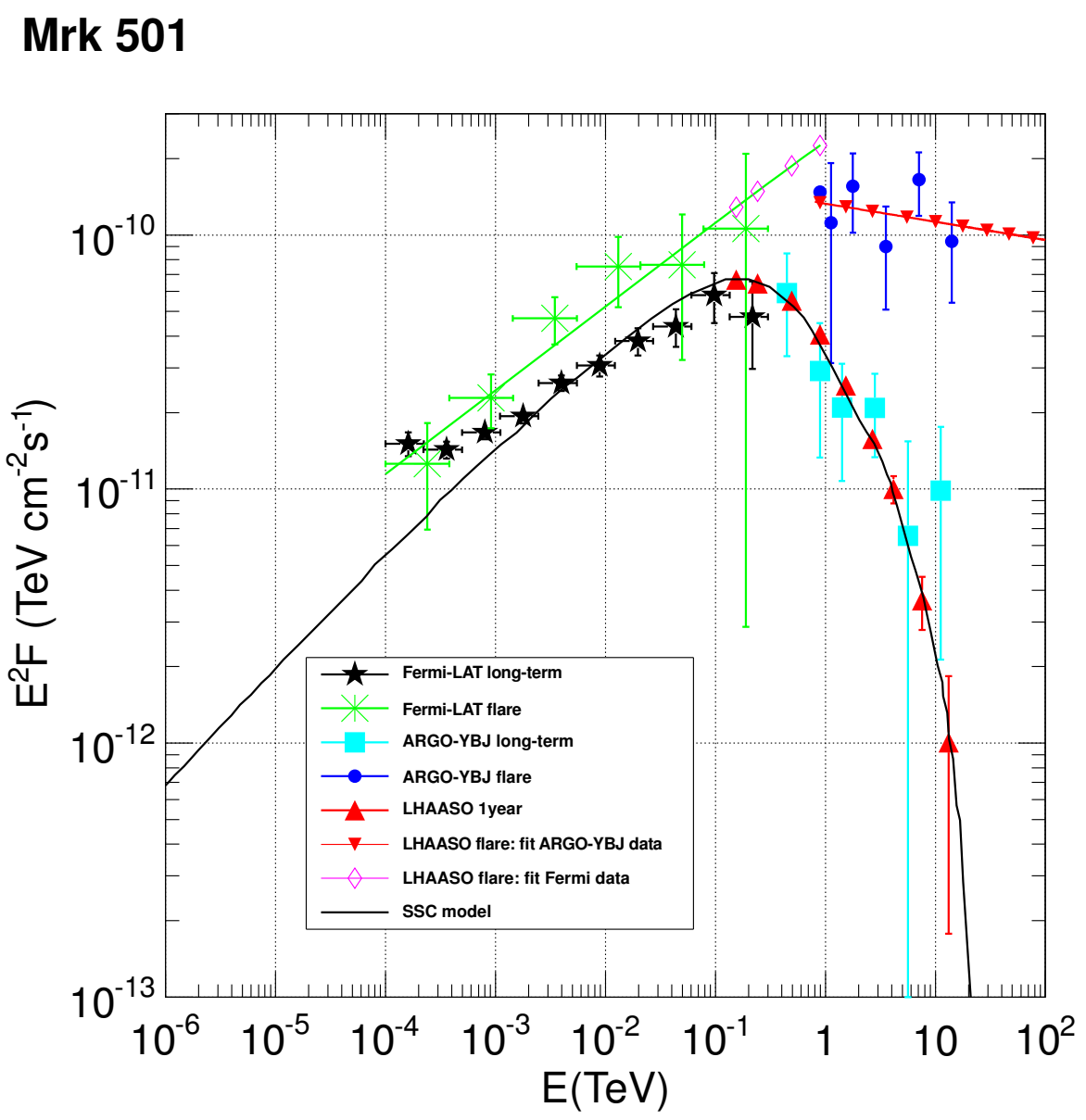

Fig. 12. Expectation of the LHAASO project on Mrk501 (open diamonds and inverted filled triangles) , comparing with the measurements by FERMI-LAT and ARGO-YBJ[24], respectively.

### 2.4 Summary

LHAASO has started its upgrading plan by replacing the 8" PMTs with 20" PMTs in 70% of area of WCDA. WCDA-1, 30% of the total area of the detector and equipped with 8" PMTs, has been put in scientific operation since April 2019. Mrk-421 has been observed with a significance of 20$\sigma$ within 173 days of observation. WCDA-2, same size as WCDA-1 but equipped with 20" PMTs, is under construction and going to be put in operation by the end of 2019. The whole array, including the scintillator counter array and muon detector array and Cherenkov telescopes, will be built up by 2021. With the upgraded configuration, the gamma ray detecting sensitivity below 300 GeV will be boosted to be about 0.2 crab unit per quarter decades of energy around 50 GeV[26] and as same sensitive as FERMI-LAT at 70 GeV. LHAASO therefore will play a significant role in surveying for new sources brighter than 7 milli-CU above 1 TeV in the northern sky and monitoring for transient phenomena in its FoV of the size of 1/7 sky at any moment. The LHAASO effective area of 1800 $m^2$ around 50 GeV is going to be a useful tool in the multi-messenger astronomy involving in ultra high energy neutrino or gravitational wave detection.

## 3 Line of shower trigger method for GRB observation by LHAASO-WCDA

Executive summary: Detection of GeV photons from GRBs is crucial in understanding the most violent phenomenon in our universe. Due to the limited effective area of space-born experiment, only 29 GRBs above 10GeV emission were detected at the time of writing. Large area EAS experiments at high altitude can reach a much larger effective area around 10 GeV, for which single particle technique is usually used to lower the threshold energy but its sensitivity is poor due to indistinguishable of direction. To reach an energy threshold as low as 10 GeV and guarantee the sensitivity at the same time, low multiplicity trigger is required, but random coincidences rather than cosmic ray showers overwhelms the signals, and it is a great challenge for traditional trigger logic and reconstruction algorithm to discriminate the signals from the noises. A new method is developed for LHAASO-WCDA(Large High Altitude Air Shower Observatory-Water Cherenkov Detector Array) to work under low multiplicity mode. With this technique, the LHAASO detector can even work below 100GeV, the sensitivity and expectation of LHAASO-WCDA with low multiplicity technique to GRBs are presented.

### 3.1 Introduction

Gamma Ray Bursts(GRBs) are among the most powerful events in the Universe, and have been the subject of observational studies from radio to multi-GeV energies. Satellites with instruments sensitive to hard gamma-rays, such as CGRO and Fermi LAT, have extended the observations from 30 MeV to tens of GeV. GRB130427A [27] that was observed up to 94 GeV, or 126 GeV once corrected for redshift, shows that GRBs are capable of producing very-high-energy photons. On the present, several GRBs have been observed above 10 GeV [12, 28–31]. It is unknown up to what energy the spectrum extends, as present-day observations are limited by effective area, in the case of space-based instruments, and by slewing constraints and energy threshold for ground-based Imaging Air Cherenkov Telescopes. Studying the spectrum beyond 10 GeV is crucial in understanding GRB mechanisms themselves, and also allows us to probe cosmological phenomena such as extragalactic background light (EBL) and it may be used to constrain Lorentz invariance violation.

Currently three major classes of high-energy detectors exist: Satellite detectors, Imaging Atmospheric Cherenkov Telescopes (IACTs) and Extensive Air Shower (EAS) particle detector arrays. Satellites can observe very wide fields of view (e.g. 2.4sr or 19% of $4\pi$ sr for Fermi LAT) and have close to a 100% operational duty cycle. On the other hand, the limited physical size of satellites prevents them from obtaining enough statistics to reach energies greater than tens of GeV. Operating above 50 GeV IACTs that have been designed for fast slewing (1 min). EAS detector arrays, such as WCDA, benefit from a very large field of view (2 sr or 16% of $4\pi$ sr) and near 100% duty cycle that will allow for observations in the prompt phase. They are also sensitive to energies beyond those covered by satellites. EAS observatories, in particular WCDA, are thus useful high-energy GRB detectors that complement the observations by satellites such as Fermi.

For EAS detectors, at present, two methods can be used to analysis the sensitivity and capabilities of detection of GRBs by WCDA: Shower mode method, Single particle technique (SPT) and Low multiplicity technique. Shower model method is a regular analysis method, threshold-energy is about 100 GeV, but GRBs with 100 GeV photons are very few. SPT can detect GRBs with 10 GeV photon but poor in sensitivity due to losing direction information. So taking advantage of characteristic of trigger mode of WCDA, a new method, low multiplicity technique is developed for GRBs detection, which can reach such energy like tens of GeV and reserve direction information at the same time. In this proceeding we will present the sensitivity and capabilities of low multiplicity technique for detection of GRBs by WCDA and show the observatory's ability to measure possible high-energy emission from GRBs.

### 3.2 WCDA experiment and trigger mode

Targeting gamma astronomy in energy band from 100 GeV to 30 TeV, the WCDA is one of the major components of the LHAASO, covering an area 90,000 $m^2$, has been proposed to be built at Daocheng County (4300 m a.s.l.), GanZizhou, SiChuan, China. Is is made of three ponds, two with area of $150{\times}150m^2$ and another with an area of $300{\times}110m^2$. Total ( $5{\times}5m^2$ each partitioned by 0.5 mm-thick black curtains made of black polyethylene lines, with a effective water depth of 4 m ) seen by one PMT located at the cell floor center and looking up to detect the direct Cherenkov light produced by the relativistic particles of the showers.

The measured counting rate was about 36 kHz for each cell, which is much higher than the noise of PMT itself. So a new trigger technique based on "Trigger-less" is proposed for large area WCDA: each PMT will output a L1 (Level 1) single trigger signal of 250 ns after hit and over threshold, and if there is another over-threshold signal in the same 250 ns period, which will be taken as a new signal and trigger signal will extend 250 ns; namely the total array is divided into 81 trigger cluster ($60m{\times}60m$ each, including $12{\times}12 = 144$ PMTs) Horizontally and vertically, and then judge whether hit multiplicity is bigger than 12 at the falling edge of the clock of each trigger cluster. When any trigger cluster satisfies this selection, then output one L2 trigger signal and produce total trigger. In this trigger system, if single counting rate produced by cosmic ray is less than 50 kHz, random coincidence trigger rate produced by single counting can be controlled smaller than 1 kHz, and trigger rate of 70 kHz produced by air shower (> 10 GeV) is expected.

### 3.3 Line of shower trigger

1. The method and feasibility analysis

   In order to lower the threshold energy and reserve the primary direction information at the same time for GRBs detection, Low multiplicity technique is produced taking advantage of LHAASO-WCDA trigger technique based on "trigger-less", wide field of view and full duty cycle, only observatory array like LHAASO-WCDA with these three characteristics can do this. For WCDA1, total array is about $10^8 Hz$ , corresponding to trigger rate of $10^6 Hz$ in the trigger window of 300ns, which is much higher than signal from cosmic ray, that means the random coincidence event rate is too high. Then how to discriminate the signals from the noise, it's a great challenge for traditional trigger logic and reconstruction algorithm, A new method is developed for LHAASO-WCDA to work under low multiplicity mode.

   Lowering down the huge background rate from single rate is crucial for this method. Three steps are take to realize this goal. firstly, we take the GRB alert as an "event" trigger for follow-up observation. As we know, typical delay of a GRB alert is about 1 minute, when a GRB alert comes, DAQ takes the GRB alert as an "event" trigger for low multiplicity technique and stores all the data in the pipeline and data of a certain time duration after the alert before reconstructing the shower core and direction for follow up observation. This implies that the GRB position is known from other observations. The time and duration of the burst are also assumed known, which allows one to efficiently reject the background by defining a restrictive time window. There is no any problem and hardware-free for WCDA. Secondly, we localize the shower, namely lower the total single rate. we can consider events with a distance from center of the array < 50 m, and with zenith angle $<40^o$, after doing so, the total number of noise hits is reduced by a factor of 10, but it's still too high for a multiplicity as low as 3. At last, we shrink the trigger time window through hit time transformation. After this transformation, the coincidence time window can be reduced by a factor of ∼30, in which the average number of noise hits can be reduced by a factor of ∼30, from 3 to 0.1, it's good for low multiplicity trigger.

In this method, all hits are to be saved, and the running time window should be applied on each hit, otherwise, shower hits can be separated into adjacent windows.

2. Signal simulation for low multiplicity technique

Gamma-ray showers are simulated with CORSIKA with an $E^{-2.0}$ spectrum at different energy. The detector response model developed for WCDA is used at an altitude of 4300 m using a GEANT4 based code. The lowest energy for primary gamma showers was set to 10 GeV. The signal rate in the low multiplicity technique is the number of PMT hit after detector response and before direction and position reconstruction.

The signal rate S is given by:

$$S(\theta)=\int dE\frac{dN}{dE}A_{eff}^{\text{low multiplicity}}(E,\theta) \quad (3)$$

Where $dN/dE$ is the photon spectrum and $A_{eff}^{\text{lowmultiplicity}}$ is the detector effective area. Depends on several variables and here only energy E and zenith angle are treated. In this proceeding, only results of $\theta=0$ are presented, and other direction will be considered later. The effective area of WCDA for gamma rays for different low multiplicity is shown in the Table 1.

Table 1. Effective area at different energy for low multiplicity

| | nhit≥1(m$^2$) | nhit≥2(m$^2$) | nhit≥3(m$^2$) |
|---|---|---|---|
| 10GeV | $3.7\times10^4$ | $4.7\times10^3$ | $1.7\times10^3$ |
| 20GeV | $9.5\times10^4$ | $1.9\times10^4$ | $8.2\times10^3$ |
| 50GeV | $3.0\times10^5$ | $8.9\times10^4$ | $4.6\times10^4$ |
| 100GeV | $5.4\times10^5$ | $2.2\times10^5$ | $1.4\times10^5$ |

3. Background for low multiplicity technique

Galactic cosmic rays are simulated with CORSIKA for multiple species with an $E^{-2.62}$ spectrum: protons, He, C, O, Ne, Mg, Si, Fe. The galactic cosmic ray spectrum is re-weighted to measurement by J.R. Hoerandel [32]. Background rate mainly includes occasional noise from cosmic rays and random coincidence noise. Fig. 13 is the distribution of hit number from cosmic ray background by simulation, from this figure, we can infer the occasional noise rate is $4.3\times10^6$ Hz for $nhit\geq1$, $1.6\times10^6$ Hz for $nhit\geq2$ and $8.4\times10^5$ Hz for $nhit\geq3$ by fitting the distribution of number of hits with a power law.

For random coincidence noise, firstly, we take the pipe line with a time duration of 200 seconds, to save all the data when a GRB alert comes and after the alert. Secondly, we localize the shower, considering events with a distance from the center of < 50 m, then the total Number of noise hits is reduced by a factor of 10, i.e. from 30 hits ($10^8$ Hz $\times300\times10^{-9}$ s = 30 hits) reduced to 3 hits in trigger time window of 300 ns. The random coincidence noise rate is about 80 MHz for nhits$\geq 3$, about 300 times higher than shower rate (with direction information). Secondly, we shrink the trigger time window, namely perform hit time transformation: GRB direction cosines are (l,m,n), for each hit(x,y,z,t), we can define $tr=t-(lx+my+nz)/C$, where C is light velocity in vacuum, Automatically removed those far from GRB direction and shower direction information is obtained without direction reconstruction. After this transformation, the coincidence time window can be reduced from 300 ns to 10 ns, in which the average number of noise hits is reduced from 3 to 0.1, good for low multiplicity trigger. For low multiplicity trigger, the running time window was applied on each hit, then the random coincidence rate is about 0.5 MHz (nhits$\geq 3$ for mean noise hit of 0.1), lowering down a factor of 100. For different low multiplicities, the random coincidence noise rate and the Background rate are presented in the following Table 2.

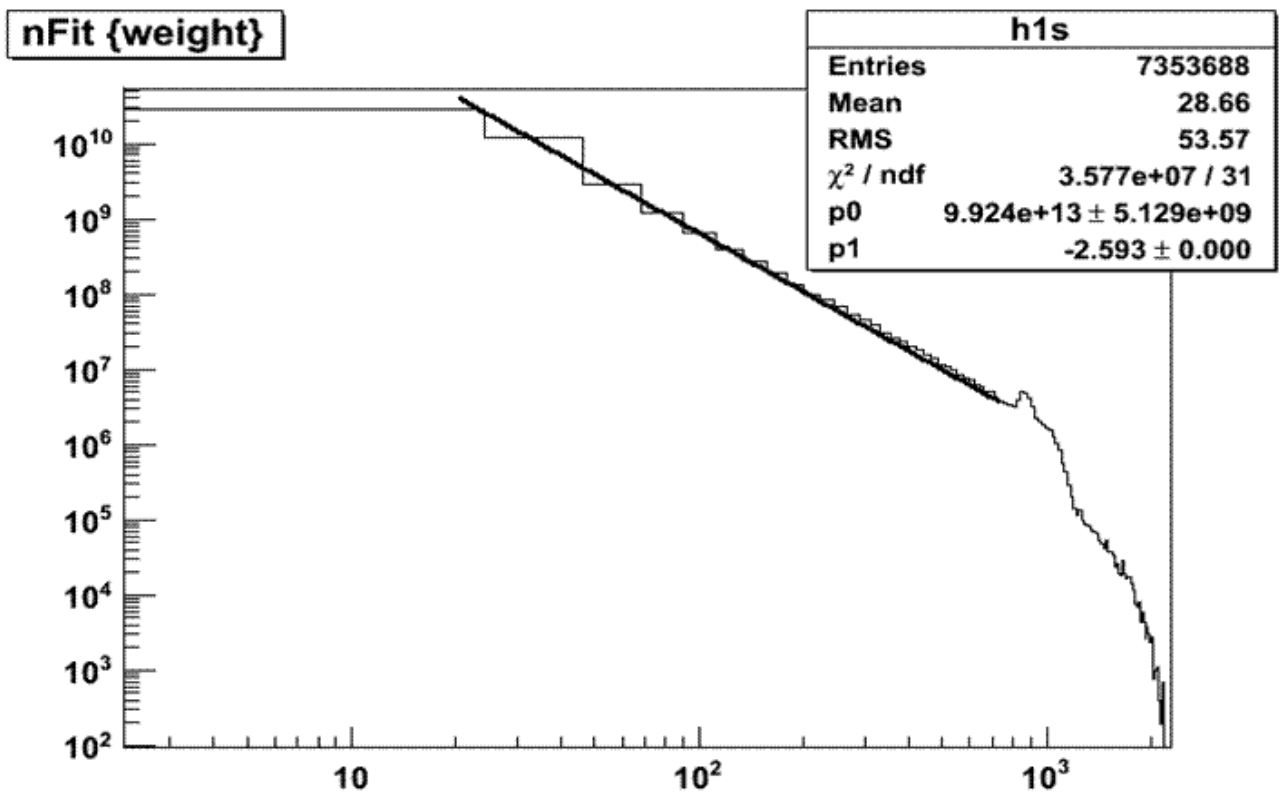

Fig. 13. Distribution of number of hit. Solid line is a power law fit to the number of hits to infer low multiplicity due to threshold effect.

| | occasional noise rate(Hz) | random coincidence noise rate(Hz) | background rate(Hz) |
|---|---|---|---|
| nhit≥1 | $4.3\times10^6$ | $10^8$ | $1.04\times10^8$ |
| nhit≥2 | $1.6\times10^6$ | $9.5\times10^6$ | $1.11\times10^7$ |
| nhit≥3 | $8.4\times10^5$ | $4.6\times10^5$ | $1.30\times10^6$ |

Table 2. Background rate at different low multiplicity

4. Sensitivity of low multiplicity technique to GRBs

For low multiplicity, the sensitivity of WCDA to GRBs depends on a number of factors, including the GRB emission time scale, emission spectrum and redshift, as well as on the signal and background estimation of the experiment. To calculate WCDA's sensitivity, we simulate gamma ray spectrum according to the power-law $dN/dE \propto E^{-2}$ with an arbitrary reference flux normalization. This injection spectrum can be weighted for any other spectral shapes. In which we take into account attenuation of VHE gamma rays due to interaction with extragalactic background light, the Franceschini et al. [33] model is used.

Given a signal rate $S(\theta)$, background rate B then the significance of a given observation is :

$$\begin{aligned} Sig &= S(\theta)T_{90}/\sqrt{BT_{90}} \\ &= \sqrt{T_{90}/B}\int dE\frac{dN}{dE}A_{eff}^{\text{low multiplicity}}(E,\theta) \quad (4) \end{aligned}$$

We have used various spectra of the type $dN/dE \propto E^{-\gamma}$ with sharp high-energy cutoffs to determine the sensitivity of the low multiplicity technique to GRBs. The sensitivity is defined as the flux detectable at $5\sigma$ significance. A range of spectral indices gamma between -3 and -1 and a range of cutoffs between 10 GeV and 10 TeV were tested, and effects of the EBL are also considered.

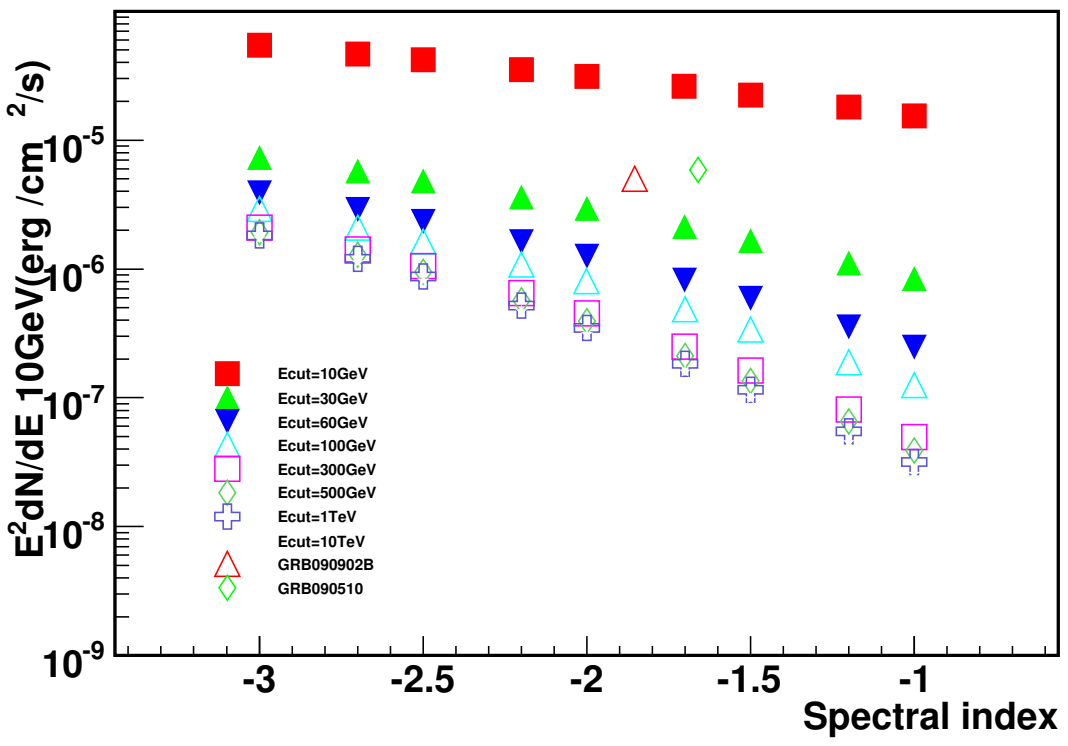

Fig. 14. Sensitivity using the low multiplicity as a function of spectral index. The $5\sigma$ discovery potential is shown as a function of spectral index for various values of a sharp high-energy spectral cutoff. The duration of the burst is fixed to 1s and the zenith angle is fixed to $0^o$. Data from 2 GRBs are corrected for duration and inserted for comparison [29] [30]

Fig. 14 shows WCDA's sensitivity curve of different GRB emission spectra on the expected sensitivity of WCDA using the low multiplicity (nhit$\geq$3) calculated with equation compared to GRBs that have been detected by Fermi LAT. Assuming that the burst occurs at a zenith angle of $0^o$ and lasts 1 second at a distance of redshift of $z=0.5$ [33]. Data for GRBs 090510 and 090902b, extracted from [29] and [30] are shown for comparison. We conclude that the most promising cases for detection with high significance are GRBs such as GRB 090510 and GRB 090902b if the high-energy cutoff is above 30 GeV. Fermi LAT observations of these two GRBs were made up to 30 GeV without any indication of a cutoff. If high-energy emission from GRBs extends beyond 30 GeV, then WCDA will become even more significant due to limited physical size of Fermi LAT.

### 3.4 Scientific prospect and conclusions

A new method was developed to detect GRBs at energy as low as 10 GeV and reserve source direction information using EAS array, like LHAASO-WCDA. From above analysis, we can conclude, WCDA, will have the capability of detecting GRBs at high energies. The simulations presented in this proceeding show that WCDA will be able to detect GRBs with characteristics similar to those of some of the brightest GRBs seen by Fermi LAT. As opposed to Fermi LAT, with a fixed physical size, the effective of the method of the low multiplicity increases with energy. Thus this method will expand upon the energy sensitivity of current detectors. Also WCDA is a wide field of view detector with near 100% duty cycle, it will be able to make GRB observation in the prompt phase. WCDA, in union with satellite or other ground based detectors, will be able to measure the high-energy GRB components including a possible high-energy cutoff. Important astrophysical information will be deduced from spectral cutoffs such as the Lorentz boost factor of GRB jets, the effects of the EBL and the maximum energy to which GRBs accelerate particles.

For low multiplicity technique, how to pick up the true events and analysis the data? This is still the question to be solved. In this proceeding, we present the result with GRBs alert information, at further step, we also can do without GRBs alert and work alone taking advantage of large field of view of WCDA, then CPU power maybe is a huge challenge.

# 4 Studies of Active Galactic Nuclei with LHAASO

Executive summary: We review the prospects for studies of active galactic nuclei (AGN) using the future Large High Altitude Air Shower Observatory (LHAASO). This review focuses on blazars, which constitute the vast majority of AGN detected at gamma-ray energies. We argue that LHAASO will enable substantial progress on searching for clear evidence of blazar releasing very high energy cosmic rays through its excellent flux sensitivity. We give two proposals: (a) searching for hard spectra >10 TeV from the extreme blazars (e.g,, 1ES 0229+200) and nearby blazars (e.g., Mrk 421); (b) searching for TeV photons from distant blazars with redshift $z \sim 1$. The surveys of LHAASO enable measurement of cosmic TeV background and construction of luminosity function of TeV blazars. These results will help us to understand the origins of Ultra-high energy cosmic rays (UHECRs). At last, we discuss the potential of LHAASO as a tool for probing new physics like Lorentz Invariance Violation (LIV) with the observatiosn of blazars. The traditional projects such as relativistic jet physics (including high-energy radiation mechanisms and acceleration of particles) and the extragalactic background light (EBL) determination are not discussed in this paper. However it should be pointed out that all these projects are interrelated.

## 4.1 Introduction

Active galactic nuclei (AGN) are the extragalactic sources of enhanced activity that are powered by the release of gravitational energy from the supermassive central black hole. Energy linked to the black hole spin [34] or rotating accretion disks [35] may be instrumental for forming prominent jets which transport the material with relativistic speed from the innermost region of the AGN to kpc-, sometimes even Mpc-scale distances. Such jets are usually identified through the detection of bright non-thermal radio emission as observed in radio-loud AGN. Only a small percentage ($\sim$10 %) of all AGN are known to be radio-loud. In the vicinity of the central region of an AGN, matter is accreted from a disk onto the black hole; line-emitting clouds (the so-called broad-line region, BLR, and narrow line region, NLR) form at pc to kpc distances from the central engine; and dusty material surrounding the accretion disk may imprint thermal signatures in the infrared part of the AGN spectrum [36] .

The radiation from the material which moves relativistically with speed $\beta_\Gamma c$ (with $\Gamma = 1/\sqrt{1-\beta_\Gamma^2}$ being the bulk Lorentz factor) along the jet axis is beamed into an angle $\sim 1/\Gamma$ around the direction of propagation. Because of this beaming effect, mostly those AGN with jets pointing towards us (i.e., blazars) are favorably detected as gamma-ray sources. However, some mis-aligned AGN (i.e., radio galaxies) can be also detected, if they are sufficiently nearby. Blazars therefore offer an excellent opportunity to study jet physics in massive black hole systems and their evolution over cosmic time through population studies.

Blazar emission is dominated by non-thermal radiation over all frequencies ranging from radio to TeV gamma-rays. Its typical multi-wavelength spectral energy distribution (SED) is characterized by two distinct humps (see Fig. 15). It is accepted that the first hump in the SED is the synchrotron emission radiated by relativistic electrons in the jet. The origin of the emission in the gamma-ray hump is still under debate.

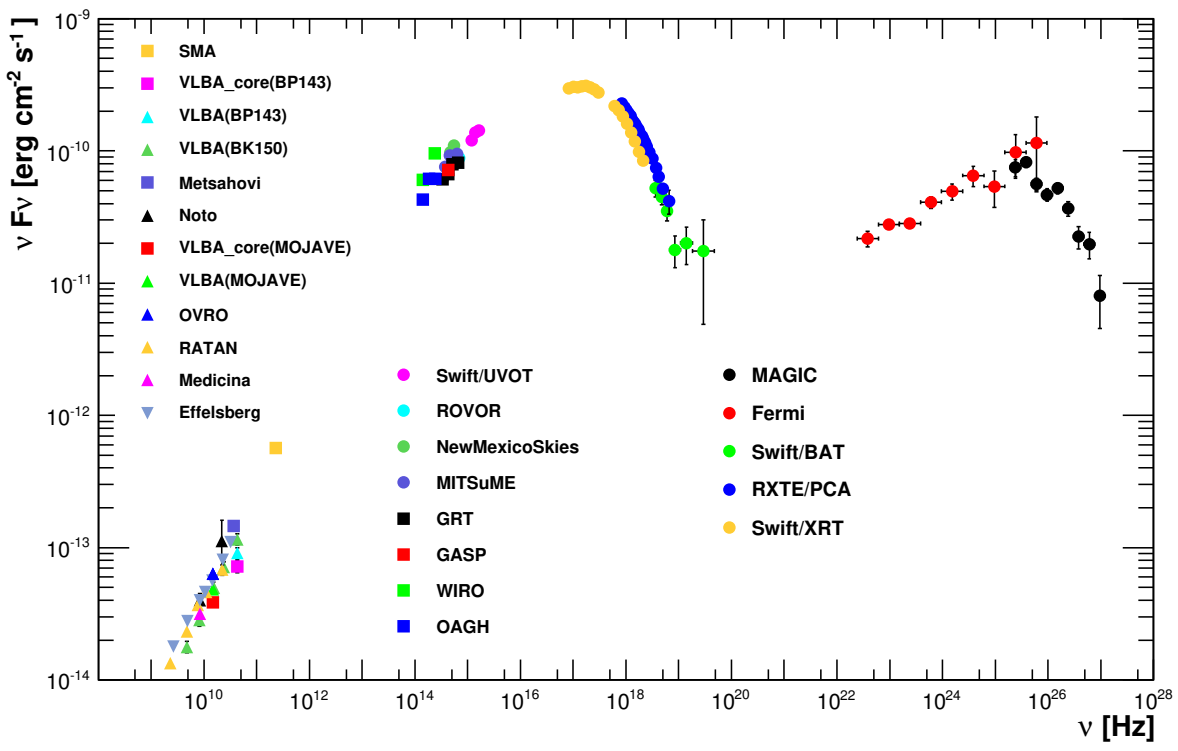

Fig. 15. One SED of Mrk 421. [from 37].

Different classes of blazars are defined according to various properties. BL Lac objects are typically defined if the equivalent width of the strongest optical emission line is < 5Å. By contrast, flat spectrum radio quasars (FSRQs) have strong optical emission lines indicating the presences of dense BLR material and strong illuminating accretion-disk radiation. Blazars also can be divided into low, intermediate, and high synchrotron-peaked sources (LSPs, ISPs, and HSPs, respectively, defined by whether the peak frequency of the synchrotron component of the SED $\nu_{\rm syn}^{\rm pk} < 10^{14}$ Hz, $10^{14} < \nu_{\rm syn}^{\rm pk}$ (Hz) $< 10^{15}$, or $\nu_{\rm syn}^{\rm pk} > 10^{15}$ Hz) [38]. Most FSRQs are LSP blazars, whereas BL Lac objects include LSP, ISP, and HSP sources. Based on blazar SED and the emission variation, the relativistic jet physics (e.g., emission mechanisms and acceleration processes) can be investigated [e.g., 39], if the gamma-ray emissions are certainly produced in the jet.

In this article, we review the prospects of LHAASO to understand the AGN high-energy phenomenon and its related physics including the origin of ultrahigh energy cosmic rays (UHECRs).

### 4.2 LHAASO and Signatures of UHECRs in Gamma-rays from Blazars

The origin of gamma-ray (GeV - TeV) emission from blazar is not resolved. Three kinds of models have been proposed to resolve this problem. In leptonic models, the gamma-ray emission is supposed to be inverse Compton (IC) emission from relativistic electrons in the jet that up-scatter either low-energy synchrotron photons emitted by the same population of electrons (synchrotron-self-Compton model, SSC), or photons originating from outside the jet (external inverse Compton, EIC). In hadronic models, the gamma-ray emission is attributed to synchrotron radiation of high-energy protons in the jet, or synchrotron radiation of secondary particles created in proton-photon interaction. In the third model, the gamma-ray emission is the secondary cascade gamma-ray photons produced in the propagation of UHECRs emitted by blazar [e.g., 40–42] . Since the gamma-ray photons in leptonic and hadronic models are produced in the jet, we classify the two kinds of models as jet model. Moreover, we refer to the model that produces gamma-ray photons in the interactions between the highest-energy cosmic rays and background photons in the Universe as cosmogenic model. The key issue of our attention is to disentangle the jet model and cosmogenic model from observations.

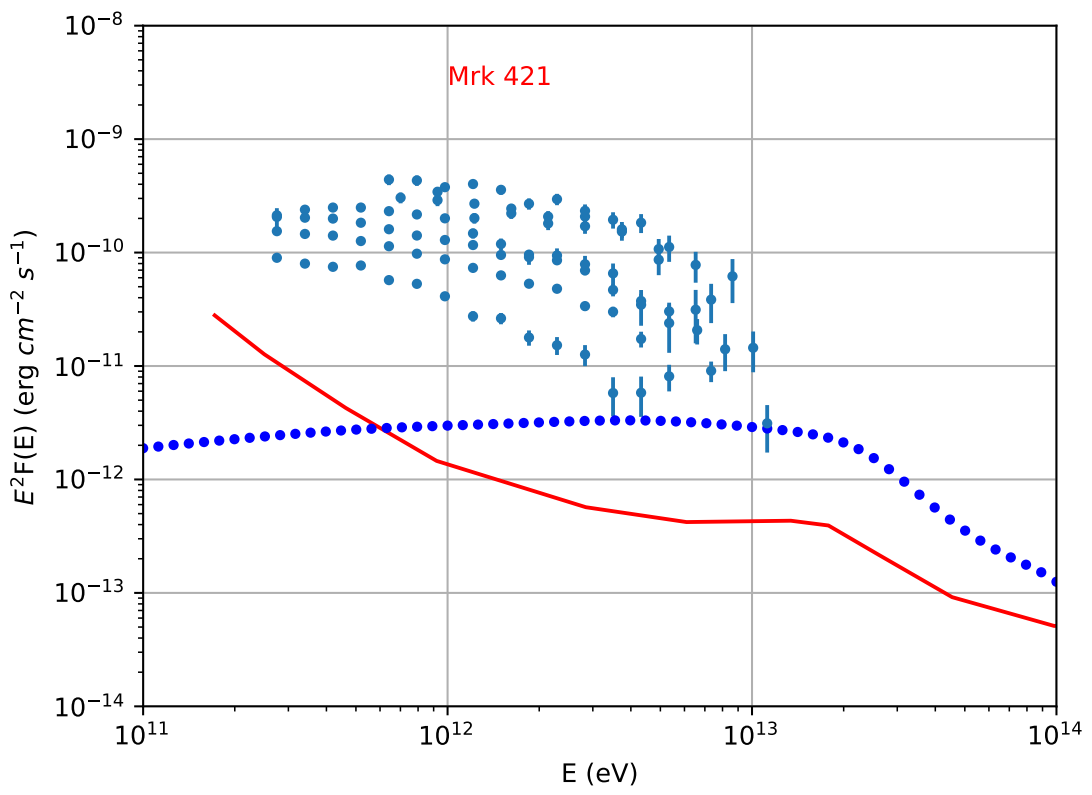

Fig. 16. A spectrum of UHECR-induced cascade gamma-rays for Mrk 421 (dotted line) and the historical TeV data of Mrk 421 (points). The data are obtained through the SED Builder of ASDC (http://tools.asdc.asi.it). The cascade gamma-rays are calculated by using the TRANSPORTCR code [43] . The injection spectrum of protons is assumed to be a power-law with an exponential cutoff. We take the index of 2.6 and the cutoff energy of $10^{19}$ eV. The EBL model of [33] is used in the calculation. The solid line is the one-year differential sensitivity of LHAASO.

The observations of HESS found the non-variable and extremely hard TeV spectra of several blazars, e.g., 1ES 0229+200 and 1ES 1101-232 [44, 45]. The classical leptonic jet model is difficult to explain such hard TeV spectra*. The hadronic jet models may account for the hard TeV spectra [e.g., 48]. Alternatively, [49] have proposed that the TeV spectrum of 1ES 0229+200 could be from the secondary gamma-ray produced in the propagation of UHECRs in the Universe. With the current TeV observation up to $\sim$ 10 TeV, we cannot disentangle the leptonic, hadronic and cosmogenic models. From [49], one can estimate that the energy flux of the UHECR-induced cascade gamma-rays calculated with a low EBL at 30 TeV is $\simeq 2\times10^{-13}$ erg cm$^{-2}$ s$^{-1}$. On the other hand, the one-year differential sensitivity of LHAASO at 30 TeV is also $\simeq 2\times10^{-13}$ erg cm$^{-2}$ s$^{-1}$. Therefore, LHAASO is capable of detecting the UHECR-induced cascade gamma-rays. By obtaining the good spectra of 1ES 0229+200 above 10 TeV, we could disentangle the jet models and cosmogenic models.

Another interesting object is Mrk 421. The current IACTs observations show that the TeV emissions from Mrk 421 are strongly variable. This suggest that the steady UHECR-induced cascade gamma-rays cannot make a significant contribution to the observed TeV emissions. In Fig. 16, we show the spectrum of UHECR-induced cascade gamma-rays constrained by the current

*Modified leptonic jet models succeed in explaining the hard TeV spectra [e.g., 46, 47]

TeV data. In this case, LHAASO can detect the cascade gamma-rays in one year. LHAASO will detect 100 TeV photons from Mrk 421 if it really emits >1 EeV protons. The observations of LHAASO for these sources could provide strong evidence for UHECR origin.

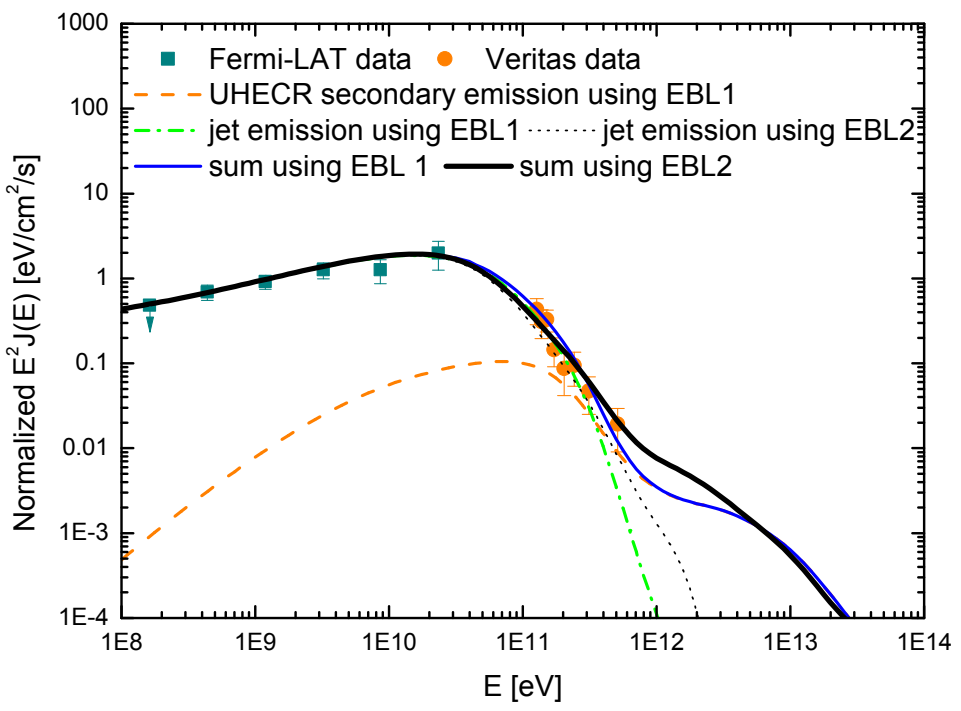

Fig. 17. Cosmogenic model of [50] for the distant TeV blazar PKS 1424+240.The resutls are obtained by assuming $z$=0.6. [from 50].

We also propose another strategy to find the clear evidence of blazar emitting UHECRs. Compared to the jet models, VHE photons produced by cosmogenic models suffer less absorption by extragalactic background light (EBL) because of the long energy-loss distance of UHECRs interactions with background lights. Therefore the VHE photons from the jet in high redshift blazars will suffer strong absorption by EBL. If LHAASO or CTA detect > 1 TeV photons from very distant blazars, the observed gamma rays are the secondary photons produced in interactions of high-energy protons originating from the blazar jet and propagating over cosmological distances almost rectilinearly [e.g., 50–52]. In Fig. 17, we show the case for the distant TeV blazar PKS 1424+240. One can see that the cosmogenic model predicts harder spectrum above one TeV than the jet model.

In the two topics mentioned above we do not seek to explain the observed cosmic ray spectrum above $10^{18}$ eV[†] . We focus on finding the evidence of UHECRs originating from the blazar jet through the observations of VHE gamma-rays by LHAASO.

### 4.3 LHAASO and Cosmic TeV Gamma-Ray Background Radiation

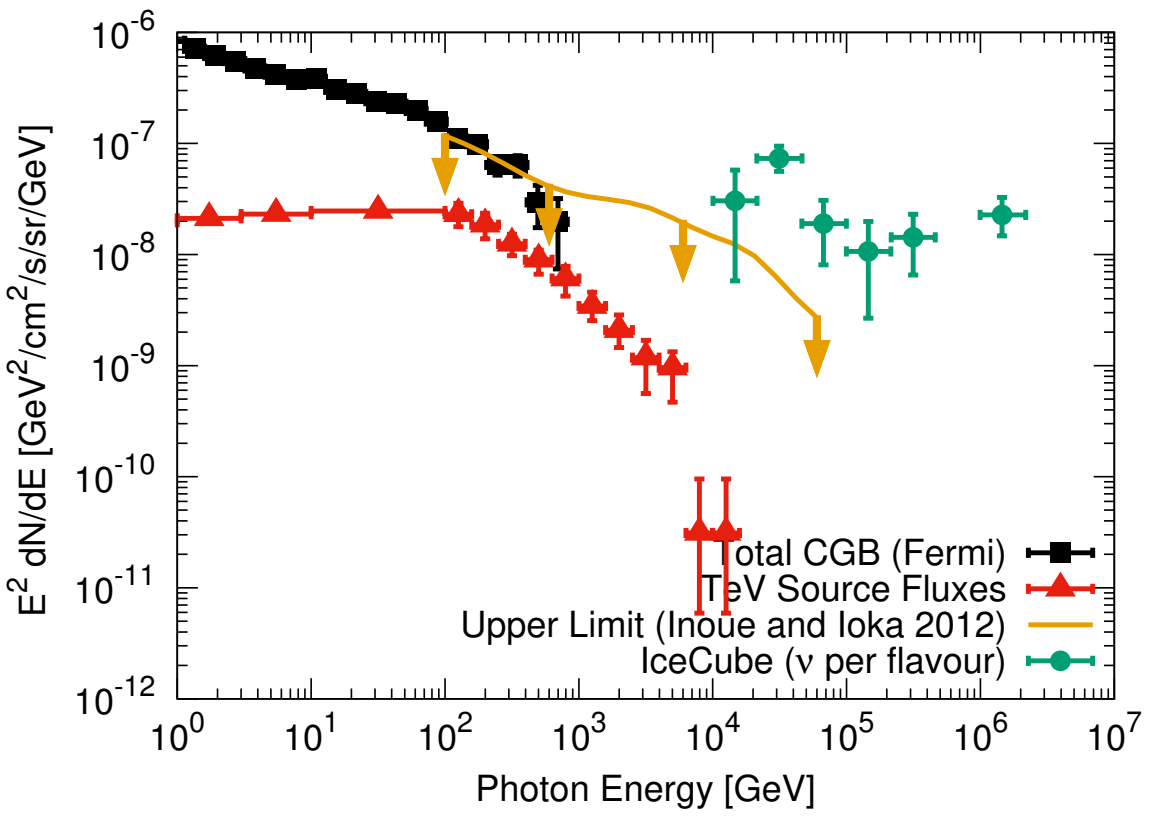

Fig. 18. GeV background radiation measured by *Fermi* gamma-ray space telescope, and the upper and lower limits on TeV background [55]. [from 55].

The *Fermi* gamma-ray space telescope has successfully measured the cosmic gamma-ray background (CGB) spectrum at 0.1 - 820 GeV [56]. It also provides an opportunity to explore and decipher the high-energy universe through a multi-messenger approach including the information from cosmic rays (CRs), gamma rays, and neutrinos [e.g., 57, 58]. However, the measurement of the cosmic TeV gamma-ray background radiation is still rare, although its upper and lower limits are given based on the current understandings of TeV sources (see Fig. 18).

Based on its good sensitivity and wide field of view, LHAASO will perform an unbiased sky survey of the Northern sky from TeV to 100 TeV. The high background rejection capability in the 10 - 100 TeV range will allow LHAASO to measure the cosmic TeV gamma-ray background radiation.

The GeV CGB are mainly contributed by the emission from blazars. Therefore, one can naturally speculate that blazars also contribute a substantial fraction of the TeV CGB. Under LHAASO extragalactic surveys, many TeV AGNs will be detected to build large and well-defined TeV AGN sample. We can construct the luminosity function of AGN at TeV band to study AGN evolution over cosmic time, and then to investigate the origin of the TeV CGB.

It is noted that the cosmogenic model mentioned in Section **4.2** predicts a flat spectrum from TeV to a few tens TeV. The signature of the flat spectrum in the cosmic TeV gamma-ray background radiation will be also a key test on the UHECR-induced cascade emissions [e.g., 57, 58].

[†]See the studies of [53] and [54] for the constraints on the origin of the observed >10 EeV cosmic rays with the non-observation of 10 PeV neutrinos by IceCube.

### 4.4 New Physics from LHAASO observations of Blazars

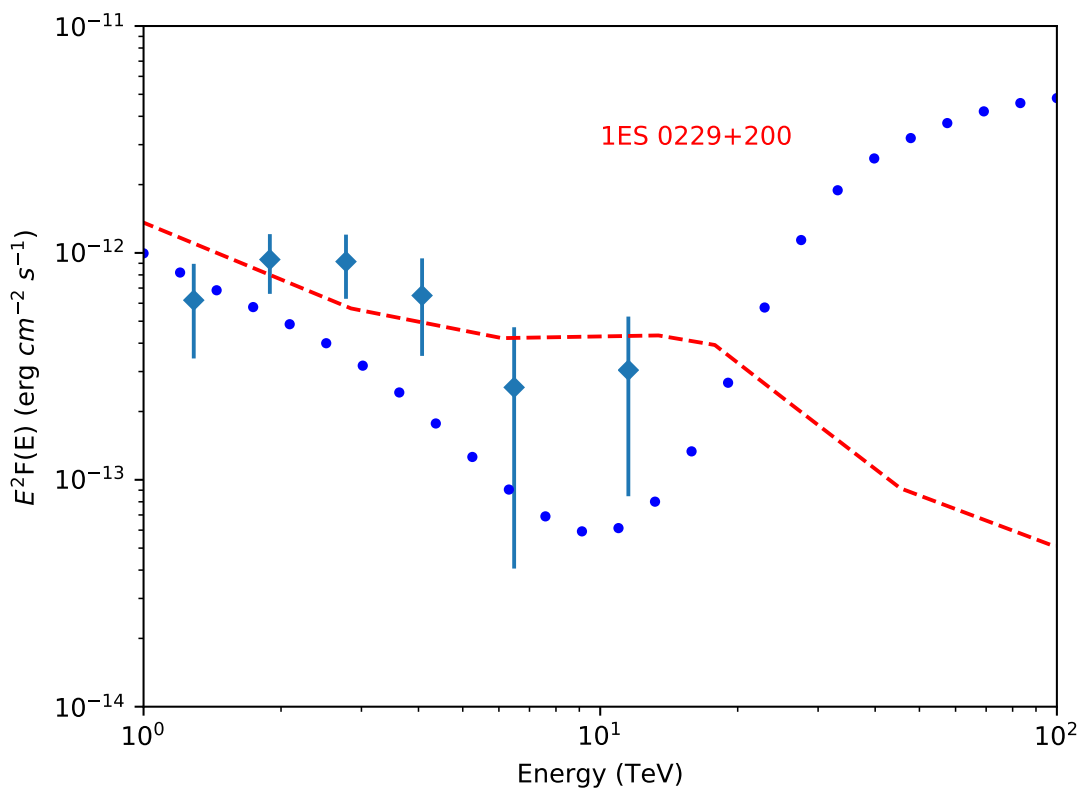

Fig. 19. TeV spectrum of 1ES 0229+200 with the consideration of LIV (dotted line). The points are the HESS data of 1ES 0229+220. The EBL model of [59] is used in the calculation. The dashed line is the one-year differential sensitivity of LHAASO.

Astrophysical observations with gamma-ray experiments have proven to be a powerful tool of searching for physics beyond the Standard Model. For example, observations at gamma-ray energies can be used to search for the traces of axion-like particles (ALPs), in which blazars are abundantly observed sources. The photon-ALP oscillations may lead to two changes in the energy spectra: a) the gamma-ray source flux can be attenuated due to pair production with low energy background photons. ALPs produced in the vicinity of the source would mitigate this attenuation, and if they reconvert to gamma rays, leading to a significant boost of the observed photon flux, and b) the oscillations of the flux should be imprinted in the spectra around $E_{\rm crit}$[‡] and $E_{\rm max}$[§] [e.g., 60]. In the analysis of *Fermi*-LAT and IACT spectra of blazars, no ALP-induced spectral signature, which is a spectral hardening at high optical depths, was found [61, 62]. The LHAASO extragalactic survey with its good sensitivity could be used to search for a spectral hardening correlated with the photon-ALP oscillations.

Lorentz Invariance (LI) is a basic component of Einstein's Special Relativity. It is strictly valid in Quantum Mechanics and has been verified in various accelerator experiments at the electro-weak scale. On the other hand, Lorentz Invariance Violation (LIV) has also been largely predicted in the framework of various classes of Quantum Gravity (QG) models. Tests of LIV with high-energy photons from distant sources have been proposed [e.g., 63, 64]. It is possible to utilize LHAASO for the detection of LIV through anomalies in the multi-TeV gamma-ray spectra of blazars [e.g, 65].

In Fig. 19, we show the predicted TeV spectrum of 1ES 0229+200 under the condition of LIV. One can see that this spectrum becomes harder at 10 TeV, and LHAASO cannot detect the concave shape around 10 TeV. But, it can detect the extremely hard spectrum (the photon index much less than 2) above 20 TeV, which is the evidence of LIV. The observed variability can also be used to probe LIV. Taking advantage of the wide energy-coverage of LHAASO, we can construct the energy-dependent light curves of blazars to search for a possible time lag between low- and high-energy photons, constraining an energy-dependent LIV [e.g., 66], i.e., an energy-dependent speed of light.

### 4.5 Concluding remarks

In this paper, we have proposed several projects for LHAASO. This is surely incomplete in the field of AGN research. The topics on the relativistic jets are not included. Although the EBL and intergalactic magnetic field (IGMF) are not specifically discussed, all the studies mentioned above are related to EBL and IGMF. All these questions are interrelated. To improve upon these constraints, we need a better understanding of the sources and emission mechanisms, including the relativistic jet physics. Actually the key issue is to determine the origin of the observed TeV photons. To better understand these questions, we need an overall emission model, e.g., a self-consistent jet+cosmogenic-propagation emission model [e.g., 50]. Combining the future measurement for the cosmic TeV background radiation and the observations on UHECRs and cosmic neutrinos, it is possible to improve the constraints on their origins. We believe that the observations of LHAASO will improve our understanding of the high-energy universe.

## 5 Measuring Extragalactic Background Light with LHAASO Observations of blazars

<u>Executive summary:</u> The extragalactic background light (EBL) contains important information about stellar and galaxy evolution. It leaves imprint on the very high energy $\gamma$-ray spectra from sources at cosmological distances due to the process of pair production. We have proposed a method to *measure* the EBL intensities by extracting the collective attenuation effects in a number of $\gamma$-ray sources at different redshifts. This method employs a Markov Chain Monte Carlo fitting algorithm to derive the EBL intensities and the intrinsic spectral parameters of $\gamma$-ray sources simultaneously, without prior assumption of the EBL shape. With larger sample of extragalactic sources, primarily blazars, and better spectral measurements by LHAASO, we expect to improve the measurement of EBL substantially.

[‡] $E_{\rm crit} = |m_a^2 - \omega_{\rm pl}^2|/2g_{a\gamma}B$, where $B$ denotes the field strength transversal to the photon propagation direction, $g_{a\gamma}$ the photon-ALP coupling, and $\omega_{\rm pl}$ the plasma frequency of the medium.

[§] $E_{\rm max} = 90\pi g_{a\gamma} B_{\rm cr}^2/7\alpha B$, with $\alpha$ the fine structure constant and the critical magnetic field $B_{\rm cr} \sim 4.4\times10^{13}$ G.

### 5.1 Introduction

The extragalactic background light (EBL) is the diffuse radiation from ultraviolet to far infrared wavelengths, spread isotropically in the universe (for a review of EBL, see [67, 68]). The EBL originates from the radiative energy releases of all the stars, other extragalactic sources and diffuse emissions since the epoch of recombination. Therefore its intensity and spectral shape hold crucial information about the formation and evolution of stellar objects and galaxies throughout the cosmic history. The EBL is one of the fundamental quantities in cosmology.

Direct measurement of EBL is, however, very difficult due to the contamination of the foreground emission from the solar system zodiacal light and the Galactic stellar and interstellar emissions [69]. Technically, it also requires the absolute calibration of the instruments, and the understanding all measurement uncertainties. Given the difficulties, direct measurements provide just lower and upper limits of EBL intensity. A strict lower limit on the EBL intensity is provided by the integrated light from resolved galaxies, e.g. in optical by the Hubble Space Telescope [70] and in infrared by the Spitzer telescope [71]. The upper limit can be derived from the uncertainties of the absolute measurement of EBL [67].

Another indirect, but effective way to study the EBL is through the observation of very high energy (VHE) $\gamma$-rays. The VHE $\gamma$-rays from extragalactic sources are attenuated by the process of electron/positron pair production, $\gamma_{\rm VHE}+\gamma_{\rm EBL}\rightarrow e^+e^-$, when propagating to the Earth [72]. With the rapid development of ground based $\gamma$-ray imagining atmospheric Cherenkov telescopes (IACT), quite a few VHE $\gamma$-ray sources from cosmological distances have been detected, most of which are blazars, a subgroup of active galactic nuclei (AGN), with relativistic jet pointing towards the observer. With reasonable assumption of the intrinsic blazar spectra we can set an upper limit of the EBL intensity by comparing the observed spectra with the intrinsic spectra [73]. The observations of blazars H 2356-309 and 1ES 1101-232 at redshifts $z=0.165$ and $z=0.186$, respectively by HESS has set a strong upper limit of EBL, close to the lower limit set by galaxy counts, at the near infrared wavelength [74]. The MAGIC observation of 3C 279 at $z=0.536$ set upper limit at the optical band [75]. In [76] Mazin and Raue gave a comprehensive study of EBL based on eleven blazars over a redshift range from $0.03-0.18$. They explored a large number of hypothetical EBL scenarios and set robust constraints on EBL over a wide wave-length range. With the Fermi observation of blazar spectra at GeV to $\sim 100$ GeV more stringent constraints on EBL are shown recently (e.g., [77–81]. These studies seem to indicate that the Universe is more transparent than we had expected.

The power of this indirect method to study EBL is limited due to the fact that the intrinsic spectrum of each blazar is unknown. Therefore it is hard to disentangle the absorption effect by EBL from the intrinsic emission nature for a specific observation. The usual practice in the literature is to reconstruct the blazar intrinsic spectrum from the observation by first assuming an EBL model. The EBL model is rejected if it results in an unphysical intrinsic spectrum, for example, the reconstructed intrinsic spectrum follows a power law with an extremely hard spectral slope or even shows an exponential rise at the high energy end. Recently with large sample of $\gamma$-ray blazars, the EBL intensities were derived through a likelihood fit with given spectral template of the EBL [82, 83].

With the fast increasing number of $\gamma$-ray sources and better measurements of their spectra, we propose to *measure* the EBL intensities through extracting the collective absorption effects in a number of sources at different redshifts, using a global fitting method [84]. The method employs the Markov Chain Monte Carlo (MCMC) global fitting algorithm to fit the intrinsic source spectra and EBL simultaneously. Different from the previous studies in the literature, we make no assumption of the EBL spectral shape in the fitting. Instead the EBL intensities are approached as free parameters in a series of discrete energy bins, which are allowed to vary during the fitting. The application to a few sources by the current IACTs illustrate that this method can give effective measurement of both the intensities and shape of the EBL [84]. The derived results are consistent with the upper limits obtained with $\gamma$-ray observations as well as the theoretical modeling from galaxy evolution.

### 5.2 Attenuation of VHE photons

The observed VHE $\gamma$-ray spectrum after absorption by the EBL is commonly expressed as

$$F_{\rm obs}(E)=e^{-\tau(E,z)}F_{\rm int}(E)\ , \tag{5}$$

where $F_{\rm int}(E)$ is the intrinsic spectrum of the source at redshift $z$. The strength of the attenuation by EBL is described by the optical depth $\tau(E,z)$ as a function of energy $E$ and the source redshift $z$. The optical depth $\tau$ is expressed as [85]

$$\tau(E,z)=\int_0^z {\rm d}l(z')\int_{-1}^{+1}{\rm d}\mu\frac{1-\mu}{2}\int_{\epsilon'_{\rm thr}}^{\infty}{\rm d}\epsilon' n'(\epsilon',z')\sigma(E',\epsilon',\mu)\ , \tag{6}$$

where the variables with prime are the quantities at redshift $z'$, $\mathrm{d}l = c\mathrm{d}t = \frac{c}{H_0}\frac{\mathrm{d}z'}{(1+z')\sqrt{\Omega_M(1+z')^3+\Omega_\Lambda}}$ is the differential path traveled by the VHE photon, $\mu = \cos\theta$ with $\theta$ the angle between the momenta of VHE and EBL photons, $n'(\epsilon', z') = n(\epsilon'/(1+z'), z=0)(1+z')^3$ is the EBL number density at redshift $z'$, and $\sigma$ is the pair production cross section. Note that the VHE measurements mainly apply for low-redshift ($z \lesssim 0.2$) sources, and we neglect the late production of EBL photons and just consider its cosmological expansion and redshift effect. $\epsilon'_{\rm thr}$ is the threshold energy for $\gamma$-ray energy $E' = E(1+z')$ with an angle $\cos\theta = \mu$ with the EBL photon. The cross section is peaked at a wavelength $\lambda/\mu{\rm m} \sim 1.24E/{\rm TeV}$ [86]. Therefore the observation of VHE $\gamma$-ray spectra can probe EBL at the wavelength from optical to far infrared, while it is not sensitive to UV band by the IACT data. The cosmological parameters used in this work are $\Omega_M = 0.274$, $\Omega_\Lambda = 1-\Omega_M$, $H_0 = 70.5$ km s$^{-1}$ Mpc$^{-1}$ [87].

### 5.3 Fitting method

We assume the intrinsic spectrum of blazar, $F_{\rm int}$, is of log-parabola shape ($F \propto E^{-\alpha-\beta\log E}$). The blazar spectrum is commonly modeled by the synchrotron-self-Compton (SSC) scenario, which shows a concave $\gamma$-ray spectrum in general. If the measured energy range is not very wide, the simple power-law can actually give a quite good description to the observations. However, at least for some sources, deviation from single power-law distribution of the spectrum, even corrected for the absorption effect, has been observed [88]. Therefore we adopt the log-parabola form of the intrinsic spectrum. It has been tested that the log-parabola assumption of the intrinsic spectrum will give robust results of the EBL [84].

No prior assumption about the EBL shape is adopted in this study. We divide the wavelength range of EBL from 0.1 $\mu$m to 100 $\mu$m, which is relevant for $\gamma$-rays between 100 GeV and 100 TeV, into 10 bins logarithmically. Within each bin the intensity $\nu I_\nu$ is assumed to be a constant $\xi_i$. Then we can fit the 10 $\xi_i$s, as well as the intrinsic source parameters of each source (nuisance parameters), from a set of observed $\gamma$-ray spectra $F_{\rm obs}(E)$.

### 5.4 Perspective of LHAASO

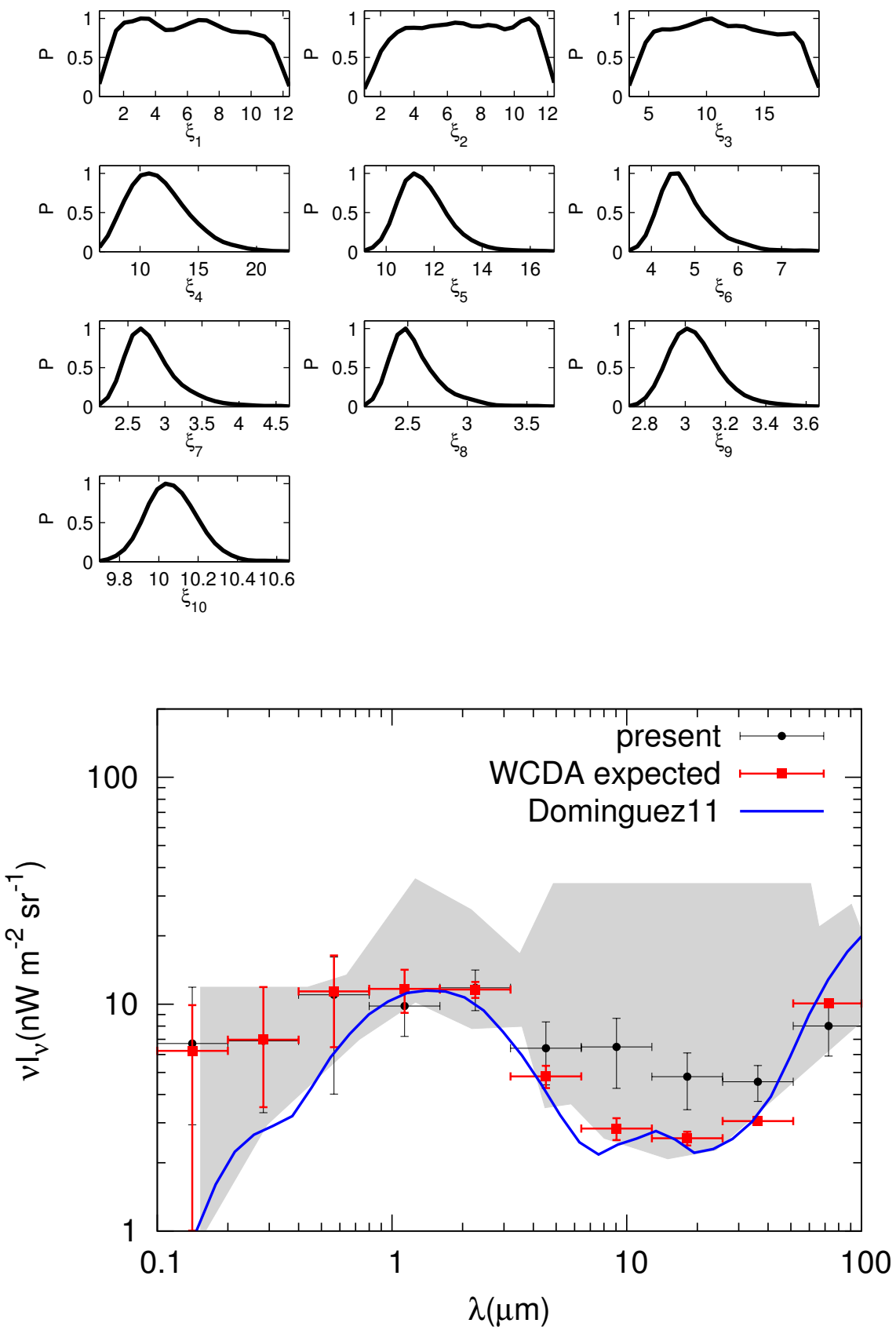

Fig. 20. Upper: 1-dimensional marginalized probability distributions of the fitting parameters $\xi_i$. Lower: Fitting results of the EBL intensities with simulated LHAASO spectra of 45 blazars, compared with the input EBL model [89]. We also show the constraints with the current blazar data as derived in [84] for comparison.

We explore the potential of LHAASO to measure the EBL intensities with this method. We first generate simulated observations of the blazar spectra with LHAASO-WCDA [90]. In [90] we studied the detectability of the blazars with LHAASO-WCDA, based on the Fermi AGN sample. The spectra of the Fermi AGNs, with known redshift measurements and within the field of view of LHAASO, are directly extrapolated to TeV energies based on the Fermi measurements. Then we apply the EBL absorption to the extrapolated spectra to derive the detected spectra of the sources. Comparing to the sensitivity of WCDA, we find that there will be about 30–40 Fermi blazars, mostly BL Lacs, could be detectable by LHAASO-WCDA in a few years' survey. The actual number of sources may be higher, due to the unexpected flaring activities of the blazars and the sources without redshift measurements. However, the sources which do not have redshift measurements will not be able to be used to constrain the EBL.

Taking the EBL model by Dominguez et al. (2011; [89]) as an example, we find that 45 sources with redshifts in the third LAT AGN catalog (3LAC; [91]) will be detectable by LHAASO-WCDA for one year sky survey. We simulate the spectral measurements of these sources following [90]. Fitting to these simulated spectra enables us to have a measurement of the EBL intensities, as shown in Figure 20. We can see that the EBL intensities above $\sim 1\ \mu$m can be well constrained with the expected LHAASO data. At shorter wavelengths the constraints become weaker, due to the relatively high energy threshold of LHAASO. The fitting results reproduce the input EBL model well, illustrating the robustness of this method. Compared with the results obtained with the present (sub-)sample of blazars [84], we find that LHAASO will have significant potential to improve the measurements of the EBL intensities.

### 5.5 Conclusion

We propose to *measure* the EBL with a global fitting methods, based on the VHE $\gamma$-ray observations of extragalactic blazars by LHAASO. This method does not assume the spectral shape of EBL, but parameterizes the EBL intensities in different wavelength bins as a constant parameter. The intrinsic spectra of the sources and the EBL intensities are fitted simultaneously using an MCMC algorithm. With simulated observations of of blazars by LHAASO, we show that the EBL intensities can be well measured. A large sample of sources with good spectral measurements, which is the object of LHAASO, is very essential for improving our understanding of the EBL.

## 6 LHAASO Science: VHE observations of star-forming/starburst galaxies

Executive summary: Detection of high-energy gamma-ray emission from star-forming and starburst galaxies by *Fermi* suggest that these galaxies are huge reservoirs of cosmic rays and these cosmic rays convert a significant fraction of their energy into gamma-rays by colliding with the interstellar medium. We propose that LHAASO observes nearby star-forming and starburst galaxies within about 20 Mpc. With its high sensitivity at energies above 10 TeV, LHAASO will be able to probe the acceleration and propagation of 0.1-1 PeV cosmic rays in these galaxies. As the processes producing VHE gamma-rays are accompanied by high energy neutrinos, the TeV-PeV gamma-ray flux of these galaxies can be used to study their contribution to the cosmic TeV-PeV neutrino background recently detected by IceCube.

### 6.1 VHE observations of star-forming/starburst galaxies

It is generally believed that Galactic cosmic rays (CR) are accelerated by supernova remnant (SNRs) shocks. CR protons interact with the interstellar gas and produce neutral pions (schematically written as $p+p \rightarrow \pi^0 + \mathrm{otherproducts}$), which in turn decay into gamma-rays ($\pi^0 \rightarrow \gamma+\gamma$). The high SN rate in star-forming and starburst galaxies implies high CR emissivities, so they are predicted to be bright gamma-ray sources. Ackermann et al. [92] examined a sample of 69 dwarf, spiral, and luminous and ultraluminous infrared galaxies using 3 years of data collected by the Large Area Telescope (LAT) on the Fermi Gamma-ray Space Telescope (Fermi). They find further evidence for quasi-linear scaling relations between gamma-ray luminosity and and total infrared luminosity which apply both to quiescent galaxies of the Local Group and low-redshift starburst galaxies. Nearby star-forming and starburst galaxies, such as M82 and NGC 253 are also detected at very high-energy (VHE) gamma-rays by e.g. HESS, VERITAS [93, 94]. But so far only quite a few galaxies have been detected. Moreover, no starburst galaxies have been detected above 10 TeV. With LHAASO, which has much higher sensitivity at energies above 10 TeV, one may expect that much more star-forming and starburst galaxies can be detected above 10 TeV and even above 100 TeV, as long as the galaxies are within the distance where VHE photons have not been absorbed by extragalactic background light (EBL) (typically within about 10-20 Mpc). At such high energy photons can only be produced by PeV cosmic rays, so LHAASO can probe the acceleration of PeV cosmic rays in these galaxies.

By extrapolating the TeV flux of M82 to the energy at 10 TeV, we find that the predicted flux at is $2{\times}10^{-13}\mathrm{TeVcm^{-2}s^{-1}(E/10TeV)^{-0.2}}$. NGC253 has a similar predicted flux above 10 TeV. These fluxes are above the sensitivity of LHAASO, so we expected that M82 and NGC 253 may be detected by LHAASO above 10 TeV. Since these two starburst galaxies are at distances of 2–3Mpc, we expected that they may be detected even above 100 TeV by LHAASO. This will be useful to study whether these starburst galaxies host extragalactic PeVatrons.

Recently, Tang et al. [95] reported the detection of gamma-ray emission above 200 MeV from a luminous infrared galaxy NGC 2146, which is at a distance of 15.2 Mpc. GeV gamma-rays have been also detected from some starburst galaxies hosting an obscured AGN, such as NGC 1068 and NGC 4945 [92], as well as from an ultra-luminous infrared galaxy, Arp 220 [96]. Using NGC 2146 as an example, we here estimate the detectability of these galaxies by LHAASO. Using the GeV flux and the spectrum of NGC 2146, we estimate the VHE flux assuming a simple power-law extrapolation. As can be seen from Figure 21, the predicted energy flux is $E^2dN/dE \simeq 10^{-13}\mathrm{TeVcm^{-2}s^{-1}}(E/1\mathrm{TeV})^{-0.1}$ at TeV energies, which is within the reach of LHAASO sensitivity. Note that at a distance of 15.2 Mpc, the absorption by EBL is not severe even for 10-100 TeV photons. Therefore, we propose that LHAASO perform systematic observations of nearby star-forming and starburst galaxies within 10-20 Mpc. The main candidates include M 82, NGC 253, NGC 2146, NGC1068, which are within the field of view of LHAASO.

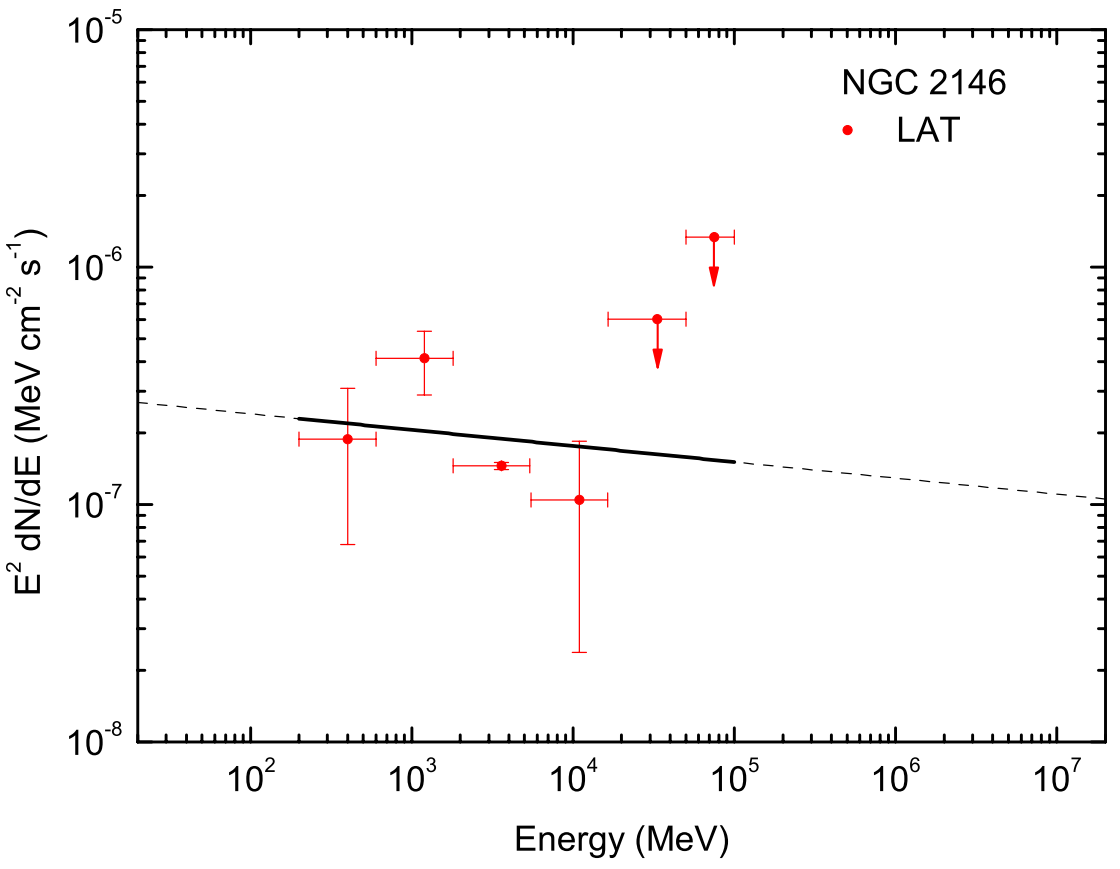

Fig. 21. Spectral energy distribution for NGC 2146 obtained in the analysis of the 68 months of Fermi/LAT data. The black solid line represents the best-fit power law in the range of 0.2-100 GeV and the dashed line is the extrapolation to higher energies.

### 6.2 Implications of VHE observations

The gamma-ray luminosity of starbursts depends not only on the CR intensity, but also on the efficiency of converting CR proton energy into pionic gamma-rays. This efficiency in turn depends on the ratio of the timescale of pion production to the escape time of protons. Protons escape by advection with galactic winds or by diffusion. The gamma-ray flux at 100 TeV will thus not only give us information about the acceleration of PeV cosmic rays, but also tell us physics about the transport of these cosmic rays in galaxies.

Proton-proton collisions in starbursts not only produce neutral pions, but also produce charged pions, which then decay and produce neutrinos. Loeb & Waxman [97] argued that supernova remnants in starburst galaxies accelerate CR protons and produce high-energy neutrinos. Chang et al. [98] calculated the accumulated neutrino flux by using the infrared luminosity function of star-forming galaxies recently obtained by the Herschel PEP/HerMES survey. Recently, the IceCube Collaboration reported 37 events ranging from 60 TeV to 3 PeV within three years of operation, corresponding to a $5.7\sigma$ excess over the background atmospheric neutrinos and muons [99]. One attractive scenario for this excess is that they are produced by cosmic rays in starburst galaxies [97]. But whether neutrinos in starburst galaxies can extend to sub-PeV/PeV energies is uncertain, given that normal supernova remnants are usually believed to accelerate protons only to PeV energy. It was suggested that hypernova remnants in starburst galaxies, by virtue of their fast ejecta, are able to accelerate protons to EeV energy [100] and produce sub-PeV/PeV neutrinos [101]. Future observations of nearby star-forming and starburst galaxies at 100 TeV will enable us to study the neutrino flux produced in such galaxies. One can then further study the total contribution to the cosmic neutrino background by all star-forming and starburst galaxies in the universe and hence pin down the starburst galaxy origin for IceCube neutrinos.

# Chapter 4 Cosmic-Ray Physics*

Benedetto D'Ettorre Piazzoli[1★¶] Siming Liu(刘四明)[2★¶†] Domenico della Volpe[3★¶] Zhen Cao(曹臻)[4,5¶]
Andrea Chiavassa[6¶] Benedetto D'Ettorre Piazzoli[1¶] Yiqing Guo(郭义庆)[4,5¶] Leonid T. Ksenofontov[7¶]
Olivier Martineau-Huynh[8¶] Diane Martraire[4¶] Lingling Ma(马玲玲)[4,5¶]
Xinhua Ma(马欣华)[4,5¶] Yuri Stenkin[9,10¶] Qiang Yuan(袁强)[11¶] Houdun Zeng(曾厚敦)[11¶]
Shoushan Zhang(张寿山)[4,5¶] Yi Zhang(张毅)[11¶] Hui Zhu(朱辉)[12¶]

[1]Dipartimento di Fisica dell'Universitàdi Napoli "Federico II", Complesso Universitario di Monte Sant'Angelo, 80126 Naples, Italy
[2]School of Physical Science and Technology, Southwest Jiaotong University, 611756 Chengdu, Sichuan, China
[3]Département de Physique Nucléaire et Corpusculaire, Facultéde Sciences, Université de Genève, 1211 Geneva, Switzerland
[4]Key Laboratory of Particle Astrophyics, Institute of High Energy Physics, Chinese Academy of Sciences, 100049 Beijing, China
[5]TIANFU Cosmic Ray Research Center, Chengdu, Sichuan, China
[6]Dipartimento di Fisica, Università degli Studi di Torino, Via Pietro Giuria 1, Torino, 10125, Italy
[7]Yu.G.Shafer Institute of Cosmophysical Research and Aeronomy SB RAS, 677027 Yakutsk, Russia
[8]Laboratoire de Physique Nucléaire et des Hautes Energies, CNRS-IN2P3, Universités Paris VI et VII, Paris, France
[9]Institute for Nuclear Research of Russian Academy of Sciences, 117312 Moscow, Russia
[10]Moscow Institute of Physics and Technology, 141700 Moscow, Russia
[11]Key Laboratory of Dark Matter and Space Astronomy, Purple Mountain Observatory, Chinese Academy of Sciences, 210023 Nanjing, China
[12]National Astronomical Observatories, Chinese Academy of Sciences, 100101 Beijing, China

**Abstract:** In the first part of this Chapter the present state of knowledge from the observations of cosmic rays between $10^{13}$ and $10^{20}$ eV is summarized. This is not intended to be a complete review, but rather a broad overview of the relevant processes involving cosmic rays, including the astrophysical environments in which they take place. This overview mainly concerns experimental results and phenomenological aspects of their interpretation, therefore experiments' description is not given but references to the vast bibliography are provided in the text. Some attempt is made to address the most popular explanations offered by theoretical models. The second part is devoted to the description of the LHAASO performance and of its capability to provide a response to several open questions, still unanswered, concerning cosmic rays above $10^{13}$ eV, highlighting which major steps forward in this field could be taken from LHAASO observations.

## I. THE COSMIC RAYS

### A. Cosmic rays as messengers of the non-thermal universe

Cosmic Rays (CRs) were discovered about a century ago. First evidence of this radiation from space came from the measurements performed by Victor Hess in 1912 during seven balloon flights at different altitudes. Before the development of accelerators in the fifties, the cosmic rays served as the main source of high-energy particles: positrons, in 1932, muons in 1937, pions and strange particles (kaons and hyperons) in the late 1940s were discovered in cosmic rays. Presently, cosmic rays are studied since they are messengers of extreme phenomena involving very high energies. Indeed, their power-law energy spectrum extending up to $10^{20}$ eV (see Fig. 1) and beyond witnesses their non-thermal origin. High-energy cosmic rays are the manifestation of the relativistic Universe involving physical processes at energies by far in excess of what could ever be achieved in man made laboratories. How cosmic accelerators can

Received 2 December 2021; Accepted 4 December 2021; Published online 18 January 2022

* Supported by the National Key R&D Program of China (2018YFA0404203), the International Scholarship Program of the MOST of China (G2021166002L), National Natural Science Foundation of China (NSFC) (12147208, U2031103, U1931204) and the Science and Technology Department of Sichuan Province (2021YFSY0031)

† E-mail: liusm@swjtu.edu.cn

★ Editors ¶ Contributors. All authors contribute equally to the work.

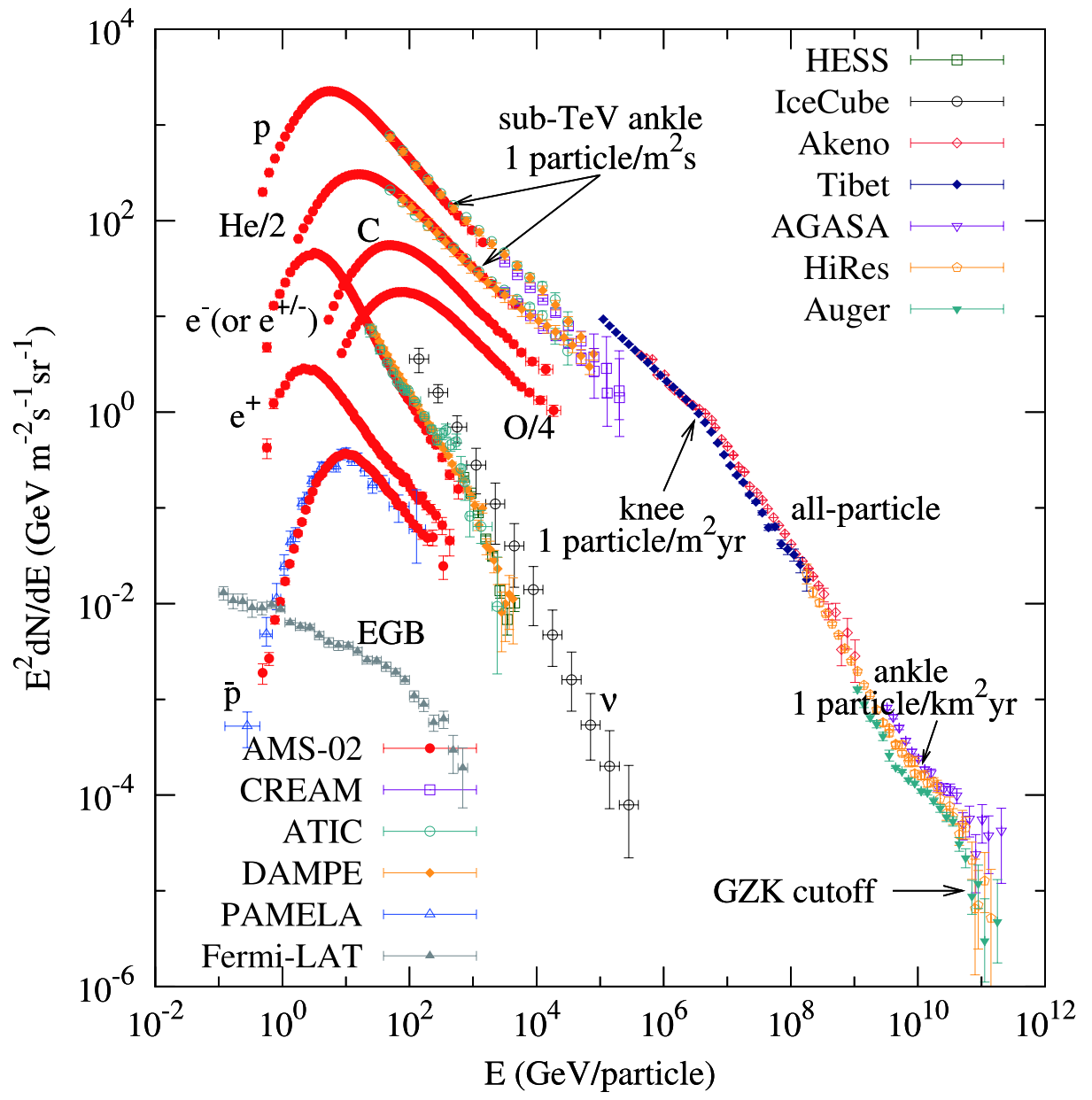

**Fig. 1.** (color online) Cosmic-ray energy spectra. For energies below ~ 100 TeV, the spectra of different species are shown, and for higher energies the all-particle spectra are plotted. References of data: CREAM [1]; ATIC [2, 3]; AMS-02 [4-8]; PAMELA [9]; DAMPE [10-12]; Fermi-LAT [13]; HESS [14]; IceCube [15]; Akeno [16]; Tibet [17]; AGASA [18]; HiRes [19, 20]; Auger [21].

boost particles to these energies and which is their nature are primary questions that yet need a firm answer. At the same time, these sources are natural laboratories to explore how the laws of physics behave at the highest energies. The center-of-mass energy of the LHC *pp* collisions is reached with the interaction on fixed target of a proton with energy $\approx 10^{17}$ eV. The highest energy cosmic rays may provide the opportunity to probe new physics beyond the Standard Model of particle physics as, for instance, a tiny violation of the Lorentz invariance at energies not reached at terrestrial accelerators. A sudden change of the hadronic interaction at the highest energies is another example. Thus, both astrophysics and particle physics are central topics of the cosmic ray research.

### B. The origin of cosmic rays : galactic or extragalactic?

The problem of the cosmic ray origin has been debated extensively. This issue concerns the nature of the cosmic ray sources, how cosmic rays are accelerated at very high energies and how they propagate in the interstellar or intergalactic medium. Two extreme models have been discussed: the galactic model, in which the cosmic ray sources are supposed being concentrated in the Galaxy, and the extragalactic models which postulate that cosmic rays occupy the extragalactic space from which they may flow into the Galaxy. Many aspects concerning the chemical composition, the energy spectrum, the anisotropy as well as the energetic requirements rule out and made unacceptable the extragalactic origin of all cosmic rays, favoring, instead, the galactic model. An exhaustive account is given in [22] and Sec. 20 of [23]. According to the hypothesis first suggested by Baade and Zwicky [24] the sources of the bulk of cosmic rays are the supernova (SN) explosions. The strongest arguments in favor of this assumption come from the energetics characterizing the SN explosions and from the models of stochastic acceleration by shock waves in SN being able to reproduce the cosmic ray power-law energy spectrum (see next Sections). However, the energy range above $10^{18}$ eV is likely dominated by extragalactic cosmic rays because of the limit of the maximum energy achievable in SN explosions, and because the Galactic magnetic field cannot confine particles of very high energy. The anisotropy of the highest energy cosmic rays also call for their extragalactic origin (see Sec. I.J).

### C. Galactic confinement and propagation

The large scale magnetic field of the Galaxy is measured in several ways. The evidence comes from the Zeeman splitting of the 21 cm line of neutral hydrogen and of molecular radio lines, from the polarization of the radio emission and of the light of nearby stars, and from the Faraday rotation of linearly polarized radio signals. All estimates of the magnetic field strength lie in the range 1-10 μG, and a typical value of 3 μG is usually assumed. The gyro-radius of a particle of rigidity $R$ in a field of strength $B$, assuming its velocity orthogonal to $B$, is $r = 1.1 \times 10^{-6} R(\mathrm{GV})/B(\mu\mathrm{G})$ pc. For protons of $10^{15}$ eV, this gives $r \approx 0.37$ pc to be compared with the galactic radius of about 15 kpc and a galactic thickness $2h \approx 400$ pc at the radius of Earth (8.5 kpc). The gyro-radius of a proton of $10^{18}$ eV is about 370 pc, comparable to the galactic thickness. One can conclude that particles up to $10^{15}$ eV are well confined within the Galaxy. At higher energies, cosmic rays start to escape more freely from the Galaxy, providing a possible explanation of the steepening of the spectrum at $10^{15}$ eV.

The steepening would then happen at different energies for each element since the gyro-radius depends on the charge $Z$ of the particle. The primary information on the propagation of cosmic rays in the galactic volume comes from the presence of secondary elements more abundant in the cosmic radiation than in solar system material and substantially absent as end products of stellar nucleosynthesis. These secondary elements are produced by fragmentation of primary cosmic rays, such as carbon and oxygen (to Li, Be, B) and iron (to Sc, Ti, Cr, Mn), in the interstellar medium. From the ratio of secondaries to primaries, decreasing as energy increases (Fig. 2), the mean amount of matter traversed is determined to be of the order of 2-10 g·cm$^{-2}$. With a nominal density in the disk of one proton per cm$^3$, the distance travelled up to the Earth is about 500-2000 kpc, far greater than the galactic dimensions. This implies that cosmic ray confinement is a diffusive process in which the particles follow a very tortuous path before escaping into the extragalactic space. Typical residence times are then in the range $10^6-10^7$ years. High energy cosmic rays diffuse out of the Galaxy faster. However, even cosmic rays with energy $\geq 10^{15}$ eV do not stream freely away from the galactic disk as proved by their low anisotropy ($\approx 10^{-3}$), and an estimate of the residence time of the order of $10^5$ years, or less, looks more appropriate. Escape times of the order of $10^7$ years are also found for GeV CRs from the study of the abundances of radioactive isotopes produced in the spallation reactions, as for instance $^{10}$Be that undergoes $\beta$-decay into $^{10}$B with a characteristic lifetime $\tau = 3.9\times10^6$ y [26].

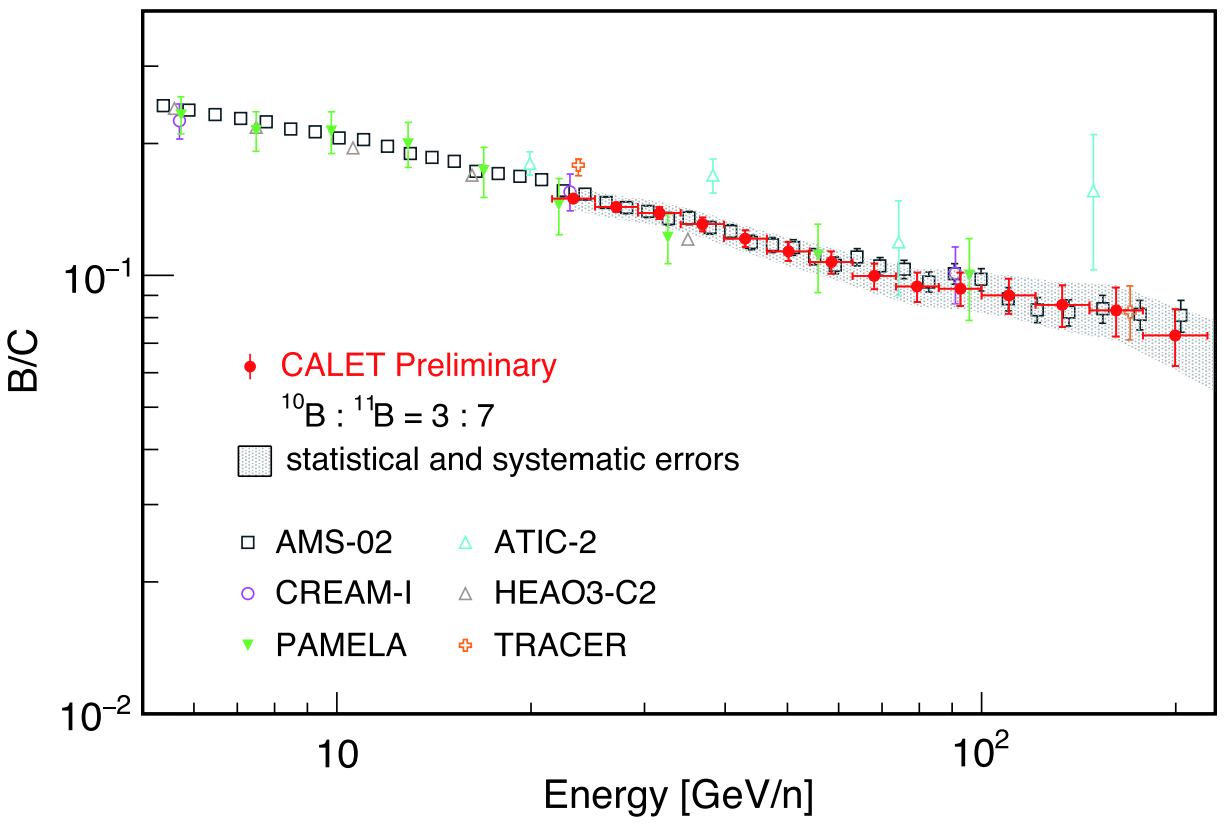

**Fig. 2.** (color online) Preliminary result of boron-to-carbon flux ratio with CALET compared with previous observations [25].

In summary, in the current propagation models, the Galaxy is assumed to be a *'leaky box'* in which cosmic rays diffuse staying confined for a long time and undergoing, before escaping, frequent scatterings either with random magnetic irregularities or with waves excited by the high energy particles themselves. A high isotropy ensues. The residence time decreases with the energy so that higher energy cosmic rays $\geq 10^{15}$ eV escape quicker and have less time to produce the lower energy secondaries. They are less confined in the galactic volume and this effect as well as the limit of the acceleration process via shock waves (see next Sections) are usually invoked to explain the steepening of the spectrum above $10^{15}$ eV (see Sec. 6 of [27]).

### D. Cosmic ray luminosity and the acceleration process

Cosmic rays constitute a remarkable component of the galactic inventory accounting for an important energetic and dynamical factor. Their energy density ($\approx$ 1 eV/cm$^{-3}$) is comparable to the magnetic energy density ($\approx$0.25 eV/cm$^{-3}$), to the energy density of the interstellar gas ($\approx$ 1 eV/cm$^{-3}$) and is also of the same order of magnitude as the energy density of the relic thermal radiation ($\approx$0.26 eV/cm$^{-3}$). Such a situation must be expected in quasi-stationary conditions and implies that cosmic rays play an essential role in the dynamics of the Galaxy. The cosmic ray luminosity at Earth, about $3\times10^{40}$ erg/s, is second only to the galactic luminosity of the optical radiation. It is generally accepted from simple energetic considerations that among the galactic sources the supernova remnants (SNRs) are the main source of cosmic rays since they may provide the total energy budget of cosmic ray in the Galaxy. Due to the falling off energy spectrum, the power required to sustain the high energy cosmic rays is considerably less, about $5\times10^{37}$ erg/s for above $10^{16}$ eV cosmic rays. The mechanical energy input to the Galaxy from each SN is about $10^{51}$ erg so that with a rate of about one explosion every 30 years the total mechanical power input from SNs is of the order of $10^{42}$ erg/s [28]. Thus SNs have enough power to drive the galactic cosmic ray acceleration if there is a mechanism for channeling about 10% of the mechanical energy into relativistic particles. An appropriate acceleration mechanism, the diffusive shock acceleration process (DSA), has been known since 1977 [29].

The diffusive acceleration process at supernovae blast waves driven by expanding SNRs can provide the spectral power-law shape of cosmic rays. In this picture, a supersonic flow, as due to the ejecta from a supernova explosion or a pulsar wind moving at speeds of about $10^4$ km/s, terminates in a shock balancing the pressure of the ambient medium. High energy particles scattered off turbulent magnetic fields on both sides of the shock may diffusively cross the shock front many times gaining each cycle an average energy $\delta E/E \approx \beta(4/3)$, where $\beta$ is the velocity of the plasma flow. A small fraction of these particles are advected downstream of the shock and may escape in the downstream flow acquiring a power-law spectrum with a spectral index mainly dependent on the ratio of the upstream and downstream gas velocity in the shock reference. This model naturally produces a power-law spectrum $\mathrm{d}N/\mathrm{d}E \approx E^{-2.0}$ at source ($\mathrm{d}N/\mathrm{d}E \approx E^{-2.3}$ in highly relativistic shocks), consistent with the locally observed cosmic ray spectrum $\mathrm{d}N/\mathrm{d}E \approx E^{-2.7}$ after correcting for the propagation effects. Indeed, in the conventional models the large-scale propagation in the Galaxy is governed by the diffusion. The spectrum becomes steeper of a quantity $\approx$ 0.3-0.6, the diffusion index, as it results from the spectra of the secondary cosmic rays produced by spallation of heavier primaries with the interstellar matter, and from the secondary-to-primary ratios, in particular the ratios Boron/Carbon (Fig. 2) and $^{10}$Be/$^9$Be.

The maximum attainable energy depends on the time the particle remains in the acceleration region before escaping. This is related to the size of the region and on the strength and structure of the magnetic field. The first es-timate of the maximum energy is due to Lagage and Cesarsky [30]. Assuming that the shock remains strong enough for about 1000 y and a typical interstellar magnetic field of a few micro-Gauss, the acceleration rates give a maximum cosmic ray energy of roughly $Z \cdot 10^{14}$ eV, ignoring any energy loss mechanism. In this simplified derivation the accelerated particles do not affect the conditions in the acceleration region. However, there are strong theoretical and observational reasons that argue for a significant amplification of the magnetic field as a result of the pressure gradient of the accelerating CRs, triggering instabilities in the precursor of the SNR shock. The most important consequence of magnetic field amplification in SNRs is the substantial increase of the maximal energy of CRs accelerated by SN shocks, which presumably provides the formation of Galactic CR (GCR) spectrum inside SNRs up to the energy $10^{17}$ eV. It is also discussed possibilities of formation GCR spectrum up to significantly higher energies $3\times10^{18}$ eV due to reacceleration of CRs generated in SNRs [31, 32], or due to contribution of more powerful type IIb supernovae [33]. The theory of particle acceleration by strong shock associated to SNRs is sufficiently well developed. Important theoretical progresses have been achieved with the development of the kinetic nonlinear theory of diffusive shock acceleration (for reviews see Refs. [34-37]). The current theoretical framework, consistently including the most relevant physical factors, allows to make quantitative predictions of the expected properties of cosmic rays accelerated in SNRs.

In Fig. 3, the calculated cosmic ray intensities of different species accelerated in SNRs are shown together with experimental data. Two different possibilities of maximal energies are shown by thin and thick curves [38]. The SNR origin model of cosmic rays via DSA has received some support from the observation of radio emissions, X rays, and GeV-TeV gamma rays from putative sources belonging to different classes of objects, such as pulsar wind nebulae, supernova remnants, compact binary systems, clusters of massive stars, and the Galactic center. These observations give direct evidence of the acceleration of electrons to beyond $10^{14}$ eV. However, the evidence that high energy cosmic rays have the same origin needs confirmation [39]. TeV gamma rays are tracer for high energy particles and can be used to search for PeV cosmic ray sources, the so-called PeVatrons. In the 'hadronic scenario' gamma rays come from the decay of neutral pions produced by the interaction of cosmic ray nuclei with the ambient matter or radiation. In the competing 'leptonic scenario' the inverse Compton scattering of ambient photons with energetic electrons may lead to the TeV gamma ray emission. Current data do not allow us to distinguish between the hadronic or leptonic origin of high energy TeV gamma rays [40]. Since the inverse Compton scattering at high energies is strongly suppressed by the Klein-Nishina effect, the observation of a gamma ray power-law spectrum extending with no break up to the 100 TeV and beyond would be a good proof of the hadronic nature of the interaction.

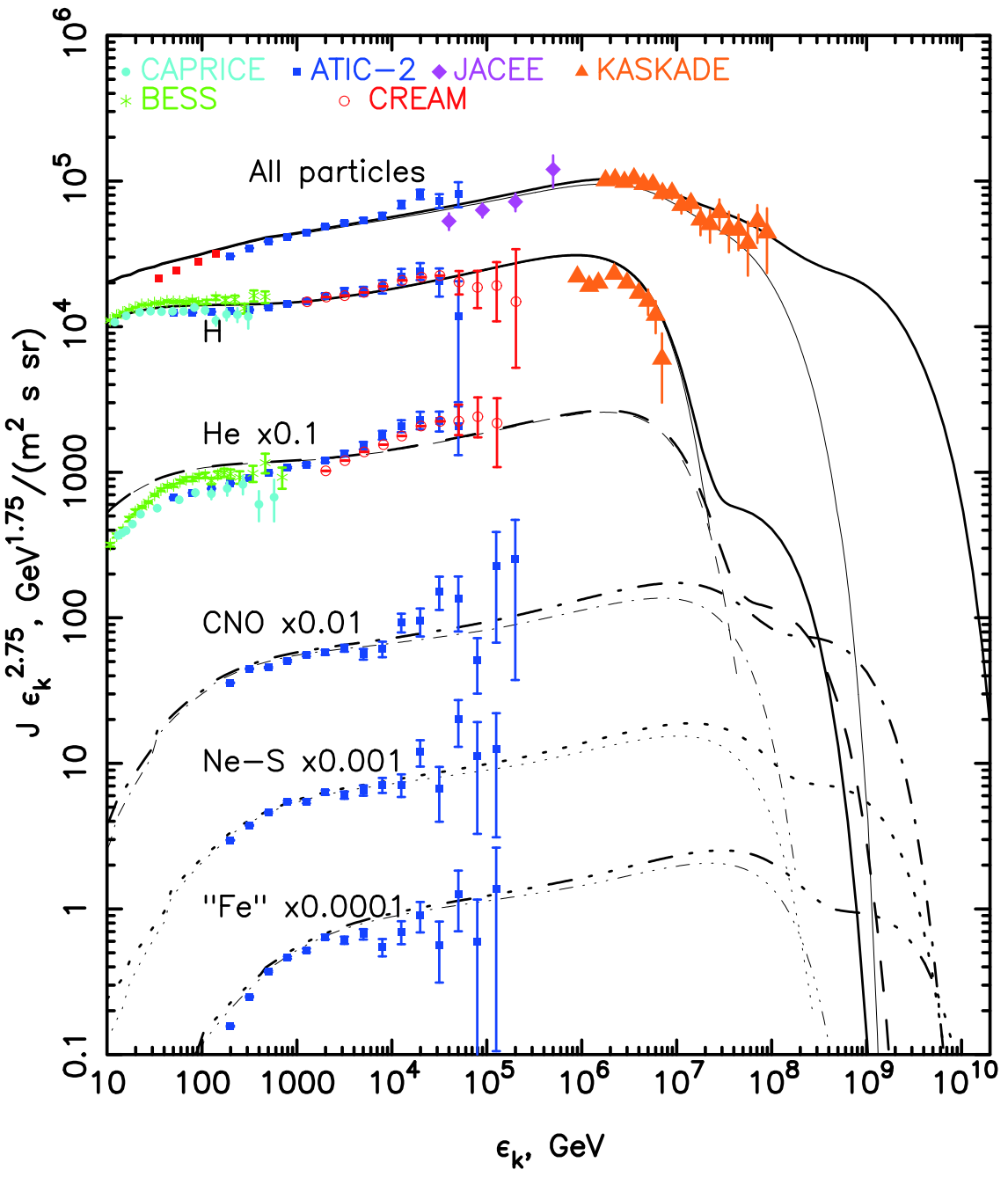

**Fig. 3.** (color online) Differential intensity $J_k$ of different species of cosmic rays as a function of the kinetic energy $\epsilon_k$. Experimental data obtained in the CAPRICE [41], BESS [42], ATIC-2 [43], CREAM [44], JACEE [45] and KASCADE [46] experiments are shown. Thin and thick curves refer to two different models of cosmic-ray acceleration in SNRs discussed in [45].

Unfortunately, current $\gamma$-ray observations show that SNRs with hard GeV spectra always have evident spectral softening in the TeV band with a potential cutoff near ~ 10 TeV. For SNRs with soft $\gamma$-ray spectra, no emission beyond 100 TeV has been detected either [39, 40]. Fig. 4 shows the multiwavelength spectra of a sample of 35 $\gamma$-ray SNRs and their distribution in the Galaxy. The only source with emission close to 100 TeV, W30, is actually associated with a pulsar wind nebula. All other sources do not have detectable fluxes above 50 TeV. Indeed, HAWC searched for TeV gamma ray emission from GeV detected SNRs. Among 9 sources with significant emission above 56 TeV, three SNRs that emit above 100 TeV have been observed. These 9 sources are close to ATNF radio-pulsars and exhibit a curved spectrum, implying a dominant leptonic origin of the emission. However, that does not immediately disqualify them from being PeVatrons [47, 48]. The half-completed LHAASO facility has reported twelve sources, including Crab Nebula, of > 100 TeV gamma rays [49]. Data collected for a livetime of 308.33 days evidence a gradual spectral steepening with energy up to 500 TeV. Apart from Crab, all sources ex-hibit an extended morphology up to one degree. The energy of the most energetic photon from an extended source positionally overlapping with the Cygnus Cocoon is about 1.4 PeV.

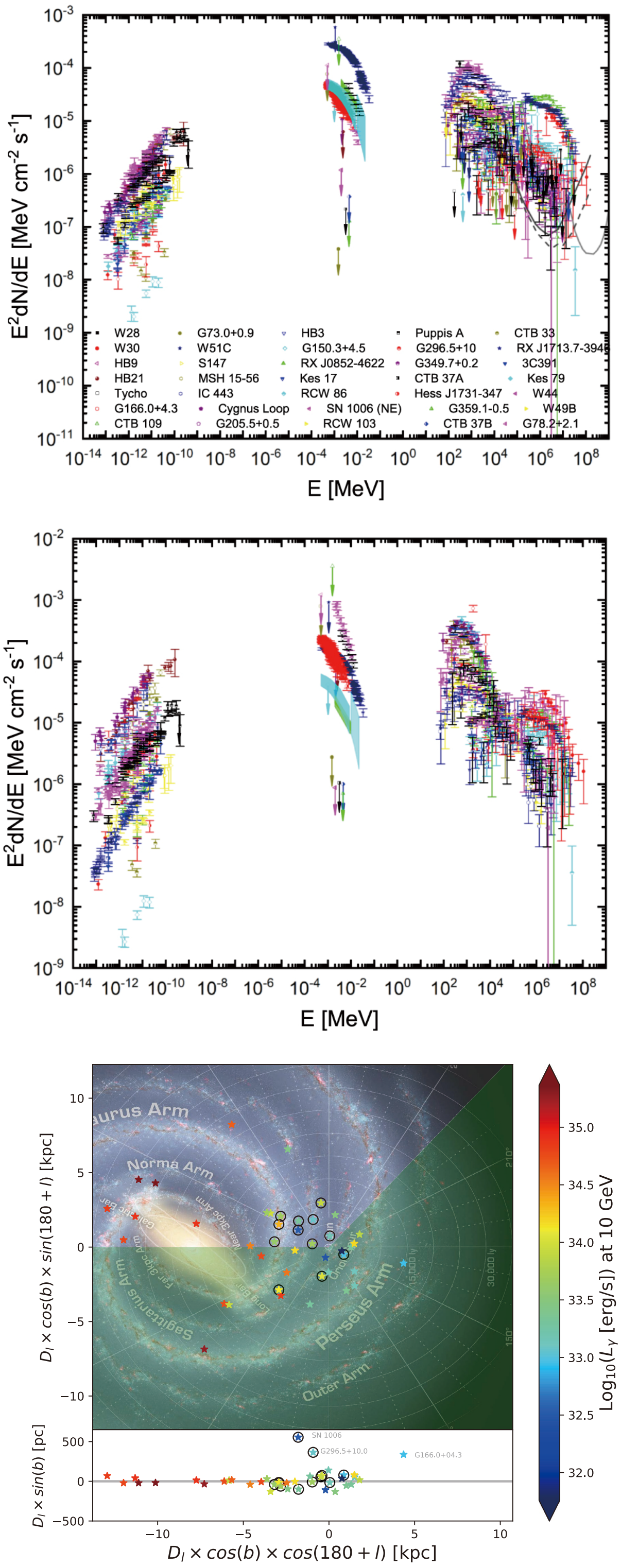

**Fig. 4.** (color online) The upper panel shows the multi-wavelength spectral data of 35 SNRs. The middle panel shows the spectra normalized at 100 GeV. Lower panel distribution of $\gamma$-ray SNRs in the Galaxy. The color indicates the luminosity at 10 GeV. Sources with hard GeV spectra are shown with open black circles [39]. The light green color indicates the region covered by molecular survey observations of the Delingha millimeter telescope.

Much larger photon statistics will allow detailed studies of the spectral and morphological features of these sources needed to assess the origin, leptonic or hadronic, of the detected radiation. The Crab Nebula spectrum measured by LHAASO [50] is shown in Fig. 5. It extends up to the PeV range including one event of ≃ 1.1 PeV energy. The spectrum shows a gradual steepening over three energy decades that can be explained by a combination of synchrotron radiation and inverse Compton scattering of relativistic electrons, accelerated at the termination shock of the pulsar wind, interacting with the ambient magnetic and radiation fields, respectively. A contribution of PeV protons to the production of the highest-energy gamma rays cannot be excluded. Gamma ray emission extending up to 50 TeV with no evidence of a cutoff has been observed by HESS from a small region surrounding the Galactic center [51]. However, the black hole Sgr A* at the Galactic center cannot be a viable alternative to SNRs as source of PeV galactic cosmic rays since at present it does not have a high enough rate of particle acceleration to substantially contribute to the population of Galactic cosmic rays [52]. Accordingly, one would expect that a few PeVatrons are currently at work in the Galaxy.

In conclusion, even if the experimental findings are not conclusive, there is a general consensus that SNRs, as well as pulsars, clusters of massive stars, and the black hole at the Galactic center or, more in general, all the high-energy systems, end-points of the stellar evolution, are able to generate the cosmic power-law spectrum and to account for the total amount of cosmic-ray energy contained in the Galaxy, at least up to $10^{17}$ eV. Galactic sources are believed to run out of power at $10^{17}-10^{18}$ eV, where the transition from galactic to extra-galactic component should take place. Higher energies beyond $10^{17}$ eV are essentially reachable by extra-galactic phenomena. The need of confining the particles in the accelerator region for long times provides a basic geometric criterion, due to Hillas [53], useful in selecting potential accelerator sites. This is a simple dimensional argument which makes it possible to identify objects that are able to accelerate particles up to a given energy. By demanding that the Larmor radius of the particle, $r = E/ZB$, does not exceed the size of the acceleration region, we obtain a limit to the maximum attainable energy $E_{\max} = \beta ZBL$ where $L$ the size of the acceleration region, and $\beta$ the speed of the magnetic scattering centers. A simplified version of the Hillas plot, not accounting for energy losses [54], is given in Fig. 6, showing, for a given maximum energy, the relation between the source magnetic field strength $B$ and its size.

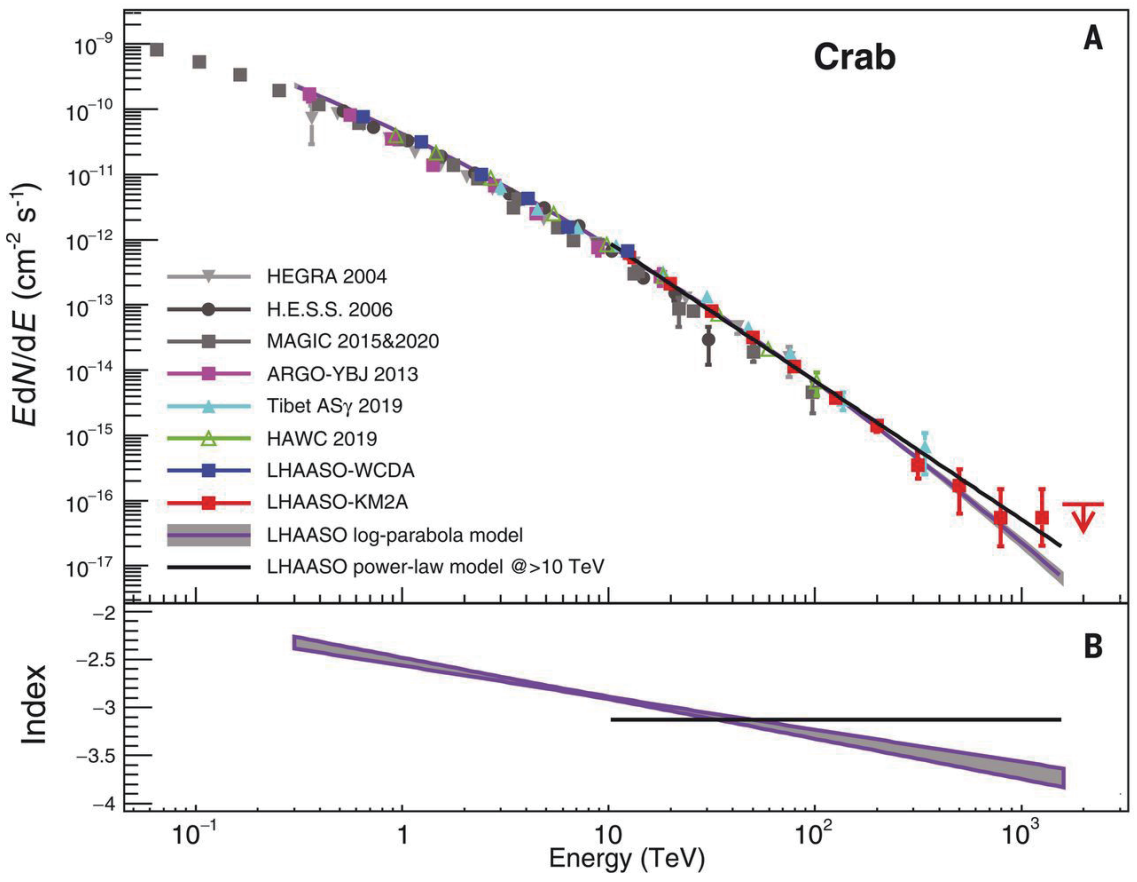

**Fig. 5.** (color online) The gamma-ray flux of the Crab measured by LHAASO (top) and the energy-dependent local power law index derived by the log-parabola model fitting as indicated by the purple band (bottom). See [50] for details.

As shown in this plot, the most luminous types of candidate sources are the Active Galactic Nuclei (AGNs) and Gamma Ray Bursts (GRBs). In many AGNs, as for instance the Fanaroff-Riley class II (FR-II) radio galaxies hosting supermassive black holes at the core, high-energy jets of radiation and relativistic material emerge along the disk's axis, with scales of pc to kpc. Shock waves propagating along these jets with Lorentz boost factors of $\Gamma = 10-30$ may be the sources of the highest

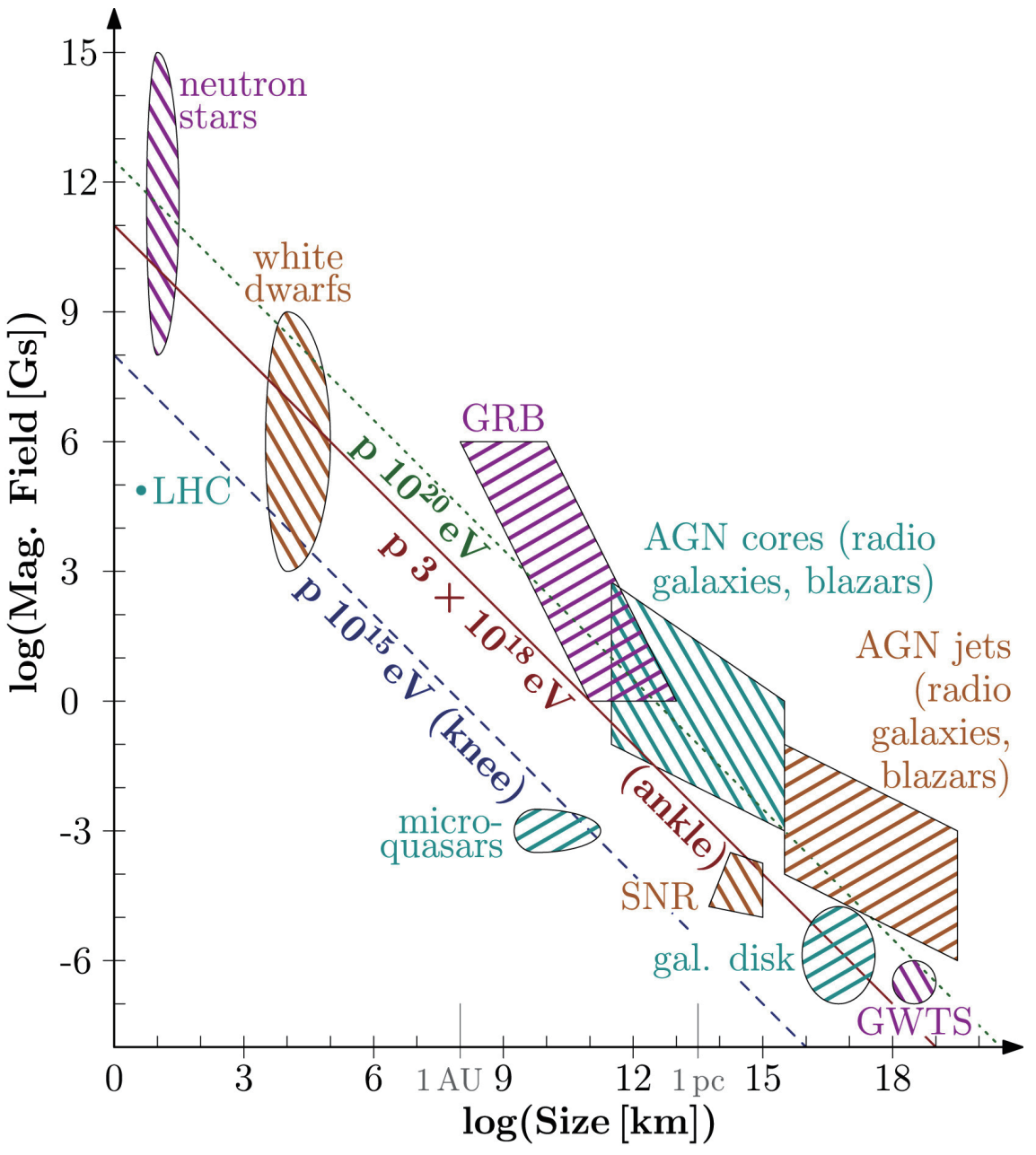

**Fig. 6.** (color online) The classical Hillas plot. The knee ($10^{15}$ eV), ankle ($3\times10^{18}$ eV) and maximum energy ($10^{20}$ eV) lines are shown (blue dashed, red solid and green dotted, respectively). See [55] for details. (Reprinted with permission from Elsevier)

energy cosmic rays [56]. GRBs, believed to occur when two neutron stars (black holes) merge or a massive star collapses into a black hole, may also provide the requisite environment for acceleration to ultrahigh energies. These events would be accompanied by jet formation where shock with speeds very close to c can occur and energies up to $10^{21}$ eV can be reached [57]. Other suggested extragalatic CR sources are the shock associated with colliding galaxies or starburst galaxies. The observed luminosity of all these systems may deliver the required flux of ultrahigh energy cosmic rays.

To be thorough, we mention the 'top-down' scenarios in which the cosmic rays are not accelerated, but are the result of the decay of supermassive 'X' particles that have been trapped in topological defects since the time of the early Universe [58]. Nevertheless, in all conceivable top-down theories photons and neutrinos should dominate at the end of the hadronic cascade, a scenario not supported by observations.

### E. The cosmic ray energy spectrum

Protons make up about 90% of all cosmic-ray components while helium nuclei amount to nearly 10% and all other nuclei comprise only about 1% of the total flux. Traditionally, the term 'cosmic rays' refers solely to the fast charged nuclei of cosmic origin. Other charged particles in the cosmic radiations are electrons, positrons and antiprotons with far lower fluxes (see Fig. 1). The abundance of nuclei in cosmic rays reflects their concentration in outer space. Indeed, the chemical composition of cosmic rays exhibits remarkable similarities to the solar system abundances. The main difference is the presence of two groups of elements, (Li, Be, B) and (Sc, Ti, V, Cr, Mn), more abundant in the cosmic radiation than in the solar system material as shown in Fig. 7.

These elements are absent as end-products of stellar nucleosynthesis and are present in the cosmic radiation as spallation products of Carbon and Oxygen and Iron nuclei, respectively. The differences can be accounted for by the collision of cosmic rays with the interstellar gas during their propagation from sources to Earth. These secondary nuclei have energy spectra steeper than the primary nuclei. At energies above a few TeV, cosmic rays are conventionally grouped into five groups: protons, helium nuclei, M(medium: $Z$ = 6-9), H(heavy: $Z$ = 10-20), VH(very heavy: $Z$ = 21-30). Another classification often used in the literature, and mainly referred to the data analysis, subdivides cosmic rays in 'light' (protons plus helium nuclei) and 'heavy' (all other nuclei) or in three mass groups ('light', 'medium', 'heavy').

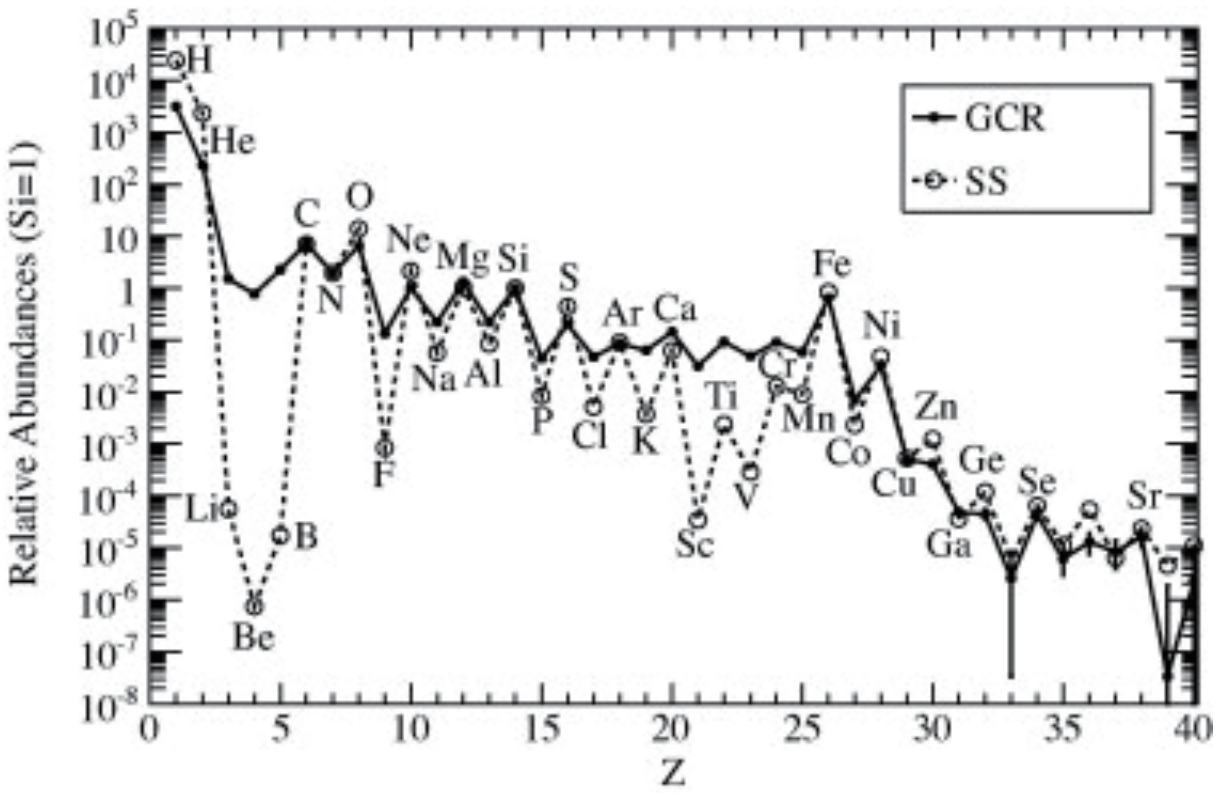

**Fig. 7.** Solar System (SS) and galactic cosmic rays relative abundances at 2 GeV/nuc normalized to $^{14}$Si = 1 [59].

Cosmic rays of energies up to about 100 TeV can be studied with telescopes on board of balloons, satellites or installed on the International Space Station (ISS). These devices achieve a reliable charge identification and a high resolution measurement of the primary energy on an event-by-event basis. Energy spectra for individual elements can be obtained, often displayed versus the energy per nucleon. At higher energies the primary radiation is studied with experiments detecting the secondary particles generated in extensive air showers (EAS). These experiments have limited sensitivity to identify the charge of the primaries, and the total flux as a function of the particle energy, the so-called all-particle spectrum, are typically displayed. As shown in Fig. 1, the all-particle spectrum follows a power-law with a spectral index of about -2.7 up to about $3\times10^{15}$ eV, where a break of the spectral index is observed ('knee'). Above this knee the spectrum becomes steeper with an index of -3.0 up to the 'ankle' around $5\times10^{18}$ eV where the spectrum becomes flatter again for about one energy decade. Above $4\times10^{19}$ eV the spectrum appears to fall-off.

Direct measurements with satellites, balloon-based detectors or detectors installed on the ISS (AMS [4], ISS-CALET [60], ISS-CREAM [61]) have provided excellent description of the evolution of each cosmic ray component up to 100 TeV, as shown in Fig. 8. These detectors consist of tracking planes and devices to measure the charge of the incoming particles. The energy measurement is accomplished with magnetic spectrometers at lower energies and calorimeters at higher energies. The measurement of the proton spectrum from GeV to TeV energies is one of the main target of these observations. Some spectral features have been observed, as shown in Fig. 9. A break in the spectrum is evident near 500 GV followed by a spectral hardening. There is also strong evidence of a softening above ~10 TV. Data from ATIC [43], CREAM [62], and JACEE [45] demonstrate a decrease of the proton-helium flux ratio at TeV energies (Fig. 10). CREAM [1, 62], DAMPE [11], and HAWC [63] report a hardening of the helium energy spectrum near 500 GV, surpassing the proton spectrum at approximately 10 TeV, a trend confirmed by the space-based observations of ISS-CALET [64]. Possible explanations to these spectral features include different source populations, nearby proton-rich source up to TeV energies or proton-poor sources in the TeV range and anomalous diffusion [65, 66].

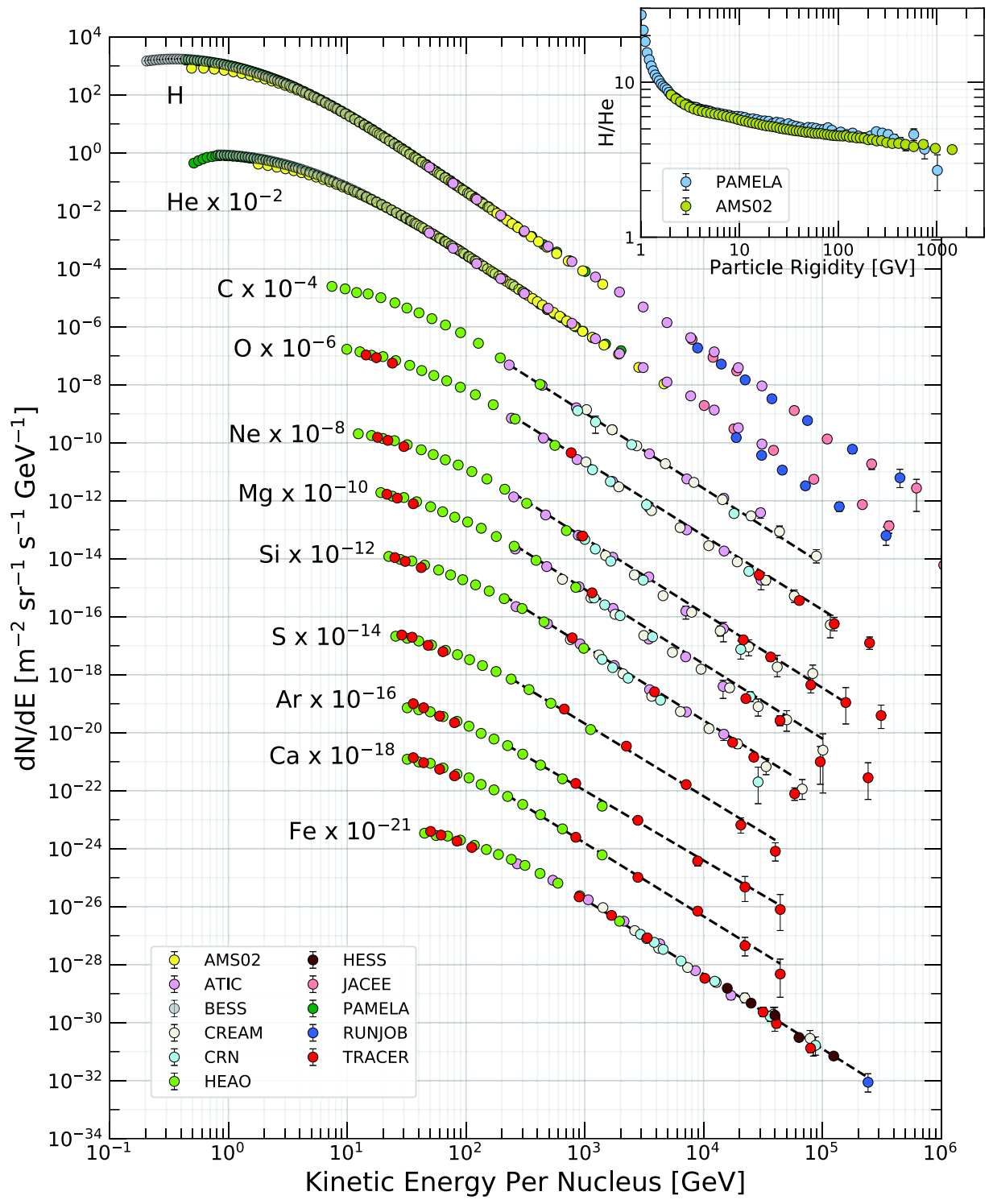

**Fig. 8.** (color online) Fluxes of nuclei of the primary cosmic radiation in particles per energy-per-nucleus PLOTTED vs energy-per-nucleus. The inset shows the H/He ratio as a function of rigidity [67].

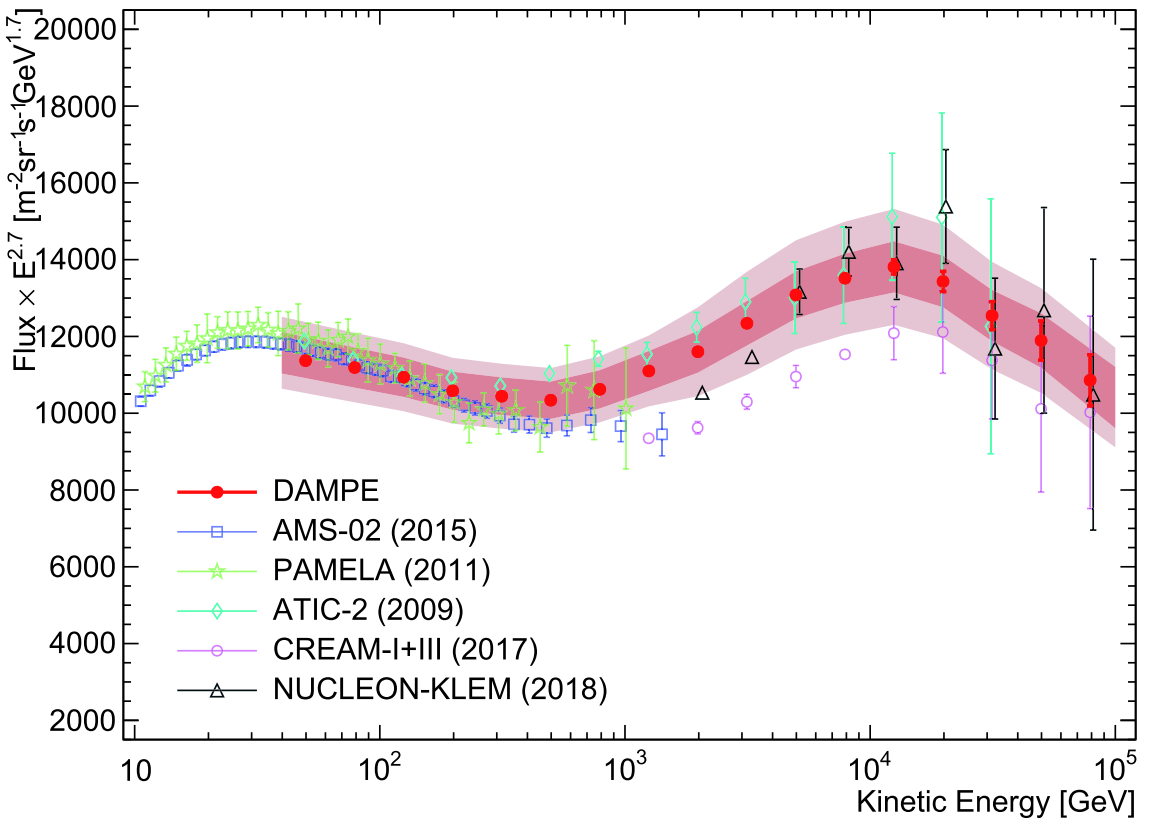

**Fig. 9.** (color online) The proton spectrum from 40 GeV to 100 TeV measured by DAMPE [10]. Similar spectral shape has also been observed by ISS-CALET [64]. (Reprinted with permission from AAAS)

In the energy region >10 TeV, air shower arrays operating at high altitude, such as Tibet AS [68], ARGO-YBJ [69] and HAWC [63], can efficiently separate the light component providing an useful complement to the direct measurements affected, due to their limited detector exposure, by large statistical uncertainties. A comparison of direct and indirect measurements shows a reasonable agreement for the light component up to 100 TeV and beyond within estimated systematic and statistical uncertainties ($\simeq$ 30%, Fig. 11).

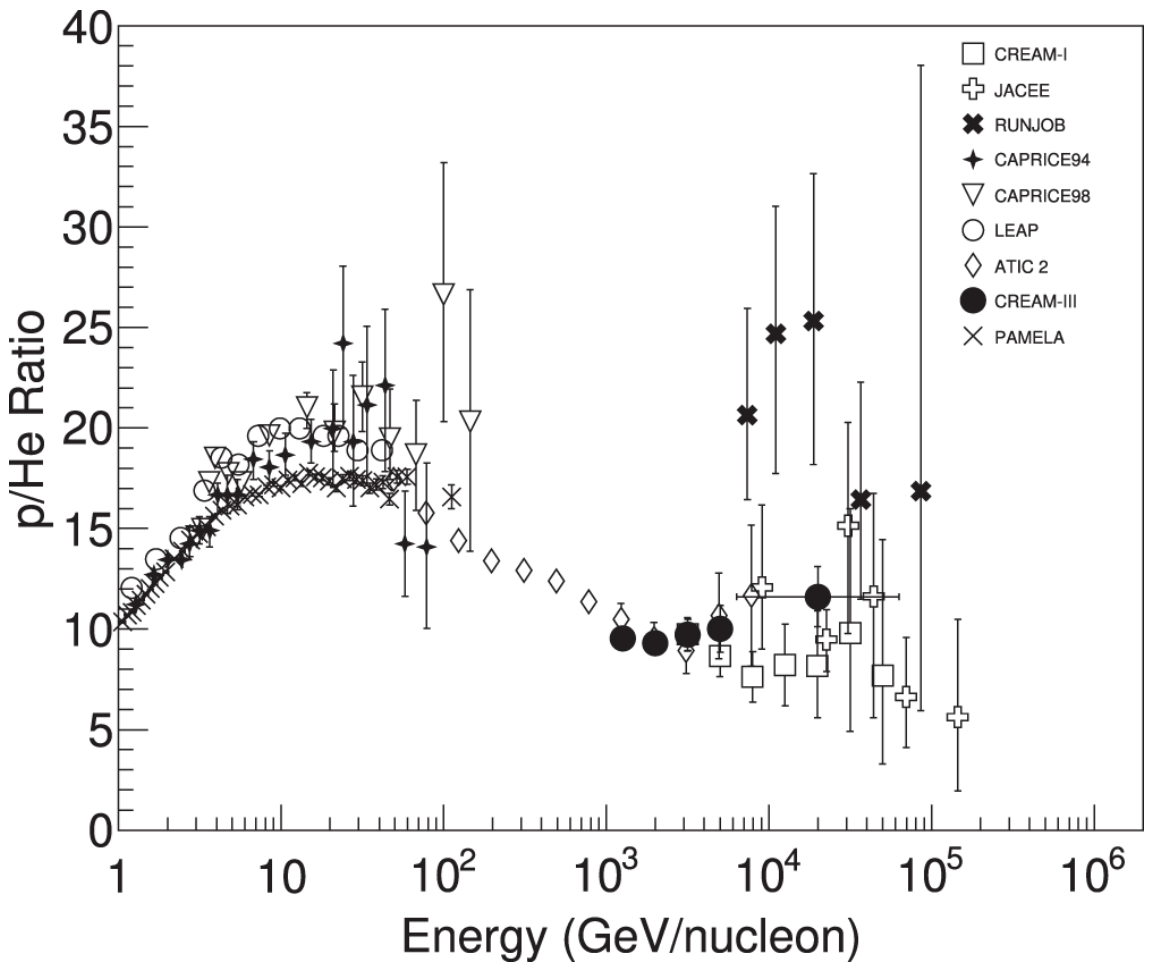

**Fig. 10.** Proton to Helium ratio in cosmic rays as measured by different experiments [62]. (Reprinted with permission from AAS)

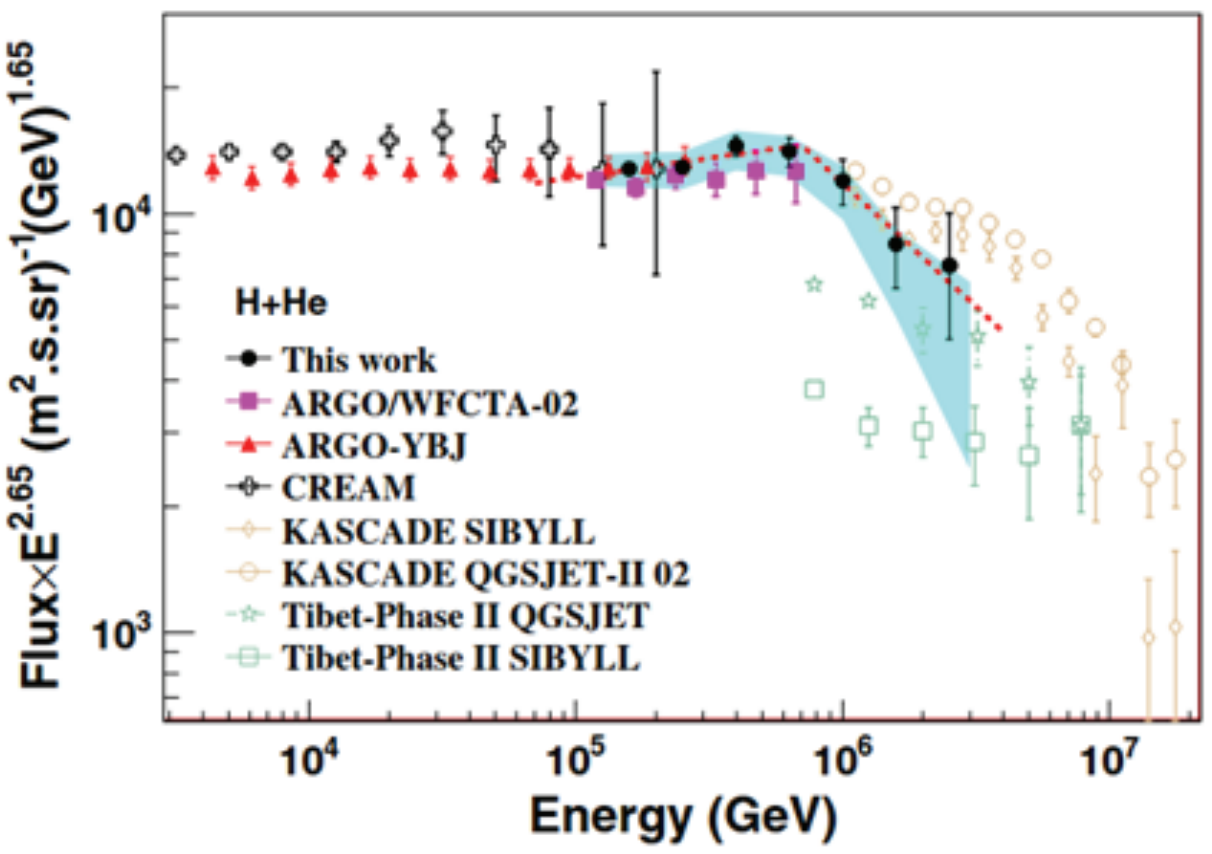

**Fig. 11.** (color online) Energy spectrum of the cosmic-ray light component ($p$+He) measured by different experiments. See [69] for details. (Reprinted with permission from the American Physical Society)

The higher energies above 100 TeV are the domain of indirect methods exploiting the observation of secondary particles in the extensive air showers (EAS) produced by the primary cosmic rays colliding with air nuclei of Earth’s atmosphere. Last generation experiments operating up to $10^{16}-10^{17}$ eV, and measuring different EAS components with good resolution (mainly the electron number $N_e$ and the muon number $N\mu$, or Cherenkov light as in TUNKA [70] and BASJE [71] experiments), have reached the sensitivity to separate into two mass groups (light and heavy) with an analysis technique not critically depending on EAS simulations, or in more mass groups using unfolding techniques heavily based on EAS simulations. Studies at the highest energies by detecting ultra-high-energy cosmic rays (UHECRs) were carried out by many pioneering experiments (Volcano Ranch, Haverah Park, AGASA, Fly’s Eyes, HiRes, see [72] for a short summary). Fly’s Eyes and HiRes explored the observation of the ultraviolet light produced by the nitrogen fluorescence, fully demonstrating the extraordinary potential of this technique. Two giant observatories (Auger in the Southern hemisphere and Telescope Array (TA) in the Northern hemisphere) are steadily observing the ultra-high energy cosmic rays whose origin and nature represent one of the most intriguing mystery of astrophysics. Both detectors combine two techniques with surface arrays of particle detectors overlooked by large field-of-view telescopes allowing for the reconstruction of the shower development in the atmosphere by imaging the ultraviolet fluorescence light produced from atmospheric nitrogen molecules excited by the EAS particles. The shower size, measured for both electrons and muons, and the distribution of the shower maximum in the atmosphere are combined to measure energy and mass composition.

In Fig. 12, where the fluxes are multiplied by $E^{2.7}$, it is shown the cosmic ray all-particle spectrum versus the energy-per-nucleus above 10 TeV, as obtained by air shower experiments. Features as the knee and the ankle are well evident. A less evident feature around $10^{17}$ eV, the so-called second knee, is also visible.

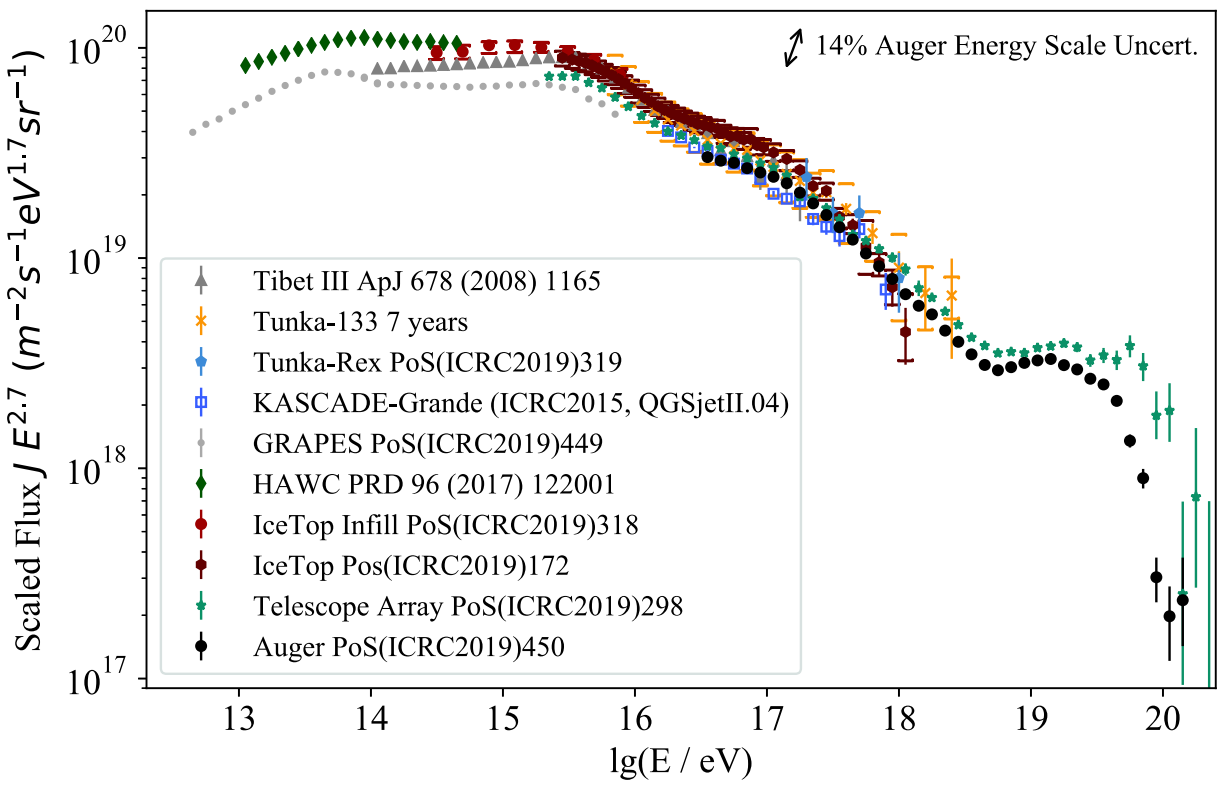

**Fig. 12.** (color online) Energy spectrum of high-energy cosmic rays obtained from air shower measurements. For details see [73].

### F. Primary composition

The origins of these spectral changes are still uncertain. Many astrophysical models assume a dependence of such features on the charge $Z$ of the primary nuclei, mainly related to the end of the acceleration mechanism or escape from the Galaxy volume. These models produce rigidity-dependent break-offs. Other mechanisms associate with new physics processes, e.g. the cannonball model of [74], may generate atomic number $A$-dependent knees.

Unfortunately, still large uncertainties affect the composition measurements preventing any conclusive assessment. According to the KASCADE measurements the knee at $3\times10^{15}$ eV is due to the steepening of the light primary masses, implying an increase of the contribution of the heavy component up to about $8\times10^{16}$ eV, where a knee-like break has been observed by KASCADE-Grande, caused by a steepening in the spectrum of this heavy component. This is the energy where the charge-dependent knee of the iron is expected, if the knee at 3 PeV is due to a decrease of the primary proton flux. However, a number of experiments, in particular those located at high altitude, seems to indicate that the bending of the light-nuclei component is well below the PeV and the knee of the all-particle spectrum can be due to heavier nuclei.

Observational results of the cosmic ray spectra up to $10^{17}$ eV can be summarized as follows:

- Some ground-based air shower arrays can extend their measurements into the energy range covered by direct experiments. The all-particle spectra in the range 10-500 TeV measured by ARGO-YBJ [75], GRAPES-3 [76], Tibet AS [17], and HAWC [77] exhibit a fair agreement within the statistical and systematic uncertainties;

- The proton + Helium spectra obtained by direct (CREAM) and indirect (ARGO-YBJ) measurements are in good agreement in the energy range covered by both experiments [69], showing the reliability of the hadronic interaction models used for the energy calibration of indirect experiments, at least until 200 TeV (Fig. 11);

- All EAS experiments detect a change of slope (known as "knee" of the primary "all-particle" spectrum at about 2-4 PeV. The "all-particle" spectrum above the knee cannot be described by a single power law (KASCADE- Grande [78], IceTop [79], TUNKA-133 [70], TALE [80]), showing a hardening around $10^{16}$ eV and a steepening at around (8-9)$\times10^{16}$ eV;

- The knee has been observed in the main EAS components at different atmospheric depths (i.e. observation height and zenith angle), including electromagnetic (EAS-TOP [81], KASCADE [82], TibetAS [68] among the others), muonic (EAS-TOP [83], KASCADE [82]), and hadronic (KASCADE [84]) components. The results obtained on every single component at different depths are in agreement with the EAS development models

- The experiments located at high altitude - EAS-TOP, CASA-MIA [85], ARGO-YBJ, Tibet-AS, and BASJE show evidence that the knee is due to primaries heavier than protons, or, more in general, than the light component. As a consequence, the composition already before the knee is expected rich of heavy nuclei;

- In addition to the knee-like features in the heavy primary spectrum at about $8\cdot10^{16}$ eV, the KASCADE-Grande experiment claims a flattening of the light component (electron-rich sample) near $10^{17}$ eV [86].

The results obtained by ground-based arrays are still conflicting with each other and the composition around the knee is being questioned. A key-point is the identification of the proton knee, at about 3 PeV as quoted by KASCADE experiment at sea level, or below 1 PeV according to the high altitude experiments. This crucial datum is the cornerstone of the interpretation of the spectral evolution at higher energies. The measurement of the helium/proton ratio, that looks increasing up to 100 TeV and beyond, brings into question the proton dominance of the spectrum. It is thus clear that a firm and precise determination of the proton knee is the key point to further improve our knowledge of the CR composition at $10^{15}$ eV and beyond. The energy range between $10^{16}$ and $10^{18}$ eV, which encompasses the transition region from galactic to extragalactic components is of crucial importance to determine origin and propagation of cosmic rays. Open questions concern the mass composition, presence of sources other than SNRs, the onset of the extragalactic component. The KASCADE-Grande results provide a general picture of the evolution of the mass composition: from lighter mean mass at $10^{16}$ eV to a heavier mean mass at $10^{17}$ eV to lighter again at $10^{18}$ eV. This spectral behavior, observed also by other experiments, is likely suggesting the contribution of different components. For instance, in ref. [87] a three component model is proposed including contributions from ‘regular’ SN explosions, exploding Wolf-Rayet stars, and extragalactic sources.

Different models have been proposed to explain the ankle, in terms of source characteristics or propagation effects. The ankle could reflect the transition between the galactic component at low energies towards a harder extragalactic component at higher energies. In this scenario extragalactic cosmic rays dominate only above $10^{19}$ eV, and one needs some kind of process providing the extension of the galactic component beyond $10^{18}$ eV [88]. Possible solutions imply re-acceleration of the most energetic cosmic rays or cosmic rays due to massive stars exploding into their own wind (e.g. Wolf Rayet stars) or cosmic rays accelerated at a Galactic wind termination shock. In an alternative scenario the ankle is formed as a dip in the extra-galactic proton spectrum, dominating above $10^{18}$ eV, from the energy loss of the protons via $e^+-e^-$ pair production in interactions with the CMB. The cosmic ray chemical composition is expected to be very different at energies from $10^{17}$ to $10^{19}$ eV in these two cases. It is also not clear whether the suppression of the energy spectrum above $4\cdot 10^{19}$ eV is due to the interaction of protons with the CMB via pion production (the Δ-resonance reaction $p+\gamma\rightarrow\Delta^{+}$), the so-called GZK cutoff, or to the photo-disintegration of heavy nuclei like the iron, or simply to the maximum energies achievable at the sources. The GZK cutoff could imply a nearby distribution of sources within 50-100 Mpc. In this case UHECRs are expected to propagate with less deflection by magnetic fields and their arrival directions should be correlated with the direction of powerful extragalactic sources. From the energy spectrum alone it is not possible to distinguish among these different scenarios whose assessment would be made easier by the measurement of the composition and anisotropy. The most recent results on the cosmic ray composition above $10^{17}$ eV come from the Auger experiment. Data point to a composition getting lighter up to about $2.1\times 10^{18}$ eV, in agreement with the KASCADE-Grande findings, and going towards intermediate-heavy masses above, implying a non-negligible fraction of heavier elements at the highest energies [89].

### G. Disentangling energy and mass

All the features observed in the cosmic-ray spectra are naturally related to different types of sources, to acceleration mechanisms, propagation effects, and interactions 'en route' from sources to Earth. Generally, apart from some tension between the flux measured by Auger and Telescope Array above $10^{19}$ eV, the measurements of the energy spectrum from different experiments are in agreement on the form of the major features in the spectrum when taking into account statistical, systematic, and energy scale uncertainties (see Fig. 12). From the spectral results alone, however, it is not possible to understand the causes for the observed features. Indeed, many underlying source, acceleration and propagation models, though to predict similar energy spectra, differ considerably in composition. To gain insight into this issue it is important to measure the composition by disentangling this degeneracy. The traditional approach, consisting in comparing data with the expected results according to different mass composition models (for instance: all-proton, all-iron, mixed 'light' plus 'heavy' as suggested by some theoretical model) may provide partial solutions but fails to give a firm and definitive answer to this question. Measuring the energy spectra of individual components is a challenging task. The mentioned goal could be achieved by selecting, on an event-by-event basis, the single elements or the main mass groups. It is a common believing that an efficient sampling of many shower components, as for instance electrons, muons, Cherenkov and/or fluorescence light and also hadrons and radio output, allowing the simultaneous reconstruction of the lateral and longitudinal shower distributions, could provide a reasonable identification of mass and energy of the primary inducing each observed events. The study of mass-sensitive parameters is carried out in Section II. Taking advantage of the high altitude site and of the hybrid detection approach by a large deployment of different kind of detectors, the LHAASO experiment is expected to be highly sensible to the individual cosmic ray components.

### H. The hadronic interaction models

The interpretation of shower data relies on the output of hadronic Monte Carlo models. High energy hadronic interaction models such as QGSJET or SYBILL are used, while the FLUKA or GHEISHA codes have been adopted to simulate low energy interactions. In recent years interaction models tuned to LHC data, such as EPOS-LHC, have been developed. Indeed, all LHC experiments feature detection capabilities with a wide phase-space coverage, in particular in the forward direction that drives the shower development. EPOS event generator is based on a combination of Gribov-Regge theory and perturbative-QCD, and can be used to simulate $pp$, $pA$, and $AA$ collisions. The use of different models may introduce some residual uncertainty in addition to the experimental systematic and statistical uncertainties. These uncertainties are commonly quoted in summarizing the experimental results. Many results are derived, by unfolding analysis techniques, from the two-dimensional $N_e-N_\mu$ correlation. The electromagnetic component is less dependent on the choice of the hadronic interaction model, while the muon content is more affected. A test of the hadronic interaction models has been carried out by the Auger experiment by comparing the measurement of a number of air shower parameters to the predictions of Monte Carlo simulations [90]. None of the currently used interaction models provides a consistent description of air showers at energies above $10^{17}$ eV, in particular of the muon production profile. A muon deficit in simulations between 38% and 53% is found. A comprehensive study shows that all these models give reasonable description, within a few percent, of experimental data up to a few $10^{16}$ eV. At higher shower energies a growing muon deficit in simulations is observed in all models [91]. No clear explanation of this effect is given so far.

### I. Cosmic ray anisotropy

The measurement of the anisotropy in the arrival direction distribution of cosmic rays is a complementary way to gather information on the origin and propagation of this radiation. Indeed, cosmic ray anisotropy is expected to reflect the source distribution and the structure of the galactic magnetic field. Furthermore, heliosphere and local magnetic fields may represent a significant source of perturbation for low energy particles. The gyro-radius of cosmic rays of $10^4$ GV rigidity (10 TeV protons) is about 0.0037 pc, comparable to the heliosphere width. The anisotropy study is thus an important tool to trace the potential sources and to probe the structure of the magnetic fields through which cosmic rays travel.

In the last few decades, the anisotropy in the cosmic-ray (CR) arrival direction distribution has been observed from tens of GeV to tens of PeV by a number of experiments located in the northern and southern hemispheres, with an intensity of $10^{-4}-10^{-3}$ with respect to the isotropic background. Before the late 1990s, the anisotropy was usually measured as a variation of the cosmic-ray flux over the sidereal day and local solar day based on harmonic analysis in one dimension. With the development of ground-based and underground/under-ice experiments, two-dimensional map of the CR arrival directions distribution was provided [93-99], thanks to the long-term stable observations and large statistic. Detailed morphological studies of the anisotropy structures became then possible. In the two-dimensional anisotropy map at multi-TeV, a few large scale structures have been identified, as shown in Fig. 13. One is the "tail-in" [100], an excess confined in a narrow cone with a half opening angle of 68° from the direction (right ascension $\alpha \approx 90°$, declination $\delta \approx -24°$) coincident with the heliospheric magnetotail direction (right ascension $\alpha \approx 90°$, declination $\delta \approx -29°$). Another feature is the so-called "loss-cone", a broad deficit in the direction of the Galactic North Pole. A small diffuse excess around $\alpha = 310°$ and $\delta = 40°$, corresponding to the Cygnus region direction, has been reported by Tibet AS and ARGO-YBJ experiments, likely due to a large contribution of gamma rays.

The maximum of the dipole anisotropy, about $10^{-3}$, is reached around 10 TeV, above which the amplitude begins to decrease with the phase gradually shifting. As the energy increases a major change in the morphology of the anisotropy is observed in several experiments in the

Northern hemisphere as well as in the Southern hemisphere (EAS-TOP, IceCube, Tibet AS [101], and ARGO-YBJ [102]). Data collected by Tibet AS and ARGO-YBJ over 7 and 5 years, respectively, of data taking show that at energies above 50 TeV the 'tail-in' and the 'loss-cone' features gradually fade away, while above 100 TeV a sudden change of the phase is observed, and the amplitude begins to increase with energy with an evident new pattern. At energies above 150 TeV the 'tail-in' and 'loss-cone' features completely disappear and the 2D map is dominated by an excess in the interval $\alpha = 20°$-300° and a deficit around $\alpha = 0°$-100°. The direction of the new excess is very close to the direction of the galactic center ($\alpha = 268.4°$), suggesting this region as a possible source of cosmic rays. Similar results are obtained by the IceCube collaboration in the Southern hemisphere up to 5 PeV [103]. The 2D maps of relative intensity for 8 median energies, from 4 to 520 TeV, as measured by ARGO-YBJ, are shown in Fig. 14.

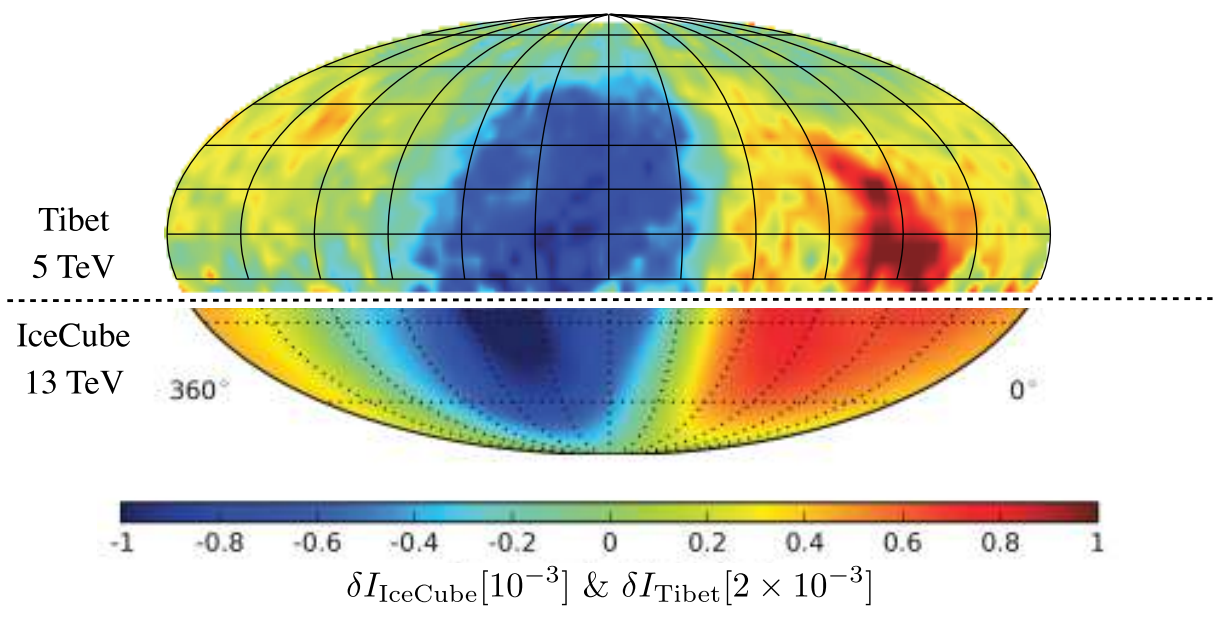

**Fig. 13.** (color online) Combined cosmic ray anisotropy of Tibet-AS and IceCube experiments in the equatorial coordinate system [92].

The cosmic ray anisotropy is not well described by a simple dipole moment which is, however, commonly used to illustrate the energy dependence of the phase and strength of the anisotropy. The amplitude and phase of the first harmonic measured by many experiments, obtained by a projection of the 2D map onto the right ascension axis, is shown in Fig. 15.

A time variation of the anisotropy in association with the 11 year solar cycle could be evidence for a heliospheric influence. Milagro [95] reported a steady increase of the 'loss-cone' amplitude at a mean energy of 6 TeV over a period of seven years as the solar activity varied from near maximum to minimum. This challenging result is not confirmed by Tibet AS [104], ARGO-YBJ [102], and IceCube [103] experiments. A comparison of data collected by the experiments located in the Northern hemisphere is shown in Fig. 16. This result implies that the anisotropy of multi-TeV cosmic rays essentially reflects the structure of the interstellar magnetic field.

By filtering out the large scale structure from the CR anisotropy, several localized regions of significant cosmic-ray excess have been observed. The Milagro cosmic-ray sky map [105] first indicated two excess regions at ($\alpha \sim 69.4°$, $\delta \sim 13°$) and ($\alpha \sim 130°$, $15° < \delta < 50°$), the relative intensity of them being $6\times10^{-4}$, and $4\times10^{-4}$, respectively. Both regions were confirmed by Tibet AS$\gamma$ and ARGO-YBJ experiments. The ARGO-YBJ experiment [106] presented evidence for two additional excess regions at ($\alpha \sim 240°$, $15° < \delta < 55°$) and around ($\alpha = 210°$, $\delta = 30°$) with a maximum relative intensity of $2.3\times10^{-4}$ and $1.6\times10^{-4}$, respectively. The last excess region has been confirmed by HAWC [107]. In the southern sky, an excess localized at ($\alpha = 122.4°$, $\delta = -47.4°$) were suggested by IceCube experiment [108].

The study of the anisotropy at high energies above $10^{16}$ eV is a very difficult and challenging task because of the low CR rate. Thus a very long duration of data taking is needed to reach a good statistical sensitivity, implying stability and a strict control of the detector perform-ance. For instance, the Auger results come from the analysis of 14 years data collected with the surface array. Anisotropies are expected to be stronger at the higher energies since both source distance and magnetic deflections are reduced. The amplitude and the phase of the first harmonic measured by KASCADE-Grande, IceCube, IceTop, and Auger at energies above $10^{15}$ eV are shown in Fig. 17.

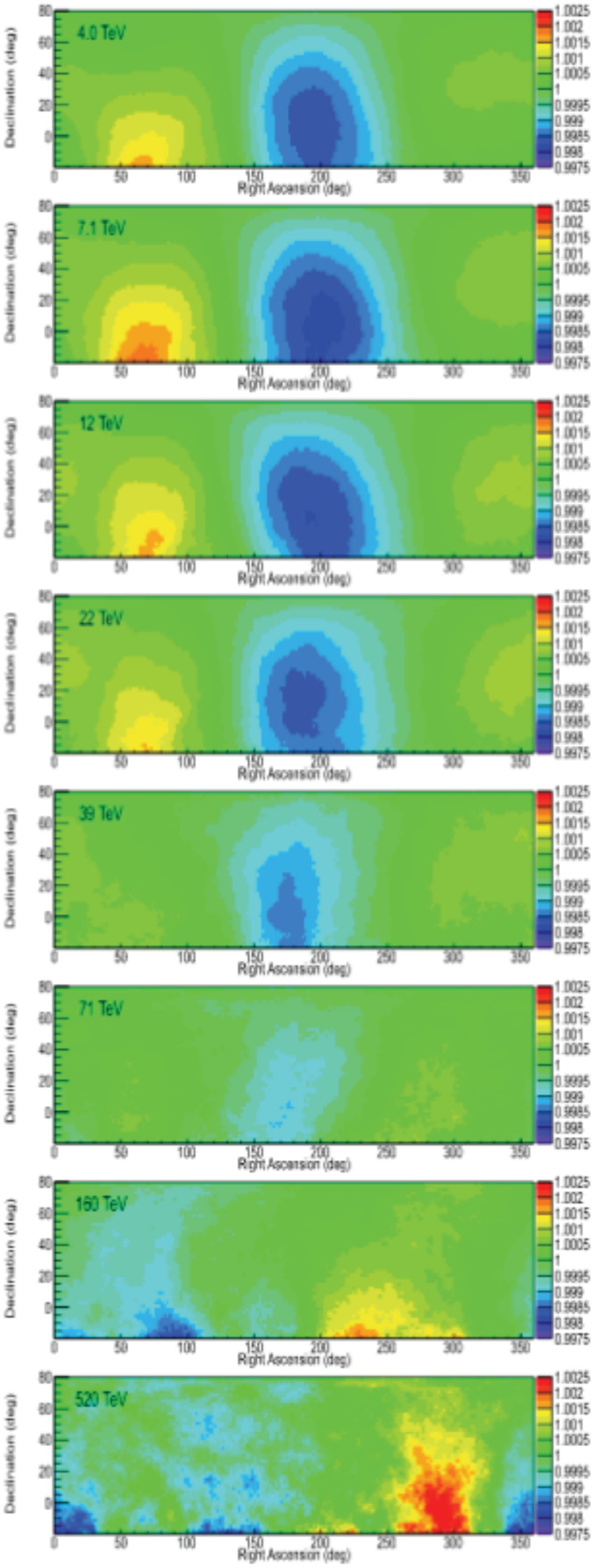

**Fig. 14.** (color online) The 2D maps of the large-scale anisotropy measured by ARGO-YBJ for eight energy bins of median energies (from top to bottom) 4, 7, 12, 22, 39, 71, 160, 520 TeV [102]. (Reprinted with permission from AAS)

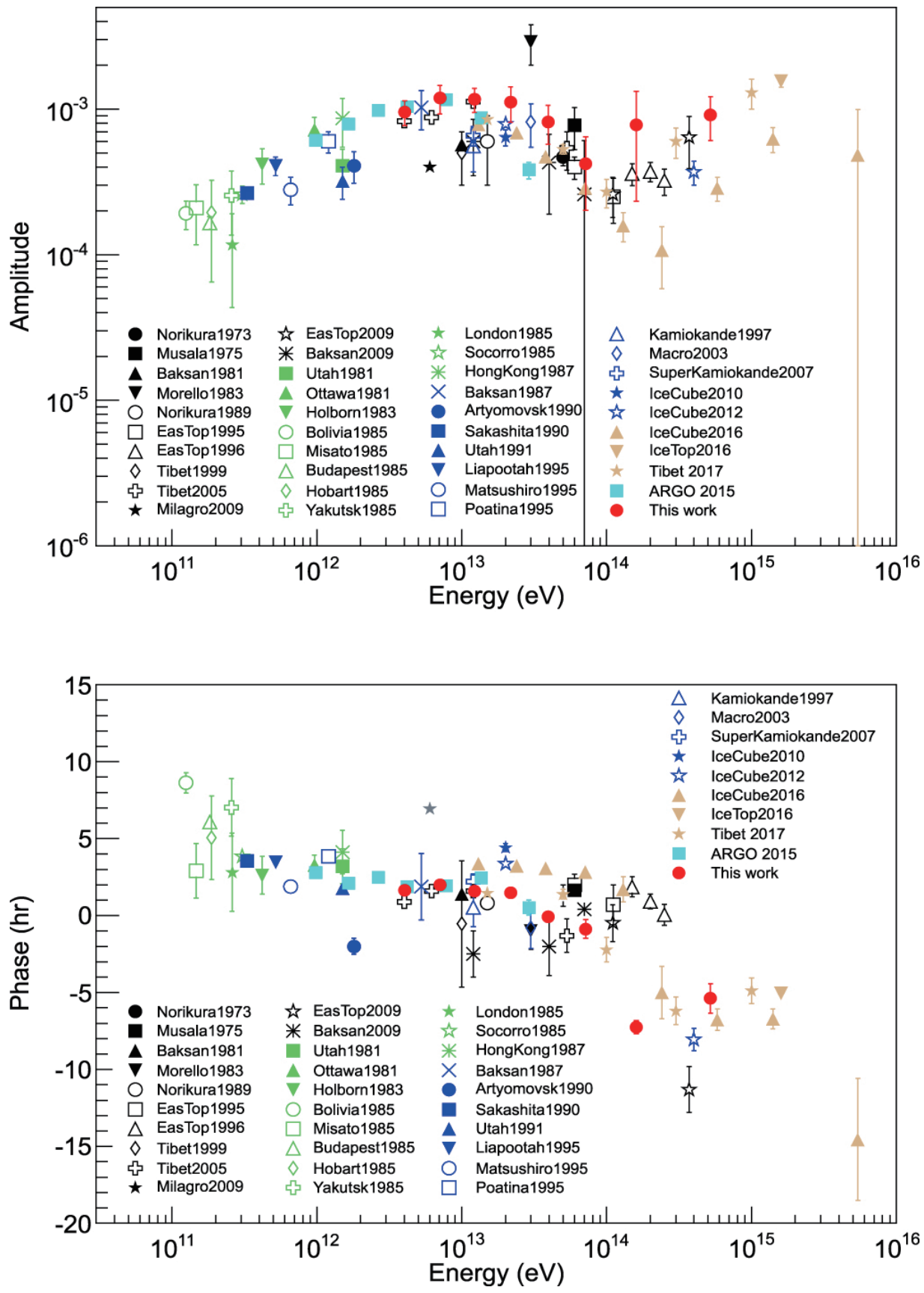

**Fig. 15.** (color online) The amplitude (top panel) and the phase (bottom panel) of the first harmonic of the sidereal anisotropy as a function of the cosmic-ray energy measured by many experiments. For details and references see [102]. (Reprinted with permission from AAS)

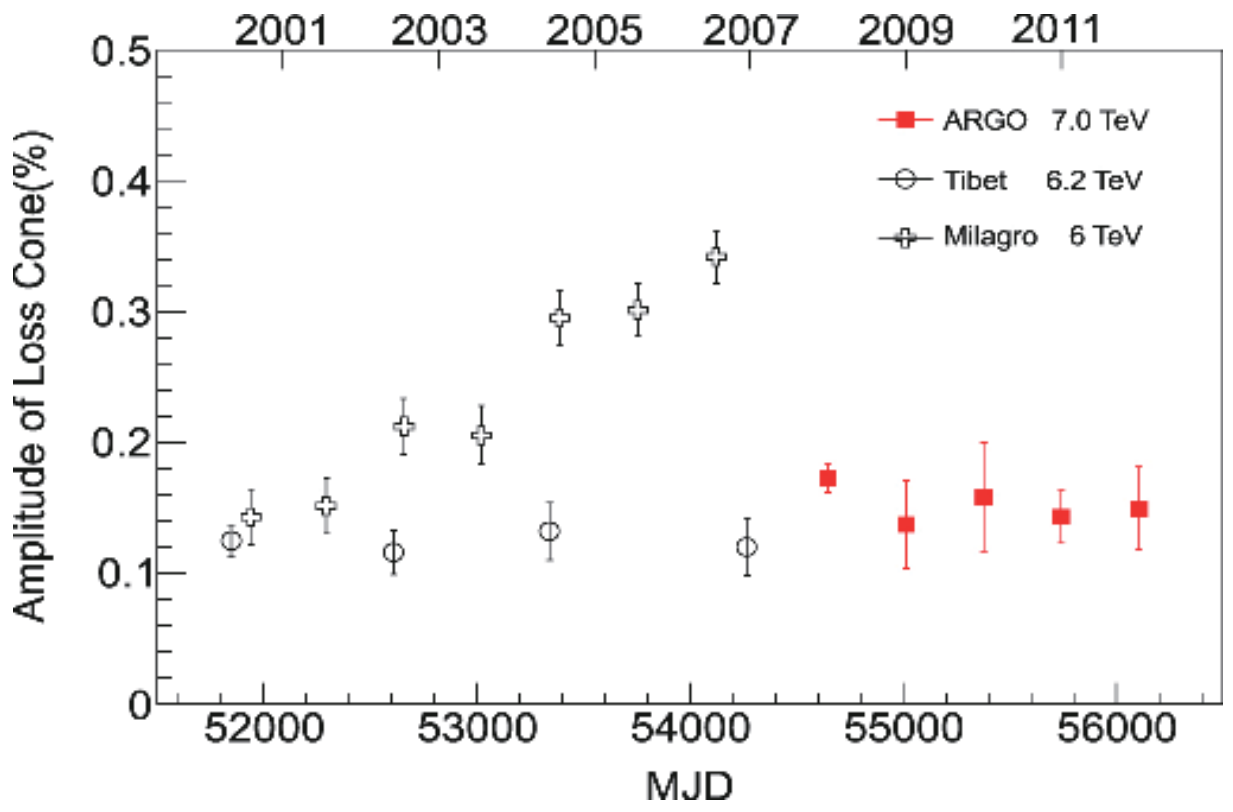

**Fig. 16.** (color online) Temporal variation of the 'loss-cone' amplitude measured by ARGO-YBJ, Tibet-AS, and Milagro experiments [102]. (Reprinted with permission from AAS)

The measured amplitudes increase from $10^{-3}$ to $10^{-2}$ up to an energy of about $8\times10^{18}$. Below $10^{18}$ eV all the measured phases point near the Galactic center ($\alpha\approx$ -94° in this plot). An amplitude of about $6.6\times10^{-2}$ and pointing $\approx125°$ away from the Galactic center is instead observed by Auger above $8\times10^{18}$ eV. The 2D map of the anisotropy is shown in Fig. 18. Taken together, these results suggest a transition from a predominantly Galactic to an extra-galactic origin of cosmic rays somewhere between 1 and a few EeV.

A number of explanations for the CR anisotropy have been discussed, pertaining to uneven distribution of CR sources in the galaxy, propagation effects, the galactic magnetic field and the local magnetic field. The global anisotropy changes with energy, however, are intriguing. According to numerical studies of CR propagation in a scenario of homogeneous and isotropic diffusion in the galaxy, the small magnitude and the energy dependence of the anisotropy amplitude can be explained with nearby and recent SN explosions [111-113], mixed mass composition, the smoothing effect of the galactic halo and the position of the Sun on the inner edge of the Orion arm [114].

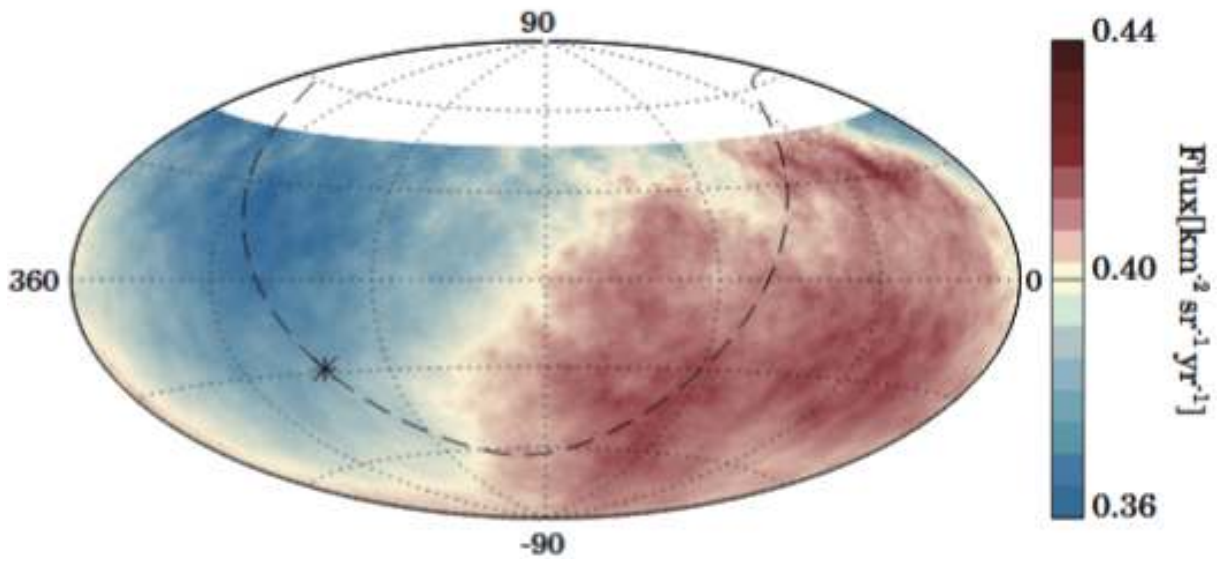

**Fig. 18.** (color online) Smoothed cosmic ray flux for $E$ >8 EeV measured by Auger. The dashed line and the star indicate the Galactic plane and the center, respectively. The direction of the dipole lies 125° from the Galactic center, disfavoring a galactic origin for these cosmic rays [109].

The local origin model of the anisotropies was proposed by Amenomori *et al.* (2010) [115], Zhang *et al.* (2014) [116] and Schwadron *et al.* (2014) [117]. In their model, the global anisotropy may be generated by galactic cosmic rays interacting with the magnetic field in the local interstellar space of scale ~ 2 pc surrounding the heliosphere. In addition, Qu *et al.* (2012) [118] proposed a global galactic CR stream model by extending the observed CR anisotropy picture from the solar system to the whole galaxy, connecting GCR streaming and the structure of the galactic halo magnetic field. On the other hand, a dipolar anisotropy is expected due to the motion of the observer relative to the CR plasma by Compton & Getting [119]. Such CR anisotropy due to the solar system rotation around the Galactic center at a speed of ~ 220 km s$^{-1}$ was excluded [93], implying that GCRs corotate with the local galactic magnetic field environment. The models for mid-scale anisotropies have been also discussed since 2008. The excess could be due to the magnetic mirror effect on CRs from a local source [120], or could be related to the Geminga pulsar [121]. Cosmic ray acceleration from magnetic reconnection in the magnetotail has been proposed as a possible source [122]. The reconstruction errors in the presence of a large angular gradient in the cosmic ray flux could also cause small-scale features [123]. Recently it has been argued that small-scale structures could be due to cosmic ray scattering in local turbulent magnetic fields [124]. Models proposed for explaining the small-scale anisotropies are reviewed in [92].

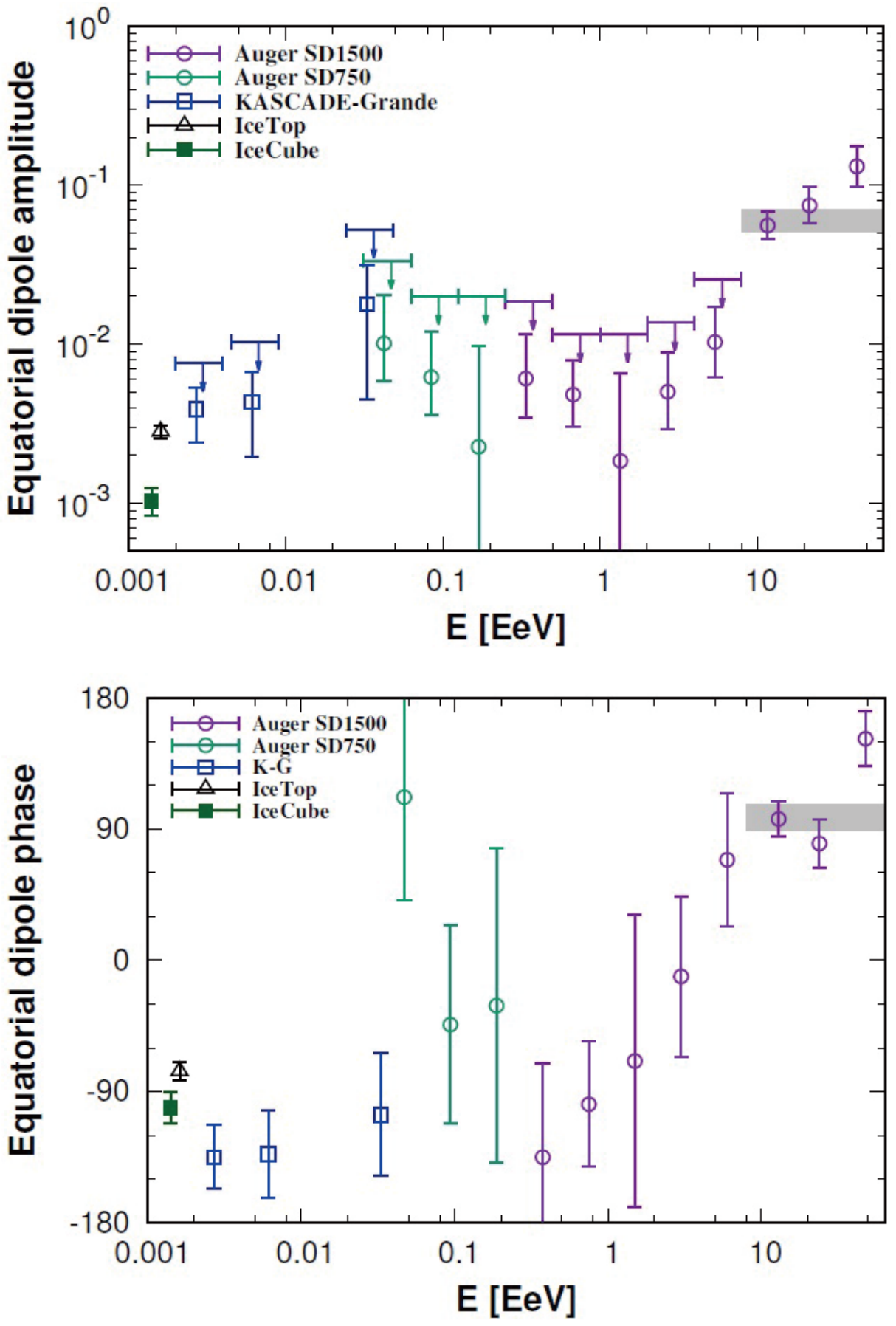

**Fig. 17.** (color online) Amplitude (top) and phase (bottom) of the equatorial dipole of the cosmic-ray large-scale anisotropy at ultra-high energies [109].

### J. Cosmic ray electrons, positrons and antiprotons

The CR electrons (the total electrons + positrons) and antiprotons are less abundant species in the CR family. Nevertheless, they are crucial for studies of various fundamental problems in physics and astrophysics, includ-ing the propagation of CRs, the search for anti-matter Universe, and the detection of particle dark matter (DM). Usually these particles need to be detected in space, by e.g. magnetic spectrometers, in order to be robustly identified from the very high proton and nuclei background.

The extension of the measurements to high energies by space detectors is difficult due to the limited counting statistics of high energy particles. Therefore, the ground-based experiments are expected to play important roles in the study of high energy ($\gtrsim$ TeV) electrons and antiprotons. Due to improving separation power between hadronic and electro-magnetic showers, high energy electrons can be detected by the ground-based experiments. Using the Earth's magnetic field as a huge magnetic spectrometer, the antiproton-to-proton ratio can also be measured via the observations of the moon shadow by air shower arrays (Tibet AS, ARGO-YBJ, HAWC).

### 1. *Electrons and positrons*

The propagation of electrons in the Milky Way is very different from that of nuclei. The radiative cooling of electrons, with cooling time scales proportional to $E^{-1}$, becomes very important and dominates the propagation processes at high energies. Therefore high energy electrons can only retain for a limited time scale and propagate within a limited distance range. The detection of high energy electrons and the precise measurement of their spectral features provide unique diagnostic of local sources of CR electrons.

The electron spectra have been measured to TeV energies with very high precision by space- and balloon-borne detectors in the past few years [3, 125-128]. The ground-based Imaging Cherenkov Telescope Arrays (IACTs) extended such measurements to about 5 TeV with relatively large systematic uncertainties [14, 129, 130]. Figure 19 summarizes the current measurements of the total $e^+ + e^-$ spectra. Data up to a few TeV have been provided by ISS-CALET [64] and the DAMPE instrument on board a satellite [12]. DAMPE is operating with high energy resolution and effective rejection of the hadronic cosmic-ray background. The observed energy spectrum is shown in the right panel. A spectral hardening near 50 GeV and a spectral break near 0.9 TeV have been reported after 530 days of operation. In the energy range of 55 GeV to 2.36 TeV the spectrum is well represented by a smoothly broken power-law model shown as a dashed line in Fig. 19.

The CALET results exhibit a lower flux than those of DAMPE from 300 GeV up to near 1 TeV, likely indicating the presence of unknown systematic effects. The energy spectrum measured by CALET above 1 TeV suggests a flux suppression consistent, within the errors, with the DAMPE results [131]. Thus both experiments confirm the change of the spectrum slope above 1 TeV firstly

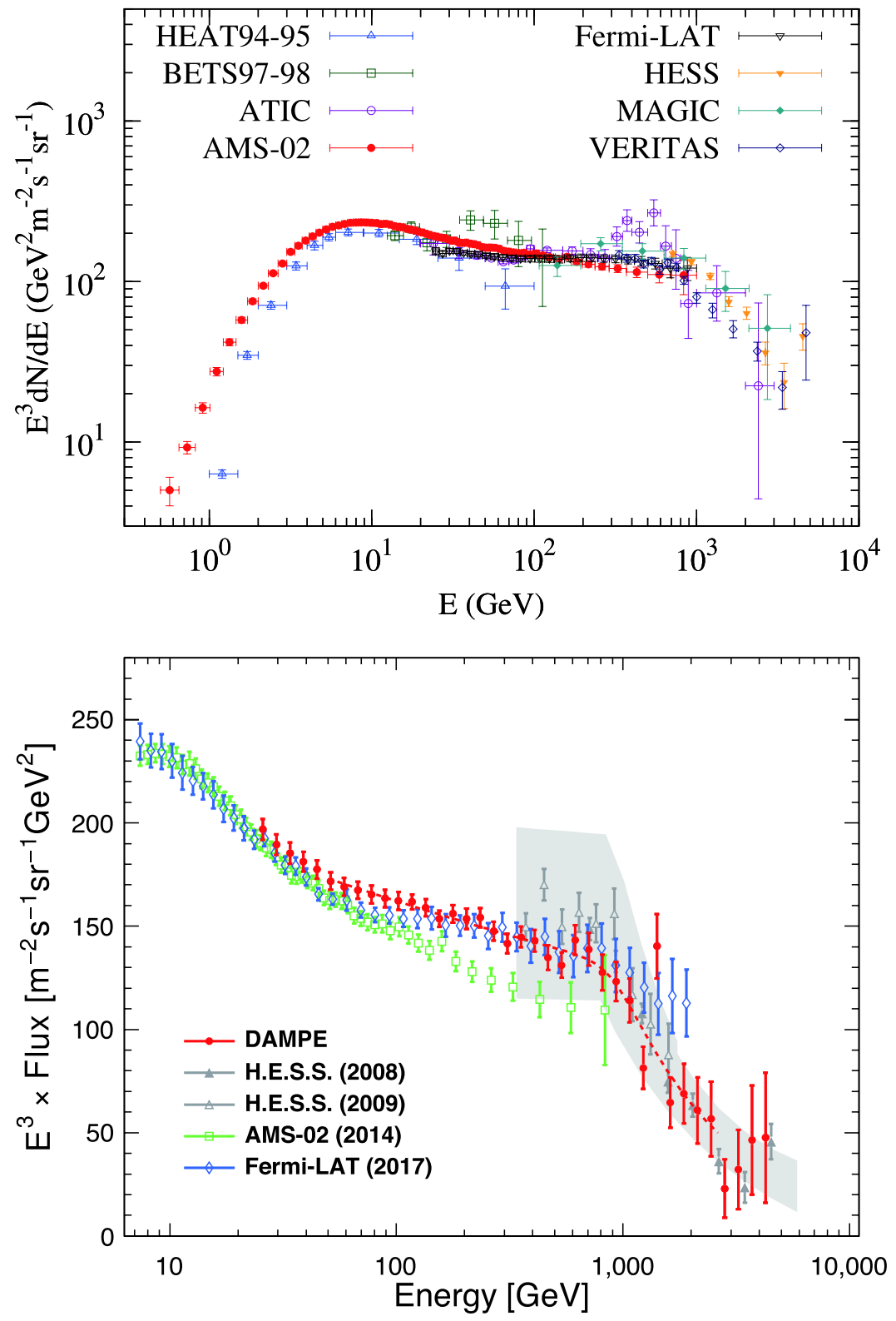

**Fig. 19.** (color online) Top: Fluxes (weighted by $E^3$) of the total $e^+ + e^-$ measured by balloon- and space-borne experiments [3, 125-128], and the ground-based IACTs [14, 129, 130]. The bottom panel shows the electron + positron spectrum measured by DAMPE [12]. The grey band represents the systematic uncertainty affecting the HESS data.

observed with large systematic uncertainties by the ground-based IACTs.

Theoretically, CR electrons include the primary component which may be accelerated simultaneously with nuclei, by e.g., supernova remnants (SNRs) and many other types of CR sources, and the secondary component from inelastic collisions between CR nuclei (mostly protons and Helium) and the interstellar medium (ISM). The secondary contribution is only a small fraction (~ 10%) of the total electrons. In addition, there might be leptonic sources which produce dominantly electron/positron pairs. Candidate sources include pulsars [132] and sources with $e^+e^-$ pair production via $\gamma\gamma$ interaction [133] or photo-nuclei interaction [134]. The existence of such primary $e^+e^-$ pair sources has been suggested by the observations of remarkable excesses of the positron fraction above ~10 GeV [135-138]. The simultaneous fitting to the positron fraction and total electron spectra further indicates the existence of high energy spectral hardenings of the primary electrons [139-141]. Such a result implies nearby sources of CR electrons [142, 143].

The annihilation or decay of DM particles can produce electrons and positrons, which may also explain the electron/positron excesses [144-146]. However, observa-tions of $\gamma$-rays strongly constrain such models, leaving only a class of decaying DM models with leptonic decaying products [147-150]. The precise measurement by CALET and DAMPE of the electron/positron spectrum up to 10 TeV is expected to narrow down the parameter space of all proposed models.

### 2. *Antiproton-to-proton ratio*

Antiprotons are important probe of the propagation of CRs and the search for DM particles. It is generally accepted that galactic antiprotons are produced via the inelastic collisions between high energy CRs and the ISM during their propagation in the Galaxy. Such a production mechanism is similar to the process of nucleus fragmentation which produces secondary nuclei (such as Lithium, Beryllium, and Boron) in CRs.

Figure 20 shows the current measurements of the $\bar{p}/p$ flux ratio by balloon- and space-borne experiments from 0.1 to 500 GeV [7, 9, 151-153]. The $\bar{p}/p$ ratio shows an increase from the lowest energy up to ~10 GeV, which reflects the antiproton production threshold of the $pp$ collision. The high energy behavior of the $\bar{p}/p$ ratio is expected to follow a power-law declination due to the energy-dependent diffusion of CRs. The current data from AMS-02 is consistent with either a flat or a shallow decline at high energies [7]. It is still unclear whether there are antiproton excesses for $E \gtrsim 100$ GeV. Based on a semi-analytical model of CR propagation [158], it has been shown that the observed $\bar{p}/p$ ratio is well consistent with the model inferred from the Boron-to-Carbon ratio [159, 160], leaving limited room for contribution from the particle DM annihilation[1)].

The extension of the measurements to higher energies is very difficult. For the direct detection in space, the event statistics is very challenging due to the extremely low fluxes of antiprotons. The ground-based experiments are, however, lack of effective charge-sign separation capability. A smart idea is to use the deflection of the moon shadow by the Earth's magnetic field to separate antiprotons and protons [165]. Analyses from a few experiments have been done, and no significant deviation of the moon shadow's deflection from Monte Carlo expectation based on pure positive CRs was found [154-156]. The corresponding upper limits of the $\bar{p}/p$ ratio were derived, as shown by arrows in Fig. 20. These upper limits are $10^2-10^3$ times higher than the direct measurements at lower energies and can only constrain few exotic theoretical models such as the antimatter model [166].

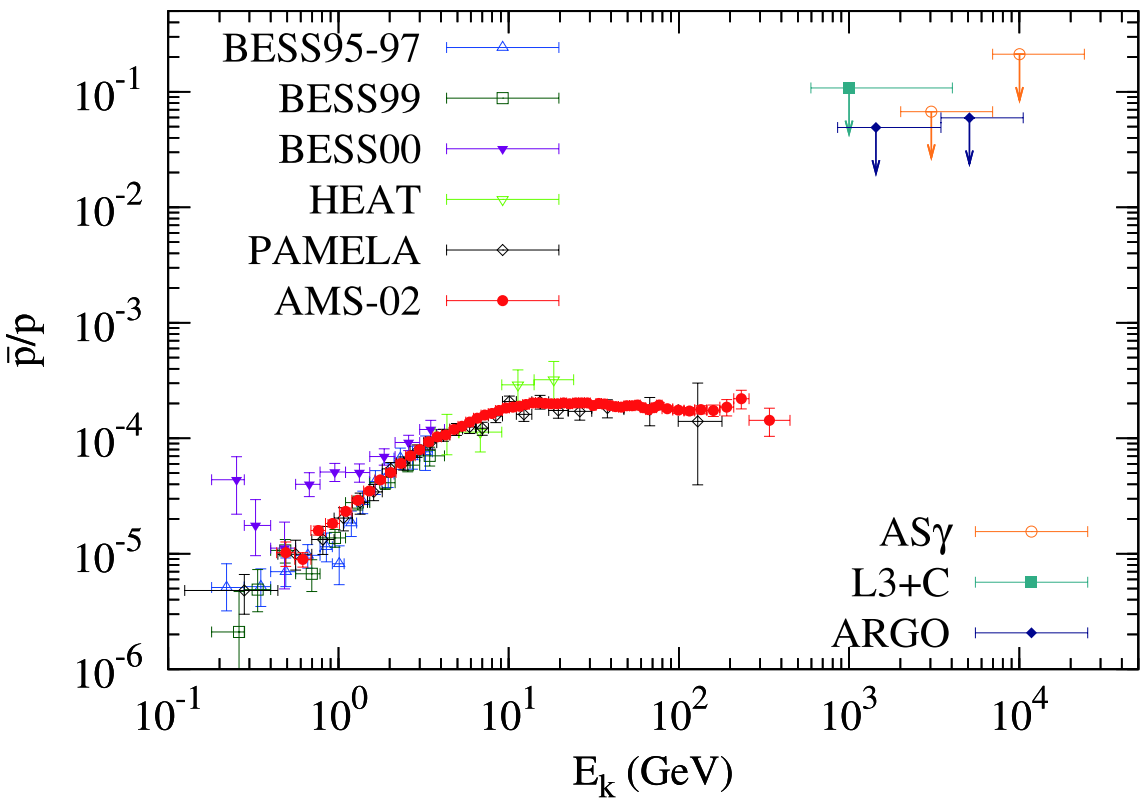

**Fig. 20.** (color online) The antiproton to proton flux ratio measured by different experiments [7, 9, 151-156]. Comparable upper limits in the TeV region have been obtained by the HAWC experiment [157].

### K. The Multi-messenger approach

Since CRs are electrically charged, their paths are bent by the cosmic magnetic fields. As a consequence, they do not point back to their sources. In contrast to cosmic rays, gamma rays produced at these sources propagate along a straight line.

They are produced in the interaction of accelerated particles, cosmic rays or electrons, with ambient matter or radiation fields, thus TeV gamma rays do trace the emission sites with the acceleration sites nearby. Since the photon production processes, hadronic interaction or inverse-Compton, are well known, TeV gamma-ray observations provide crucial information about the intensity and the spectrum of the accelerators. After the pioneering observation of the Whipple Telescope that detected in 1989 the first source, the Crab Nebula, the gamma-ray astronomy above a few tens of GeV by ground-based detectors developed rapidly thanks to the IACT techniques. Nowadays more than 200 TeV sources are reported in the TeV-cat [167] catalog. Wide-angle devices like MILAGRO, ARGO, HAWC may complement the observations by IACT due to their better survey capabilities and better sensitivity to extended sources. The status of this field and the LHAASO performance in the gamma-ray observations are discussed in Chapter 2.

In the context of the search of cosmic-ray sources, high energy neutrinos play a crucial role being able to provide an uncontroversial proof of their hadronic character; moreover, they can reach us from cosmic regions which are opaque to other type of radiation, including high-energy gamma rays. High energy neutrinos, as well as photons, are expected to be produced by UHECRs interacting with extragalactic background photons during intergalactic propagation. An attractive plot, see Fig. 21, has been proposed by the IceCube collaboration [168]. The energy flux of diffuse extragalactic gamma rays, ultra-high energy cosmic rays, and high energy neutrinos have similar spectra despite their disparate energy re-

1) Note, however, that slightly different conclusions about the low energy ($\lesssim$ tens of GeV) part were obtained with more detailed numerical models [161-164].

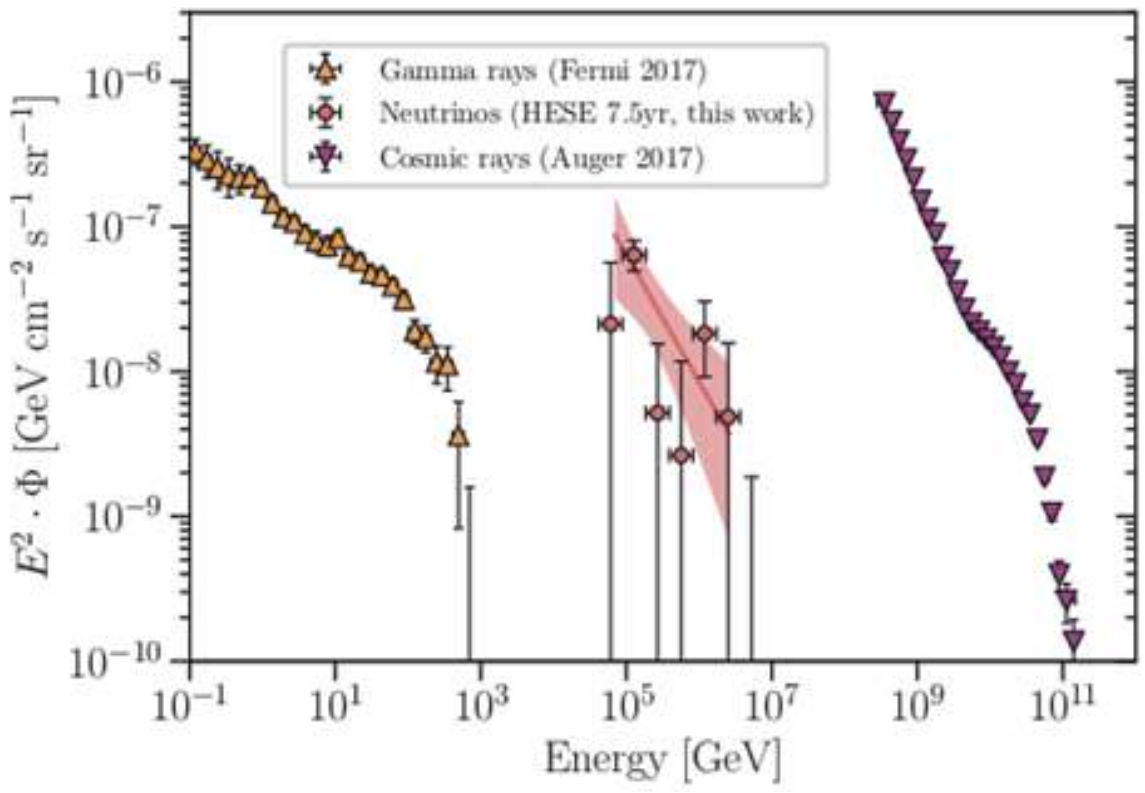

**Fig. 21.** (color online) Energy flux of extragalactic diffuse gamma rays, high-energy astrophysical neutrinos, and UHECRs [168]. (Reprinted with permission from the American Physical Society)

gimes, implying a common origin of these radiations. The expected neutrino fluxes from sources are of the same order of magnitude as the gamma-ray fluxes, however the km$^3$-scale neutrino detectors currently in operation have a sensitivity far lower than the LHAASO sensitivity ($\approx 10^{-14}$ erg cm$^{-2}$ s$^{-1}$) in the most relevant, 100 TeV to 1 PeV, energy range, envisaging at this time only a marginal contribution to the LHAASO search for cosmic-ray sources. A dedicated study of the Multi-messanger physics with LHAASO is given in Chapter 6.

### L. Future prospects

Cosmic rays are the most energetic particles in nature and their studies address fundamental issues as the nature of the non-thermal Universe and the violent cosmic processes there at work. In 1938 Pierre Auger and his colleagues detected extensive air showers with two Geiger-Muller counters operated in coincidence 300 m apart. The energy of the primary particle inducing these events was estimated to be about $10^{15}$ eV. In 1989 the Whipple telescope discovered TeV gamma rays from the Crab Nebula. That represents the first direct evidence of a cosmic source accelerating particles at multi-TeV energies. Since then a wealth of information about putative sources of high energy cosmic rays has been produced by the observations of a remarkable suite of experiments. The emerging general picture evidences that cosmic rays are a varying mixture of protons and nuclei of galactic origin up the energies below $10^{17}$ eV, the second knee, and that a transition from the galactic to an extragalactic component takes place in the energy region between the second knee and the ankle. The energy range above a few $10^{18}$ eV is likely dominated by extragalactic cosmic rays. These findings point out the existence of many types of candidate sources as anticipated by theoretical models. However, *"neither the most energetic galactic nor the extragalactic sources of cosmic rays have been discovered, yet... "* [73]. To shed light on this fundamental question, information about the spectrum and anisotropy of each cosmic ray component is of paramount importance. Indeed, understanding the cosmic ray origin and propagation is made difficult by the poor knowledge of the elemental composition of the radiation as a function of the energy. Direct measurements by above-the-atmosphere detectors may provide the energy spectrum of each cosmic ray elements. Probing these spectra via direct detection becomes a challenge beyond 100 TeV, due to the limited detector exposure. Present ground-based air shower arrays, which exploit the atmosphere as target and a calorimeter, do not suffer of this limitation, but have limited sensitivity to the charge of the primaries. Separating as much as possible the mass groups is thus a mandatory goal of new generation experiments.

In summary, high energy spectral features such as breaks, bumps or cut-offs of the all particle spectrum may reflect the superposition of many contributions that could be solved only by comparing the proposed models to the measured energy dependence of the single components or mass groups. In this respect, the most fundamental topics concern

1. the measurement of the single component, or mass groups, energy spectrum through the knee at 3 PeV, allowing the determination of the energy of the knees of each element or mass groups;

2. the detailed study of the energy range $10^{16}-10^{18}$ eV, where the shape of the mean of the logarithmic mass $\langle \ln A\rangle$ measured by various experiments (see Fig. 22) cannot be explained by a single galactic component with ri-

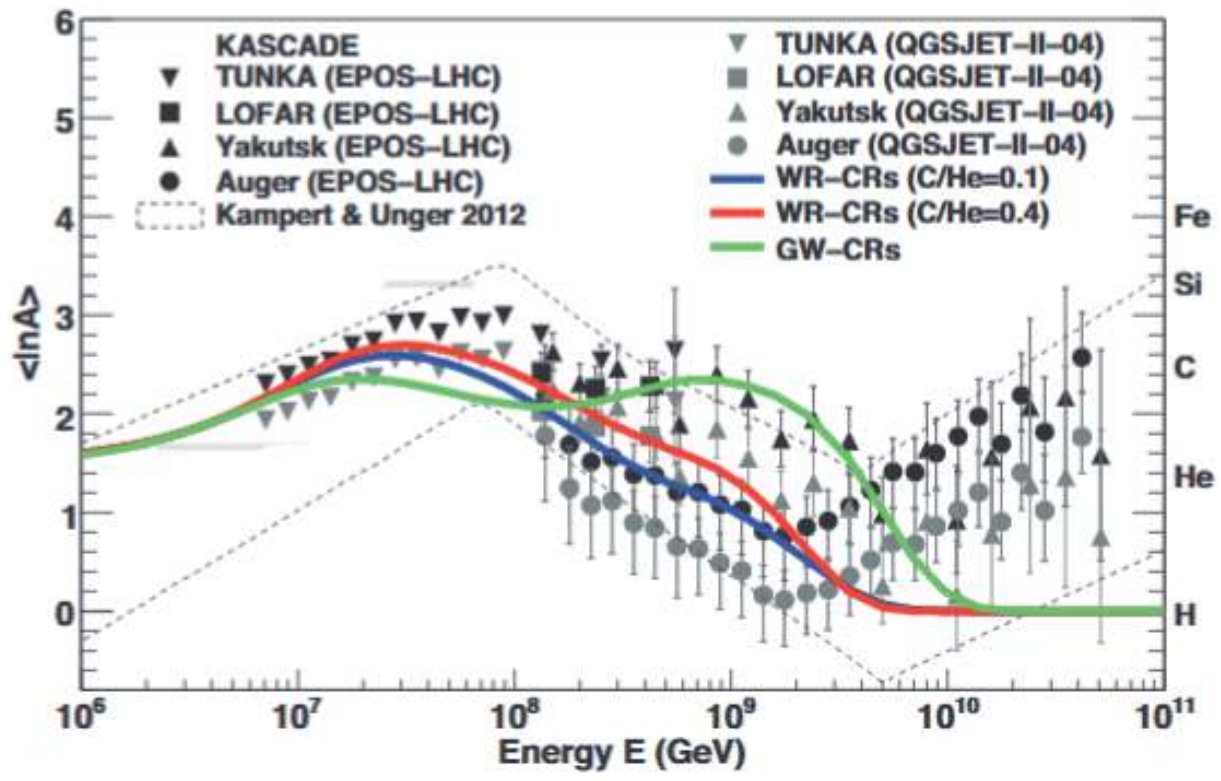

**Fig. 22.** (color online) Mean of the logarithmic mass $<\ln A>$ as measured by various experiments interpreted with two hadronic interaction models (EPOS-LHC and QGSJET-II-04). Predictions are shown for three different models of the additional Galactic component: cosmic rays from Wolf-Rayet stars (C/He = 0.1 and C/He = 0.4) and cosmic rays being re-accelerated by the galactic wind. For details and references see [87]. (Reproduced with the permission @ ESO)

gidity-dependent energy cut-offs in the individual spectra of different elements;

3. the measurement of the anisotropy for different mass groups.

How LHAASO may accomplish this task is discussed in the second part of this Chapter. In the search for cosmic sources LHAASO will greatly benefit of its remarkable sensitivity in gamma astronomy. Indeed, magnetic fields prevent the direct identification of the sources by simply detecting cosmic rays. Observation of high energy gamma rays produced by the interaction of PeV protons or nuclei is widely recognized as a very powerful tool to identify the cosmic ray accelerators. The need of ruling out their possible leptonic origin requires a proper sensitivity to survey the sky at above 100 TeV photon energies. In this respect, LHAASO's sensitivity and wide field of view may provide a unique discovery potential, as discussed in Chapter 2.

By combining the study of the cosmic ray elemental composition and the observation of very high energy gamma rays, LHAASO will become one of the most powerful instrument to search for PeVatrons.

Important contributions to this search may come from a correlated study with low frequency components of the electromagnetic spectrum. Radio data at the GHz scale are produced at the source sites as well as by the interaction of cosmic rays with the ISM. Results from LHAASO gamma-ray observations in the multi-TeV range may be usefully combined with data from the next generation SKA observatory to deepen the problems related to the cosmic ray acceleration and propagation in the Galaxy. An overview of this topic is hereinafter reported.

## M. Multi-wavelength studies of cosmic ray acceleration and transport in the Galaxy

Multiwavelength observations can be used to study the Galactic cosmic rays' origin, propagation, and distribution in the Galaxy. With the next generation telescopes, such as LHAASO and SKA, we may make one giant leap for understanding GCRs by finding PeVatrons, measuring the magnetic field amplification, examining the energy conversion rate and nonlinear effect, increasing evidence for TeV CRs diffusive propagation and studying their distribution in our Galaxy.

### 1. *Background*

Supernova remnants (SNRs) are known as the best origination candidate for GCRs (other candidates include pulsar wind nebular, X-ray binaries, Galactic center, superbubbles, and so on). Multi-wavelength observations have provided lots of evidence supporting SNRs as the origin of GCRs: (1) Radio observations display bright filaments and twisty structures of SNRs which are predicted by DSA. (2) The average spectral index, $\alpha$, of SNRs is about 0.5 ($S_\nu \propto \nu^{-\alpha}$) indicating a particle energy index, $\gamma$, of about 2 ($\gamma$ = 1+2$\alpha$). (3) The magnetic fields derived from observing OH 1720 MHz masers in the SNRs shocked regions are significantly amplified to magnitude of mG. (4) X-ray observations detect synchrotron emissions from young SNRs showing that electrons have been accelerated up to 100 TeV and the magnetic fields are amplified to 100-600 μG. (5) Molecular spectral line observations detect enhanced ionization rate surrounding SNRs implying efficient low energy CR acceleration. (6) Many SNRs interacting with molecular clouds or neutral hydrogen clouds, which are identified by infrared, centimeter, millimeter and sub-millimeter observations, are also GeV and/or TeV emitting objects. (7) The discovery of two components of optical H$\alpha$ line supports the existence of CRs induced shock precursor. (8) *Fermi* satellite has detected the pion bump feature from SNRs IC443 and W44 giving the first direct evidence that both SNRs accelerate CRs to GeV.

A combination of DSA and CRs propagation in our Galaxy is usually referred as the SNR paradigm. The theoretical and observational works mentioned above are in favor of this paradigm. However, many questions in the paradigm are still open. Multi-wavelength observations from next generation telescopes especially LHAASO and SKA should play a key role in solving the problems in DSA theory, CRs diffusive propagation and distribution.

### 2. *The diffuse shock acceleration theory*

Some key predictions or requirments of DSA are that: SNRs could accelerate CRs to the knee, i.e., about 4 PeV; magnetic field amplification is needed to accelerate CRs; the energy conversion rate should be high, i.e., larger than 10%, and CRs should have important nonlinear effect on the structure of the shock [169, 170].

The CRs are usually traced by 4 emission processes. For electrons, the tracers are synchrotron radiation ($\propto N_{\rm CRe}B^2$, where $N_{\rm CRe}$ is the number density of electron), bremsstrahlung ($\propto N_{\rm CRe}N_{\rm H}$, where $N_{\rm H}$ is the number density of ionized, neutral and molecular hydrogen) and inverse Compton (IC) scattering ($\propto N_{\rm CRe}N_*$, where $N_*$ mean the number density of background photon). For protons, the tracer is neutral pion decay ($\propto N_{\rm CRp}N_{\rm H}$, where $N_{\rm CRp}$ is the number density of protons). The first process usually dominates in the radio band and sometimes appears in the X-ray band. The last three processes produce radiation in the $\gamma$-ray band. The key to illustrate SNRs as the origin of GCRs is to separate the hadronic process from the leptonic processes. Since both bremsstrahlung and pion decay are proportional to $N_{\rm H}$, their relative intensity is determined by the density ratio between electrons and protons ($K_{\rm ep}$). Be-cause $K_{\rm ep}$ is usually smaller than 0.01, bremsstrahlung is much less efficient than pion decay (see the estimation from [171]). The main confusion is from IC.

Multi-wavelength observations are so far the best way to solve the problem. From synchrotron radiation (radio and X-ray bands), we could investigate the electron energy index which can be used to restrict the IC radiation. Furthermore, the ratio of electron energy loss between synchrotron radiation and IC is $P_{\rm sy}/P_{\rm IC} = U_{\rm B}/U_{\rm ph}$, where $U_{\rm B}$ and $U_{\rm ph}$ are the energy densities of magnetic field and background photon field, respectively. Higher magnetic field strength will lead to less IC radiation. The OH 1720 MHz maser (centimeter band), X-ray synchrotron radiation (X-ray band) can be used to estimate the magnetic field strength. The background photon field includes the 3 K cosmic background radiation. However, for some SNRs, the infrared radiation from dust (infrared band) also has a great contribution to the photon field. For pion decay, it depends on the material distribution which can be inferred by the molecular lines observation (centimeter, millimeter/submillimeter band), dust observation (infrared band) and X-ray observation.

1. PeVatrons

In the $\gamma$-ray band, there are two crucial spectral windows to distinguish pion decay from leptonic processes. The first one is the sub-GeV window. In this window, the spectrum of pion decay is characterized by the pion bump---rises steeply below ~ 200 MeV. This feature has been observed as the first direct evidence for SNR accelerating protons at GeV. Since the current ongoing $\gamma$-ray satellites are not sensitive at this band, further MeV-GeV telescopes, such as PANGU [172], may have a large sample investigation. Another window is the band well beyond 10 TeV, such as 100 TeV. In this band, the $\gamma$-ray contribution from the IC component is greatly suppressed due to the Klein-Nishina effect. The hadronic origin could be established through detailed modeling with multi-wavelength information. So far, LHAASO has the best sensitivity at the energy above 10 TeV (see Fig. 23).

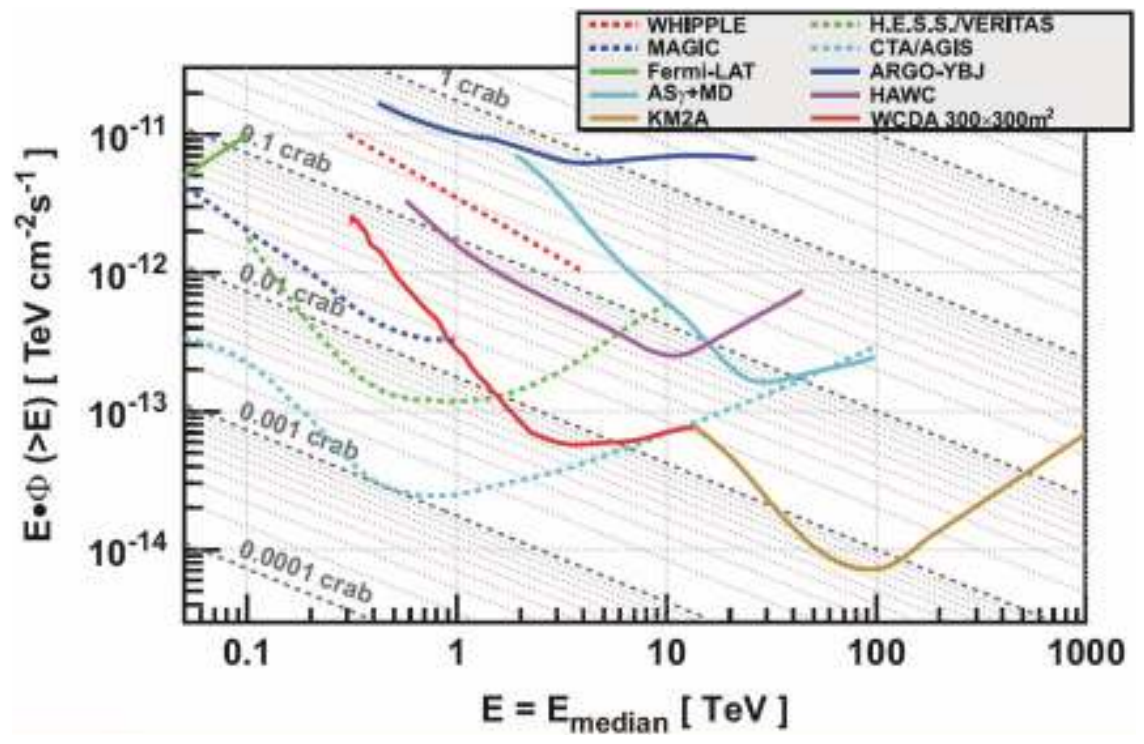

**Fig. 23.** (color online) The sensitivity of LHAASO-WCDA and LHAASO-KM2A [173].

It will not only give the first SNR observation above 30 TeV, but also greatly reduce the error bar of the data which is critical to distinguish different models. The SNR paradigm predicts that young SNRs should be PeVatrons, which can be verified with LHAASO observations.

2. Magnetic field amplification

Magnetic filed amplification is a prediction of sufficient CRs acceleration and is also required if SNRs can indeed accelerate CRs to PeV. LHAASO could give direct evidence of PeVatrons, but the process of how the CRs are accelerated to PeV is not within its reach. As mentioned above, previous magnetic field strength studies are mainly based on OH 1720 MHz maser and X-ray synchrotron radiation observations. Both studies indicate significant magnetic filed amplification. The OH 1720 MHz masers only appear in shocked molecular cloud with density of about $10^5$ cm$^{-3}$. That means the magnetic field strength measurement is constrained to a compact region. For most parts of an SNR, OH maser observation is not able to measure the magnetic field strength. For young SNRs, X-ray synchrotron emission is only identified in narrow regions close to shock front. So, does magnetic field amplification really appear in the whole region of an SNR?

The Zeeman effect of neutral hydrogen has been used to measure the magnetic field strength of the interstellar medium. The difficulty of this method is the superposition of different hydrogen clouds similar line-of-sight velocities. Recently, observations have shown that some SNRs are associated with high velocity neutral hydrogen clouds [174]. Since those clouds are distinct from background ones, to measure their magnetic fields is possible. SKA with its sensitivity, angular resolution and big field of view (see Fig. 24), will bring us a chance to map the magnetic field strength with great details in the large area of an SNR. It could help to reveal where the magnetic field amplification happens and how large the amplification is.

Magnetic field amplification is believed to be associated with turbulence. This turbulence will cause scattering, scintillation of background light, and might cause the background point source to become an "extended" one. The scintillation of pulsars has been widely used to detect interstellar cloud physical properties to study the Kolmogorov spectrum. When a pulsar is located behind an SNR, even behind the shock region, we could use it to detect the turbulence in the shock region with the same method used to study the interstellar electron clouds. Since most pulsars are faint (previous studies usually use pulsars with flux larger than 20 mJy at 400 MHz), a more sensitive telescope like SKA is needed to do this work.

3. Energy conversion rate and nonlinear effect

To explain the observed CRs energy density, an en-

Table xxx: Parameters for Comparable Telescopes

| | | eMERLIN | JVLA | GBT | GMRT | Parkes MB | LOFAR | FAST | MeerKAT | WSRT | Arecibo | ASKAP | SKA1-survey | SKA1-low | SKA-mid |
|---|---|---|---|---|---|---|---|---|---|---|---|---|---|---|---|
| $A_{eff}/T_{sys}$ | $m^2/K$ | 60 | 265 | 276 | 250 | 100 | 61 | 1250 | 321 | 124 | 1150 | 65 | 391 | 1000 | 1630 |
| FoV | $deg^2$ | 0.25 | 0.25 | 0.015 | 0.13 | 0.65 | 14 | 0.0017 | 0.86 | 0.25 | 0.003 | 30 | 18 | 27 | 0.49 |
| Receptor Size | m | 25 | 25 | 101 | 45 | 64 | 39 | 300 | 13.5 | 25 | 225 | 12 | 15 | 35 | 15 |
| Fiducial frequency | GHz | 1.4 | 1.4 | 1.4 | 1.4 | 1.4 | 0.12 | 1.4 | 1.4 | 1.4 | 1.4 | 1.4 | 1.67 | 0.11 | 1.67 |
| Survey Speed FoM | $deg^2\ m^4\ K^{-2}$ | $9.00\times10^2$ | $1.76\times10^4$ | $1.14\times10^3$ | $8.13\times10^3$ | $6.50\times10^3$ | $5.21\times10^4$ | $2.66\times10^3$ | $8.86\times10^4$ | $3.84\times10^3$ | $3.97\times10^3$ | $1.27\times10^5$ | $2.75\times10^6$ | $2.70\times10^7$ | $1.30\times10^6$ |
| Resolution | arcsec | $10\text{-}150\times10^{-3}$ | 1.4 - 44 | 420 | 2 | 660 | 5 | 88 | 11 | 16 | 192 | 7 | 0.9 | 11 | 0.22 |
| Baseline or Size | km | 217 | 1 - 35 | 0.1 | 27 | 0.064 | 100 | 0.5 | 4 | 2.7 | 225 | 6 | 50 | 50 | 200 |
| Frequency Range | GHz | 1.3-1.8, 4-8, 22-24 | 1 - 50 | 0.2 - 50+ | 0.15, 0.23, 0.33, 0.61, 1.4 | 0.44 to 24 | 0.03 – 0.22 | 0.1 – 3 | 0.7 - 2.5, 0.7 - 10 | 0.3 – 8.6 | 0.3 - 10 | 0.7-1.8 | 0.65-1.67 | 0.050 – 0.350 | 0.35-14 |
| Bandwidth | MHz | 400 | 1000 | 400 | 450 | 400 | 4 | 800 | 1000 | 160 | 1000 | 300 | 500 | 250 | 770 |
| Cont. Sensitivity | $\mu Jy\text{-}hr^{-1/2}$ | 27.11 | 3.88 | 5.89 | 6.13 | 16.26 | 266.61 | 0.92 | 3.20 | 20.74 | 0.89 | 28.89 | 3.72 | 2.06 | 0.72 |
| Sensitivity, 100 kHz | $\mu Jy\text{-}hr^{-1/2}$ | 1714 | 388 | 373 | 411 | 1029 | 1686 | 82 | 320 | 830 | 89 | 1582 | 263 | 103 | 63 |
| SEFD | Jy | 46.0 | 10.4 | 10.0 | 11.0 | 27.6 | 45.2 | 2.2 | 8.6 | 22.3 | 2.4 | 42.5 | 7.1 | 2.8 | 1.7 |

Notes to Table

| | | | | |
|---|---|---|---|---|
| eMERLIN | Frequencies non-contiguous | | | |
| JVLA | Multiple antenna configurations | | | |
| GBT | Single dish | | | |
| GMRT | Frequencies non-contiguous | | | |
| Parkes MB | Multi-beam (13) | Frequencies non-contiguous | | |
| LOFAR | Parameters for all NL stations | Frequencies non-contiguous | | |
| FAST | Single dish | Under construction | | |
| MeerKAT | SKA Precursor | Under construction | | |
| WSRT | Frequencies non-contiguous | | | |
| Arecibo | Single dish | | | |
| ASKAP | SKA Precursor | Multi-beam (36) | Under construction | |
| SKA1-survey | Multi-beam (36) | Mixed 12-m & 15-m dishes | FoV based on 15-m dishes | Planned |
| SKA1-low | | | | Planned |
| SKA-mid | | Mixed 13.5-m & 15-m dishes | FoV based on 15-m dishes | Planned |
| Notes: All | Fiducial frequency: Most Parameters | $\Omega_{FoV} = (\pi/4)(66\lambda/Ddish)^2$ | Gray shading: <400 MHz capable | SEFD: System Equivalent Flux Density |
| (cont'd) | SEFD derived from $A_{eff}/T_{sys}$ | Sensitivity derived from SEFD & BW | System efficiency assumed 100%. | |

**Fig. 24.** (color online) The basic parameters for SKA [175].

ergy conversion rate of about ~10% is needed. In the nonlinear DSA theory, the conversion rate in effective CRs acceleration shock can reach up to 50%. However it is not true for some SNRs. One case is Cas A. Abdo *et al.* (2010) claimed [176] that only less than 2% of the total energy is used to accelerate CRs. LHAASO may push this study further by measuring and modeling many SNR energy spectra with high sensitivity and broad energy coverage and give more accurate conversion rate estimates to a sample of SNRs.

A general condition for the 10% conversion rate is a Galactic supernova explosion rate of 2-3 per century. Considering the typical life time of about $10^5$ years for an SNR, the total number of Galactic SNRs should be more than 1000. This is much larger than ~300 SNRs currently detected in our Galaxy. Is this gap real or just because we miss lots of SNRs due to observation selection effects? For the first one, we need reconsider the theory of SNR paradigm. For the second one, we need to find the missing ones. Previous Galactic radio surveys are usually sensitivity limited or resolution limited which lead to the failed detection of old, faint, large remnants or young, small remnants. The ability of SKA (high resolution, sensitivity, and big field of view) gives us a chance to discover the missing SNRs in our Galaxy. It will answer how many SNRs are in our Galaxy and even tell us how the SNRs are distributed in our Galaxy. The total number of SNRs is critical to answer whether they are the main accelerator of Galactic CRs. The distribution of SNRs affect the CRs injection model which is important when modeling the diffuse $\gamma$-ray emission of our Galaxy.

Another way to find SNRs is to identify the lower energy counterparts of unidentified GeV/TeV sources. One example is the discovery of SNR G353.6-0.7 which is the first SNR discovered at TeV band and then identified at radio band [177]. Till now, more than 200 TeV sources have been discovered, however, more than 1/3 of them have no lower energy counterparts [178]. It is undoubted that LHAASO will find more TeV sources and some of them should be SNRs. The combination of SKA and LHAASO, will identify those missing SNRs, which allows us a compelling population study of the conversion rate problem.

When the energy is effectively converted into CRs, the shock structure will be modified that will lead to a curvature of electron spectrum with spectral hardening toward high energies. It has been claimed that this effect was detected for a few SNRs [179], but there are still lack of a large sample and spatially detailed studies, e.g. more convincing nonlinear effect towards TeV SNRs. To do this study, the SKA and LHAASO need work together.

### 3. *CRs diffusive escape and distribution near SNRs*

When CRs are accelerated to high energies in SNRs and the shock also slows down, the CRs will propagate diffusively from SNRs to the Galaxy. These CRs interact with the interstellar medium forming the non-thermal diffuse background emissions from radio to $\gamma$-ray bands.

#### 1. Escape

The escaped CRs take energy away from their mother SNRs. Therefore it is a possible explanation why some TeV bright SNRs have a very low energy conversion rate.

One case for CRs escape is from the *Fermi* observation of W44 in the GeV band [180]. While, for young SNRs like Cas A, the escaped CRs should have very high energy. These CRs can interact with surrounding materials and produce TeV emission. Compared with CTA, LHAASO has higher sensitivity to extended sources, which makes it a perfect facility to detect the TeV halo surrounding young SNRs. Since pion decay also depends on the material density, the infrared, centimeter or millimeter observations are also needed to derive the density distribution surrounding SNRs. High energy CRs will escape earlier and faster than lower energy CRs, so the halo may also have a GeV/TeV ratio change with distance away from the SNR. But the angular resolution of LHAASO is low, so CTA is more suitable for this kind of studies.

2. Distribution

The energy distribution of electrons can be obtained via modeling the diffuse emission from radio to $\gamma$-ray bands. Their spatial distribution can be obtained via measuring the emissivity of electrons from radio observations with the help of absorption from HII regions and planetary nebulae. The first one will only give two dimensional information and the second one may map the three dimensional electron distributions.

By employing the 21 months *Fermi* data, [181] used the GALPROP software to analyse the Galactic diffuse $\gamma$-ray emission. Their work successfully reproduces the observed $\gamma$-ray emission and gives the $\gamma$-ray composition and contributions from electrons and protons, respectively. However, they do not consider whether the electrons and protons, which are used to model $\gamma$-ray emission, could produce the observed radio emission or not. A combination modeling of radio and $\gamma$-ray is necessary. However, the angular resolution of current radio surveys in frequency of a few tens MHz to a few hundreds MHz is poor (usually worse than 1 degree) and can not effectively separate point sources from diffuse emission. SKA can provide the needed high resolution low frequency radio data and LHAASO will supplement the high energy TeV data.

Generally speaking, if we could get the synchrotron emission and magnetic field information at each position, it is possible to give a three dimensional model of electron distribution in our Galaxy. The only problem is how to get the distance information for synchrotron emission. A long time ago, people have noticed that HII regions can absorb the background low frequency radio emission though free-free absorption. This gives us a chance to estimate averaged foreground synchrotron emissivity as the background emission has been screened. Furthermore, if there are many HII regions distributed close to one line of sight, we could even estimate the emissivity between those HII regions. Figure 25 displays the relation between optical depth and frequency for typical HII regions and planetary nebulae [182].

Figure 26 shows the spatial distribution of HII regions and planetary nebulae [182]. For HII regions, they are big, so easy to be detected. Their distances are also easily determined. However, the total number of known

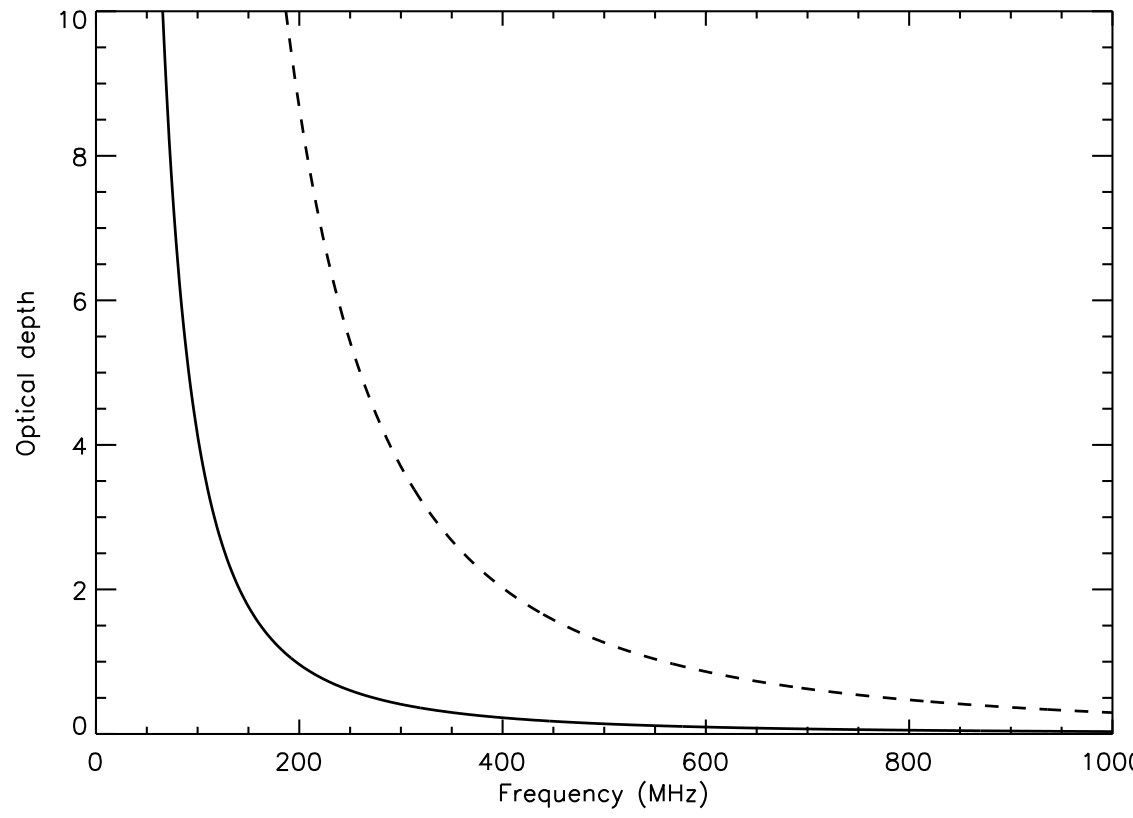

**Fig. 25.** Optical depth vs observation frequency of typical HII region and planetary nebulae. The solid line is for HII with temperature $T_e$ = 10000 K, electron density $n_e$ = 100 cm$^{-1}$, size $\Delta l$ = 10 pc. The dashed line is for planetary nebulae with temperature $T_e$ = 10000 K, electron density $n_e$ = 3000 cm$^{-1}$, size $\Delta l$ = 0.1 pc.

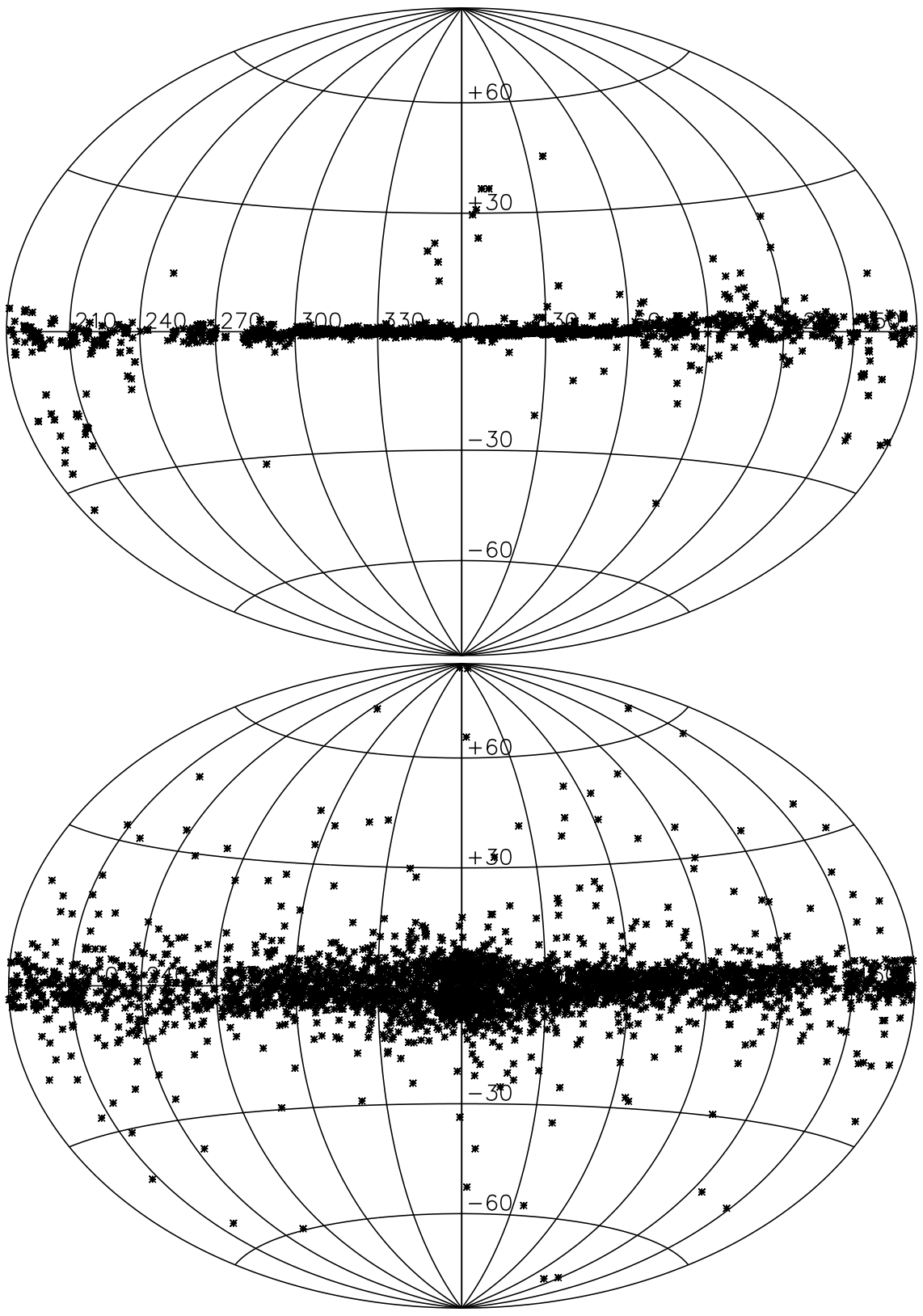

**Fig. 26.** The distribution of HII regions (upper) and planetary nebulae (lower) in Galactic coordinates.

HII regions is small and they are mainly located on the Galactic plane. For planetary nebulae, their distribution is wider than HII regions, so they can be used to estimate the emissivity of middle latitude regions. The total number of planetary nebulae is big and planetary nebulae become optical thick at higher frequency which mean they can measure the emissivity at broader region and dynamical range. The disadvantages are that their sizes are small and measuring the distances are not easy for most of them.

Currently, only a few tens absorption features from HII regions have been detected and no absorption detections for planetary nebulae. The main problem is due to the poor angular resolution and sensitivity of current low frequency radio surveys. SKA has enough resolution (such as a few arcsecond) and sensitivity to carry out this study. This will be a great step to know the CRs-electron's distribution in our galaxy.

### 4. *The propagation of cosmic rays in the Galaxy*

It has been long considered that SNRs are the origin of primary Galactic Cosmic Rays, and the diffusive shock acceleration is regarded as the main acceleration mechanism. CRs are accelerated at SNRs and then diffuse in the Galaxy due to magnetic turbulence, suffering from fragmentation and energy losses in the interstellar medium (ISM) and interstellar radiation field (ISRF), and other processes. Considering those processes, the CRs propagation equation can be written as

$$\begin{aligned}\frac{\partial\psi(\vec{r},p,t)}{\partial t} =& Q(\vec{r},p,t)+\nabla\cdot(D_{xx}\nabla\psi-V_c\psi)\\ &+\frac{\partial}{\partial p}p^2D_{pp}\frac{\partial}{\partial p}\frac{1}{p^2}\psi-\frac{\partial}{\partial p}\left[\dot{p}\psi-\frac{p}{3}(\nabla\cdot V_c\psi)\right]\\ &-\frac{\psi}{\tau_f}-\frac{\psi}{\tau_r},\end{aligned} \tag{1}$$

where $\psi(\vec{r},p,t)$ is the density of CR particles per unit momentum p at position $\vec{r}$[1)], $Q(\vec{r},p,t)$ is the source distribution, $D_{xx}$ is the spatial diffusion coefficient, $V_c$ is the convection velocity, $D_{pp}$ is the diffusive reacceleration coefficient in momentum space, $\dot{p}\equiv\frac{\mathrm{d}p}{\mathrm{d}t}$ is momentum loss rate, and $\tau_f$ and $\tau_r$ are the characteristic time scales for fragmentation and radioactive decay respectively.

In the conventional model (CM), CR diffusion is assumed to be uniform in space and only energy-dependent, and the diffusion coefficient is parametrized as $D_{xx}=\beta D_0(\rho/\rho_0)^\delta$, a function where $\rho$ is the rigidity and $\delta$ reflects the property of the ISM turbulence. The reacceleration can be described by the diffusion in momentum space and the momentum diffusion coefficient $D_{pp}$ is coupled with the spatial diffusion coefficient $D_{xx}$ as [183].

$$D_{pp}D_{xx}=\frac{4p^2v_A^2}{3\delta(4-\delta^2)(4-\delta)\omega}, \tag{2}$$

where $v_A$ is the Alfvén speed, and $\omega$ is the ratio of magnetohydrodynamic wave energy density to magnetic field energy density, which can be fixed to 1 for strong turbulence. In this formula, the observed CRs are modulated by three main processes: injection, propagation and solar modulation.

It is generally believed that SNRs are the sources of Galactic CRs. The spatial distribution of SNRs is usually described by the following empirical formula:

$$f(r,z)=\left(\frac{r}{r_\odot}\right)^a\exp\left(-b\cdot\frac{r-r_\odot}{r_\odot}\right)\exp\left(-\frac{|z|}{z_s}\right), \tag{3}$$

where $r_\odot=8.5$ kpc is the distance from the Sun to the Galactic center, $z_s\approx0.2$ kpc is the characteristic height of the Galactic disk, $a$ = 1.25 and $b$ = 3.56 are adopted from [184], which are suggested from *Fermi* studies on diffuse $\gamma$-ray emission in the 2nd Galactic quadrant [185]. The accelerated spectrum of primary CRs at source region is assumbed to be a broken power law function:

$$q(p)=q_0\times\begin{cases}(p/p_{\rm br})^{-\nu_1} & \text{if}(p<p_{\rm br}),\\ (p/p_{\rm br})^{-\nu_2}\cdot f(\hat{p}) & \text{if}(p\geq p_{\rm br})\end{cases} \tag{4}$$

where $p$ is the rigidity, $q_0$ is the normalization factor for all nuclei, relative abundance of each nuclei follows the default value in GALPROP or DRAGON package. $p_{\rm br}$ is the broken energy and $\nu_1,\nu_2$ are the spectrum indexes before and after the broken energy $p_{\rm br}$, respectively. $f(\hat{p})$ is used to describe the high energy cut-off.

Recent measurements have shown a spectral hardening for the CR proton and He at several hundreds GeV, which suggests a new component in the CR spectra. By assuming that this spectral hardening origins from the propagation process, a spatial dependent diffusion is introduced and attributed to different magnetic turbulence between the galactic plane and the outer galaxy region.

As described in [186], a 1D spatial-dependent diffusion model, the two halo model (THM), has been studied and can give a good description of the primary proton and He spectra. As described by the THM, the diffusion process of the CRs can be separated into two regions, the inner halo and the outer halo. The inner halo contains the galactic plane with a lower diffusion coefficient owing to a higher magnetic confinement, and the outer halo is the

1) Note that it has been assumed that the CR distribution is nearly isotropic, which is appropriate away from the source regions.

rest region with a higher diffusion coefficient. The scenario is shown in Fig. 27. Based on the discussion about the THM, the spatial-dependent diffusion coefficient, $D_{xx}$, depends on both the spatial coordinates $(r,z)$ and the particle rigidity, which is parameterized as [187]

$$D_{xx}(r,z,\rho)=\begin{cases}\eta(r,z)\beta\left(\dfrac{\rho}{\rho_0}\right)^{\varepsilon(r,z)}, & |z|<\xi z_h \quad \text{(disk)}\\[2ex] D_0\beta\left(\dfrac{\rho}{\rho_0}\right)^{\delta_0}, & |z|>\xi z_h \quad \text{(halo)}\end{cases} \qquad (5)$$

where $\beta$ is the velocity of the particle in unit of the light speed $c$, $D_0$ represents the normalization of the halo diffusion efficient at $\rho_0=4$ GV, $\delta_0$ characterizes the rigidity dependence of the diffusion coefficient, $\xi z_h$ denotes the thickness of the disk, $\eta(r,z)$ and $\varepsilon(r,z)$ describes the spatial dependence of the diffusion coefficient in the disk. $\eta(r,z)$ and $\varepsilon(r,z)$ can be related to the source distribution.

The numerical code DRAGON is used to calculate the spatial dependent diffusion of CRs [194]. Figure 28 shows the model calculation results for the $Z$-dependent scenario to compare with the data. In this figure, panels (a)-(c) are the spectra of protons, Helium, and $p$+He, corresponding to three different data set of the light component knee as described above. The model parameters are also different. Panel (d), (e), and (f) is for C and O, for Mg, Al, and Si, and for Fe, respectively. Panels (g)-(i) are for the all-particle spectra. It is obvious that CREAM data prefers a relatively lower energy knee of the light components, which under-shoots the all-particle spectra. The KASCADE data gives the highest energy of the knee, which slightly over-shoots, but nevertheless is roughly consistent with the all-particle spectra. The Tibet AS data, taken at high altitude close to the shower maximum depth, are expected to provide the best reference for the all-particle spectrum. In all these fittings, it seems that there is some tension between the CREAM data and the ground-based measurements. However a good agreement is found between CREAM and ARGO-YBJ measurements concerning the spectrum of the light component (see Fig. 11).

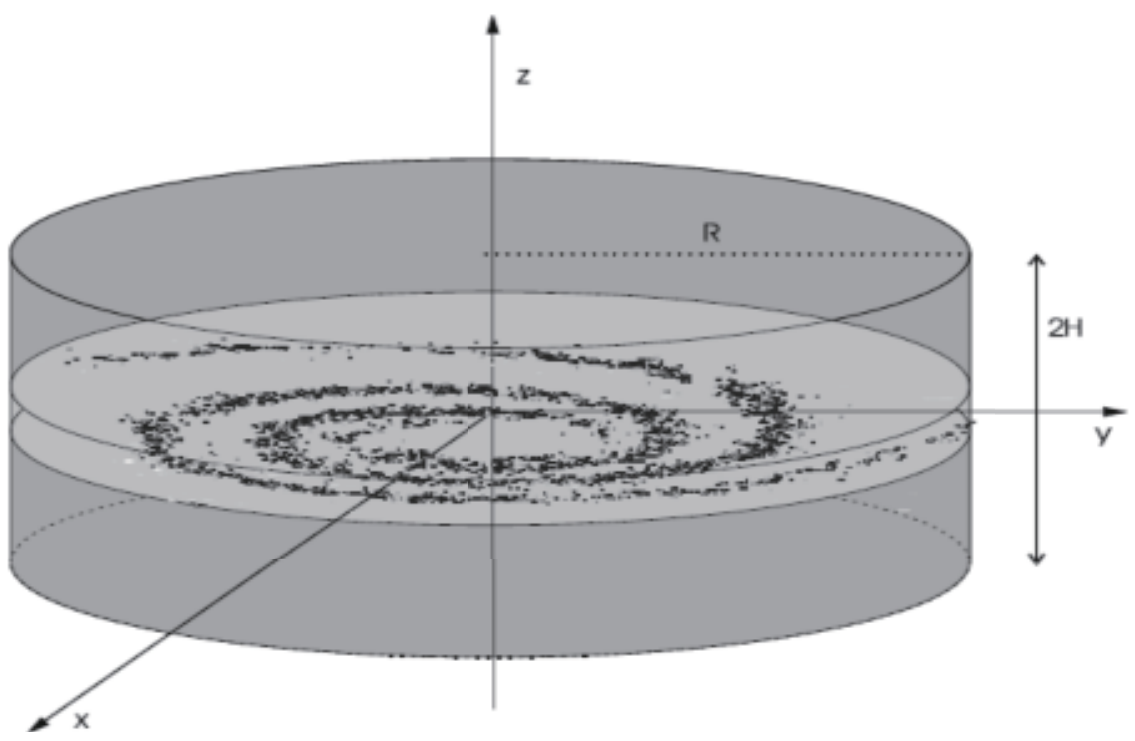

**Fig. 27.** The cartoon of two halo physical scenario.

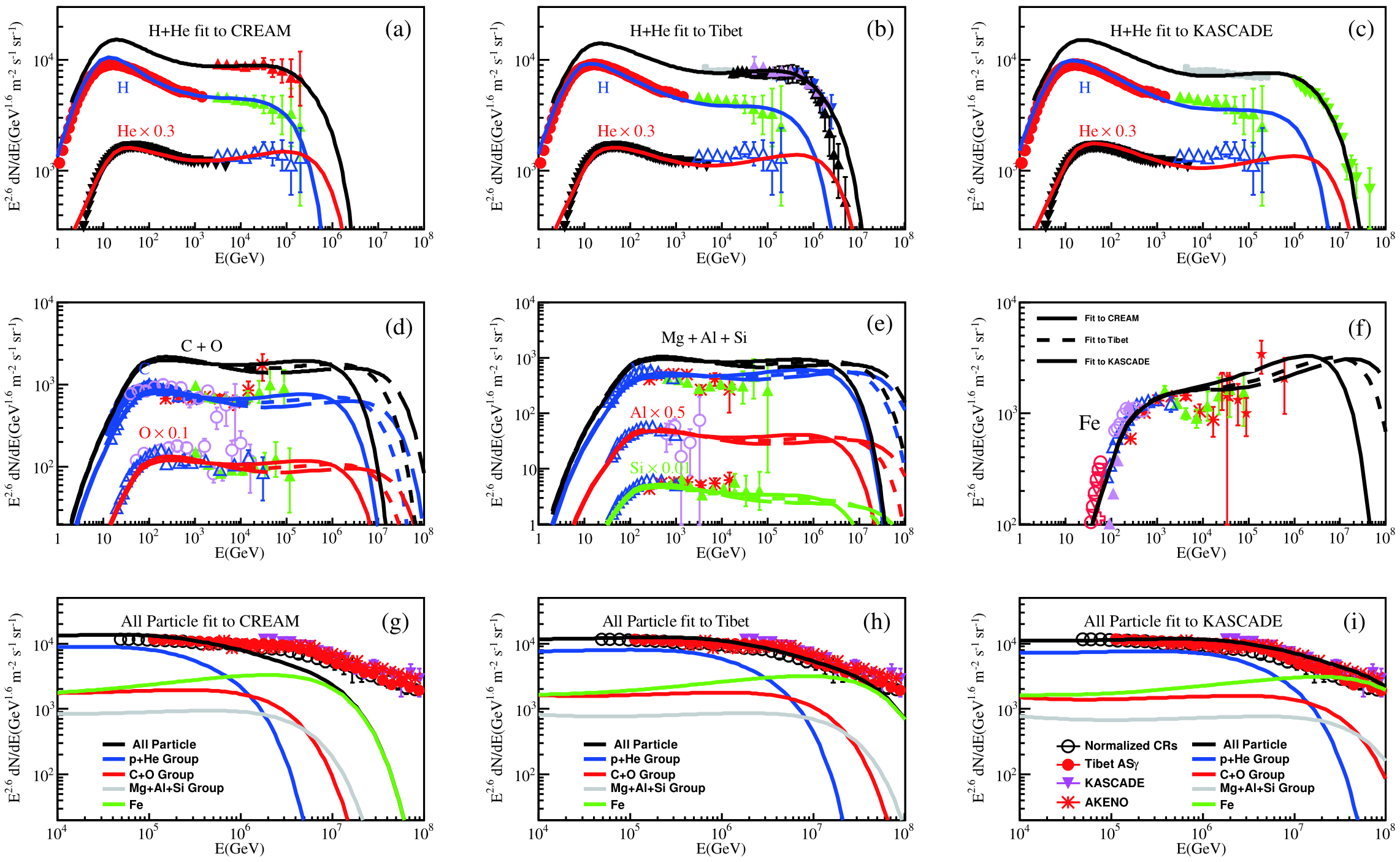

**Fig. 28.** (color online) The comparison between the model calculations and the experimental data for the Zdependent cases. The proton data are from: AMS-02 [4], CREAM [1], ATIC-2 [43]; the Helium data are from: AMS-02 [5], CREAM [1]; ATIC-2 [43]; the Carbon, Oxygen, Magnesium, Aluminium, Silicon, and Iron data are from: HEAO-3 [188], the proton + Helium data are from: KASCADE [189], Tibet-AS$\gamma$ [190], ARGOWFCTA [69], ARGO-YBJ [191, 192]; the allparticle data are from: Tibet-AS$\gamma$ [17], KASCADE [189], Akeno [16], and the normalized allparticle spectrum is from [193].

In general, as shown in Figs. 9 and 10, discrepancies between the direct measurements of different detectors may become increasingly larger at the highest energies approaching 100 TeV, due to both statistical and systematic uncertainties. For example, the proton and Helium spectra by CREAM [1] differ much from that by ATIC-2 [43]. A direct comparison of the AMS-02 measured fluxes of Helium and that by CREAM shows that the CREAM ones are higher by about 20% at 1 TeV/nucleon [5]. The spectrum measured by AMS-02 is also softer than that by CREAM. Further more precise measurements of the energy spectra of various species, by e.g., ISS-CALET [195], DAMPE [196], and LHAASO [197] will be very important to address this issue and better determine the model parameters.

## II. COSMIC RAY PHYSICS WITH LHAASO

### A. Selecting mass groups: the strategy

As illustrated in the previous Section, the energy spectrum of cosmic rays follows roughly a power-law shape from about $10^{10}$ eV to $10^{20}$ eV and is characterized by many features. The slope changes at least at three points, one around 3 PeV energy where the spectral index steepens from -2.7 to -3.1, the so-called knee, another around 80 PeV, the second knee, and then around $5\times10^{18}$ eV, the so-called ankle, where the spectrum again flattens. Any viable model of origin and propagation of cosmic rays has to explain all these features that contain the imprints of the energy spectra of each cosmic ray component. The measurement of spectra of five mass groups ($p$, He, CNO, MgSi, Fe) up to $10^{17}$ eV should be a high priority for LHAASO. The approach based on an event-by-event classification, which has a low dependence on hadronic interaction models and can be used for anisotropy studies, appears very promising with respect to the traditional reconstruction of the cosmic ray elemental composition carried out by means of complex unfolding techniques, procedures that heavily depend on the hadronic interaction model. Indeed LHAASO consists, in its basic configuration, of four types of detectors which may provide the lateral distribution and the size of the soft and muon shower components, the particle density near the shower core, and the image of the Cherenkov light produced along the longitudinal development of the shower. A large area densely instrumented enabling sensitive measurements up to $10^{17}$ eV, and a high altitude site approaching the atmospheric depth of maximum development of showers, where fluctuations are smaller and all nuclei produce the same electromagnetic size in a large energy range, are prominent figures of merit of LHAASO. The envisaged strategy to exploit, event-by-event, the experimental observables may be implemented through the following steps:

- combining the experimental observables in order to obtain an energy estimator and a set of mass sensitive parameters.

- use of selection criteria or multivariate analysis to separate the event samples according to the nature of the primary cosmic ray or mass group.

- estimate of the purity of the selected samples.

This program has been pursued using data obtained from the simulation of the shower development and detector response, based on CORSIKA and dedicated codes for each detector. The performance of the 'simulated LHAASO' is then applied to determine the spectrum of the light components in a wide energy range up to 10 PeV including the knee of the all-particle spectrum, and to reconstruct the heavy nuclei spectrum above 10 PeV (see Secs. II.D, II.E, and II.F). The envisaged implementation of the basic layout with neutron detectors and antenna for radio-detection will add further EAS observables sensitive to the nature of primary particles with energies above $10^{15}$ eV. The expected results based on preliminary studies are reported in Sec. II.g and Sec. II.H.

### B. The LHAASO layout

The LHAASO layout has been extensively described in Chapter 1. However we find convenient to shortly recall here some of its main features.

The Large High Altitude Air Shower Observatory (LHAASO) consists of a 1.3 km$^2$ EAS array (KM2A), a water Cherenkov detector array (WCDA), a wide field of view Cherenkov/fluorescence telescope array (WFCTA). KM2A includes 5195 scintillator detectors, with 15 m spacing, for electromagnetic particle detection and 1188 underground water Cherenkov tanks (36 m$^2$ each), with 30 m spacing, for muon detection.

A comparison with other large experiments concerning the number and active area of the deployed electron and muon detectors is reported in Table 1, singling out the relevance of this installation to reconstruct the lateral distribution of the main shower components.

WCDA consists of two 150 m × 150 m water pools plus one of dimensions 300 m × 110 m, all filled to a depth of 4.5 m. The total area is about 78000 m$^2$ divided in 3120 cells of size 5 m × 5 m each. The first pond of 150 m × 150 m (WCDA-1) is equipped with 900 pairs of 8-inch and 1.5-inch of PMTs to enhance the dynamical range. The other ponds are equipped with pairs of 20-inch and 3-inch PMTs.

WFCTA is composed of 18 telescopes each consist-

**Table 1.** LHAASO vs other EAS arrays.

| Experiment | depth/(g/cm)$^2$ | Detector | $\Delta E$/eV | e.m. Sensitive area/m$^2$ | Instrumented area/m$^2$ | Coverage |
|---|---|---|---|---|---|---|
| ARGO-YBJ | 606 | RPC/hybrid | $3\times10^{11}-10^{16}$ | 6700 | 11,000 | 0.93 (central carpet) |
| BASJE-MAS | 550 | scint./muon | $6\cdot10^{12}-3.5\cdot10^{16}$ | | $10^4$ | |
| TIBET AS$\gamma$ | 606 | scint./burst det. | $5\times10^{13}-10^{17}$ | 380 | $3.7\times10^4$ | $10^{-2}$ |
| CASA-MIA | 860 | scint./muon | $10^{14}-3.5\cdot10^{16}$ | $1.6\times10^3$ | $2.3\times10^5$ | $7\times10^{-3}$ |
| KASCADE | 1020 | scint./mu/had | $10^{15}-10^{17}$ | $5\times10^2$ | $4\times10^4$ | |
| KASCADE-Grande | 1020 | scint./mu/had | $10^{16}-10^{18}$ | 370 | $5\times10^5$ | $7\times10^{-4}$ |
| Tunka | 900 | open Cher.det. | $3\cdot10^{15}-3\cdot10^{18}$ | – | $10^6$ | – |
| IceTop | 680 | ice Cher.det. | $10^{15}-10^{18}$ | $4.2\times10^2$ | $10^6$ | $4\times10^{-4}$ |
| LHAASO | 600 | Water C scint./mu/had Wide FoV Cher.Tel | $3\times10^{11}-10^{18}$ | $5.2\times10^3$ | $1.3\times10^6$ | $4\times10^{-3}$ [KM2A] |

Muon detectors

| Experiment | m asl | $\mu$ Sensitive area/m$^2$ | Instrumented area /m$^2$ | Coverage |
|---|---|---|---|---|
| LHAASO | 4410 | $4.2\times10^4$ | $10^6$ | $4.4\times10^{-2}$ |
| TIBET AS$\gamma$ | 4300 | $4.5\times10^3$ | $3.7\times10^4$ | $1.2\times10^{-1}$ |
| KASCADE | 110 | $6\times10^2$ | $4\times10^4$ | $1.5\times10^{-2}$ |
| CASA-MIA | 1450 | $2.5\times10^3$ | $2.3\times10^5$ | $1.1\times10^{-2}$ |

ing of a segmented spherical mirror of 4.7 m$^2$ with a SiPM based camera installed at its focal plane. Each camera has a 16° × 16° FoV and is equipped with 1024 pixels each with a size of about 0.5° × 0.5°. The telescopes are arranged in a mobile design allowing the array to cover a wide patch of the sky for diffused cosmic ray measurements in both Cherenkov and fluorescence mode. They can work also with full Moon light, except the case of Moon directly in their FoV, with increased energy threshold, achieving a large duty cycle.

More details concerning construction, calibration and performance of these detectors can be found in Chapter 1. A schematic drawing of the LHAASO layout is shown in Fig. 29. Combining and integrating the information delivered by all detectors WCDA, KM2, WFCTA, it is possible to explore the energy range up to about 10 PeV selecting showers with core inside KM2A or WCDA. At low energies below 100 TeV, the LHAASO measurements will thus overlap with those from space detectors. At higher energies above 10 PeV WCDA cannot be used to sample the shower core due to saturation problems, and the mass sensitive parameters will be mainly obtained from the KM2 and WFCTA data. Figure 30 shows a simulated shower event induced by a 20 PeV iron nucleus as imaged by the LHAASO detectors. The map of hits in the scintillator array, muon counter array and the Cherenkov image in the cameras are shown (from left to right). According to the shower geometry determined by the scintillator counter array the impact parameter $R_p$, the distance between the Cherenkov image and the core location on the ground, is about 200 m.

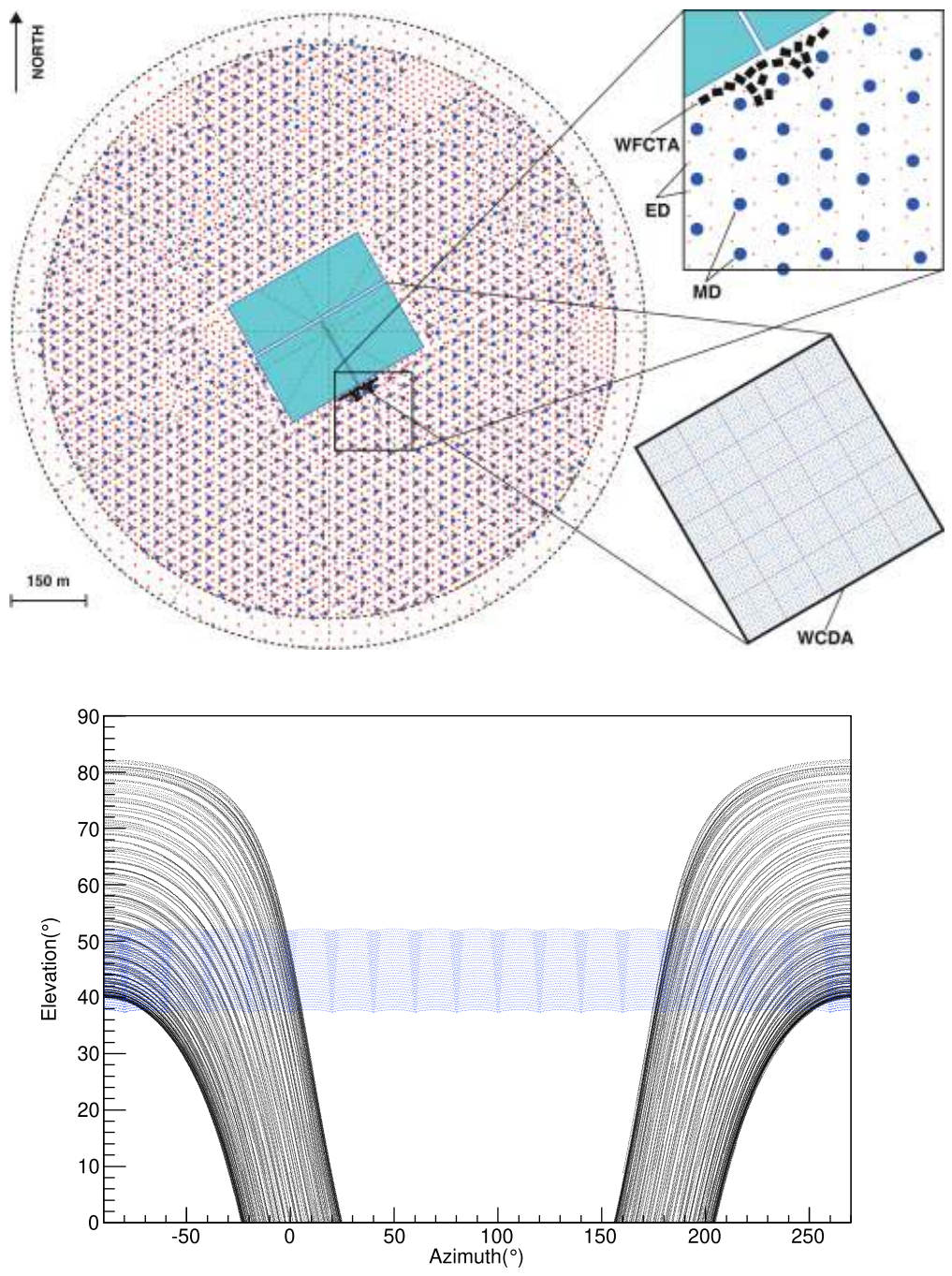

**Fig. 29.** (color online) The layout of the scintillator counter (small dots) array, muon counter (big dots) array, water Cherenkov detector (rectangle in the center) array and the location of the wide field of view (FoV) Cherenkov telescope (small squares) array in the LHAASO experiment (upper panel). The FoV of the telescopes in the northern sky (lower panel). The azimuth angle 90° is the north direction. Curves in the sky indicate the trajectories of the moon in one year. The FoV of the array covers a ring in the sky with a 16° elevation width, from 37° to 53° as shown in the figure.

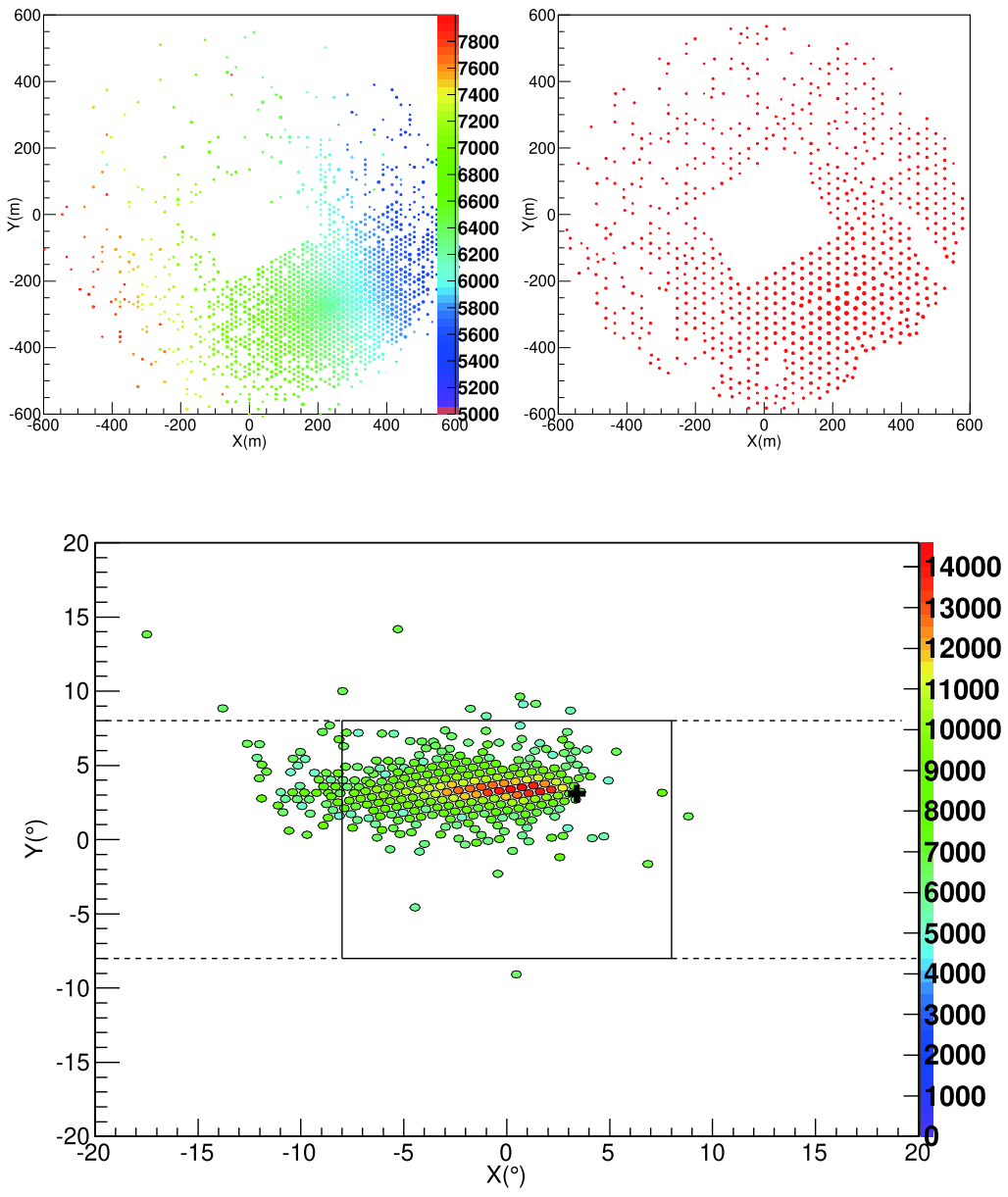

**Fig. 30.** (color online) A 20 PeV iron induced shower as imaged by the LHAASO detectors. The maps of the scintillation counters and of the muon counters of KM2A are shown in the upper two panels. The dimensions of the points are proportional to the logarithm of the particle number, the color scale indicates the time (in $10^{-2}$ ns) of each hit. The image of the shower taken by the Cherenkov telescopes at a distance $R_p \simeq$ 200 m is shown in the lower panel. The color scale indicates the number of photoelectrons

For observations above 100 PeV WFCTA will be operated to detect fluorescence light.

## C. Energy estimation and mass sensitive parameters

The basic information concerning the detector set up and operation is reported in Chapter 1. Here we shortly summarize the detector performance and describe the combination of the experimental observables currently studied to build up an energy estimator independent of composition and a set of mass sensitive parameters.

### 1. *Energy estimation*

Showers are sampled by WCDA and KM2A at a fixed depth ($600/\cos\theta$, $\theta$ being the zenith angle of the arrival direction) of their development in atmosphere. These detectors provide the basic information, namely the shower core position, the shower lateral profile, the arrival direction, the electron and muon content ($N_e$ and $N_\mu$ respectively) from which an energy estimate is obtained, as well as a first indication of the nature of the primary particle based on the ratio $N_\mu/N_e$. The Cherenkov image carries additional information about the shower energy, mainly from the total number of photons.

KM2A. The charge output of each scintillation counter is used to measure the number of crossing shower particles with excellent linearity up to more than $10^4$ particles. The shower trigger logic allows high efficient detection of shower events at an energy threshold of about 10 TeV. The signals from the scintillation counters are used to determine the impact point of the shower axis, the arrival direction, and fitting particle densities by a N-K-G like function, the shower profile. The electron size $N_e$, the muon number $N_\mu$ and/or the particle density at a fixed distance may be used as energy estimators. The weighted combination of $N_e$ and $N_\mu$ is a robust way to determine an energy estimator insensitive to composition. Work to fix the best combination is in progress. An energy resolution of about 15%-40% is expected, depending on the shower energy and zenith angle of the arrival direction [198].

WCDA. Each cell of the detector collects the Cherenkov light signals generated by the shower particles in the water. The total amount of Cherenkov photons is proportional to the energies carried by the particles, except muon traversing the water. The energy resolution depends on the shower energy and core location, with typical values in the 20%-35% range [199].

WFCTA. The Cherenkov image is an integration of the light yield along the longitudinal shower development in atmosphere and turns out to be an excellent estimator of the shower energy after corrections to account for the impact parameter $R_p$, that is the distance between the telescope and the core location on the ground, and at a minor extent, for the angular offset $\delta\theta$ of the centroid of the shower image from the shower arrival direction. Thus the shower energy is basically a function of the number of detected photoelectrons $N_{\rm pe}$ and the impact parameter $R_p$ (Fig. 31 upper). The energy resolution depends on these parameters, being about 20% with a bias less than 3% (Fig. 31 lower) [200].

### 2. *Mass sensitive parameters*

The muon content, the particle density in the shower core, the depth in atmosphere of the shower maximum, and the length to width ratio of the Cherenkov image are experimental observables that can be correlated and combined to define a set of mass sensitive parameters. A preliminary study, based on Monte Carlo simulations (by Corsika version 74005 with EGS4 to simulate electromagnetic processes, and QGSJET-II04 and FLUKA to model high and low energy hadronic interactions respectively), has been carried out to identify the most suitable parameters allowing an efficient selection of showers generated by light primaries (protons and Helium nuclei) and by nuclei of the iron group. Indeed, an accurate measurement of the proton and iron spectra can allow to set the relevant energy scale for all the other components.

Two different cases have been envisaged in this study corresponding to events with shower core impacting on WCDA or on the surface instrumented with the KM2A detectors. They address different energy intervals, below 10 PeV in the first case and up to 100 PeV for showers with core landing on the large area of KM2A. The aim of the first selection is the study of the light component and the measurement of its spectrum across the knee of the all particle spectrum. Selecting showers with core on KM2A allows the collection a sufficient number of events covering the relevant energy range from 10 to 100 PeV where the knee of the galactic iron component is expected.

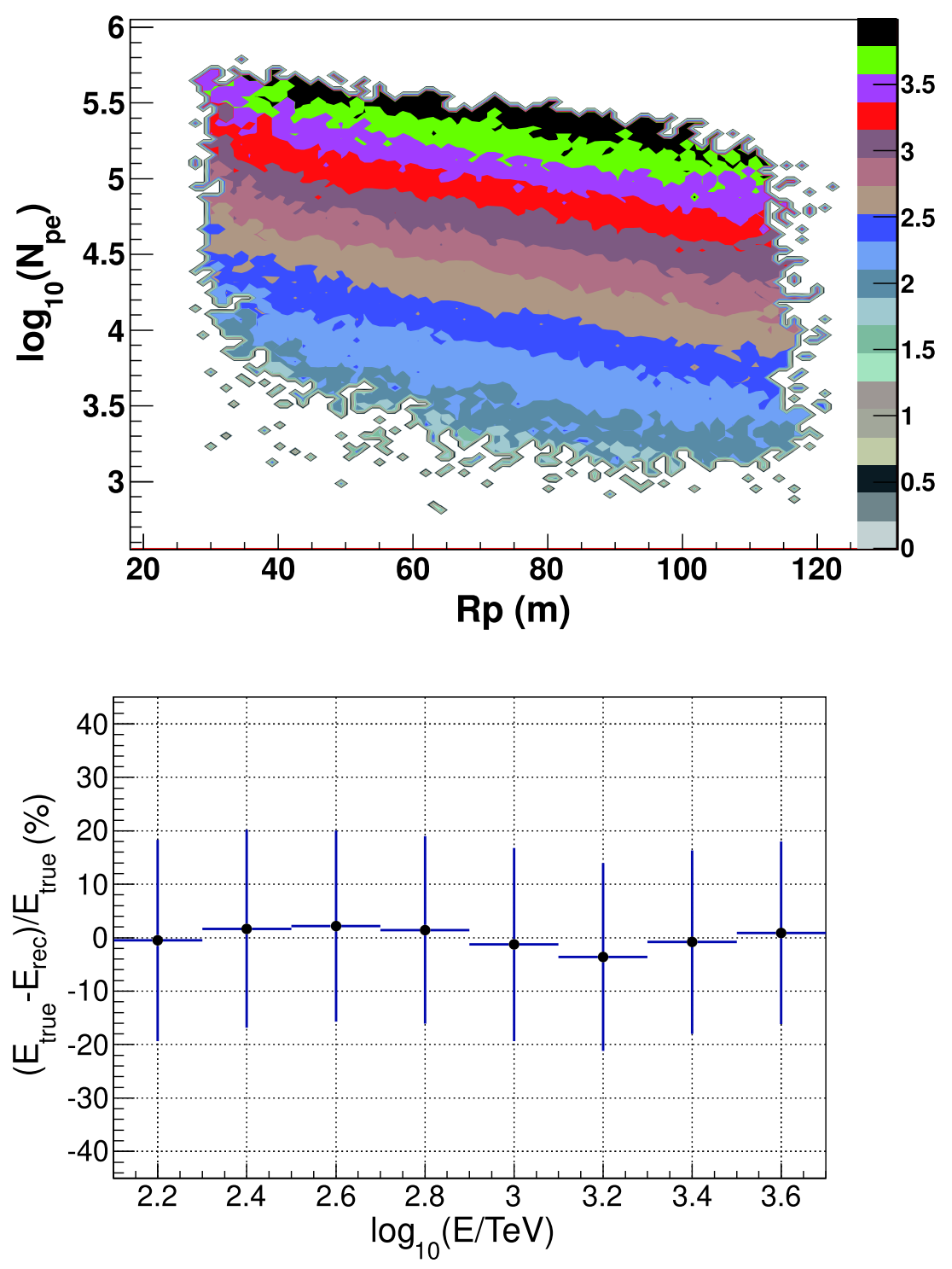

**Fig. 31.** (color online) The total number of photoelectrons $N_{pe}$ as a function of the impact parameter $R_p$ for primary protons. The color scale represents the shower energy in bins of $\Delta\log_{10}E=0.2$, covering primary energies from 30 TeV to 10 PeV (upper). The shower energy resolution of the Cherenkov telescopes (vertical bars) as a function of the energy of primary protons. $E_{true}$ and $E_{rec}$ are the simulated and reconstructed energies, respectively. A bias less than 3% is visible (lower).

Mass sensitive parameters are obtained by a hybrid approach, simultaneously using Cherenkov light and particle data. What is crucial for the development of an air shower is the distribution $\mathrm{d}P/\mathrm{d}x=\exp[-x/\lambda]$ of the first interaction point of the incoming cosmic ray, where $x$ is the slant depth measured from the top of the atmosphere, and $\lambda$ is the interaction length.

At TeV-PeV energies the proton interaction length is about 80 g/cm$^2$, while the interaction length for an iron nucleus is about 2.3 g/cm$^2$. Thus showers initiated by light elements, such as protons and helium nuclei, penetrate more deeply into the atmosphere than those from heavier nuclei. That means that the particle density nearby the core is higher in proton-showers than that of iron-showers whose lateral extension is more spread. Thus the measurement of the particle density distribution around the core of showers landing on WCDA may provide a sensitive mass parameter.

Usually, the cells with the largest number of particles recorded, $N_{max}$, are the closest to the core of the event. The value of $N_{max}$ in cores due to a heavy nucleus is lower than that due to a light nucleus. Obviously, $N_{max}$ is energy dependent and, therefore, a normalization procedure is necessary before it can be used to assess the composition. According to the simulation, $N_{max}$ is proportional to $(N_0^{pe})^{1.44}$, where $N_0^{pe}$ is the total number of photoelectrons measured by WFCTA normalized for $R_p=0$ and $\alpha=0°$.

A good indicator of the shower composition is the reduced parameter, $p_{max}$, defined as

$$p_{max}=\log_{10}(N_{max})-1.44\cdot\log_{10}(N_0^{pe}). \tag{6}$$

The other mass sensitive parameter is the total number of photoelectrons measured by the WCDA, $N_{WCDA}^{pe}$. Obviously, $N_{WCDA}^{pe}$ is primary energy dependent. The reduced parameter

$$p_{Npe}^{WCDA}=\log_{10}(N_{WCDA}^{pe})-1.18\cdot\log_{10}(N_0^{pe}), \tag{7}$$

may serve as a good indicator of the shower composition.

Another consequence is that the proton initiated showers develop to their maximum deeper in atmosphere than ones from heavy nuclei. The atmospheric slant depth of shower maximum, $X_{max}$, is the traditional mass sensitive parameter, which can be reconstructed from the Cherenkov image with a resolution of about 50 g/cm$^2$ (see Sec. II.F). Since $X_{max}$ depends on the primary energy, a good indicator of the primary mass is the reduced parameter $p(X_{max})=X_{max}-k\cdot\log_{10}N_0^{pe}$ where $k=49$ is a fitting parameter.

The Cherenkov images look like an ellipse and are more stretched,i.e. narrower and longer, for showers more deeply developed in atmosphere. The ratio of the length to the width ($L/W$) is therefore a good parameter that is sensitive to the nature of the primary particle. It is also known that the image is more elongated when the shower is farther away from the telescope, i.e. the image becomes longer and narrower for showers located farther away. Before they are used as indicators of the composition, images must be normalized to showers with different impact parameters, $R_p$. Furthermore, the images are also more stretched for the more energetic showers. According to simulation, the ratio $L/W$ of images is linearly proportional to $R_p$ and $N_0^{pe}$. The reduced parameter $L/W - 0.018R_p + 0.28\log_{10}N_0^{\rm pe}$, denoted as $p_C$, is related to the primary cosmic ray mass.

Muons in air are mostly from decay of charged pions and kaons. They do not multiply but only lose energy by ionization. The muon content of a shower builds up to a maximum then attenuates very slowly. On the contrary, the electron component attenuates relatively rapidly after maximum. At a fixed observation level the electron number $N_e$ is smaller for a heavy nucleus initiated shower than for a proton-shower of the same energy, whereas the low energy muon number $N_\mu$ (> 1 GeV) is larger approximately of a factor $A^{1-p}$ with $p \approx 0.86-0.93$ according to different hadronic interaction models [27].

The ratio $N_\mu/N_e$ for vertical shower at the LHAASO altitude (4410 m a.s.l.), obtained by CORSIKA, is shown in Fig. 32.

The low energy muon size $N_\mu$ is itself a mass sensitive parameter, once the dependence on the energy is taken into account. From simulations we find that the parameter

$$p_\mu = \log(N_\mu) + 0.001 \cdot R_p - 0.86 \cdot N_0^{\rm pe} \qquad (8)$$

is strongly related to the nature of the primary.

Many other parameters have been studied. The quoted parameters are not independent, there is some correlation between them. The correlations between mass sensitive parameters $p_{\rm max}$ and $p_c$, $p_{\rm max}$ and $p_\mu$, $p_{X_{\rm max}}$ and $p_{\rm max}$ are shown in Figs. 33, 34, 35, respectively.

These five parameters combine information from different detectors and turn out to be well suitable to implement the separation between cosmic ray components. Two different studies have been carried out concerning the identification of proton and proton+Helium nuclei around the knee of the all-particle spectrum, and the separation above 10 PeV of the iron nuclei from the other cosmic ray elements in order to identify the knee of this component. In this preliminary study only the correlation of two parameters are used for particle identification. The use of more parameters can be accomplished by means of a Multivariate Analysis currently in progress.

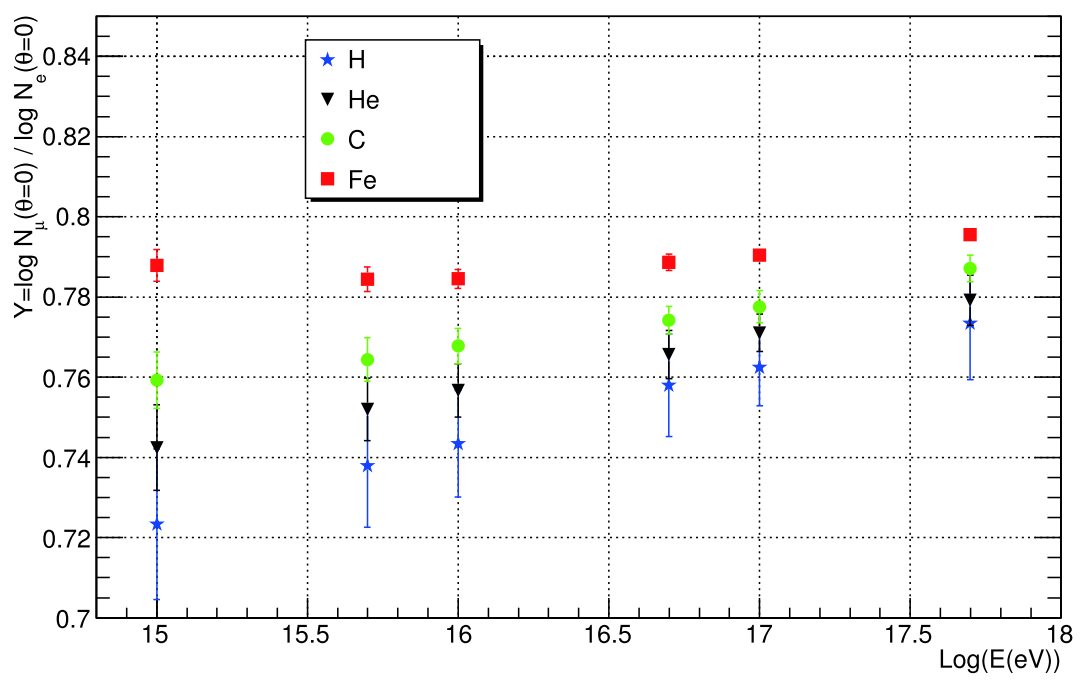

**Fig. 32.** (color online) $N_\mu/N_e$ calculated for EAS observed at the LHAASO altitude (4410 m above sea level) in the ideal case of a full coverage experiment without detection errors.

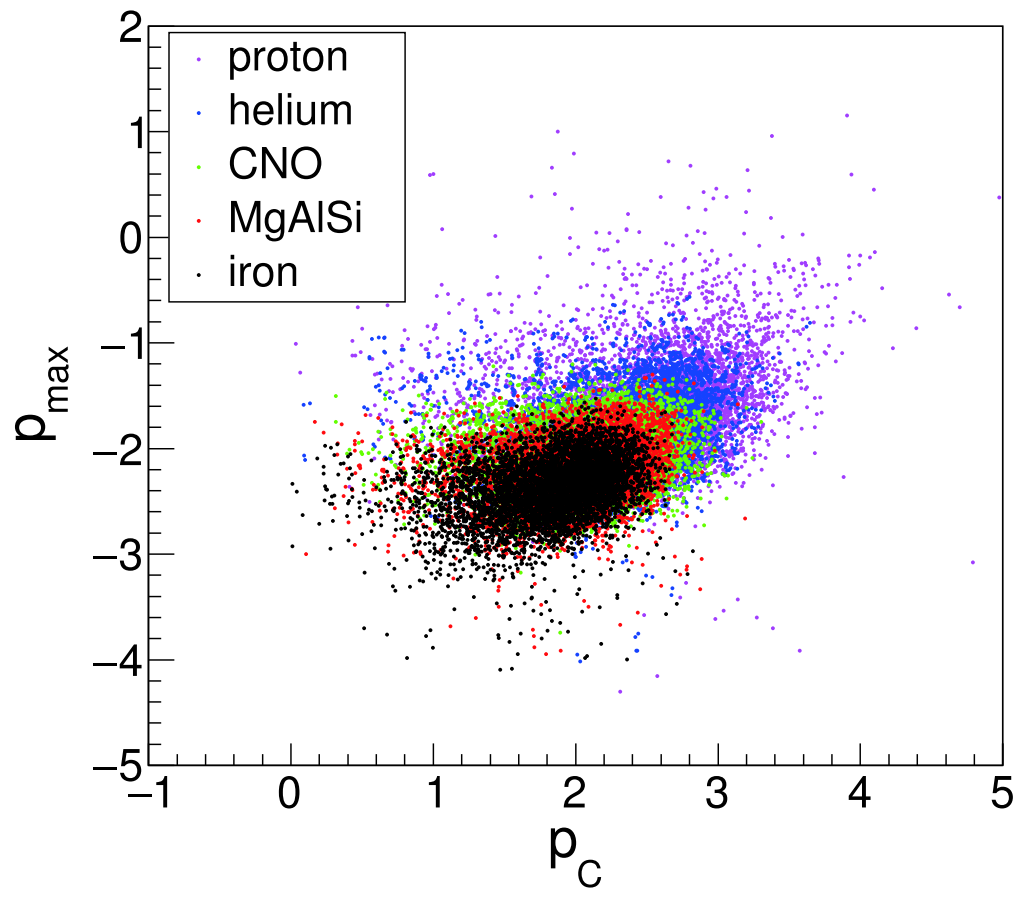

**Fig. 33.** (color online) Composition-sensitive parameters $p_{\rm max}$ and $p_{\rm c}$ for each primary cosmic ray element/group.

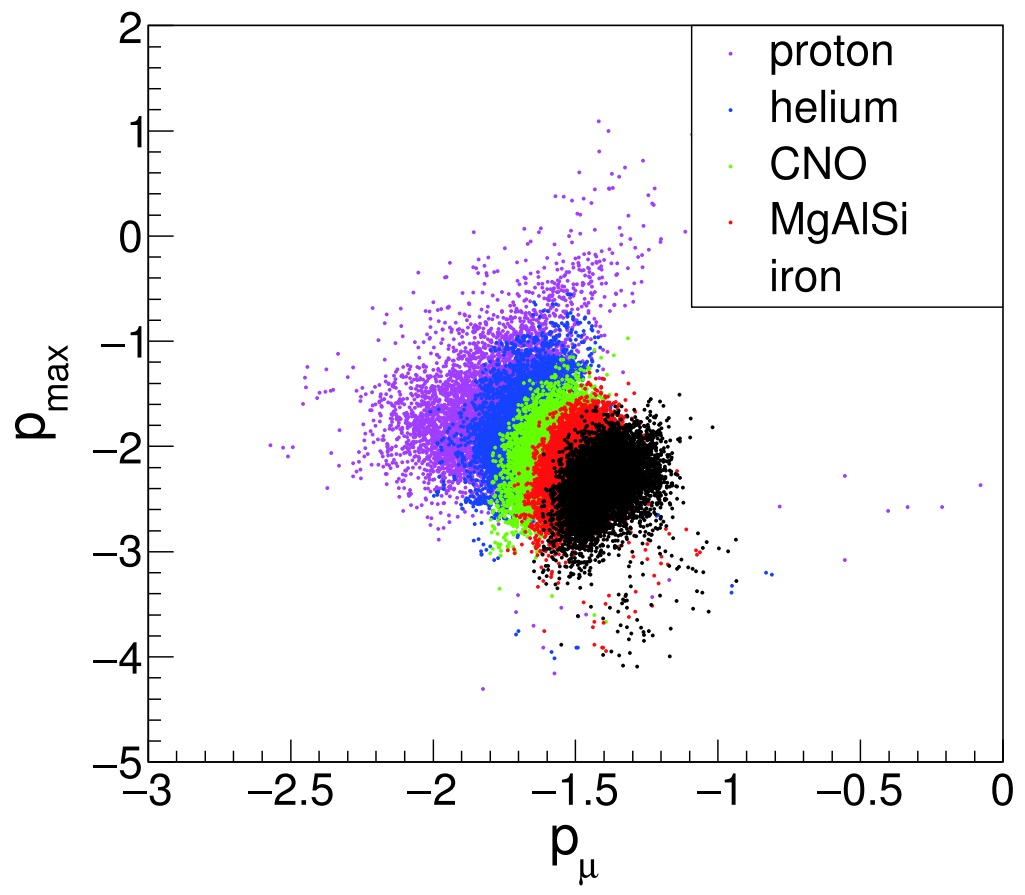

**Fig. 34.** (color online) Composition-sensitive parameters $p_{\rm max}$ and $p_\mu$ for each primary cosmic ray element/group.

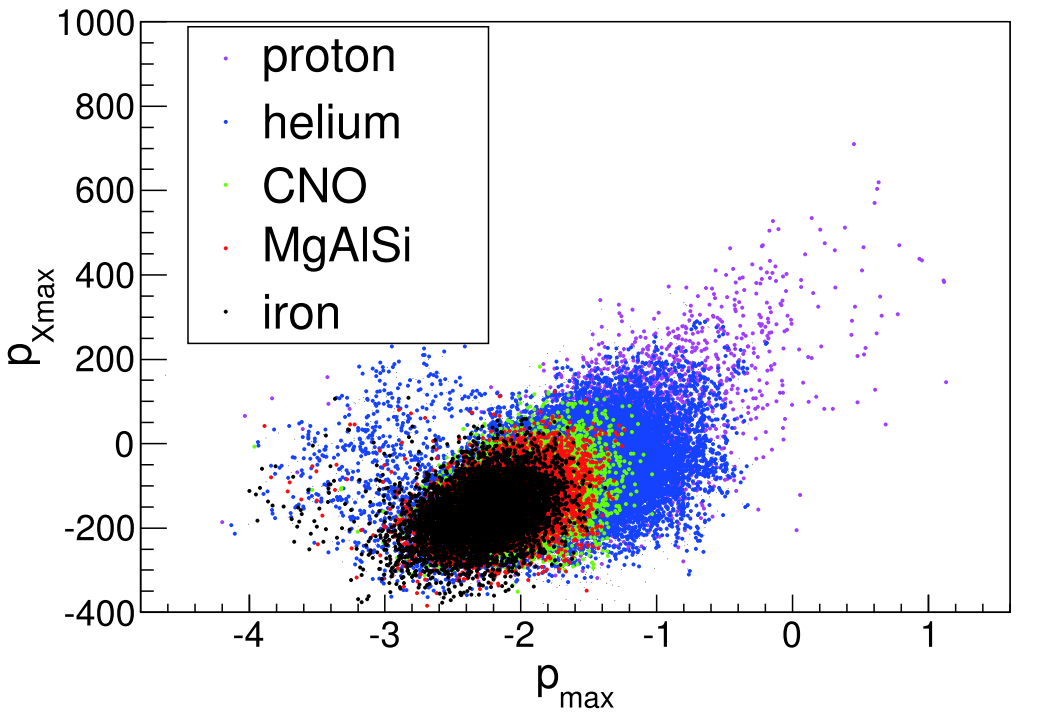

**Fig. 35.** (color online) Composition-sensitive parameters $p_{X_{\rm max}}$ and $p_{\rm max}$ for each primary cosmic ray element/group.

### D. Selection of Proton induced showers

The sample of proton (H event) induced showers is selected from the coincident events by combining the two composition-sensitive parameters $p_{\rm max}$ and $p_\mu$. This sample can be statistically separated on the $p_{\max} - p_\mu$ map shown in Fig. 34.

The cuts $p_{\max} \geq -1.0$ and $p_\mu \leq -1.9$ result in a selected sample of proton showers with a purity of 85% assuming the Hörandel composition models [193]. The aperture, defined as the geometrical aperture times the selection efficiency, gradually increases to 2600 m$^2$ sr at 100 TeV and remains nearly constant at higher energies (see Fig. 36). The selection efficiency is defined as the ratio of the selected number of proton events to the total number of injected proton events in the simulation.

In the selected sample, the contamination from the heavy species (CNO, MgAlSi, Iron groups) depends on the composition. Assuming the Hörandel composition [193], the contamination of heavy species is found to be less than 15% at energies ranging from 100 to 3 PeV, as shown in Fig. 37. After the composition selection, *H* like events from 100 TeV to 10 PeV are selected. The total exposure time of $1\times10^6$ seconds per year (~3.2% duty cycle) is assumed. The number of events in the each energy bin ($\Delta\log_{10}(E_0/\mathrm{TeV}) = 0.2$) is shown in Fig. 38. About 1000/year proton-like events around 1 PeV can be measured after the composition selection.

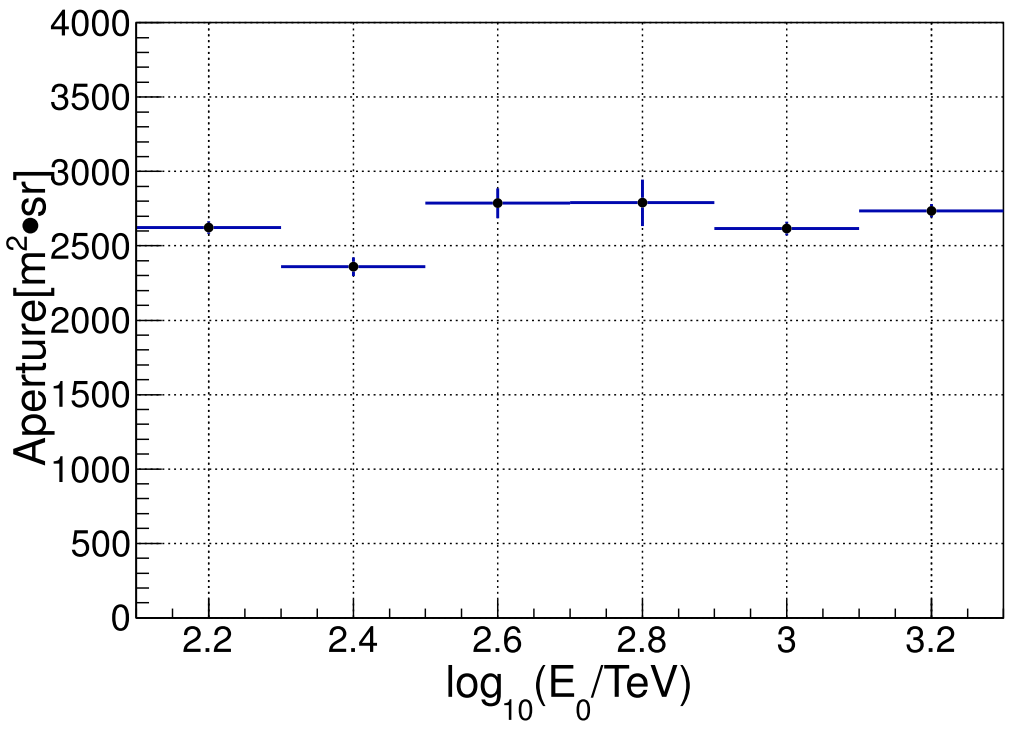

**Fig. 36.** (color online) The aperture for proton selected events as a function of the primary energy $E_o$.

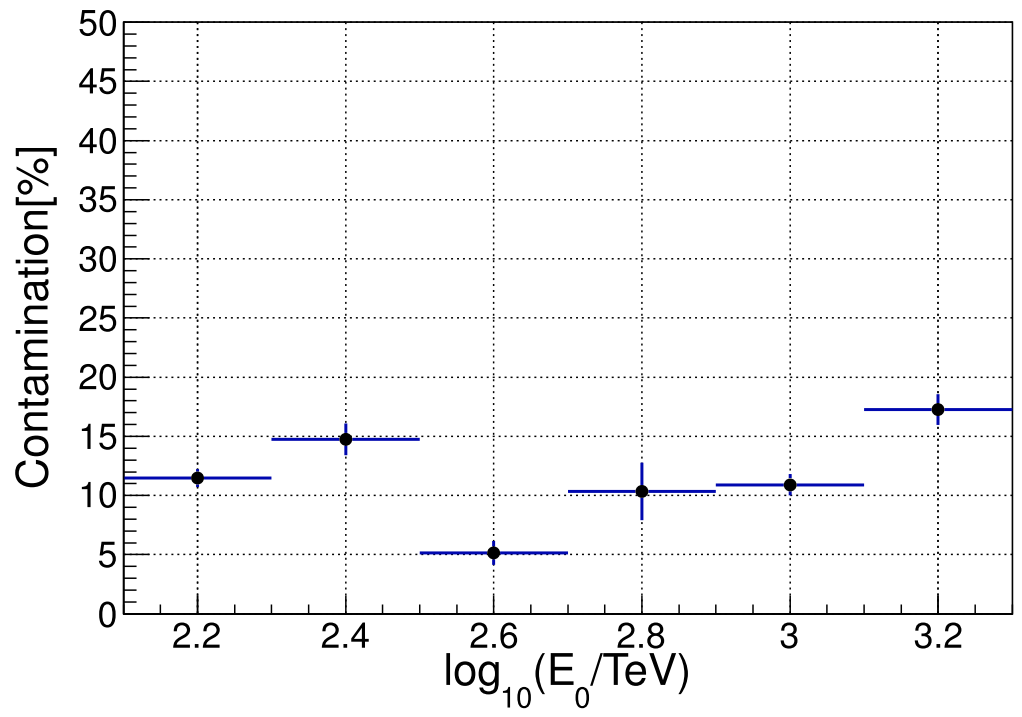

**Fig. 37.** (color online) The fraction of heavy nuclei that contaminates the proton sample as a function of the primary energy $E_o$. The Horandel composition model is assumed.

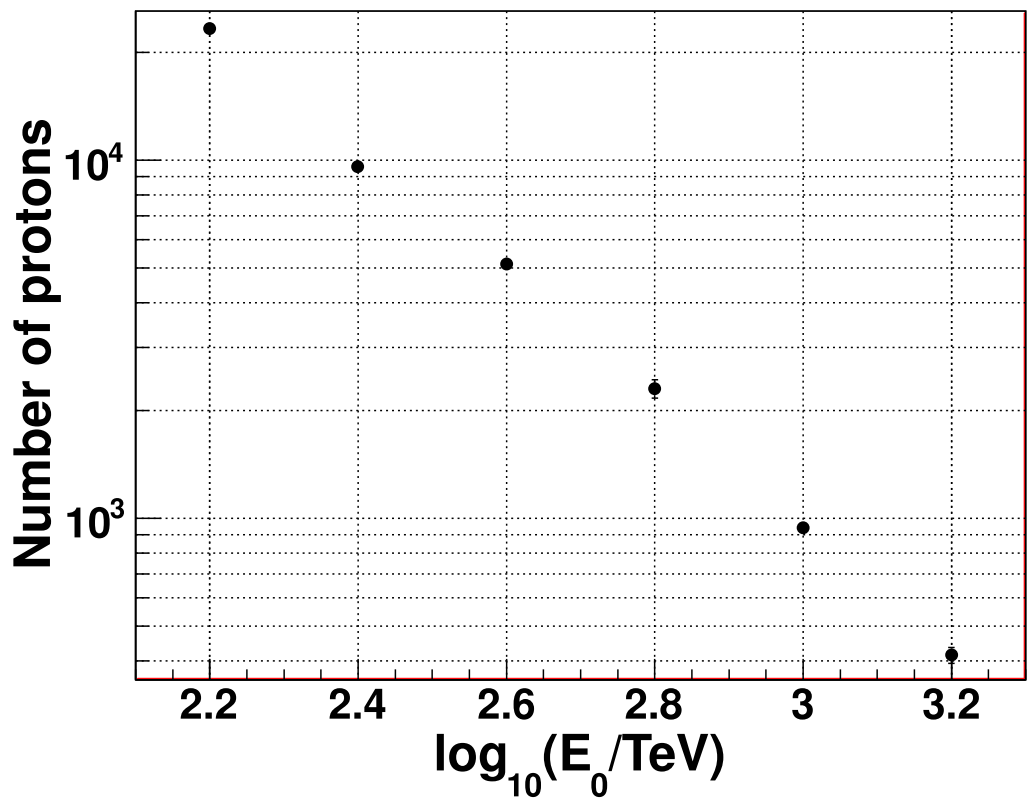

**Fig. 38.** (color online) The number of proton-like events in each energy bin measured per year after the composition selection. The Hörandel model is assumed in the simulation.

## E. Selection of proton plus Helium nuclei induced showers

The selection of showers generated by the light components, i.e. protons plus Helium nuclei (He), can be obtained from the same $p_{\max} - p_\mu$ map in Fig. 34.

The cuts $p_{\max} \geq -1.3$ and $p_\mu \leq -1.7$ result in a selected sample of p+He showers with a purity of 96% assuming the Hörandel composition models [193]. The aperture gradually increases to 4500 m$^2$ sr at 100 TeV and remains nearly constant at higher energies (see Fig. 39).

In the selected sample, the contamination from the heavy nuclei depends on the composition. Assuming the Hörandel composition [193], the contamination of heavy species is found to be less than 5% at energies ranging from 100 TeV to 3 PeV, as shown in Fig. 40.

After the composition selection, *p*+He like events from 100 TeV to 10 PeV are selected. The total exposure time of $1\times10^6$ seconds per year (3.2% duty cycle) is assumed. The number of events in each energy bin is shown in Fig. 41. About 3000/year *p*+He like events around 1 PeV can be measured after the composition selection.

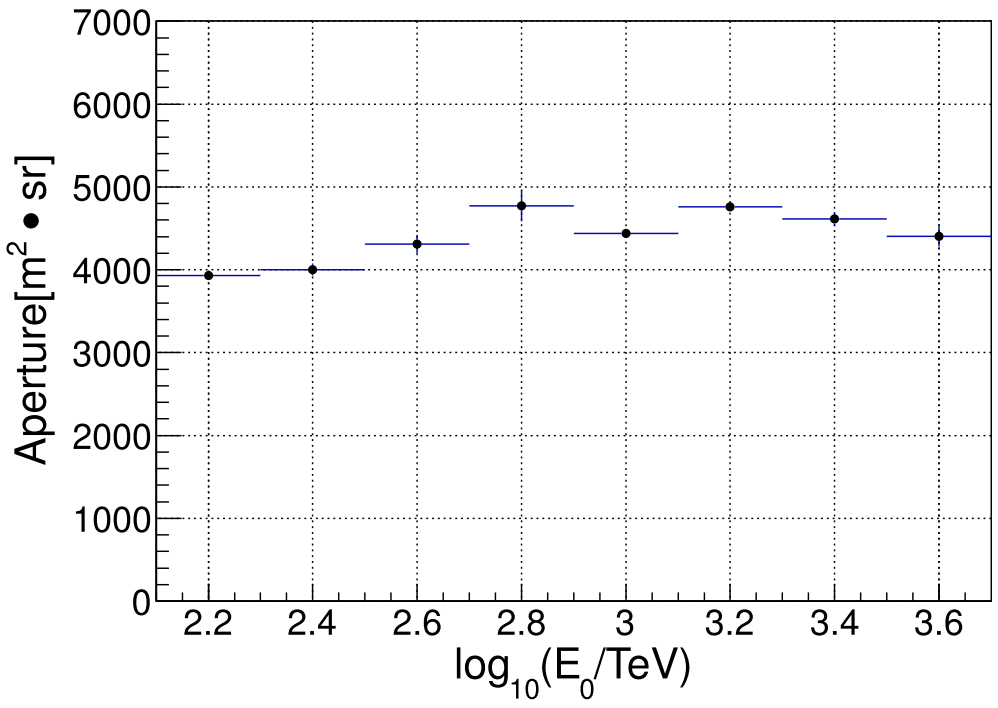

**Fig. 39.** (color online) The aperture for *p*+He selected events as a function of the primary energy $E_0$.

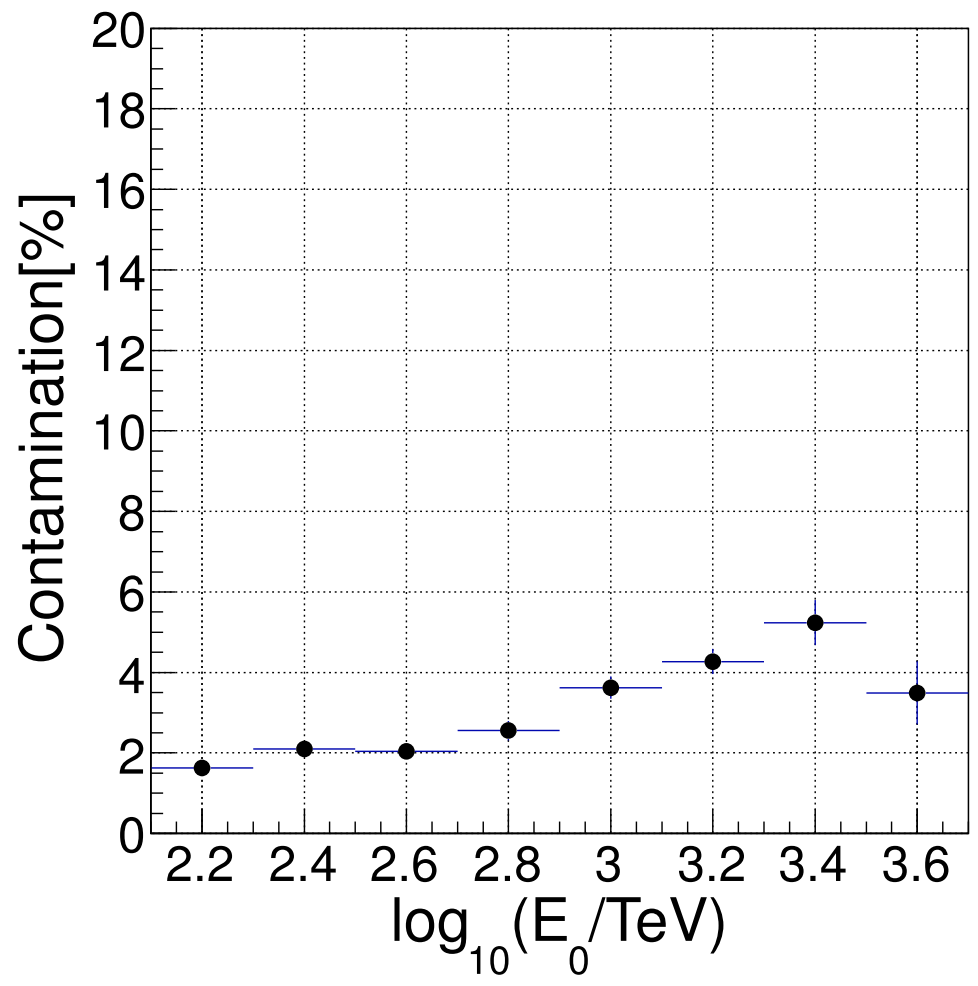

**Fig. 40.** (color online) The fraction of heavy nuclei that contaminate the proton+He sample as a function of the primary energy $E_0$. The Horandel composition model is assumed.

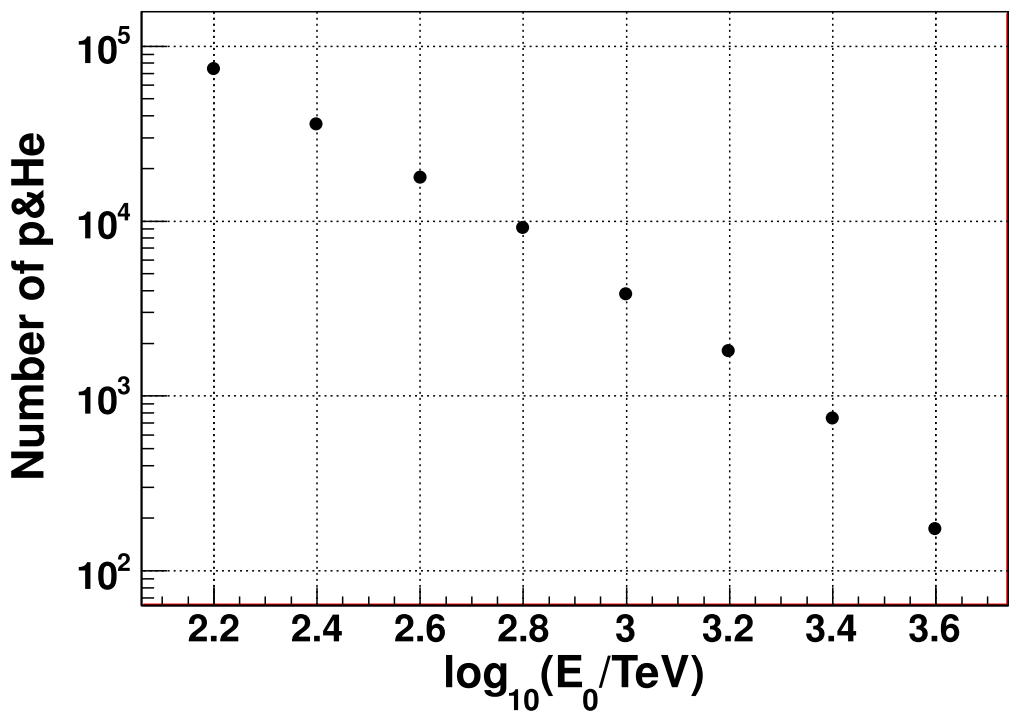

**Fig. 41.** (color online) The number of ($p$+He)-like events measured per year in each energy bin after the composition selection. The Horandel composition model is assumed.

### F. Measuring the spectrum of the heavy components above 10 PeV

Measuring the knees of the energy spectrum of the single components would imply a significant improvement in understanding the origin and propagation of the galactic cosmic rays. This is of particular importance in the energy range above 10 PeV where, according to some proposed models, a second galactic component is expected in addition to cosmic rays from SNRs, for instance re-accelerated particles at a galactic wind termination shock or cosmic rays from Wolf-Rayet stars exploding in the wind of the massive progenitor [87]. In the most accepted interpretation of the KASCADE and KASCADE-Grande data, the so-called second knee at 80 PeV is attributed to the bending of the iron component, assuming that the knee of the all-particle spectrum at 3 PeV is due to the decrease of the proton flux. On the other hand, high altitude experiments as Tibet As and ARGO-YBJ find the bending of the light component below 1 PeV, thus implying the knee of the iron component being at an energy below 80 PeV for both the plausible assumptions of a bending rigidity or mass A dependent.

An example of the LHAASO capability of imaging high energy iron induced showers is displayed in Fig. 30, where the imprint of a 20 PeV iron nucleus event is simulated.

One of the peculiar aspects of the LHAASO experiment is that the atmospheric depth of the experimental site (4410 m.a.s.l.) is close to the maximum development of $10^{16}-10^{17}$ eV air showers, with the electron size almost independent of the masses of the primary cosmic rays. On the contrary, the low energy muon number $N_\mu$ as mentioned before, depends on the mass of the primary particle. Accordingly, the ratio $N_\mu/N_e$ between the total number of low energy muons and the shower size is expected to be a mass sensitive parameter as shown in Fig. 32. The large electron and muon arrays of LHAASO may reconstruct these sizes with excellent resolution. A shower of 10 PeV typically generates more than 5000 hits in the ED array and about 50-100 hits in MD. The angular resolution is about 0.3° and the core location is measured with a resolution ≤ 2 meters. In the reconstruction of the muon lateral distribution, the information from the muon detectors close to the core is not taken into account to avoid the punch-through effect due to high energy electrons near the shower core. The ratio of the reduced muon number $N_\mu$ to the shower size $N_e$, $C_\mu = N_\mu/N_e$, can be exploited to selected heavy nuclei induced showers. The distribution of $1/C_\mu$ is displayed in Fig. 42 for all mass groups.

The plot in lower panel of Fig. 42 shows that, assuming a realistic composition, the parameter $1/C_\mu$ may be used to discriminate iron-induced events keeping at a low fraction the contamination from other nuclei. More efficient selection can be achieved by combining this parameter with the measurement of the shape of the Cherenkov images provided by the Cherenkov telescopes. Showers from heavy nuclei develop higher and faster (and with less shower to shower fluctuations) than shower initiated by lighter nuclei of the same energy. In the framework of the superposition model we have $X_{\max,A} = X_{\max,p} - \lambda_r \ln(A)$, where $A$ is the atomic mass and $\lambda_r$ is the radiation length in air (≃ 37 g/cm$^2$). All detailed simulations and hadronic models share this result predicting that iron showers have a smaller average $X_{\max}$ and less fluctuations on $X_{\max}$ than proton ones.

When the arrival direction of the shower and the impact parameter $R_p$ are known, simple geometry can be used to reconstruct the amount of light received from each altitude of the shower, light which is proportional to the number of electrons.

The angular offset ($\delta\theta$) of the centroid of the Cherenkov image to the arrival direction of the shower is related to $X_{\max}$, though $R_{\rm p}$ dependent due to the elongation of the geometry. The distributions of angular offset after $R_{\rm p}$ correction for the five mass groups are displayed in Fig. 43 with the assumption of primaries evenly distributed or distributed according to the Horandel model.

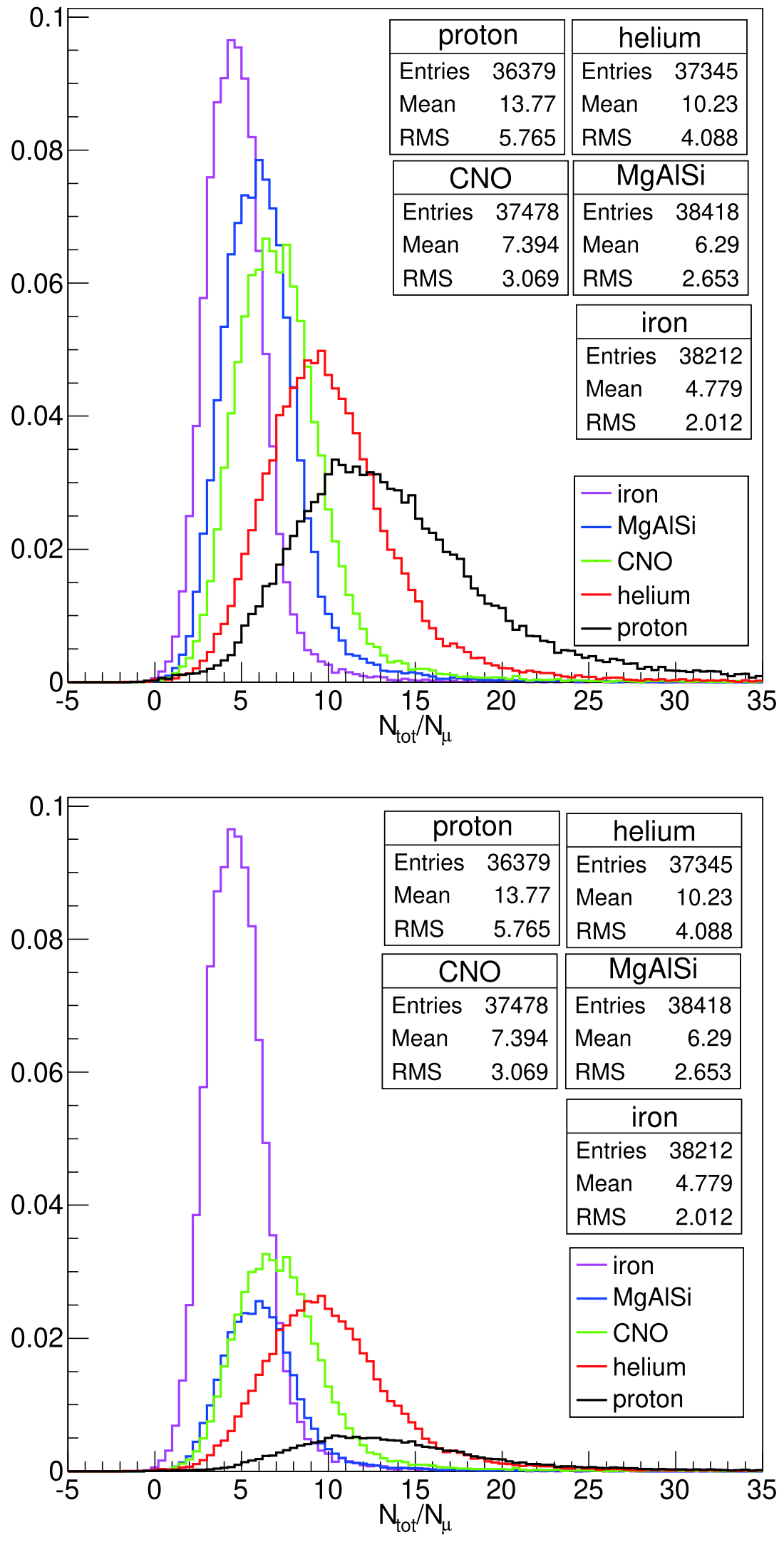

**Fig. 42.** (color online) Upper: Distributions of the inverted muon content $1/C_{\mu}$ for 5 cosmic-ray groups, i.e. proton, helium, CNO, MgAlSi, Iron (left). Lower: assuming the composition of the Horandel model [193].

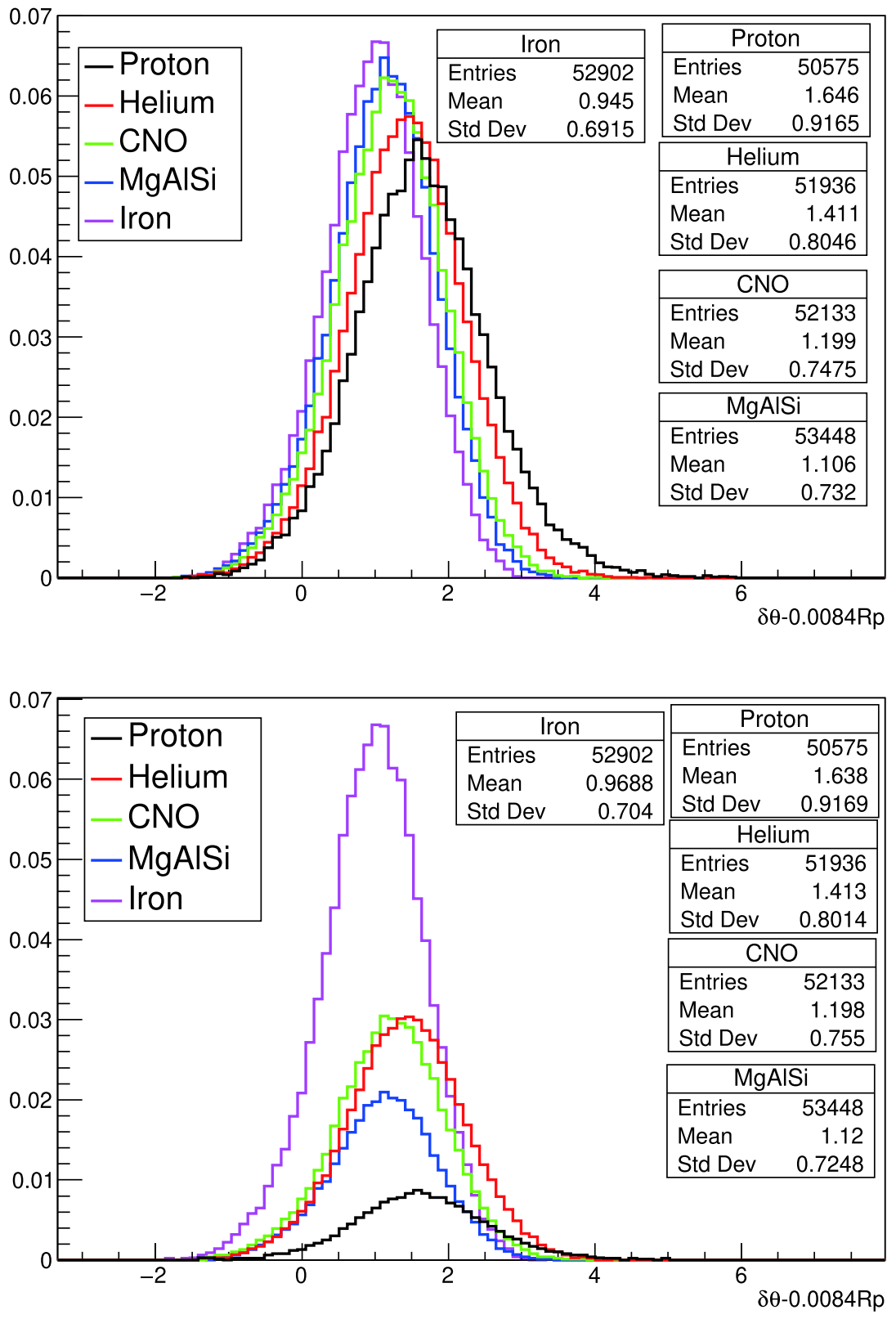

**Fig. 43.** (color online) (upper) Distributions of angular offset with $R_{\rm p}$ correction for 5 cosmic-ray groups, i.e. proton, helium, CNO, MgAlSi, Iron (left). (lower) Assuming the composition of the Horandel model [193].

### 1. *Selection of Iron induced showers at energies above 10 PeV*

The cuts $1/C_{\mu} < 6$ and $\delta\theta - 0.0084R_{\rm p} < 1.8°$ result in a selected sample of iron showers with a purity of 70% at 10 PeV and 85% at 100 PeV. The effective aperture, defined as the geometrical aperture times the selection efficiency, is about $3.4 \times 10^5$ $\rm m^2$ sr, allowing a collection of about 16000 /year iron showers above 10 PeV assuming the Horandel composition model. The shower energy is measured by both detectors KM2A and WFCTA with an expected resolution of about 20 % over a wide energy range (see Sec. II.C.1).

Given a sample of a single cosmic ray element with a purity of 70% or better, the energy reconstruction of the shower is rather straightforward by using the total number of Cherenkov photons in the shower image. This minimizes the uncertainty due to the unknown composition. The total number of photons has been proved to be a good energy estimator because the resolution function is symmetric Gaussian with the bias less the 5%. This is a good feature of the Cherenkov technique in the power-law-like spectrum measurement with minimized distortion. The other good feature of the technique is that the energy resolution is almost a constant better than 20% over a wide energy range. This is very important in finding the structures of the spectrum if there are, such as the knee. Every part of the spectrum is equally measured with consistent resolution. Both the resolution and the reconstruction bias as functions of the shower energy are shown in Fig. 44. With the reconstructed energy of selected showers, the expected spectrum of iron showers is shown in Fig. 45 by solid red squares [201]. The flux in the last bin near 100 PeV corresponds to about 164 events/year. The knee, if it is there, will be discovered with high significance in one year operation of the hy-brid observation using LHAASO instruments.

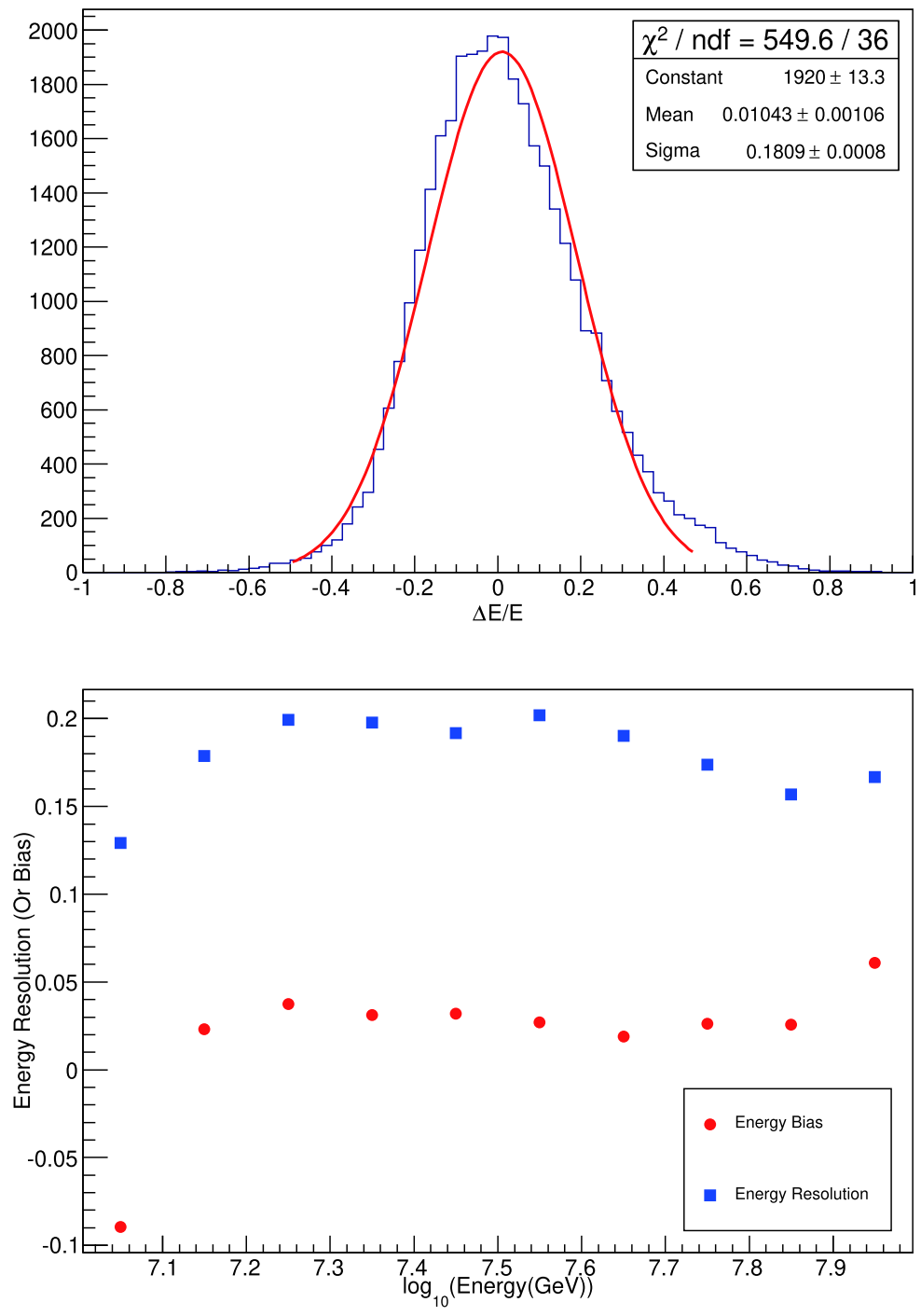

**Fig. 44.** (color online) The energy resolution of 30 PeV pure iron showers using the total numbers of Cherenkov photons in shower images well contained in the telescope FOV (upper). Resolution and bias as a function of the shower energy in the 10-100 PeV energy range (lower).

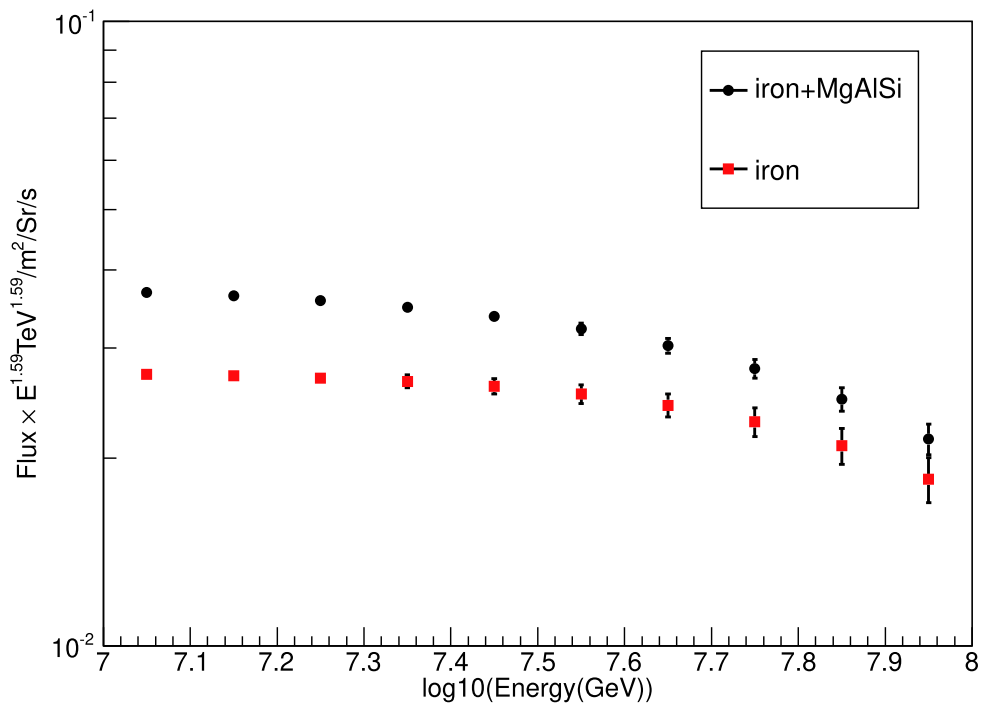

**Fig. 45.** (color online) The expected spectra of pure irons and of the mixed heavy nuclei group over the energy range from 10 to 100 PeV in one year of data taking. The Horandel model is assumed. The iron knee-like feature expected below 100 PeV should be readily observed.

### 2. *Selection of Iron and MgAlSi induced showers above 10 PeV*

Assuming that the iron spectrum above 10 PeV has a constant index below its knee, it is possible to observe the bending of the spectrum of mixed irons and Mg, Al, Si nuclei, if we can separate the combined event group out of all events. It turns out an easier job with a purity better than 70% and even 90% around 100 PeV. The gain is that the effective aperture increases, and reaches to $4.2\times10^5\,m^2sr$ due to relaxing the cut on the angular offset measured by the Cherenkov telescopes. The total number per year of selected showers is about 26000 above 10 PeV and 200 events in the last bin near 100 PeV. In fact, the difference between the two types of showers is not very significant. The down side is the energy resolution becoming slightly worse due to the mixing of composition. From 18% for pure iron showers, the resolution worsens to 20% for the mixed heavy samples. The expected spectrum of the mixed sample is plotted in Fig. 45 by solid black circles. The Horandel composition model is assumed.

### 3. *Summary*

The LHAASO capability to measure the spectra of the heavy components beyond 10 PeV as well as the spectra of the proton and Helium nuclei at PeV energies (Secs. II.D, II.E) is a powerful tool to ascertain the evolution of the cosmic-ray spectra and clarify the phenomena associated to their detailed structures. This will greatly enhance our knowledge of the mechanisms governing production and propagation of galactic cosmic rays.

## G. Implementing LHAASO with neutron detectors

When arriving at Earth, high energy cosmic rays interact with the air nuclei producing extensive air showers (EAS). They consist of a core of high energy hadrons that continuously feed the electromagnetic part of the shower, mainly with photons from neutral pion, kaon and eta particle decays. Nucleons and other high energy hadrons contribute to the hadronic cascade. High energy hadrons, which constitute the EAS skeleton, may carry important information for multi-parameter correlation studies, since some hadronic observables, primarily the hadron number/electron number correlation, depend on the nature of the particle inducing the shower [202, 203]. Thus, the detection of high energy hadrons, designed to improve the discrimination power in these analysis, is highly advisable. A way to deal with this problem avoiding the use of huge and expensive HCALs was brought out in [203-205]. In these papers the detection of thermal neutrons generated by EAS hadrons is proposed.

Indeed, due to the tight correlation between the air showers hadrons and the thermal neutrons, this technique can be envisaged as a simple way to estimate the number of high energy hadrons in EAS.

It is well known that hadrons interacting with ambient matter (air, building, ground, etc.) produce evaporation neutrons due to nuclei disintegration. The neutrons have no charge and lose energy only by scattering. If the medium is a good moderator, i.e., the absorption cross section is much less than the scattering cross section, the neutrons lose energy via scattering down to the thermal ones (moderation process) and then live in the matter until capture. Evaporation neutrons need about 0.5 ms to thermalize in rock (concrete). Neutrons are generated abundantly, up to 2 orders of magnitude more than parent hadrons. The mean number of evaporation neutrons <n> produced by hadrons in a 120 cm layer of surrounding soil (about 3 hadron interaction lengths) and/or construction materials can be estimated using the empirical relationship

$$< n > \approx 36 \times E_{\rm h}^{0.56}, \tag{9}$$

where $E_h$ is the hadron energy in GeV [206]. A large fraction of the evaporation neutrons thermalize, so that recording thermal neutrons can be exploited to reconstruct the hadron content in the shower. This approach looks very promising for measurements carried out at high altitude. Indeed, since the hadron content in EAS increases with the altitude, an abundant production of thermal neutrons can be predicted for experiments at 4 (or more) km a.s.l., about a factor 10 higher than that at sea level [206]. These considerations suggested the development of a simple and cheap thermal neutron detector, to be deployed over a large area, as 'hadron counter' in EAS experiments at mountain level. This idea led to the development of the EN-detector, made of a mixture of the well-known inorganic scintillator ZnS(Ag) with $^6$LiF, capable of recording both thermal neutrons and charged particles [204, 207].

Thermal neutrons are detected via the capture reaction (940 barns)

$$^6{\rm Li}+n\rightarrow{}^3{\rm H}+\alpha+4.78\ {\rm MeV}, \tag{10}$$

producing a light yield in the scintillator of about 160,000 photons per neutron capture.

However the light output is different for different types of particles. Charged particles produce weak and fast signals comparing with the high amplitude, slow and delayed signals from thermal neutron capture. The first big and fast peak is generated by the large amount of charged particles in the shower front while the smaller delayed signals are due to the thermal neutron capture. Thus these peculiar features make this detector well suitable for operation in the framework of EAS experiments. The amplitude of the fast signal can be used to mea-sure the charged particle density while the delayed sig-nals recorded in a time gate of 10 ms give the number of captured thermal neutrons.

All details concerning construction, operating principles and performance are described in Chapter 1 , Sect. 1.6. In order to check the performance of this detector at a high altitude site, a small array composed of four EN-detectors (PRISMA-YBJ) has been installed inside the hall hosting the ARGO-YBJ experiment at the Yangbajing Cosmic Ray Observatory (Tibet, China, 4300 m a.s.l., 606 g/cm$^2$). The two arrays operated together, and coincident events have been analyzed to gather information on the PRISMA-YBJ performance [208].

The results of the combined operation ARGO-YBJ/PRISMA confirmed the excellent linearity of the EN-detector whose fast output is found proportional to the number of shower particles measured by ARGO-YBJ. Thermal neutrons are found distributed around the shower core with a very narrow lateral distribution (Fig. 46) and their total number is linearly correlated with the truncated shower size (particles within 10 m from the shower axis) measured by ARGO-YBJ. The three $N_{\rho 10}$ selections characterize the cosmic ray energy range from 0.5 PeV to 2 PeV. Both features are consistent with the ones characterizing the EAS high energy hadrons. Indeed, electrons and hadrons are closely related to each other, many experiments proving that the number of hadrons $N_h$ in a shower is almost proportional to the shower size $N_e$, that is $N_h = kN_e^{\alpha}$ with $\alpha$ varying between 0.9 and 1.0 [209]. The analysis of more than two thousands EAS events confirmed that the EN-detectors worked properly at high altitude in combination with an array of particle detectors.

To overcome the problem of $^6$Li procurement, a novel type of ZnS(Ag) scintillator alloyed with $B_2O_3$, with the $^{10}$B isotope about 20%, has been developed to build up the ENDA array (see Chapter 1, Sect. IV.B).

Up to now, ENDA has totally 66 detectors (ENDA-64 and the other two as backup), ready for deploying inside LHAASO to make a hybrid detection of cosmic ray spectrum from 100 TeV to 50 PeV. ENDA will be extended up to 400 detectors with an array area of $10^4$ m$^2$ deployed inside LHAASO to provide additional in-formation for the study of energy spectrum and mass composition of cosmic rays up to 300 PeV. The expected capability to address the mass of the primary cosmic rays is presented in Chapter 1, Sect. IV.C.

### H. Implementing LHAASO with radio detectors

Here we discuss the opportunity to perform radio-detection of extensive air showers (EAS) in combination with LHAASO measurements. In Sec. II.H.1 we present a brief status of EAS radio-detection. We then study in Sec. II.H.2 the possible benefit of radio measurements for LHAASO and finally (Sec. II.H.3) evaluate how the LHAASO detector could be instrumental in the perspective of the foreseen Giant Radio Array for Neutrino Detection [210].

#### 1. *Status of extensive air shower detection*

Creation and acceleration of charges during the devel-opment of EAS induced by high energy cosmic rays naturally generates electromagnetic radiations. The dominant effect is the so-called *geomagnetic* effect [211], corresponding to the drift in opposite directions of positive and negative charges from the shower because of the Lorentz force associated with the Earth magnetic field $\boldsymbol{B}_{\rm geo}$. The resulting charge current produces brief flashes (≤50 ns) of coherent radio emission in the ~10-200 MHz frequency range, linearly polarized along the $\boldsymbol{B}_{\rm geo}\times\boldsymbol{v}$ direction.

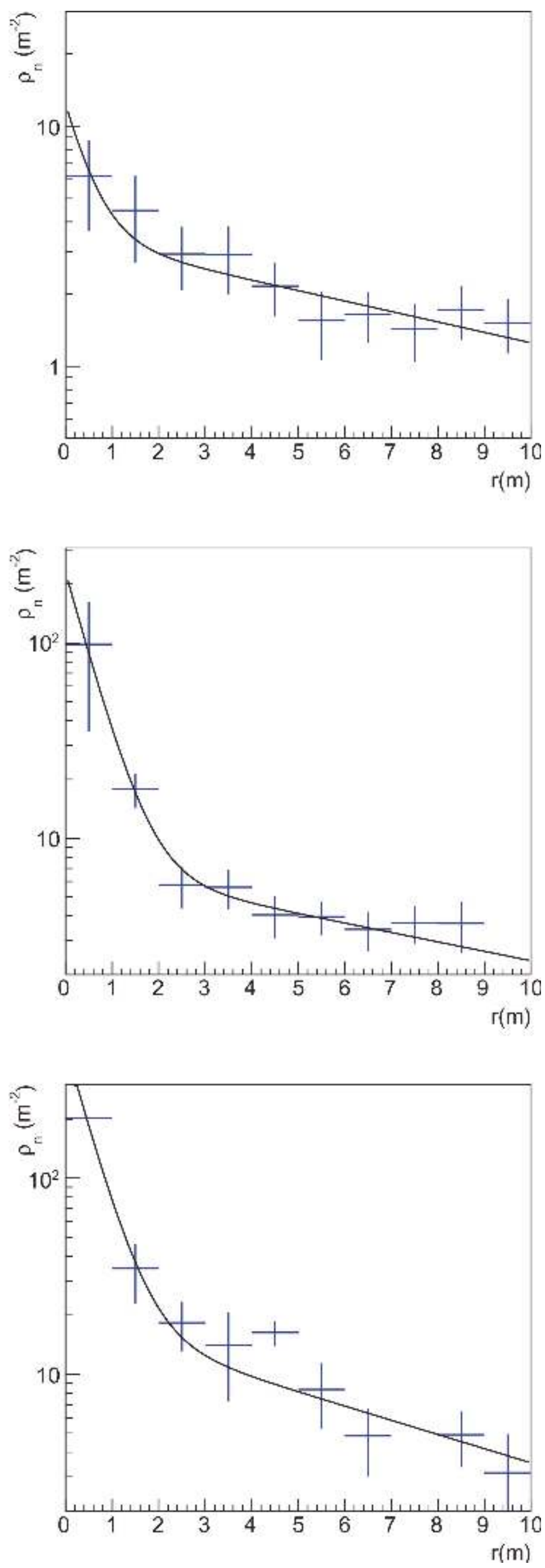

**Fig. 46.** Lateral distributions of thermal neutrons detected at Yangbajing (4100 m a.s.l.) fitted by the double exponential function. $r$(m) is the distance from the shower core, $\rho_n$(m$^{-2}$) is the neutron density. (upper) $\lg(N_{\rho 10}) < 4.8$; (middle) $4.8 < \lg(N_{\rho 10}) < 5.4$; (lower) $\lg(N_{\rho 10}) > 5.4$, where $N_{\rho 10}$ is the truncated size [208], see text. (Reprinted with permission from Elsevier)

Radio emission by EAS was experimentally observed as soon as 1966 [212], but it was not before the new millennium that extensive experimental efforts were carried out in order to establish the radio technique as a valid tool for the study of high energy cosmic rays.

● CODALEMA and LOPES were the two pioneering experiments in the early 2000, with radio arrays composed of few tens antennas deployed over areas ≤ 1 km$^2$, and triggers provided by ground arrays (the KASKADE-GRANDE experiment in the case of LOPES).

● LOFAR is a radio telescope deployed over several countries in Europe. Among other science goals, LOFAR aims at detecting cosmic rays with the central part of the telescope, composed of ~2400 antennas clustered on an area of ~10 km$^2$. This high density of antennas makes LOFAR the perfect tool to study features of the radio emission created by extensive air showers. Air-shower measurements are conducted based on a trigger received from an array of scintillators (LORA). LOFAR comprises two types of antennas, recording radio emission in low-frequency band from 10 to 90 MHz and also in the high-frequency band (110-190 MHz) [213].

● The members of these three collaborations later joined efforts with others to develop the Auger Engineering Radio Array (AERA), with the explicit goal to test if radio-antenna arrays could eventually replace the standard technics (ground arrays or fluorescence detectors) for future UHECRs detectors. This was motivated by the fact that radio antennas were suspected to be cheaper, easily deployable and would require minimal maintenance and would thus be potentially well suited to the giant detector surfaces required for the detection of UHECRs. AERA is an array of 150 radio antennas working in the 30-80 MHz frequency range and deployed over ~17 km$^2$ with array stepsize between 150 and 350 m. AERA is located in a region with a higher density of water-Cerenkov detectors (on a 750 m grid) and within the field of view of the HEAT fluorescence telescope, allowing for the calibration of the radio signal using super-hybrid air-shower measurements, i.e., recording simultaneously the fluorescence light, the particles at the ground and the radio emission from extensive air showers [213].

● Tunka-Rex is the radio extension of the Tunka observatory for cosmic-ray air showers. Its main detector, Tunka-133, is an array of non-imaging photomultipliers detecting the Cherenkov light emitted by the air-showers in the atmosphere in the energy range of $10^{16}$ to $10^{18}$ eV. Tunka-Rex is composed of 25 antennas deployed over 1 $km^2$ [214].

• TREND [215] (Tianshan Radio Experiment for Neutrino Detection) is a setup composed of 50 self-triggered antennas running in the 30-100 MHz frequency range deployed over 1.5 $km^2$ on the site of the 21 CMA radio-interferometer in the Tianshan mountains, Xinjiang Autonomous Province, China. Compared to the above-mentioned projects, all triggered by other types of detectors, TREND specifically focuses on autonomous detection and identification of EAS with radio signals only.

A decade of efforts by these various experiments brought some significant results :

• As the geomagnetic effect is the dominant contribution to the radio signal of air showers, its strength strongly depends on its direction of origin, and more precisely on the geomagnetic angle ($\boldsymbol{v}$, $\boldsymbol{B}_{geo}$). For air showers developing in a direction perpendicular to the geomagnetic field, energies down to few $10^{16}$ eV could be detected by dense arrays like CODALEMA or LOFAR [216]. An efficiency larger than 80% is reached by CODALEMA for energies above $10^{17}$ eV [217]. Detection at low energies is limited by the sky background noise, due in particular to Galactic emission, which significantly affects the signal-to-noise ratio. It should be noted however that, to our knowledge, no specific signal treatment was ever applied to identify low amplitude radio pulses in the data. As both noise (from measurements) and air-shower induced radio waveforms (from simulations) can be determined, a dedicated filtering treatment might allow to dig out EAS-induced radio signals from noise for primary energies down to $10^{16}$ eV.

• LOPES, LOFAR and AERA were able, thanks to their ~ns timing resolution, to reconstruct the direction of origin of the incoming cosmic particle from the radio data with a precision of a fraction of a degree typically [218], using a conical parameterization of the shower front [219].

• As the strength of the electromagnetic field is directly related to the number of particles in the shower (coherent radio emission), it is possible to estimate the energy of the primary cosmic particle from the radio signal in a rather straightforward way. A 17% precision was achieved by AERA [220] and 20% by Tunka [214] (see Fig. 47).

• The radio signal pattern at ground depends on the longitudinal development of the shower, and in particular

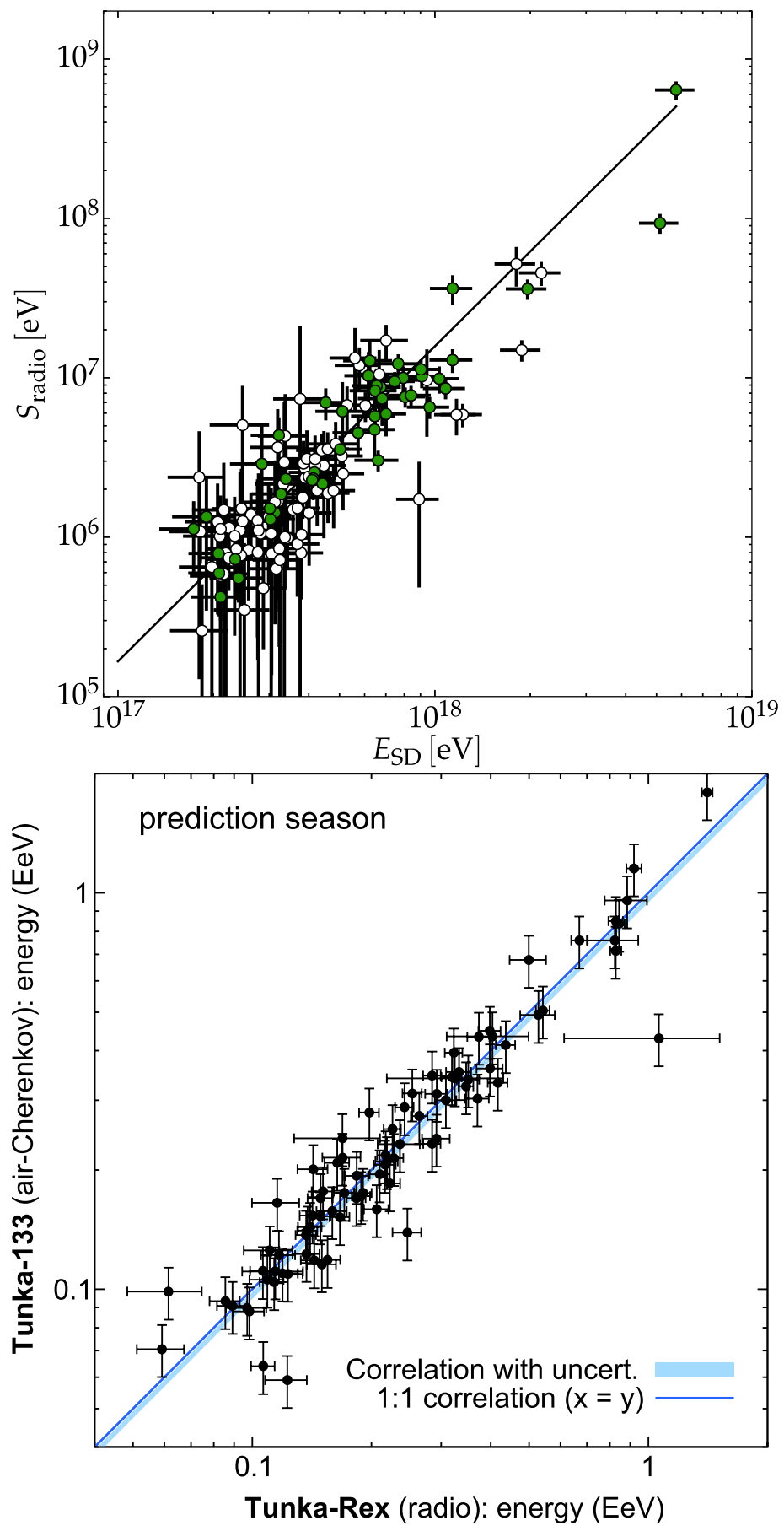

**Fig. 47.** (color online) (upper) Radio-energy estimator $S_{radio}$ as a function of the cosmic-ray energy measured with the Auger surface detector. Green filled circles denote air showers with at least five stations with signal. Open circles denote events with less than five stations with signal and use the surface detector core position. A 17% energy resolution could be achieved for events with 5+ stations triggered. Taken from [220]. (lower) Correlation of the shower energy reconstructed with Tunka-Rex radio and Tunka-133 air-Cherenkov measurement. Taken from [214].

on the position of its maximum of development $X_{max}$, as can be seen from Fig. 48. It is therefore possible in principle to perform a measurement of $X_{max}$ and hence determine the nature of the primary from the radio data. Various technics were used: LOPES used the information on the shape of the radio wavefront (with a smaller curvature radius for showers developing deeper in the atmosphere) to achieve a 140 g/$cm^2$ resolution on $X_{max}$, while simulation indicate that precision as good as 30 g/$cm^2$ may be achieved for denser and/or more extended arrays deployed in quieter radio environment [218]. Tunka-Rex estimated $X_{max}$ with a ~40 g/$cm^2$ accuracy by measuring the slope of the lateral intensity profile of radio footprint at ground (steeper for showers developing deeper in the atmosphere) [214]. LOFAR took advantage of it high-density array to reach a 17 g/cm$^2$ using a similar technique [221]. AERA developed very recently a method based on the measured frequency spectrum (flatter for showers developing higher in the atmosphere), allowing in principle to measure $X_{max}$ from a single antenna only, and reaching a ~20 g/cm$^2$ resolution for a subset of AERA events [222].

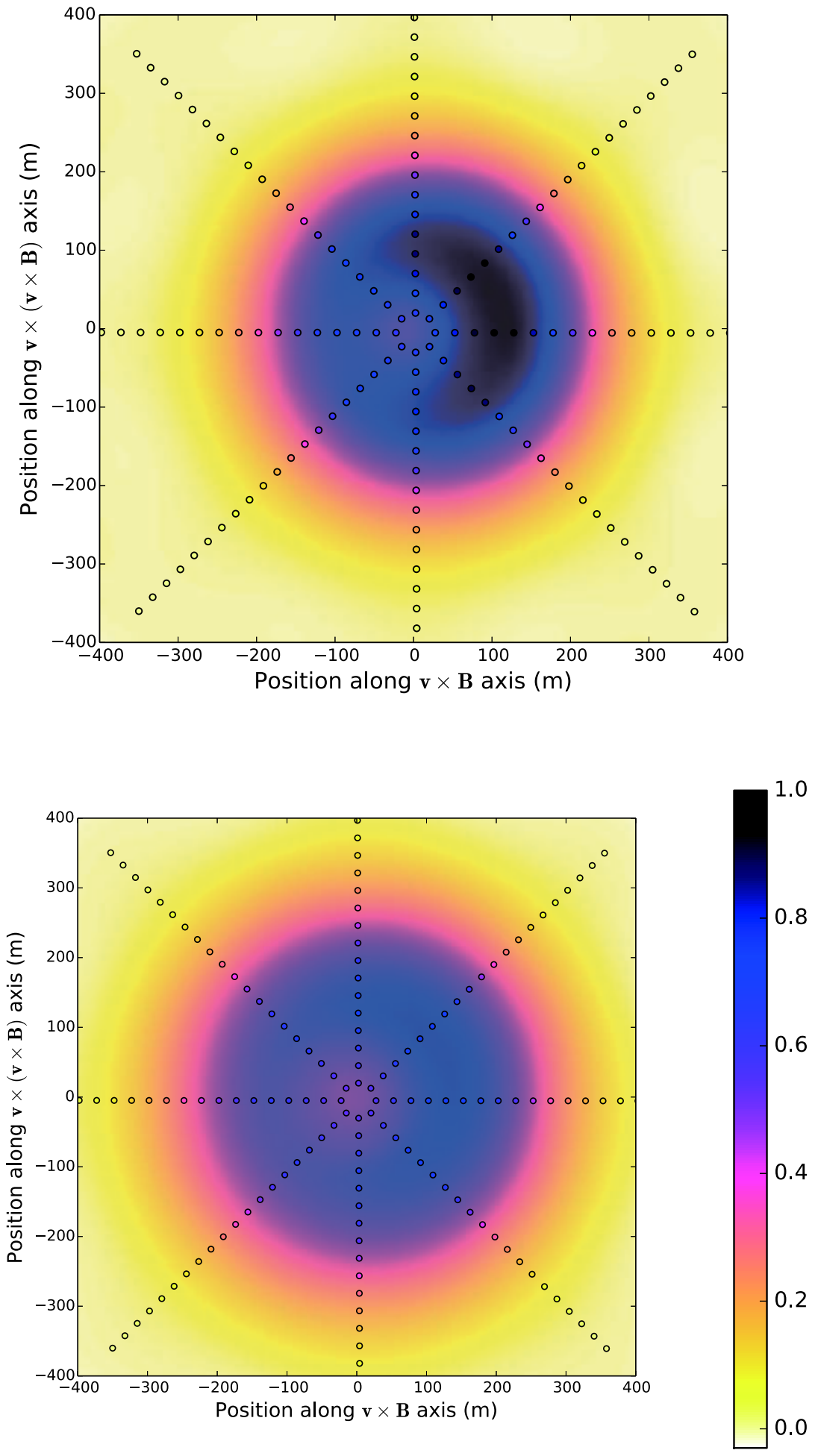

**Fig. 48.** (color online) Radio profiles in arbitrary units for a proton shower with $X_{max}$ = 794 g/cm$^2$ (upper) and an iron shower with $X_{max}$ = 573 g/cm$^2$ (lower). Both showers have an energy of 2.3 10$^{17}$ eV and a zenith angle of 49 degrees. Taken from [221].

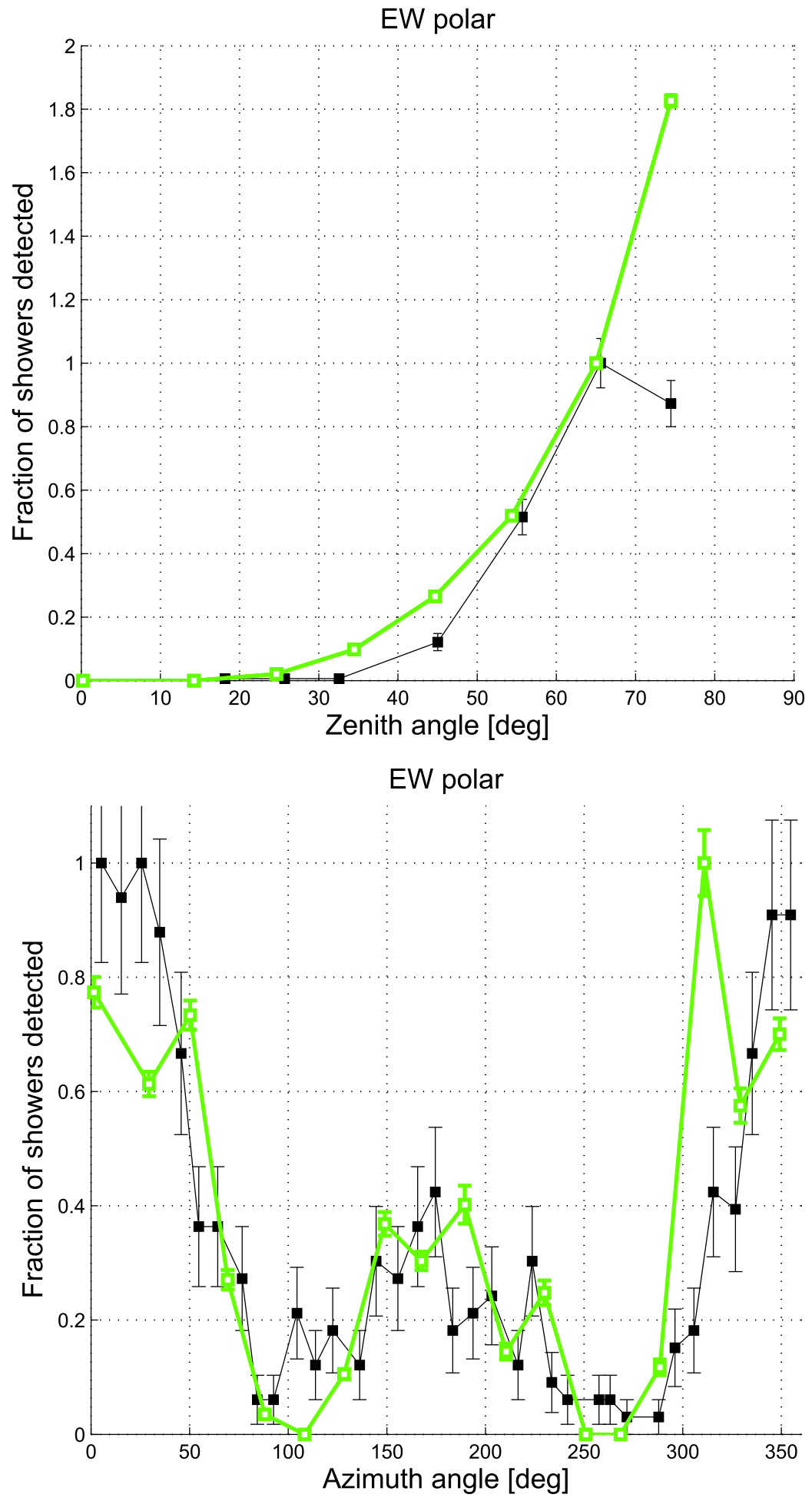

**Fig. 49.** (color online) Distribution of reconstructed zenith (upper) and azimuth (lower) angles for the 465 EAS radio candidates selected in the 317 live days of TREND data (black squares). Also shown are the expected distributions for air showers initiated by protons with $E$ = 10$^{17}$ eV (green empty squares).

● The TREND experiment focused on the detection and identification of air showers based on their radio signals only. To achieve this result, TREND developed a DAQ system allowing for a ~200 Hz trigger rate for each antenna and performed an offline identification of air shower signals based on their specific characteristics, following an algorithm detailed in [223]. TREND could select 465 EAS candidates for 317 live days of data. According to simulations, the distribution of the direction of arrival of these events follows rather well that expected for EAS with energies of 10$^{17}$ eV for zenith angles $\theta \leq 70°$ (see Fig. 49). This result, still to be refined, indicates that it is possible to trigger and identify EAS with a self triggered radio array, with a limited contamination by background events. However TREND detection efficiency was estimated to be around 10% only because of the background rejection cuts applied. Other EAS selection procedures may have to be found to improve the EAS detection efficiency.

The experimental developments above detailed allowed a better understanding of EAS radio emission, thus feeding the various simulation codes [224-226] developed and refined in that period of time, which now fit very well the experimental data. These codes in turn constitute a very valuable tool to further develop the air-shower radio detection technique.

If nice results were achieved by EAS radio detection, some limitations were reached as well. We may stress in particular the fact that the radio emission is very much beamed around the shower axis, with an abrupt exponential drop when moving away from the shower core (signal typically divided by 10 between 100 and 200 m from the shower core for a vertical shower). This feature does not significantly depend on the energy, which implies that arrays of very high density (detector spacing ~50 m) would be necessary to perform EAS radio-detection and reconstruction. This is not realistic for UHECRs detection, which requires huge detection areas. This statement however has to be mitigated by the observation that the EAS radio footprint at ground is much larger for inclined showers [227], as the zone of main electromagnetic emission (mostly around $X_{max}$) is in that case much more distant from ground, and also because the projection of the radio emission cone on a flat ground is, by construction, more elongated for inclined trajectories. Giant radio arrays might therefore be able to perform a competitive study of UHECRs by selected inclined trajectories. This is presently being studied in the framework of the GRAND project [210].

Another major issue for EAS radio-detection is the high rate of background events. Even in remote areas like the TREND site, background radio sources (trains, planes, cars, but even more frequently HV lines and electric transformers) generate event rates that surpass the EAS flux by orders of magnitudes [223]. The DAQ system of a radio array has to take into account this constraint in order to perform autonomous triggering successfully. GRANDproto should allow to determine the EAS detection efficiency and background rejection potential achievable for an autonomous radio array. GRANDproto [228] is an hybrid setup composed of 35 radio-antennas with a DAQ guaranteeing a 0% dead time for an individual antenna trigger rate up to 5 kHz, running in parallel to a cosmic ray detection array of 21 scintillators. EAS radio-candidates will be selected based on the events polarization information measured by the triggered antennas, while the scintillator array will be used as a cross-check to the EAS nature of the selected radio candidates, thus allowing a quantitative determination of the background rejection potential of the array. GRANDproto will be fully deployed in summer 2016.

### 2. *Benefit of radio-measurements for LHAASO*

Here we only give some hints on the potential added value of EAS radio measurements for LHAASO, in the light of the status presented in Sec. II.H.1. We should stress however that a rigorous response to this issue would require a dedicated study based on detailed simulations taking into account the specifics of LHAASO (altitude, magnetic field at the detector location, electromagnetic background,...) in order to determine what goals and performances would be actually achievable.

In the light of LOPES, Tunka or AERA results for example, it seems realistic to think that a radio array deployed at the LHAASO location could provide an independent measurement of cosmic ray parameters (energy and $X_{max}$ in particular) with good precision, provided the electromagnetic background level is low enough at the LHAASO site, and that other detectors (PMTs in particular) are well shielded. There is no reason to think that performances similar to present arrays (energy resolution of 15%-20%, $X_{max}$ resolution in the range of 20 to 40 g/cm$^2$) should not be achievable. An external trigger could be provided by LHAASO detectors to circumvent the challenges of radio autonomous trigger mentioned in the previous section. We shall stress however that the threshold for radio is presently ~$10^{17}$ eV for the energy measurement, and even higher for $X_{max}$. It is possible that a very dense array (~50 m detector spacing), and a dedicated signal treatment to improve signal-to-noise ratio could lower this threshold, but this is hard to assess *a priori*. We suggest that a radio array may be interesting as a complement to the high energy end of the KM2A array measurements ($X_{max}$), or as a complement to WFCTA in order to better constrain the shower geometry through the measurement of the shower core position.

### 3. *LHAASO and GRAND*

GRAND [210] is a proposal to build a giant radio array (total area of 200000 km$^2$) primarily aiming at detecting cosmic neutrinos. The project is still at a very early stage, and many issues have to be studied and solved before the project comes to reality. Preliminary sensitivity studies are however extremely promising, with an expected sensitivity guaranteeing -even for the weakest expected fluxes [229]- the detection of the so-called *cosmogenic neutrinos* produced by the interaction of UHECRs with CMB photons during their cosmic journey [230, 231].

Among the many steps to be completed before GRAND comes to life, an important one will consist in deploying a ~1000 km$^2$ engineering array (GRAND-EA) composed of ~1000 antennas in order to validate the technological choices defined for GRAND. This array will obviously be too small to perform a neutrino search, but cosmic rays should be detected above $10^{18}$eV. Their reconstructed properties (energy spectrum, directions of arrival, nature of the primaries) will enable us to validate this stage, if found to be compatible with the expectations. Even if the two detectors areas differ a lot, it may be interesting to consider in more details a deployment of GRAND-EA around the LHAASO experimental site. An independent detection by the 2 setups of a statistically significant number of cosmic ray events would indeed be very valuable for the evaluation of GRAND-EA perform-ances. Given the present status of the GRAND proposal, GRAND-EA could not be deployed before 3 or 4 years.

# Chapter 5 Dark Matter and New Physics Beyond the Standard Model with LHAASO

Xiaojun Bi(毕效军)[1★¶†] Andrea Addazi[2,3¶] Konstantin Belotsky[4¶] Vitaly Beylin[5¶] Marco Cirelli[6¶]
Arman Esmaili[7¶] Nicolao Fornengo[8,9¶] Qingyu Gan(甘庆余)[2¶] Michael Kachekriess[10¶] Maxim Khlopov[11,4,5¶]
Vladimir Korchagin[5¶] Alexander Korochkin[11¶] Vladimir Kuksa[5¶] Antonino Marciano[12,3¶] Andrei Neronov[11,13¶]
Paolo Panci[14¶] Roman Pasechnick[15¶] Alexander Sakharov[16,17¶] Filippo Sala[6¶] Giuseppe Di Sciascio[18¶]
Dimiri Semikoz[11,4¶] Pasquale Dario Serpico[19¶] Nikolay Volchanskiy[5,20¶] Pengfei Yin(殷鹏飞)[1¶]

[1]Key Laboratory of Particle Astrophysics, Institute of High Energy Physics, Chinese Academy of Sciences, Beijing 100049, China
[2]Center for Theoretical Physics, College of Physics Science and Technology, Sichuan University, Chengdu 610065, China
[3]Laboratori Nazionali di Frascati INFN, Frasacati (Rome), Italy, EU
[4]National Research Nuclear University MEPHI (Moscow State Engineering Physics Institute), 31 Kashirskoe chaussee, Moscow 115409, Russia
[5]Institute of Physics, Southern Federal University Stachki 194, Rostov on Don 344090, Russia
[6]LPTHE, CNRS and Sorbonne Universit'e, 4 Place Jussieu, F-75252, Paris, France
[7]Departamento de Fisica, Pontificia Universidade Catolica do Rio de Janeiro, Rio de Janeiro 22452-970, Brazil
[8]Dipartimento di Fisica, Universita di Torino, Via P. Giuria 1, 10125 Torino, Italy
[9]Istituto Nazionale di Fisica Nucleare, Sezione di Torino, Via P. Giuria 1, 10125, Torino, Italy
[10]Institutt for fysikk, NTNU, 7491 Trondheim, Norway
[11]Universite de Paris, CNRS, Astroparticule et Cosmologie, F-75013 Paris, France
[12]Department of Physics and Center for Field Theory and Particle Physics, Fudan University, Shanghai 200433, China
[13]Astronomy Department, University of Geneve, Ch. d'Ecogia 16, 1290, Versolx, Switzerland
[14]Dipartimento di Fisica, Universita di Pisa and INFN, Sezione di Pisa Largo Pontecorvo 3, 56127 Pisa, Italy
[15]Department of Astronomy and Theoretical Physics, Lund University, 221 00 Lund, Sweden, EU
[16]Experimental Physics Department, CERN, Geneva 23 CH 1211, Switzerland
[17]Department of Physics, New York University, New York NY; United States of America
[18]INFN sezione Roma Tor Vergata, I-00133 Rome, Italy
[19]LAPTh, Univ. Grenoble Alpes, USMB, CNRS, F-74000 Annecy, France
[20]Bogoliubov Laboratory of Theoretical Pjysics, JINR, Dubna, Russia

**Abstract:** In order to reveal the nature of dark matter, it is crucial to detect its non-gravitational interactions with the standard model particles. The traditional dark matter searches focused on the so-called weakly interacting massive particles. However, this paradigm is strongly constrained by the null results of current experiments with high precision. Therefore there is a renewed interest of searches for heavy dark matter particles above TeV scale. The Large High Altitude Air Shower Observatory (LHAASO) with large effective area and strong background rejection power is very suitable to investigate the gamma-ray signals induced by dark matter annihilation or decay above TeV scale. In this document, we review the theoretical motivations and background of heavy dark matter. We review the prospects of searching for the gamma-ray signals resulted from dark matter in the dwarf spheroidal satellites and Galactic halo for LHAASO, and present the projected sensitivities. We also review the prospects of searching for the axion-like particles, which are a kind of well motivated light pseudo-scalars, through the LHAASO measurement of the very high energy gamma-ray spectra of astrophysical sources.

## I. THEORETICAL MOTIVATION AND BACKGROUND OF HEAVY DARK MATTER

Over the past century, a growing number of astrophysical and cosmological observations has led to overwhelming evidence for the so-called "dark matter" (DM), which is non-luminous and non-relativistic (hence dubbed *cold*). While it is not unusual in astrophysics to detect unseen objects gravitationally first or exclusively, the existence of DM cannot be explained in the current standard model (SM) of particle physics plus general relativity.

† E-mail: bixj@ihep.ac.cn

★ Editors ¶ Contributors. All authors contribute equally to the work.

Given the universality of the gravitational force, in order to identify the nature of DM, it is required to detect other interactions of DM. Candidates for cold DM have been proposed in an extremely wide range of masses and interaction strengths. These parameters notably control its main observable: the DM average cosmological abundance, now measured to percent precision thanks to Cosmic Microwave Background (CMB) data, which must be correctly reproduced within a specific production mechanism.

Despite the huge parameter space, over the past few decades the dominant DM searches have overwhelmingly focused on the so-called Weakly interacting massive particles (WIMPs) paradigm, which features stable particles with masses in the electroweak scale region (give or take one order of magnitude) and couplings to the SM of electroweak strength, produced as thermal relics in the early universe. This has been promoted especially by

(1). The rather predictive thermal freeze-out production mechanism.

(2). A widespread theoretical expectation of new particle physics (such as Supersymmetry, Extra Dimensions etc.) at or slightly above the weak scale in order to fix the hierarchy problem of the Higgs mass. Such new physics produces (or can easily accommodate) also a WIMP DM particle as a byproduct.

(3). The numerous approaches that such a possibility would open for experimental searches, with technologically accessible and mature technologies: via nucleus recoils induced by DM scattering in underground experiments (direct detection), via direct searches at colliders, or indirectly detecting the SM byproducts of DM annihilation (the same process setting their relic abundance in the early universe) in space or ground-based observations of astrophysical fluxes (indirect detection).

Null results of all the searches at point (3). have often pushed the surviving WIMP candidates to heavier masses, notably above the kinematical production threshold of collider searches. There is thus a renewed interest of traditional DM searches above TeV scale, where both forthcoming Imaging Cherenkov Telescopes (notably CTA) and the Large High Altitude Air Shower Observatory (LHAASO) may contribute.

On the other hand, in recent years the trend has been to tackle the DM problem on its own, exploring alternatives to the baseline production mechanism (point (1).), to break more and more free from the "theoretical injunction" (point (2).) in building DM models. More and more studies explore alternative detection strategies, channels, and energy windows, compared to the limited framework of point (3). above. In that sense, the window roughly spanning the energies 0.1-10 PeV has become an interesting new target for DM modeling, for a variety of reasons. We can loosely classify them as:

1. Models explicitly built to account for a DM candidate, with mass and interaction scales emerging rather from the broader possibilities allowed for a viable production than from theoretical bias. In the standard freeze-out picture, an upper limit on the DM mass, of order 100 TeV, follows from the unitarity (i.e. the sum of probabilities equals one) upper limit on the DM annihilation cross section [1], see [2] for a recent appraisal. For the sake of logical organisation, let us classify the models of heavy DM in terms of the assumptions they drop, in the derivation of that result.

1.1 A straightforward option is that DM was never in chemical equilibrium with any bath. In this case DM can be produced in the right abundance by a multitude of mechanisms [3], just to name a few via freeze-in [4], via (p)reheating and/or inflaton decays (see the recent [5] and references therein), at phase transitions [6, 7], at inflation production and decay of heavy Gravitinos or Polonyi fields into the lightest supersymmetric particles (LSP) [8– 10] etc. Such DM candidates can easily have the masses needed to give signals at LHAASO, e.g. via their decays.

1.2 An option that allows to keep DM at equilibrium with some (SM or 'dark') bath is to modify the standard radiation-dominated cosmology. A first possibility to achieve this is by an early matter-dominated phase, where the decay of massive particles dilute the density of preexisting relics, thus allowing for very large DM masses [11]. Models where the DM abundance is set by annihilations into lighter mediators ('Secluded DM' [12]) realise the above picture if mediators are long-lived enough [13, 14], and allow for DM masses as heavy as a few EeV [15] ('Homeopathic DM' [15]). Importantly for LHAASO, they have been shown to give indirect signals at high-energy telescopes [15]. Early vacuum-energy domination offers a second possibility to dilute the over-abundance of heavy DM, via the resulting period of exponential expansion of the universe ('Supercool DM' [16]).

1.3 One can even not depart from radiation domination and keep DM at equilibrium with a bath, provided it is a dark one. If that bath is at a temperature much smaller than the SM one, then the unitarity limit on the cross section can allow for DM masses as heavy as tens of PeV if DM freezes-out while relativistic [17].

1.4 Models of heavy DM have also been proposed where both DM is at chemical equilibrium with the SM bath, and radiation domination is not modified. They require the DM abundance to be set either by long chains of nearest-neighbour interactions [18], allowing for DM as heavy as $\sim 10^{14}$ GeV, or by processes like DM $\zeta^+ \to \zeta\zeta$ with $\zeta$ any slightly lighter particle, allowing for DM masses as large as $\sim 10^9$ GeV ('Zombie DM') [19].

2. Models at some high energy scale aimed at solving other issues of the Standard Model, that can accommodate DM at similar scales via one of the production mechanisms mentioned at point (1). "Traditional" constructions like supersymmetry (SUSY) and confinement constitute an only-partly explored playground in this respect, as we now sketch with a few examples. Models with a large SUSY-breaking scale can predict heavy DM in the form of a Gravitino, see e.g. [20] and references therein. Moving to lower SUSY-breaking scales, in gauge mediation new stable baryon-like composite particles, belonging to the strongly-interacting SUSY breaking sector, provide viable DM candidates [21–23]. SUSY also naturally accommodates dilution of relics via early matter domination, see e.g. [23]. Confining theories predict stable particles that are viable heavy DM candidates [24], and likewise can feature an early matter dominated phase associated to glueballs, which are naturally long-lived [25, 26]. If the confining phase transition is supercooled, then these models realise instead the picture of Supercool Dark Matter [27].

3. Models inspired by phenomenological puzzles in the 0.1-10 PeV energy range, notably the origin of the astrophysical neutrino flux discovered by IceCube [28–33]. If interpreted as a byproduct of DM, it is more straightforward to reconcile them with a *decaying* DM [34–51]. In these scenarios, DM is a non-thermal relic, whose abundance is set by particle decays, freeze-in, or inflationary-related mechanisms. Often, it is linked to the neutrino mass generation mechanism. It is important to remark that the sensitivity of LHAASO telescope, for large angular diffuse gamma-ray flux, will be an important discriminator to distinguish among astrophysical or heavy DM decay explanations for IceCube neutrinos [52, 53].

Similar to the WIMP paradigm, it seems to us that the PeV regime may establish itself as one of the new paradigms in the future, and therefore enjoy the synergy of theory and experiment that has made WIMP searches so prominent in the recent past. LHAASO offers a particularly important opportunity, in that respect. Note that many heavy DM models predict signals also at other experiments, from telescopes to colliders, from gravitational wavedetectors to cosmological surveys, thus widening the context of relevance for the LHAASO physics program.

## II. PHENOMENOLOGY AND NUMERICAL EXPECTATIONS OF DARK MATTER SIGNALS

The number of the particles of type $i$ per unit energy $E$ per unit volume per unit time injected at position $\mathbf{x}$ and time $t$ via annihilation of the self-conjugated DM particles writes

$$Q_i(\mathbf{x},t,E) = \frac{\langle\sigma v\rangle}{2}\frac{\rho^2_{\rm DM}(\mathbf{x},t)}{m^2_{\rm DM}}\frac{{\rm d}N_i(E)}{{\rm d}E}\,. \tag{1}$$

For the decaying DM particles with lifetime $\tau_{\rm DM} = \Gamma^{-1}_{\rm DM}$, the number of injected particles writes

$$Q_i(\mathbf{x},t,E) = \frac{\rho_{\rm DM}(\mathbf{x},t)}{\tau_{\rm DM}\, m_{\rm DM}}\frac{{\rm d}N_i(E)}{{\rm d}E}. \tag{2}$$

In general, knowing the signal at the emission point is not enough, since we need the flux detectable at the Earth. For charged particles, one key difficulty is that they do not retain directionality due to deflections in interstellar magnetic fields: CR trajectories are similar to random-walks typically described via a diffusion-loss equation. Although we will not cover this channel in detail, note that it is of some interest for LHAASO as well, since at very least energetic $e^\pm$ final states are responsible for *secondary* gamma-rays, notably via Inverse-Compton up-scattering of background photons, i.e. galactic radiation fields and the ubiquitous CMB.

Even limiting oneself to prompt photons, it is important to notice that absorption is significant at LHAASO energies. In particular, LHAASO will be the *first experiment to be sensitive to gamma-ray absorption even for Galactic sources*. If we define the optical depth to photon-photon pair production $\tau$, the differential flux (number of particles per unit time, energy, surface, and solid angle) writes as an integral of the above term $Q$ along the line of sight (below, $b$ and $l$ indicate latitude and longitude in Galactic coordinates, respectively)

$$\frac{{\rm d}\Phi_i}{{\rm d}E}(E,b,l) = \frac{1}{4\pi}\int_0^\infty Q_i(\mathbf{x},t,E)\,{\rm e}^{-\tau(E,s,b,l)}\,{\rm d}s. \tag{3}$$

For a spherically symmetric DM distribution (as typically assumed for the halo of our Galaxy) one has $\rho_{\rm DM}(s,b,l) = \rho_{\rm DM}[r(s,b,l)]$ where

$$r(s,b,l) = \sqrt{s^2 + R^2_\odot - 2sR_\odot\cos b\cos l}, \tag{4}$$

and $R_\odot = 8.249 \pm 0.045_{\rm syst} \pm 0.009_{\rm stat}$ kpc [54] is the distance of the Sun from the Galactic Center. A benchmark density profile is the Navarro-Frenk-White (NFW) density profile

$$\rho_{\rm DM} = \frac{\rho_0}{(r/r_0)(1+r/r_0)^2} \tag{5}$$

with typical range of the scale radius $10\,{\rm kpc} \lesssim r_0 \lesssim 20$ kpc and the $\rho_0$ corresponding to a local abundance of $\rho_{\rm DM}(R_\odot) \simeq 0.4$ GeV/cm$^3$ (for recent determinations, see e.g. [55, 56].) The profile becomes more uncertain towards the inner Galaxy. Yet, along most directions, the uncertainties on the flux for the case of decaying DM are within a factor two.

The best sensitivity for indirect searches of DM is naively expected by telescopes with the largest "grasp" $G = A\Omega$ [53, 57–59], with $A$ being the area and $\Omega$ the solid angle field of view. The current HAWC and, especially, the forthcoming LHAASO are the detectors with the largest grasp in the very-high-energy $\gamma$-band and are therefore well suited for the DM search.

### A. Signals from dwarf spheroidal satellites

Dwarf spheroidal satellites (dSphs), which are large Galactic DM substructures with high DM densities, are ideal targets for detecting the gamma-ray signals induced by DM annihilation or decay. This search is almost background-free, because of the lack of gamma-rays from astrophysical processes in dSphs. Since LHAASO has strong background rejection power (~1%) and large field of view (FOV) (~ 2 sr), it is promising for LHAASO to investigate the properties of massive DM particles via the gamma-ray observations of dSphs. In this subsection, we discuss the LHAASO sensitivities to gamma signals induced by DM annihilation from dSphs, which are investigated in Ref. [60].

For DM annihilation in a point-like source, the expected gamma-ray flux in an energy bin is given by

$$\Phi_\gamma = \frac{1}{4\pi}\frac{\langle\sigma v\rangle}{2m_{\rm DM}^2}\int_{E_{\rm min}}^{E_{\rm max}}\int \frac{{\rm d}N_\gamma}{{\rm d}E}{\rm e}^{-\tau(E,s)}{\rm d}s\,{\rm d}E \times J. \tag{6}$$

For a certain annihilation channel, the initial photon spectrum $\frac{{\rm d}N_\gamma}{{\rm d}E}$ can be obtained using the `PPPC4DM` package [61, 62]. The $J$-factor is defined as the integral of the DM density squared along the line of sight within a solid angle of $\Delta\Omega = 2\pi(1-\cos\alpha_{\rm int})$. Using the results of the kinematic observation of dSphs, their $J$-factor can be derived from Jeans analysis [63–65].

19 dSphs with large $J$-factors are considered in the analysis. These dSphs are located in the LHAASO FOV with favored declination angles. Compared with the HAWC investigation [66], four more dSphs, namely Draco II, Leo V, Pisces II, and Willman 1, are taken into account. Notice that Refs. [67, 68] provide two sets of $J$-factors with different choices for the integration angle $\alpha_{\rm int}$. One set is derived with a constant $\alpha_{\rm int} = 0.5°$, while the other is derived with varying $\alpha_{\rm int} = \theta_{\rm max}$. $\theta_{\rm max}$ is the maximum angular radius of a certain dSph and can be determined by the observation of the outermost member star. To get a large signal-to-background ratio, the $J$-factor of each dSph is taken to be that derived with a smaller $\alpha_{\rm int}$ as $\min\{\theta_{\rm max}, 0.5\}$.

The main backgrounds in the analysis result from the mis-identified cosmic-ray particles. The expected background number in an energy bin can be calculated as

$$B = \int_{E_{\rm min}}^{E_{\rm max}}\int_{\Delta\Omega}\int_0^T \zeta_{cr}\cdot\frac{{\rm d}\Phi_p}{{\rm d}E}\cdot A_{\rm eff}^p(E,\theta_{\rm zen}(t))\cdot\varepsilon_p(E){\rm d}t{\rm d}\Omega{\rm d}E, \tag{7}$$

where $\frac{{\rm d}\Phi_p}{{\rm d}E}$ is the flux of primary cosmic-ray proton and is assumed to be a single power-law, the observational time $T$ is taken to be one year, and $E_{\rm max}/E_{\rm min}$ is assumed to be 3. In order to include the contributions from heavier nuclei in the primary cosmic-rays, an additional factor $\zeta_{cr} = 1.1$ is introduced. The background number is calculated within a cone around the direction of dSph with a solid angle $\Delta\Omega = 2\pi\times[1-\cos(\max\{\alpha_{\rm int},\theta_c\})]$, where $\theta_c$ is the angular resolution of LHAASO. With increasing photon energy, $\theta_c$ varies from 2° to 0.1°. The effective area of LHAASO $A_{\rm eff}^p$ is taken from Ref. [69], which depends on energy and zenith angle $\theta_{\rm zen}(t)$. In order to show the visibility of dSphs, the values of $r_{\rm eff}$, which is defined as the ratio of the observation time with $\theta_{\rm zen} < 60°$ to total time, are listed in Table 1.

Since the main backgrounds arise from cosmic-ray particles, it is crucial to improve the $\gamma/p$ discrimination in the analysis. Ref. [70] provides the energy-dependent quality factor $Q \equiv \varepsilon_\gamma/\sqrt{\varepsilon_p}$ for WCDA, where $\varepsilon_\gamma$ and $\varepsilon_p$ are survival ratios of gamma-rays and primary protons, respectively. For $\varepsilon_\gamma$~50% and $E > 0.6\,{\rm TeV}$, $\varepsilon_p$ varies within a range of 0.04%-0.11%. In Here $\varepsilon_p$ is taken to be 0.278% for $\varepsilon_\gamma$~40.13% for a conservative calculation.

In order to calculate the LHAASO sensitivity, the mimic observations with only backgrounds are assumed. The event count $N$ in each mimic observation is randomly generated around the expected background number $B$ by Poisson sampling. The impact of the statistical uncertainty of the $J$-factor is also considered as Refs. [71, 72]. For one mimic observation, the likelihood is defined as

$$\mathcal{L}_j = \prod_i \mathcal{L}_{ij}(S_{ij}|B_{ij},N_{ij}) \times \frac{1}{\ln(10)J_{{\rm obs},j}\sqrt{2\pi}\sigma_j} \times\exp-[\log_{10}({\rm J}_j)-\log_{10}({\rm J}_{{\rm obs},j})]^2/2\sigma_j^2, \tag{8}$$

**Table 1.** The astrophysical properties of the 19 selected dSphs. The columns for each dSph denote the name, right ascension (RA.), declination (DEC.), effective time ratio ($r_{\rm eff}$), maximum angular radius ($\theta_{\rm max}$), and the median value and standard deviation of the $J$-factor. The $J$-factor and $\theta_{\rm max}$ of most of the dSphs are taken from Ref. [67]. The $J$-factors of four dSphs marked with asterisks are not provided by Ref. [67]; they are taken from Ref. [68]. From [60].

| Source | RA. (deg) | DEC. (deg) | $r_{\rm eff}$ (year) | $\theta_{\rm max}$ (deg) | $\log_{10} J_{\rm obs}$ (GeV$^2$cm$^{-5}$) |
|---|---|---|---|---|---|
| Boötes I | 210.02 | 14.50 | 0.352 | 0.47 | $18.2\pm0.4$ |
| Canes Venatici I | 202.02 | 33.56 | 0.398 | 0.53 | $17.4\pm0.3$ |
| Canes Venatici II | 194.29 | 34.32 | 0.399 | 0.13 | $17.6\pm0.4$ |
| Coma Berenices | 186.74 | 23.90 | 0.377 | 0.31 | $19.0\pm0.4$ |
| Draco | 260.05 | 57.92 | 0.442 | 1.30 | $18.8\pm0.1$ |
| Draco II* | 238.20 | 64.56 | 0.451 | – | $18.1\pm2.8$ |
| Hercules | 247.76 | 12.79 | 0.348 | 0.28 | $16.9\pm0.7$ |
| Leo I | 152.12 | 12.30 | 0.346 | 0.45 | $17.8\pm0.2$ |
| Leo II | 168.37 | 22.15 | 0.372 | 0.23 | $18.0\pm0.2$ |
| Leo IV | 173.23 | −0.54 | 0.303 | 0.16 | $16.3\pm1.4$ |
| Leo V | 172.79 | 2.22 | 0.314 | 0.07 | $16.4\pm0.9$ |
| Pisces II* | 344.63 | 5.95 | 0.327 | – | $16.9\pm1.6$ |
| Segue 1 | 151.77 | 16.08 | 0.357 | 0.35 | $19.4\pm0.3$ |
| Sextans | 153.26 | −1.61 | 0.299 | 1.70 | $17.5\pm0.2$ |
| Triangulum II* | 33.32 | 36.18 | 0.403 | – | $20.9\pm1.3$ |
| Ursa Major I | 158.71 | 51.92 | 0.432 | 0.43 | $17.9\pm0.5$ |
| Ursa Major II | 132.87 | 63.13 | 0.449 | 0.53 | $19.4\pm0.4$ |
| Ursa Minor | 227.28 | 67.23 | 0.455 | 1.37 | $18.9\pm0.2$ |
| Willman 1* | 162.34 | 51.05 | 0.430 | – | $19.5\pm0.9$ |

where $i$ and $j$ denote the $i$-th energy bin and $j$-th dSph, respectively, $\mathcal{L}(S|B,N)$ is the Poisson distribution with expected signal number from DM annihilation $S$, $\log_{10}(J_{{\rm obs},j})$ is the observed mean value of the $J$-factor, and $\sigma_j$ is the corresponding standard deviation. The value of $\log_{10}(J_j)$ is adjusted to maximize $\mathcal{L}_j$ for given $\langle\sigma v\rangle$ and $m_{\rm DM}$. In order to derive the upper-limit on $\langle\sigma v\rangle$ at 95% C.L., it is required that the log-likelihood including the contribution of DM annihilation with increasing $\langle\sigma v\rangle$ decreases by 2.71/2 from its maximum. Furthermore, a combined likelihood $\mathcal{L}^{\rm tot}=\prod_j \mathcal{L}_j$ can be used to derive an improved upper-limit on $\langle\sigma v\rangle$ in the joint analysis with many dSphs. As an example, the individual sensitivity for single dSph and the combined sensitivity for all the 19 dSphs for the $b\bar{b}$ annihilation channel from one mimic observation are shown in Figure 1.

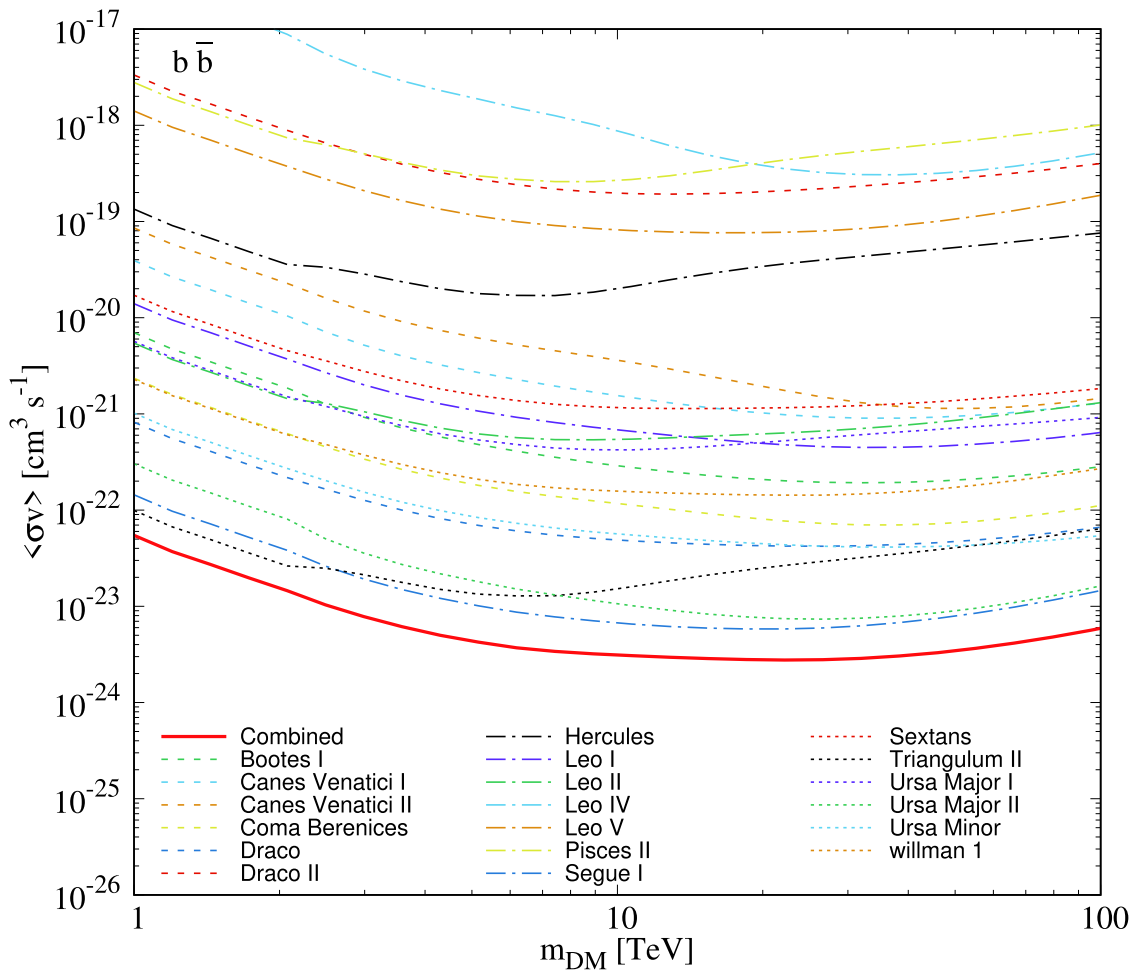

**Fig. 1.** (color online) The expected one-year sensitivities to the DM annihilation cross section $\langle\sigma v\rangle$ at 95% C.L. for the $b\bar{b}$ annihilation channel in one mimic observation. The solid red line represents the combined sensitivity derived in the joint analysis for the 19 selected dSphs; while other lines represent the sensitivities for single dSph. From [60].

It is found that the sensitivities derived from the observations of three dSphs, namely Segue 1, Ursa Major II, and Triangulum II, are much better than other selected dSphs. Therefore, the combined sensitivity in the joint analysis is dominantly determined by these three dSphs, which have large $J$-factors and favorable locations in the LHAASO FOV. Note that Triangulum II has almost the largest $J$-factor among 19 dSphs, but the statistical uncertainty of its $J$-factor is also large. This is because that Triangulum is an ultra-faint dSph and the corresponding kin-ematic data is not sufficient. As a consequent, although Triangulum is located very close to the center of LHAASO FOV, it cannot utterly determine the combined sensitivity. This means that the uncertainty of the *J*-factor should be including in the analysis; otherwise, the sensitivity given by some dSphs may be overestimated.

Since the event number of very high energy photon is small, a series of mimic observations should be performed to reduce the impact of statistical fluctuation. The median values and the two-sided 68% and 95% containment bands of the combined sensitivities derived from 500 mimic observations for five typical DM annihilation channels, including $b\bar{b}$, $t\bar{t}$, $\mu^+\mu^-$, $\tau^+\tau^-$, and $W^+W^-$, are shown in Figure 2. In order to compare with the available able limits from other gamma-ray observations of dSph, the results of the HAWC combined limit [66], Fermi-LAT combined limit [73], HESS combined limit [74], VERITAS Segue 1 limit [75], and MAGIC Segue 1 limit [76] are slso shown.

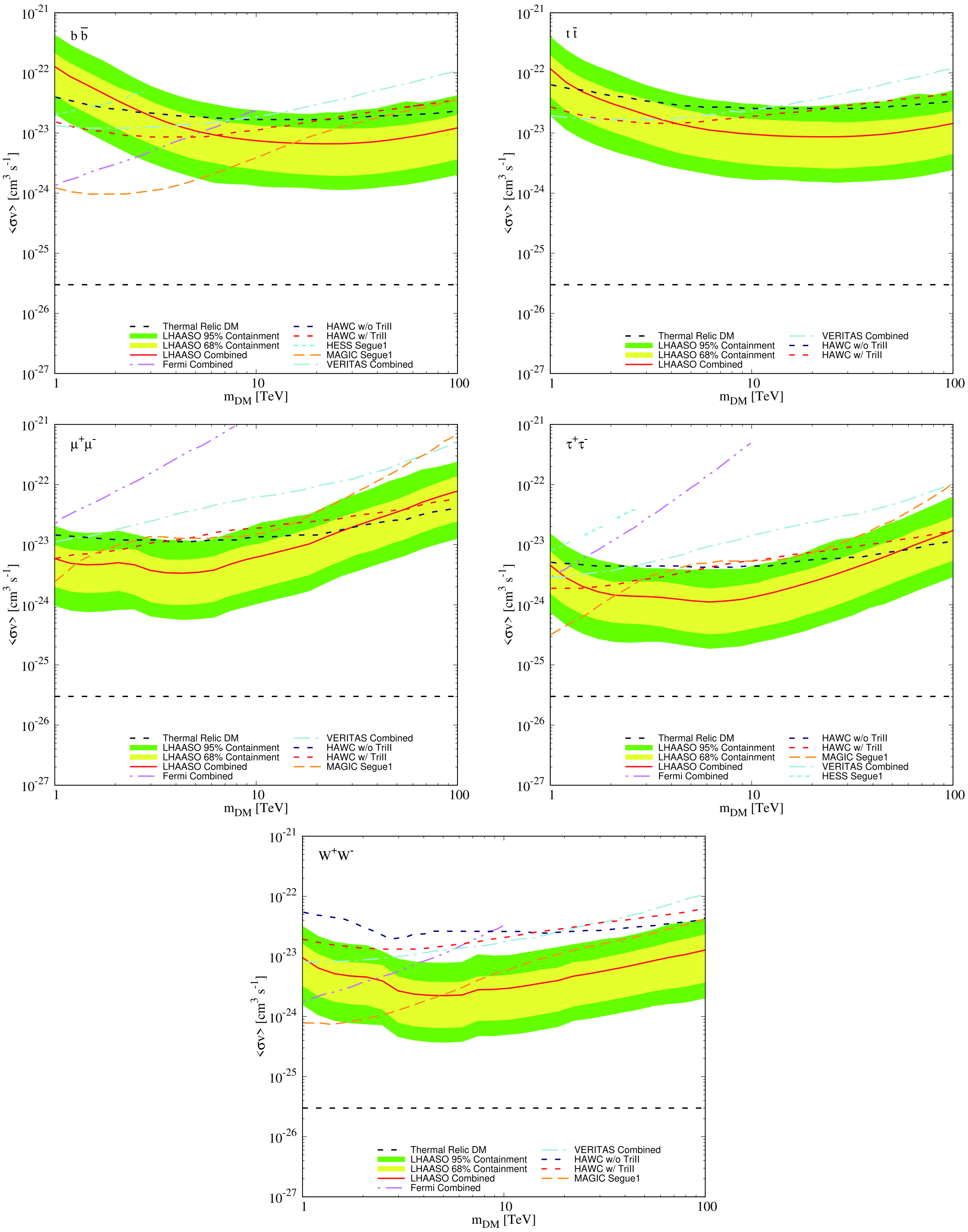

**Fig. 2.** (color online) The combined one-year LHAASO sensitivities at 95% C.L. for five DM annihilation channels, including $b\bar{b}$, $t\bar{t}$, $\mu^+\mu^-$, $\tau^+\tau^-$, and $W^+W^-$. The red solid lines represent the median values. The yellow and green bands represent the corresponding two-sided 68% and 95% containment bands, respectively. The HAWC combined limits [66], Fermi-LAT combined limit [73], VERITAS Segue 1 limit [75], HESS combined dSph limit [74] and MAGIC Segue 1 limit [76] are also shown for comparison. From [60].

Since the initial photon spectrum of the $\tau^+\tau^-$ annihilation channel is hard, the sensitivity for this channel, which can reach $\sim 10^{-24}$ cm$^3$ $s^{-1}$ for $m_{\rm DM} > 1$ TeV, is better than other channels. For the $\tau^+\tau^-$, $W^+W^-$, and $b\bar{b}$ channels, the LHAASO sensitivities are better than the current limits for $m_{\rm DM}$ larger than $\sim 2$ TeV, $\sim 3$ TeV, and $\sim 8$ TeV, respectively. For the $\mu^+\mu^-$ and $t\bar{t}$ channels, LHAASO has good sensitives to explore the DM signals for from $m_{\rm DM}$ in the range of $\sim$ 1 -100 TeV.

The similar analysis procedures can be applied for decaying DM. The flux of gamma-ray signals from DM decays depend on the lifetime of DM $\tau_{\rm DM}$ and the $D$-factor, which is the integral of the DM density along the line of sight. The sensitivities to $\tau_{\rm DM}$ for five DM decay channels from the LHAASO gamma-ray observation of dSphs are investigated in Ref. [77]. 19 dSphs within the LHAASO FOV are studied in the individual and combined analyses. Two dSphs, namely Draco and Ursa Major II, would significantly affect the combined sensitivity, due to their large $D$-factors and suitable locations in the FOV of LHAASO. For $m_{\rm DM} \sim 100$ TeV, the LHAASO sensitivities to $\tau_{\rm DM}$ can reach $\sim 10^{-27}$ s.

### B. Signals from the Galactic halo

In this subsection, we focus on the gamma-ray signals from DM decays in the Galactic halo. For particles with negligible absorption, such as neutrinos, a further, quasi-isotropic contribution to the flux is due to annihilations or decays in the whole universe. This term is in general negligible for annihilating DM (unless one assumes rather extreme DM halo clumpiness), but it is definitely comparable to the Galactic term for decaying DM, and should be taken into account. In particular, this component reduces the scale of variations of the signal across the sky in the neutrino channel. For gamma-rays in the range of LHAASO, however, the extragalactic sky is fully opaque (see e.g. [78]), hence the extragalactic DM contribution is degraded in energy below the pair-production threshold on the EBL, at $E \lesssim 10^2 - 10^3$, GeV. Still, the diffuse gamma ray background measured by Fermi-LAT provides an upper limit to the gamma-ray flux at TeV and PeV energies.

Different strategies are possible for the search of the DM decay signal. The analysis by the HAWC collaboration [79] has adopted an approach in which a signal from the direction around the Galactic Center (more precisely, the region of the Fermi Bubble) is searched for, and the rest of the sky is considered for the background estimate. An alternative possibility is to search for a somewhat weaker (by a factor of two, on average) signal, but extending across the entire sky. An advantage of the latter approach is the larger exposure available for the full-sky search, while a disadvantage is the stricter requirements on the charged-particle vs. gamma-ray separation, due to the modest if not negligible angular variation of the signal. A simple rescaling suggests that a full-sky exposure would provide an increase of the DM signal-to-noise ratio by a factor of $\simeq 2$ on one-year observation time span, compared to the Fermi Bubble region exposure, in spite of the lower average flux. The use of the full (or large) sky exposure, rather than of limited sky region around the Galactic Centre, is important also in the view of uncertainties on the Galactic diffuse $\gamma$-emission unrelated to the DM decay flux. This diffuse emission provides a background on top of which the DM decay signal is searched for. Even if it is possibly sub-dominant compared to the residual charged particle background in $\gamma$-telescopes, it might still be stronger than the DM decay signal.

Intriguingly, the birth of high-energy neutrino astronomy provides a benchmark region in parameter space to search for a possible DM decay signal. The IceCube experiment, completed in 2011, continues observing a flux of high energy ($\gtrsim 10$ TeV) neutrinos significantly in excess with respect to the expected background from atmospheric neutrinos and muons [28–33]. The source(s) of these neutrinos is yet unknown, although based on their almost uniform angular distribution an extragalactic origin or a galactic halo origin is favored. Directional analyses with various classes of astrophysical objects and catalogs are not showing any correlation leading to the conclusion that the contribution of well-known objects, such as blazars, to the observed diffuse neutrino flux is $\lesssim O(10\%)$ [80]. In this context, it has been quite natural to consider unconventional sources for these neutrinos. Also, since neutrinos provide for the first time a window on the 0.1-10 PeV Universe, it may not be so bizarre that new classes of sources can pop up. The potential to answer long-standing problems such as the nature of DM by investigating this energy regime has only recently been entertained.

A decaying DM scenario has gained some attention, mainly due to its minimal assumptions and its testability in future gamma-ray experiments. Interestingly, not only the first PeV-scale events discovered [34], but the whole observed flux of neutrinos by IceCube can be interpreted in this scenario [35], although a multi-component flux arising from both the conventional astrophysical sources and DM also has been investigated, starting from [38]. In a phenomenological approach, the properties of the required DM particle can be deduced from a fit to the neutrino data. In this case, the free parameters are the decay lifetime, the mass and the branching ratios of the DM decay to various standard model particles. The ballpark lifetime is $\sim 10^{27} - 10^{28}$ s and the mass has to be $\gtrsim$ few PeV in order to interpret the highest energy observed events in IceCube (obviously, assuming the multi-component neutrino flux, the DM mass can take any value in our range of interest $\gtrsim 10$ TeV). The highest energy events are typically accounted for via 'hard' leptonic final states, while lower energy events are fitted via soft channels including e.g. gauge bosons and quarks. Part of the flux can also be accommodated via some astrophysical component.

For any decay channel of PeV-scale DM particles explaining or contributing to the IceCube neutrino flux, gamma rays are unavoidably associated decay products, and their Galactic fraction can be observed by LHAASO. The following processes contribute to the expected gamma ray flux at Earth: i) A prompt flux is at very least due to the electroweak corrections ii) A secondary flux is induced by the unavoidable prompt (as well as secondary) charged leptons, via the Inverse-Compton process onto the CMB and star-light which lead to a spectrum of high energy gamma rays where the Galactic part of it contribute to the total flux. Of course the exact spectral shape of the flux depends on the magnitude and profile of the magnetized halo in our Galaxy, which are yet not known very well.

Figure 3, updated from Ref. [52], shows the spectrum of gamma ray yield from the decay of DM with mass 4 PeV and final state branching ratios given by: $\ell^{\pm}W^{\mp} : \overset{(-)}{\nu_\ell} Z : \overset{(-)}{\nu_\ell} h = 1:2:2$. The solid curves show the prompt flux accounting for gamma-ray absorption; different colors represent different directions in the sky. Even this Galactic flux suffers from absorption due to the pair

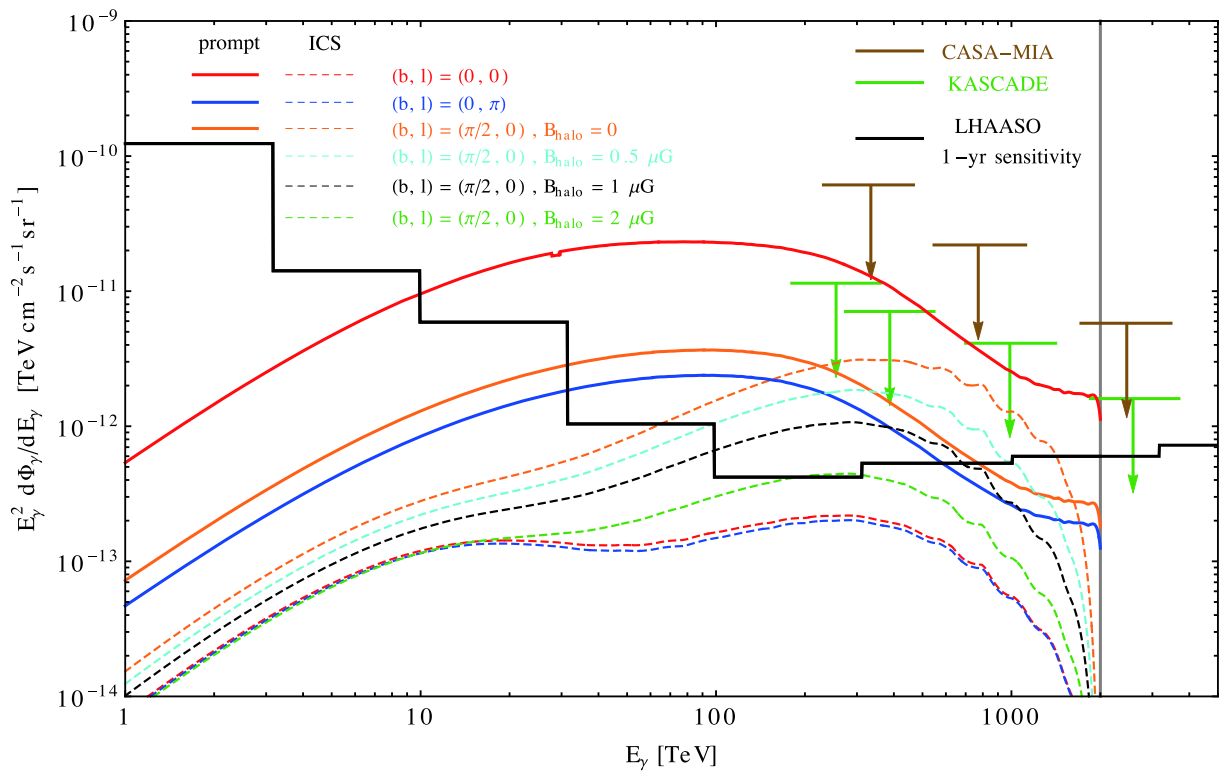

**Fig. 3.** (color online) The gamma-ray flux from DM decay from various directions, with $m_{\rm DM} = 4$ PeV and $\tau_{\rm DM} = 10^{28}$ s, and branching ratios reported in the text. The solid colored curves show the prompt flux, including the absorption of gamma-rays; different colors represent different directions in the sky. The dashed curves show the IC flux, for various assumptions for the constant halo magnetic field, $B_{\rm halo}$, possibly pervading the thick diffusive halo of the Galaxy up to large distances. The green and brown bar lines show the upper bound on gamma-ray flux from CASA-MIA [81] and KASCADE [82], respectively. The black line is an indicative 1 yr LHAASO sensitivity.

production on CMB and star-light, with the suppression reaching ~70% for the Galactic center line of sight. Dashed curves show the flux due to inverse-Compton photons, for various assumptions for the constant halo magnetic field, $B_{\rm halo}$, possibly pervading the thick diffusive halo of the Galaxy up to large distances. While uncertain, it is particularly important in the range above 100 TeV. Note how upper bound from two-decade old experiments CASA-MIA [81] and KASCADE [82] are within one order of magnitude of the expected flux (even less, if they had been sensitive to regions closer to the Galactic center), while LHAASO should provide a definite test of this scenario (the black line indicates its 1 yr nominal sensitivity).

Meaningful bounds can also be obtained thanks to diffuse gamma-ray data by Fermi-LAT in the GeV band, see e.g. [83], exploiting the cascading effect on the extragalactic part of the flux previously mentioned. However, such constraints are rather indirect, depending on the datasets used, the final state considered, and the different assumptions for the contributions to the astrophysical background. LHAASO would allow one to probe the scenario directly and unambiguously, achieving great sensitivity also to sub-leading DM contributions. This was explicitly illustrated in Ref. [84], where the authors estimated the LHAASO sensitivity reach for the decaying DM search, following the approach of Ref. [79]: In each energy bin they compared the DM decay flux levels for different values of $m_{\rm DM}, \tau_{\rm DM}$ with the residual charged particle background levels and calculate by how much is the $\chi^2$ of the fit of the signal+background data is inconsistent with the backgorund-only model in all energy bins. In this way they found the minimal detectable DM decay flux as a function of the DM mass for the model of Ref. [84] of DM decaying into quark-antiquark pair, in turn converted into a maximal measurable DM decay time. The results are shown in Fig. 4. It clearly illustrates how LHAASO will explore DM lifetimes up to $\tau_{\rm DM} \sim 3\times10^{29}$ s over a wide DM mass range $m_{\rm DM} > 10$ PeV. In the mass range 10 TeV$< m_{\rm DM} < 10$ PeV LHAASO will provide a factor of 3-to-10 improvement of sensitivity compared to HAWC. In any case, LHAASO will fully test models where a non-negligible fraction of the IceCube astrophysical neutrino flux is generated by DM decays.

## III. SEARCHES FOR AXION-LIKE PARTICLES

Let us remind that the LHAASO sensitivity to new physics is not limited to DM candidates. As an example, light pseudo-Goldstone bosons (PGBs) can efficiently be converted by the interaction with high energy photons that belong to a magnetic field background, according to the Primakoff effect [86, 87]. PGBs arise ubiquitously in physics, in particular from several possible spontan-eously broken global symmetries beyond the SM. Notable examples are provided by the QCD axion, the familon/archion [88, 89], the majoron [90, 91], the baryo-majoron [92] and and the exoticon [93]. All PGBs interact with the SM particles through terms of the Lagrangian that are suppressed by an energy scale related to the spontaneous symmetry breaking and the decay constant $f$, as $f^{-1}J^{\mu}\partial_{\mu}a$, $J$ standing for the Noether current related to the spontaneously broken global symmetry, and $a$ denoting the light PGB field.

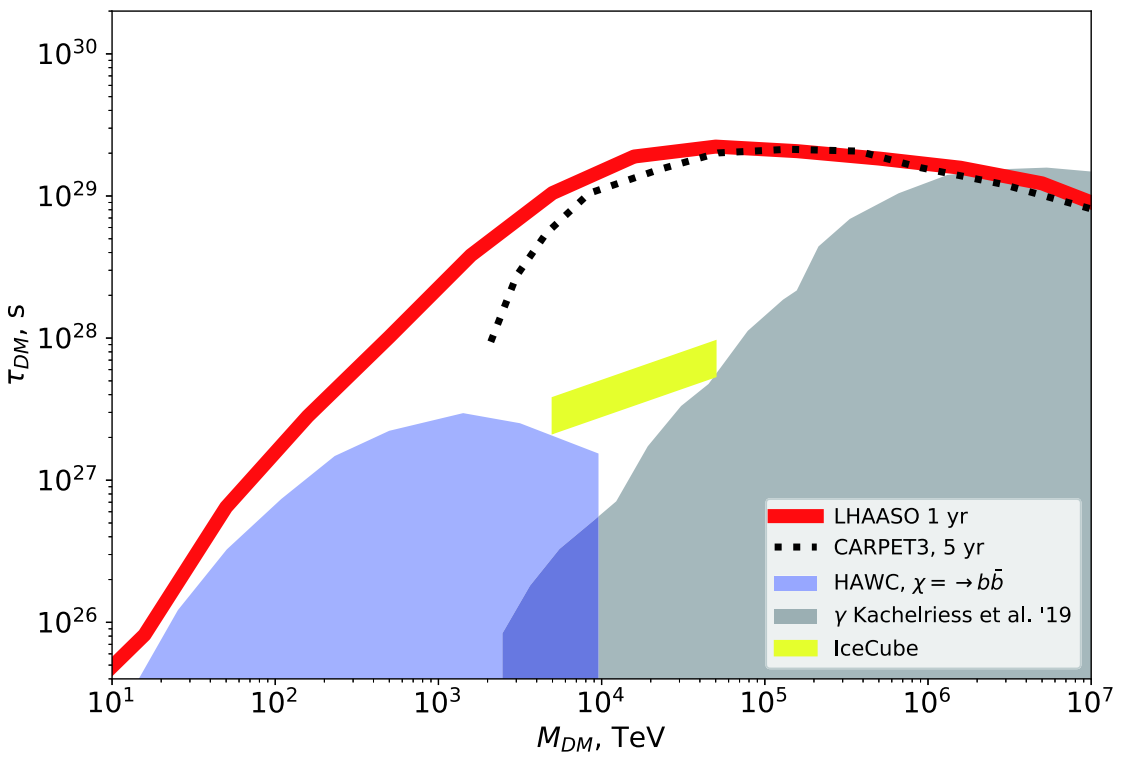

**Fig. 4.** (color online) Sensitivity of LHAASO for the measurement of dark matter decay time (for DM decaying into quarks). Yellow band shows the range of decay times for which DM decays give sizeable contribution to the IceCube neutrino signal [84]. Blue and grey shaded regions show the existing bounds imposed by HAWC [79] and ultra-high-energy cosmic ray experiments [85]. and dashed cureves are from the HAWC search of the DM decay signal in the Fermi Bubble regions [79]. From [53].

Similar light particles are also suggested from string-inspired phenomenology: moduli scalar fields from string compactifications, as well as anomalous Peccei-Quinn symmetries $U_{P.Q.}(1)$ from intersecting D-branes models, provide a rich spectrum of different axion-like particles (ALPs) beyond QCD axion mass and couplings [94–96]. PQ symmetries, generated from string theory, are explicitly violated by worldsheet instantons, brane instantons, new gauge instantons and gravitational instantons, which are beyond QCD instantonic effects [94–96].

The QCD axion is proposed as a solution of the strong CP problem, dynamically shifting out the $\theta$-parameter in $\mathcal{L}_{\theta} = -\theta(\alpha_s/8\pi)G^{\mu\nu a}\tilde{G}^{a}_{\mu\nu}$, where $G^{a}_{\mu\nu}$ is the gluon field strength and $\tilde{G}^{\mu\nu a} \equiv \epsilon^{\mu\nu\lambda\rho}G^{a}_{\lambda\rho}/2$. The QCD axion is related to a $U_{P.Q.}(1)$, which is broken through anomalous triangle couplings to gluons: $\mathcal{L} = (a/f-\theta)(\alpha_s/4\pi)G^{\mu\nu a}\tilde{G}^{a}_{\mu\nu}$. Non-perturbative topological fluctuations of the gluon field induce a periodic potential for $a$-field, the minimum of which is set at $a = f\theta$, screening the CP violating $\theta$-term [97–100]. It is worth to note that axions may be also thought as composite states of neutrinos, through new fundamental fifth force interactions [101] or chiral symmetry breaking topological terms sourced by non-perturbative gravitational effects [102, 103].

### A. Searches for the axion-photon oscillation effect

The ALP-photons interaction can be generated by anomalous triangle diagrams as $\mathcal{L}_{a\gamma\gamma} = g_{a\gamma\gamma}aF_{\mu\nu}\tilde{F}^{\mu\nu}/4$, where $F^{\mu\nu}$ is the electromagnetic field strength. In term of electric and magnetic fields, $\mathcal{L}_{a\gamma\gamma}$ can be rewritten as $g_{a\gamma\gamma}a\vec{E}\cdot\vec{B}$. This interaction implies that an external magnetic field background may source photon-ALPs mixings, through the aforementioned Primakoff effect [86, 87]. This motivates searches for ALPs in laboratory experiments with high magnetic fields, including the CAST experiment in CERN facility [104, 105].

On the other hand, possible hints of $a-\gamma$ oscillations may be observed from VHE gamma-ray sources, triggered by cosmic magnetic fields along the sight line [106–110]. An amplifier for this effect would be provided by the distance of sources, such as blazars. These are active galactic nuclei (AGN) with gamma beam pointing toward our line of sight, and representing the most distant and energetic gamma-ray sources observed. This certainly individuates blazars as natural candidates for the searches of ALP's effects. The mean free path of VHE photons above 100 GeV energy is related to the interactions with radiation backgrounds and pair production $\gamma\gamma \to e^{+}e^{-}$. It is well known that VHE gamma rays have a non-negligible probability of disappearing into electron-positron pairs, because of the annihilation with extragalactic background light (EBL) [111–114].

For gamma-ray energies below 100 GeV, the pair production effect is completely negligible, since the mean free path length $\lambda_{\gamma}(E)$, with $E$ the VHE photon energy, is comparable with the Hubble radius. On the other hand, for $E > 100$ GeV, the EBL effects quickly increase and $\lambda_{\gamma}(E)$ rapidly decreases with the energy. The $\gamma \to a \to \gamma$ oscillations can occur in the intergalactic space from the source to the Earth, turning the average $\lambda_{\gamma}$ into an higher value than expected, competitively with electron-positron pair production [106–108].

Another possibility is that $\gamma - a$ conversion happens inside or near blazar sources, and axions are converted back into photons inside the Milky Way [109, 110, 115–120]. This mechanism induces VHE gamma path length to be become larger than expected. Thus the Universe would appear more transparent in the very high energetic spectrum. A hardening of the observed blazar spectra, parametrically controlled by the $a\gamma\gamma$ coupling, would provide a hint of a photon-ALP oscillation in $1-100$ TeV range. The future LHAASO measurements of blazar gamma-ray spectra in this energy range will impose stronger limits on the ALP parameters than current results. For the blazars with variations in the spectra, such as Markarian 421, the combined analysis of observations in different periods can future improve the constraints [120].

The ALP interpretations of such observations depend on some astrophysical features. If the blazar is located at a rich galaxy cluster, the $\gamma-a$ conversion could happen in the turbulent inter cluster magnetic field of $O(1)\mu$G. Since there is no concrete model for this magnetic field, a series of random configurations is introduced in the analysis [115]. For blazars that are not resides in such environment, the $\gamma-a$ conversion dominantly happen in the blazar jet magnetic field, whose strength can be determined by the multi-wave observations [115, 118]. The ALP interpretations also depend on the EBL model. The predicted blazar spectra with the ordinary EBL models such as given by Ref. [78], can well fitted most of current measurements. Thus the measurements of blazar gamma-ray spectra impose constraints on the ALP parameters. On the other hand, if there exist some exotic EBL contents at $\sim O(1)\mu$m as reported by some recent measurements such as CIBER [121] the additional EBL absorption effect would reduce the observed VHE photons. In this case, the $\gamma-a$ oscillation compensates the lost of VHE photons and would provide better fitting to the measurements [116, 119].

The similar analysis can be applied for the VHE gamma-ray sources in the Galaxy [122, 123]. An advantage of such analysis is that the model of the Galactic magnetic field is more clear than those in the extragalactic space. The AS$\gamma$ observation of sub-PeV gamma-rays from the Crab Nebula has been used to set limits on the ALP parameter space [123]. The future LHAASO detection for the photons with higher energies and larger statistic will improve the current limits.

### B. Searches for infrared photons from ALP decays

Line emission from ALPs decaying everywhere across the Universe also contributes to the EBL [133]. Superposition of the line emission accumulated at different redshifts leads to a broad "bump-like" feature in the EBL spectrum with the width $\Delta E \sim E$. Such feature could be directly measured by dedicated observations probing the EBL. Such measurements [121, 134– 137] are, however, challenging in the eV photon energy range because of the presence of strong Zodiacal light background. The direct measurements typically find an excess EBL flux above the estimate from direct galaxy counts in the near-infrared [138–141]. Recent measurements by CIBER [121] estimate the excess flux at the level 29 to 42 nW/m$^2$/$sr$, depending on assumptions about Zodiacal flux model. Recent results from AKARI [136] between $\sim 2\,\mu$m and $\sim 5\,\mu$m also demonstrate an increase in EBL intensity at shorter wavelengths.

Attenuation of gamma-ray signal from distant sources by the effect of pair production on the EBL photons provides an alternative way of probing the EBL [142–146]. This method has been used to set limits on the normalisation of the EBL, under certain assumptions about its spectral shape. However, such an analysis does not account for possible unforeseen spectral features in the EBL, such as e.g. the ALP dark matter decay bump. On the other hand, presence of the bump leads to a characteristic dip in the observed gamma-ray spectra of distant sources, thus making this feature potentially observable.

In a recent study [147] combined Fermi/LAT and Cherenkov telescopes gamma-ray spectra of a large number of blazars were used to derive tight constraints on possible EBL bump features by fitting the gamma-ray spectra with models which include additional attenuation of the gamma-ray flux by absorption on the EBL bump. These constraints were then used to explore the parameter space of ALPs. Fig 5 shows constraints on the mass and two photon coupling constant obtained with current data and predictions for the sensitivity of LHAASO.

The existing data from Imaging Atmospheric Cherenkov Telescopes and the direct measurements of the EBL flux indicate that a small bump in the EBL spectrum, which might be due to the ALP decays, exists in the wavelength range around 1 micron [132]. The range of ALP parameters which corresponds to this hint is shown by the pink hatched region in Fig. 5. The "dip" in the gamma-ray spectra imprinted by the ALP-induced EBL bump is only marginally detectable by the current generation telescopes. Next generation telescope CTA will be able to test the hypothesis of the EBL excess produced by the ALP decays and pinpoint the parameters of ALP particles. Red colored region in Fig. 5 shows the precision of measurement of ALP paeramenters which can be achieved with CTA [132].

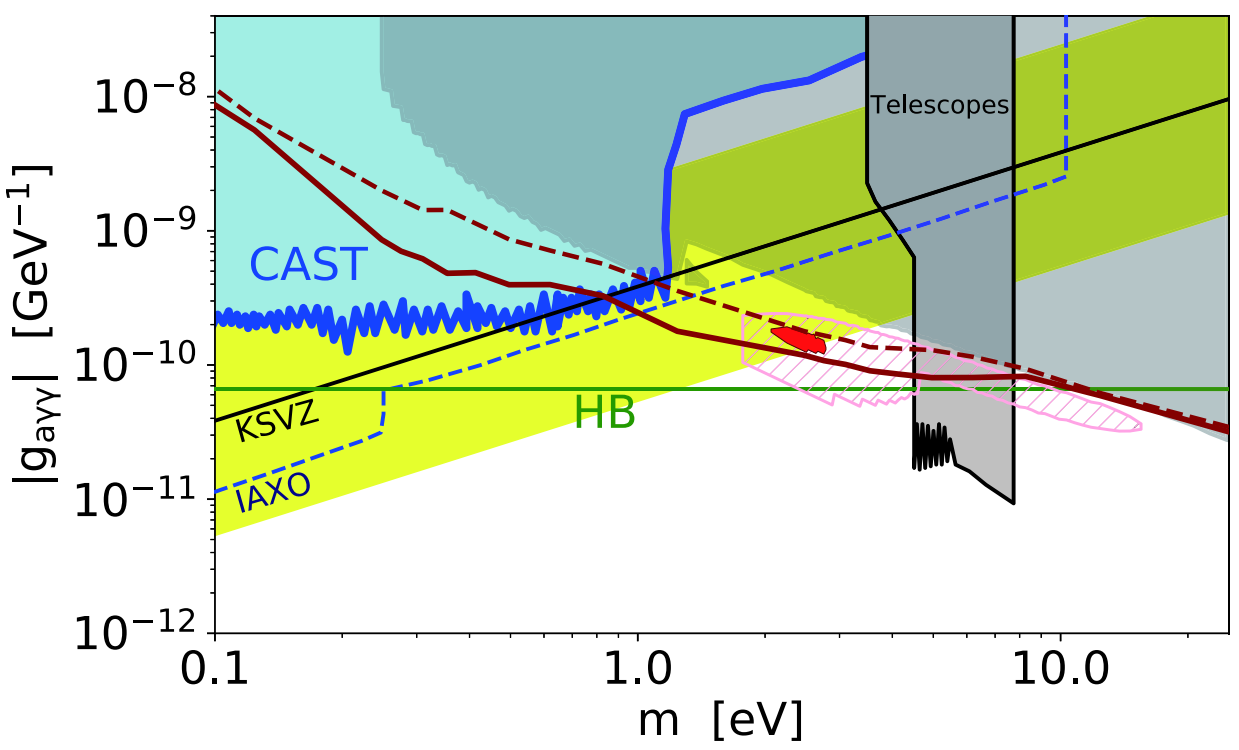

**Fig. 5.** (color online) Constraints of ALP parameters from laboratory and astrophysical probes. Yellow band and black solid line corresponds to the QCD axion models [124–127]. Blue region is the limit impost by CAST non-observation of solar axions [128]. Grey vertical band comes from high-resolution spectroscopy of intracluster medium in by optical telescope [129]. Black dashed line is the EBL limit estimate of [130]. Green horizontal line is the limit imposed by non-observation of excess energy loss in Horizontal Branch stars [131]. Light grey shaded region shows 95% confidence level excluded range of parameters from non-observation of absorption feature in the Fermi/LAT+VERITAS spectrum of 1ES 1218+304 [132]. Pink hatched and region show 68% confidence level preferred range of the parameters obtained by the spectral analysis of 1ES 1218+304. Red region is the same as pink but for the simulated CTA spectrum. Brown dashed and solid lines show the precision of the ALP mass and two-photon coupling which can be reached with 1- and 5-years of observations with LHAASO (95% confidence level).

Non-observation of axion flux from the Sun by CAST experiment constrains $g_{a\gamma\gamma}$ from above in the mass range below 1 eV, as shown in Fig. 5 [128]. The CAST limit will be improved by the next-generation facility IAXO [148], also shown in Fig. 5. Another strong constraint on the ALP parameters is imposed by non-observation of the effect of energy loss through emission of ALPs on stellar evolution [149, 150]. Recent constraints on two-photon coupling [131, 151] are shown by the horizontal line in Fig. 5, stem from non-detection of excessive energy loss through ALP emission which would affect evolution of the Horizontal Branch (HB) stars in globular clusters. Similarly to most of the astrophysical constraints, the stellar evolution argument relies on the assumption of validity of complex stellar evolution models controlled by large number of parameters which are not directly measured [152, 153]. Direct searches for ALP decays into two photons were performed using highresolution spectroscopy of galaxy clusters in the range of masses between 4.5 and 7.7 eV [129]. Non-observation of non-identified spectral emission lines from ALP decays imposes a constraint on $g_{a\gamma\gamma}$ shown by grey shading in Fig. 5.

To estimate the sensitivity reach of LHAASO for the ALPs parameters we performed the same analysis as proposed in [132]. Six sources, which satisfy the criteria from [132] and located in LHAASO field of view were selected: Markarian 501, Markarian 421, 1ES 1218+304, 1ES 1215+303, 1ES 1959+650, 1ES 1011+496. Using gamma-ray sensitivity, effective area and gamma/hadron separation efficiency we estimated background rate and simulated spectra for the blazars in the sample for 1- and 5-years of observations with LHAASO in the energy range between 200 GeV and up to 20 TeV. One can see that 5-years LHAASO sensitivity is sufficient to explore potentially interesting area of ALPs masses around 1 eV and $g_{a\gamma\gamma} \sim 10^{-10}\,\mathrm{GeV}^{-1}$ indicated from previous observations. In the case of absence of the signal LHAASO will improve current limitation on $g_{a\gamma\gamma} \sim 10^{-10}\,\mathrm{GeV}^{-1}$ by a factor of 3-5 in the mass range from 1 eV to 5 eV. From Fig. 5 one can see that LHAASO will have enough sensitivity to cross-check the CTA measurement of the ALP parameters (if the existence of a bump in the EBL spectrum will be confirmed).

# Chapter 6 Multimessenger Physics

Editor: Zhuo LI

Contributors: Yiqing GUO[1], Haoning HE[2], Zhuo LI[3,4], Ruo-Yu LIU[5,6], Kai WANG[7]

[1] Key Laboratory of Particle Astrophysics, Institute of High Energy Physics, Chinese Academy of Sciences, Beijing 100049, China
[2] Key Laboratory of Dark Matter and Space Astronomy, Purple Mountain Observatory, Chinese Academy of Sciences, Nanjing 210023, China
[3] Department of Astronomy, School of Physics, Peking University, Beijing 100871, China
[4] Kavli Institute for Astronomy and Astrophysics, Peking University, Beijing 100871, China
[5] School of Astronomy and Space Science, Nanjing University, Nanjing 210023, China
[6] Key laboratory of Modern Astronomy and Astrophysics (Nanjing University), Ministry of Education, Nanjing 210023, China
[7] Department of Astronomy, School of Physics, Huazhong University of Science and Technology, Wuhan 430074, China

**Abstract:** Combining observations of multi-messengers help in boosting the sensitivity of astrophysical source searches, and probe various aspects of the source physics. In this chapter we discuss how LHAASO observations of very high energy (VHE) gamma rays in combination with telescopes for the other messengers can help in solving the origins of VHE neutrinos and galactic and extragalactic cosmic rays.

## 1 Search for Galactic Cosmic-Ray PeVatron Source

Executive summary There are many candidate sources of PeV cosmic rays in the Milky Way. Gamma-ray detection of these sources at the ultrahigh-energy band proves their potential of acceleration of PeV protons as particle accelerators. High-energy neutrino detection would serve as a smoking gun to identify that acceleration of PeV protons are indeed proceeding in these sources.

### 1.1 Introduction

Detection of the so-called 'knee' at ∼PeV in the cosmic ray proton spectrum implies that there are petaelectronvolt accelerators ('PeVatron') residing in our Galaxy. Power-law spectra of gamma-rays extending to at least several tens of TeV without a cut-off has been suggested as the identifier of such kind of powerful proton accelerators. However, despite of a large amount of cosmic ray accelerators has been found at TeV in our Galaxy, none of them has shown the unequivocal feature of PeVatron, except that HESS experiment has discovered a likely existence of proton PeVatron within the central 10 parsecs of the Galaxy at roughly $2\sigma$ confidence level [18]. LHAASO has a good sensitivity for gamma-ray with $>10$TeV, in particular, reaching an unprecedented level around 100 TeV and hence can serve as an efficient PeVatron detector. Among 78 VHE sources in the HESS Galactic Plane Survey catalog[19], 21 of them are in the field of view of LHAASO. By extrapolating the spectrum of these sources with the best-fitting spectral model (power-law model or exponential cutoff power-law model) to 1 PeV, 19 of them is beyond the $5\sigma$ detection limit of LHAASO with one-year observation. There are many more PeVatron candidates beyond the field of view of HESS but in that of LHAASO. For example, the star-forming region Cygnus Cocoon is one of the proposed cosmic-ray accelerator [20]. Gamma-ray emission from this region has been detected by Fermi-LAT[21], Milagro[22], ARGO-YBJ[23], Veritas[24] and HAWC[25] in GeV–TeV band with a quite complex morphology, implying contributions from multiple sources. In fact, some of these sources have already been detected by half of LHAASO-KM2A at the ultrahigh-energy (UHE) gamma-ray band ($E \geq 100$ TeV, [26]). Among 12 UHE gamma-ray sources detected by LHAASO, most of them are associated with supernova remnants (SNRs), pulsar wind nebulae (PWNe) and star-forming regions, implying these sources as proton PeVatron candidates.

To put it shortly, LHAASO has revealed many sources as potential proton PeVatrons and is promising to discover more new PeVatron candidates in the future. Based on the current observation, however, it is difficult to differentiate the hadronic origin and the leptonic origin for most of LHAASO detected sources. High-energy neutrinos from these sources would be a smoking gun for the hadronic origin, since high-energy gamma rays and neutrinos are produced associately in hadronic interactions of cosmic rays with comparable flux. However, no astrophysical neutrino sources had been clearly identified by IceCube yet and hence only neutrino flux upper limit are available [27]. Even so, the upper limits can already put interesting constraints on the origin of UHE gamma rays of some LHAASO sources [28]. We note that using the LHAASO measured spectrum as a prior in IceCube's analysis may enhance the post-trial significance of the sources and in turn leads to more stringent neutrino flux upper limits.

### 1.2 Supernova Remnants

SNRs, especially young SNRs with age less than one thousand years such as Tycho, Cas A and so on, are believed to be able to accelerate CRs up to PeV energies, and contribute to the Galactic CRs. The interaction between accelerated CRs and the surrounding matters would produce $\gamma$-rays and neutrinos. $\pi^0$ bumps are observed in the $\gamma$-ray spectra of SNRs W44 and IC443 as the indications of hadronic interactions [29]. Previous studies on the TeV gamma-ray emissions associated with SNRs W28, W41, W51C and CTB37A also suggested that these TeV emissions are possibly powered by the hadronic interactions. In particular, a middle-aged SNR G106.3+2.7 has been extensively studied in GeV band [30], TeV band[31–33], and X-ray band[34, 35]. Theoretical studies have also been carried out to investigate its possibility as a proton PeVatron [34, 36]. LHAASO's measurement extended its spectrum up to about 600 TeV. It is predicted that IceCube can detect 0.4 muon neutrinos above 50 TeV for the ten-year operation in the hadronic model. Future observation of LHAASO and IceCube (as well as the next-generation neutrino telescope) should be able to test the hadronic model.

### 1.3 Pulsar wind Nebulae

PWNe are also believed to be a kind of Galactic cosmic ray sources. The pulsar wind interacting with the ambient medium around a pulsar forms a terminal wind shock, which will accelerate particles to high energy. The accelerated CRs interact with matters or photons in the nebulae would produce gamma-rays and neutrinos. More than 30 PWNe have been detected at TeV energies. A stacking analysis to search for neutrino emission from 35 TeV PWNe using 9.5 years of all-sky IceCube data finds no significant correlation between PWNe and neutrinos [37]. Extended TeV $\gamma$-ray emission are detected from nearby sources Geminga pulsar and B0656+14 by HAWC [38]. More extended TeV images of PWNe might be discovered by LHAASO. The spectral features and the gamma-ray profiles of PWNe detected by the LHAASO would help IceCube to improve the neutrino searches. It is interesting to note that the latest observation of LHAASO on the Crab Nebula has extended its spectrum up to 1.1 PeV and reveal a possible hardening of its spectrum above several hundred TeV [39]. This might be interpreted as an additional hadronic component at the highest energies.

### 1.4 Star-forming Regions

Star-forming regions are factories of stars/star clusters, and usually associated with molecular clouds, such as W51A, W51B, Cygnus region, W49A and W43. The young OB star clusters, supper bubbles supplied by supernova explosions or collective stellar winds, or SNRs/PWNe in the star-forming regions could accelerate CRs to high energy. The high energy CRs that are confined in the star-forming regions would interact with molecular clouds, and produce gamma-rays and neutrinos.

We assume that, in a star-forming region in the Galaxy, a PeV CR accelerator are accompanied with molecular clouds, and the accelerated PeV CRs escape from the accelerator and are confined in the molecular clouds with the total energy of $E_{\rm inj}$. Gamma-rays and neutrinos are produced via the interaction between CRs and molecular clouds. The corresponding extended sources might be observed with the gamma-ray and neutrino profiles following the distribution of the molecular clouds. The gamma-ray flux at the energy of 100 TeV is about $f_\gamma = \frac{1}{3} E_{\rm inj} n_{\rm H} \sigma_{\rm pp}^{\rm inel} c/(4\pi D_{\rm L}^2)$, i.e.,

$$f_\gamma = 3\times10^{-12}\ {\rm TeV cm^{-2} s^{-1}} \times \left(\frac{E_{\rm inj}}{10^{50}\ {\rm erg}}\right)\left(\frac{n_{\rm H}}{1\ {\rm cm^{-3}}}\right)\left(\frac{D_{\rm L}}{10\ {\rm kpc}}\right)^{-2} \quad (1)$$

where $\sigma_{\rm pp}^{\rm inel} \simeq 50$ mb is the approximated cross-section for inelastic $pp$ collision [40], and $n_{\rm H}$ is the density of the molecular clouds. The muon neutrino flux at 50 TeV is about $10^{-12}\ {\rm TeV cm^{-2} s^{-1}} (n_{\rm H}/1\ {\rm cm^{-3}})$ for $E_{\rm inj} = 10^{50}$erg and $D_{\rm L} = 10$kpc, and the integrated counts of muon neutrinos for ten year operation of IceCube can be estimated to be $N_\mu = 4(A_{\rm eff}/1\ {\rm m^2})(E_{\rm inj}/10^{50}{\rm erg})(n_{\rm H}/1\ {\rm cm^{-3}})(D_{\rm L}/10\ {\rm kpc})^{-2}$. IceCube reported the results of searching for extended sources of neutrino emission with 7 years of IceCube data, with the discovery potential flux at 50% confidence level for the Northern Hemisphere about $\sim 10^{-12} - 10^{-11}$ TeV cm$^{-2}$ s$^{-1}$ [41]. Therefore, a PeVatron associated with molecular clouds in the Galac-tic star-forming region might be a possible extended gamma-ray source candidate for LHAASO, and also a possible extended neutrino source candidate for IceCube, depending on the total injected energy, the source distance, and the mass of molecular clouds. LHAASO's future discovery on new extended sources in the Galaxy will improve the discovery potential of IceCube.

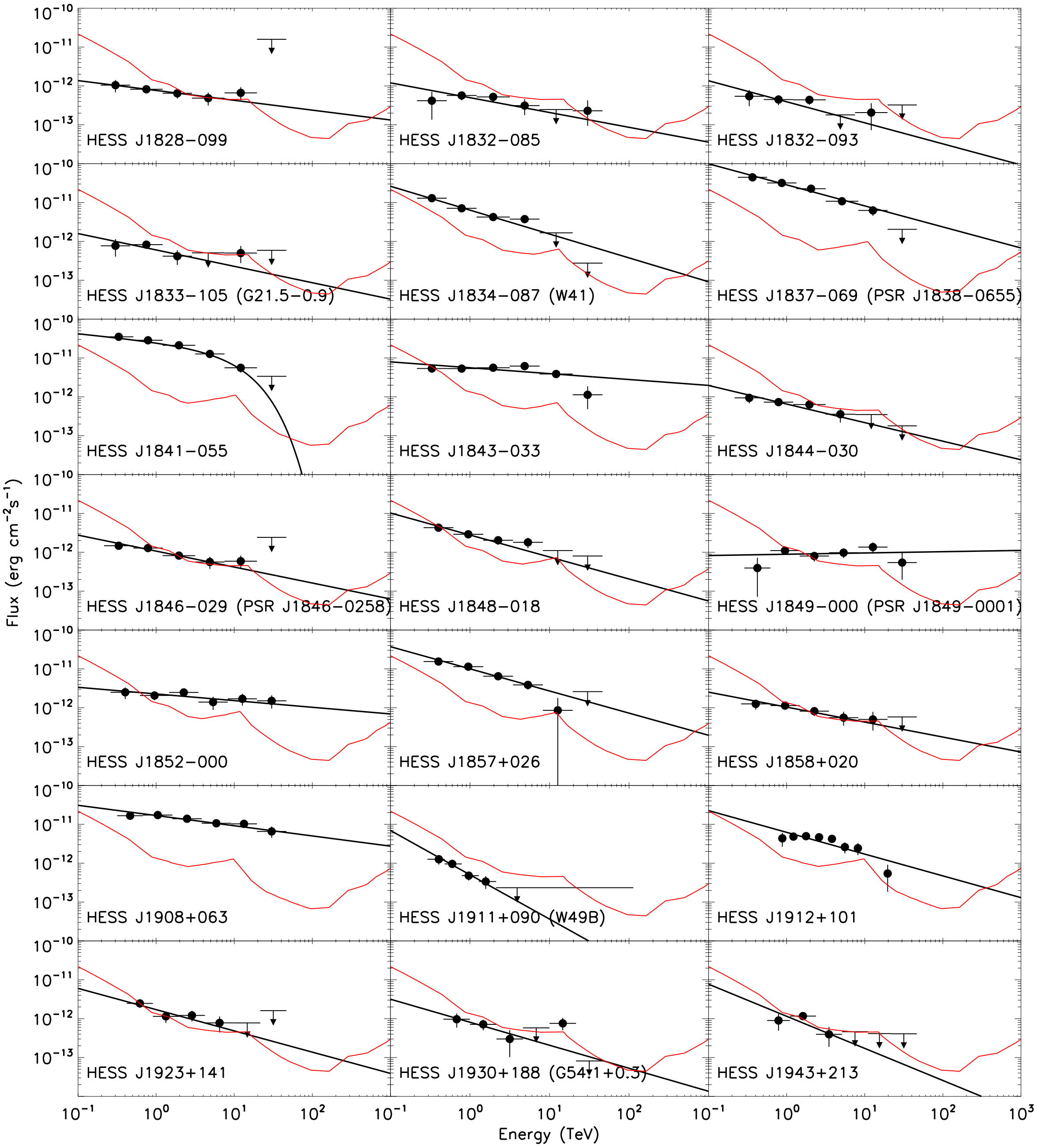

Fig. 1. 21 sources from HGPS catalogue that are in the field of view of LHAASO. Black data points are the measurement by HESS, and black curves show the best-fit spectrum model [19]. Red curves show the sensitivity of LHAASO for each source with one-year observation. The sensitivity is estimated by multiplying a factor of $\max(1,\theta_s/\theta_{\rm PSF})$ to the sensitivity for Crab, where $\theta_s$ is the angular size of the source measured by HESS while $\theta_{\rm PSF}$ is the size of PSF of LHAASO.

### 1.5 Gamma-ray Binary System

In $\gamma$-ray binary systems, such as LS 5039, HESS J0632+057, and LS I+61°303 [42], particles might be accelerated to high energy via the pulsar wind interacting with the strong wind of massive star, the jet activities in micro-quasars, or relativistic outflow interacting with the ISM, and further produce gamma-rays and neutrinos. In particular, HAWC detected TeV emission above 25 TeV from the jet's lobe of microquasar SS433 [43], demonstrating that the binary system could be efficient particle accelerator. The searches for time dependent neutrino sources with IceCube data from 2008 to 2012, find no significant time dependent point sources of neutrinos, but the most significant neutrino excess from the binary system HESS J0632+057 with pre-trial p-value of 0.087 [44]. Chances are that LHAASO would discover more nearby gamma-ray binaries, and provide a gamma-ray binary catalog for the IceCube to search for more neutrino sources.

## 2 Follow-up Study on IceCube Neutrino Sources

Executive summary The IceCube neutrino observatory is sending public real-time alerts on single muon neutrino-induced track events with a high probability of being of astrophysical origin. Since 2019, more than 40 neutrino singlet alerts located within LHAASO's field of view (FOV) were reported. If the source is nearby with no photon attenuation inside the source and along the propagation path, there is a high chance that LHAASO can observe high energy photons associated with the IceCube announced single muon neutrino alerts. Besides performing observations following up the IceCube neutrino alerts , LHAASO can also provide public alerts for the follow-up neutrino detections. What's more, LHAASO's observations on nearby blazars and starburst galaxies can provide more details on the nonthermal processes operating in the neutrino source candidates.

### 2.1 Introduction

The IceCube neutrino observatory, located under the Antarctic ice, is the largest neutrino detector to date. Since 2016, IceCube started to send public real-time alerts on single-muon neutrino-induced track events with a high probability of being of astrophysical origin, based on the real-time, online event reconstruction, through Astrophysical Multimessenger Observatory Network (AMON) and Gamma-ray Coordination Network (GCN) [45]. In 2019 June, the HESE (High Energy Starting Events) notices and EHE (Extremely High Energy) notices are replaced by so called "ICECUBE ASTROTRACK GOLD notices" and "BRONZE notices", with the rate about 12/yr and 16/yr, and the chance of $>50\%$ and $>30\%$ to be astrophysical [46], and the position error of $0.2°-0.75°$. Since 2019 September, more than 40 neutrino singlet alert's within LHAASO's field of view were reported. In addition to muon neutrino-induced event alerts, IceCube has a real-time program to search for muon-neutrino multiplets. In 2016, the IceCube real-time neutrino search identified a muon-neutrino multiplet, with no likely electromagnetic counterpart detected [47].

### 2.2 Follow-up Observations of the IceCube Neutrino Alerts

The large FOV, high duty cycle and high sensitivity make LHAASO a perfect detector on searching for very high energy gamma-ray transients or steady sources associated with the IceCube neutrino alerts. What's more, IceCube is more sensitive to sources in the Northern hemisphere, since events from the Southern hemisphere are highly contaminated by the muon backgrounds [48], and most of the LHAASO's FOV is in the Northern hemisphere. This fact also makes LHAASO a suitable detector to do the follow-up observations the IceCube neutrino alerts. Since most of IceCube neutrino alerts locate to the direction of high galactic latitude, if they are of astrophysical origins, they are likely to be from extragalactic sources. Due to the extragalactic background light (EBL) absorption, the flux of $\gtrsim$10 TeV photons from sources with distance larger than a few Mpc will be suppressed. Therefore, the detection horizon of LHAASO KM2A is limited to be a few Mpc. The EBL absorption is weak for TeV photons from sources within a few hundred Mpc, thus, the LHAASO WCDA are able to detect photons from sources within a few hundred Mpc.

For steady neutrino sources, the duration of the neutrino emission at the source can be as high as the IceCube operation time. Let us denote the energy of a neutrino by $E_{\nu_\mu}^{\rm obs}$, the effective area of the IceCube by $A_{\rm eff}$, the duration of neutrino emission at the source by $T_\nu$. Assuming the neutrino spectrum $\frac{dN_\nu}{dE_\nu} \propto E_\nu^{-2}$ with $E_{\nu,\rm min} = 1$ TeV and $E_{\nu,\rm max} = 10$ PeV, the flux of a single muon neutrino-induced event is estimated as

$E^2_{\nu_\mu} dN_{\nu_\mu}/dE_{\nu_\mu} = E^{\rm obs}_{\nu_\mu}/(\ln(E_{\nu,\rm max}/E_{\nu,\rm min}) A_{\rm eff} T_\nu)$, i.e.,

$$E^2_{\nu_\mu}\frac{dN_{\nu_\mu}}{dE_{\nu_\mu}} \simeq 4\times10^{-12}\ {\rm TeV\ cm^{-2}\ s^{-1}} \times\left(\frac{E^{\rm obs}_{\nu_\mu}}{100\ {\rm TeV}}\right)\left(\frac{A_{\rm eff}}{1\ {\rm m}^2}\right)^{-1}\left(\frac{T_\nu}{10\ {\rm yr}}\right)^{-1}. \quad (2)$$

Since the gamma ray flux is about 2 times larger than the muon neutrino flux considering the equipartition among the three neutrino flavors after their oscillations during the propagation, the gamma-ray flux at 100 TeV is about $\frac{dN_\gamma}{dE_\gamma} \sim 8 \times 10^{-16}\ {\rm TeV^{-1}\ cm^{-2}\ s^{-1}} \left(\frac{E^{\rm obs}_{\nu_\mu}}{100\ {\rm TeV}}\right)^{-1} \left(\frac{A_{\rm eff}}{1\ {\rm m}^2}\right)^{-1}\left(\frac{T_\nu}{10\ {\rm yr}}\right)^{-1}$, if there is no photon attenuation within the source and along the propagation path. Therefore, if a single neutrino with energy about $E^{\rm obs}_{\nu_\mu}=100$ TeV is observed from a nearby steady source, the expected gamma-ray flux at 100 TeV with no absorption is larger than the sensitivity of LHAASO at 100 TeV for one year operation, as shown in Fig. 1. To be noticed here, the effective area of IceCube is a function of declination. Therefore, if there is no photon attenuation inside the source and along the propagation path, there is a high chance that LHAASO KM2A can observe high energy photons associated with the IceCube announced single muon neutrino alerts and muon neutrino multiplets.

Besides performing observations following up the IceCube neutrino alerts , LHAASO can also provide public alerts for the follow-up neutrino detections. LHAASO are able to search for the hotspot with a cluster of events above the estimated cosmic-ray background level with an excess significance above 2.75 $\sigma$, and provide the information of the hotspot for the IceCube collaboration to search for neutrinos associated with the LHAASO hotspot temporally and spatially. Then, the IceCube and LHAASO collaborations can issue the IceCube LHAASO alert based on gamma rays and neutrino sub-threshold detections, similar to the IceCube HAWC alert. The combined analysis on the observed gamma-ray photons and neutrinos would help to improve the discovery potential of the neutrino sources.

### 2.3 Blazars and Starburst Galaxies

Blazars and starburst galaxies are two major source candidates for the IceCube diffuse neutrino background.

**Blazars** are relativistic jets driven by supermassive black holes with directions aligned with the observer's line of sight. They have been proposed as the high-energy neutrino sources for decades[49–52]. Ref.[53] found 11 significant neutrino flares using a sample of muon track neutrino events from 2012 April to 2017 May, associating with 10 AGN counterparts, including FSRQs, BL lac and radio galaxies. In addition, 9 blazars are summarized associated with single high-energy neutrino events, including both archival neutrino events and neutrino alert events[54].

One of the most promising high-energy neutrino candidates to date is blazar TXS 0506+056, which is detected in spatial and temporal coincidence with a $\sim$ 300TeV neutrino event IC-170922A at $3\sigma$ level during its flaring state[55]. A later analysis on the archival data revealed 13$\pm$5 additional neutrino events from the same direction during about 4 month in 2014-2015[56]. This may be an evidence of blazars as cosmic ray proton accelerators at least to $\sim$10 PeV. If TXS 0506+056 is truly a neutrino emitter, it will be also a multi-TeV gamma-rays producer through the same hadronic interactions. However, due to the large distance of the blazar (at a redshift of 0.34) to Earth, the produced multi-TeV gamma-ray photons will be severely absorbed during their propagation in the intergalactic space, even if the internal absorption due to the emission of the blazar itself is not important.

Nevertheless, it'd be worth monitoring the closest blazars, such as Mrk 421 and Mrk 501 with distance 126 Mpc and 157 Mpc away from Earth, respectively, at which distance the attenuation for multi-TeV flux is not very strong. Another interesting source in the field of view of LHAASO is the radio galaxy M87, which is considered as the misaligned counterpart of blazars. It was found associated with a short neutrino flare in 2016 of a duration about 3.9 minutes with a p-value of $1.91\times10^{-3}$[53]. Given the distance of M87 to be about 16.4 Mpc, the detection of 100 TeV gamma-ray photons is in principle possible, if the intrinsic flux is high and the internal absorption is not strong. Nevertheless, even if only upper limits of multi-TeV gamma-ray fluxes are obtained, the results may be useful to constrain the radiation model of blazars and relevant physical quantities, such as the size/locations of the emitting region, particle acceleration capability and the composition of the jet.

**Starburst galaxies** are another major candidates of high-energy neutrino sources. Theoretical studies have shown that starburst galaxies are able to contribute to, at least a considerable fraction of, the diffuse neutrino background detected by IceCube [e.g. 57–60]. In the ten-year search for steady point-like * neutrino sources by IceCube, the hottest spot with a post-trial significance of $2.9\sigma$ [27] is in the direction of a nearby starburst galaxy NGC 1068 (M77) (14.4Mpc from Earth), which shows starburst activity and an active galactic nucleus.

There are some other nearby starburst galaxies located in the LHAASO's FOV. For example, M82 (also

*Given different angular resolutions between IceCube and LHAASO, a point-like source viewed by IceCube could be an extended source viewed by LHAASO.

known as Cigar galaxy) is one of the closest starburst galaxies (3.5 Mpc) and is usually considered as a prototype of starburst galaxy. It has been observed by VERITAS [61] up to a few TeV with a power-law spectrum showing no clear cutoff. The TeV emission is believed to arise from the *pp* collision between cosmic rays in the galaxy and the interstellar medium (ISM), so high-energy neutrino emission from M82 is naturally expected. Although the IceCube has not found any neutrino excess in the direction of M82, the multi-TeV gamma-ray observation would help to determine the hadronic interaction efficiency in starburst galaxies at the energy regime interesting for high-energy neutrino astronomy.

# 3 High-Energy Neutrinos and Gamma Rays from Core-Collapse Supernovae

Executive summary The interaction between the core-collapse supernova (CCSN) ejecta and the dense circumstellar medium (CSM) could generate the shock waves and then accelerate the cosmic rays beyond PeV energies. Multi-messenger signatures, e.g., the high-energy gamma-rays and the high-energy neutrino emission, could rise from the inelastic *pp* collision between the high-energy cosmic rays and the gas therein. Such high-energy gamma-rays from the interaction of the CCSN ejecta and the CSM could reach the sensitive energy range of LHAASO. The joint observations and studies of the simultaneous gamma-ray emission and neutrino emission could help us to find out the nature of CCSNe, e.g., the properties of the progenitor, the circumstellar environment, the acceleration of cosmic rays and so on.

## 3.1 Introduction

A high-density circumstellar wind environment in the immediate vicinity of the progenitor can be caused by the sustained mass-loss of the progenitor before the explosion of CCSNe, and after the explosion, the interaction of supernova ejecta with the optically thick wind could result in a bright, long-lived wind breakout (i.e., CSM breakout) event, which may also make the usual envelope breakout delay. There are increasing evidences that large mass loss episodes closely preceding the CCSN explosion are not uncommon. The early lightcurves of Type IIn supernovae (SNe IIn) are consistent with the predictions of wind breakouts [62–64] with inferred mass loss rates $\dot{M} > 10^{-3}\,M_\odot\,yr^{-1}$ [65–70]. Suggested by some literature [71–78], the month-scale pre-SN "precursors" in SNe IIn before the explosions are common, providing an independent evidence for the existences of intense mass loss episodes in most SNe IIn preceding the explosion. Besides, for superluminous SNe (SLSNe), type II or type I, may originate from the interactions between the SN ejecta and the extended CSM [79–91]. Finally, the rapid follow-up observations suggests that SN 2013fs, a Type IIp SN (SN IIp), went through a mass-loss of the progenitor prior to the explosion at a high rate of $\sim 3\times10^{-3}\,M_\odot\,yr^{-1}$ [92]. What's more, the further early observations for dozens of rising optical lightcurves of Type II SNe candidates indicate that the SN 2013fs is not a special case and the mass loss episodes should be ubiquitous for regular Type II SNe [93].

The dense CSM would be formed due to continuous mass loss before the SN explosion. After the explosion, the shock waves generated by the interaction of the SN ejecta with the CSM could accelerate protons beyond PeV energies, and subsequently the inelastic *pp* collision between the accelerated protons and the shocked CSM can give rise to interesting multi-messenger signatures, such as high-energy gamma-ray emissions, neutrino productions, as well as X-ray, optical and radio emissions[94–99]. Such high-energy gamma-rays could reach a energy of $\sim 100\,$TeV, locating the sensitive energy range of LHAASO.

## 3.2 The multi-messenger signatures of the regular Type II SN ejecta-CSM interaction

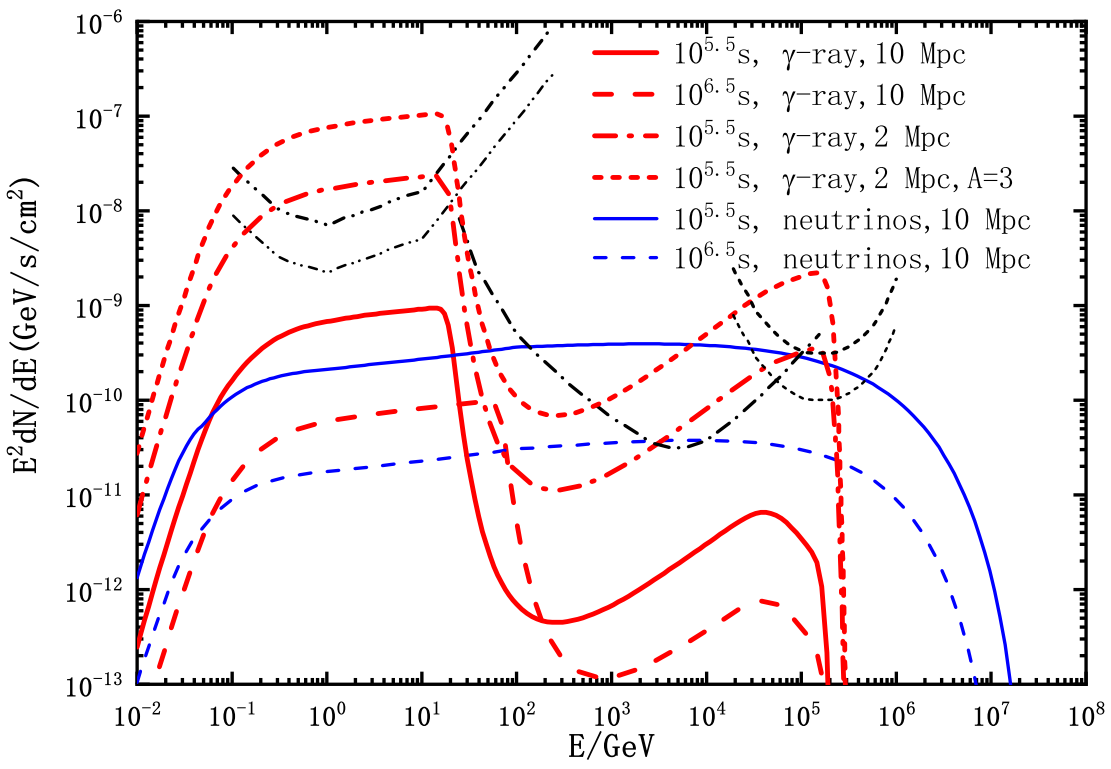

Fig. 2. The fluxes of produced gamma-rays (red thick lines) and neutrinos (blue thin lines) for a typical interaction of the SN ejecta with the CSM. The red solid, red dashed and red dash-dotted lines are shown for two different time windows, i.e., $10^{5.5}$s and $10^{6.5}$s and two representative source distances i.e., 2Mpc and 10Mpc by assuming that the CSM is SN 2013fs-like. Besides, the red short dashed line for the $A = 3$ times denser CSM environment is presented as well. The black thick and thin dot-dot-dashed lines represent respectively the differential sensitivity of *Fermi*-LAT for a observational time $10^{5.5}$s and $10^{6.5}$s, the black dot-dashed line indicates the 50hr differential sensitivity of CTA, and the black short dashed lines indicate the differential sensitivity of LHAASO for a observational time $10^{5.5}$s and $10^{6.5}$s. For more details, please

see [99].

For regular Type II SNe, productions of gamma-rays and neutrinos for the typical interaction of the SN ejecta and the SN 2013fs-like CSM is presented in Fig. 2, as in Ref. [99]. Due to the absorption of low-energy photon field in the emission region, the spectrum gets a significant suppression at the energy range of $\sim$10 GeV–100 TeV. Besides, the gamma-ray flux above tens of TeV will be reduced during the propagation to Earth as well, due to the absorption by the EBL and CMB photons. In Fig. 2, the sensitivities of *Fermi*-LAT, Cherenkov Telescopes Array (CTA) and LHAASO are shown to compare with the gamma-ray emissions. As we can see, for typical values of parameters, at 10Mpc, high-energy gamma-rays are hard to be observed by all the three experiments for the case with the 2013fs-like CSM. However, for a time window of $\sim 10^{5.5}$s, at a distance $\lesssim$2−3Mpc, gamma-rays could be detected around GeV by *Fermi*-LAT, around few−100 TeV by the CTA and above $\sim$100 TeV by LHAASO. For a denser CSM environment (e.g., the red short dashed line in Fig. 2), the flux of gamma-rays is significantly enhanced and it could be still detectable for a further distance of source. The cooperative observations to gamma-rays from a ejecta-CSM interaction by the experiments in the future could be expected and a broad spectrum of gamma-rays from GeV to hundreds of TeV may be approached. Especially, LHAASO, which has a competitive sensitivity around 100 TeV, can provide the unique information of very-high-energy gamma-rays and can help us to expose the origin of cosmic rays more directly.

In contrast to gamma-rays, which may be absorbed via photon-photon attenuation in the shocked region and during their propagation in the universe, neutrinos are more effective information carriers as they can escape from the source and propagate to Earth unimpeded. For the scenario with the SN CSM breakout as the point source of neutrinos, at 10Mpc, the flux is $\sim 3\times10^{-10}\,\mathrm{GeVcm^{-2}s^{-1}}$ (see Fig. 2), which could reach the sensitive level of future IceCube Gen2 [100]. If the source is located at a closer distance or with the Galaxy, those neutrinos could be detected by the current IceCube observatory [101]. The joint observations of high-energy gamma-rays by LHAASO and neutrinos by IceCube could become a reality in the future and then help us to determine the nature of these explosive phenomena, e.g., properties of the progenitor and the acceleration mechanism of cosmic rays [102]. For the high-energy gamma-rays, LHAASO is able to detect signatures of CSM breakouts of regular Type II SNe at $\lesssim$2–3Mpc for a time window of several days. Such a size is comparable with the size of a local galaxy cluster. The expected regular Type II SN event rate in local galaxy cluster is about a few in ten years [103].

### 3.3 The multi-messenger detection probability of other types of CCSN

The SNe IIn may have a denser and more extended CSM than that of SNe IIp, resulting in a larger detectable distance of $\sim$10Mpc and longer time window of month-scale for high-energy gamma-rays [95]. However, SNe IIn is relatively rare with a event rate of 7–9% of all CCSNe [104, 105], and therefore the occurrence rate of nearby SNe IIn is $\sim 1\mathrm{yr}^{-1}$ within 30 Mpc and $\sim 0.03\mathrm{yr}^{-1}$ within 10 Mpc. In addition, during the propagation, the EBL attenuation is significant at $\gtrsim$ 100 TeV for a distance of $\gtrsim$10Mpc. As a result, high-energy gamma-rays observed by LHAASO can be operated only within $\sim$ 10Mpc and with a event rate of $\sim 0.03\mathrm{yr}^{-1}$. However, this event rate is conservative since the SN rate density within 10Mpc is higher than the global one [106] and more importantly, due to the detectable distance of high-energy neutrinos from SNe IIn up to $\sim$ 10Mpc [95, 97], SNe IIn could be promising sources for multi-messenger observations. For Type Ibc SNe, whose rate is $\sim$20% of CCSNe [104, 107], it may be difficult to implement such a joint multi-messenger owing to the lower expected flux, but it could be also possible if the CSM is as dense as that for low-luminosity gamma-ray bursts [108] or it takes place in a closer distance.

### 3.4 Summary

In summary, LHAASO can play an important role to participate in these multi-messenger observations for the CCSN ejecta-CSM interaction specifically by focusing on the very-high-energy gamma-ray observations. As the above estimation, LHAASO is able to detect high-energy gamma-rays of CSM breakouts for regular Type II SNe at $\lesssim$2−3Mpc with a event rate around a few in ten years, and for SNe IIn at $\lesssim$10Mpc with a event rate of $\sim 0.03\mathrm{yr}^{-1}$.

## 4 Extended/Diffuse Sources of High-Energy Neutrinos

Executive summary The origins of astrophysical neutrinos could be either galactic or extra-galactic. The galactic origins include the TeV $\gamma$-ray point sources, the galactic center and the diffuse one in Fermi bubble region, galactic halo and galactic plane. According to the production site, two regions as galactic disk and halo are discussed in this section. The diffuse $\gamma$-rays observed by LHAASO in those regions can play very important roles to constrain the neutrino flux from the galactic contribution.

### 4.1 Contribution from the galactic plane

Besides various potential point-like neutrino sources, there are also possible extended neutrino sources. Galac-tic plane is actually a promising one because the hadronic interaction between cosmic rays and gas in ISM has been demonstrated by Fermi-LAT [109], ARGO-YBJ[110] and Tibet AS+MD array [33]. There has been analysis on the diffuse neutrino flux from the Galactic plane, and the upper limit is found to be $\sim 10\%$ [41, 111]. The diffuse gamma-ray emission from the Galactic plane at multi-TeV energy would be a direct indicator of Galactic high-energy neutrino flux, although the emission by unresolved faint gamma-ray sources may also contribute [112, 113]. Figure 3 shows the expected Galactic diffuse neutrino spectrum under the conventional propagation model. These neutrinos originate from the collisions of the diffuse cosmic rays with the interstellar medium nearby the Galactic disk. The solid and dashed lines represent the contributions from the inner Galactic plane ($-30^\circ < l < 30^\circ$, $|b| < 5^\circ$) and outer Galactic plane ($90^\circ < l < 270^\circ$, $|b| < 5^\circ$) respectively. The Galactic diffuse neutrinos intensity is about $\sim 10\%$ of all-sky one, which is consistent with the IceCube's observation.

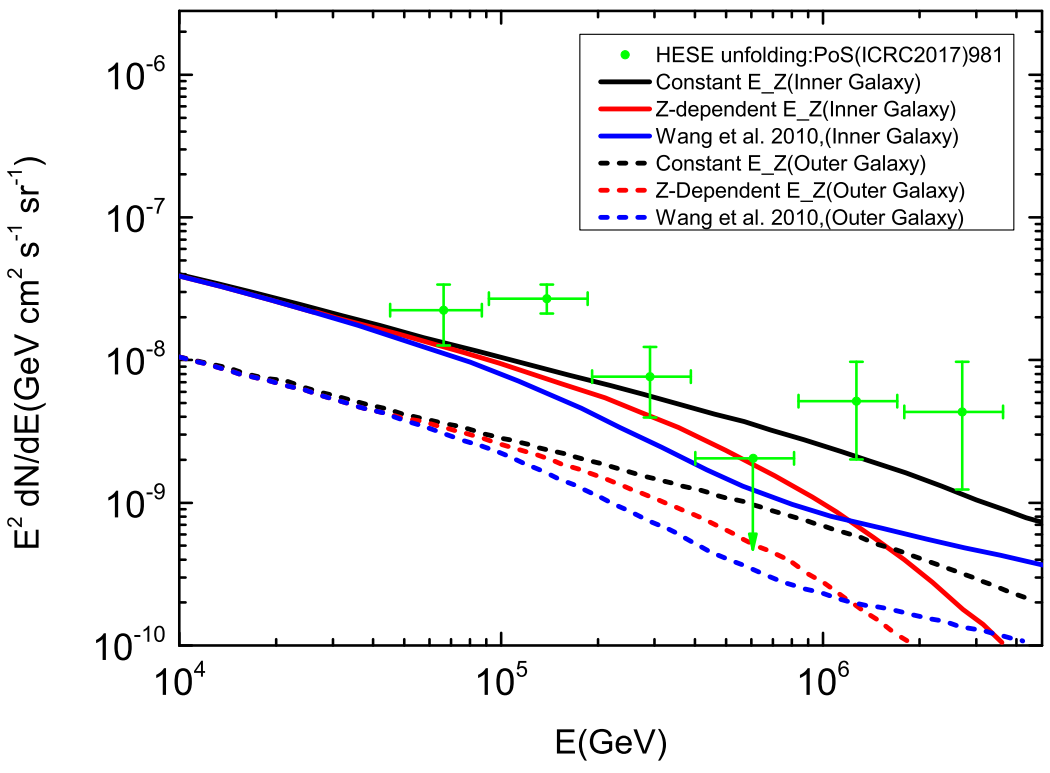

Fig. 3. Expected diffuse neutrino spectrum from the collision of the CRs with the ISM in the inner ($-30^\circ < l < 30^\circ$, $|b| < 5^\circ$, upper solid) and outer ($90^\circ < l < 270^\circ$, $|b| < 5^\circ$, lower dashed) Galactic plane regions, respectively. Also shown are the atmospheric muon neutrino background observed by IceCube and the recent IceCube results of possible astrophysical neutrinos.

In addition, fresh cosmic rays injected recently from the sources could be detained by local magnetic fields. Recent observations revealed a universal hardening of $\gamma$-rays, $e^+/e^-$ and B/C ratio from $\sim$10 GeV, which indicate an additional hard component of cosmic rays and can be ascribed to these detained fresh cosmic rays. By interacting with local gases, these cosmic rays can generate additional gamma rays and neutrinos in the Galactic plane. Figure 4 shows the expected diffuse neutrino spectrum by the collisions between a hard galactic plane component and ISM. As can be seen, the theoretical calculation of all-sky flux including both conventional model and hard component is lower than the experimental observation, which contribute to $\sim 60\%$ of the IceCube observation [114]. Therefore the Galactic neutrinos could contribute between $\sim$10% and $\sim$60% of total neutrino flux observed by IceCube experiment. But as shown in the right panel of Fig. 4, the Galactic neutrino flux mainly comes from galactic plane by the black solid line. This significantly contradicts with the observations of the IceCube collaboration, which claimed an isotropic distribution based on the current number of neutrino events. It is possible that the extrapolation of the hard component to $\sim$ PeV is unlikely to be the right approach, which may overestimate the galactic contribution. The diffuse gamma-ray observations at $\sim$100 TeV by the LHAASO experiment could constrain the Galactic neutrino contribution.

### 4.2 Contribution from Fermi-Bubbles and Galactic halo

In addition to the Galactic plane, the Fermi bubble could be another potential emitter of TeV-PeV gamma rays and neutrinos, via the *pp* collisions of protons or nuclei accelerated in or transported to the bubble region. Recently HAWC reported an absence of gamma-ray excess from the Northern bubble at $b \gtrsim 6^\circ$ Galactic latitude, posing a flux upper limit in the energy range of $1.2-126\,$TeV [115]. The upper limit is consistent with the gamma-ray spectrum measured by Fermi-LAT at $|b| \geq 10^\circ$, where an exponential cutoff at energies $\gtrsim$100 GeV is evident. However, the gamma-ray spectrum at $b \leq 10^\circ$ does not show any cutoff feature up to around 1 TeV, and hence it remains a potential source for TeV-PeV gamma rays and neutrinos. It has been shown that LHAASO may constrain emission in the $0.1-100\,$TeV range if $\leq$10% systematic uncertainties can be achieved [116].

Furthermore, the extended halo of our Galaxy as a diffuse neutrino sources has also been studied by different groups [117–119]. It is suggested that Galactic cosmic rays that escape from the Galactic plane can interact with the diffuse hot gas in the halo, producing gamma rays and neutrinos. The large FOV of LHAASO is beneficial for the measurement of such a diffuse flux. After a long-term operation, either detection or nondetection of the diffuse sub-PeV photon from high Galactic latitude will be useful to evaluate neutrino flux originating from the extended halo. The high-latitude diffuse gamma-ray flux may also shed some light on the "missing" Galactic baryon content in the halo [119].

These result would also be important to understand the distribution and transport of sub-PeV/PeV cosmic rays in our Galaxy.

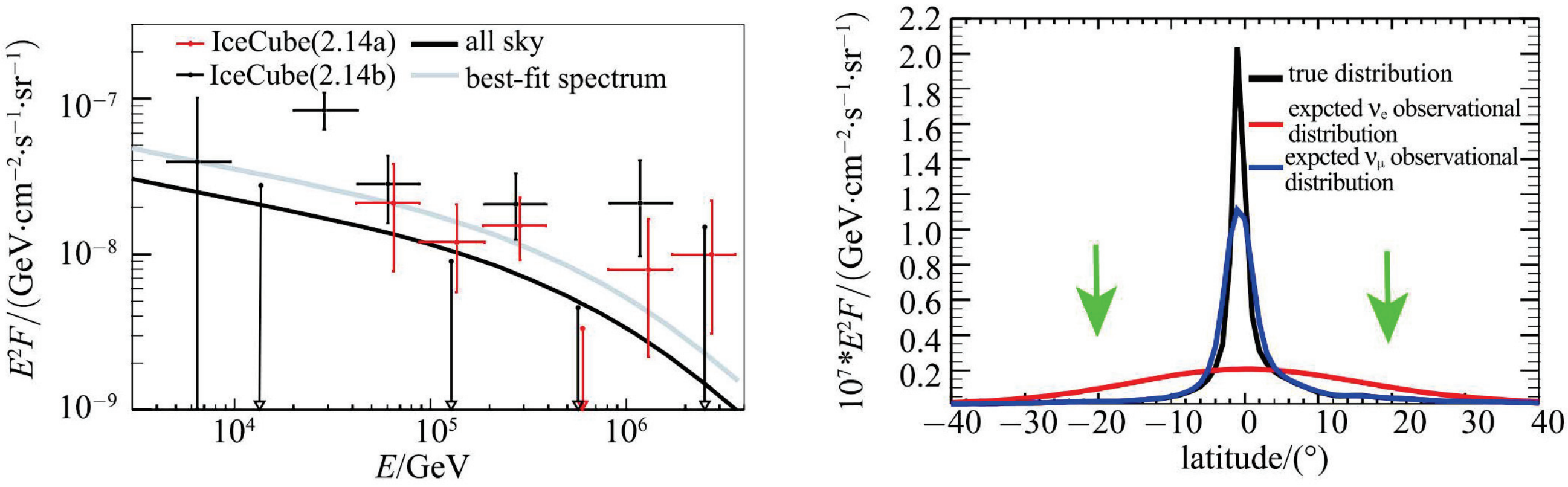

Fig. 4. The left panel shows the calculated diffuse neutrino spectrum from collision of CRs with ISM. The data is astrophysical neutrino observations. The right panel is the integrated neutrino flux for $E > 30$ TeV along $l = 0$. The black line is the true distribution in our model and the red is the reconstructed distribution for $\nu_e(\nu_\mu)$ after considering the angular resolution of $15°(1.5°)$.

## 5 LHAASO Probes the Origin of Ultra-High Energy Cosmic Rays

Executive summary The origin of ultrahigh energy cosmic rays (UHECRs) is not solved yet. The sources of UHECRs may be also producing high energy gamma-rays and neutrinos via hadronic interactions. If the diffuse TeV-PeV neutrinos and UHECRs are produced in related processes, then LHAASO may be expected to identify TeV gamma-ray sources from a fraction $\sim 0.1(n_s/10^{-5}\mathrm{Mpc}^{-3})^{-1/2}$ of the UHECR positions (with $n_s$ being the source density). Even if the UHECR sources are too numerous and weak to detect, the non-detection can still put stringent constraint on the source density and hence the origin of UHECRs.

The origin of the observed UHECRs, $>10^{19.5}$eV, are still unknown (see review [120]). Because of the Greisen-Zatsepin-Kuzmin (GZK) effect, the effective propagation length of cosmic rays with $>$50EeV is only $d_{GZK}\lesssim$1Gpc. The UHECRs detected on the Earth should be originated from sources with a distance of $d<d_{GZK}$. Cosmic rays are deflected by magnetic field during propagation, but UHECRs of $>50$EeV are expected to be deflected by only $<2°$, assuming UHECRs are protons [see, e.g., 121]. Thus their arrival directions may trace back to the sources. Cosmic ray sources may also produce high energy gamma-rays and neutrinos by the hadronic interactions of cosmic rays. Using LHAASO to observe the positions of UHECRs may enhance the chance of finding the UHECR sources. Note, within 100 Mpc the gamma-gamma absorption due to the extra-galactic background lights may not be very important for TeV gamma rays. Therefore, high energy gamma-ray observations by LHAASO must be very helpufl to probe UHECR origin.

The IceCube-detected TeV-PeV neutrino flux is comparable to the Waxman-Bahcall bound, which is derived from observed UHECR flux. This may indicate that the origin of TeV-PeV neutrinos is related to the origin of UHECRs, i.e., the cosmic rays that result in the TeV-PeV neutrinos is with the same origin as the UHECRs [122]. If so, a TeV-PeV gamma-ray flux comparable to the TeV-PeV neutrino flux should also be accompanying the production of UHECRs. We here suggest LHAASO to search the high energy gamma ray signals from the UHECR positions.

Let us estimate the possible observational results by LHAASO. Derived from the IceCube detection, the gamma-ray emissivity, i.e., the energy production rate density in the universe, at $\sim 10$ TeV should be about $\dot\rho \sim 10^{43}\dot\rho_{43}\mathrm{erg Mpc^{-3} yr^{-1}}$ [123]. If the source density is $n_s = 10^{-5}n_{-5}\mathrm{Mpc}^{-3}$, then the (average) gamma-ray luminosity of a single source is $L = \dot\rho/n_s$, and the maximum distance that the sources can be detected for a telescope of given sensitivity $S$ is $d_M = (L/4\pi S)^{1/2}$. Since the integrated sensitivity of LHAASO at 3-TeV is $S_3{\sim}6{\times}10^{-14}\mathrm{TeV cm^{-2}s^{-1}}$ (for 1-yr exposure), the maximum distance is given by

$$d_M = 54\dot\rho_{43}^{1/2} n_{-5}^{-1/2}(S_3/6\times10^{-14}\mathrm{TeV cm^{-2}s^{-1}})^{-1/2}\mathrm{Mpc}. \quad (3)$$

For comparison, the mean free path of 3-TeV gamma-rays in the intergalactic medium is larger, $d_\tau \sim 100$Mpc. The number of sources that are within a distance of $d_M$ is $N_s \approx \frac{4}{3}\pi d_M^3 n_s$, i.e.,

$$N_s \sim 6.6\dot\rho_{43}^{3/2} n_{-5}^{-1/2}(S_3/6\times10^{-14}\mathrm{TeV cm^{-2}s^{-1}})^{-3/2}. \quad (4)$$

As the exposure time increases, the sensitivity goes as $S \propto t^{-1/2}$, thus the number of observed sources for 10 yrs is

$$N_s \sim 40 n_{-5}^{-1/2}(t/10\mathrm{yr})^{3/4}. \quad (5)$$

Here the source density has been normalized to a value typical for starburst galaxies, whereas the number of observed sources will decreases if the sources are more numerous, and hence weaker, $N_s \propto n_s^{-1/2}$.

These gamma-ray sources should lie in the directions of the observed UHECRs. For a certain UHECR experiment, the fraction of its detected UHECRs that are originated within a distance $d_M$ is about $f_s \sim d_M/d_{GZK}$ ($d_M < d_{GZK}$), i.e., $f_s \sim 0.1 n_{-5}^{-1/2}(t/1\mathrm{yr})^{1/4}$ for UHECRs of $>50$EeV, which does not vary much with exposure time, $\propto t^{1/4}$. LHAASO can search for these gamma-ray sources by carrying out observations at a few TeV at the positions of the UHECRs that are detected by the Telescope Array (TA) experiment and Pierre Auger Observatory (PAO). A fraction $f_s$ of these UHECR positions are expected to be identified with TeV gamma-ray sources. If the sources are too weak (i.e., $n_s$ is large) to detect, we can carry out stacking analysis of all the TA and/or PAO detected UHECR positions. Due to the wide field of view and high sensitivity of LHAASO, even non-detection of signals will put stringent limit on the source density and hence the origin of UHECRs.

## 6 Search for Neutrinos with Horizontal Air Showers from LHAASO

The observation of PeV cosmic neutrinos is an essential tool for understanding cosmic ray acceleration, composition and source evolution. These particles are produced either within astrophysical sources or when ultra-high energy cosmic rays interact in transit through the cosmic background radiation. Neutrino signals may accompany electromagnetic and gravitational messengers, or they may be the primary signal at the site of cosmic ray acceleration. Neutrino messengers from transient or flaring sources (AGN, GRB) are a topic of intense study. In 2018, the IceCube detection of a neutrino event coincident with a gamma-ray flaring blazar started a multi-messenger era that includes neutrinos [1]. The detection of cosmic neutrinos above energies of $10^{16}$ eV (= 10 PeV) has not yet been achieved.

The observation of neutrinos through the detection of Horizontal Air Showers (HAS) with the LHAASO experiment is the subject of this paper. In fact, the only way for LHAASO to observe such events is to look for HAS, i.e., showers with zenith angles >70° [2, 3]. The study of HAS is an important tool for UHE cosmic neutrinos measurement, in particular for the quest for prompt $\nu_e$ produced in GRBs [3].

The cosmic ray (CR) flux is a steeply falling function of zenith angle because the depth of atmosphere traversed by a shower reaching the ground rises rapidly from 1030 to about 36000 g/cm$^2$ as the zenith angle varies from zero to 90°. Thus near the horizon the interaction point is separated by about 1000 radiation lengths of matter from the detector. Most secondaries such as electrons, pions and kaons are absorbed in the dump and only penetrating particles, such as muons and neutrinos produced in the initial interaction, are able to reach the detector. Therefore, to CRs incident near the horizon the Earth's atmosphere represents a beam dump.

The observation of extensive air showers (EAS) in nearly horizontal directions provides a *"well shielded laboratory"* for the detection of penetrating particles: high energy muons, cosmic neutrinos, possible weakly interacting particles produced in the decays of cosmological superheavy particles, will leave a clear signature in this dump.

Measurements of the CRs rate at different zenith angles give information on the relative number of muons in a shower, which is dependent on the CR elemental composition, thus providing an important tool to probe the CR mass distribution [5]. In addition, for very high energy interactions the decay of charm particles is the dominant source of high energy secondary muons. So counting high energy muons at large zenith angles determines the charm cross section [6–8]. There is no background from the semi-leptonic decay of pions and kaons which, as a result of time dilation, interact and lose energy rather than decay into high energy muons.

The detection of EAS at large atmospheric zenith angles (Horizontal Air Showers, HAS) has been firstly reported in 1965 at an energy above $10^{14}$ eV [9]. In the seventies their origin has been studied by Bohm and Nagano [10], but their interpretation was not straightforward, due to the contradiction between the expected and detected muon contents. The EAS-TOP experiment studied in detail the phenomenology of HAS, finding that they are mainly due to muon-dominated showers produced by Ultra High Energy (UHE) cosmic rays interacting at very large distance in the atmosphere [11]. In the last years a big effort has been made to study in detail the phenomenology of these events with accurate MC simulations (see, e.g., [12–15]).

HAS are believed to be mainly due to the atmospheric muons and their interactions, as example:

- (a) high energy single muons can interact through bremsstrahlung or deep inelastic scattering and initiate showers at the depth appropriate for detection. Such showers are essentially electromagnetic, since the remnant muons from the initial shower (whose typical primary energy is not much larger than the muon one) are dispersed over a very large area.

- (b) UHE CRs interacting at the top of the atmosphere, at very large zenith angles, produce a "large" amount of muons through the pion decays (favoured, at large angles, with respect to pion in-teractions due to the low atmospheric density at the interaction altitude). Such showers are therefore composed essentially of muons since the e.m. component is completely absorbed.

Neutrino induced showers have some intermediate typology, being more similar to conventional CR air showers or to events (a), when a large amount of their energy is transferred to the electromagnetic cascade. EAS arrays must have the possibility of discriminating between the different typologies of events through $\mu$/e identification. The LHAASO experiment, with an array of about 40,000 $m^2$ muon detector, is the most suitable apparatus for such studies.

### 6.1 Atmospheric neutrino induced showers

Fig. 5 summarizes the weak interaction channels of neutrinos [4]. In all cases around 20% of the energy of the primary neutrino is transferred to the hadronic jet which results from the nucleon debris. These particles initiate cascades very similar to those produced by protons. The remaining 80% of the energy of the primary particle is contained in a ultra-energetic lepton. The actual energy that is transferred to the shower depends on the interaction channel and neutrino flavour.

If the shower is initiated by a $\nu_e$ through charged current (CC), the resulting electron initiates an electromagnetic shower overlapping the hadronic one produced by the jet. In this case 100% of the energy is transferred to the shower. On the contrary, neutral current interactions (NC) produce a secondary neutrino instead of an electron. This neutrino escapes and does not contribute to the process of multiplication, carrying around 20% of the energy of the primary neutrino.

Inclined showers initiated by a $\nu_\mu$ through CC are very similar to the ones initiated via NC.

The $\nu_\tau$ via CC presents an interesting characteristic. In the same way as the muon, the $\tau$ lepton is a very penetrating particle which can travel an important distance from the point at which it was produced. On the other hand, its lifetime is seven orders of magnitude lower so it can decay before reaching the surface producing a secondary shower that is added to the one initiated by the hadronic jet (see third panel in Fig. 5). This kind of showers are commonly known as *"Double Bang" (DB)*. Depending on the decay channel of the $\tau$, the second shower will be of hadronic or electromagnetic nature.

The mean free path for neutrinos is much higher than the atmospheric depth, therefore, neutrinos can interact at any point in the atmosphere with almost the same probability. In particular, neutrinos can initiate an inclined shower deep in the atmosphere, inside a fiducial volume in which the electromagnetic component reaches the surface. This characteristic distinguishes neutrinos from other possible particles like protons, nuclei or photons which interact in the first hundred grams of the atmosphere with a probability close to 1.

Tau neutrinos are not produced in the decay of charged pions, therefore they are suppressed in the neutrino production relative to $\nu_e$ and $\nu_\mu$. Nevertheless, after travelling cosmological distances, because of neutrino flavour mixing, the usual 1:2 of $\nu_e$ to $\nu_\mu$ ratio at production is altered to approximately equal fluxes for all flavours [16].

A VHE $\nu_\tau$ entering in the Earth or mountain can undergo charged-current interactions producing a tau lepton $\tau$ that can emerge into the atmosphere and decay in flight producing a nearly horizontal extensive shower [17]. Subsequently, it decays and produces an air shower. Particles or Cherenkov photons from the air shower are detected. Due to the separation of the first interaction where $\nu_\tau$ produces tau and $\tau$ decay generating air shower, air shower observation becomes possible while preserving the huge target mass required to compensate for the low cross section of the first interaction. For this detection method, it is crucial for the $\tau$ to go through the Earth and/or mountain before its decay, and to develop the air-showers in the atmosphere in front of the detector after its decay. The Earth crust, with a density 1000 times greater than the air density ($\rho_{Earth}$ = 2.65 g/cm$^2$), is a target much more massive than the atmosphere. Under a spherical Earth approximation the distance the neutrino has to go through is about 220 km at 89°. This means that 30% of these neutrinos should interact at this zenith angle. However, one needs also to consider the probability of the resulting $\tau$ escaping the Earth and decaying in a fiducial volume to be detected.

The tau decay probability depends on the tau energy $E_\tau$. For reference, the tau decay length, with the $\gamma$-factor $\gamma = E_\tau/(m_\tau c^2)$, is $\gamma c\tau = 5$ km$\times E_\tau/10^8$ GeV.

The Earth-skimming process (neutrino interaction in the Earth crust) only applies to tau neutrinos. The detection of electron neutrinos when interacting in the Earth is very suppressed as the resulting high energy electron will give rise to a shower in the Earth. The problem with muon neutrinos is that muons have a lifetime $7.5\times10^6$ longer than taus so even if they can escape the Earth they decay very high in the atmosphere and the particles from this up-going shower never reach the ground.

### 6.2 Search for 100 TeV $\nu_e$ from high-energy transients

High-energy transients, including gamma-ray bursts (GRBs), supernovae, and blazars, are potential sources of high-energy cosmic rays. Neutrinos are penetrating neutral particles, and among the four traditional messengers (charged cosmic rays, photons, gravitational waves and neutrinos), they maybe the best probe of the origin

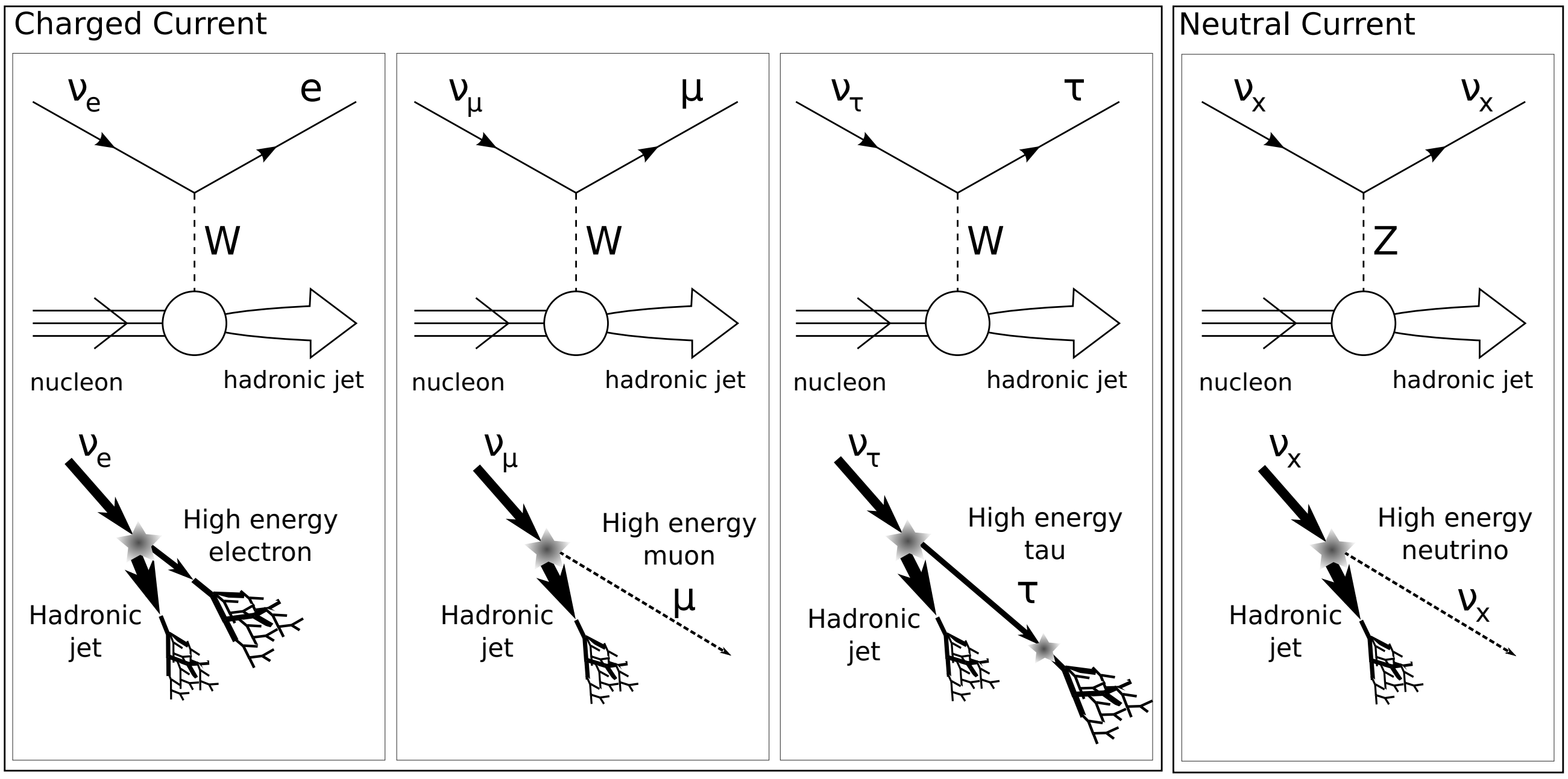

Fig. 5. Neutrino interaction channels (reprinted from [4]).

of cosmic rays.

The expected neutrino flux can vary by order of magnitude between GRBs due to the fluctuations in the burst parameters and several GRBs in the LHAASO data set will be not in the IceCube list. Moreover, core-collapse supernovae (SNe), which are believed to be the origin of long-duration (>2 s) GRBs, could produce neutrino bursts in events in which mildly relativistic jets interact inside the stellar envelope. The neutrino spectrum produced in these events (as, for instance, the "choked GRBs") is expected softer but very high in fluence.

This technique could represent a complementary methodology to search with large area arrays for $\nu_e$ associated to buried GRB jets which are predicted to generate a soft neutrino spectrum with a fluence at TeV energies a few orders of magnitude greater than that characterizing the GRBs observed in gamma-rays.

The search relies on the directional and temporal information coming from satellite observations. In the angular and time windows of the prompt emission we look at an excess of events with respect to the background air showers induced by high energy atmospheric muons which undergo catastrophic energy losses due to radiative processes. Electrons produced in the CC interaction can carry a significant fraction of the neutrino energy, about 50% at low energies, rising to 75% above 100 TeV. These electrons promptly initiate an electromagnetic cascade that can be detected if the neutrino interaction point is at the appropriate distance from the LHAASO array (inside a fiducial volume). Due to the limited longitudinal development of the shower, the target thickness is by far smaller than the one obtained by observing long-range high energy muons in IceCube or Cherenkov photons from tau showers.

Cascade events are also produced in CC interactions when the generated muon radiates knock-on electrons, bremsstrahlung photons or electron pairs, and in $\nu_\tau$ CC interactions when the resulting tau decays into an electron (three body decay, about 18% branching ratio) or into mesons (about 64% branching ratio, but only 12% for two body decays). At any given neutrino energy all these showers have much less energy than that induced by the electrons generated in the $\nu_e$ interactions, and the contribution of these process can be at first neglected.

All the neutrino flavours can generate cascades via the NC interaction, producing hadronic jets. The contribution of these events is expected to be quite low, reflecting the combination of a smaller cross section (about a factor 3 at energies <100 TeV) and the decrease of the neutrino flux with the energy. Moreover, the energy deposited in the hadronic jet, less than 50% that of the neutrino, can be shared by more than one particle.

The LHAASO experiment, due to different detectors (water ponds, dense scintillator array, muon detectors and wide field of view Cherenkov telescopes) is well suited to measure HAS and study neutrino-induced events in an unprecedented energy range, from TeV up to about $10^{16}$ eV. In particular, the unprecedented muon detection area (more than 40,000 $m^2$) will allow to discriminate neutrinos from the cosmic ray background in an excellent way.

### 6.3 Observation of HAS with LHAASO

At zenith angles $\theta > 60°$ an excess of events is observed above the rate of EAS as expected from the exponential absoprtion (with $\Lambda_{EAS} \approx 220$ g/cm$^2$) of the air shower electromagnetic component in the large atmospheric depth (see Fig. 6), which implies a decrease of the EAS counting rate with $\Lambda_c \approx 130$ g/cm$^2$.

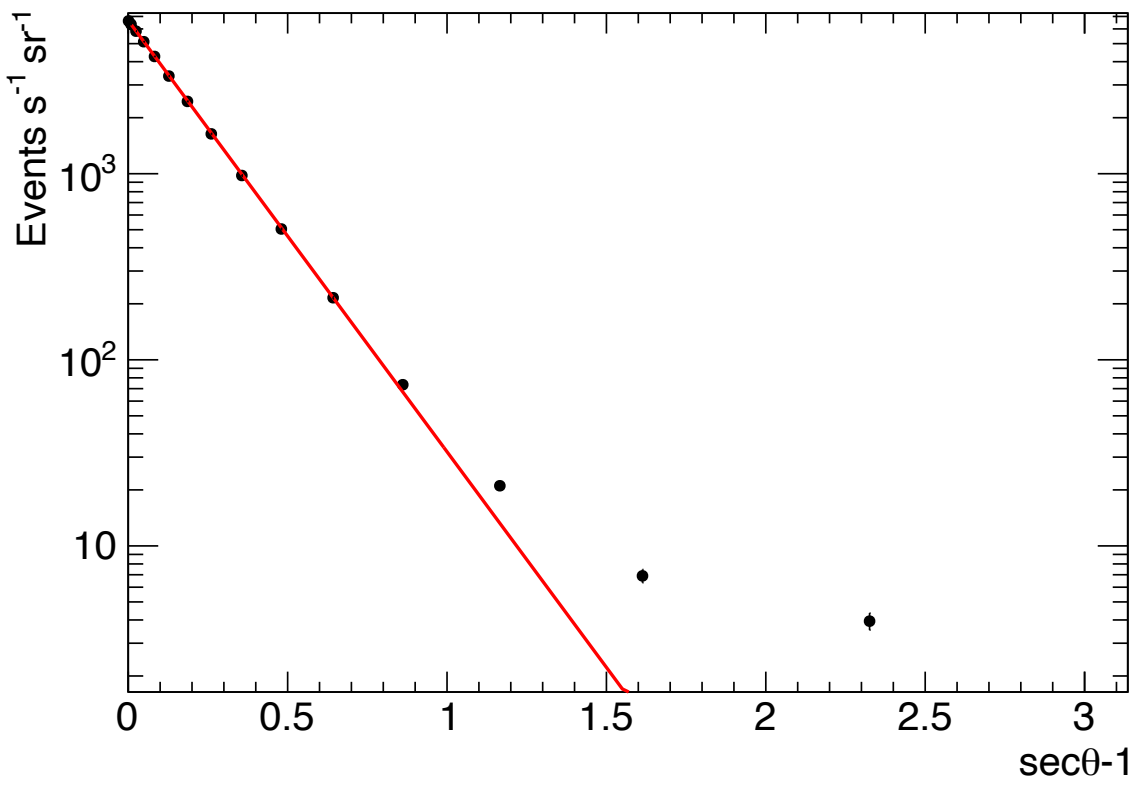

Fig. 6. The zenith angle distribution of EAS measured with LHAASO. The best fit out to ∼60° is also shown.

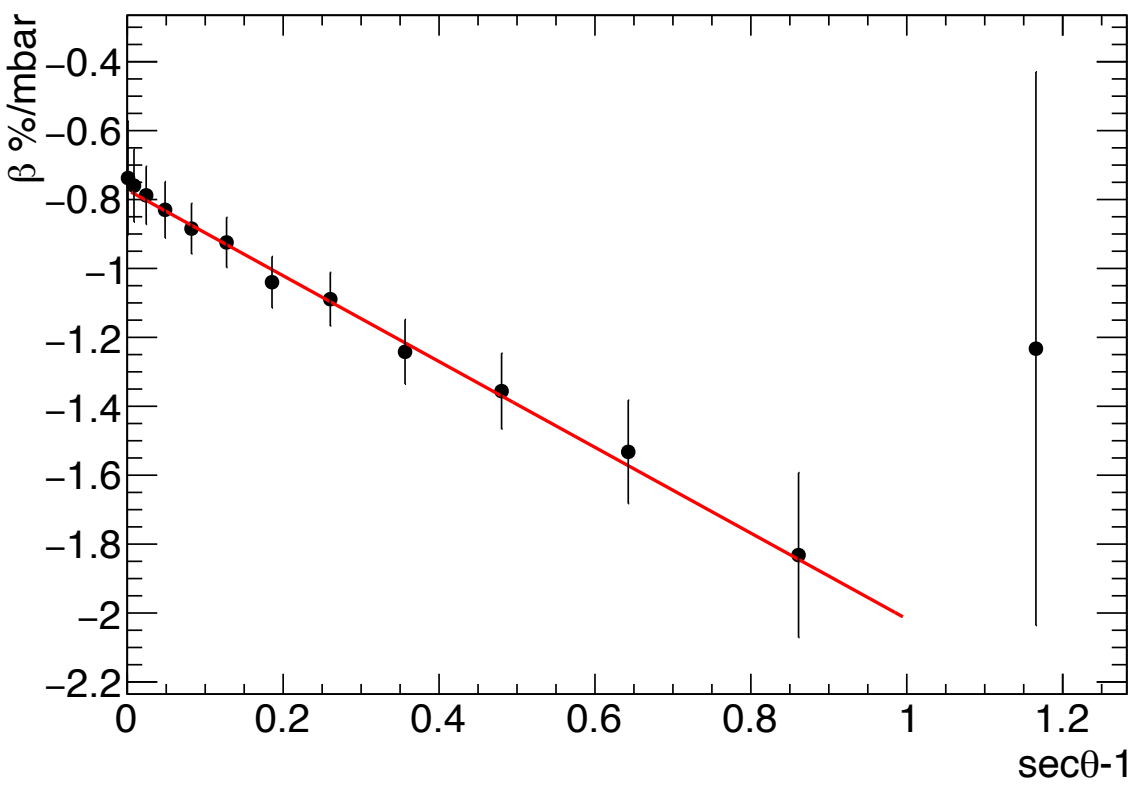

Fig. 7. The barometric coefficient for different zenith angles as measured by LHAASO.

The dependence of the barometric effect on the zenith angle, shown in Fig. 7, clearly shows a deviation from the sec$\theta$ behaviour for sec$\theta$>2. In fact, the barometric coefficient $\beta = \frac{1}{n}\frac{dn}{dx}$ ($n$ = counting rate, $x$ = atmospheric pressure) is related to zenith angle as: $\beta(\theta) = \beta(0°) sec\theta$. This can be explained by the presence of a "non-attenuated" EAS component that dominates for angles larger than 70°.

Finally, the physical nature of the HAS observed by LHAASO is demonstrated by the observation of showers shown in Fig.**??**

A particle (muon or neutrino) with $\theta$>70° interacting deep will present a young shower front.

The fair agreement with the expected spectral index makes us confident that the bulk of HAS observed by LHAASO are due to muon-induced showers.

### 6.4 Conclusions

The study of HAS has been long recognised as a useful tool to investigate the interactions of high energy muons and to detect ultra high energy neutrinos.

The LHAASO experiment, due to different detectors (water ponds, dense scintillator array, muon detectors and wide field of view Cherenkov telescopes) is well suited to measure HAS and observe neutrino-induced events in an unprecedented energy range, from TeV up to about $10^{16}$ eV. In particular, the unprecedented muon detection area (more than 40,000 m$^2$) will allow to discriminate neutrinos from the cosmic ray background in an excellent way.

# Chapter 7 Solar and Heliospheric Physics and Interdisciplinary Research with LHAASO*

Huanyu Jia(贾焕玉)[1★¶†] Li Feng(封莉)[2★¶] D. Ruffolo[3¶] Yu. V. Stenkin[4,5¶] Xunxiu Zhou(周勋秀)[1¶]

[1]School of Physical Science and Technology, Southwest Jiaotong University, Chengdu, China
[2]Key Laboratory of Dark Matter and Space Astronomy, Purple Mountain Observatory, Chinese Academy of Sciences, Nanjing, China
[3]Department of Physics, Faculty of Science, Mahidol University, Bangkok, Thailand
[4]Institute for Nuclear Research of Russian Academy of Sciences, Moscow, Russia
[5]Moscow Institute of Physics and Technology, Moscow, Russia

**Abstract:** In the following sub-sections, studies of solar-heliospheric effects on cosmic rays, investigating a possible link between cosmic ray flux and Earth's climate, and detection of MeV-range $\gamma$-rays from thunderstorms with the data from LHAASO will be discussed; geophysical research with environmental neutrons will be introduced, and some Monte Carlo simulation results about effects of thunderstorm electric fields on LHAASO observations of cosmic rays will be given.

## I. SOLAR–HELIOSPHERIC EFFECTS ON COSMIC RAYS

The LHAASO team will also perform interesting studies, both for basic science and for applications, of solar and heliospheric effects on the cosmic ray flux. These are basically the effects of the solar wind and solar storms, and as such they are directly related to so–called "space weather" effects of the solar wind and solar storms on human activity. LHAASO will obtain unique information on the magnetic fields between the Sun and the Earth, which are moving toward Earth with the solar wind. Indeed, LHAASO will obtain information on the direction of the interplanetary magnetic field, which is a key determinant of whether a solar wind disturbance or solar storm will result in reconnection with Earth's magnetic field and trigger a geomagnetic storm. Thus the real–time data from LHAASO will complement other information for space weather forecasting. In addition, LHAASO will perform numerous other studies of solar, heliospheric, and geomagnetic effects on cosmic rays.

### A. Types of data

Solar storms and the solar wind, as they propagate throughout the heliosphere, have a profound effect on cosmic rays at GeV–range energy, leading to a wide variety of signatures in cosmic ray flux variations with time. These have mostly been studied with detection thresholds up to 10 GeV or slightly higher. With its tremendous count rate at high altitude, LHAASO will obtain sufficient statistics to open the gateway to study many of these phenomena at even higher energy, thus providing new information on these processes. In addition, some phenomena, such as the Sun shadow and Moon shadow, are more profitably examined at TeV energies, where LHAASO will again provide improved statistical accuracy. In studies of time variations, improved statistics allow the possibility of studies at finer time resolution. As we shall describe, LHAASO can even provide useful real–time data for space weather forecasting.

LHAASO will generate two types of data of interest for solar–heliospheric studies: reconstructed shower rates (as a function of energy, direction, and time) and scaler rates (as a function of threshold energy and time).

One of the design goals of LHAASO is to reconstruct showers for gamma rays and cosmic rays down to tens of GeV in energy. Thus reconstructed showers with information on the arrival direction of primary cosmic rays will allow the study of the Sun and Moon shadows over a

Received 2 December 2021; Accepted 4 December 2021; Published online 20 January 2022

* Supported in China by National Natural Science Foundation (NSFC) (12047576, U2031101, 11475141), and the National Key R&D Program of China (2018YFA0404201, 2018YFA0404202), and in Thailand by the Thailand Science Research and Innovation (RTA6280002)

† E-mail: hyjia@swjtu.cn

★ Editors ¶ Contributors. All authors contribute equally to the work.

wider energy range, as well as loss–cone anisotropy to provide advance warning of the arrival of some interplanetary shocks, which can lead to various space weather effects. Directional shower data down to tens of GeV will allow a better determination of the diurnal anisotropy as well [1].

LHAASO will also produce scaler rates (also referred to as the single particle technique or SPT), in which the shower is not reconstructed but count rates are collected for various threshold numbers of "hits" in the detectors. The rate for each threshold has a different response as a function of the primary cosmic ray energy. This opens a possibility to obtain higher rates (and better time resolution) and information on lower energies below the shower reconstruction threshold. (Note, however, that the LHAASO site has a cutoff rigidity of about 13 GeV for protons, so cosmic rays below this energy cannot be examined.) Examples of existing detectors that have examined scaler rates are ARGO–YBJ [2–4] and Auger [5]. To make proper use of scaler rates, we will need to correct for environmental factors. This will require careful monitoring of the weather, atmospheric conditions, and local temperature at each detector.

### B. Sun shadow and advance warning of the interplanetary magnetic field for space weather forecasting

#### 1. *Information from the Sun shadow*

The shadow of the Sun in TeV–range cosmic rays directly relates to solar–terrestrial relations, i.e., how solar phenomena affect the Earth and its immediate environment. The solar wind is a radial flow of plasma out from the Sun at supersonic speeds, which comes out at all times and in all directions (Figure 1). The solar wind drags out the complex coronal magnetic field to become the interplanetary magnetic field. However, both the solar wind plasma flow and the interplanetary magnetic field are highly turbulent, and magnetic fluctuation amplitudes are of the same order as the mean field. Roughly speaking, an interplanetary field line connects parcels of plasma that came from the same region of the solar corona, and because of the solar rotation, its shape is curved into an Archimedean spiral. The solar wind plasma and magnetic field usually take about 4 days to come from the Sun to the Earth.

The arrival direction distribution from shower reconstruction of TeV–range cosmic ray trajectories shows deficits corresponding to the locations of the Sun and Moon [7]. The solar, interplanetary, and terrestrial magnetic fields deflect the particle paths and shift the shadow of the Sun from its actual location, as first reported by the Tibet AS experiment [8]. In other words, the measured deflection of cosmic rays is a cumulative effect of magnetic fields along the whole path from the Sun to the Earth.

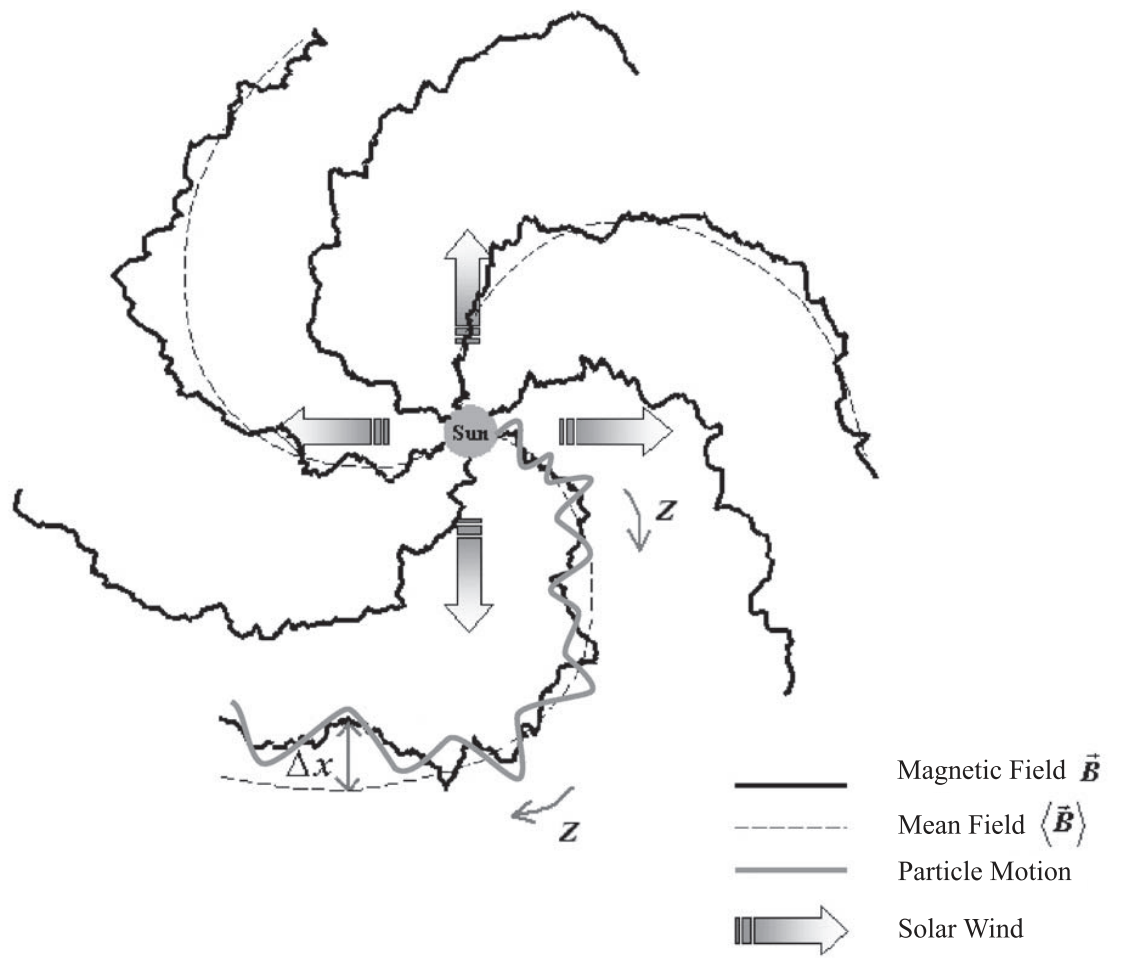

**Fig. 1.** (color online) Illustration of the solar wind and interplanetary magnetic field. The solar wind is emitted radially from the Sun in all directions at all times. The spiral magnetic field lines connect plasma that originated from the same location on the rotating solar surface. Note that the turbulent magnetic field lines (solid lines) do not coincide with the mean magnetic field lines (dashed lines). The Earth might be located near the bottom of the figure. [6]

This experiment also observed the effect of the interplanetary magnetic field (IMF) [9, 10] a solar cycle variation [11], and coronal mass ejections [12], and evaluated the effects of two coronal magnetic field models [13]: the potential field source surface (PFSS) [14, 15] and current sheet source (CSSS) models [16, 17]. Further experiments of LHAASO can also be used to evaluate more advanced coronal magnetic field models, e.g., the nonlinear force–free field (NLFFF), magneto–hydrostatic (MHS), and magneto–hydrodynamic (MHD) magnetic field models [18]. On the other hand, the *electromagnetic fields investigation* (FIELDS) aboard the *Parker Solar Probe* (PSP) offers the first–ever magnetic field measurements in the interplanetary space as close as 10 solar radii from the Sun. Together with the solar wind speed measurements by the *Solar Wind Electrons Alphas and Protons Investigation* (SWEAP), PSP can add further valuable observational constraints to the aforementioned magnetic field models [19]. The recently launched Solar Orbiter mission is equipped with in–situ magnetic field and solar wind measurements in interplanetary space as well. Together with the measurements close to Earth, the optimal magnetic field models can hopefully be obtained.

#### 2. *Solar cycle variation of the Sun shadow*

Solar activity, including the likelihood of solar storms and space weather effects on human activity, is positively associated with the sunspot number, which varies with a cycle of roughly 11 years, known as the "sunspot cycle" or "solar cycle" (see Figure 2). The ARGO–YBJ collaboration also found that the deficit of cosmic ray flux in the Sun shadow is reduced with increasing solar activity [22], as shown in Figure 3. To understand the shadow effect, it is useful to imagine trajectories of antiparticles traveling backward from Earth to intersect the Sun's surface, which are equivalent to the forward trajectories that are blocked by the Sun, causing the shadow. One possible explanation of the weaker Sun shadow with increasing solar activity is that if the solar coronal magnetic fields are very irregularly distributed, the cosmic ray deflections could be so randomized that backward trajectories over a wider range of angles can intersect the Sun. Ref. [22] considers another mechanism: variation and frequent reversals of the IMF during each three–month observation period causes a superposition of Sun shadows with different shifts and leads to an observed shadow that is wider and weaker.

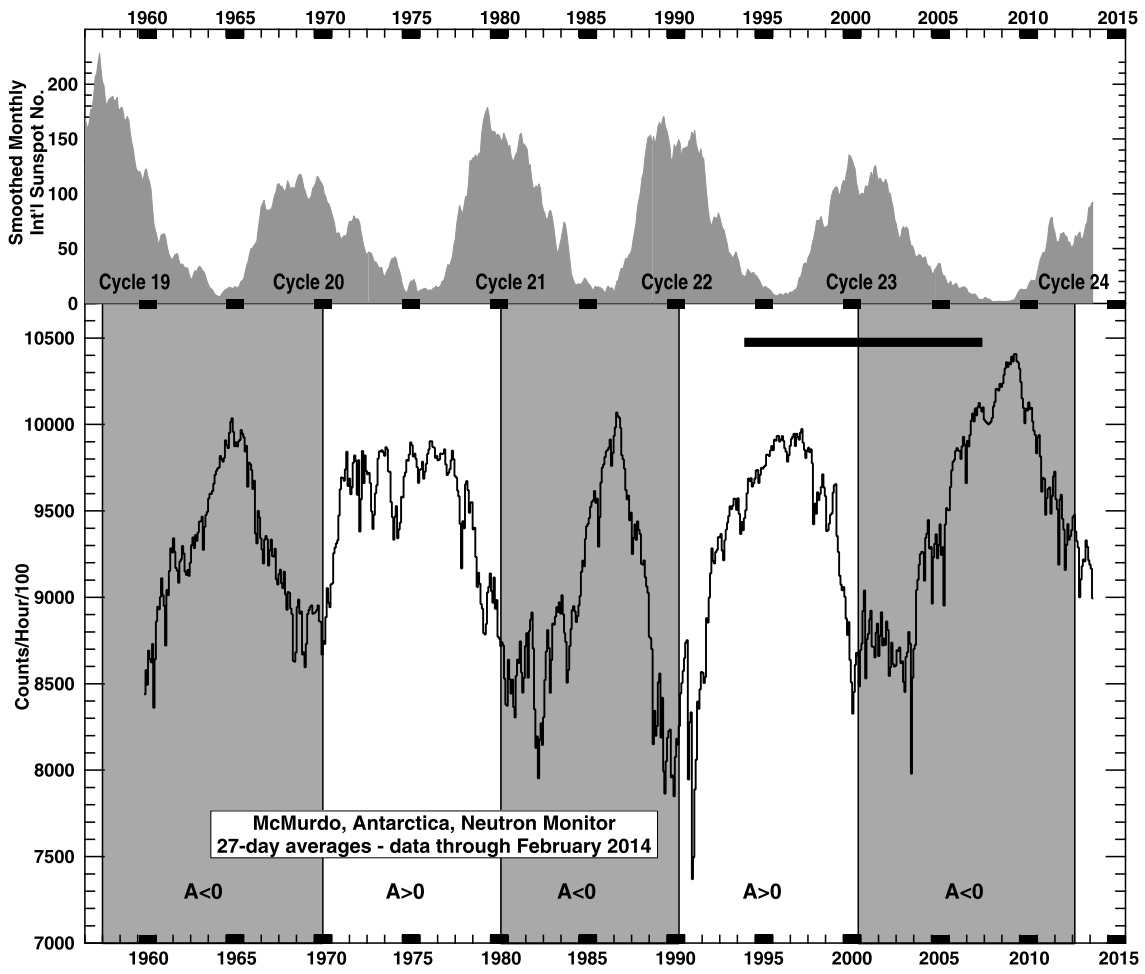

**Fig. 2.** (color online) Smoothed monthly international Sunspot number (using 5 –month boxcar smoothing) and McMurdo neutron monitor count rate as a function of time. The long–term drift at McMurdo has been corrected following [20]. A neutron monitor count rate indicates the Galactic cosmic ray flux, which undergoes "solar modulation" in association with solar activity. Solar modulation includes dramatic 11–year variations with the sunspot cycle, and a 22–year variation with the solar magnetic cycle, seen here in changes in the solar modulation pattern between positive ($A$>0) and negative ($A$<0) magnetic polarity. [21].

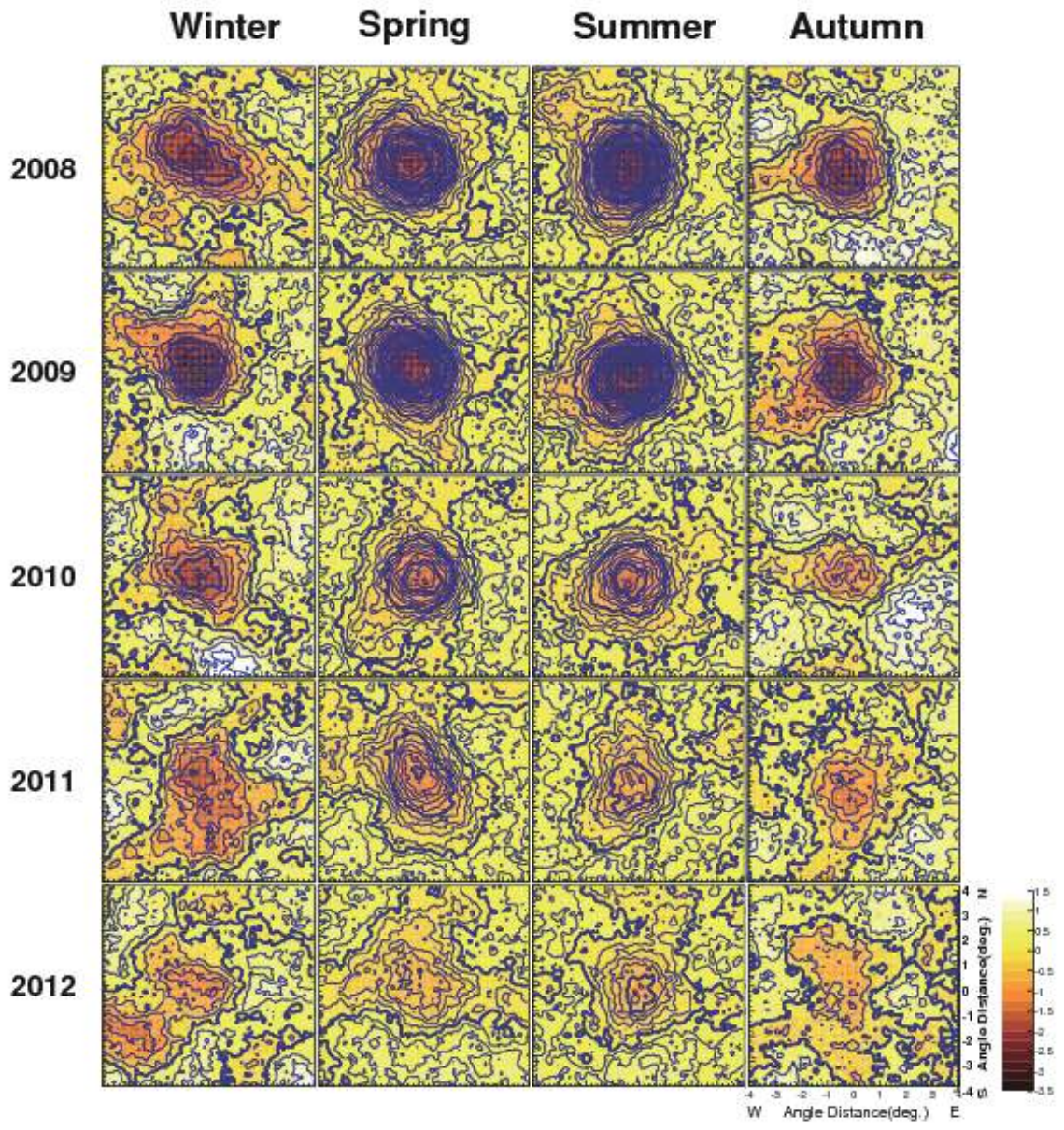

**Fig. 3.** (color online) Seasonal variation in the Sun shadow observed by the ARGO–YBJ experiment in cosmic rays at median energy 5 TeV. The observation period for each map is one astronomical season in the Northern Hemisphere. The smoothing radius is 1.2° and the pixels are 0.1° × 0.1°. Each map shows the fractional change in the cosmic ray flux (color scale) and the statistical significance of the change (contours). Each contour represents an integral value of the significance (in units of the standard deviation), with darker contours every 5 units. Maps for the Spring and Summer seasons show stronger significance because the Sun was higher in the sky at the ARGO–YBJ site in Tibet. The fractional change suddenly weakened in Spring 2010, in association with a sudden increase in IMF variability, whereas the sunspot number and some other generic indicators of solar activity started to increase rapidly only in Spring 2011 [22].

### 3. *Relevance to space weather forecasting*

The Sun produces energetic particles due to occasional sudden explosions at its surface, called solar storms, which can accelerate particles to relativistic energies (ions up to tens of GeV, electrons up to tens of MeV) for durations up to about an hour. Furthermore, a type of storm called a coronal mass ejection (CME) can drive an interplanetary shock that accelerates ions up to tens of MeV (called "energetic storm particles") over several days. The particles due to solar storms, collectively called solar energetic particles (SEPs), pose a radiation hazard to astronauts and high –altitude passenger aircraft for short but unpredictable time periods, as well as damaging expensive satellites and spacecraft (at least fifteen have been disabled by solar storms to date). Strong UV and X–ray fluxes lead to increased ionospheric ionization and disturb human communications and navigation signals. The shock and CME carry particularly strong magnetic fields, and they can significantly disturb the Earth's magnetosphere. In particular, a strong southward magnetic field can lead to magnetic reconnection and a strong inflow of solar wind plasma and energetic particles into the Earth's magnetosphere, which can also damage satellites. A disturbed magnetosphere can lead to geomagnetically induced currents and power outages. All these effects on human activity can collectively be called "space weather" effects.

There is great practical interest in space weather pre-diction, but current prediction capabilities are very limited. The situation is analogous to long–term weather forecasting some decades ago, when the best results were based on prior experience and qualitative concepts. For modern space weather forecasting, even after a CME has been observed at the Sun, it remains difficult to predict when an interplanetary shock and CME will arrive at Earth (which can be ~1–4 days, depending on the CME speed), and very difficult to measure or infer the orientation of the magnetic field of the CME; a southward field would result in particularly strong space weather effects.

The ARGO–YBJ experiment first used the Sun shadow displacement in the south–north direction to measure the intensity of the magnetic field between the solar wind from the Sun to the Earth, during the previous period of minimum solar activity [23] (see also [10]). This capability could also be used to determine the mean magnetic field orientation between the Sun and Earth, before the field arrives at Earth. At present, the best reported time resolution for Sun shadows – that of the ARGO–YBJ group [22] – is three months, which is not of practical use for space weather forecasting. However, because of its much greater size, LHAASO is expected to produce a statistically meaningful Sun shadow every 1–2 days. This is then directly useful for space weather forecasting. For example, a Sun shadow determined for time *t* can be compared with the previous Sun shadow, observed 1–2 days earlier, and the difference is due to new magnetic fields that have emerged from the Sun minus old fields that have passed the Earth. When making use of *in situ* spacecraft observations of the interplanetary magnetic field, we can determine the new magnetic field that emerged from the Sun during that time interval. Then we can infer the orientation of the magnetic field that will pass by Earth in the next few days, including whether the field will have a strong southward component. This will be important information to complement existing data for space weather forecasting. On the other hand, we can perform state–of–the–art MHD simulations of the CME propagation in solar–terrestrial space, e.g., with the SWMF code, in the solar corona (SC) and inner heliosphere (IH) regimes to derive the evolution of the interplanetary magnetic field associated with the CME transition [24]. A next step is to include the CME disturbed solar–terrestrial magnetic field in order to simulate the resulting Sun shadow. The interlink between CME simulations and Sun shadow simulations can shed light on the capability of the Sun shadow as an approach for space weather forecasting.

### C. Loss–cone anisotropy and advance warning of shock arrival for space weather forecasting

The loss cone anisotropy is another type of measurement of Galactic cosmic rays that is directly relevant to space weather forecasting because it could provide advance warning of the arrival of an interplanetary shock, and could also indicate an expected time of arrival. This would be useful because a shock arrival often coincides with a sudden storm commencement as determined by ground–based geomagnetic observatories, i.e., it marks the start of a geomagnetic storm and the associated space weather effects.

The loss cone anisotropy is a decrease in GeV–range cosmic ray density in only a narrow range of directions, found within 1–2 days before the arrival of an interplanetary shock at Earth. Note that after the shock arrives, there is a decrease in the cosmic ray flux from all directions, known as a Forbush decrease [25], because the high plasma speed and strong magnetic fields inhibit access of cosmic rays to the region downstream of the shock. Now the Forbush decrease itself is not directly useful for space weather forecasting because it arrives after the shock, i.e., after the geomagnetic storm commencement. However, the "loss cone" is a range of angles close to the interplanetary magnetic field direction toward the shock, and particles from these directions came from downstream of the shock where the particle flux is lower. Hence a loss cone anisotropy is an indicator of the approach of an interplanetary shock.

A loss cone anisotropy was first reported in data from neutron monitors, in 1992 [26]. Later there were numerous other reports of loss cone decreases in neutron monitor data, as well as an enhancement of cosmic ray flux in a ring of directions surrounding the loss cone, which was attributed to reflection from the shock. The first theoretical description of the anisotropy and its spatial distribution was provided by [27]. Further computer simulations [28] provided a basis for comparison, so that an observed loss cone angle can be used to infer the shock–field angle. That work has been used to parameterize more recent determinations of loss cone shock precursors by the Global Muon Detector Network (GMDN) with fine directional resolution [29, 30].

LHAASO's reconstructed showers will have excellent directional precision and a huge count rate, over a cosmic ray energy range similar to that of GMDN, so LHAASO will provide improved measurements of the loss cone anisotropy, including a possible discovery of fine directional structure beyond the axisymmetric fits performed with presently available data. This could be used in real time to provide advance warning of impending shock arrivals and geomagnetic storm onsets. According to [28], in this energy range the loss cone feature can provide warning up to 12 hours in advance. With a single detector facility, loss cone precursors can be seen when the interplanetary magnetic field direction rotates into view, which will often but not always occur during that 12–hour window. Therefore, a more comprehensive warning system could be obtained by teaming up with GMDN, other air shower arrays, or neutron monitors worldwide to continuously monitor loss cone features along the interplanetary magnetic field direction.

### D. Forbush decreases due to solar storms

There are also transient cosmic ray flux and anisotropy variations due to major solar storms. The main effect is the so–called Forbush decrease [25]. The first stage of the decrease occurs at a shock driven by a CME. There may be a second stage associated with the arrival of the CME ejecta, with a further decrease that lasts while the ejecta pass the observer [31]. After that, the flux returns to normal over the next few days. It is common to observe interesting anisotropy patterns during a Forbush decrease, often indicating interesting directional distributions of particles following the magnetic structures of the CME. The mechanism for the Forbush decrease is not clear, and the energy distribution of the decrease could provide important clues.

Air shower arrays can play a role by determining the Forbush decrease at high energies, where the energy and time dependence have not been systematically studied. For example, the ARGO–YBJ Collaboration reported the detection of a Forbush decrease on 2005 Jan 18 [32]. LHAASO could provide a great increase in statistics, though it will be necessary to understand and correct for environmental effects on the count rate as a function of time. Together with time–dependent cosmic–ray transport models with a three–dimensional diffusion barrier [33], we can simulate the Forbush–decrease events in different rigidity ranges and for different particle species to understand their possible mechanisms.

### E. Modulation of the cosmic ray flux with the solar cycle

The longest–period cosmic ray variations that are directly measured are related to the 11–year sunspot cycle and the 22–year solar magnetic cycle. The number of sunspots typically varies over 11 years. There were sunspot maxima in 1989, 2000, and 2014, and sunspot minima in 1996, 2008, and 2019. Because magnetic fields and solar storms are concentrated near sunspots, numerous solar phenomena vary with the solar cycle. They do not precisely depend on the sunspot number, so we tend to speak of "solar maximum" as a period of several years around solar maximum, and "solar minimum" as a period of several years with very few sunspots. Because of the higher solar wind speed (on average) and stronger magnetic fields during solar maximum, the transport of cosmic rays into the inner heliosphere is inhibited. Thus the flux of cosmic rays is observed to have an inverse association with the sunspot number, with the most cosmic rays during solar minimum, and the fewest during solar maximum [34]. The amplitude of variation is $\sim 30\%$ at an energy of 1 GeV. This roughly 11–year variation is called the solar modulation of cosmic rays (Figure 2).

The Sun's magnetic field is much more complex than the Earth's, and magnetic fields are highly concentrated at the sunspots, typically directed outward at one and inward at another. Nevertheless, there is an overall preponderance of one polarity on one hemisphere and the opposite polarity on the other. Every 11 years or so, at solar maximum, there is a magnetic reversal in which the preponderance reverses sign. Therefore, a complete magnetic cycle requires two sunspot cycles, i.e., about 22 years. Charged particle orbits undergo drift motions that depend on the charge sign and the sign of the magnetic field. The drifts therefore reverse every 11 years and repeat every 22 years. The same holds for the effect of magnetic helicity on the particle scattering. Therefore, there is also a roughly 22–year cosmic ray flux variation corresponding to the solar magnetic cycle [35]. In other words, 11–year periods with opposite magnetic polarity exhibit distinct cosmic ray variations. These effects are associated with a variety of interesting phenomena, such as guiding center drifts, cosmic ray gradients with helio–latitude, particle charge sign dependence, and changing diffusion coefficients [36–38]. These phenomena depend on the sign of $qA$, where $q$ is the particle charge and $A$ is the solar magnetic polarity.

With stable, long–term operation, LHAASO will provide important information to further explore solar modulation as a function of energy and time, if its threshold can be reduced to $\lesssim 50$ GeV. There may even be sufficient compositional information to discern the individual modulation of protons and alpha particles as a function of energy throughout the solar cycle. This is information that is not available from traditional ground–based observatories of GeV–range cosmic rays, such as neutron monitors and muon detectors.

### F. 27–day variations

Roughly speaking, a faster solar wind speed can inhibit the entry of cosmic rays to the inner heliosphere, and is typically associated with a reduced cosmic ray flux. Thus co–rotational variations in solar wind speed (which rotate with the Sun) are associated with well–known "synodic" or "27 day variations" in the cosmic ray flux [39], which have sometimes been called "recurrent Forbush decreases." It frequently happens that a region of the solar corona that produces fast solar wind, e.g., a coronal hole, lies eastward of a region that produces slow solar wind. Then as the Sun rotates, the source region of fast solar wind moves underneath the region where slow solar wind came out previously, and the fast solar wind will collide with the slow solar wind that lies in front of it. Such a collision region is called a co–rotating interaction region (CIR), The CIR also has a spiral shape, and it represents a region where solar wind is suddenly compressed by the collision. An observer near Earth sees the solar wind speed suddenly increase when the faster solar wind arrives.

It is common to see the cosmic ray flux suddenly decrease at the time of the CIR, either as part of the inverse relationship with solar wind speed or because the compressed plasma and increased magnetic field serve as a barrier to hinder access to cosmic rays (see Figure 4, and note the reversed time axis). These jumps are a key component of the 27–day variations in cosmic ray flux with solar rotation. Note that CIRs also cause geomagnetic storms and space weather effects, so there is some practical interest in what cosmic rays can tell us about the physical properties of CIRs. LHAASO data could provide further insight, especially with regard to the energy dependence, which may allow us to clarify and quantify the association with solar wind parameters. It will be challenging for LHAASO to reduce its threshold to sufficiently low energy to observe cosmic ray flux variations, but it is perhaps more likely for 27–day effects to be observed in the solar diurnal anisotropy (see below). And promisingly, future coordinated observations by solar space missions, e.g., NASA's Parker Solar Probe, ESA's Solar Orbiter and the Chinese Advanced Space–based Solar Observatory will provide better solar wind parameters and better Carrington maps of photospheric magnetic field to calculate interplanetary magnetic field.

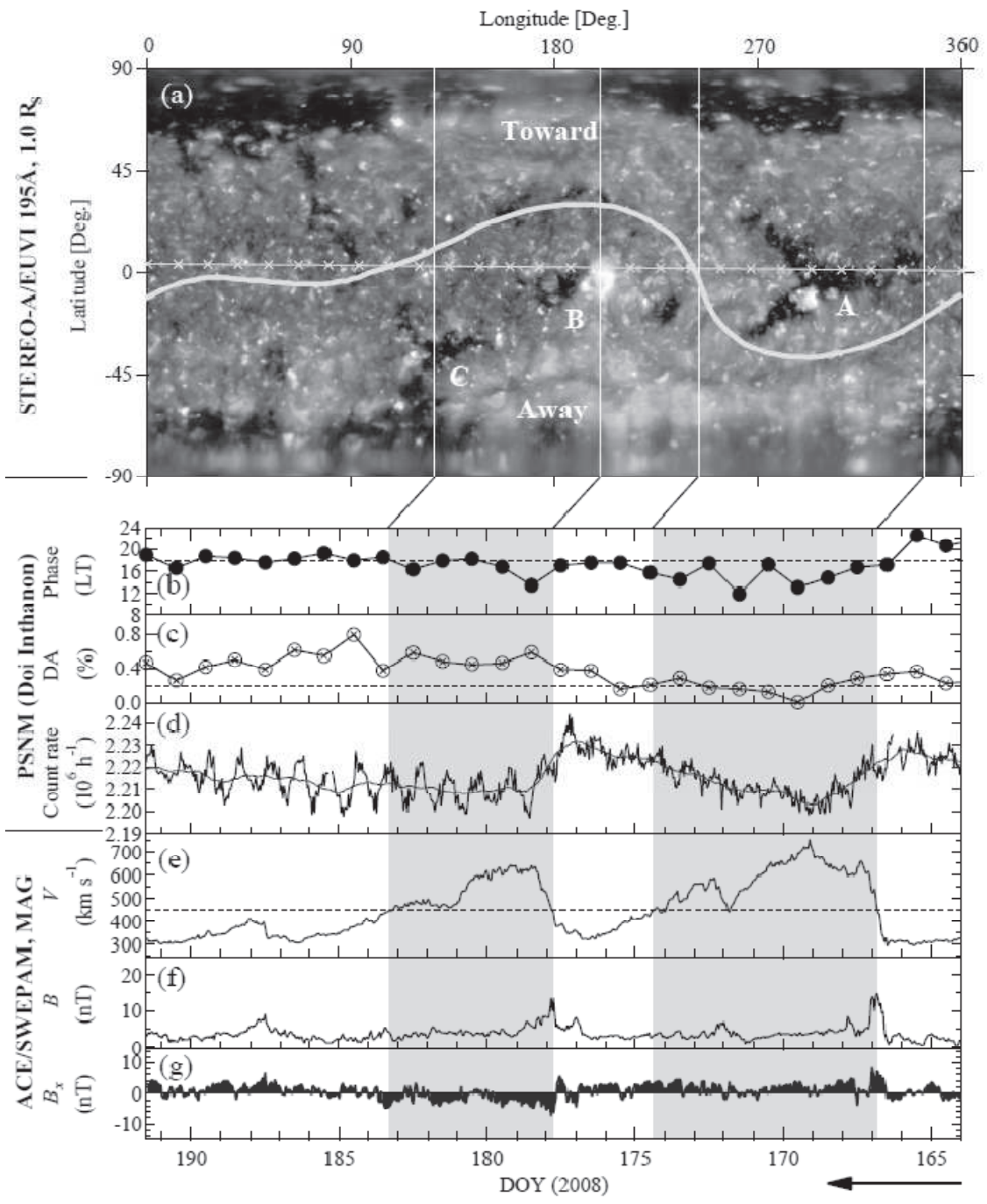

**Fig. 4.** (color online) Reversed time plots in day of year (DOY) for Carrington (solar) rotation 2071 (between 2008 June 12 and 2008 July 10). Top panel: Synoptic map of the solar corona as observed by the EUVI–A imager in the Fe XII 195–Å bandpass. Upper three graphs: Data of the diurnal anisotropy and flux of Galactic cosmic rays (GCRs) as measured by the Princess Sirindhorn Neutron Monitor at Doi Inthanon, Thailand. Lower graphs: Hourly interplanetary plasma parameters from the ACE and OMNIWeb databases in GSE coordinates. When the high speed solar wind streams pass the Earth they reduce the cosmic ray flux. After the rapid solar wind speed increase of DOY 177, there was a strong, long–lasting enhancement in the diurnal anisotropy of GCRs. This is attributed to an extra $\mathbf{B}\times\nabla n$ anisotropy with a latitudinal gradient in association with the coronal hole (dark region) morphology [40].

### G. Sidereal anisotropy

The sidereal anisotropy (also called the sidereal diurnal anisotropy) refers to the difference in cosmic ray flux from different directions in space, ideally averaged over the Earth's yearly orbit of the Sun. For scaler rates from a ground–based detector rotating with Earth, the sidereal anisotropy is related to the data organized as a function of sidereal time, as opposed to solar time. The sidereal anisotropy of TeV–range cosmic rays has been a very exciting topic of study, since the initial discoveries of the "loss–cone" deficit from a direction close to the Galactic center and a "tail–in" enhancement from the direction of an assumed heliotail [41]. Further measurements have produced sky maps of the large–scale anisotropy, e.g., [42, 43] (see Figure 5). With better statistics and detector sensitivity, more and more structures have been found at medium and small scales [44, 45]. The possible explanations include large–scale flows in the galaxy, nearby sources of cosmic rays, and/or the fingerprint of interstellar turbulence [46]. For more details, see the section on Cosmic Ray Measurement and Physics.

To some extent, the cosmic ray anisotropy pattern must be affected and distorted by heliospheric magnetic fields [47], so solar and heliospheric phenomena are relevant. Because these magnetic fields vary strongly with the (roughly) 11–year solar cycle, various air shower experiments are looking for such a time dependence in the anisotropy pattern. Other possible types of time variation are a difference between patterns for opposite polarities of the interplanetary magnetic field, and a dependence on the location of the Sun in the sky. Results published to date are consistent with a time–independent anisotropy, but when LHAASO takes data with improved statistics over a substantial portion of a solar cycle, the imprint of heliospheric magnetic fields should be found.

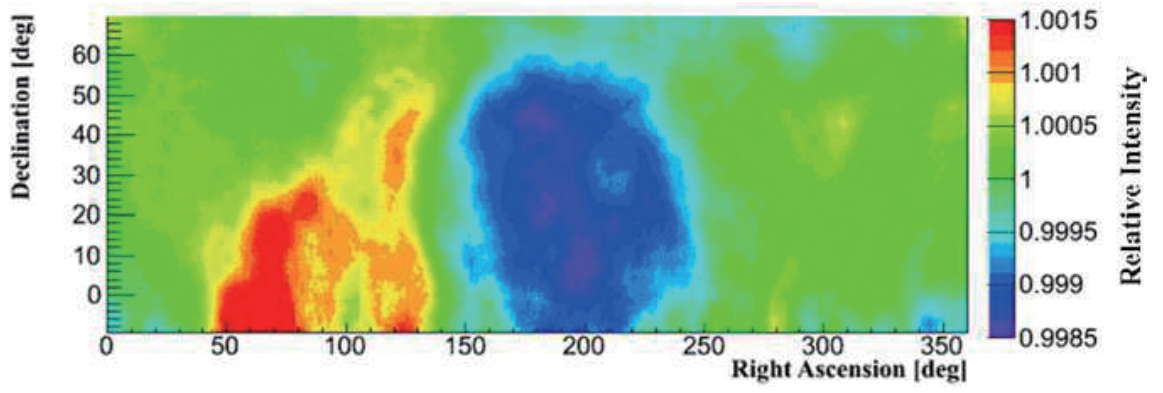

**Fig. 5.** (color online) Relative intensity map of TeV cosmic rays as measured by ARGO–YBJ, showing the sidereal anisotropy [43].

The first impact of such a ground–breaking measurement would be to help determine the heliospheric magnetic field, and indeed the shape of the heliosphere itself. The large–scale structure of the heliosphere, and the shape of its boundary, the heliopause, are still hotly debated. The traditional view is that the interstellar medium (ISM), which moves relative to the heliosphere, pushes past the heliosphere to create a bullet–shaped nose to the heliopause on its upstream side and an extended tail on its downstream side. Others contend that there is no heliotail and that the solar wind instead flows as jets along the poles of solar rotation, with the jets bent downstream by the ISM [48].

The sidereal anisotropy in GeV–range cosmic rays is also of substantial scientific interest. The anisotropy decreases in amplitude with decreasing energy [43], presumably due to solar modulation. However, in data from the Matsushiro underground muon detector at ~0.6 TeV, there was at most a minor solar cycle dependence [49]. This is surprising, because solar modulation had apparently reduced the amplitude by a factor of 3 at that energy. With greater statistics and improved resolution of time variations, LHAASO data may help shed light on this mystery.

### H. Diurnal anisotropy

The diurnal anisotropy (also called the solar diurnal anisotropy) refers to the difference in cosmic ray flux from different directions in space relative to the Sun, e.g., as expressed in geocentric solar ecliptic (GSE) coordinates. This is an anisotropy related to solar phenomena, or the Earth's orbit around the Sun. For scaler rates from a ground–based detector rotating with Earth, the diurnal anisotropy is related to the data organized as a function of local solar time. For GeV–range cosmic rays, it is also necessary to account for significant deflection of the cosmic ray direction by Earth's magnetic field.

The basic physical explanation of the diurnal anisotropy is very different for TeV–range and GeV–range cosmic rays. In the TeV range, the cosmic ray distribution is almost isotropic in an inertial frame, so the diurnal anisotropy is dominated by the Compton–Getting effect from Earth's orbital motion. The greatest flux arrives at ~0600 local time. In contrast, cosmic rays of energy up to ~100 GeV are affected by the Sun and the interplanetary magnetic field (IMF), which introduces an energy–dependent anisotropy [50]. The average diurnal anisotropy (DA) vector has been explained as a consequence of the equilibrium established between the radial convection of the cosmic ray particles by solar wind and the inward diffusion of GCR particles along the IMF. In a reference frame co–rotating with the Sun, convection and parallel diffusion (i.e., diffusion parallel to the large–scale magnetic field) can nearly cancel and the GCR distribution has almost no net flow. Then in Earth's reference frame, there is a net flow as the co–rotating GCR distribution impinges on Earth from the dusk sector, i.e., ~1800 local time. Transient variations are superimposed on the steady state co–rotational anisotropy, and they sometimes form "trains" of enhanced diurnal variation that persist for several consecutive days (see Figure 4). Thus the long–term variation in GeV–range diurnal anisotropy provides information on solar modulation and cosmic ray gradients [1, 38], while the changes on shorter time scales tell us about the changing structure of the heliosphere [39].

LHAASO will examine the diurnal anisotropy of cosmic rays over a wide range of energies, using both scaler data and shower data. Consistency between those two data sets, and also with previous reports, will provide a demanding test that the flux data are properly corrected for environmental effects. Even in the TeV range, previous experiments apparently disagree about whether there is a strong deviation from the expected Compton–Getting effect [43, 51]. Then there is an interesting transition in the TeV range from Compton–Getting to mostly co–rotational anisotropy. Finally, in the GeV range, LHAASO results can be compared with results from neutron monitors (~10–35 GeV median energy) and muon detectors (~60–110 GeV median energy for surface detectors, and higher for underground detectors), and we expect to find interesting structure in the diurnal anisotropy as a function of energy and time.

### I. Short–time variations

Clearly there are sharp decreases in cosmic ray flux associated with discrete structures that accompany an interplanetary coronal mass ejection (ICME) and accompanying shock, as discussed in the section on Forbush decreases. Here we consider the slightly different issue of variations in cosmic ray flux, over times shorter than one day, due to fluctuations in the IMF, the solar wind, or possibly the magnetosphere. It has become clear from observational and theoretical work that apparently homogeneous regions of the solar wind are really permeated by flux–tube like structures that can guide the motion of energetic particles, leading to non–uniform spatial distributions [52–58], including clear observational confirmation at MeV energies or lower. For decades, there has been a notion in the cosmic ray community that there should be local fluctuations in the GeV–range cosmic ray rate in concert with small–scale turbulent fluctuations or coherent structures in the IMF [59]. However, the correlations obtained are somewhat weak, and instrumental and environmental fluctuations could be important. Furthermore, for a ground–based detector with no directional informa-tion for individual cosmic rays, and a directional acceptance that rotates with Earth, it is often unclear whether short–term variations in the cosmic ray flux are due to temporal changes in the IMF or due to the cosmic ray directional distribution, i.e., structure in the diurnal anisotropy.

LHAASO data from shower reconstruction at tens of GeV could be a game–changer for this type of study, providing an ability to distinguish between temporal effects and changes in the directional distribution over its wide field of view. There is some reason to expect temporal changes in the Galactic cosmic ray flux in association with interplanetary structures, based on successful observations at MeV energies [60, 61]. Furthermore, neutron monitors in Antarctica had a rare opportunity to observe minute–scale fluctuations in GeV–range solar particles during the giant solar event of 2005 Jan 20 [62] (the Galactic cosmic ray flux does not provide sufficient statistics to study minute–scale fluctuations in such detectors). That study found huge variations in flux with periods of 2 to 4 minutes, which they attributed to fluctuations in the beaming direction of the particle distribution. Thus LHAASO data, with excellent statistics and direction information, will provide a means to seriously search for short–term variations in Galactic cosmic rays at tens of GeV and above and to identify their nature and origin. Environmental stability of the LHAASO detectors will be crucial for this work.

### J. Moon shadow and geomagnetic field variations

The moon shadow in TeV cosmic rays is a very important tool for calibrating the resolution and absolute energy scale of an air shower array [7], because the Moon has a known size and the observed shadow has an energy–dependent deflection due to the known geomagnetic field; see also the section on Cosmic Ray Measurement and Physics. Usually time variations in the moon shadow are not expected, and indeed the constancy of the moon shadow is an important test of an air shower detector's stability. However, there has been a suggestion of a possible so–called day/night effect, because the solar wind continually impinges upon Earth's magnetosphere and compresses the dayside magnetosphere, while the nightside magnetosphere is elongated into the magnetotail. Thus the moon shadow deflection due to the geomagnetic field could be different during different phases of the Moon's orbit, depending on whether it is on Earth's dayside or nightside. There have been previous reports of no day/night effect [7, 63], and also a claim of such an effect [64]. LHAASO should be able to check for this possible effect with better statistics and over a wider energy range. If successfully detected, this could provide a valuable magnetospheric database of measurements of the integrated geomagnetic field along the line of sight to the Moon as it orbits the Earth.

## II. INVESTIGATING A POSSIBLE LINK BETWEEN COSMIC RAY FLUX AND EARTH'S CLIMATE

Earth's climate change, including global warming, is one of the most important scientific issues of our time. It has been suggested that solar activity has historically played an important role in governing Earth's temperature [65], and one possible mechanism for such a connection involves cosmic rays. In this scenario, increased solar activity leads to decreased cosmic ray flux (see Figure 2), cosmic ray showers are the main cause of atmospheric ionization a few kilometers above ground level, and decreased atmospheric ionization leads to decreased cloud formation, stronger sunlight at Earth's surface, and an increased surface temperature.

The controversy is whether cosmic ray variations really lead to significant changes in cloud cover. Some researchers have claimed a correlation between temporary Forbush decreases and cloud cover [66, 67], while others claim there is no significant effect of cosmic ray variations on cloud cover or on Earth's temperature [68–70]. It should be possible to improve upon the methodology used by [66]. For example, they model the effect of a Forbush decrease on the GCR spectrum using a function that gives a nonsensical decrease of > 100% at a rigidity of 1 GV, and they treat each detection rate as a differential flux at the median rigidity rather than an integral flux. With LHAASO data, we can estimate the GCR spectrum with greater accuracy, and we can also use Monte Carlo simulations based on the inferred spectrum to estimate atmospheric ionization and its dependence on geomagnetic cutoff rigidity (or roughly speaking, on geomagnetic latitude), altitude, and time. We should be able to address the issue of a possible effect of cosmic ray variations on cloud cover variations with much greater accuracy.

In the big picture, the world's experts on climate change, through the Intergovernmental Panel on Climate Change, have reached a consensus that solar and volcanic variations can account for Earth's surface temperature changes before 1951 and that anthropogenic effects have dominated thereafter.[1)] We do not intend to challenge that expert consensus. We can address the specific question of whether (and how) cloud cover changes are associated with cosmic ray variations, but we would not interpret a positive association as indicating the dominance of solar effects over anthropogenic effects.

1) https://www.ipcc.ch/report/ar6/wg1/downloads/report/IPCC_AR6_WGI_Chapter_03.pdf

## III. DETECTION OF MEV–RANGE $\gamma$–RAYS FROM THUNDERSHOWERS

The scalar rates at LHAASO may be able to detect MeV–range $\gamma$–rays from thunderstorms, which have previously been detected by ground–based $\gamma$–ray detectors [71] and the solar neutron telescope and neutron monitor at Yangbajing, China [72]. The latter reference contends that the signals in neutron detectors were due to $\gamma$–rays. For this purpose, it will be useful to have electric field measurements at the LHAASO site, to corroborate an association with lightning activity. Measurements of the time profile of $\gamma$–ray emission, as indicated by increased scaler rates in LHAASO's electromagnetic detectors, in conjunction with the electric field data, may help clarify the physical mechanism causing this mysterious emission from thunderstorms.

## IV. GEOPHYSICAL RESEARCH WITH ENVIRONMENTAL NEUTRON FLUX

Environmental neutrons are produced by two natural sources: by cosmic rays in air and in upper layers of soil by natural radioactivity (mostly due to $(\alpha,n)$–reactions on light nuclei) throughout the Earth's crust. Being produced as fast the neutrons are moderated by media down to thermal energy and persist until nuclear capture. The neutron lifetime depends on media chemical composition, temperature and water (or any hydrogenous material). A natural radioactivity chain daughter product, the inert gas radon–222 with 3.8–d half–life can migrate in air and in soil (rock, concrete, etc.) to a long distance and even accumulate in some places, thus changing the neutron flux in underground locations. It is also sensitive to local seismic activity. Therefore, the flux (or concentration) of thermal neutrons in the media is sensitive to the media parameters such as its temperature, humidity, porosity (seismic activity), etc. Measurement of neutron flux time variations for a long time could thus be used to control the above media parameters.

We plan to use the EN–detectors of the ENDA array for continuous environmental thermal neutron flux monitoring and its variation study, needed not only for EAS experiment background estimation but also for some geophysical applications. We already have some results [73] of this study and it has a promising future. The following geophysical phenomena will be investigated through thermal neutron study:

- Neutrons during thunderstorms (surface)
- Lunar tidal effects in Earth's crust (surface and underground)
- Seasonal radon–neutron waves at high altitude (surface)
- Free Earth oscillations (underground)
- Forbush effect and Sun–Earth interconnections (surface and underground)
- Ground Level Enhancement (GLE) effect (surface)

The additional geophysical studies on the Earth's surface could be performed cost–free using ENDA detectors. The items needed in an underground detector location could use existing underground or basement rooms to decrease the cosmic ray source and to emphasize the radon–neutron source. Otherwise, additional investment will be required.

1. Development and construction of a prototype array (PRISMA–YBJ) which consisted of 4 en–detectors in the ARGO hall at high altitude in Tibet, which operated continuously from August 30, 2013 to March, 2017. Some results are already published and some are in preparation. Through long–term stable observations of environmental thermal neutron flux variations, periodic changes in the thermal neutron flux caused by the gravitational tidal effect of the Sun–Moon–Earth system have been observed [74], which fully proves that this detector can be used to monitor weak changes in the Earth's crust. Near the strong (M = 7.8) earthquake on April 25, 2015 in the Gorkha (Nepal) region, the thermal neutron flux monitored by PRISMA–YBJ exhibited a number of anomalies. A series of strong aftershocks followed the main earthquake, with magnitudes > 6 and as high as 7.3. Close to the main earthquake and strong aftershocks, PRISMA–YBJ twice detected significantly changed thermal neutron flux diurnal wave shape [75].

2. Coincidence run of PRISMA–YBJ and ARGO in 2013. The results are partially published.

3. Autonomous running accumulated up to date 2 years of data taking. Results on thermal neutrons lateral and temporal distributions in EAS were published at 33rd ICRC, 34th ICRC and TAUP2015 conferences.

4. Monte–Carlo simulations based on CORSIKA and GEANT were performed to simulate the PRISMA–YBJ experiment configuration. Now we have very good agreement between the simulations and experiment and we need not make any normalization. The program code is ready now to simulate the ENDA–LHAASO configuration.

5. Search for new cheap scintillator for thermal neutron detection has been done. As a result we found scintillator producer in Russia and have developed together a novel technology for scintillator compound based on ZnS(Ag) with natural boron addition as a target for neutron capture. Resulting thermal neutron recording efficiency of the compound is close to 20% at the compound thickness of 50 mg/cm$^2$. The price for the compound is now a factor of 5 lower than that a $^6$Li enriched compound.

6. Data acquisition system has been developed and it has been tested at an expanded up to 16 en–detector prototype in Yangbajing

## V. EFFECTS OF THUNDERSTORM ELECTRIC FIELD ON COSMIC RAYS

Studying the effects of thunderstorm electric fields on cosmic rays is very useful in understanding the acceleration mechanism of secondary charged particles caused by an atmospheric electric field. In this work, Monte Carlo simulations are performed with CORSIKA to study the intensity and energy changes of positrons and electrons due to such electric fields at the observatory of LHAASO. The variations of the secondary cosmic ray intensity are found to be highly dependent on the strength and polarity of the electric field. The energy distributions of positrons and electrons have also changed. In the low energy region, the total number of positrons and electrons increases significantly, while at high energies, it does not change obviously. Key components of LHAASO, the electromagnetic particle detectors (ED) in the kilometer–square array (KM2A) and the water Cherenkov detector array (WCDA), are sensitive to the secondary positrons and electrons in extensive air showers. Thus our simulation results could also be helpful in understanding the experimental data of LHAASO during thunderstorms, and may provide important information for studying the acceleration mechanism of secondary charged particles caused by electric fields and the possible physical mechanism of lightning triggered by high energy cosmic rays.

### A. Introduction

During thunderstorms, the maximum strength of electric fields has been found up to 2000V/cm [76, 77]. In such strong fields, by accelerating or decelerating the charged particles in extensive air showers, the intensity and energy of secondary cosmic rays will be considerably affected. In 1924, Wilson [78] first pointed out that the electron with a very small mass can be accelerated to a very high energy by the thunderstorm electric field, and additionally proposed the concept of "runaway" electrons. Gurevich *et al.* [79] put forward a new breakdown mechanism in 1992, and suggested that the secondary electrons can be accelerated in such fields. When the particles acquire energies greater than the energies lost in ionization and bremsstrahlung, they will ionize the atmospheric molecules to produce new free electrons. These new free electrons are accelerated again by the electric field, resulting in an avalanche process in which the number of electrons increases exponentially. Marshall, Dwyer and Symbalisty *et al.* [80–82] built upon this theory, now commonly called a relativistic runaway electron avalanche (RREA).

For years, scientists have carried out lots of ground–based experiments to detect the thunderstorm ground enhancements (TGEs) [83] and masses of satellite–borne experiments to investigate the terrestrial gamma flashes (TGFs) [84, 85], trying to find the high–energy electrons accelerated by the thunderstorm electric fields or the high–energy gamma rays radiated by bremsstrahlung. This indicates that the RREA process is believed to be a reasonable explanation of these phenomena.

The correlations between the intensity of the ground cosmic rays and the thunderstorms electric field were measured at YBJ [86, 87]. They found that the particle count rates did not always increase in the field, and in some cases would decline. The intensity decreases cannot be explained by the RREA mechanism. What is the acceleration mechanism for these abnormal results? In order to learn more about the acceleration mechanism and the intensity change, more theoretical, experimental and careful simulation results are needed.

The LHAASO station is located at a high altitude in southwest China. The thunderstorms are frequent in the summer season and beneficial to studying the effects of thunderstorm electric fields on the secondary cosmic rays. In this work, Monte Carlo simulations are performed with CORSIKA to study the effects of thunderstorm electric fields on the secondary positrons and electrons at LHAASO.

### B. Simulation Setup

CORSIKA is a detailed Monte Carlo program to study the evolution and properties of extensive air showers in the atmosphere [88]. The code of CORSIKA7.5700 is used in this work. The hadronic interaction models used are QGSJETII–04 for high energy particles and GHEISHA in the low energy range. In simulations, we apply a vertical uniform electric field from 6400 m to 4400 m above sea level. Here we define the positive electric field as one that accelerates positrons downward in the direction of the earth. In view of the acceleration of the field, the energy cutoff is set to 0.1 MeV, below which value positrons and electrons are discarded from the simulation.

### C. Simulation results and discussion

Since positrons and electrons predominate in the secondary charged particles of cosmic rays, the effects of the electric field on electrons and positrons are properly taken into account in this work.

### 1. *The intensity variations of positrons and electrons in fields*

The electric fields are chosen as a series of values in the range of –1000 to 1000 V/cm. Fig. 6 shows the percent change of the average numbers of positrons and electrons and the sum of both in different electric fields at LHAASO.

As shown in Fig. 6, in a negative electric field (accelerating electrons), the number of electrons increases, while the number of positrons decreases, and the total number increases with the increasing strength of electric field. When the electric field is positive (accelerating positrons), the number of electrons decreases, while the number of positrons increases. In the range 0 to 600 V/cm, the total number declines and the maximum decrease is about 2.5 percent. In the positive field greater than 600 V/cm, the sum of positrons and electrons increases with the increasing strength of the field.

Why does an electric field have a stronger effect on electrons? And what causes the decline in the positive field? In order to answer these questions, the ratio of electrons to positrons and the energies of positrons and electrons are analyzed. As show in Fig. 7, there are more electrons at lower energies than positrons, and the electric field accelerates the particles with lower energy more obviously. From Fig. 8, we can see the number of electrons is greater than the number of positrons (which is mainly due to the Compton effect). So the accelerating effects on electrons are more obvious than the decelerating effects on positrons in negative fields, resulting in a total number of positrons and electrons that goes up with increasing electric field strength. For the same reason, the deceleration of electrons is more important than the acceleration of positrons in a positive field with small strength, and the increase of positrons cannot compensate for the decrease of electrons. As a result, the total number will decrease in the field 0 to 600 V/cm. However, if the positive field strength becomes larger, the accelerating effects on positrons are dominant. The increase of positrons could compensate for the decrease of electrons and the total number starts to increase. This mechanism, explained in detail in [89], produces the asymmetric behavior shown in Fig. 6.

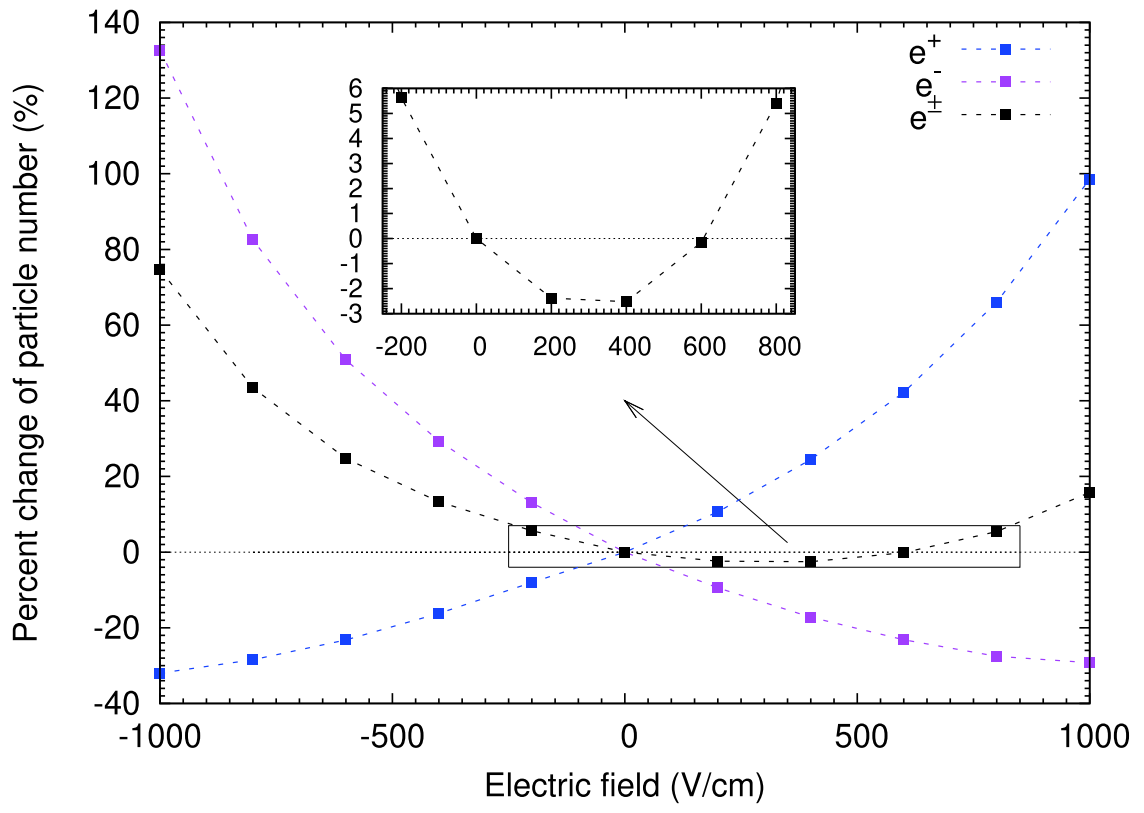

**Fig. 6.** (color online) Percent change of particle number as a function of electric field at LHAASO.

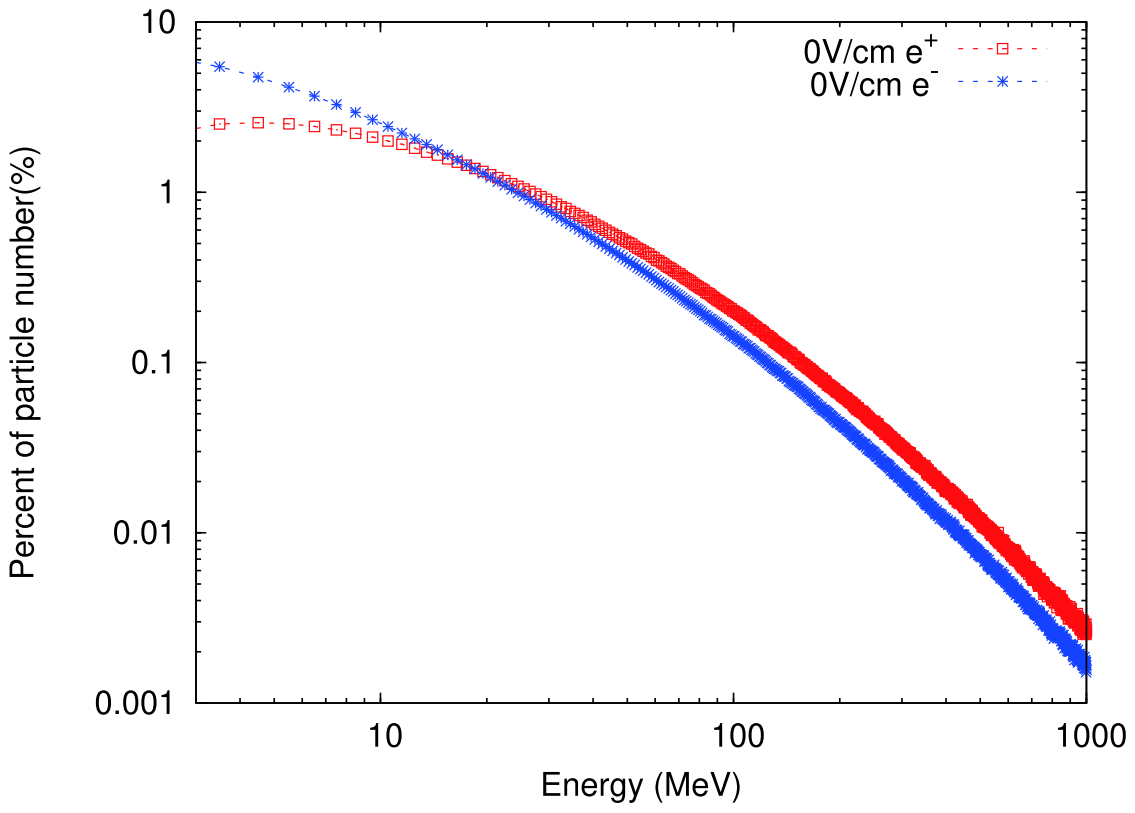

**Fig. 7.** (color online) Percent distributions of positrons and electrons as a function of energy at LHAASO.

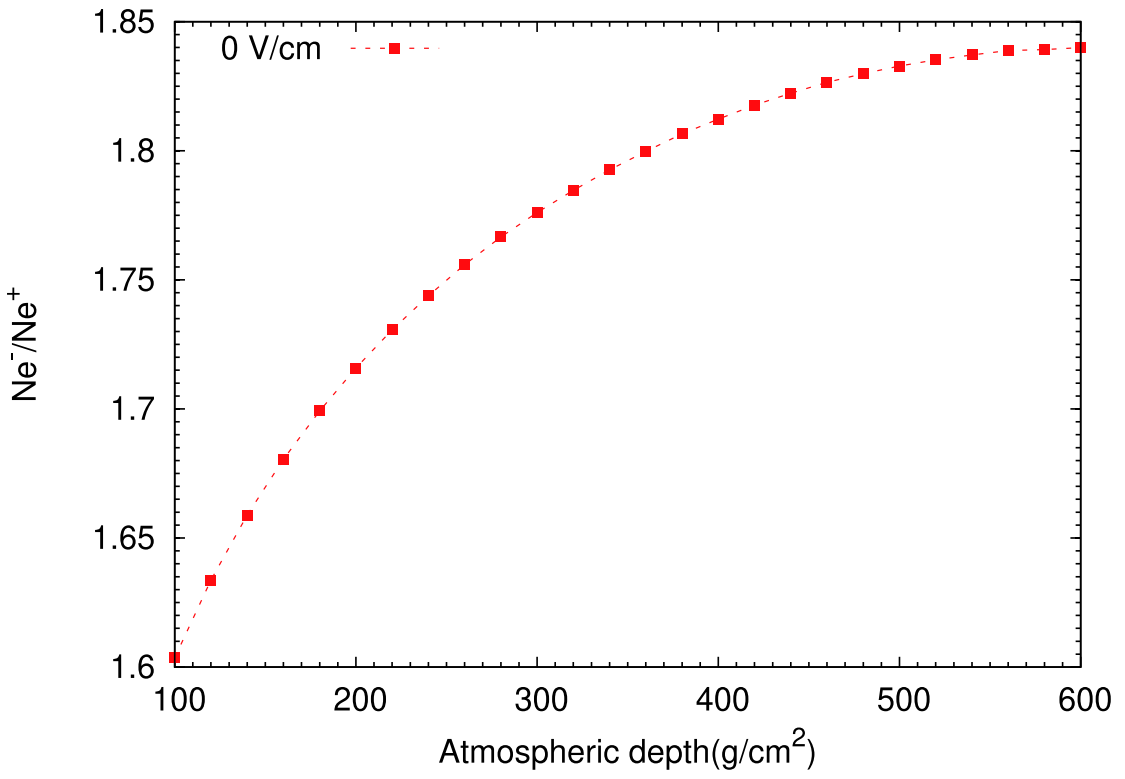

**Fig. 8.** (color online) Ratio of electrons to positrons as a function of the atmospheric depth.

### 2. *The field effects on the energy of positrons and electrons*

From the evaluations of Dwyer and Symbalisty et al., the threshold field strength for the development of the RREA process is about 1650 V/cm at LHAASO (4400 m a.s.l.). In order to get clues to the mechanism of electron acceleration in the thunderclouds, two typical strengths of thunderstorm electric field are selected, namely the field strength 1000 V $cm^{-1}$ (smaller than the threshold strength for RREA) and the field strength 1700 V $cm^{-1}$ (greater than the threshold strength for RREA).

Fig. 9 gives the number distributions of positrons in positive fields (the results in 0 V/cm are plotted just for comparison). The increase appears in positive electric fields, especially in the low energy band. With energy lower than 60 MeV, the number of positrons increases obviously. And the enhancement in +1700 V/cm (about 153%) is greater than that in +1000 V/cm (about 17%). As the energy increases, the influence of the electric field on particles gradually decreases. That is, the energy spectrum will become soft.

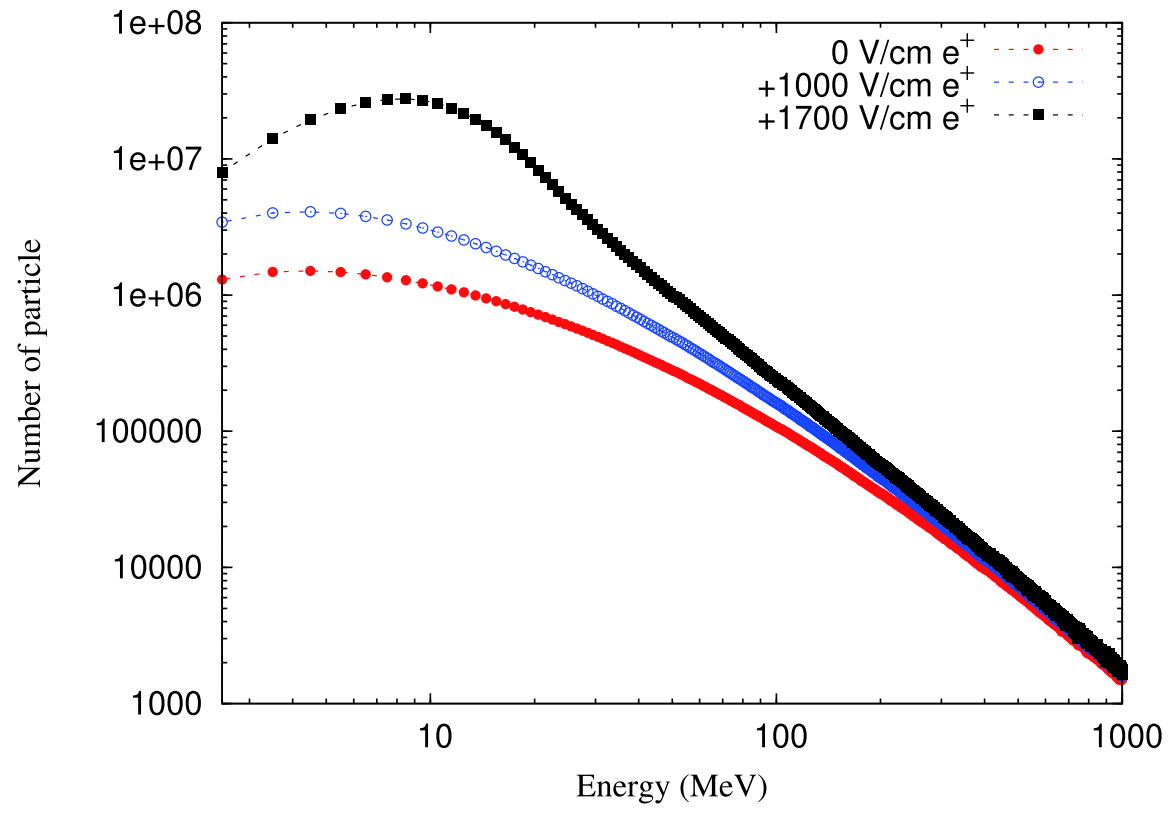

**Fig. 9.** (color online) The number of positrons as a function of energy in positive fields.

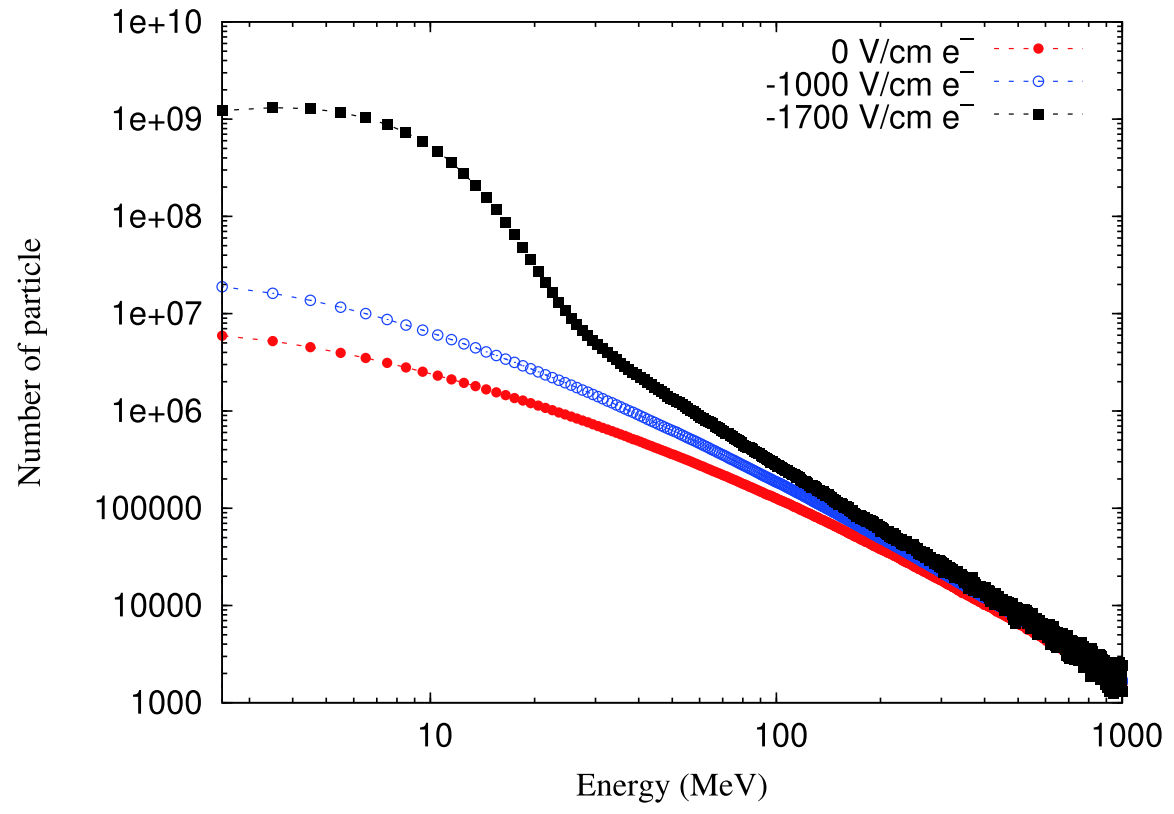

**Fig. 10.** (color online) The number of electrons and as a function of energy in negative fields.

Fig. 10 describes the variations of electron number with different energies in negative fields (the results in 0 V/cm are plotted just for comparison). When the electric field strength is 1700V/cm, the electron number exhibits an exponential increase at the low energy region. With energy less than 60 MeV, the increase is up to 14,310%. This is consistet with the RREA theory. From this figure, we can see that when the energy is lower than a few tens of MeV, the increase of electrons in 1000 V/cm (about 42%) is far less than the increase in 1700 V/cm. Based on the results of the above two figures, we also can conclude that the negative electric field has a more obvious effect on particles than positive fields. The acceleration processes in two typical strengths of fields were discussed in detail in Ref. [90].

### D. Conclusion

Monte Carlo simulations are performed to study the electric field effects on the secondary positrons and electrons at LHAASO. The intensity and energy of the particles change significantly, and the variation amplitudes are closely related to the strength and polarity of the electric field. These results can explain some of the existing experimental phenomena involving the intensity increases and decreases. Now, LHAASO experimental facilities have been fully built and been operational and acquired data normally. The experimental data of KM2A and WCDA will be analyzed, and the correlations between the variations of cosmic rays and thunderstorm electric fields will be studied. With the simulation and experimental results, we can better understand the data changes in LHAASO experiments during thunderstorms. At the same time, these results may provide important information for studying the acceleration mechanism of secondary charged particles by electric fields and the possible physical mechanism of lightning triggered by high energy cosmic rays.